%% file: nuctag.tex
\documentclass[letterpaper,twoside,12pt,
cleardoubleempty,openright,pagesize,abstractoff,
bibtotoc]{scrreprt}
\addtocounter{secnumdepth}{1}
\usepackage{textcomp}
\usepackage[adobe-utopia,sfscaled=true]{mathdesign}
\usepackage{helvet}
\usepackage{courier}
\usepackage{amsmath}
\usepackage{floatflt}
\usepackage{tabularx,multirow}
\usepackage[square,numbers,sort&compress,nonamebreak]{natbib}
\usepackage{hypernat}
\usepackage[linkbordercolor={0 .5 0},
            citebordercolor={.8 0 0},
            urlbordercolor={0 0 .5},
            raiselinks=true,
            bookmarks=false,
            pdfborder={0 0 1 [2]}]{hyperref}
\usepackage{doc}
\usepackage{url}
\usepackage[tight]{subfigure}
\usepackage[printonlyused,smaller]{acronym}
\usepackage{fancyvrb}
\usepackage{arydshln}
\usepackage{feynmf}		%Package for feynman diagrams. 

\newif\iffinal
\finaltrue

\newif\ifabstract
\abstracttrue

\newif\ifindex
\newif\iftocs
\tocstrue

\newif\ifbib
\bibtrue

\newif\ifappendix
\appendixtrue

\input{srcs/commands}

\newif\ifpdf
\ifx\pdfoutput\undefined
   \pdffalse
\else
   \pdftrue
\fi

\ifpdf
   \usepackage[pdftex]{graphicx}
   
   \iffinal
   \else
      \usepackage{pdfdraftcopy}
      \draftstring{Draft \today}
   \fi
   
   \pdfoutput=1        % we are running pdflatex
   \pdfcatalog{/PageMode(/UseOutlines)} % or UseNone
\else %pdf=false
   \usepackage[dvips]{graphicx}
   \usepackage{rotating}

   \iffinal
   \else
    \usepackage[dvips,english,all,light]{draftcopy}
   \fi

   \hypersetup{
     pdfauthor   = {\myname},
     pdftitle    = {\thesistitle},
     pdfcreator  = {\myname},
     pdfproducer = {},
     pdfsubject  = {},
     pdfkeywords = {}
    }
\fi %ifpdf

\graphicspath{ {./img/} }
\ifindex
   \usepackage{makeidx}
   \makeindex
\fi

\typearea[.9in]{14}

\begin{document}
\pagestyle{empty}
\thispagestyle{empty}

\addtolength{\topmargin}{.23in}
\addtolength{\textheight}{.2in}
% TITLE
\maketitle

\cleardoublepage

% ABSTRACT
\ifabstract
   \begin{abstractpage}
   \input{srcs/abstract}

   \end{abstractpage}
\fi

\addtolength{\topmargin}{.25in}
\addtolength{\textheight}{-.2in}

\begin{dedication}
This work is dedicated to the memory of\\
Ann-Marie, Donna and Theodore Elias.\\
\end{dedication}

% TOC
\iftocs
   %\addtocontents{toc}{\relax \protect \pagestyle{empty}}
   %\addtocontents{toc}{\relax \protect \thispagestyle{empty}}
   %\addtocontents{toc}{\relax \protect \enlargethispage{0.5cm}}
   %\addtocontents{lof}{\relax \protect \pagestyle{empty}}
   %\addtocontents{lof}{\relax \protect \thispagestyle{empty}}
   %\addtocontents{lot}{\relax \protect \pagestyle{empty}}
   %\addtocontents{lot}{\relax \protect \thispagestyle{empty}}
   \tableofcontents
   \thispagestyle{empty}
   %\listoffigures
   %\listoftables
\fi

% ****************************************
% ** Chapter content is in separate files
% ****************************************
\pagestyle{headings}
\thispagestyle{headings}
\input{chaps}

% APPENDIX
\ifappendix
   \begin{appendix}
   \input{srcs/appendix}

   \input{srcs/acronyms}
   \end{appendix}
\else
   \input{srcs/acronyms}
\fi

\ifbib
   %\setbibpreamble
   \bibliographystyle{unsrtnat}
   \bibliography{nuctag}
\fi

\ifindex
   \IndexPrologue{\chapter*{Index}}
   %\setindexpreamble
   \printindex
\fi

\pagestyle{plain}
\input{srcs/acknowledge}

\end{document}

%% file: srcs/commands.tex
\def\phob         {PHOBOS}
\def\prap         {pseudorapidity}
\def\Prap         {Pseudorapidity}
\def\mrap         {mid-rapidity}
\def\cheren       {\v{C}erenkov}

\def\MIT          {MASSACHUSETTS INSTITUTE OF TECHNOLOGY}
\def\Mit          {Massachusetts Institute of Technology}
\def\thesistitle  {Studies of Nucleon-Gold Collisions at 200~GeV per Nucleon Pair Using Tagged d+Au Interactions}
\def\myname       {Corey Reed}
\def\department   {Department of Physics}
\def\thesisdate   {September 5, 2006}
\def\degreeword   {degree}
\def\degree       {Doctor of Philosophy}
\def\prevdegrees  {B.S., University of California (1999)}
\def\degreeyear   {2006}
\def\degreemonth  {September}
\def\copyrightnotice {{\copyright} {\Mit} {\degreeyear}.  All rights reserved.}
\def\supervisor         {George S. F. Stephans}
\def\supervisortitle    {Senior Research Scientist}
\def\cosupervisor       {Wit Busza}
\def\cosupervisortitle  {Professor of Physics}
\def\chairmanname  {Thomas J. Greytak}
\def\chairmantitle {Associate Department Head for Education}

\def\signature#1#2{\par\noindent#1\dotfill\null\\*
  {\raggedleft #2\par}}
\def\abstractsupervisor{\vskip\baselineskip\par\noindent%
Thesis Supervisor: {\supervisor} \\
Title: {\supervisortitle}}

\def\titlepage{\cleardoublepage\centering
  \thispagestyle{empty}
  \parindent 0pt \parskip 10pt plus 1fil minus 1fil
  \def\baselinestretch{1}\normalsize\vbox to \vsize\bgroup\vbox to 9in\bgroup}
\def\endtitlepage{\par\kern 0pt\egroup\vss\egroup\newpage}
\def\maketitle{\begin{titlepage}
\large
{\def\baselinestretch{1.2}\Large\bf {\thesistitle} \par}
by\par
{\Large {\myname}}
\par
{\prevdegrees}
\par
Submitted to the {\department} \\
in partial fulfillment of the requirements for the
{\degreeword}
of
\par
{\degree}
\par
at the
\par\MIT\par
{\degreemonth} {\degreeyear}
\par
\copyrightnotice
\par
\vskip 3\baselineskip
\signature{Author}{{\department} \\ {\thesisdate}}
\par
\vfill
\signature{Certified by}{{\supervisor} \\ {\supervisortitle} \\
Thesis Supervisor}
\par
\vfill
\signature{Certified by}{{\cosupervisor} \\ {\cosupervisortitle} \\
Thesis Co-Supervisor}
\par
\vfill
\signature{Accepted by}{{\chairmanname} \\ {\chairmantitle}}
\vfill
\end{titlepage}}

\def\abstract{\subsection*{Abstract}\small\def\baselinestretch{1}\normalsize}

\def\abstractpage{\cleardoublepage
\begin{center}{\large{\bf {\thesistitle}} \\
by \\
{\myname} \\[\baselineskip]}
\par
\def\baselinestretch{1}\normalsize
Submitted to the {\department} \\
on {\thesisdate}, in partial fulfillment of the \\
requirements for the {\degreeword} of \\
{\degree}
\end{center}
\par
\begin{abstract}}
\def\endabstractpage{\end{abstract}\noindent
{\abstractsupervisor} \newpage}

\newcommand{\dedicationtop}{\vspace*{\stretch{1}}}
\newcommand{\dedicationbottom}{\bigskip\bigskip\vspace*{\stretch{1}}}
\renewenvironment{dedication}{%
  \cleardoublepage
  \thispagestyle{empty}
  \centering
  \begingroup
  \begin{itshape}\dedicationtop
}{\dedicationbottom\end{itshape}\par\endgroup}%dedication

\DefineVerbatimEnvironment{CodeBox}{Verbatim}
   {frame=single, fontsize=\small, samepage=true, commandchars=\\\@\#}
\newcommand {\code}[1]{\texttt{\small{#1}}}
\def\ttree        {\code{TTree}}
\def\tclon        {\code{TClonesArray}}
\def\tobjar       {\code{TObjArray}}
\def\tsel         {\code{TSelector}}
\def\tbrow        {\code{TBrowser}}
\def\tamm         {\code{TAModule}}
\def\tams         {\code{TAMSelector}}
\def\tamo         {\code{TAMOutput}}

\newcommand {\snn}      {\sqrt{s_{\scriptscriptstyle{{\rm NN}}}}}
\newcommand {\dndy}     {\ensuremath{d\!N/\mathit{dy}}}
\newcommand {\dnchdy}   {\ensuremath{d\!N_{\mathrm{ch}}/\mathit{dy}}}
\newcommand {\dndeta}   {\ensuremath{d\!N/\mathit{d\eta}}}
\newcommand {\dnchdeta} {\ensuremath{d\!N_{\mathrm{ch}}/\mathit{d\eta}}}
\newcommand {\ave}[1]   {\ensuremath{\left< #1 \right>}}
\newcommand {\abs}[1]   {\ensuremath{\left| #1 \right|}}
\newcommand {\prn}[1]   {\ensuremath{\left( #1 \right)}}
\newcommand {\sqb}[1]   {\ensuremath{\left[ #1 \right]}}
\newcommand {\lsim}     {\,{\buildrel < \over {_\sim}}\,}
\newcommand {\gsim}     {\,{\buildrel > \over {_\sim}}\,}
\newcommand {\cov}      {\ensuremath{\mathrm{cov}}}
\newcommand {\hpos}     {\ensuremath{\mathrm{h}^+}}
\newcommand {\hneg}     {\ensuremath{\mathrm{h}^-}}
\newcommand {\have}     {\ensuremath{(\mathrm{h}^+ + \mathrm{h}^-) / 2}}

\newcommand {\hrefurl}[1]{\href{#1}{\mbox{#1}}}
\newcommand {\fig}[1]{Fig.~\ref{#1}}
\newcommand {\figs}[2]{Fig.~\ref{#1} and~\ref{#2}}
\newcommand {\Fig}[1]{Figure~\ref{#1}}
\newcommand {\eq}[1]{Eq.~\ref{#1}}
\newcommand {\eqs}[2]{Eq.~\ref{#1} and~\ref{#2}}

\newcommand {\Eqs}[2]{Equations~\ref{#1} and~\ref{#2}}
\newcommand {\sect}[1]{Sect.~\ref{#1}}

\newcommand {\chap}[1]{Ch.~\ref{#1}}
\newcommand {\chaps}[2]{Ch.~\ref{#1} and~\ref{#2}}

\newcommand {\tab}[1]{Table~\ref{#1}}
\newcommand {\tabs}[2]{Tables~\ref{#1} and~\ref{#2}}
\newcommand {\Tab}[1]{Table~\ref{#1}}
\newcommand {\peq}[1]{Eq.~\ref{#1} on page~\pageref{#1}}

\newcommand {\pfig}[1]{Fig.~\ref{#1} on page~\pageref{#1}}

\newcommand {\ptab}[1]{Table~\ref{#1} on page~\pageref{#1}}

\newcommand {\stolenfig}[2]{\fig{#1}~\cite{#2}}

\newcommand {\appndx}[1]{Appendix~\ref{#1}}

\newcommand {\pt}       {\ensuremath{p_{\mathrm{T}}}}
\newcommand {\mt}       {\ensuremath{m_{\mathrm{T}}}}
\newcommand {\mpi}      {\ensuremath{m_\pi}}

\newcommand {\Npard}    {\ensuremath{N_{\rm part}^{d}}}
\newcommand {\NparA}    {\ensuremath{N_{\rm part}^{Au}}}
\newcommand {\Ncoll}    {\ensuremath{N_{\rm coll}}}
\newcommand {\Npart}    {\ensuremath{N_{\rm part}}}
\newcommand {\Nptbs}    {\ensuremath{N_{\rm part}^{\rm biased}}}
\newcommand {\spcal}    {\ensuremath{S_{\rm pcal}}}
\newcommand {\scorl}    {\ensuremath{S_{\rm corl}}}
\newcommand {\cpar}     {\ensuremath{P_{\rm coll}}}
\newcommand {\dcab}     {\ensuremath{\mathit{DCA}_{\rm beam}}}
\newcommand {\aveeta}   {\ensuremath{\ave{\smash[b]{\eta}}}}
\newcommand {\avenpt}   {\ensuremath{\ave{\smash[b]{\Npart}}}}
\newcommand {\avencl}   {\ensuremath{\ave{\smash[b]{\Ncoll}}}}
\newcommand {\pbarprat} {\ensuremath{\bar{\rm p}/{\rm p}}}
\newcommand {\pbarpirat}{\ensuremath{\bar{\rm p}/{\pi^{-}}}}
\newcommand {\phikrat}{\ensuremath{\bar{\phi}/{K^{-}}}}
\newcommand {\mub}      {\ensuremath{\mu_{B}}}
\newcommand {\pttrue}   {\ensuremath{\pt^{\text{true}}}}
\newcommand {\ptrecon}  {\ensuremath{\pt^{\text{recon}}}}

\newcommand {\rad}   {\mbox{${\rm rad}$}}
\newcommand {\mrad}  {\mbox{${\rm mrad}$}}
\newcommand {\dg}    {\mbox{$^\circ$}}
\newcommand {\ev}    {\mbox{${\rm eV}$}}
\newcommand {\tev}   {\mbox{${\rm TeV}$}}
\newcommand {\gev}   {\mbox{${\rm GeV}$}}
\newcommand {\mev}   {\mbox{${\rm MeV}$}}
\newcommand {\kev}   {\mbox{${\rm keV}$}}
\newcommand {\mom}   {\mbox{\rm GeV$\kern-0.15em /\kern-0.12em c$}}
\newcommand {\mmom}  {\mbox{\rm MeV$\kern-0.15em /\kern-0.12em c$}}
\newcommand {\mass}  {\mbox{\rm GeV$\kern-0.15em /\kern-0.12em c^2$}}
\newcommand {\mmass} {\mbox{\rm MeV$\kern-0.15em /\kern-0.12em c^2$}}
\newcommand {\gm}    {\mbox{${\rm g}$}}
\newcommand {\km}    {\mbox{${\rm km}$}}
\newcommand {\m}     {\mbox{${\rm m}$}}
\newcommand {\mps}   {\mbox{${\rm m/s}$}}
\newcommand {\cm}    {\mbox{${\rm cm}$}}
\newcommand {\mm}    {\mbox{${\rm mm}$}}
\newcommand {\um}    {\mbox{$\mu{\rm m}$}}
\newcommand {\nm}    {\mbox{${\rm nm}$}}
\newcommand {\fm}    {\mbox{${\rm fm}$}}
\newcommand {\fmc}   {\mbox{${{\rm fm}/c}$}}
\newcommand {\amps}  {\mbox{${\rm A}$}}
\newcommand {\vdc}   {\mbox{${\rm VDC}$}}
\newcommand {\volts} {\mbox{${\rm V}$}}
\newcommand {\mvolt} {\mbox{${\rm mV}$}}
\newcommand {\Mvolt} {\mbox{${\rm MV}$}}
\newcommand {\pF}    {\mbox{${\rm pF}$}}
\newcommand {\fC}    {\mbox{${\rm fC}$}}
\newcommand {\tesla} {\mbox{${\rm T}$}}
\newcommand {\us}    {\mbox{$\mu{\rm s}$}}
\newcommand {\ns}    {\mbox{${\rm ns}$}}
\newcommand {\ps}    {\mbox{${\rm ps}$}}
\newcommand {\kg}    {\mbox{${\rm kg}$}}

\newcommand {\mhz}   {\mbox{${\rm MHz}$}}

\newcommand {\kB}    {\mbox{${\rm kB}$}}

\newcommand {\mbyte} {\mbox{${\rm MB}$}}

\newcommand {\gbyte} {\mbox{${\rm GB}$}}
\newcommand {\mbyteps}{\mbox{${\rm MB/s}$}}

\newcommand {\kelvin}{\mbox{${\rm K}$}}
\newcommand {\inbarn}{\mbox{${\rm nb}^{-1}$}}
\newcommand {\dens}  {\mbox{${\rm gm}/{\rm cm}^{3}$}}
\newcommand {\mohm}  {\mbox{${\rm M}\Omega$}}
\newcommand {\mb}    {\mbox{${\rm mb}$}}

\newcommand {\pp}    {\mbox{p+p}}
\newcommand {\pbarp} {\mbox{p+\={p}}}
\newcommand {\Au}    {\mbox{Au}}
\newcommand {\AuAu}  {\mbox{Au+Au}}
\newcommand {\CuCu}  {\mbox{Cu+Cu}}
\newcommand {\dAu}   {\mbox{d+Au}}
\newcommand {\pAu}   {\mbox{p+Au}}
\newcommand {\nAu}   {\mbox{n+Au}}
\newcommand {\NAu}   {\mbox{N+Au}}

\newcommand {\stype}[1] {\mbox{#1-type}}
\newcommand {\pnjnc}    {\mbox{p-n}~junc\-tion}
\newcommand {\naive}    {na\"{\i}ve}
\newcommand {\myupdate}[1] {\iffinal\else\begin{center}\textbf{#1}\end{center}\fi\noindent}

\newcommand {\cpartab}[6]{%
   \begin{table}[#6]
      \begin{center}
         \begin{tabular}{|llllll|}
\hline
\multicolumn{6}{|c|}{#1 $#2$ with #3 event selection} \\
\hline
         &            & \multicolumn{4}{c|}{$\ave{#2} \pm \text{Sys. Err.}$} \\
Variable & Simulation & 70-100\% & 40-70\% & 20-40\% & 0-20\% \\
\hline
#4
\hline
         \end{tabular}
      \end{center}
      \caption{   \label{data:tab:#5}
         #1 $#2$ obtained using different centrality
         measures and the #3 event selection.}
    \end{table}
}

%% file: srcs/abstract.tex
% $Id: abstract.tex,v 1.6 2006/09/03 19:33:04 cjreed Exp $
%
% Abstract

The spectra of charged hadrons produced near {\mrap} in {\dAu}, {\pAu}
and {\nAu} collisions at $\snn=200~\gev$ are presented as a function of
transverse momentum and centrality. These measurements were performed
using the {\phob} detector at the \acf*{RHIC}. Nucleon-nucleus
interactions were extracted from the {\dAu} data by identifying the
deuteron spectators. The deuteron spectators were measured using two
calorimeters; one that detected forward-going single neutrons and a
newly installed calorimeter that detected forward-going single protons. 

The large suppression of high-$\pt$ hadron production in central {\AuAu}
interactions relative to a {\naive} superposition of {\pbarp} collisions
has been interpreted as evidence of partonic energy loss in a dense
medium. This interpretation is founded upon the absence of such
suppression in the yield of {\dAu} collisions. The validity of using
{\dAu} interactions in place of a nucleon-nucleus reference is tested.
It is shown that hadron production in {\dAu} agrees with a simple binary
collision scaling of hadron production in {\pAu}. An ideal reference for
{\AuAu} collisions is constructed using a weighted combination of {\pAu}
and {\nAu} yields and is found to be similar to the {\dAu} reference.
Further, hadron production in {\pAu} interactions is compared to that of
{\nAu} interactions. The single charge difference between a {\pAu} and a
{\nAu} collision allows for a unique study of the ability of the
interaction to transport the proton from the initial deuteron to
{\mrap}. However, no asymmetry between the positively and negatively
charged hadron spectra of {\pAu} and {\nAu} interactions is observed at
$\aveeta=0.8$.

Collision centrality was determined using several different observables,
including those based on the multiplicity in different regions of
{\prap} and those based on the amount of nuclear spectator material. It
is shown that measurements made on small collision systems in the
{\mrap} region are biased by centrality variables based on the {\mrap}
multiplicity. Despite this bias, a smooth evolution with centrality is
observed in the Cronin enhancement of hadrons produced in {\dAu}
collisions. It is shown that this smooth progression is independent of
the choice of centrality variable when centrality is parametrized
by the multiplicity measured near {\mrap}.

%% file: chaps.tex
% $Id: chapters.tex,v 1.10 2006/08/17 17:52:29 cjreed Exp $

\input{srcs/intro}
\input{srcs/chapter1}
\input{srcs/chapter2}
\input{srcs/reconChapter}
\input{srcs/trackChap}
\input{srcs/anaChap}

\input{srcs/data}

\input{srcs/results}
\input{srcs/conc}

%% file: srcs/intro.tex
%---------------------------------------------------------------
\chapter{Strongly Interacting Matter}
\label{intro}
%---------------------------------------------------------------

\myupdate{$*$Id: intro.tex,v 1.35 2006/09/05 01:18:31 cjreed Exp $*$}%

%---------------------------------------------------------------
\section{Forces of Nature}
\label{intro:forces}
%---------------------------------------------------------------

A successful description of the motion of matter is of fundamental
importance to the understanding of nature. Classical descriptions of the
mechanics of motion, developed during the Scientific Revolution around
1500-1700, postulated that objects tend to move in a straight line at a
fixed speed. It followed that changes in the speed or direction of an
object's motion did not occur spontaneously, but were forced upon the
object. The concept of forcing motion on an object through contact with
the object is fairly intuitive. What is perhaps less obvious is how a
force could act on an object at a distance and without contact. An
example of such a force would be the ability of the Earth to accelerate
flakes of snow to the ground (particularly in the Boston area).

\begin{figure}[t]
   \begin{center}
      \includegraphics[width=0.5\linewidth]{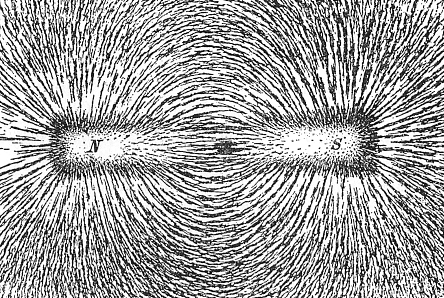}
   \end{center}
   \caption{\label{intro:fig:Magnet0873}
      Iron filings reveal the field generated by a bar
      magnet~\cite{Image:Magnet0873}.}
\end{figure}

Modern descriptions of the dynamics of objects have expanded upon these
early concepts. Forces are no longer thought to act instantaneously at a
distance. Instead, they are presumed to be mediated by fields, which
describe physical quantities at every point in space. Force fields, for
example, describe the magnitude and direction of the force that would be
applied to an object at any particular point in space. Variations in a
field progress at a finite speed. A visualization of a field is
presented in \stolenfig{intro:fig:Magnet0873}{Image:Magnet0873}, which
shows how iron filings orient themselves in a magnetic field.
Indeed, studies of electricity and magnetism led to the foundation of
theories of field dynamics. These studies not only showed that
electricity and magnetism are two aspects of the same field, but that
variations in this electromagnetic field travel through otherwise
vacuous regions at the speed of light, $c \approx 3\times10^{8}~\mps$.
The latter observation led to the postulate that light itself is
composed of waves propagating through the electromagnetic field.

These ideas were refined with the advent of quantum mechanics and
special relativity. Relativity postulated that the laws of physics are
the same in all inertial frames of reference; that is, to any observer
not accelerating. It followed that, since the speed of light is an
integral part of the descriptions of the electromagnetic field, light is
always observed to be moving at a speed $c$ -- regardless of how fast
the light source or the observer may be moving. An important consequence
of relativity was the equivalence of mass and energy. Expressed in what
is likely the most famous equation of physics, $E = m c^2$, this concept
reveals that an object at rest and under the influence of no forces
still possesses a finite amount of energy proportional to its mass.

Quantum mechanics held that the motion of an object is not determined
uniquely. Instead, a property of an object, such as its position, can be
observed to take on a particular value only with some probability. That
is, it is not possible to predict with certainty the outcome of a single
measurement. However, predictions can be made regarding the possible
outcomes of a measurement and the likelihood with which each outcome
could occur. It was possible to interpret quantum mechanics as
describing the dynamics of a wavefunction, rather than the dynamics of
an object. The value of this wavefunction at any point in space is
related to the probability of observing the object at that point.

The current understanding of the dynamics of objects combines all of the
concepts discussed above into a relativistic quantum field theory. This
description of nature is built upon the existence of fields which
describe the possible states of a physical system (the universe) and the
probability with which the system might be found in each state. When
such an observation is performed, the field will be measured not as a
continuous wave, but rather as indivisible, dimensionless packets called
`quanta' or `particles.' Because the fields are consistent with
relativity, energy in a field can be converted to mass, and vice-versa.
Thus, the number of particles observed in the field can change;
increasing as energy is converted to mass, and decreasing as mass is
converted to energy. The dynamics of the particles are driven by
interactions between various fields.

\begin{table}[t]
   \begin{center}
      \begin{tabular}{|rrccc|}
\hline
Generation & Particle & Symbol & Charge & Mass
({\mmass})~\cite{Yao:2006} \\
\hline
\multirow{2}{*}{First} & Electron Neutrino & $\nu_e$ & $0$ & $< 2\times10^{-6}$\\
 & Electron & $e^{-}$ & $-1$ & $0.511$\\
\hdashline[1pt/4pt]
\multirow{2}{*}{Second} & Muon Neutrino & $\nu_\mu$ & $0$ & $< 0.19$\\
 & Muon & $\mu^{-}$ & $-1$ & $105.7$\\
\hdashline[1pt/4pt]
\multirow{2}{*}{Third} & Tau Neutrino & $\nu_\tau$ & $0$ & $< 18.2$\\
 & Tau & $\tau^{-}$ & $-1$ & $1777$\\
\hline
      \end{tabular}
   \end{center}
   \caption{   \label{intro:tab:leptons}
      The leptons. For each particle, there is an oppositely charged
      antiparticle (not shown). The neutrino masses listed here are
      effective masses, see~\cite{Yao:2006}.}
\end{table}

\begin{table}[t]
   \begin{center}
      \begin{tabular}{|rccc|}
\hline
Generation & Particle & Symbol & Charge \\
\hline
\multirow{2}{*}{First} & Up & $u$ & $+2/3$ \\
 & Down & $d$ & $-1/3$ \\
\hdashline[1pt/4pt]
\multirow{2}{*}{Second} & Charm & $c$ & $+2/3$ \\
 & Strange & $s$ & $-1/3$ \\
\hdashline[1pt/4pt]
\multirow{2}{*}{Third} & Top & $t$ & $+2/3$ \\
 & Bottom & $b$ & $-1/3$ \\
\hline
      \end{tabular}
   \end{center}
   \caption{   \label{intro:tab:quarks}
      The quarks. Each quark listed here can carry one of the three
      types of color charge: red, blue or green (see \sect{intro:qcd}).
      For each of these 18 quarks, there is an antiparticle with an
      opposite electric and color charge.}
\end{table}

In the Standard Model, nature is characterized by a set of fields that
describe the fundamental building blocks of matter and by another set of
fields that mediate their interactions: the forces. The building blocks
of matter are divided into two distinct groups known as quarks and
leptons. For example, an everyday object such as a book is composed of
an unimaginably large number of atoms. Each atom consists of a tiny
nucleus surrounded by a cloud of electrons (the electron is the most
famous lepton). The nucleus is made up of a collection of tightly bound
protons and neutrons, each of which is itself composed of quarks. All of
the fundamental particles, quarks and leptons, are quanta of their
respective fields. For each particle, there exists an antiparticle. An
antiparticle is also a quantum of its field, so it shares certain
properties with its corresponding particle, such as mass. However, other
properties of an antiparticle known as additive quantum numbers, such as
electric charge, are exactly opposite to those of its corresponding
particle~\cite{Cristinziani:2003xu}. Currently, it is believed that
there are three \emph{generations} of both leptons and quarks. The
particles in a given generation share certain properties that influence
how they react with the quanta of other fields~\cite{Rolnick:particles}.
The current understand of the properties of elementary particles can be
found in~\cite{Yao:2006}. A list of the lepton generations is presented
in \tab{intro:tab:leptons} and a list of the quark generations is shown
in \tab{intro:tab:quarks}. Because free quarks are not observed, as will
be discussed in \sect{intro:qcd}, determining the mass of the quarks is
the subject of study; see~\cite{Yao:2006} for current estimates.

The interactions between particles of matter are thought to be governed
by four forces. Gravity is likely the most familiar force,
yet a description of gravity in terms of relativistic quantum field
theories has proven extraordinarily difficult~\cite{Ashtekar:2004eh}, to
the extent that gravity is not included in the Standard Model. However,
gravity is so incredibly weak\footnote{If this seems surprising,
consider the fact that a small magnet can easily hold a paper-clip off
the ground, despite the pull of gravity exerted on the paper-clip by the
entire planet Earth.} compared to the other fundamental forces -- in the
physical phenomena discussed in this thesis -- that it can be safely
ignored.

The second least powerful force is known as the weak force. Unlike
gravity, the weak force has a very small range, about $10^{-18}~\m$, and
is not observed to bind objects together. Instead, the weak force
mediates interactions between the quanta of matter fields, both quarks
and leptons, in such a way as to allow a change in the type, or flavor,
of the initial particle(s). For example, the decay of a neutron to a
proton, known as beta decay, is governed by the weak force. In this
interaction, a down quark changes flavor to an up quark through the
process \mbox{$d \rightarrow u e^{-} \bar{\nu}_e$}.

The electromagnetic force mediates the interaction between electrically
charged particles. The charge of any particle can be expressed in
discrete units of the magnitude of an electron's charge, as is done in
\tabs{intro:tab:leptons}{intro:tab:quarks}. The electromagnetic force is
the most prevalent force in everyday experience, and consequently is the
most well understood force~\cite{Escribano:1996hf}. All molecular
binding and chemical reactions are the result of electromagnetic
interactions. In addition, contact forces in the macroscopic world can
be attributed to the repulsion of atoms near the surface of opposing
objects.

Such repulsive electromagnetic forces also cause protons in an atomic
nucleus to repel each other. Yet despite this fierce repulsion, the
protons are bound together, along with neutrons, into a very small
volume -- the radius of a nucleus is over ten thousand times smaller
than the radius of an atom. Thus the protons and neutrons attract each
other through some force that is far stronger than the electromagnetic
repulsion of the protons. This force is aptly, if not creatively, named
the strong force. The strong force is thought to be so powerful that the
binding of protons and neutrons in a nucleus is merely a residual effect
of the interactions between quarks in the nucleons (a nucleon being
either a proton or a neutron). This residual force is analogous to the
residual electromagnetic force that binds atoms into molecules;
see~\cite{LandauLifshitz:qm} for a discussion of atomic molecules
and~\cite{Isgur:1985vp} for a discussion of the residual nuclear force.
The strong force is not experienced by leptons, nor by the quanta of the
other force fields. It is described in the Standard Model by the
relativistic quantum field theory known as \acf{QCD}.

%---------------------------------------------------------------
\section{Quantum Chromodynamics}
\label{intro:qcd}
%---------------------------------------------------------------

\acf{QCD} is a relativistic quantum field theory that is believed to
describe the strong force. The theory describes the dynamics of six
quark fields, the quanta of which are listed in \tab{intro:tab:quarks}.
Each quark carries an electric charge as well as a different kind of
charge known as color. There are three types of color charge: red, blue
and green. A quark carries one of the three colors, red, blue or green,
while an antiquark carries one of the three anticolors. For example, the
antiparticle of an up quark carrying a red color charge, $u_r$, would be
an antiup quark carrying an ``antired'' charge, $\bar{u}_{\bar{r}}$.
This is analogous to electromagnetism, in which there is a single type
of electric charge that can take on two values, positive or negative.
The interactions of particles carrying color charge are mediated by
eight force fields, the quanta of which are massless particles known as
gluons. The rationale for eight gluon fields is rooted in the fact that
gluons themselves carry color: each gluon carries both a color charge
and an anticolor charge. The various color states of the gluons can be
written as

\begin{equation}
   \label{intro:eq:gluoncolor}
r\bar{b}, r\bar{g}, b\bar{g}, b\bar{r}, g\bar{b},
\frac{r\bar{r} - b\bar{b}}{\sqrt{2}},
\frac{r\bar{r} + b\bar{b} - 2g\bar{g}}{\sqrt{6}}
\end{equation}

\noindent%
where the latter two gluons can be understood through a quantum
mechanical description. For example, the \mbox{$(r\bar{r} - b\bar{b}) /
\sqrt{2}$} gluon can be thought of as being at once \emph{both} a red,
antired gluon and a blue, antiblue gluon that has a 50/50 chance of
being observed as a $G_{r\bar{r}}$ or as a $G_{b\bar{b}}$. The ninth
possible combination of colors,

\begin{equation}
   \label{intro:eq:ninthgluon}
\frac{r\bar{r} + b\bar{b} + g\bar{g}}{\sqrt{3}}
\end{equation}

\noindent%
is not a part of \ac{QCD}, nor of nature. Such a gluon would be what
is known as a \emph{color singlet}, that is, it would carry no net
color. This would allow other colorless particles (such as protons) to
exchange the gluon, and would facilitate strong interactions over
macroscopic distances, assuming this gluon is also massless. Clearly,
there is no evidence that such a gluon exists.

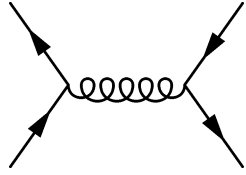
\begin{figure}[t]
   \begin{center}
\begin{fmffile}{gluonexchange}
  \fmfframe(1,7)(1,7){ 	%Sets dimension of Diagram
   \begin{fmfgraph*}(110,62) %Sets size of Diagram
    \fmfpen{thin}
    \fmfleft{i1,o1}	%Sets there to be 2 sources 
    \fmfright{i2,o2}    %Sets there to be 2  outputs
    \fmflabel{$u_g$}{i1} %Labels one of the left sources
    \fmflabel{$\bar{u}_{\bar{g}}$}{i2} %Labels one of the left sources
    \fmflabel{$u_b$}{o1} %Labels one of the right outputs
    \fmflabel{$\bar{u}_{\bar{b}}$}{o2} %Labels one of the right outputs
    \fmf{fermion}{i1,v1,o1} %Connects the sources with a vertex.
    \fmf{fermion}{o2,v2,i2} %Connects the outputs with a vertex.
    \fmf{gluon,label=$G_{g\bar{b}}$,label.dist=0.09w}{v1,v2}
       %Labels the conneting line.
   \end{fmfgraph*}
  }
\end{fmffile}
   \end{center}
   \caption{\label{intro:fig:gluonexchange}
      An example of a quark-antiquark interaction through the exchange
      of a gluon. The flow of time in the diagram is up-wards; the
      horizontal direction represents spatial separation. Note that the
      quarks change color after interacting.}
\end{figure}

\begin{table}[t]
   \begin{center}
      \begin{tabular}{|lccc|}
\hline
Particle [Quark Content] & Symbol & Mass (\mmass) & Mean Lifetime (s) \\
\hline
Proton $[uud]$ & p & $938.3$ & stable \\
Neutron $[udd]$ & n & $939.6$ & $898$ \\
\hdashline[1pt/4pt]
Positive Pion $[u\bar{d}]$ & $\pi^{+}$ & $139.6$ &
   $2.603\times10^{-8}$\\
Negative Pion $[d\bar{u}]$ & $\pi^{-}$ & $139.6$ & $2.603\times10^{-8}$\\
Neutral Pion $[(d\bar{d} + u\bar{u}) / \sqrt{2}]$ & 
   $\pi^{0}$ & $135.0$ & $0.83\times10^{-16}$\\
Positive Kaon $[u\bar{s}]$ & $K^{+}$ & $493.7$ & $1.237\times10^{-8}$\\
Negative Kaon $[s\bar{u}]$ & $K^{-}$ & $493.7$ & $1.237\times10^{-8}$\\
\hline
      \end{tabular}
   \end{center}
   \caption{   \label{intro:tab:hadrons}
      Properties of some common hadrons~\cite{Perkins:HEP}. The mean
      lifetimes shown are for free particles. The stability of the
      proton is the subject of active experimental and theoretical
      study~\cite{Shiozawa:1998si, Friedmann:2002ty}.}
\end{table}

An example of an interaction between an up quark and an antiup quark in
one dimension is shown in \fig{intro:fig:gluonexchange}. A study of such
simple gluon exchanges  shows that the lowest energy, and therefore most
stable, bound states of quarks -- called hadrons -- are (a)~a
combination of three quarks each having a different color, called a
baryon, and (b)~a combination of a quark and its antiquark, called a
meson; see~\cite{Perkins:HEP} for a discussion. This is in good
agreement with observation. The proton and neutron are the most common
baryons observed in nature. Two mesons that are abundantly produced in
high energy hadron collisions are the pion and the kaon. The quark
content of these particles is listed in \tab{intro:tab:hadrons}.

\begin{figure}[t]
   \centering
   \parbox{0.95\linewidth}{%
      \subfigure[Quark Exchange]{
         \label{intro:fig:ggqq}
         \begin{fmffile}{ggqq}
           \fmfframe(1,7)(1,14){ 	%Sets dimension of Diagram
            \begin{fmfgraph*}(100,62) %Sets size of Diagram
             \fmfpen{thin}
             \fmfleft{i1,o1}	%Sets there to be 2 sources 
             \fmfright{i2,o2}    %Sets there to be 2  outputs
             \fmflabel{$G_{r\bar{g}}$}{i1} %Labels one of the left sources
             \fmflabel{$G_{g\bar{r}}$}{i2} %Labels one of the left sources
             \fmflabel{$\bar{u}_{\bar{g}}$}{o1} %Labels one of the right outputs
             \fmflabel{$u_g$}{o2} %Labels one of the right outputs
             \fmf{fermion}{o1,v1} %Connects the sources with a vertex.
             \fmf{fermion}{v2,o2} %Connects the sources with a vertex.
             \fmf{fermion,label=$u_r$}{v1,v2}
             \fmf{gluon}{i1,v1} %Connects the sources with a vertex.
             \fmf{gluon}{i2,v2} %Connects the sources with a vertex.
            \end{fmfgraph*}
           }
         \end{fmffile}
      }
      \hfill
      \subfigure[Gluon Exchange]{
         \label{intro:fig:ggggexg}
         \begin{fmffile}{ggggexg}
           \fmfframe(1,7)(1,14){ 	%Sets dimension of Diagram
            \begin{fmfgraph*}(100,62) %Sets size of Diagram
             \fmfpen{thin}
             \fmfleft{i1,o1}	%Sets there to be 2 sources 
             \fmfright{i2,o2}    %Sets there to be 2  outputs
             \fmflabel{$G_{r\bar{g}}$}{i1} %Labels one of the left sources
             \fmflabel{$G_{g\bar{r}}$}{i2} %Labels one of the left sources
             \fmflabel{$G_{b\bar{g}}$}{o1} %Labels one of the right outputs
             \fmflabel{$G_{g\bar{b}}$}{o2} %Labels one of the right outputs
             \fmf{gluon}{v1,o1} %Connects the sources with a vertex.
             \fmf{gluon}{v2,o2} %Connects the sources with a vertex.
             \fmf{gluon,label=$G_{r\bar{b}}$,label.dist=0.09w}{v1,v2}
             \fmf{gluon}{i1,v1} %Connects the sources with a vertex.
             \fmf{gluon}{i2,v2} %Connects the sources with a vertex.
            \end{fmfgraph*}
           }
         \end{fmffile}
      }
      \hfill
      \subfigure[Direct Interaction]{
         \label{intro:fig:gggg}
         \begin{fmffile}{gggg}
           \fmfframe(1,7)(1,14){ 	%Sets dimension of Diagram
            \begin{fmfgraph*}(75,62) %Sets size of Diagram
             \fmfpen{thin}
             \fmfleft{i1,o1}	%Sets there to be 2 sources 
             \fmfright{i2,o2}    %Sets there to be 2  outputs
             \fmflabel{$G_{r\bar{g}}$}{i1} %Labels one of the left sources
             \fmflabel{$G_{g\bar{r}}$}{i2} %Labels one of the left sources
             \fmflabel{$G_{b\bar{g}}$}{o1} %Labels one of the right outputs
             \fmflabel{$G_{g\bar{b}}$}{o2} %Labels one of the right outputs
             \fmf{gluon}{v1,o1} %Connects the sources with a vertex.
             \fmf{gluon}{v1,o2} %Connects the sources with a vertex.
             \fmf{gluon}{i1,v1} %Connects the sources with a vertex.
             \fmf{gluon}{i2,v1} %Connects the sources with a vertex.
            \end{fmfgraph*}
           }
         \end{fmffile}
      }
   }
   \caption{\label{intro:fig:gginteract}
      Examples of gluon-gluon interactions.
      \subref{intro:fig:ggqq}~Interaction via quark exchange.
      \subref{intro:fig:ggggexg}~Interaction via gluon exchange.
      \subref{intro:fig:gggg}~Direct interaction.}
\end{figure}

One of the most striking differences between \ac{QCD} and
electromagnetism is the fact that the quanta of the \ac{QCD} fields
carry color charge. Consequently, gluons can directly interact with
other gluons, as shown in \figs{intro:fig:ggggexg}{intro:fig:gggg}.
There is no analog for this in electromagnetism; the quantum of that
force, the photon, does not carry electric charge and therefore does not
interact directly with other photons. This direct interaction between
the gluon fields has severe effects on the dynamics of colored
particles. An example of this is the coupling strength between colored
particles, parametrized by $\alpha_s = g_{s}^2 / 4\pi$, where $g_{s}$ is
related to the amount of color charge carried by the particles.
Fluctuations of the quark and gluon fields in the vicinity of a colored
particle are capable of altering the effective coupling strength,
$\alpha_{s}^{\rm eff}$.

There is an electromagnetic analogy to this, in which fluctuations of
the electron field result in the production of ``virtual''
electron-positron pairs. These fluctuations are just that: pairs of
particles which are produced and annihilated, in otherwise empty space,
in such a short amount of time that their presence is not directly
observable, hence the label `virtual.' In the presence of a
(non-virtual) electron, these virtual pairs can be pictured as forming a
cloud around the electron. Pairs in this cloud become polarized by the
electron and result in an effective reduction, or screening, of the
electric charge felt by a test particle at some distance from the
electron~\cite{1995JPhB...28.1105C}. As the test particle approaches the
electron and pierces the cloud of virtual particles, the strength of
this screening is reduced.

In \ac{QCD}, on the other hand, there is evidence that just the opposite
happens. While fluctuations in the quark fields surrounding a colored
particle may screen its color charge, in a perfect analogy to
electromagnetism, fluctuations in the gluon fields can \emph{enhance}
the particle's color charge. This enhancement is due entirely to the
self-interaction of the gluon fields. The effective coupling strength of
the strong force felt by a test particle at short distances has been
calculated~\cite{Peskin:QFT}, and is presented in
\eq{intro:eq:couplingconst}. In experimental observation, however, it is
not possible to measure the distance between a test particle and the
colored object -- nor is it very meaningful. Thus instead of distance,
the effective charge seen by a test particle is parametrized by the
amount of momentum transfered between the test particle and the colored
object. Large momentum transfers correspond to small distances.

\begin{equation}
   \label{intro:eq:couplingconst}
\alpha_{s}^{\rm eff}\prn{Q^2} \approx 
   \frac{4\pi}{\prn{11 - \frac{2}{3}n_f}\ln\prn{Q^2/\Lambda^2}}
\end{equation}

\noindent%
in natural units (\mbox{$c=\hbar=1$}), where $Q$ is the momentum
transfer, $n_f$ is the number of quark flavors (six) and $\Lambda$ is a
constant of nature, roughly 200~{\mmom}. That is, for \mbox{$Q \gg
\Lambda$}, \eq{intro:eq:couplingconst} gives a reasonable approximation
of the effective coupling strength. It is clear from this equation that
as the momentum transfer increases, and shorter distances are probed,
the effective coupling strength approaches zero. This property of the
strong force is known as asymptotic  freedom~\cite{Gross:1973id,
Politzer:1973fx, Gross:1973ju, Politzer:1974fr}. On the other hand, for
small momentum transfers, the effective coupling strength increases
without bound. Of course, \eq{intro:eq:couplingconst} is not thought to
be a good approximation of the \ac{QCD} coupling strength, or of nature,
for very small values of momentum transfer. Nevertheless, the
implication is that while the strong force is weak at very small
distances, it is strong at large distances.

\begin{figure}[t]
   \begin{center}
      \includegraphics[width=0.5\linewidth]{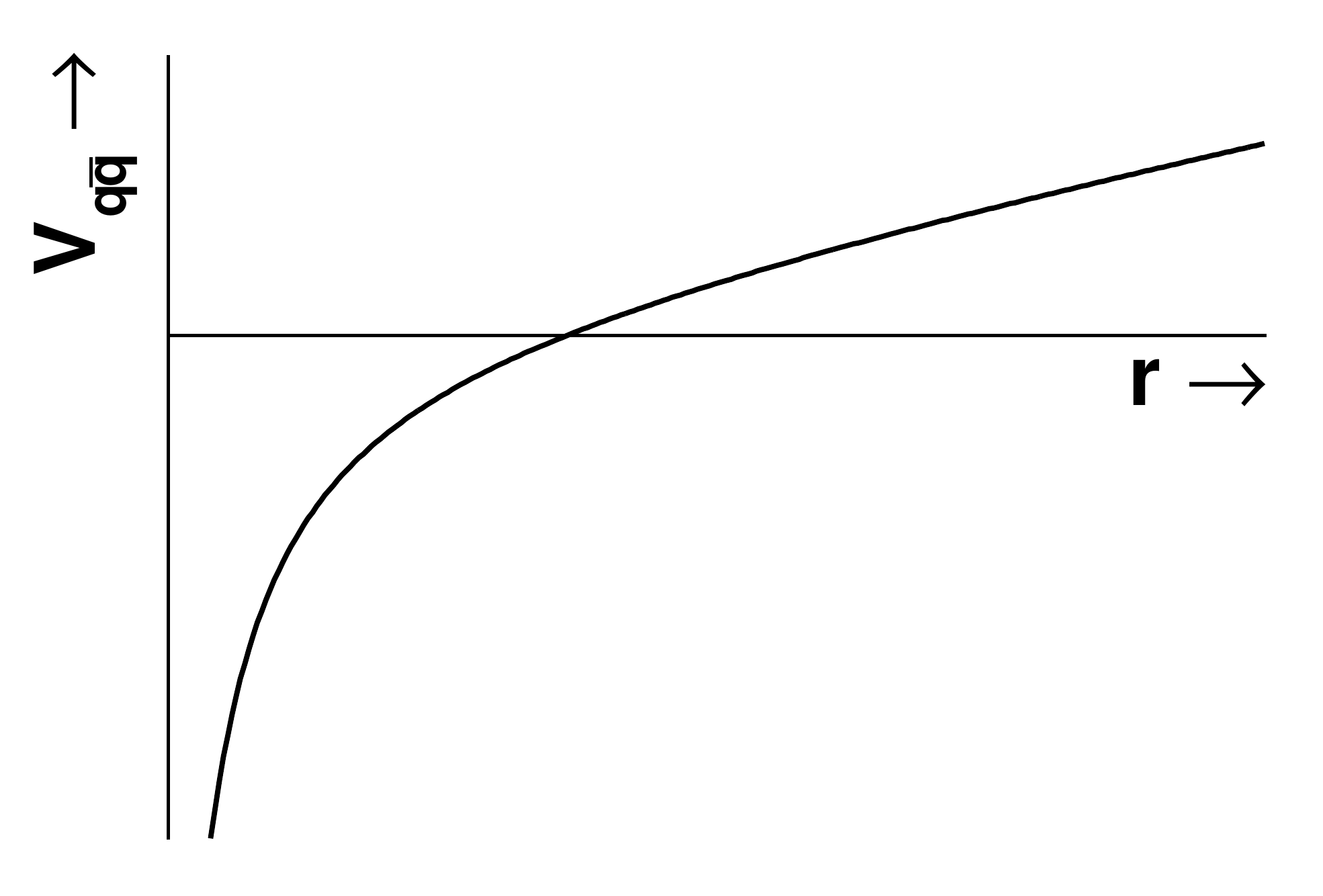}
   \end{center}
   \caption{\label{intro:fig:quarkPotential}
      The qualitative form of the potential between a (very heavy) quark
      and its antiquark, as estimated in~\cite{Olsson:1985ch}. For
      large separation, the potential becomes linear, and for closer
      separation, it falls as $-1/r$.}
\end{figure}

This might seem to suggest that the strong force should be observable,
and indeed quite powerful, on macroscopic scales. After all, distances
on the order of centimeters are certainly large distances compared to
the relevant length scale $\Lambda^{-1}$ -- by about thirteen orders
of magnitude\footnote{This estimate is made using the uncertainty
principal of quantum mechanics, $\Delta{x}\Delta{p}\ge\hbar/2$. Taking
$\Delta{p}\sim\Lambda$, $\Delta{x}\gtrsim1~\fm$; which is roughly the
size of a hadron.}. This hypothesis can be explored by studying the
static, one dimensional potential between a quark and its antiquark
(with anticolor). This potential has been estimated for very heavy
quarks, and is thought to be valid for quark-antiquark separations
larger than \mbox{$\sim0.1~\fm$}~\cite{Olsson:1985ch, German:1989vk}.
For relatively small $q-\bar{q}$ separations, the potential has a $-1/r$
dependence, analogous to the static electric potential between two
charged particles. However, at larger separations, the potential becomes
\emph{linear}, as shown in \fig{intro:fig:quarkPotential}. Thus, at
large distances, the force between a quark and an antiquark is the same
regardless of how far apart the quarks are separated. This implies that
the quarks are always bound, as it would take an infinite amount of
energy to separate the quarks completely. Notice further that the
$q-\bar{q}$ system in question is colorless; thus this potential is
relevant to the study of mesons. For a discussion of the static
potential between three quarks, thus relevant for baryons,
see~\cite{Alexandrou:2002sn}. This property of \ac{QCD}, in which
particles that interact via the strong force (i.e.~``strongly
interacting'' particles) form colorless bound states from which they
cannot escape, is known as confinement.

\begin{figure}[t]
   \begin{center}
      \includegraphics[width=0.8\linewidth]{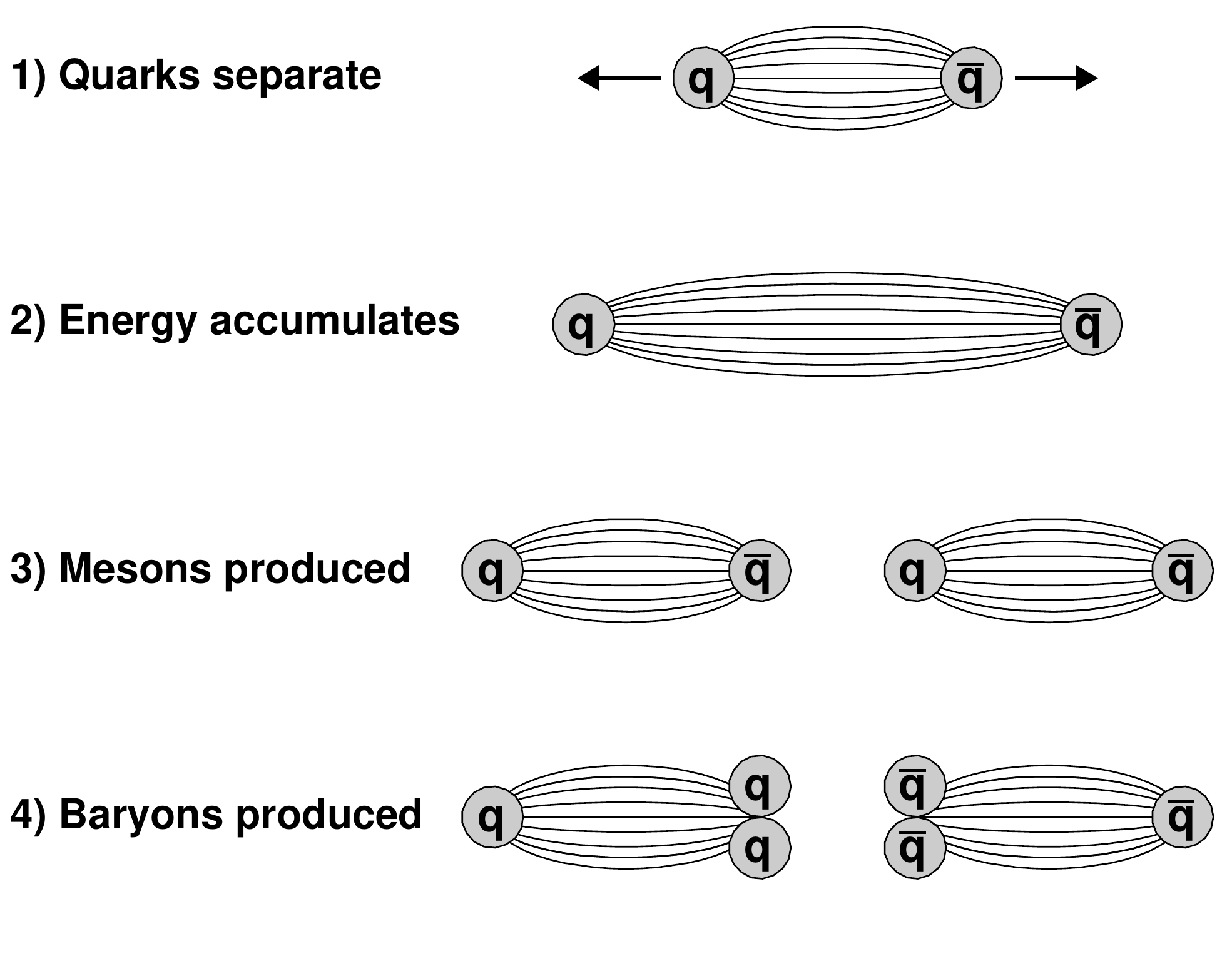}
   \end{center}
   \caption{\label{intro:fig:stringBreak}
      A cartoon of hadron production in the color flux tube, or string
      breaking, model~\cite{Casher:1978wy}. As the quarks separate,
      energy accumulates in the field. When the energy is more than
      double the quark mass, it becomes possible for a $q-\bar{q}$ pair
      to be produced. The produced quarks will bind to the original
      quarks to form two colorless mesons. Baryon production can happen
      in an analogous fashion, through the production of a diquark,
      antidiquark pair~\cite{Schierholz:1981en}.}
\end{figure}

A further implication of the linear form of the potential between two
quarks is revealed by the consideration of the quarks being drawn apart
by some external agent. Because the force between the quarks is
constant, the external agent will need to continuously add energy into
the system to separate the quarks. Eventually, the amount of energy
added will be greater than the rest mass of a quark-antiquark pair. It
would then be possible for this energy to be converted into mass. The
probability of this happening increases with the amount of energy added
to the system~\cite{Casher:1978wy}. Thus, as shown in
\fig{intro:fig:stringBreak}, new quarks would be produced, and all
quarks in the system would remain confined to colorless bound states.
This is an example of \emph{hadronization}, the process by which the
fundamental particles of \ac{QCD} are formed into the hadrons that are
observed in nature.

The confinement of colored objects into colorless hadron states sheds
light on the question of strong forces at macroscopic distances. It can
be shown that the fields generated by a single, free color source would
have an infinite amount of energy~\cite{Stuller:1975im}; thus free
colored states do not exist on their own. Further, as suggested by the
example described above, colored particles cannot be freed from their
colorless bound states; attempts to do so simply result in the
production of more colorless hadrons. Thus, all colored particles remain
confined to colorless bound states, and of course, (color) neutral
particles do not interact when they are separated by distances much
larger than their size. Interactions occur only when these objects are
close enough that their internal composition becomes discernible. For
example, colorless protons and neutrons in the nucleus of one atom of a
molecule, where the separation between nucleons in the nucleus is 
\mbox{$\sim1.5~\fm$~\cite{Vanttinen:1998iz}}, are bound together by the
residual strong force. However, these nucleons do not experience any
appreciable strong force interactions with nucleons in neighboring
atoms, as the separation between atoms is roughly one hundred thousand
times larger than the size of a nucleon.

Thus, in cold nuclear matter, quarks and gluons are confined to
colorless protons and neutrons. It should be noted, however, that the
structure of hadrons is far more interesting than the quark content
shown in \tab{intro:tab:hadrons} might suggest, and is the subject of
ongoing study~\cite{Adams:2003fx} (see \cite{Vetterli:1998tk,
Drechsel:1999qy} for more information). These composite particles in
turn settle into bound states to form nuclei. This matter is ``cold'' by
definition, since it is in its ground state, the state in which the
system stores the least amount of kinetic and potential energy. The
behavior of strongly interacting matter under very different conditions
-- high temperature and high density -- can be studied experimentally by
colliding highly energetic nuclei.

%---------------------------------------------------------------
\section{Heavy Ion Collisions}
\label{intro:hic}
%---------------------------------------------------------------

The theory of \ac{QCD} is thought to provide a valid description of
strongly interacting matter. Unfortunately, calculating the dynamics of
realistic systems using the full theory of \ac{QCD} has proven
exceedingly difficult. Consequently, experimental studies of strongly
interacting matter are valuable not only in their ability to test the
predictions of \ac{QCD}, but in their ability to explore more exotic and
less well understood systems. One of the most effective experimental
techniques that enables such studies is the analysis of collisions of
highly energetic nuclei, i.e.~heavy ions. As two relativistic nuclei
(relativistic in that their kinetic energy is much greater than their
rest mass) collide, a system of strongly interacting matter may be
produced. The dynamics of this system is a subject of intense study. It
is expected that the produced system will consist of densely packed
quarks, antiquarks and gluons.

%---------------------------------------------------------------
\subsection{The Density of Produced Matter}
\label{intro:hic:density}
%---------------------------------------------------------------

The energy density of the system produced in {\AuAu} collisions, in
which each nucleus carries 19.7~{\tev} of kinetic energy\footnote{To put
this in perspective, 19.7~{\tev} is roughly one hundred trillion times
more kinetic energy than that of the average molecule at room
temperature, but is about one hundred billion times \emph{less} kinetic
energy than that of a typical car on a highway.}, can be estimated using
the so-called Bjorken estimate~\cite{Bjorken:1982qr}. The energy density
will, of course, be the amount of energy contained in some volume. The
volume of the system produced in a head-on, or ``central,'' collision
can be approximated as a cylinder, with a transverse area equal to that
of the nuclei, $\mathscr{A}$, and some reasonable length. The length of
the cylinder will clearly increase with time as the produced system
expands, and it is reasonable to believe that the length of the cylinder
will expand much more rapidly than the radius, due to the large initial
longitudinal momenta of the nuclei. Thus, the estimation of the length
of the cylinder is rather an estimate of the \emph{time} at which it is
appropriate to consider the energy density of the system. Since it is
clear that typical length scales of the strong force must be on the
order of the size of the proton, it is natural to estimate the energy
density at \mbox{$\tau_0\sim1\to2~\fmc$}. Finally, it is necessary to
estimate the amount of energy contained in the produced system. This
energy should be directly related to the energy of hadrons emitted by
the system. Since it is the energy in the produced system that is of
interest, and not the energy carried by the initial nuclei, hadrons
produced with a momentum that is roughly transverse to the initial
nuclei should provide the best estimate of the energy of this matter.
Thus, the energy density can be estimated as follows.

\begin{equation}
   \label{intro:eq:bjorked}
\epsilon_0 \approx \frac{\ave{E_T} \dndy}{\tau_0 \mathscr{A}}
   \approx \frac{\sqrt{(0.500~\gev)^2+m_{\pi}^2}\; (700)\; (3/2)}{
      (1 \to 2~\fm)\; \pi\; (6.5~\fm)^2} \approx 4 \to 2~\gev/\fm^3
\end{equation}

\noindent%
where $\ave{E_T}$ is the average transverse energy of emitted hadrons
and $\dndy$ is the number of hadrons in the {\mrap}\footnote{Rapidity
is a variable that measures the longitudinal velocity of a particle
and is convenient for relativistic velocities; see \peq{ana:eq:rap}. A
particle that has a rapidity close to zero in the center of mass frame
of the system is said to be at \emph{\mrap}.} region. It was assumed
that the vast majority of produced particles are
pions~\cite{Veres:2004kc} having an average transverse momentum of
500~{\mmom}~\cite{phobWhitePaper}. It was further assumed that all
three pion species -- positive, negative and neutral -- are produced
in equal number, so that the total number of particles is \mbox{$(3/2)
\times \dnchdy$}, where $\dnchdy$ is the multiplicity (number of
produced particles) of charged hadrons. Using the {\mrap} multiplicity
measured in~\cite{Back:2002uc}, the total number of particles is
\mbox{$(3/2) \times700$}. The radius of the nucleus was taken to be
about 6.5~{\fm}; see \pfig{recon:fig:nucProfs}.  While this estimate
of the energy density of the matter produced in a high energy {\AuAu}
collision is fairly rough, it does imply that the matter is an order
of magnitude greater than that of a gold nucleus,
\mbox{$\sim160~\mev/\fm^3$}, and at least five times as dense as that
of a proton. Approximating the volume of a proton by a sphere with a
radius of 0.8~{\fm}, $\epsilon_{\rm p}~\approx~0.938~\gev /
\sqb{(4/3)\pi(0.8~\fm)^3} \approx~0.450~\gev/\fm^3$. This high density
suggests that the relevant constituents of the matter produced in
these collisions are likely to be the quarks, antiquarks and gluons
themselves, rather than the hadrons which are subsequently
observed. It is an exciting prospect that this loss of ``hadronic
degrees of freedom'' may mean that the strongly interacting matter
produced in these collisions is not confined (at least for some amount
of time) to hadrons, but only to the larger volume of the system as a
whole. However, no direct evidence has been presented, and the topic
of deconfinement remains a subject of extensive
inquiry~\cite{Satz:2000bn, Gazdzicki:2003fj}.

The energy density is an informative property, as it expresses the
density of the system in a way that is independent of the specific
make-up of the matter. However, it is also important to estimate the
relative abundance of baryons and antibaryons in the produced medium.
This is typically done experimentally by measuring the ratio of
antiprotons to protons produced in the {\mrap} region. Antiprotons
generated by the collision will be produced together with protons, via
pair production~\cite{Libby:1978qf}. Because the initial nuclei
consist solely of baryons, and because the number of baryons minus
antibaryons is unchanged in any interaction, the antiproton to proton
ratio is an indication of how many of those initial protons get
``transported'' to {\mrap} during the collision, relative to the number
that are produced.

\begin{equation}
   \label{intro:eq:pbarprat}
\frac{\bar{\rm p}}{\rm p} = \frac{\bar{\rm p}_{\rm produced}}{
   {\rm p}_{\rm produced} + {\rm p}_{\rm transported}}
\end{equation}

Thus, not only is the ratio a measure of how baryon rich the produced
medium is, but also of how effective the collision is at changing the
momenta of the initial nucleons. It is instructive to examine the
extreme values this ratio may assume. A $\pbarprat$ ratio of one implies
that the produced medium is completely baryon free and none of the
original nucleons are transported to the {\mrap} region. The term
`baryon free,' refers to the \emph{net} number of baryons in the system:
the number of baryons in excess of the number of antibaryons. Such a
baryon free system could be created by very distinct scenarios. In the
first, the initial nuclei pass through each other completely, none of the
nucleons are stopped, and the medium produced in the central region is
the result of an excited vacuum caused by the
collision~\cite{Lee:1995cm}. In the second scenario, the nuclei stop
completely upon colliding, creating an extremely dense medium that comes
to equilibrium and then explodes, sending the nuclei receding in
directions opposite to their initial momenta. The third scenario is a
combination of the other two, in which the quarks of the initial nuclei
pass through each other completely, while the gluons stop completely and
form a very dense medium that then explodes, as described
in~\cite{Carruthers:1983xk}. The $\pbarprat$ ratio is not able to
distinguish between these very different models of particle production.
At the opposite extreme, a $\pbarprat$ ratio near zero implies that the
produced medium is intensely baryon rich. That is, essentially no pair
production of baryons occurs, so any protons observed near {\mrap} must
have been transported from the initial nuclei.

\begin{figure}[t]
   \begin{center}
      \includegraphics[width=0.6\linewidth]{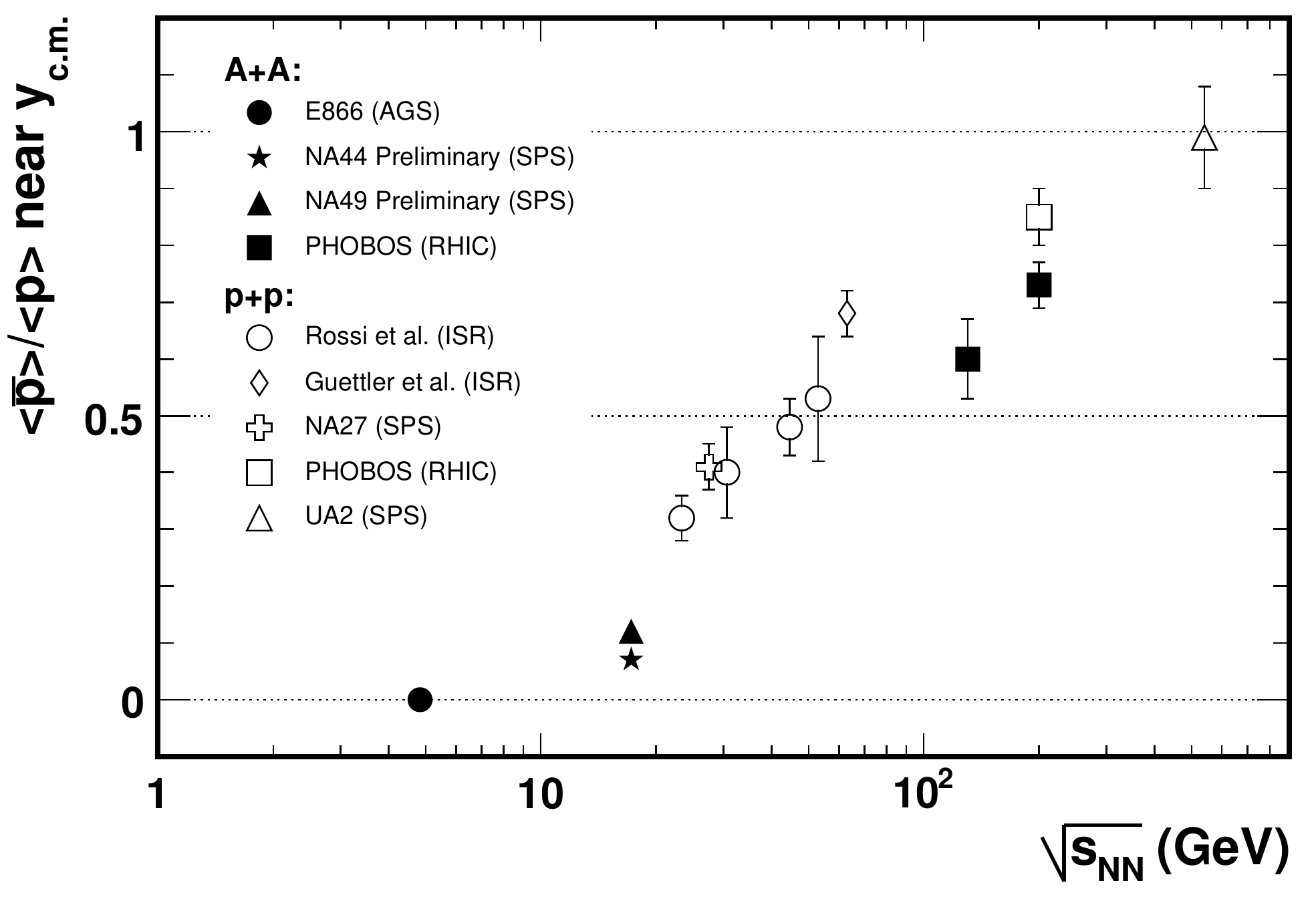}
   \end{center}
   \caption{\label{intro:fig:pbarpRatioSqrtS}
      The $\pbarprat$ ratio measured near {\mrap} in nucleus-nucleus and
      nucleon-nucleon collisions, as a function of energy per nucleon
      pair in the center of mass frame. The plot is
      from~\cite{phobWhitePaper, abbysThesis}. The data are
      from~\cite{Back:2004bk, Back:2001qr, Ahle:1998xz, Ahle:1999in,
      Banner:1983jq, Aguilar-Benitez:1991yy, Bearden:1996qb,
      Bachler:1999hu, Rossi:1974if, Guettler:1976ce}.}
\end{figure}

The ratio of antiprotons to protons observed in the {\mrap} region of
both nucleus-nucleus and nucleon-nucleon collisions is presented as a
function of the collision energy per nucleon pair in
\stolenfig{intro:fig:pbarpRatioSqrtS}{phobWhitePaper, abbysThesis}. This
shows that the $\pbarprat$ ratio rises with collision energy in both
systems. Thus, the higher the kinetic energy of the initial hadrons, the
more important pair production processes become compared to baryon
transport. Additionally, these results show that the net baryon content
of the matter produced in the central region of such collisions
decreases with increasing collision energy.

%---------------------------------------------------------------
\subsection{The Temperature of Produced Matter}
\label{intro:hic:temp}
%---------------------------------------------------------------

It is clear that heavy ion collisions are capable of producing matter
that is both far more dense and far less baryon rich than ordinary
nuclear matter. This begs the question of whether this dense matter,
which is presumably expanding rapidly, has sufficient time to achieve
equilibrium, and if so, does it have a high temperature. This is an
important question, since studies of \ac{QCD} have suggested that a
system of strongly interacting matter that is in equilibrium and is
sufficiently hot may form a phase of matter known as a
\ac{QGP}~\cite{Harris:1996zx, Rischke:2003mt}. Conceptually, this phase
of matter would consist of a soup of quarks, antiquarks and gluons that
are not confined to hadrons. Due to the extremely large temperature,
these elementary particles would have large kinetic energies on
average.  Thus, interactions between them would occur with large
momentum transfers -- or equivalently, at short distances -- and
according to \eq{intro:eq:couplingconst}, such interactions would be
governed by weak coupling. Thus the constituents a \ac{QGP} should be
relatively mobile.  The presence of mobile color charges, quarks, would
be expected to screen any long distance interactions; hence the term
`plasma.' It is this state of matter that is thought to have been
predominant in the very early universe, during the first 10~{\us} after
the big bang, prior to the formation of
hadrons~\cite{Boyanovsky:2006bf}. Whether such a weakly interacting
\ac{QGP} is produced in heavy ion collisions is the subject of ongoing
investigation, and will be discussed in \chaps{rslt}{conc} of this
thesis.

Regardless of the phase of matter produced in heavy ion collisions, it
is valuable to estimate how hot the matter is so that its dynamics can
be described in terms of its temperature. However, direct measurements
of the temperature of the produced matter are not possible. In
macroscopic systems, such a measurement is typically performed by
allowing the system in question to come to thermal equilibrium with a
second system whose behavior as a function of temperature is well
understood, i.e.~a thermometer, and observing the temperature of the
second system. Clearly, such a technique cannot be applied to the matter
produced by a heavy ion collision. However, inspiration can be drawn
from astronomy, where the spectrum of light radiated by a star is used
to determine its surface temperature~\cite{Arny:stars}.

In an analogous fashion, the spectrum of particles emitted by the
matter produced in a heavy ion collision can be used to estimate its
temperature. Unfortunately, the particles measured by a heavy ion
experiment will not be the same particles that were radiated by the
initially produced medium. This fact can be understood by following
the evolution of the heavy ion collision. The beginning of the
collision can be taken to be the point at which the distance between
the nuclei is much smaller than their longitudinal size, but before
any particle production has occurred. The energy density is at its
maximum at this point, but is not particularly interesting, since it
does not describe any produced matter. During some time following
this, substantial particle production will occur. Given what is known
about the typical strength of \ac{QCD} interactions, the time for this
particle production is expected to be roughly $\lsim1~{\fmc}$. How
particles are actually produced is in principle described by \ac{QCD},
however the difficulty of carrying out calculations in the full theory
has lead to the creation of a host of more simplified models. After
this early period, the produced quarks, antiquarks and gluons may come
to thermal equilibrium. It is at this point that the energy density
estimate of \eq{intro:eq:bjorked} is applicable, and at which a
\ac{QGP} may be formed. As this dense system of strongly interacting
matter expands, it will cool and the elementary particles will form
hadrons. Again, \ac{QCD} should in principle describe the process of
hadronization, but as it involves long-range interactions,
calculations using the full theory have not yet been developed and
simplified hadronization models are preferred. At some point during
the expansion of this hadron gas, inelastic collisions between hadrons
will cease, and the relative yield of particle species, such as the
$\pbarprat$ ratio, will be completely determined. This stage of the
collision is therefore known as chemical freeze-out. Finally, as this
gas of hadrons expands further, even elastic collisions will cease.
At this point, the momentum spectrum of each particle species is
determined, since the particles will not interact further before being
detected by an experimental apparatus. Therefore, this stage of the
collision is known as thermal freeze-out.

From this picture, it is clear that a temperature which is {\naive}ly
extracted from the spectrum of hadrons observed in an experiment will
not describe the temperature of the very dense medium formed early in
the collision. Instead, this temperature would describe the system as it
was at thermal freeze-out. Therefore, to estimate the temperature of
interest, it would be necessary either (a)~measure the spectra of
particles that reach thermal freeze-out at much earlier times in the
collision, such as photons, or (b)~to model the full collision process
such that the final measured spectra could be described in terms of the
properties of the initial dense medium. The spectra of penetrating
probes such as photons are the subject of study,
see~\cite{Adler:2005ig}, for example. However, there is presently no
model of heavy ion collisions which is able to predict all aspects of
the collision in a coherent manner. Instead, simpler models that
describe certain aspects of the collision can be used to extract the
temperature of the produced particle system during a certain stage of
the collision.

These models are typically based on hydrodynamics or statistical
mechanics. One such hydrodynamical model is known as a
blast-wave~\cite{Retiere:2003kf}, since it assumes thermal emission of
particles from a rapidly expanding shell. This model can be used to
extract the temperature of the system at the point of thermal
freeze-out. Typical freeze-out temperatures given by this model for
central {\AuAu} collisions at a center of mass energy of 39.4~{\tev}, or
200~{\gev} per nucleon pair, are ${\rm T_{\rm
f.o.}}\sim110~{\mev}$~\cite{Trzupek:2005ef, KiyomichiThesis}. It should
be noted that the temperature parameter of this model is highly
anti-correlated with the velocity at which the shell is expanding.

\begin{figure}[t]
   \begin{center}
      \includegraphics[width=0.6\linewidth]{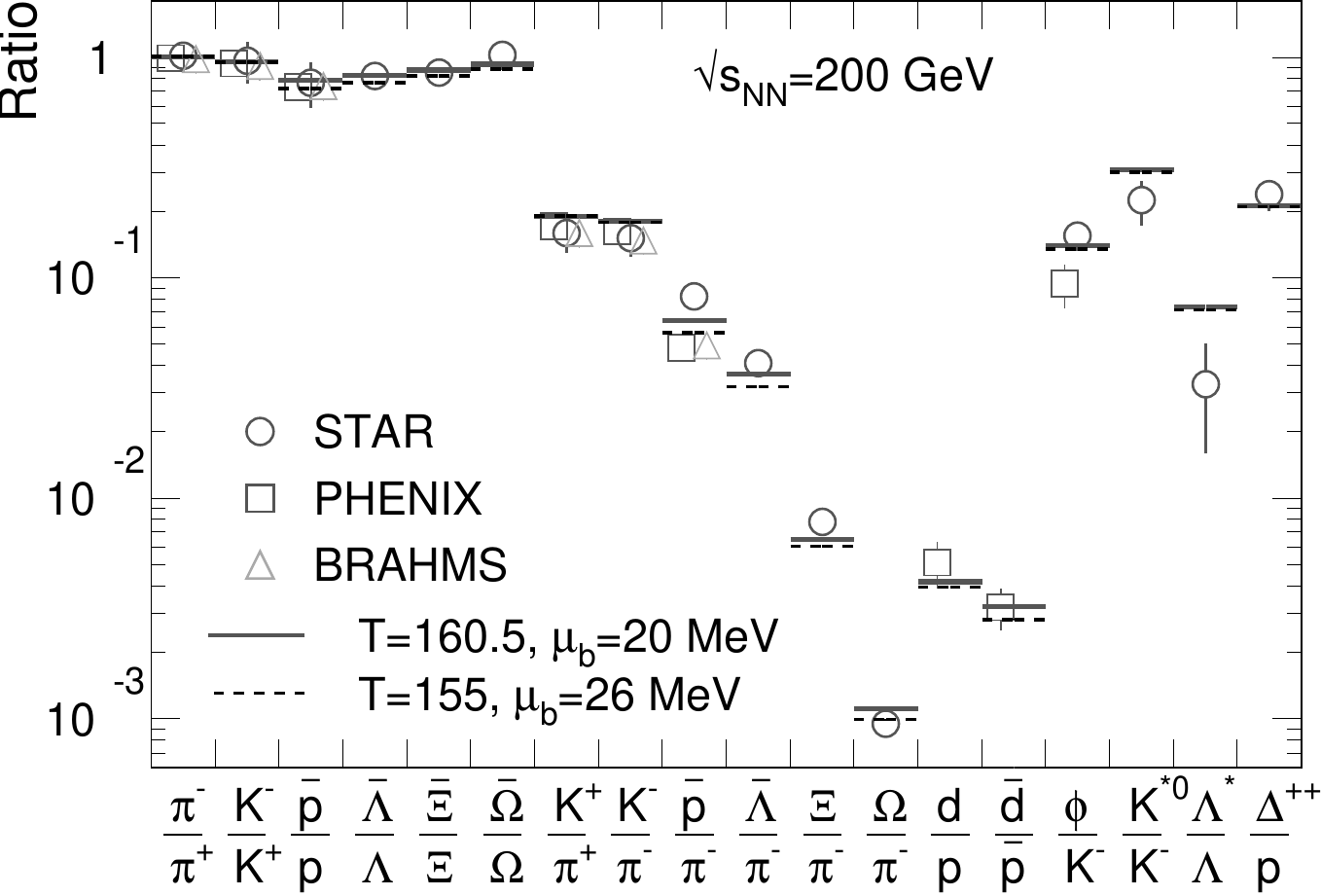}
   \end{center}
   \caption{\label{intro:fig:ratios_s200all}
      Relative hadron yields measured in {\AuAu} collisions at a center
      of mass energy of 200~{\gev} per nucleon pair. The statistical
      model was used to fit these ratios to extract the temperature at
      chemical freeze-out. The three right-most ratios, involving
      resonances, were not included in the fits. The dashed line shows
      the fit to the data excluding the $\pbarpirat$ and $\phikrat$
      ratios~\cite{Andronic:2005yp}.}
\end{figure}

Statistical models are better suited to describe the collision at the
chemical freeze-out stage~\cite{Becattini:2002en}. One such model uses
the grand canonical ensemble to estimate the yield of various particle
species. In this statistical ensemble, one envisions having a large set
of systems, each of which is identically prepared and is in equilibrium
with some external bath of energy and particles. These systems would
then be allowed to evolve, and the properties of the \emph{average}
system are determined. Thus, in the grand canonical ensemble, the number
of particles of each type is allowed to vary. Physical rules that
constrain the number of particles, for example the conservation of the
number of net baryons in the system, are obeyed only on average. These
models display impressive success at fitting the relative yields of a
wide variety of particle species with a single set of parameters. The
temperature of the matter produced in {\AuAu} collisions at a center of
mass energy of 200~{\gev} per nucleon pair at chemical freeze-out, as
estimated by one such model, is shown in
\stolenfig{intro:fig:ratios_s200all}{Andronic:2005yp}.

While the success of these models seems to suggest that the matter
produced in high energy heavy ion collisions was able to reach
equilibrium, it should be noted that statistical models are also able to
describe hadron production in more elementary collisions, where no
thermalized medium is expected~\cite{Becattini:1995if,
Becattini:1997rv}. Nevertheless, the results of these models seem to
present a consistent picture. At the latest stage of the collision,
thermal freeze-out, the system seems to have a temperature of
\mbox{$\sim110~\mev$}. Before the system had ceased inelastic
interactions, at chemical freeze-out, its temperature was higher,
$\sim160~\mev$. These results imply that the medium produced in the
early stages of the collision, prior to chemical freeze-out, should be
at an even higher temperature. 

\begin{figure}[t]
   \centering
   \subfigure[Statistical Models]{
      \label{intro:fig:temperatureRootS}
      \includegraphics[width=0.4\linewidth]{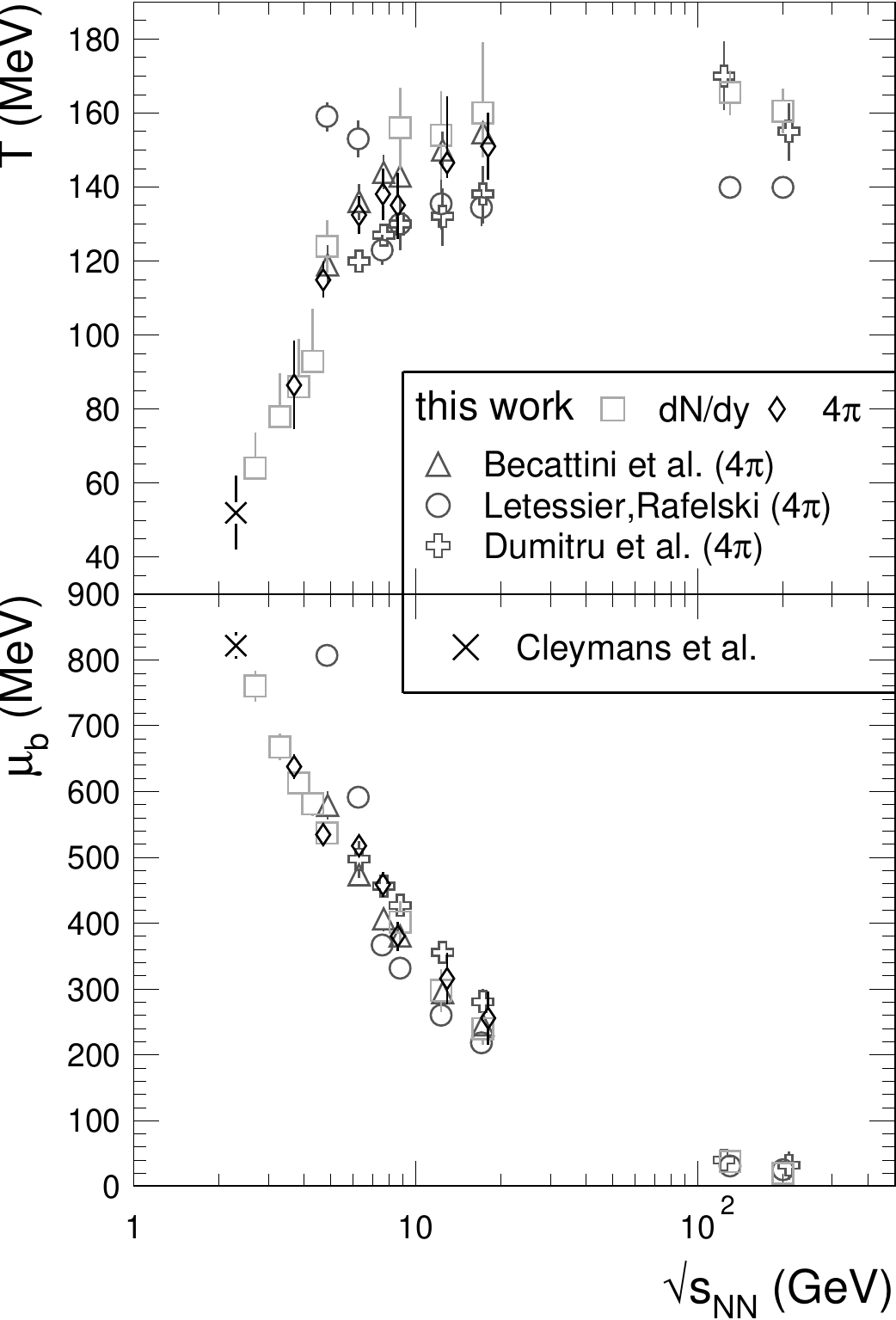}
   }
   \hspace{0.05\linewidth}
   \subfigure[Parametrization]{
      \label{intro:fig:temperatureRootSfit}
      \includegraphics[width=0.4\linewidth]{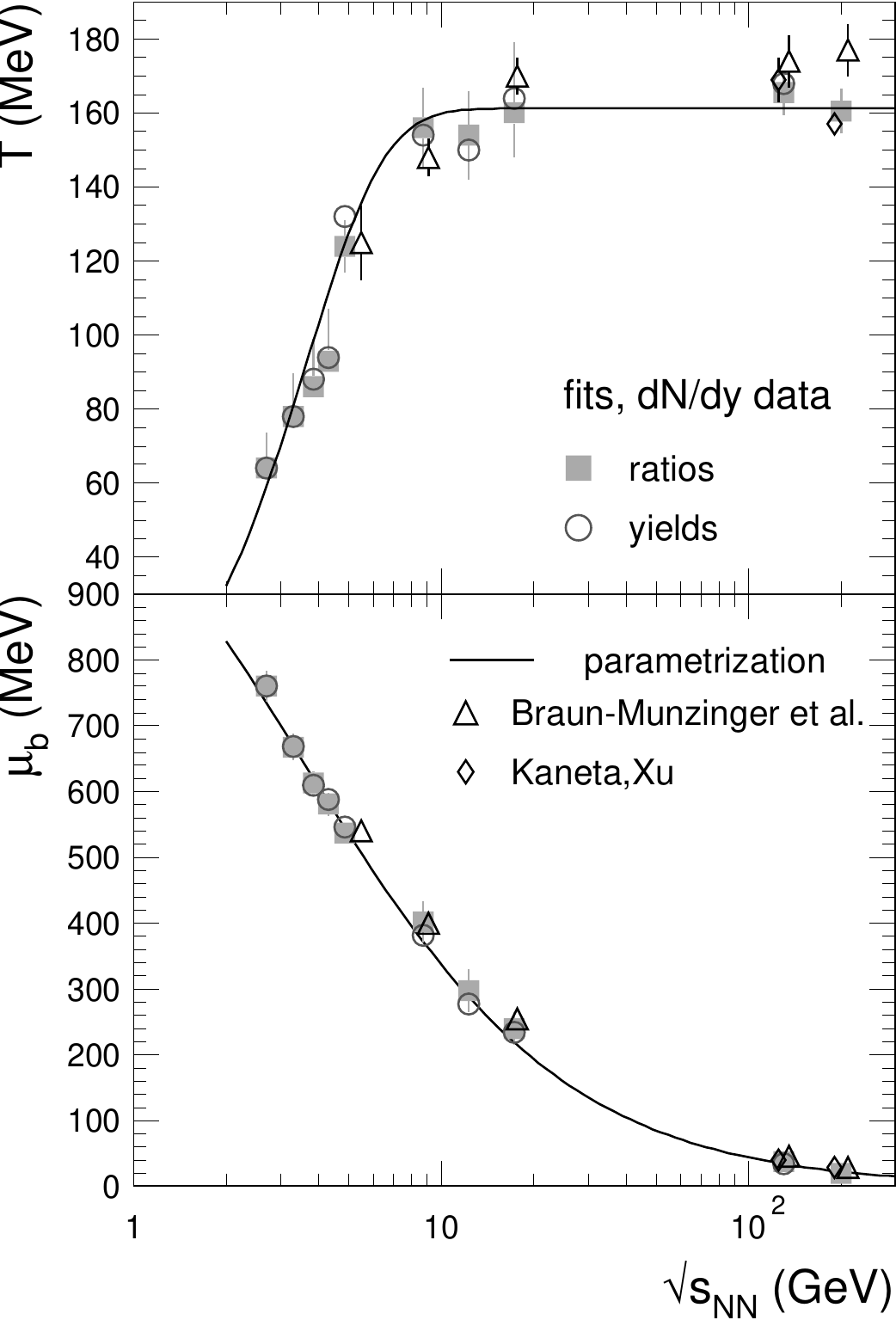}
   }
   \caption{   \label{intro:fig:tempRootS}
      \subref{intro:fig:temperatureRootS}~The chemical freeze-out
      temperature and baryon chemical potential (see text) of heavy ion
      collisions extracted using statistical models are shown as a
      function of collision energy.
      \subref{intro:fig:temperatureRootSfit}~The energy dependence is
      parametrized for the models described in~\cite{Andronic:2005yp}.
      Figures taken from~\cite{Andronic:2005yp}.}
\end{figure}

The temperature and baryon chemical potential of heavy ion collisions,
extracted using statistical models, is presented as a function of
collision energy per nucleon pair in
\stolenfig{intro:fig:tempRootS}{Andronic:2005yp}. The baryon chemical
potential, $\mub$, is related to the net baryon density of the produced
matter. It gives the amount by which the energy of the system would
change if a baryon were added. In this way, it describes the tendency of
the system to favor baryon production over antibaryon production; a
system with more baryons than antibaryons would have a large $\mub$, as
it would cost more energy to add a baryon than an antibaryon. The
temperature at chemical freeze-out is seen to increase with collision
energy sharply at first, and then very moderately for
$\snn\gsim10~\gev$, as shown in \fig{intro:fig:temperatureRootSfit}. The
parameterizations used to fit the temperature and $\mub$ dependence are
given in~\cite{Andronic:2005yp}. A limiting temperature of
$161\pm4~\mev$ was extracted from the fit. See~\cite{Andronic:2005yp}
and~\cite{Braun-Munzinger:2001as} for a discussion of the energy
dependence of the temperature extracted from statistical models.

%---------------------------------------------------------------
\subsection{Some Collision Models}
\label{intro:hic:models}
%---------------------------------------------------------------

As seen in \figs{intro:fig:pbarpRatioSqrtS}{intro:fig:tempRootS}, the
matter produced in heavy ion collisions at high energy has a large
energy density, low net baryon content and a high temperature compared
to the matter produced in lower energy collisions. To learn more about
this matter, it is necessary to have theoretical calculations of the
collisions dynamics to which experimental results can be compared. Due
to the difficulty of calculations in \ac{QCD}, it is more practical to
build a model from a set of coherent assumptions, which can be used to
simulate the collision dynamics. Two such models were used for the
analysis presented in this thesis.

The \acf{HIJING}~\cite{Gyulassy:1994ew} model is built upon the idea
that in high energy heavy ion collisions, multiple \emph{minijet}
production will be important. A jet is a collection of closely
correlated particles produced in elementary
collisions~\cite{Webber:1983if}. The hadrons in a jet are thought to be
generated from a single quark or gluon that was produced with a large
transverse momentum. A minijet is a jet whose transverse momentum is not
extremely large compared to the average $\pt$ of produced particles. In
the presence of a hot, dense medium, the width of a jet may be
broadened, and a minijet may be completely
absorbed~\cite{Kajantie:1987pd}. \ac{HIJING} can be used to test this
idea, as it includes both minijet production and energy loss of
particles in the produced medium. The production of low $\pt$ particles,
and the production of hadrons from quarks and gluons, are modeled via
string breaking pictures (see \fig{intro:fig:stringBreak}). Some nuclear
effects, such as shadowing~\cite{Kumano:1992ef}, are also included.

The other collision simulation package used extensively in this analysis
is known as \acf{AMPT}~\cite{Zhang:1999bd} model. This model uses
\ac{HIJING} to calculate the position and momentum of each parton (a
parton is either a quark, antiquark or gluon) immediately following the
collision. \ac{AMPT} does not include the continuous minijet energy loss
modeled in \ac{HIJING}; instead \ac{AMPT} models the individual
re-scatterings of partons, using \acf{ZPC}~\cite{Zhang:1997ej} model. The
partons are then hadronized using a modified version of the string
fragmentation model in \ac{HIJING}. Finally, interactions between
hadrons are simulated using \ac{ART}~\cite{Li:1995pr} model, which is
modified to include inelastic interactions between nucleons and
antinucleons, as well as between kaons and antikaons.

%---------------------------------------------------------------
\subsection{Overview}
\label{intro:hic:overview}
%---------------------------------------------------------------

Models such as \ac{HIJING} and \ac{AMPT} provide a phenomenological
description of heavy ion collisions under a certain set of assumptions.
How valid these assumptions are, and how accurate the descriptions turn
out to be, can only be determined by observing nature. The {\phob}
experiment at the \acf*{RHIC} used heavy ion collisions to shed light on
the dynamics of strongly interacting matter under extreme conditions. It
has been shown that this matter is both far more dense and far more hot
than normal nuclear matter. The analysis presented in this thesis will
examine the production of charged hadrons by both {\dAu} and
nucleon-nucleus collisions, to serve as a reference for nucleus-nucleus
interactions.

In the absence of any nuclear or produced medium effects, a
nucleus-nucleus collision may be interpreted as a superposition of
independent binary nucleon collisions. As will be discussed in
\chap{rslt}, a significant suppression relative to binary collision
scaling was observed in the high-$\pt$ hadron production of central
{\AuAu} collisions at $\snn=200~{\gev}$. This result led to two
competing hypothesis. One held that the initial nuclei were modified
such that high-$\pt$ hadron production was reduced. The other maintained
that hadron production was unchanged, but that particles produced in the
collision lose momentum as they travel through a dense, strongly
interacting medium. These ideas were tested using {\dAu} interactions,
which were expected to include any initial modification of the gold
nucleus, but were not expected to produce a dense medium. The absence of
any significant suppression of hadron production in {\dAu} interactions
led to the acceptance of the hypothesis of partonic energy loss.

The analysis presented in this thesis will examine the validity of using
{\dAu} collisions in place of nucleon-nucleus interactions as a
reference for {\AuAu}. This study was performed by measuring the charged
hadron spectra of {\dAu}, {\pAu} and {\nAu} collisions. The
nucleon-nucleus interactions were extracted from the {\dAu} data by
identifying the deuteron spectators. Two calorimeters were used to
measure the deuteron spectators. One detected forward-going single
neutrons and the other, installed just prior to the {\dAu} physics run
at \acs*{RHIC}, detected forward-going single protons. It will be shown
that the hadron production of {\pAu} and {\nAu} collisions can be
combined to form an ideal reference for {\AuAu} interactions. In
addition, the yield of positive and negative hadrons in {\pAu}
collisions will be compared to that of {\nAu}. This comparison allows
unique study of the ability of a nucleon-nucleus interaction to
transport the initial ``projectile'' nucleon to {\mrap}.

Finally, the hadron production of {\dAu} collisions will be examined as
a function of centrality (i.e.~impact parameter). Different measures of
centrality will be employed, and the effects of centrality determination
on the final measurement will be explored. These centrality measures
will be used to study the shape of the charged hadron spectrum in {\dAu}
collisions. As will be discussed in \sect{rslt:shape}, the production of
hadrons having a transverse momentum of \mbox{$\sim2.5~\gev$} is known
to be enhanced in nucleon-nucleus collisions over nucleon-nucleon
interactions. The centrality dependence of this enhancement will be
studied, and an intriguing scaling behavior will be revealed.

%% file: srcs/chapter1.tex
% $Id: chapter1.tex,v 1.50 2006/09/05 02:20:50 cjreed Exp $
%

%---------------------------------------------------------------
\chapter{The {\phob} Experiment}
\label{exp}
%---------------------------------------------------------------

\myupdate{$*$Id: chapter1.tex,v 1.50 2006/09/05 02:20:50 cjreed Exp $*$}%
The analysis presented in this thesis used data obtained by the {\phob}
experiment. {\phob} was one of four heavy ion experiments installed at
the \acf{RHIC}. It studied the strong interaction by observing
nucleon-nucleon, nucleon-nucleus and nucleus-nucleus collisions. The
collisions were provided by \acs{RHIC} over a broad range of
center-of-mass energies.

%---------------------------------------------------------------
\section{The Relativistic Heavy Ion Collider}
\label{exp:RHIC}
%---------------------------------------------------------------

The \acf{RHIC} at \acf{BNL} was designed to collide heavy ions,
primarily gold nuclei, at energies up to $200~\gev$ per nucleon pair in
the center of mass frame ($\snn$). Four experiments were built at
\ac{RHIC} to study collisions of nuclei: BRAHMS, PHENIX, {\phob} and
STAR. The BRAHMS experiment used its movable spectrometer arm to study
particle production at different values of {\prap}. The PHENIX detector
measured hadrons, photons and electrons and included two large muon
hodoscopes. The STAR detector featured a large barrel \ac{TPC} in a
solenoidal magnet which could observe almost all charged particles at
{\mrap}. The {\phob} experiment is described in \sect{exp:phobdet}.

\begin{figure}[t]
   \begin{center}
      \includegraphics[width=0.5\linewidth]{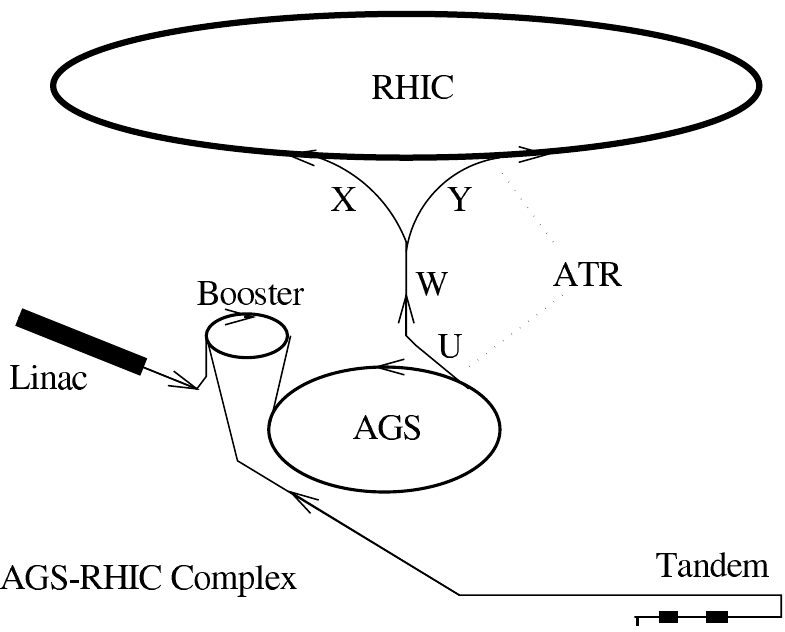}
   \end{center}
   \caption{\label{exp:fig:rhic}
      A schematic of the \protect\acs{RHIC} layout. $\Au^{+77}$ ions exit
      the \protect\acs{AGS} at \emph{U}, are stripped of electrons at
      \emph{W} and $\Au^{+79}$ ions are sent into two separate rings at
      \emph{X} and \emph{Y}~\cite{Wei:1997fq}.}
\end{figure}

Generating collisions of heavy ions required sophisticated machinery to
(a) produce ions by stripping electrons from atoms, (b) accelerate the
ions to nearly the speed of light and (c) steer the beams of ions to
cross paths and collide.  \acs{RHIC} made use of much of the existing
accelerator infrastructure at \ac{BNL} to strip electrons and accelerate
ions. A new facility was constructed to complete the electron stripping,
accelerate the ions to the final desired momentum and to collide the ion
beams. A diagram of the complete \acs{RHIC} facility can be seen in
\stolenfig{exp:fig:rhic}{Wei:1997fq}.

Gold ions were first put into motion at the Tandem Van~de~Graaff
accelerator. A pulsed sputter ion source produced gold ions with a net
charge of negative one. The Tandem generated two static electric
potential differences in sequence, one of $+15~{\Mvolt}$ and one of
$-15~{\Mvolt}$. $\Au^{-1}$ ions were first accelerated by the positive
potential at the center of the machine and then passed through a thin
foil which removed on average 13 electrons. The $\Au^{+12}$ ions were
then repelled and thereby accelerated by the positive potential and
passed through another stripping foil.

Positively charged $\Au^{+32}$ ions were then sent into the \ac{AGS}
complex. They were first accelerated by the Booster synchrotron from
1~{\mev} per nucleon up to 95~{\mev} per nucleon. As the ions exited the
Booster, they were further stripped of electrons. $\Au^{+77}$ then
entered the \ac{AGS}, where they were accelerated to \acs{RHIC}
injection energy of 10.8~{\gev} per nucleon. Beams were stored in the
\acs{AGS} until the \acs{RHIC} rings could be filled. The ions were then
stripped of the final two electrons, and $\Au^{+79}$ ions entered the
\acl{RHIC}.

To store two beams of positive ions, \acs{RHIC} consisted of two
independent rings. The rings had a circumference of 3.8~{\km} and
provided six interaction regions where the beams crossed. \acs{RHIC}
performed the final acceleration of the ions up to the full collision
energy, a maximum of 100~{\gev} per nucleon for gold nuclei. At this
energy, a magnetic field of 3.458~{\tesla} was required to keep the ions
traveling in a circle. This field was provided by superconducting dipole
magnets, each of which was 9.46~{\m} long and was cooled by liquid
Helium to a temperature of 4.2~{\kelvin}.

Particles in the beams traveled through the rings in bunches of
\mbox{$\sim10^9$} ions. This allowed the bunches to be accelerated by
successive ``kicks'' from radio-frequency electromagnetic fields. The
longitudinal positions at which a bunch could ride the r.f. wave were
referred to as buckets. Since not all buckets contained a bunch of ions,
the \ac{RHIC} facility kept track of which buckets were filled. A
crossing clock was used to inform experiments of the times at which
filled buckets would collide.

The beams were brought into alignment for collisions by two types of
\acs{RHIC} magnets, the D0 and DX dipoles. Four D0-magnets, two on each
side of the \ac{IP}, brought the beams close enough that they could
share a single beam pipe. The D0-magnets provided a field of
3.52~{\tesla} for this purpose. The beams were then steered to collide
by two DX-magnets, one on each side of the \ac{IP}, each of which
generated a field of 4.3~{\tesla}.

During the 2003 physics run, \acs{RHIC} delivered deuteron-gold
collisions with an integrated luminosity of 75~{\inbarn}. The {\phob}
experiment recorded a total of 146~million {\dAu} collisions to tape.
\acs{RHIC} also provided {\AuAu} collisions at $\snn=$19.6~{\gev},
55.9~{\gev}, 62.4~{\gev}, 130~{\gev} and 200~{\gev}; polarized {\pp}
collisions at $\snn=$200~{\gev} and 410~{\gev}; and {\CuCu} collisions
at $\snn=$22.5~{\gev}, 62.4~{\gev} and 200~{\gev}. The energy and
species scans that \acs{RHIC} provided were a striking display of the
machine's capabilities, and have been an invaluable source of
information in the quest to understand how strongly interacting matter
behaves under varying conditions. See~\cite{Harrison:2002es} for an
in-depth discussion of the \ac{RHIC} facility.

%---------------------------------------------------------------
\section{{\phob} Detector Setup}
\label{exp:phobdet}
%---------------------------------------------------------------

\begin{figure}[t]
   \begin{center}
      \includegraphics[width=\linewidth]{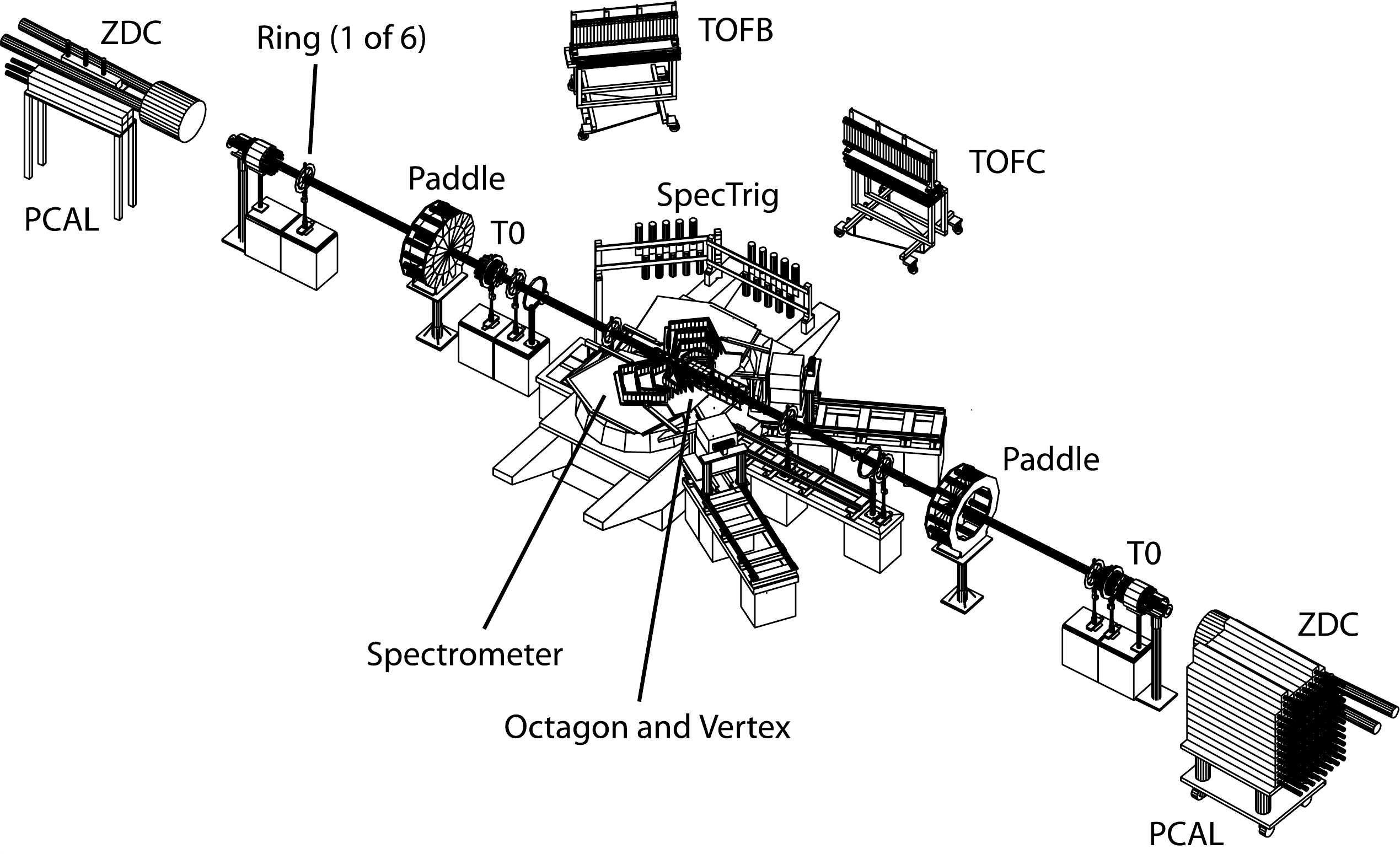}
   \end{center}
   \caption{   \label{exp:fig:phobos}
      The {\phob} detector setup used to take deuteron-gold collision
      data. The proton calorimeters (see \sect{exp:phobdet:calo:pcal})
      are drawn to scale, but are located roughly 3 times further from
      the interaction point than shown.}
\end{figure}

The {\phob} experiment was designed to study global properties of the
collisions and to search for previously unknown signals of new physics.
These design goals established the need for several essential detector
pieces. To observe global properties of collisions, a multiplicity
detector array was constructed which could observe nearly all particles
produced in a collision. To investigate the collision dynamics and
search for signals of new physics, a two-arm spectrometer was assembled
which could study a relatively small number of particles in great
detail. To help determine the impact parameter of collisions, detectors
were added to measure the amount of nuclear material which did not
participate in the collision. Finally, to ensure that new and rare
physics signals were not overlooked, a triggering and data acquisition
system was built which could collect collision data at a high rate. A
diagram of the {\phob} detector used during deuteron-gold ({\dAu})
collision data taking is shown in \fig{exp:fig:phobos}.

%---------------------------------------------------------------
\subsection{The Multiplicity and Vertex Detectors}
\label{exp:phobdet:mult}
%---------------------------------------------------------------

The {\phob} multiplicity detector covered nearly the full 4$\pi$ solid
angle.  The bulk of produced particles were measured by the Octagon -- a
single layer of silicon pad sensors which form an octagonal barrel
around the beam pipe. Particles emerging from the interaction point at
very high {\prap} were measured by the Rings -- six separate rings of
silicon pad detectors oriented perpendicular to the beam. These
detectors were used to count the number of produced particles, to
classify collisions according to their centrality and to determine the
collision vertex.

In addition to the silicon detectors described below, the {\phob}
experiment benefited greatly from its beam pipe. Of the four heavy ion
experiments at \acs{RHIC}, {\phob} was unique in its use of 12~meters of
beryllium beam pipe. Three sections of beryllium beam pipe were
installed, each of which was 4~{\m} in length, 76~{\mm} in diameter and
about 1~{\mm} in thickness. The use of a thin beryllium pipe over the
entire length of the {\phob} multiplicity detector allowed a clean
multiplicity measurement even at high values of {\prap}, where low
particles with low transverse momentum would have been much more likely
to scatter off of a steel beam pipe.

%---------------------------------------------------------------
\subsubsection{The Octagon}
\label{exp:phobdet:mult:oct}
%---------------------------------------------------------------

The Octagon was used to detect particles over a wide range of azimuthal
angle\footnote{Azimuthal angle refers to the angle in the plane
transverse to the beam line.} and {\prap} values. It consisted of a
light-weight aluminum frame which could support eight rows of up to
thirteen silicon pad sensors. Each row of sensors formed one face of
the octagonal barrel detector. In four of the faces, sensors were
removed such that the acceptance of the Octagon did not overlap the
acceptance of the Vertex or Spectrometer detectors. The Octagon was
1.1~{\m} long with a diameter of 90~{\mm} and could measure particles
having {\prap} $\abs{\eta}<3.2$ produced by collisions within 10~{\cm} of
the nominal interaction point. In addition to the silicon sensors
themselves, the frame supported the electronics used to readout signals
from the sensors, including a water cooling system that removed heat
generated by the electronics.

\begin{figure}[t!]
   \begin{center}
      \includegraphics[width=0.8\linewidth]{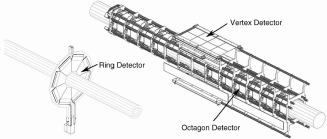}
   \end{center}
   \caption{   \label{exp:fig:multdiag}
      The {\phob} silicon multiplicity detectors~\cite{phobosNIM}.}
\end{figure}

All silicon sensors used in the Octagon were 84~{\mm} long by 36~{\mm}
wide, with a grid of four by thirty active pads. Like all silicon
sensors in \phob, sensors in the Octagon had a thickness that was 
within 5\% of 300~\um. Each pad was 2.708~{\mm} long (in the beam
direction) and 8.710~{\mm} wide. Despite the relatively large pads,
sensors in the Octagon achieved a signal-to-noise ratio of
$S/N\approx13/1$. The Octagon is shown in
\stolenfig{exp:fig:multdiag}{phobosNIM}.

%---------------------------------------------------------------
\subsubsection{The Rings}
\label{exp:phobdet:mult:rings}
%---------------------------------------------------------------

Particles emitted at very forward angles were detected by the Rings.
Each individual Ring was a set of eight trapezoidal silicon sensors,
placed side-by-side to form a disk. The Rings were oriented transverse
to the beam direction, so that the trajectory of particles with large
longitudinal momentum would be nearly perpendicular to the face of the
detector. The Rings and readout electronics were supported by
carbon-fiber frames, to ensure that inactive material around the
detectors was low-Z and would not be a large source of secondary
particles. There were six Ring detectors in all, placed $\pm1.13~\m$,
$\pm2.35~\m$ and $\pm5.05~\m$ from the \ac{IP}. These detectors observed
particles having {\prap} $3\leq\abs{\eta}\leq4$, $4\leq\abs{\eta}\leq4.7$
and $4.7\leq\abs{\eta}\leq5.4$, respectively. The inner radius of the
Rings was 10~\cm, with each Ring sensor extending out 12~{\cm} radially.

All silicon sensors used in the Rings had 64 active pads, arranged in
eight radial rows and eight azimuthal columns. Unlike the rectangular
sensors in other silicon detectors, pads in a single Ring sensor were
not equally sized. Rather, each pad had the same acceptance in {\prap},
$\Delta\eta\approx0.1$, and azimuth, $\Delta\phi\approx2\pi/64$. Thus,
pad sizes ranged from about 3.8~{\mm} in the azimuthal direction by
5.1~{\mm} in the radial direction for pads near the beam pipe, to about
10.2~{\mm} by 10.2~{\mm} for pads at the outer edge of the Ring. The
average signal-to-noise ratio of Ring sensors was comparable to that of
the Octagon. A diagram of a ring detector is shown in
\stolenfig{exp:fig:multdiag}{phobosNIM}.

\begin{figure}[t!]
   \begin{center}
      \includegraphics[angle=270,width=0.7\linewidth]{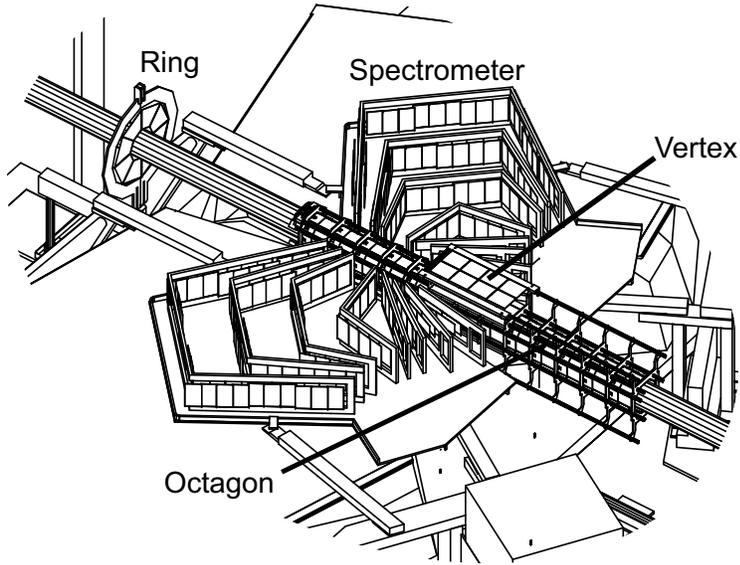}
   \end{center}
   \caption{   \label{exp:fig:specdiag}
      The {\phob} silicon detectors located close to the nominal
      interaction point. The top of the {\phob} magnet has been removed
      to make the detector pieces visible~\cite{phobWhitePaper}.}
\end{figure}

%---------------------------------------------------------------
\subsubsection{The Vertex Detector}
\label{exp:phobdet:mult:vtx}
%---------------------------------------------------------------

The Vertex detector was designed to provide accurate vertex resolution,
better than 0.2~{\mm} along the beam direction for collisions within
10~{\cm} of the \ac{IP}, in the high-multiplicity environment of a
central {\AuAu} collision. To achieve this, two double-layer detectors
were constructed, one placed above the beam line and one below, both
centered around the nominal interaction point. This arrangement allowed
the collision position to be reconstructed by identifying two-point
tracks that point to a common vertex in each of the double-layer
detectors. While this was the preferred method for {\AuAu} collisions, in
low-multiplicity environments such as \dAu, a different vertex
reconstruction method was used (see~\sect{recon:vertex}).

Two types of silicon sensors were used in the Vertex detector. The
layers closest to the beam line (56~{\mm} in the vertical direction),
known as the Inner Vertex, consisted of four sensors placed side-by-side
in the beam direction. Each sensor had a grid of 128 pads in the beam
direction by 4 pads in the transverse direction. Pads in the Inner
Vertex were 0.473~{\mm} long in the beam direction and 12.035~{\mm} wide.
The layers further from the beam line (118~{\mm} in the vertical
direction), known as the Outer Vertex, consisted of two rows of sensors,
each row having four sensors placed side-by-side in the beam direction.
Each of these sensors had a grid of 128 pads in the beam direction by 2
pads in the transverse direction. Pads in the Outer Vertex were
0.473~{\mm} long in the beam direction and 24.070~{\mm} wide. Sensors in
the Vertex detector achieved a better signal-to-noise ratio than those
in the Octagon or Rings. The top Outer Vertex can be seen in
\stolenfig{exp:fig:multdiag}{phobosNIM}.

%---------------------------------------------------------------
\subsection{The Spectrometer Detectors}
\label{exp:phobdet:spec}
%---------------------------------------------------------------

Whereas the multiplicity and vertex detectors were used to observe the
bulk of particles produced in a collision, {\phob} employed its
spectrometer to study in detail a small number of produced particles.
Particle momentum was determined by tracking the particle's motion
through a magnetic field. Tracking was performed using a two-arm silicon
pad spectrometer that was situated in a 2~T magnetic field. Particle
mass could then be determined using energy loss in the silicon and/or
time-of-flight information. A \acf{TOF} wall was used to obtain the
particle's speed.

%---------------------------------------------------------------
\subsubsection{The {\phob} Magnet}
\label{exp:phobdet:spec:mag}
%---------------------------------------------------------------

The {\phob} magnet made it possible to identify the sign of the electric
charge and the momentum of particles produced near {\mrap}. For purposes
of particle tracking, it was desirable to have a low-field region close
to the interaction and a high-field region further from the collision.
Particle trajectories in the low-field region were essentially
straight. These straight tracks were used as seeds from which the full
path taken by a particle through the high-field region could be
reconstructed. In addition, a near-zero field very close to the beam
helped to ensure stable storage of the beams inside the collider.

\begin{figure}[t!]
   \centering
   \begin{minipage}[c]{0.5\linewidth}
      \centering
      \subfigure[Physical Magnet]{
         \label{exp:fig:physmagnet}
         \includegraphics[width=0.95\textwidth]{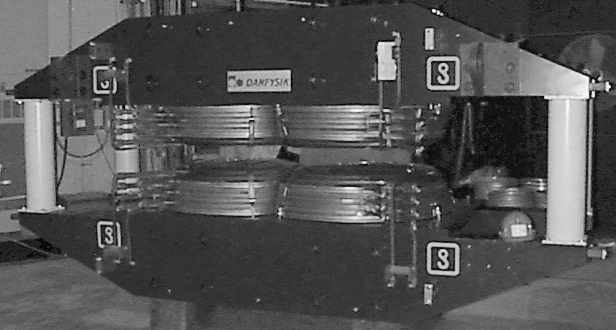}
      }
   \end{minipage}%
\hspace{0.05\linewidth}%
   \begin{minipage}[c]{0.4\linewidth}
      \centering
      \subfigure[Magnetic Field]{
         \label{exp:fig:magfield}
         \includegraphics[width=0.95\textwidth]{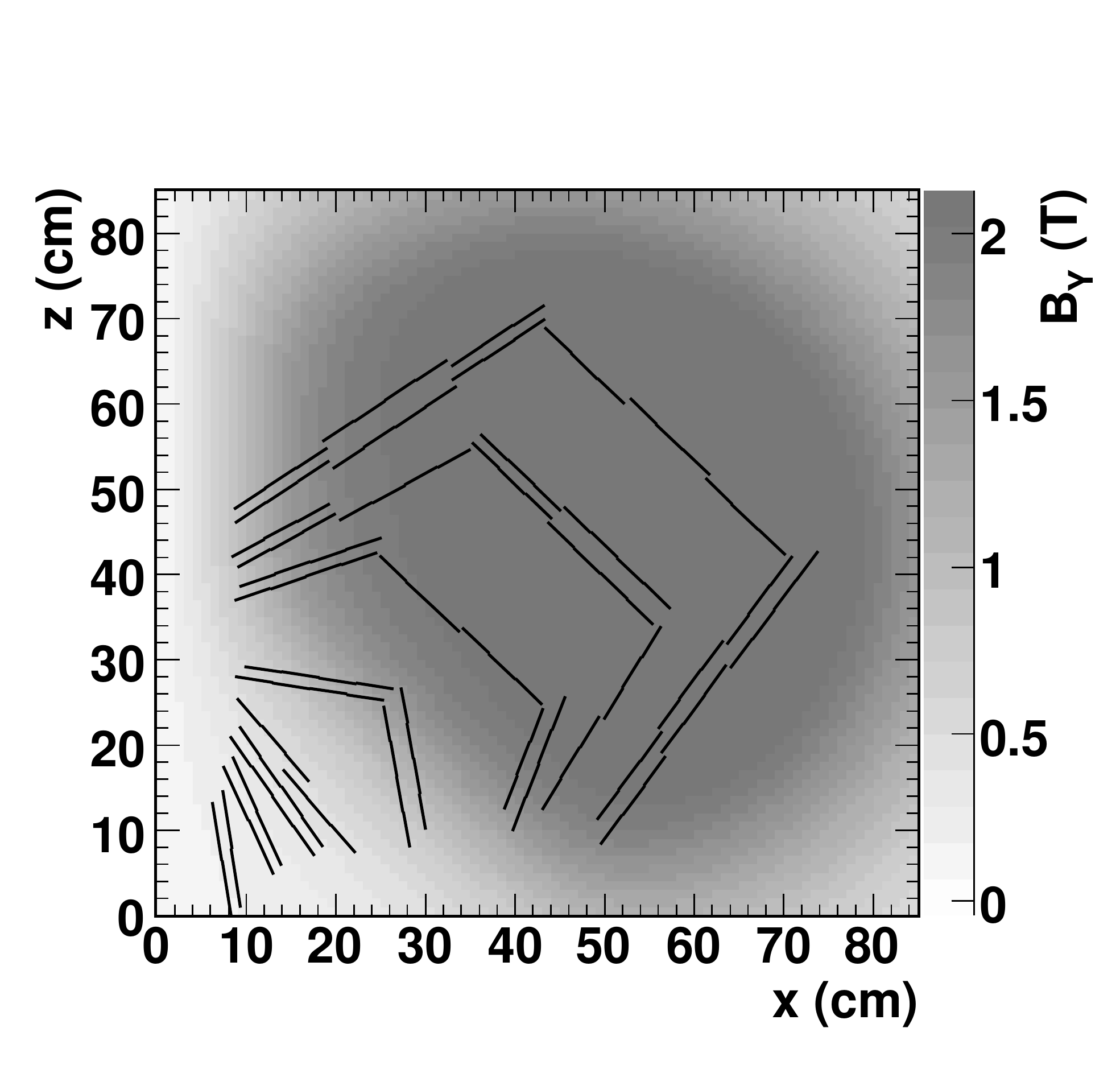}
      }
   \end{minipage}%
   \caption{   \label{exp:fig:magnet}
      The {\phob} magnet. \subref{exp:fig:physmagnet}~A picture of the
      magnet before installation. \subref{exp:fig:magfield}~The
      vertical component of one of the dipole fields, with Spectrometer
      shown, as viewed from the bottom-up.}
\end{figure}

Such a field was achieved by a double-dipole magnet that functioned at
room temperature~\cite{patricksThesis}. The two dipole fields were
produced using four copper coils, two just above each Spectrometer arm
and two just below. The vertical gap between the pole tips, into which
the Spectrometer was placed, was 158~{\mm} when the magnet was off. The
four coils were of cylindrical ``double-taper'' design, but with two
cuts to provide the desired field shape. One cut was vertical and one
was at 12{\dg} to vertical. The coils were supported by a steel flux
return yoke, two pole support plates, four support columns and an
adjustable magnet stand. Under full power and full magnetic field, the
yoke deflection was about 2~{\mm}. The physical design of the {\phob}
magnet is shown in \fig{exp:fig:physmagnet}.

The coils were energized using a refurbished \ac{AGS} power supply that
provided up to 3600~{\amps} at 115~{\vdc}. The coils were driven in a
series-parallel configuration, with the top two coils in one series
electrical circuit and the bottom two coils in another circuit. This
produced two dipole fields with opposite polarity. The voltage drop
across each coil was $0.0264~\Omega\times1800~\amps=47.5~\volts$, so the
total voltage drop for the series circuit was 95~{\volts}. Heat
generated by resistance in the conductors was dissipated by water
cooling. The maximum field strength in the vertical direction was
2.18~{\tesla}, while the field components in the two horizontal
directions were less than 0.05~{\tesla}. A map of the vertical field
strength is shown in \fig{exp:fig:magfield}. It was possible to invert
the vertical component of the magnet field by reversing the flow of
electric current through the coils. The field orientation was said to be
positive, ``B$+$,'' when the field covering the outer-ring Spectrometer
arm was directed upward. This is the orientation shown in
\fig{exp:fig:magfield}. During data taking, {\phob} attempted to record
an equal number of collisions with each magnet polarity.

%---------------------------------------------------------------
\subsubsection{The Silicon Spectrometer}
\label{exp:phobdet:spec:si}
%---------------------------------------------------------------

The motion of particles produced near {\mrap} was tracked by the
Spectrometer, a two-arm silicon pad detector. The spectrometer arms were
centered on the beam line vertically, located on opposite sides of the
beam and supported inside the {\phob} magnetic field
(see~\sect{exp:phobdet:spec:mag}). Each spectrometer arm consisted of 16
layers of silicon sensors, as shown in
\stolenfig{exp:fig:specdiag}{phobWhitePaper}. The arms were not centered
at {\mrap}, but rather extended slightly forward. Sensor layers in the
Spectrometer were mounted on aluminum frames which were themselves
mounted on large carrier plates. The carbon-epoxy carrier plates were
designed to minimize vibrations caused by changes in the magnet current.
These plates were supported on rails, allowing the Spectrometer arms to
be mounted outside the magnet and then slid into place.

\begin{figure}[t]
   \begin{center}
      \includegraphics[width=0.5\linewidth]{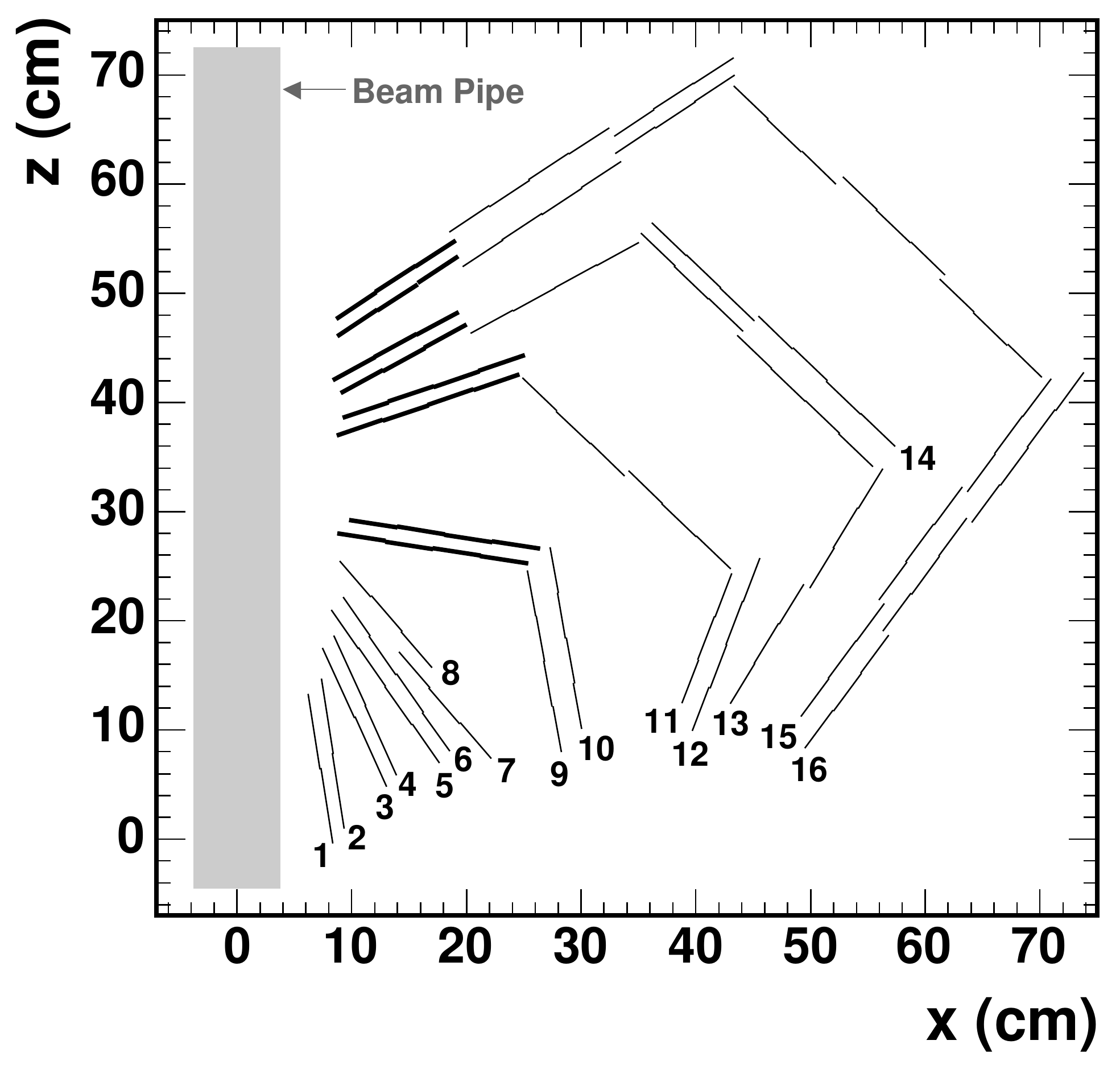}
   \end{center}
   \caption{   \label{exp:fig:specpos}
      Location of sensors in one arm of the {\phob} spectrometer, as
      viewed from the bottom-up. Layers are labeled by number and
      type~3 sensors are shown with thick lines.}
\end{figure}

The Spectrometer contained five different types of silicon sensors.
Sensors positioned closest to the interaction were the most
finely-grained sensors; in general, pad size increased as distance from
the interaction became greater. This strategy was employed to reduce the
cost (and difficulty) of silicon production at the expense of decreased
resolution in the particle's azimuthal angle. Details on the size and
placement of each sensor type used in the Spectrometer can be found in
\tab{exp:tab:siSpec}. Spacial positions of the sensors are presented in
\fig{exp:fig:specpos}. The horizontal width of pads in type~2 sensors
are smaller than those in type~1 sensors to make it possible to detect
small deflections in particle trajectory when entering the non-zero
region of the magnetic field. The width of pads in sensor types~3-5 are
smaller than those in type~1 to improve momentum resolution.

\begin{table}[t]
   \begin{center}
      \begin{tabular}{|llll|}
\hline
Sensor Type &  Number of Pads &         Pad Size  &          Sensor Placement \\
            &  (horiz.$\times$vert.) &  (\mm$\times$\mm) &   (layers) \\
\hline
1 &            $70\times22$ &           $1.000\times1.0$ &   1-4 \\
2 &            $100\times5$ &           $0.427\times6.0$ &   5-8 \\
3 &            $64\times8$ &            $0.667\times7.5$ &   9-16, near beam \\
4 &            $64\times4$ &            $0.667\times15.0$ &  9-12 \\
5 &            $64\times4$ &            $0.667\times19.0$ &  13-16 \\
\hline
      \end{tabular}
   \end{center}
   \caption{   \label{exp:tab:siSpec}
      Physical description of silicon sensors in the Spectrometer.}
\end{table}

%---------------------------------------------------------------
\subsubsection{The Time-of-Flight Wall}
\label{exp:phobdet:spec:tof}
%---------------------------------------------------------------

\begin{figure}[t!]
   \begin{center}
      \includegraphics[width=0.5\linewidth]{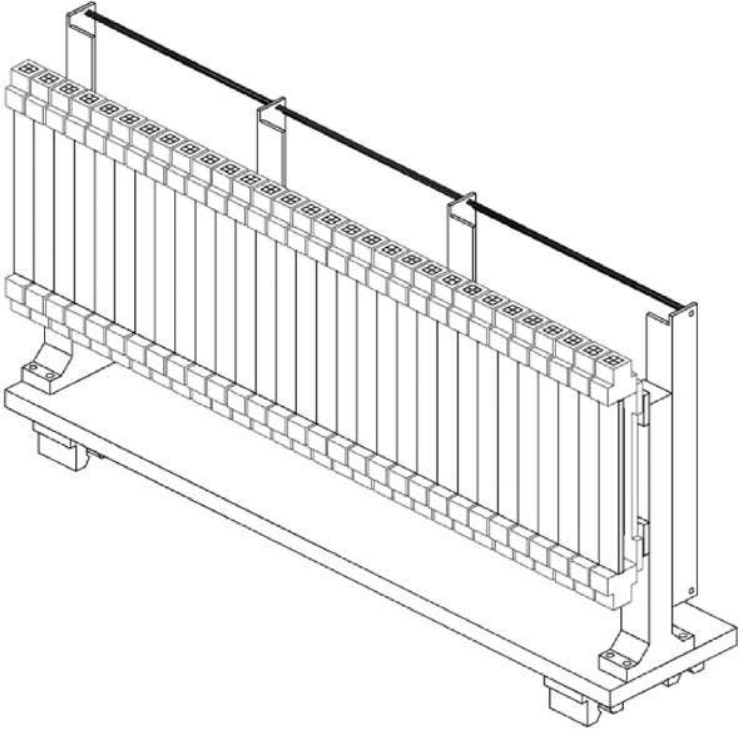}
   \end{center}
   \caption{   \label{exp:fig:tofwall}
      The physical layout of one \protect\acs{TOF} wall~\cite{phobosNIM}.}
\end{figure}

{\phob} used a \acf{TOF} wall composed of two sections to extend its
particle identification capability. Each section of the wall consisted
of 120~scintillators, giving the complete wall a {\prap} coverage of
$0<\eta<1.24$. As the wall was located on the inner-ring side of the
beam, particle identification using the \ac{TOF} wall was only possible
for particles passing through the corresponding spectrometer arm. Prior
to the {\dAu} physics run, the \ac{TOF} wall was moved away from the beam
as far as possible, to extend particle identification capabilities
for particles with higher transverse momentum. A schematic of one
\ac{TOF} wall is shown in \stolenfig{exp:fig:tofwall}{phobosNIM}.

The scintillator material was chosen to have good timing (1.8~{\ns} decay
constant), a reasonable attenuation length (8 times the scintillator
length) and an emission wavelength (408~\nm) that would ensure good
response from the \acp{PMT}. Each scintillator had a cross sectional
area of $8~\mm\times8~\mm$ and a height of 200~\mm. Scintillator light
was detected by \acp{PMT} with a fast rise time of 1.8~{\ns} and a high
gain of $10^{6}$. Anode signals from the \acp{PMT} were split; one
signal was sent to a leading-edge discriminator located close to the
tubes while the other was sent to an \ac{ADC} module after passing
through a 400~{\ns} delay cable. The discriminator signal was sent to a
\ac{TDC} module that featured a sensitivity of 25~{\ps} per digital
channel. Pulse height and timing information were used together to
perform slewing corrections and improve the overall timing resolution of
the \ac{TOF} wall.

The \acp{PMT} were attached at both the top and bottom of the \ac{TOF}
wall, which allowed the vertical position at which a particle struck the
wall to be determined. The position resolution, as measured using cosmic
rays and radioactive sources, was found to be 10~{\mm} for reconstruction
based on time difference and 37~{\mm} for reconstruction based on the
ratio of pulse heights. Timing resolution for the \ac{TOF} wall in data
was better than 100~{\ps}.

%---------------------------------------------------------------
\subsection{The Calorimeters}
\label{exp:phobdet:calo}
%---------------------------------------------------------------

Since nuclear collisions in the collider occurred with varying impact
parameter, it was necessary to have physical observables which were
correlated with the impact parameter of a collision. The number of
nucleons leaving the collision with near-beam rapidity is an example of
one such observable, since such nucleons should not have directly
participated in the collision. {\phob} measured the energy, rather than
the number, of such nucleons. The energy of these \emph{spectator}
neutrons was measured using the \acf{ZDC}. The energy of spectator
protons was measured using the \acf{PCAL}.

%---------------------------------------------------------------
\subsubsection{The Zero-Degree Calorimeters}
\label{exp:phobdet:calo:zdc}
%---------------------------------------------------------------

The \acf{ZDC} detectors were used to detect free neutrons (those not
bound in a nuclear fragment such as an alpha particle) emerging from a
collision with near-beam rapidity. While such neutrons were not
deflected by the collision itself, their trajectories could be affected
by the breakup of the nucleus. At \acs{RHIC}, such evaporation neutrons
diverged from the beam axis by less than 2~\mrad. The \acp{ZDC} were
10~{\cm} wide and located roughly 18~{\m} from the \ac{IP}, so that they
covered a horizontal deflection of about $2.7~\mrad$. There were two
\acp{ZDC}, one on each side of the collision. Each calorimeter consisted
of 3~\ac{ZDC} modules.

\begin{figure}[t!]
   \includegraphics[width=0.9\linewidth]{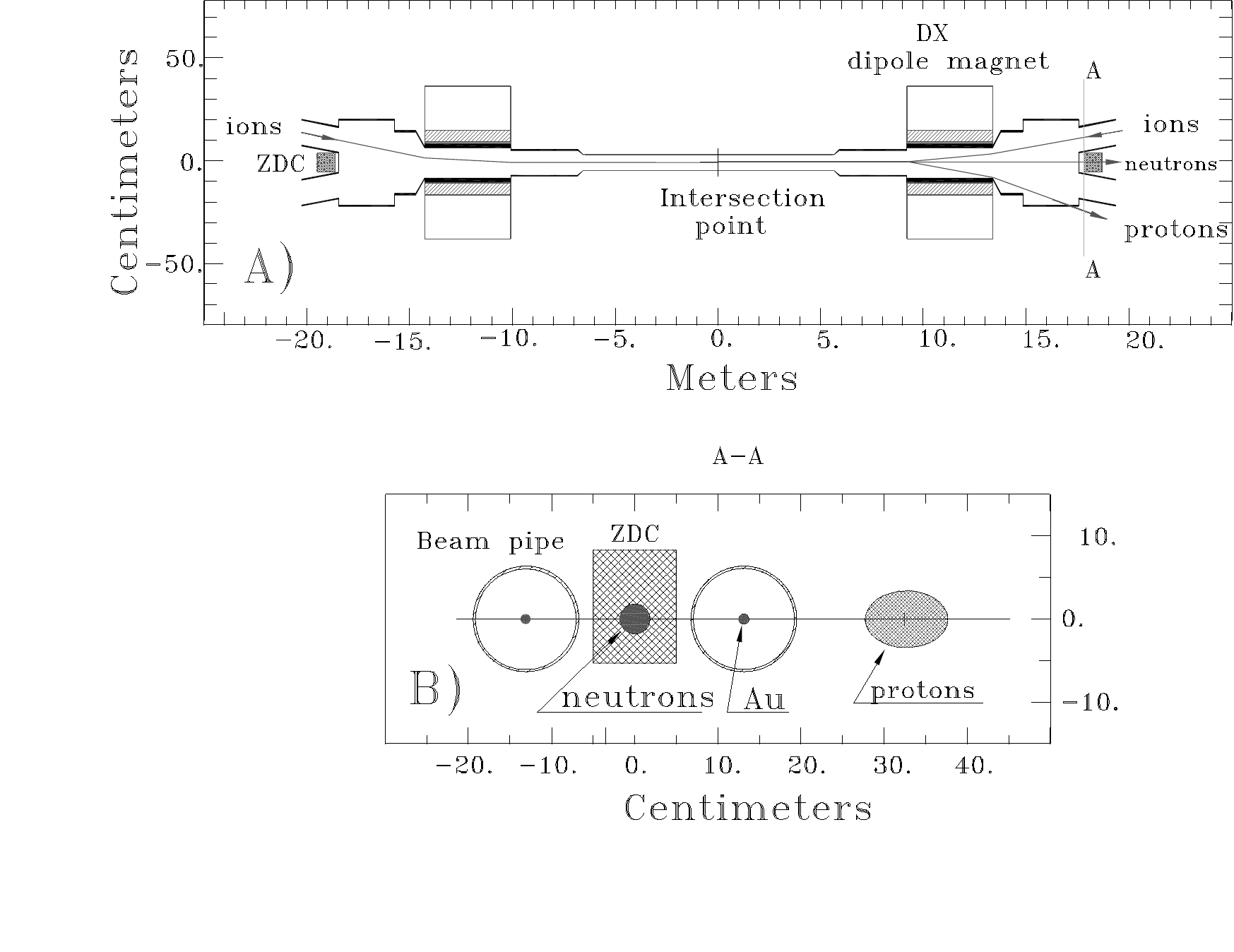}
   \caption{   \label{exp:fig:zdcPos}
      Bird's-eye view (top) and ``beam's-eye'' view (bottom) of the
      collision region, showing the location of the \protect\acsp{ZDC}
      and the paths taken by gold ions, protons and
      neutrons~\cite{Adler:2000bd}.}
\end{figure}

The \ac{RHIC} accelerator magnets were exploited to remove all charged
particles from the acceptance of the \ac{ZDC}. As the beams emerged from
the interaction region, they passed through the \acs{RHIC} DX-magnets,
which were used to bend the two beams back into their separate pipes.
These magnets have the desirable side-effect of causing the region
between the \acs{RHIC} beam pipes, past the DX-magnets, to be
inaccessible to charged particles emerging from the interaction region.
The \ac{ZDC} detectors were placed in this ``zero-degree'' region, where
any produced or secondary particles would deposit negligible energy
compared to that of spectator neutrons. The design of the \ac{ZDC}
modules was restricted by the space limitations imposed by their
position in the experiment.
See~\stolenfig{exp:fig:zdcPos}{Adler:2000bd} for a schematic of the
\ac{ZDC} positioning and the paths taken by charged particles through
the DX-magnets.

\begin{figure}[t!]
   \begin{center}
      \includegraphics[width=0.7\linewidth]{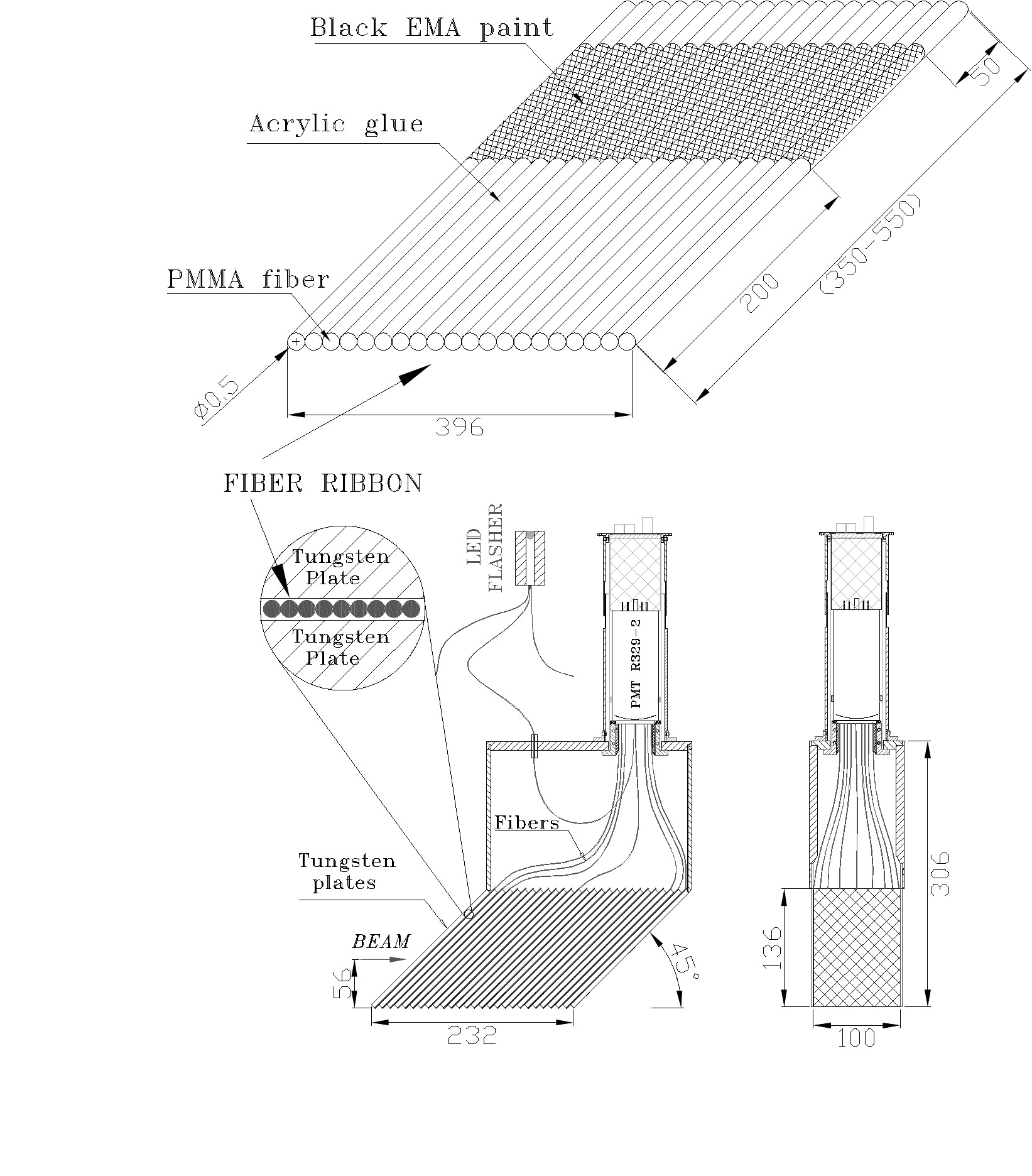}
   \end{center}
   \caption{   \label{exp:fig:zdcDiag}
      Construction design of a \protect\acs{ZDC} module. Dimensions
      shown are in millimeters~\cite{Adler:2000bd}.}
\end{figure}

The calorimeters were designed to minimize the loss in energy resolution
due to shower leakage. They were not designed to provide information
about the transverse position of the neutrons. The \acp{ZDC} did not use
scintillating material to generate light, rather optical fibers were
employed in which shower secondaries produced {\cheren} light. Early
simulations had shown that a detector using {\cheren} light could
observe more of the shower signal than a detector using scintillator.
Each \ac{ZDC} module consisted of optical fibers sandwiched between
27~Tungsten alloy plates, with each plate having dimensions of
$100~\mm\times187~\mm\times5~\mm$. Tungsten alloy was chosen as the
stopping material since simulations showed that it allowed more of the
shower signal to be detected than Lead would have. The plates were
oriented 45{\dg} from vertical, to roughly coincide with the {\cheren}
angle of light emitted by particles going near the speed of light in the
fibers. The {\cheren} light from one module was detected by a single
photomultiplier tube, which gave both signal strength and timing
information. The energy resolution of a full \ac{ZDC} (3~modules) was
found to be $\sigma_{E}/E=19\%$ and the timing resolution was found to
be better than 200~{\ps}. The timing resolution was sufficient to allow
these detectors to be used as a minimum bias collision trigger during
{\AuAu} physics runs. A diagram of the \ac{ZDC} construction is shown
in~\stolenfig{exp:fig:zdcDiag}{Adler:2000bd}.

%---------------------------------------------------------------
\subsubsection{The Proton Calorimeters}
\label{exp:phobdet:calo:pcal}
%---------------------------------------------------------------

Before the {\dAu} physics run, {\phob} installed additional calorimeters
to compliment and extend its ability to measure forward-going nuclear
fragments. Whereas the \ac{ZDC} detectors collected energy from
spectator neutrons, the \acf{PCAL} detectors measured the energy of free
spectator protons. The \ac{PCAL} detectors were not symmetric during the
{\dAu} run; a large calorimeter was installed on the Au-exit side of the
interaction region and a small calorimeter was placed on the d-exit
side. The \ac{Au-PCAL} was used for obtaining a measurement of collision
centrality, while the \ac{d-PCAL} was used to identify collisions in
which the proton from the deuteron did not participate. A large fraction
of the design and assembly of the \ac{PCAL} detectors as well as the
entirety of the commissioning, monitoring, calibration (see
\sect{calib:pcal}), and analysis (see \chaps{recon}{ana}) was completed
as part of this thesis work.

\begin{figure}[p]
   \centering
   \subfigure[Physical Layout]{
      \label{exp:fig:pcalLayout}
      \includegraphics[angle=90,width=0.95\textwidth]{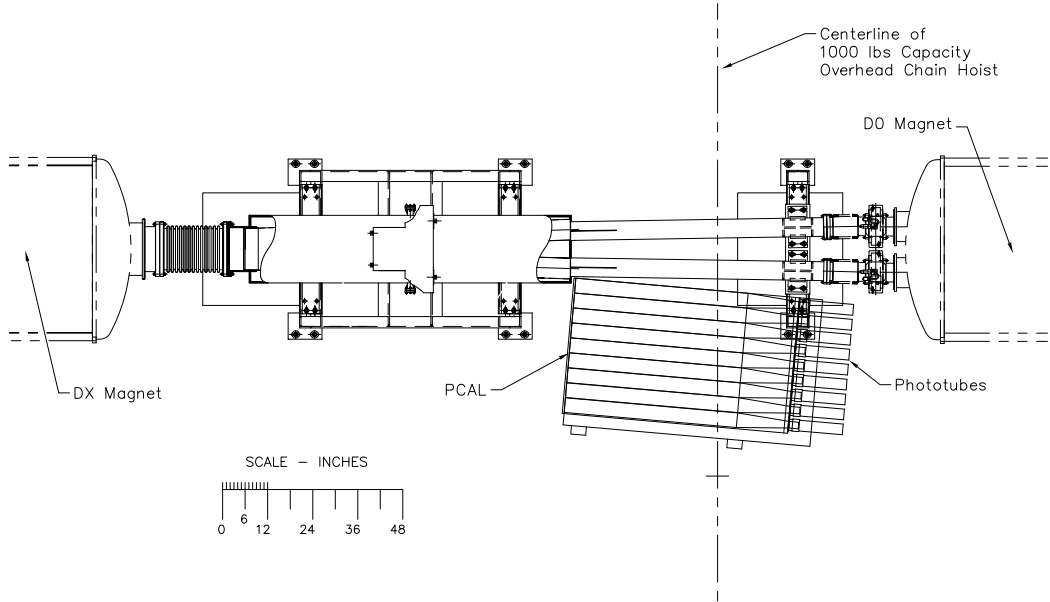}
   }
   \subfigure[Proton Trajectories]{
      \label{exp:fig:protpaths}
      \includegraphics[angle=180,width=0.95\textwidth]{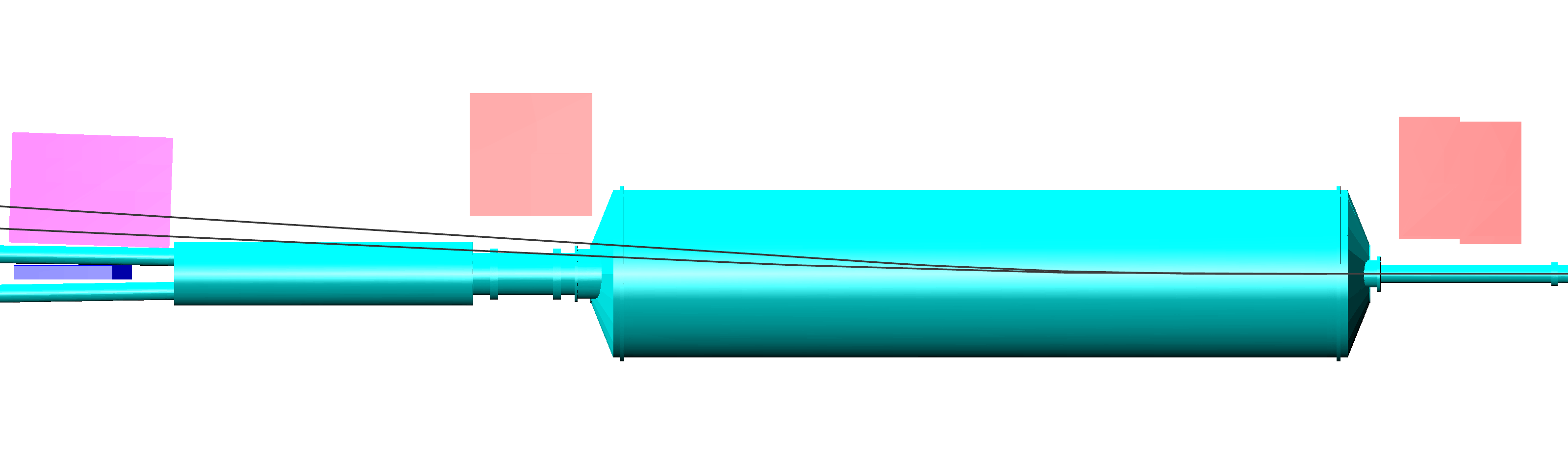}
   }
   \caption{   \label{exp:fig:pcal}
      Top views of the {\phob} Proton Calorimeter.
      \subref{exp:fig:pcalLayout}~The physical location of the
      \protect\acs{PCAL}. The \protect\acs{PCAL} Shielding and the
      \protect\acs{ZDC} are not shown. \subref{exp:fig:protpaths}~The
      black lines show the paths taken by 100~{\mom} and 50~{\mom} protons
      as they are bent by the DX~Magnet. Also shown are the
      \protect\acs{PCAL} Shielding blocks and the \protect\acs{ZDC}.}
\end{figure}

Like the \acp{ZDC}, the \ac{PCAL} detectors also exploited the \ac{RHIC}
DX-magnets. These magnets were designed to bend beams of gold nuclei
into their separate beam pipes. Since a gold nucleus has a rigidity of
$Z/A=79/197=0.4$, while a single proton has a rigidity of 1, the
DX-magnets deflected beam-rapidity protons twice as much as they
deflected gold nuclei. This extra deflection was sufficient to kick
protons completely out of the \ac{RHIC} beam pipes and into the
\ac{Au-PCAL}. See~\fig{exp:fig:protpaths} for a comparison of the
paths taken by gold nuclei and protons. A similar situation occurred on
the deuteron-exit side of the interaction, since a deuteron has a
rigidity of 0.5, comparable to that of a gold nucleus. The rough
equivalence in rigidity was a major motivation for \acs{RHIC} to
perform {\dAu} collisions rather than {\pAu}.

\begin{figure}[t]
   \begin{center}
      \includegraphics[width=0.65\linewidth]{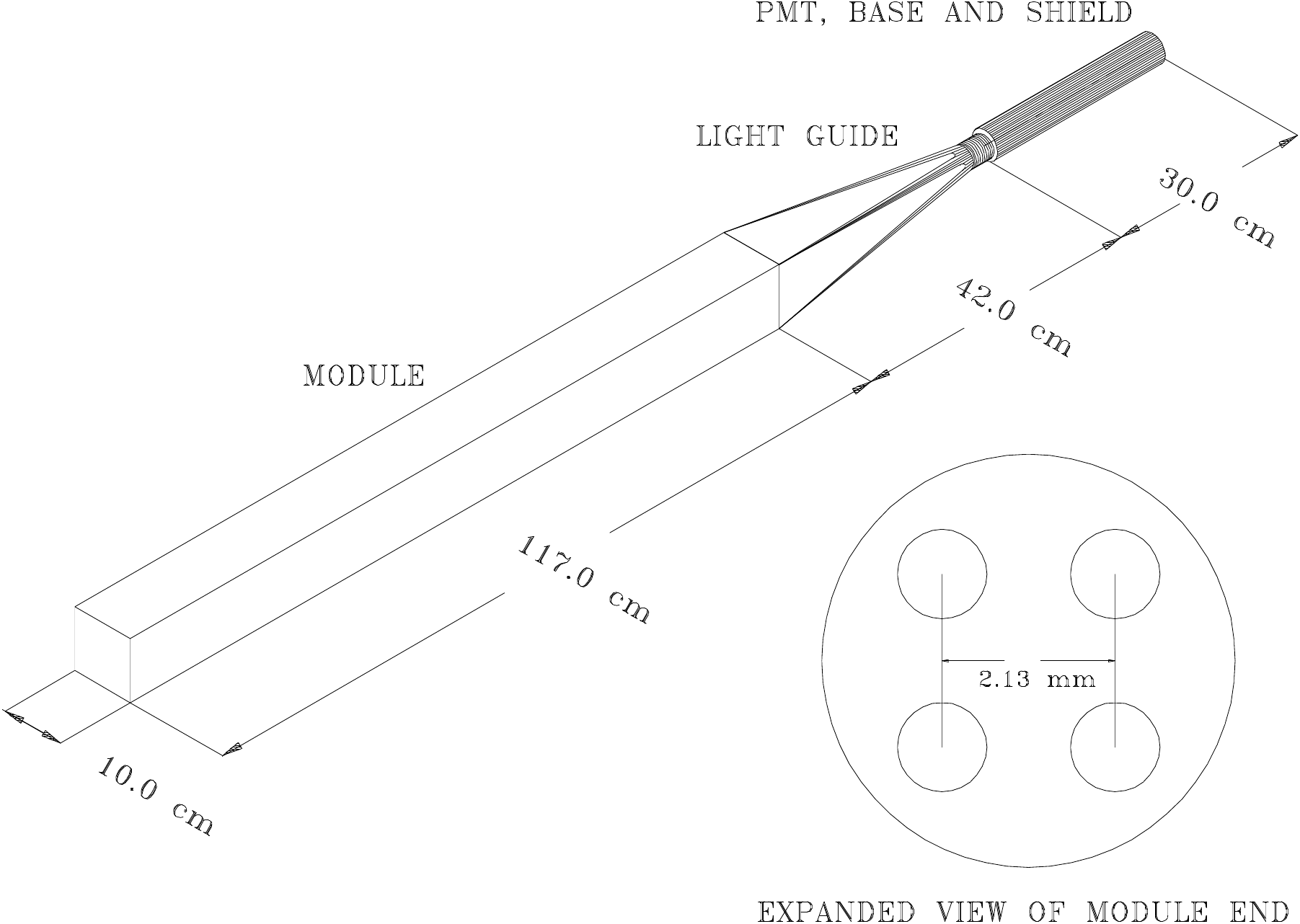}
   \end{center}
   \caption{   \label{exp:fig:pcalModule}
      Design of a \protect\acs{PCAL} module. The inset shows the spacing
      of scintillator fibers~\cite{Pruneau:1996ik}.}
\end{figure}

The \ac{PCAL} detectors were constructed using lead-scintillator
hadronic calorimeter modules that had originally been assembled for the
E864 experiment at the \ac{AGS}~\cite{Armstrong:1998qs}. Each module
consisted of a lead-scintillator brick 117.0~{\cm} in length with a
square cross-section of 10~{\cm} on each side. An array of $47\times47$
scintillator fibers was contained in each module. The  lead-scintillator
bricks were constructed by rolling thin sheets of lead (with a 1\%
antimony admixture) through a grooving machine, laminating the sheet and
then manually placing a ribbon of 47 scintillator fibers into the
grooves. This process was then repeated 46 times to complete a module,
and more than 750 modules were constructed for use in the E864
experiment. Great care was taken to ensure precise uniformity in the
inter-module fiber lattice. Attached to the brick section was an
ultra-violet absorbing Lucite light guide, that ensured clean
transmission of scintillation light ($\lambda=435~\nm$) without
contamination from any {\cheren} radiation created in the light guide.
See~~\stolenfig{exp:fig:pcalModule}{Pruneau:1996ik} for a diagram of a
\ac{PCAL} module.

Since the decommissioning of the original E864 calorimeter was not an
easy process -- the modules had been freshly painted before assembly and
had since stuck together -- and because the modules had been stored
outside, careful testing of modules was necessary before they could be
used in {\phob}. Testing was performed using cosmic rays. After the
decommissioning of E864, the modules had been cleverly stacked in rows of
ten, six layers high. Alternating layers were oriented at right angles.
Thus it was possible to use modules from three layers as cosmic ray
triggers to test both the response and attenuation length of modules in
the remaining three layers. Since \acp{PMT} were also salvaged from the
original E864 equipment, these tubes were tested at the same time as the
modules. For purposes of building the \ac{PCAL} detectors, modules with
better attenuation lengths and \acp{PMT} with desirable gains were
placed in the region of the detector expected to contain most of the
shower signal.

Assembly of the \ac{PCAL} detectors was complicated by the weight of the
modules, about 100~{\kg}, and by their fragility. In addition to the
softness of the lead and the possibility of breaking some of the
scintillator fibers, light guide detachment was a common problem. As a
result of testing and handling, more than 10\% of modules were rejected.
After having passed testing and visual inspection (to reject modules
which had bowed), modules were placed into a specially designed,
light-tight aluminum box. Mounting the modules onto the table of the box
presented its own challenge. Since the \acp{PCAL} were positioned on the
outer-ring side of the beam, and since there was no direct access to
that side of the \ac{RHIC} tunnel, the calorimeters had to be assembled
in the experimental area -- the full calorimeter was too big and too
heavy to be transported above or below the beam pipes. The modules were
mounted using a manually operated chain hoist that had been installed at
the proper location in the \ac{RHIC} tunnel for a different purpose (to
transport accelerator power supplies). As the shape of each module had
warped to some extent, great care was taken to ensure that the modules
were spaced uniformly in both the horizontal and vertical direction.
Uniformity was further necessitated by the back plate of the
\ac{Au-PCAL} box: an inch-thick aluminum plate with a grid of
$9\times12$ conical holes through which the ends of all light guides
were required to pass. The \ac{Au-PCAL} detector consisted of an array
8~modules wide and 10~modules high. On top of this array was one row of
7~modules, with the ``missing'' module away from the beam line. Two more
modules were then added to this row, one near the beam line and one
away. These modules were used to provide extra support for the top of
the calorimeter box. Light absorbing filters were then added to the
\acp{PMT} of two modules, see \sect{calib:pcal:aupcalgain}.

\begin{figure}[t!]
   \begin{center}
      \includegraphics[width=0.55\linewidth]{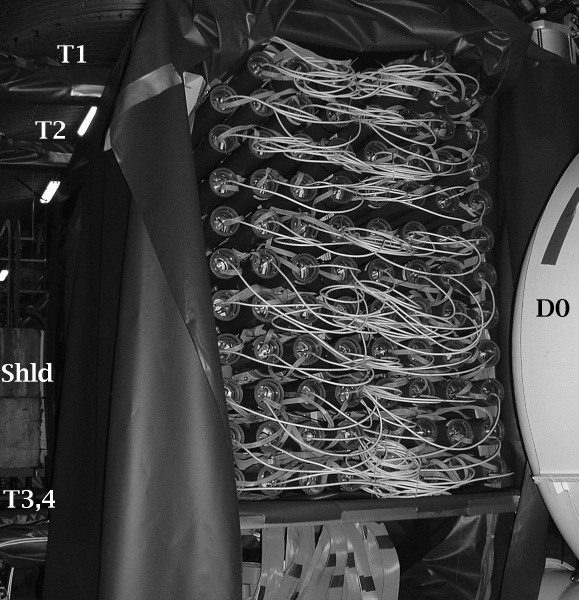}
   \end{center}
   \caption{   \label{exp:fig:fullpcal}
      A photograph of the fully commissioned \protect\acs{Au-PCAL}.
      Easily seen are the \protect\acs{PMT} mountings and readout
      cables. To guide the eye, certain features have been labeled: the
      cosmic ray trigger modules (T1-T4), the shielding (Shld) and the
      D0-magnet (D0). During normal operation, the Kevlar tarp covered
      the full box, to help prevent light-leaks.}
\end{figure}

Assembly of the small \ac{d-PCAL}, used only during the {\dAu} physics
run, was far more simple. This calorimeter consisted of only four
modules, individually wrapped in Kevlar for light-tightness and to
contain any lead-oxide, resting on an open table adjacent to the beam
pipe. After the {\dAu} physics run, the small \ac{d-PCAL} was replaced
with a full sized calorimeter. Both full sized calorimeters had
Kevlar-wrapped modules placed outside, above and below, the calorimeter
boxes to be used as cosmic ray triggers. An assembled calorimeter is
shown in \fig{exp:fig:fullpcal}.

%---------------------------------------------------------------
\subsubsection{Shielding for the Proton Calorimeters}
\label{exp:phobdet:calo:pcalshield}
%---------------------------------------------------------------

Shielding was installed to block produced particles from depositing
energy into the \ac{PCAL} detectors. While the \acp{ZDC} were
inaccessible to produced particles due to their extreme forward
position, the \ac{Au-PCAL} covered a {\prap} region of roughly
$-3.6<\eta<-5.2$. Two shields were constructed, each using four blocks
of high-density (4.0~\dens) concrete. The blocks measured
$88.9~\cm\times44.45~\cm\times44.45~\cm$ and were positioned roughly at
beam height. One shield was placed about 8~{\m} from the \ac{IP} and
moved as close to the beam as possible (about 20~{\cm}). The other shield
was situated near the calorimeter, almost 15~{\m} from the \ac{IP} and
was positioned roughly 40~{\cm} from the beam line, to allow spectator
protons to reach the calorimeter without being blocked by the shielding.
Simulations of these shielding blocks showed that even 50~{\gev} pions
(protons) would deposit less than 5\% (2.5\%) of their energy into the
calorimeter. The shield positions can be seen in
\fig{exp:fig:protpaths}.

%---------------------------------------------------------------
\subsection{The Trigger Detectors}
\label{exp:phobdet:trig}
%---------------------------------------------------------------

{\phob} used several different detectors to determine that a collision
with various properties had occurred in the experiment. For high
multiplicity {\AuAu} collisions, the Paddles served as the primary event
trigger while the {\cheren} counters could be used to trigger on the
collision vertex. Before the {\dAu} physics run, two more trigger
detectors were installed -- the \acp{T0} and the \ac{SpecTrig}. The
\acp{T0} were used to provide a more accurate vertex and collision time
than the {\cheren} counters could provide. The \ac{SpecTrig} provided a
trigger for collisions producing a high-{\pt} particle in one arm of the
Spectrometer.

%---------------------------------------------------------------
\subsubsection{The Paddle Counters}
\label{exp:phobdet:trig:paddle}
%---------------------------------------------------------------

\begin{figure}[t!]
   \begin{center}
      \includegraphics[width=0.3\linewidth]{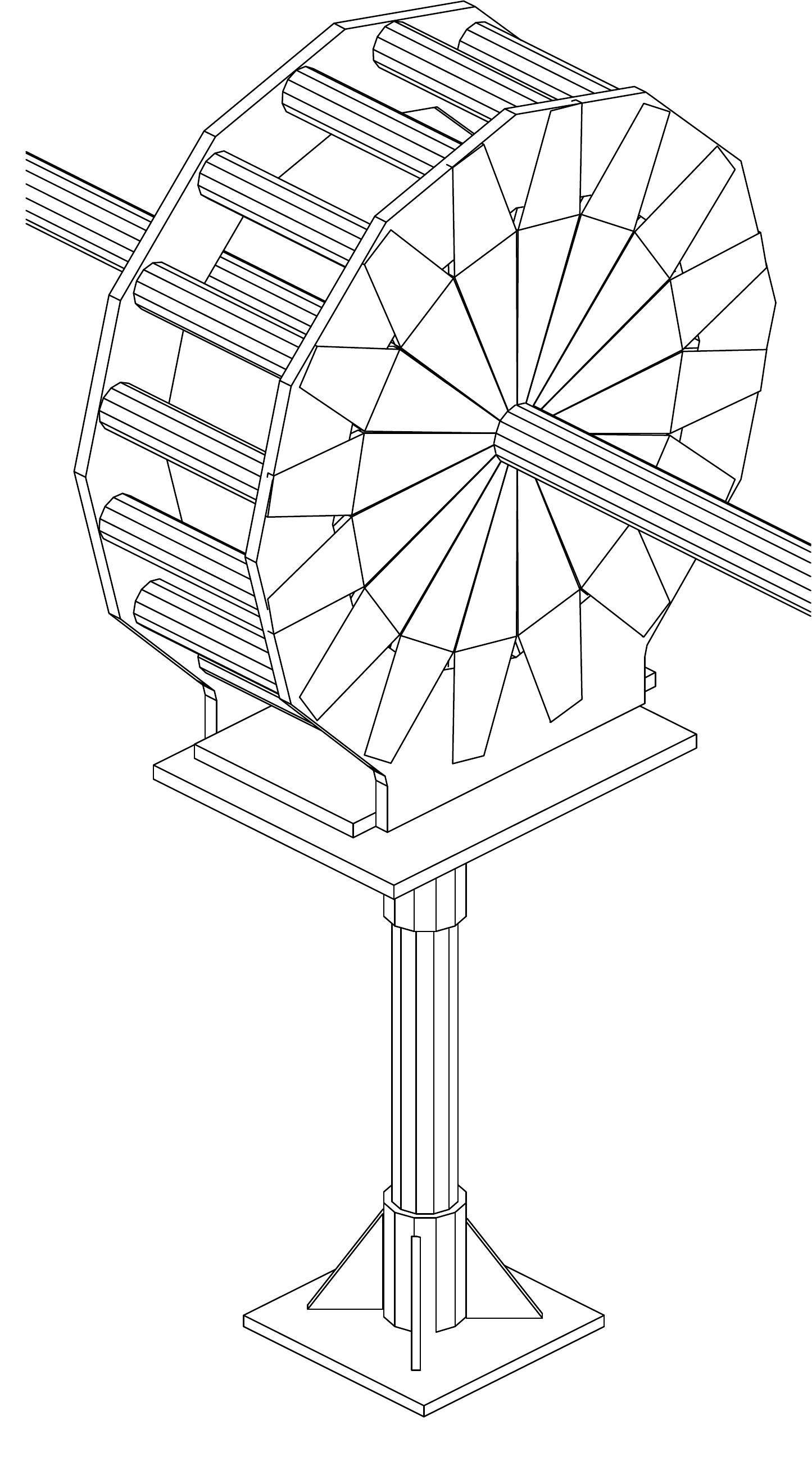}
   \end{center}
   \caption{   \label{exp:fig:pdldiag}
      Schematic diagram of a Paddle counter~\cite{Bindel:2001sp}.}
\end{figure}

The Paddle counters were used during {\AuAu} physics runs both to trigger
on collisions and to determine centrality. There were two Paddle
counters, each consisting of a circular array of 16 plastic
scintillators, positioned $\pm3.21~\m$ from the nominal interaction
point. This position gave them a {\prap} coverage of $3<\abs{\eta}<4.5$.
Plastic scintillator was chosen for its good timing resolution
(150~\ps), large dynamic range (from 1 \ac{MIP} up to 50 per collision)
and high tolerance of radiation. A diagram of a Paddle counter is shown
in \stolenfig{exp:fig:pdldiag}{Bindel:2001sp}.

Individual scintillators were 18.6~{\cm} long, 0.85~{\cm} thick,
1.9~{\cm} wide at the inner edge and 9.5~{\cm} wide at the outer
edge. Attached to the scintillator was a two-component acrylic light
guide.  While the scintillator was oriented transverse to the beam,
the \acp{PMT} were oriented longitudinally. This arrangement was
achieved by coupling one section of the light guide to the
scintillator, one section to the photomultiplier tube and using a
45{\dg} Aluminized mirror between the light guide components. Both
amplitude and timing information was available from the Paddle
signals. The full Paddle counters performed with a time resolution
of about 1~{\ns}, an energy resolution of $\sigma_{E}/\Delta E=45\%$
for a single \ac{MIP}, a signal-to-noise ratio of about 20/1 and
a triggering efficiency of 100\% for central and semi-peripheral
{\AuAu} collisions.

%---------------------------------------------------------------
\subsubsection{The {\cheren} Counters}
\label{exp:phobdet:trig:cheren}
%---------------------------------------------------------------

\begin{figure}[t!]
   \begin{center}
      \includegraphics[width=0.4\linewidth]{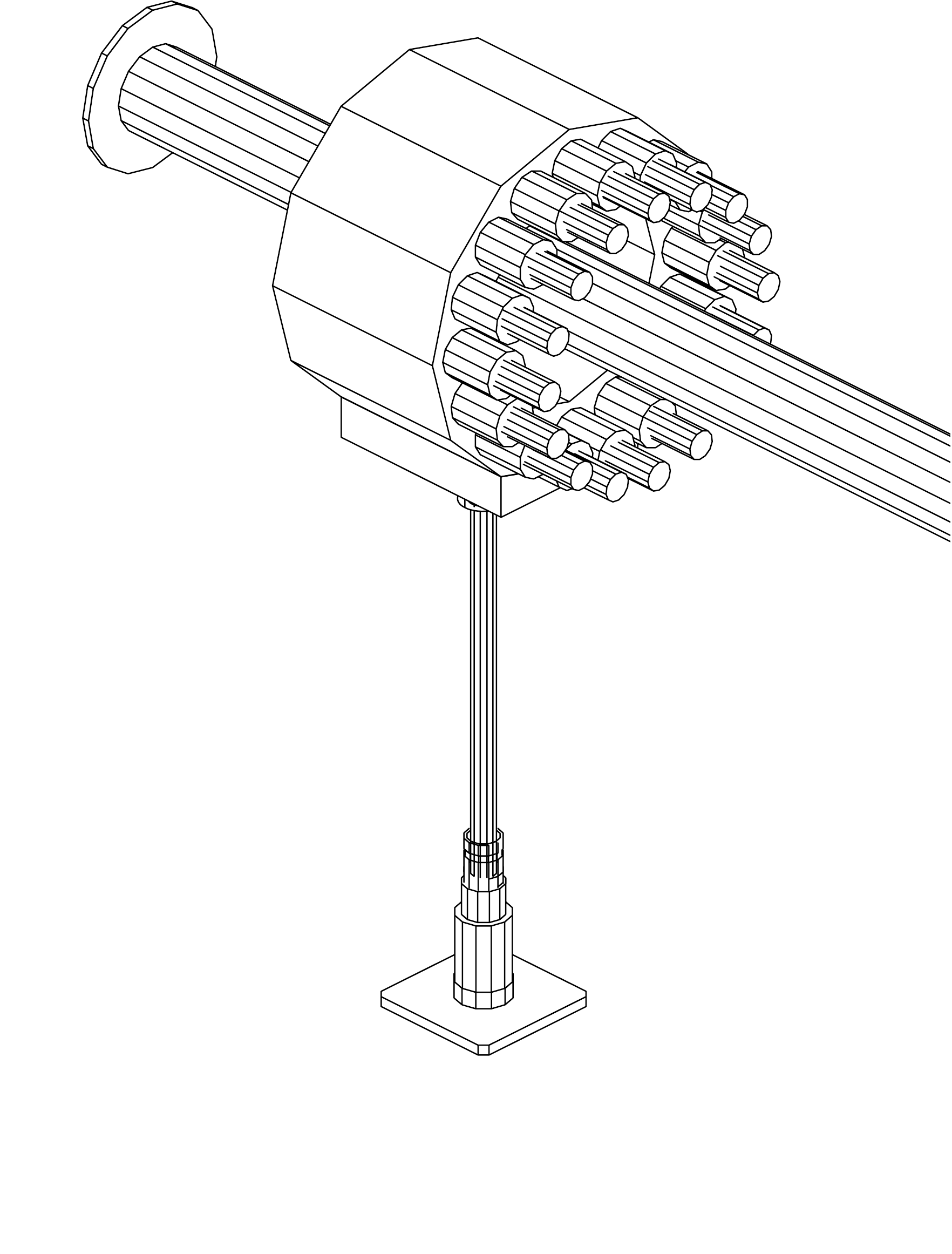}
   \end{center}
   \caption{   \label{exp:fig:cherenk}
      Schematic diagram of a {\cheren} detector~\cite{Bindel:2002up}.}
\end{figure}

The {\cheren} counters were used to provide real-time collision vertex
information. There were two {\cheren} detectors, each consisting of 16
modules, placed $\pm5.5~\m$ from the \ac{IP}. The radiators circled the
beam pipe at a radius of 8.57~{\cm} and were oriented longitudinally to
the beam. Their positioning covered 37\% of the solid angle in the
{\prap} range $4.5<\abs{\eta}<4.7$. See
\stolenfig{exp:fig:cherenk}{Bindel:2002up} for a diagram of a {\cheren}
detector.

Individual {\cheren} radiators were made of acrylic (the same type used
as a light guide for the Paddles) in the shape of a cylinder with a
4.0~{\cm} length and a 2.5~{\cm} diameter. No light guide was necessary;
\acp{PMT} were attached to the radiators with silicon elastomer. The
modules were mounted in a mechanical structure that allowed individual
modules to be moved within a 100~{\mm} range in the beam direction. This
allowed individual counters from each side of the interaction to be
matched to within 50~{\ps}. The total timing resolution achieved by the
{\cheren} counters was 380~{\ps}.

%---------------------------------------------------------------
\subsubsection{The Time-Zero Counters}
\label{exp:phobdet:trig:t0}
%---------------------------------------------------------------

Prior to the {\dAu} physics run, \acfp{T0} were installed. These
detectors were used to obtain accurate time-zero information and could
be moved to a region of {\prap} appropriate for a {\dAu} collision. 
There were two \acp{T0}, each consisting of 10 modules arranged in a
circle of 151~{\mm} in diameter about the beam pipe. One of the \acp{T0}
was located 2.6~{\m} from the \ac{IP} on the deuteron exit-side of the
collision and the other was located 5.4~{\m} from the \ac{IP} on the
gold exit-side of the collision. These locations were not fixed -- for
{\pp} collisions, for example, each module was moved closer to the
\ac{IP}. At their {\dAu} physics running positions, the \acp{T0} had a
{\prap} acceptance of $3.7<\eta<4.2$ (d-side) and $-4.9<\eta<-4.4$
(Au-side).

Acrylic {\cheren} radiators were also used for the \acp{T0}. While the
type of radiator was the same as that used for the {\cheren} counters,
the type of \acp{PMT} was not. The \acp{PMT} used for the \acp{T0} had a
rise time that was twice as fast as those used for the {\cheren}
counters. The shape of the radiators used for the \acp{T0} also differed
from those used in the {\cheren} counters. Radiators in the \acp{T0} had
a diameter of 50~{\mm} and a length of 25~{\mm}. Their acceptance, mobility
and response time made the \acp{T0} the preferred detector for
determining time-zero and for obtaining real-time vertex information for
use in collision triggers.

%---------------------------------------------------------------
\subsubsection{The Spectrometer Trigger}
\label{exp:phobdet:trig:spectrig}
%---------------------------------------------------------------

The \acf{SpecTrig} was commissioned for use in the {\dAu} physics run. It
consisted of two walls of plastic scintillators, each containing 10
modules, and was designed to have an acceptance which covered that of
the \ac{TOF}. Each scintillator module was 11~{\cm} long vertically,
7.24~{\cm} wide horizontally and 0.5~{\cm} thick. The scintillators were
connected to trapezoidal light guides, that were then glued to
\acp{PMT} using optical cement. The \ac{SpecTrig} modules were mounted
on an aluminum frame that was then attached to the {\phob} magnet yoke.

This trigger was used in low-multiplicity environments ({\dAu} and {\pp})
for the real-time selection of events containing a particle with high
transverse momentum that could be tracked using the Spectrometer and
\ac{TOF}. The selection was provided by combining real-time vertex
information from the \acp{T0} with the position of hits in the
\ac{SpecTrig} and \ac{TOF} walls. For particles with large
transverse momentum ($p_{T}\gsim2~\mom$), these three positions should
fall on a single line. Thus, the trigger selected collisions for which
some combination of a hit in the \ac{SpecTrig} and a hit in the
\ac{TOF} formed a line that pointed back to the vertex position
provided by the \acp{T0}. This trigger could only function when the
occupancy of the \ac{SpecTrig} and \ac{TOF} walls were low, so that
random combinations of hits from different particles were not likely to
be collinear with the collision vertex.

%---------------------------------------------------------------
\subsection{The Data Acquisition}
\label{exp:phobdet:daq}
%---------------------------------------------------------------

Once signals from the trigger detectors had been digitized, a decision
could be made regarding whether or not to record the full {\phob}
detector state to disk. A description of the digitizing process for
silicon signals can be found in \sect{calib:si}, while descriptions of
the various trigger criteria can be found in \sect{recon:trig}. During
the {\dAu} physics run, the {\phob} \acf{DAQ} system was used to process
signals from over 135,000 silicon channels and 1,500 photo-tube channels
at a rate of up to 280 events per second. The \ac{DAQ} resided in a
single \ac{VME} crate and consisted of a 22-node computing farm to
process the data, a computer module to assemble the processed data into
an event, a disk array for temporary data storage, a module to monitor
trigger signals and server to ship the processed data to tape
storage~\cite{Kulinich:2002qy}.

The RACEway computing farm that processed the data was provided by
Mercury Computer Systems, Inc. Each node had a 300~{\mhz} PowerPC-750
processor with 1~{\mbyte} of \ac{L2} \ac{SRAM} cache and 32~{\mbyte} of
\ac{DRAM} memory. The computing farm was mainly used to check the data
for consistency and to compress the silicon data. Because it was
necessary to make offline corrections for common-mode noise and
crosstalk among silicon channels, signals from empty channels needed to
be preserved. This was done by first subtracting the known pedestals and
then using lossless Huffman compression, which encoded the most
frequently occurring \acs{ADC} samples into short bit patterns. A
compression factor of 3.5 was achieved for {\AuAu} collision data.

After processing, the data was sent to the event builder through a
\ac{FPDP} connection at a rate of 43~\mbyteps. The event builder used
synchronization tags, which had been applied to each worker's data block
when the trigger signal was received, to assemble a complete event. The
event was then stored in memory until it could be written to the disk
cache. Files of roughly 1~{\gbyte} in size (over 10,000 {\dAu} events)
were shipped to the \ac{HPSS} at the \ac{RCF} over a Gigabit Ethernet
connection. While the {\phob} \ac{DAQ} disk cache could write events at a
rate of 50~{\mbyteps}, simultaneous writing and reading resulted in a
writing speed of 28~{\mbyteps} and a sustained throughput to \ac{HPSS} of
30~{\mbyteps}.

%% file: srcs/chapter2.tex
% $Id: chapter2.tex,v 1.41 2006/09/05 01:01:30 cjreed Exp $
%

%---------------------------------------------------------------
\chapter{Detector Calibration}
\label{calib}
%---------------------------------------------------------------

\myupdate{$*$Id: chapter2.tex,v 1.41 2006/09/05 01:01:30 cjreed Exp $*$}%
All detectors in {\phob} reported their measurements through various
electronic read-outs. Once these raw signals had been recorded, several
steps were required to determine the physical properties of the
particles that produced them. The first step was to determine the amount
of energy the particle had deposited into the detector as it passed
through.

%---------------------------------------------------------------
\section{Silicon Signal Processing}
\label{calib:si}
%---------------------------------------------------------------

The purpose of a {\phob} Silicon detector was to measure both the
position at which particles passed through the detector as well as the
amount of energy they deposited in doing so. This information was
obtained by correcting the raw signal for (a)~an overall offset,
(b)~fluctuations in the read-out electronics, (c)~the amount it had
been amplified and (d)~the possibility that the deposited energy was
not due to a particle traveling transverse to the detector plane.

%---------------------------------------------------------------
\subsection{Semiconductor Detectors}
\label{calib:si:semicond}
%---------------------------------------------------------------

As charged particles pass through a semiconductor material,
electromagnetic collisions excite electrons out of the valence energy
band of the crystal, leaving behind an effective positively charged
hole. The goal of such a detector is to measure the number of
electron-hole pairs created due to the ionizing particle. This
information can then be related back to the amount of energy the
particle deposited into the detector.

Silicon semiconductor devices are desirable due to the small amount of
energy (\mbox{$\approx1~\ev$}) required to excite an electron from the
valence to the conduction band. This allows an ionizing particle to
create a large number of electron-hole pairs, which in turn allows for a
precision measurement to be made of the deposited energy. However, the
gap between the valence and conduction bands is so small that, at finite
temperatures, thermal energy is sufficient to create electron-hole
pairs. These electron-hole pairs lead to a current through the
semiconductor, even in the absence of an ionizing particle. The ideal
semiconductor detector would be a material which could freely conduct
any electron-hole pairs produced by an ionizing particle, but would have
no such leakage current. While  the ideal situation cannot be realized,
the leakage current of semiconductor detectors is minimized by creating
a large insulating region in the material known as a depletion region.

A depletion region arises when a crystal with an excess of electrons
above the valence band is brought into equilibrium with a crystal that
has an excess of holes above the valence band. Such materials are
obtained by doping a semiconductor with impurities. When the dopant atom
has more electrons than the base semiconductor atoms, the resulting
crystal has an energy level between the valence and conduction bands in
which the extra electrons would sit at zero-temperature. At room
temperatures, on the other hand, nearly all of the electrons in this
energy band are excited into the conduction band by thermal energy.
However, these electrons do not leave a hole in the lattice, so such
doped crystals have an excess of negative charge carriers and are thus
known as \emph{\stype{n}} doped semiconductors. Similarly,
\emph{\stype{p}} doped semiconductors are those which have an excess of
positive charge carries (holes) resulting from dopant atoms that have
fewer electrons than the base semiconductor. A \emph{\pnjnc} is formed
when an \stype{n} semiconductor is brought in contact with a \stype{p}
semiconductor. Excess electrons in the \stype{n} material are attracted
to the excess holes in the \stype{p} material, resulting in a current
across the junction. As electrons fall into holes and become stuck at
crystal lattice points, the dopant atoms are left lacking an electron
(or hole) and become exposed charged ions. The charged ions create an
electric potential that is positive on the \stype{n} side of the
junction and negative on the \stype{p} side. This potential tends to
prevent the flow of charges and allows the material to reach
equilibrium. The region at the junction is thus a good region for
detecting ionizing particles: it has effectively no leakage current but
would conduct excited charge carriers. The extent of this depletion
region can be increased by applying an external electric field across
the junction to increase the potential in the \stype{n} side. This is
known as reverse-biasing the \pnjnc. For more information on
semiconductor detectors, see~\cite{Leo94:semiconductors}.

\begin{figure}[t]
   \begin{center}
      \includegraphics[width=0.7\linewidth]{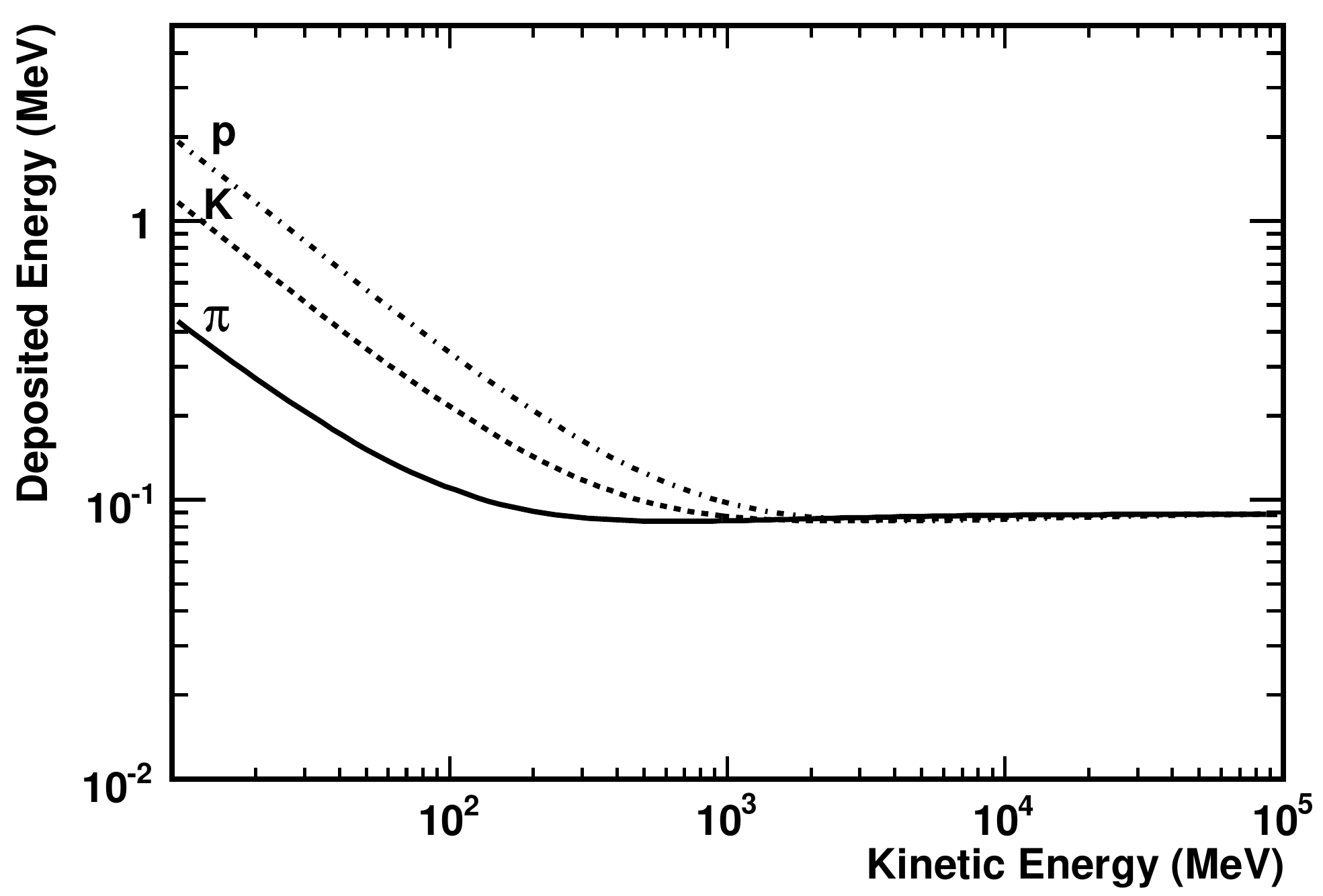}
   \end{center}
   \caption{\label{calib:fig:betabloch}
      The average amount of energy deposited by pions, kaons and
      protons into 300~{\um} of silicon.}
\end{figure}

The average amount of energy an ionizing particle loses as it travels
through material can be estimated using the Bethe-Bloch formula. This
formula has the form

\begin{equation}
   \label{calib:eq:bb}
-\frac{\mathit{dE}}{\mathit{dx}} = (0.1535~{\mev {\cm^2}/\gm})~\rho
   \frac{Z}{A} \frac{z^2}{\beta^2} 
   \sqb{\ln\prn{\frac{2 m_e \gamma^2 \beta^2 c^2 W_{\rm max}}{I^2}}
   - 2\beta^2}
\end{equation}

\noindent%
where $\rho$~is the density of the material in {\dens}, $Z$~is the
atomic number of the material, $A$~is the atomic weight of the material
in a.u., $m_e$ is the mass of the electron (0.511~{\mmass}), $z$~is the
charge of the ionizing particle in units of $e$, $\beta$~is the speed of
the particle in units of $c$, $\gamma=1/\sqrt{1-\beta^2}$, $W_{\rm
max}$ is the maximum energy transfer in a single collision and $I$~is
the average energy required to excite a bound electron from an atom in
the material. Discussions of this equation and common corrections to
the energy loss formula can be found in the literature, see for
example~\cite{Leo94:bb}. The average amount of energy deposited by
several different types of particles in a typical {\phob} Silicon
detector is shown in \fig{calib:fig:betabloch}. Notice that the minimum
amount of energy deposited by pions, kaons and protons is roughly
equivalent. Particles depositing this minimum amount of energy are known
as \acfp{MIP}.

\acsp{MIP} are important in high-energy physics, since most of the
particles produced in a collision will deposit nearly the minimum amount
of energy into a detector. For a very thick detector, an ionizing
particle would experience a very large number of roughly independent
collisions, while losing a negligible amount of energy in each
collision. After measuring many ionizing particles, the distribution of
energy deposited into the thick detector by the particles would be a
Gaussian whose mean was equivalent to the energy deposited by one
\ac{MIP}. For a thin detector, such as the Silicon detectors used in
\phob, the number of collisions would be much smaller. The energy
distribution observed by such a detector would not be Gaussian, but
would have a high-energy tail due to (rare) Collins in which the
ionizing particle loses a significant amount of energy. This
distribution is known as a Landau; examples can be seen in
\figs{calib:fig:rawadc}{calib:fig:gauslan}.

%---------------------------------------------------------------
\subsection{The {\phob} Silicon Read-out}
\label{calib:si:readout}
%---------------------------------------------------------------

\begin{figure}[t]
   \begin{center}
      \includegraphics[width=0.8\linewidth]{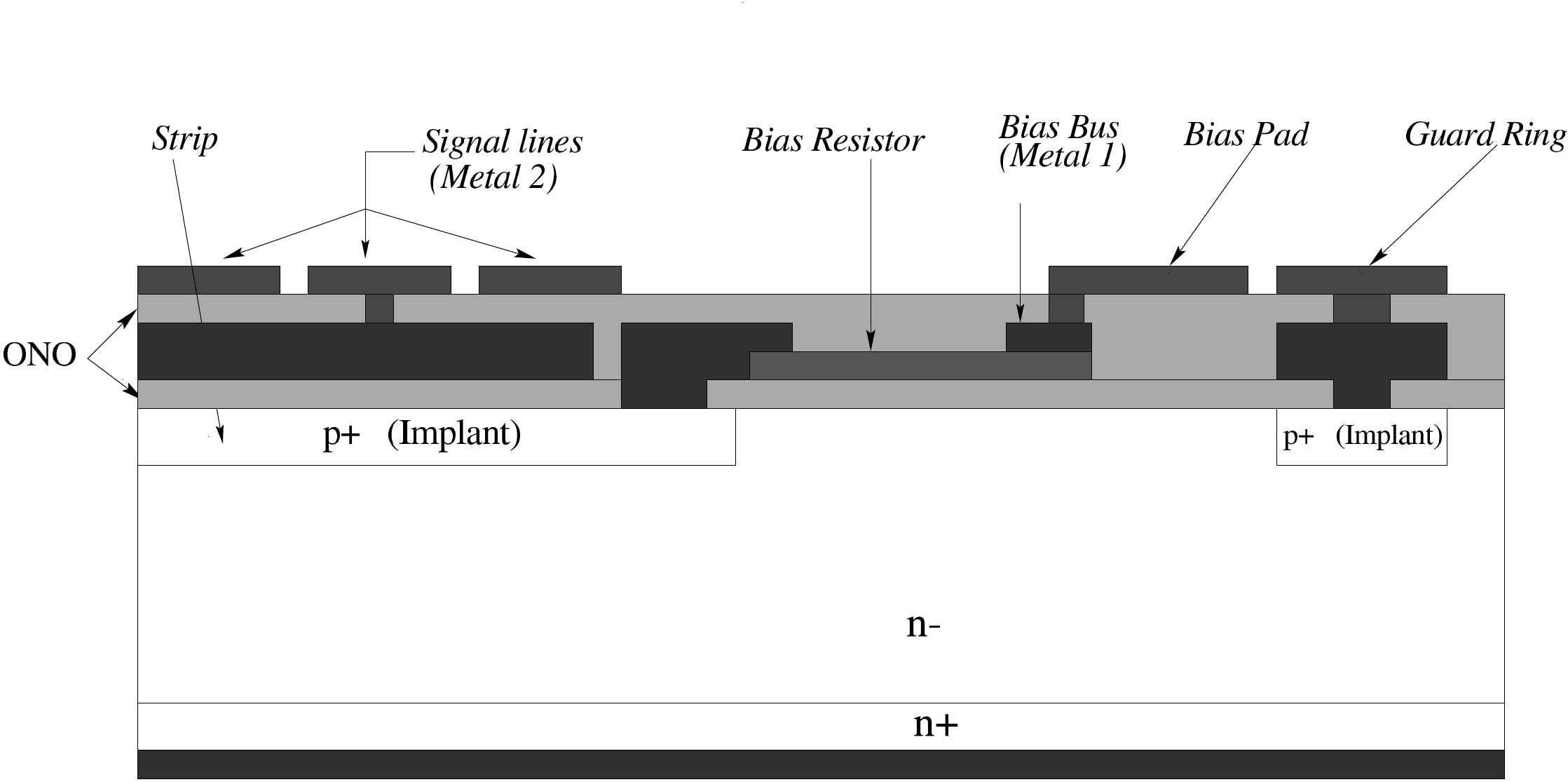}
   \end{center}
   \caption{\label{calib:fig:phobsiGray}
      Diagram of the cross section of a {\phob} silicon
      pad~\cite{Nouicer:2001qs}.}
\end{figure}

{\phob} Silicon detectors were constructed using a ($\approx300~\um$)
thick \stype{n} layer of silicon, with \stype{p} silicon read-out pads.
The pads were reverse-biased by applying positive voltage to the back
plane of the \stype{n} wafer and grounding the \stype{p} pad via a
\mbox{$5.5\pm0.2~\mohm$} resistor. This design was based on the silicon
detectors developed for the WA98 experiment~\cite{Lin:1997qx}. While the
reverse-biasing voltage served to create the depletion region, it also
pulled apart electron-hole pairs created by ionizing particles.
Electrons drifted through the \mbox{$\sim70~\volts$} potential and were
collected at the \stype{p} silicon pad. A coupling capacitor was formed
by the \stype{p} silicon and a layer of aluminum, separated by a
$0.2~\um$ thick insulating layer of silicon \ac{ONO}. The aluminum layer
of this capacitor was connected to a metal read-out line that ran along
the surface of the sensor to a bonding pad at the edge of the sensor.
Another layer of silicon \ac{ONO} served to insulate the read-out lines.
A guard ring was added to protect the silicon pads from edge effects. A
cross section of one Silicon sensor pad can be seen
in~\stolenfig{calib:fig:phobsiGray}{Nouicer:2001qs}.

Voltage built up in the capacitor of a silicon pad was sent into a
pre-amplifier and signal shaping chip that was bonded directly to the
sensor. The chip was a 64-channel or 128-channel \mbox{VA-HDR-1} chip,
manufactured by the IDEAS company (\hrefurl{http://www.ideas.no/}). The
pre-amplified signals were read out by \acp{FEC}, located in the
experimental hall, that digitized the signals using 12~bit \acp{ADC}.
These signals were then sent to the {\phob} \ac{DAQ} to be recorded (see
\sect{exp:phobdet:daq}).

%---------------------------------------------------------------
\subsection{Pedestal and Noise Correction}
\label{calib:si:pedcor}
%---------------------------------------------------------------

While the leakage current of the Silicon was minimized by the
reverse-biased \pnjnc, it was not zero. This current caused an offset
to the signal read out from a Silicon channel, even if that signal
would otherwise have been zero. The leakage current was roughly
constant during one running of the silicon (during the time the
reverse-biasing voltage was turned on), but could change by a small
amount from one running to the next. Hardware offsets in the read-out
system also contributed an overall offset to the Silicon signals. In
addition, electronic noise in the system contributed an offset that
would fluctuate from one event to the next.

\begin{figure}[t]
   \begin{center}
      \includegraphics[width=0.7\linewidth]{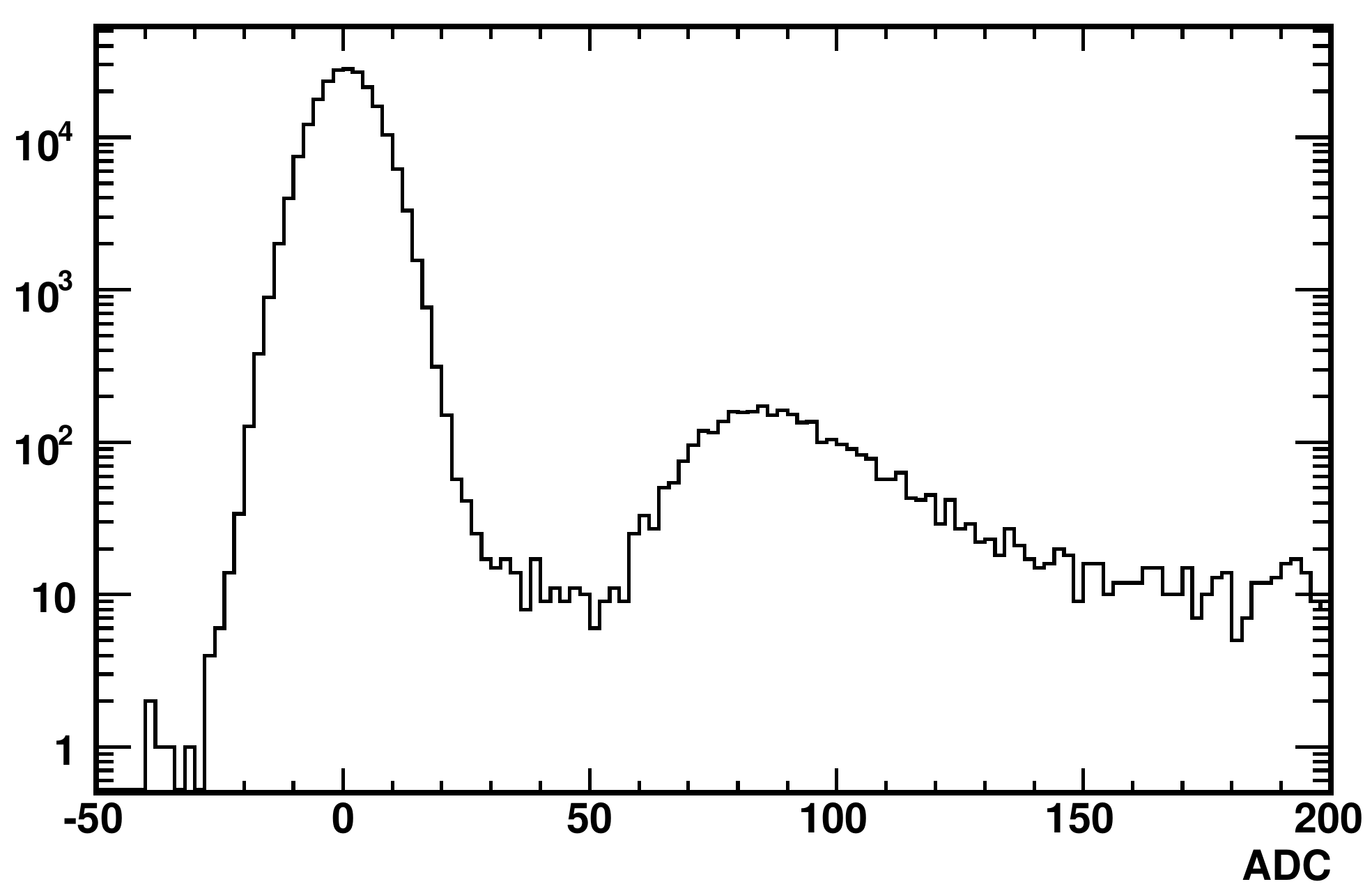}
   \end{center}
   \caption{\label{calib:fig:rawadc}
      Pedestal-subtracted \protect\acs{ADC} distribution of a typical
      channel in the Ring detector. Note that the width of the Gaussian
      peak is roughly 5~\protect\acsp{ADC} channels.}
\end{figure}

The effect of these offsets could be measured by histogramming the
signal from one Silicon channel over many events. In a typical collision
event, it was unlikely that an ionizing particle had passed through any
given silicon channel -- in other words, the occupancy of the Silicon
detectors was low. Thus, the lower part of the raw \ac{ADC} distribution
of a Silicon channel would provide a measure of the various offsets.
This distribution was a Gaussian, the mean of which, known as the
pedestal, reflected the overall offset due to electronic offsets and
leakage current. The width of the Gaussian, referred to as the noise in
the channel, was due to electronic noise and small fluctuations of the
leakage current. \Fig{calib:fig:rawadc} shows the distribution of
\ac{ADC} values of a Silicon channel from many events, where the
channel's pedestal has been subtracted from the raw \ac{ADC} signal in
each event. The Gaussian pedestal/noise peak is clearly visible. 

The pedestal and noise were calculated in several steps. First, the raw
\ac{ADC} signals of each channel were averaged over 200~low multiplicity
events. This \emph{pre-pedestal} was then used to find a window in which
to calculate the pedestal and noise. Next, \ac{ADC} values within
100~\ac{ADC} channels of the pre-pedestal from 600~low multiplicity
events were histogrammed for each Silicon channel. The  actual pedestal
and noise were calculated in an even smaller window, found by locating
the most populated bin of the histogram, and then stepping out in each
direction until consecutive empty bins were encountered. Note that such
bins are not seen in \fig{calib:fig:rawadc}, since this figure shows the
distribution from a large number of events, many of which had a high
multiplicity. Finally, the peak in this small window was used to find
the pedestal and noise. Silicon channels in the Octagon had the highest
occupancy, and for those channels, the peak was fit with a Gaussian to
extract the pedestal and noise values without contributions from actual
deposited energy signals. For all other Silicon channels, the pedestal
(mean) and noise (\ac{RMS}) were calculated explicitly to save computing
time. The pedestal and noise were then assumed to remain constant for
about an hour of Silicon running time. Pedestal and noise values would
be recalculated each time the Silicon bias was turned on and at least
once an hour during Silicon data-taking.

\begin{figure}[t]
   \centering
   \subfigure[Pedestal Values]{
      \label{calib:fig:sipeds}
      \includegraphics[width=0.4\linewidth]{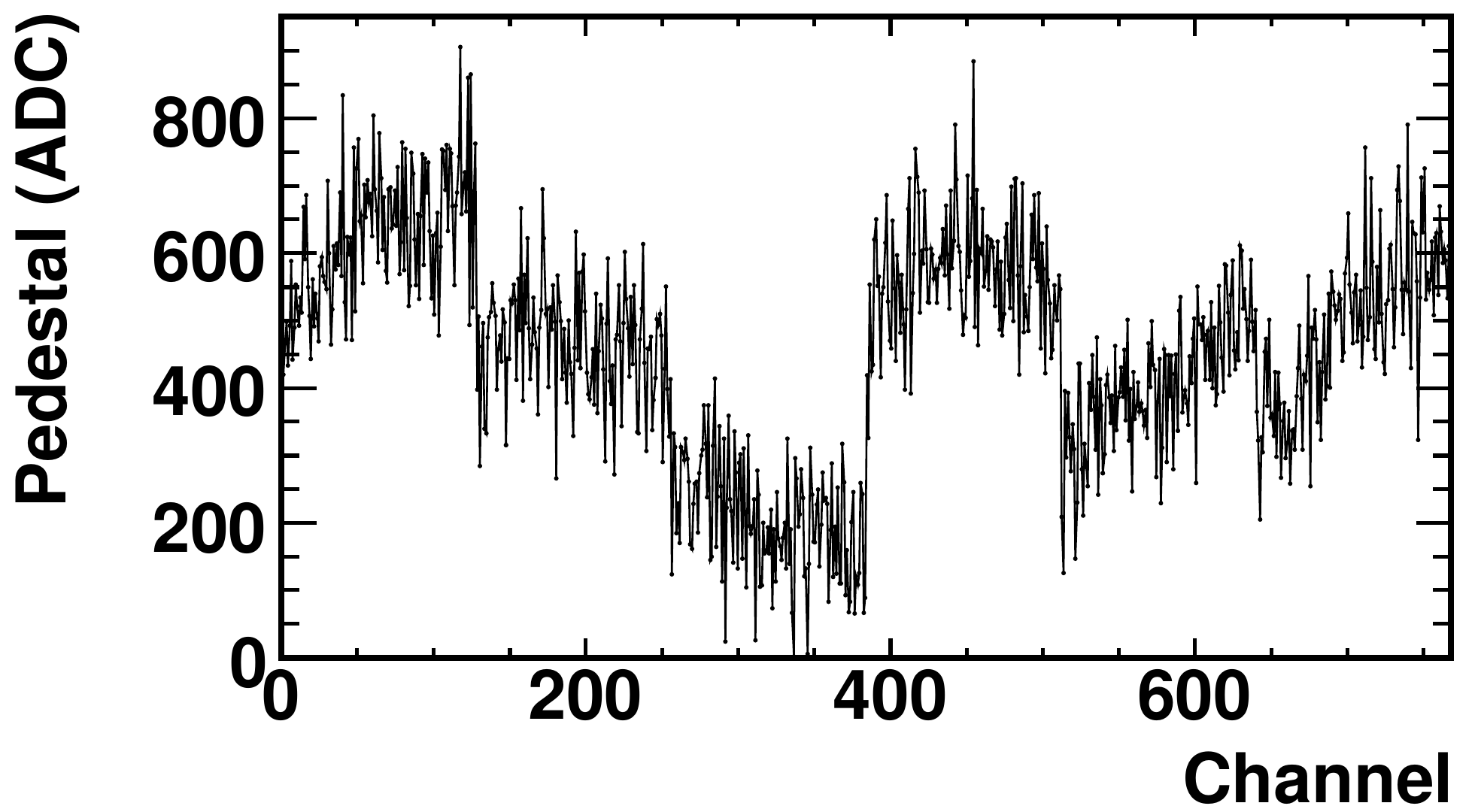}
   }
   \subfigure[Noise Values]{
      \label{calib:fig:sinoises}
      \includegraphics[width=0.4\linewidth]{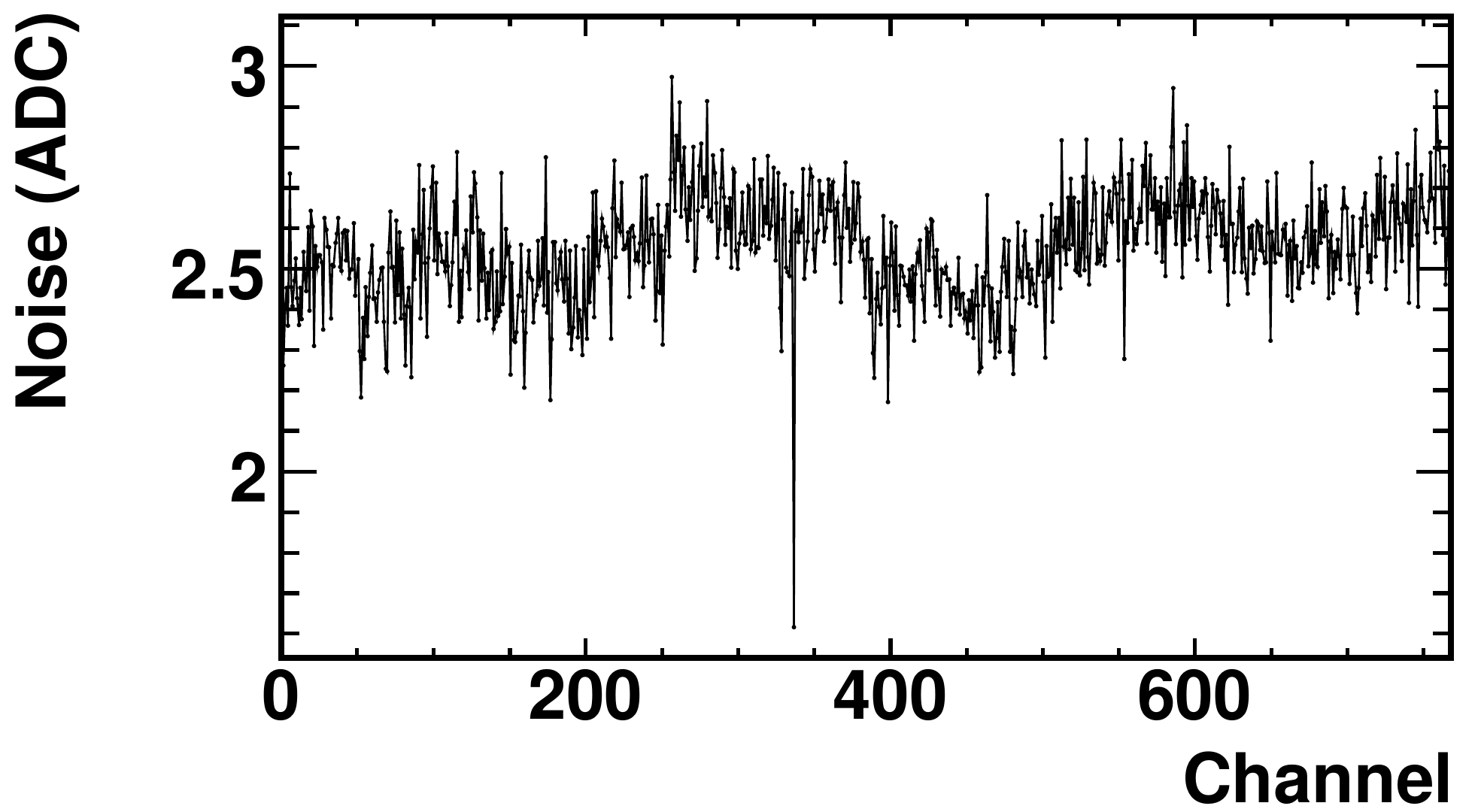}
   }
   \subfigure[\protect\acs{CMN}-Corrected Noise Values]{
      \label{calib:fig:sicmns}
      \includegraphics[width=0.4\linewidth]{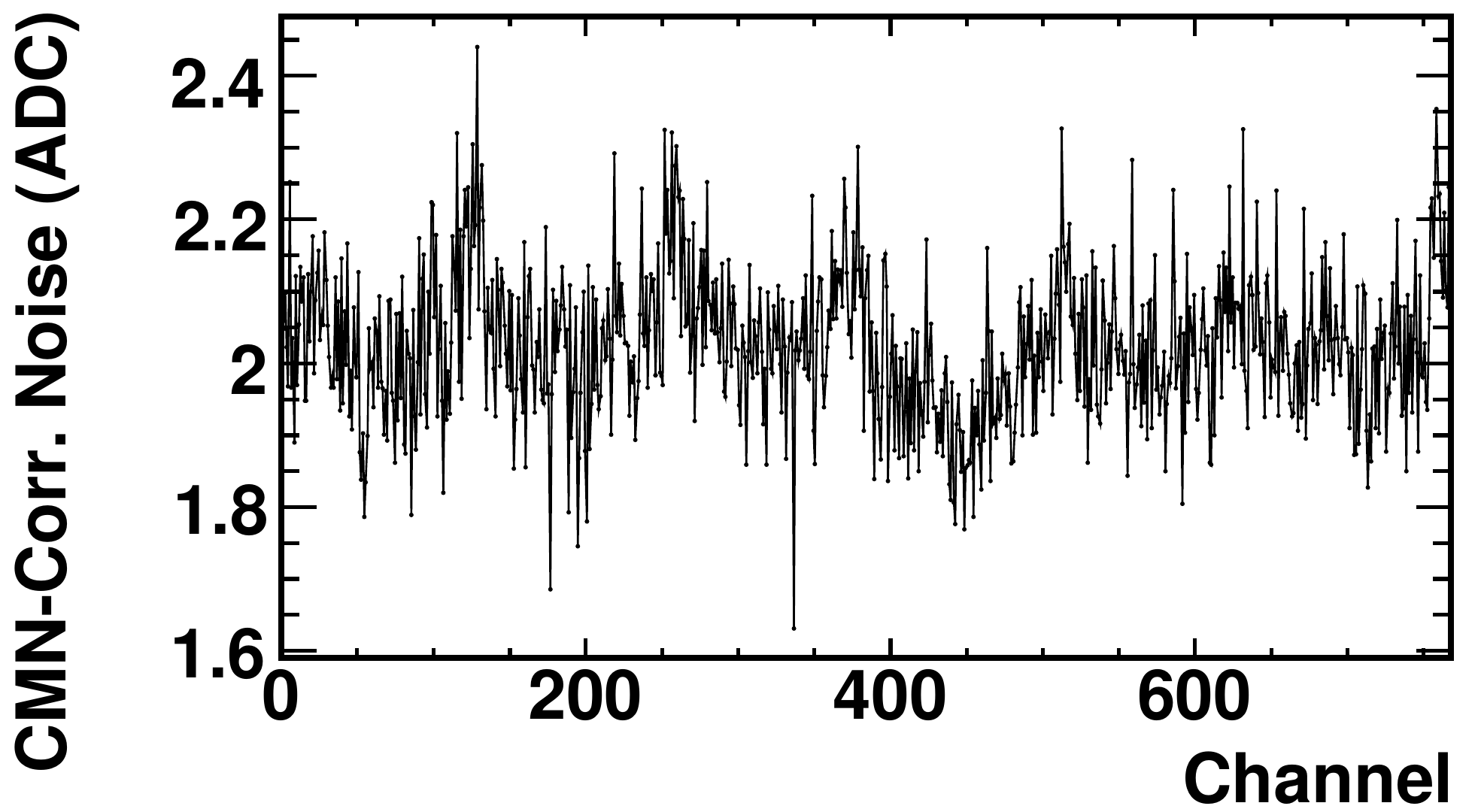}
   }
   \caption{   \label{calib:fig:sipednoise}
      Typical values of the pedestal and noise in Silicon channels.
      \subref{calib:fig:sipeds}~Pedestal values. The \mbox{128-channel}
      chips can be seen, as electronic offsets were implemented on
      a chip-by-chip basis. \subref{calib:fig:sinoises}~Noise values
      (the width of the pedestal peak). \subref{calib:fig:sicmns}~The
      noise of each channel after \protect\acs{CMN}-correction.}
\end{figure}

After pedestal subtraction, the signal was further corrected for an
offset effect known as \acf{CMN}. This type of noise was the result of
electronic cross-talk in the Silicon read-out chips. It caused an
offset to the \ac{ADC} which was different in each event, but in any
given event the offset was the same for all channels being read out by
the same chip. The amount of \ac{CMN} was found by filling the
pedestal-subtracted \ac{ADC} of every channel in a given read-out chip
into a histogram. This was done separately for every chip. The mean of
this distribution, which would be close to zero in the absence of
\ac{CMN}, was taken to be the value of the \ac{CMN} of that chip in the
event. The mean was found by the same procedure used to find the
pedestal: by calculating the mean in a small window about the most
populated bin. However, the number of entries in the window was
required to be above a certain threshold before a valid \ac{CMN}
correction could be performed. The \ac{CMN} offset was then subtracted
from the pedestal-subtracted \ac{ADC} value. Typical values of the
pedestal, noise and \ac{CMN}-corrected noise are shown in
\fig{calib:fig:sipednoise}. Notice that the \ac{CMN} correction reduced
the noise of the channels by roughly 20\%.

An additional \ac{CMN} effect occurred in the Ring detectors. The 
capacitance of pads in the Ring detectors varied due to the changing pad
size in the radial direction. This led to \ac{CMN} that was not due to
the read-out chip, but rather some Silicon effect, such as leakage
current. The value of this \ac{CMN} was found by assuming that all
channels in a Ring sensor could have some constant offset and that the
channels could have an offset that scaled with capacitance. Since the
capacitance scaled with radial distance, the \ac{CMN} offset was assumed
to have two components: one that was a constant offset and one that
varied linearly with radial distance.

\begin{figure}[t]
   \begin{center}
      \includegraphics[width=0.7\linewidth]{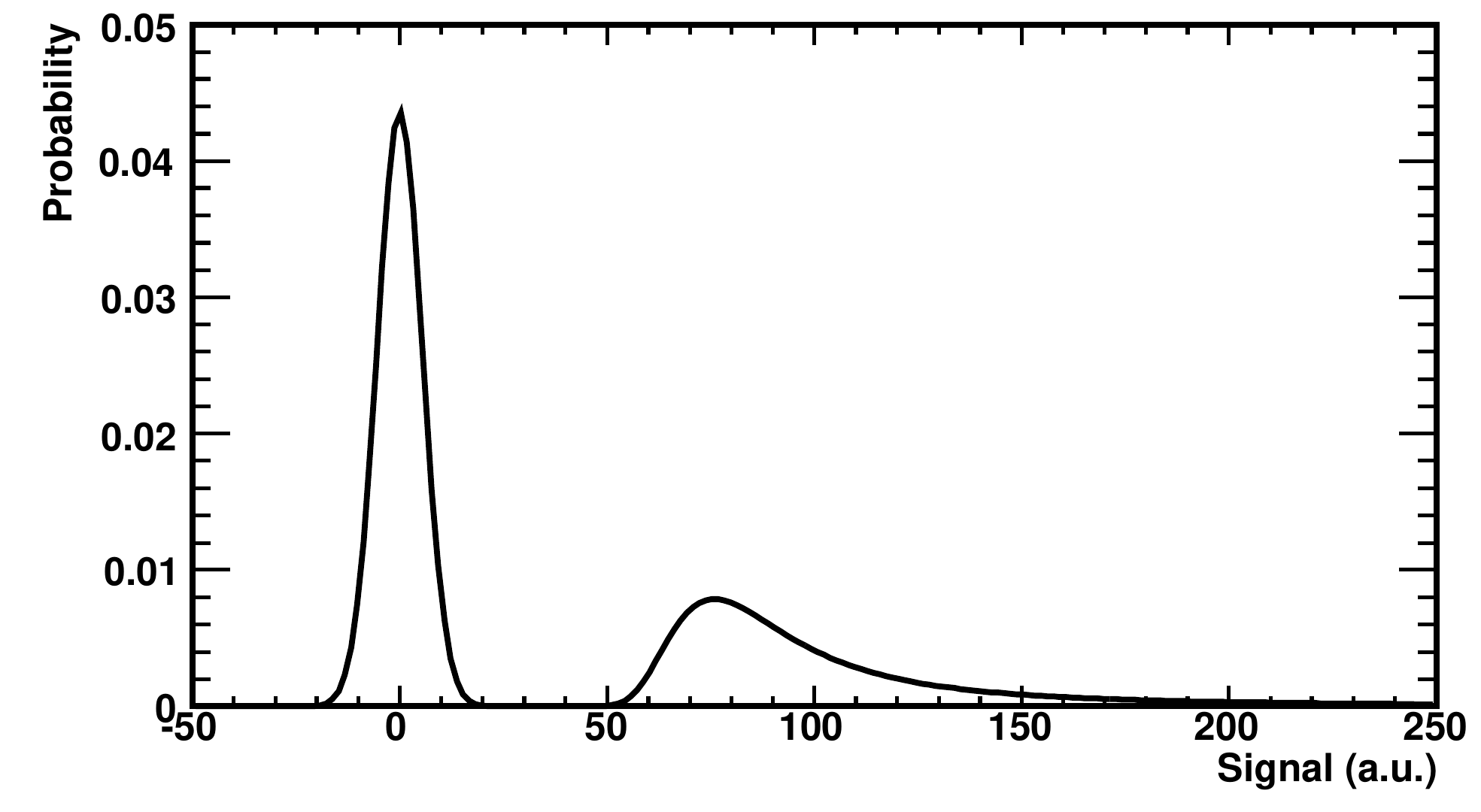}
   \end{center}
   \caption{\label{calib:fig:gauslan}
      A typical model distribution (with no baseline shift) used in the
      maximum-likelihood fit to correct \protect\acs{CMN} in the Ring 
      detectors.}
\end{figure}

The value of the baseline offset and slope were found on an
event-by-event basis. The correction had to be done for every event, but
occupancy in the Ring detectors could be very high in central
collisions. This meant it was not possible to find the baseline using
only information from channels with no signal from an ionizing particle.
However, a baseline offset would shift the signal of channels at a
particular radius by the same amount, whether the signal was due to a
hit or not. This fact was exploited by the use of a maximum-likelihood
fit. At each iteration of the fit, test values of the offset and slope
were used to find a test baseline. This baseline was then used to shift
a model signal distribution composed of a Gaussian pedestal and a Landau
\ac{MIP} peak, such as the one shown in \fig{calib:fig:gauslan}. Eight
model distributions were used, one for each radial position of pads in
the Ring detectors. The likelihood that the signal of one channel came
from the shifted model distribution was taken as the height of the
distribution at the location of the signal. Thus, the most likely
baseline would shift the model distributions such that most of the
signals of all channels in a Ring sensor were under a peak -- either the
pedestal peak or the \ac{MIP} signal peak. Once the best baseline offset
and slope parameters were obtained, the  baseline was calculated for
each radial position and subtracted from the signal in each channel of
the Ring detectors.

%---------------------------------------------------------------
\subsection{Energy Calibration}
\label{calib:si:gain}
%---------------------------------------------------------------

Once the offset of the signal in a Silicon channel had been corrected,
it was possible to extract the amount of energy deposited into a pad
using the \ac{ADC} signal in the channel. For this to be possible, it
was necessary to measure the factor by which the shaper and
pre-amplifier chips amplified the signal created by charge collected in
a Silicon pad. This factor is known as the gain of the chip, but the
term `gain' is also commonly used by physicists to refer to the ratio
between deposited energy and \ac{ADC} value. 

\begin{figure}[t]
   \begin{center}
      \includegraphics[width=0.5\linewidth]{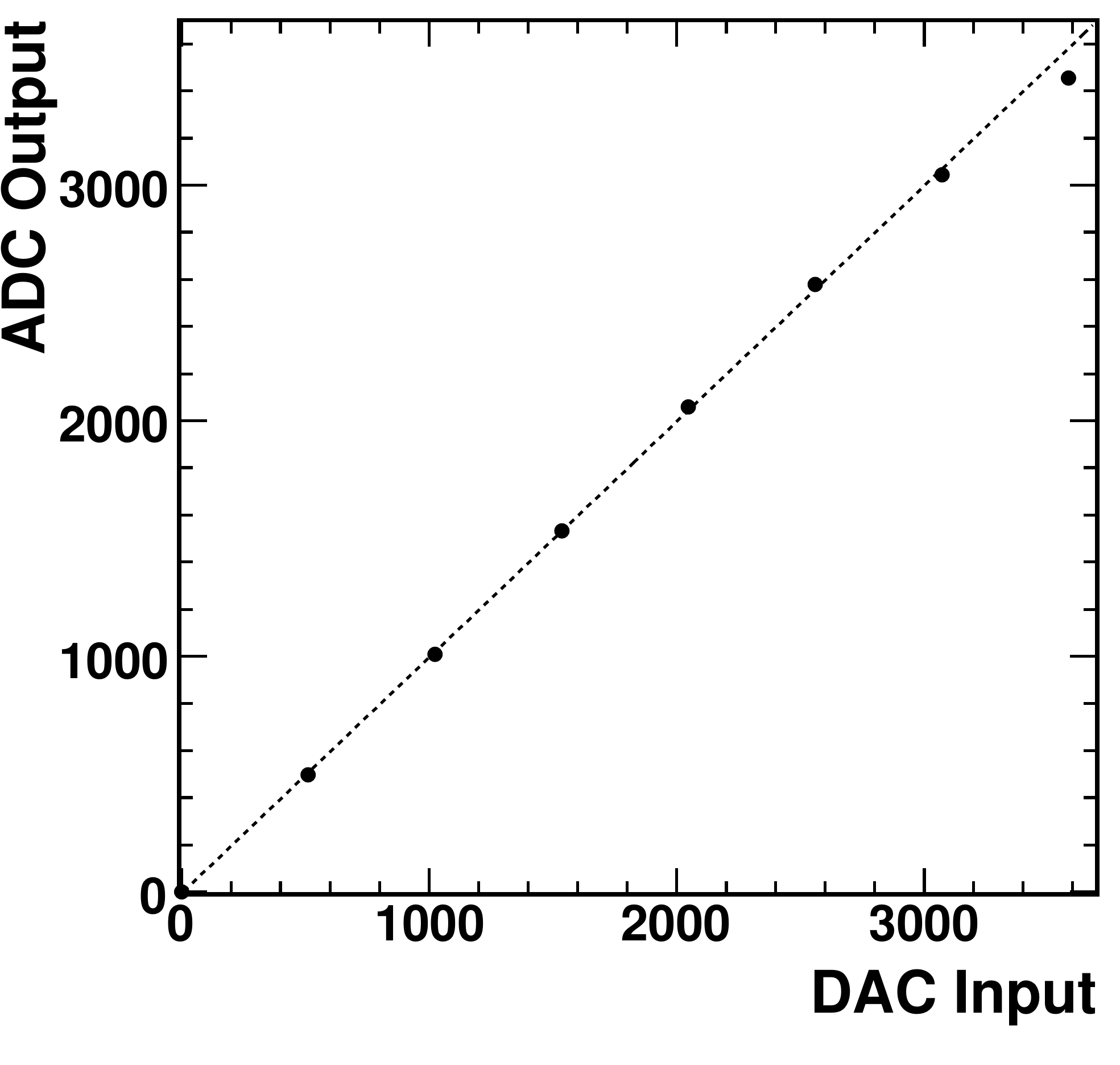}
   \end{center}
   \caption{\label{calib:fig:gainslope}
      The \protect\acs{ADC} output signals of a typical Silicon read-out
      chip generated by eight different \protect\acs{DAC} input signals.
      The slope of the line (1.00) is the gain of the chip.}
\end{figure}

Calibration circuitry was built-in to each read-out chip to allow the
electronic gain of the chip to be measured. This was done by sending
known \ac{DAC} signals into the chip and measuring the output \ac{ADC}
signals. Special calibration runs were taken by the experiment to
measure the gains of all Silicon read-out chips. In these runs, eight
different \ac{DAC} signals were used to measure the gain of the chip as
a function of input signal. As shown in \fig{calib:fig:gainslope}, the
relationship between input \ac{DAC} and output \ac{ADC} was linear over
almost the entire dynamic range of the read-out chips (the 12~bit
\ac{ADC} had values ranging from 0 to $2^{12}-1=4095$). The slope of
this line was the gain of the read-out chip. Typically, the gain of a
chip was very close to one.

The gain of the chip related the digital output \ac{ADC} value to the 
digital input \ac{DAC} value, but more information was required to use
this gain to extract the amount of deposited energy from a measured
\ac{ADC} value. The input \ac{DAC} signal was generated by charging a
capacitor, with capacitance $C_{t}$, such that it stored enough charge
to produce a signal of $V_{c}$ volts per \ac{DAC}. Typical values for
$C_{t}$ were around $2.14~\pF$, while the calibration electronics
typically produced a signal with
$V_{c}\approx4.33\times10^{-2}~\mvolt/\protect\rm\ac{DAC}$. These
parameters, combined with the gain of each chip, were sufficient to
relate an output \ac{ADC} value to the amount of charge that created
the signal sent into the chip. Of course, this relation was the same
for charge produced by the calibration circuitry as it was for charge
created by an ionizing particle. At room temperature, the average
energy required to create an electron-hole pair in silicon is
3.62~{\ev} (it increases as the temperature is decreased, to 3.72~{\ev}
at 77~Kelvin~\cite{1966PhRvL..17..653D}). Combining this with the
charge of an electron, $1.6022\times10^{-4}~\fC$, it was possible to
relate the output \ac{ADC} signal, $A_{s}$, to the amount of energy
deposited into the Silicon pad

\begin{align}
   \label{calib:eq:depE}
E_{dep}(A_{s}) &= \frac{A_{s}}{\rm gain} \: V_{c} \: C_{t} \:
                 \frac{3.62\ev/e}{1.6022\times10^{-4}\fC/e}\\
\notag &\approx \frac{A_{s}}{1 \protect\rm\ac{ADC}/\protect\rm\ac{DAC}} 
                 \prn{0.0433~\frac{\mvolt}{\protect\rm\ac{DAC}}}
                 (2.14~\pF)
                 \prn{22.6~\frac{\kev}{\fC}}\\
\notag &\approx A_{s} \times 2.09\kev/\protect\rm\ac{ADC}
\end{align}

%---------------------------------------------------------------
\subsection{Event-by-event Energy Correction}
\label{calib:si:corrections}
%---------------------------------------------------------------

\begin{figure}[t]
   \begin{center}
      \includegraphics[width=0.7\linewidth]{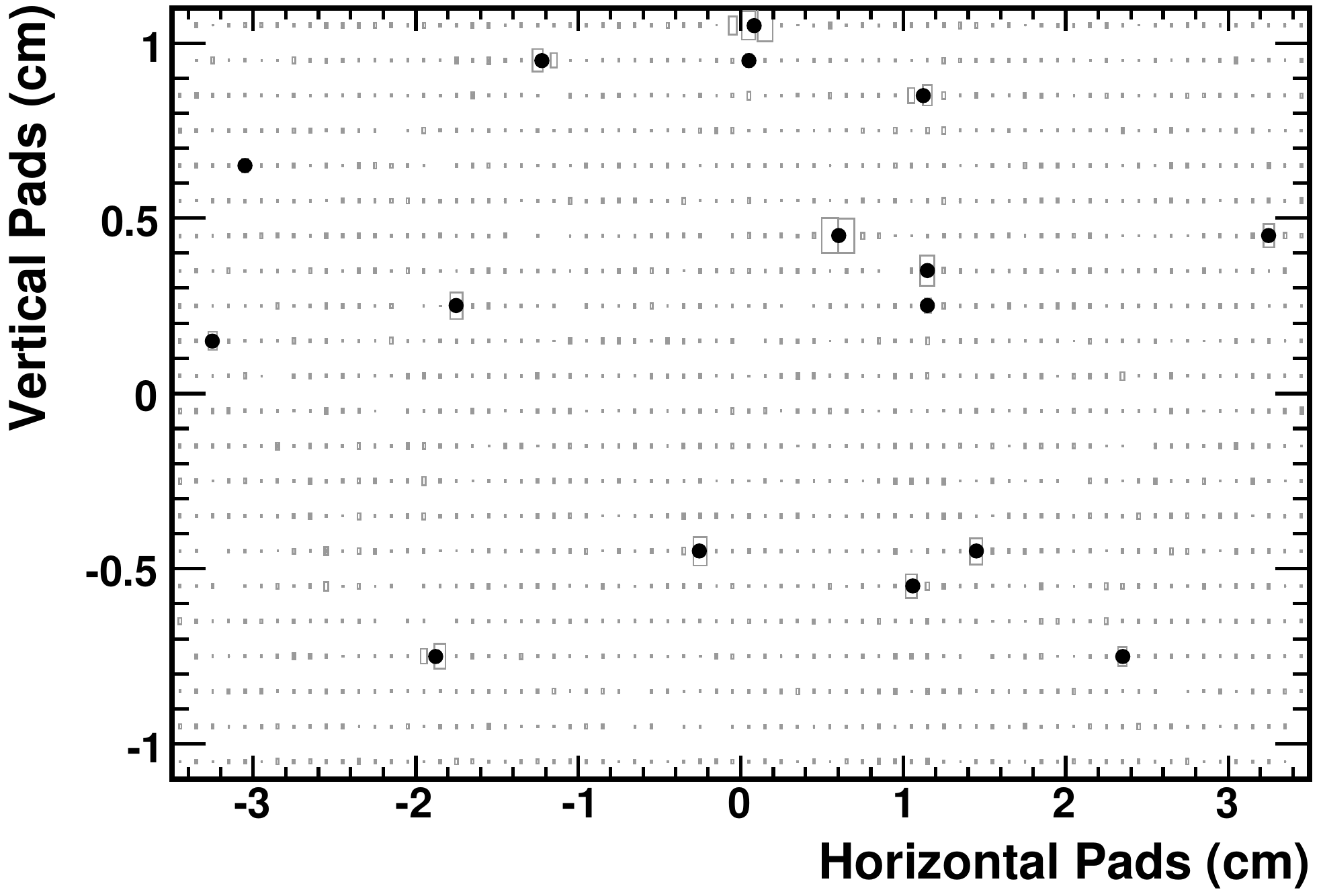}
   \end{center}
   \caption{\label{calib:fig:specHitMerge}
      An example of merged hits in a Spectrometer sensor. The size of
      the grey boxes shows the amount of energy deposited in each pad of
      the sensor. The black circles show the resulting merged hit
      locations.}
\end{figure}

With the amount of energy deposited into each Silicon pad known, it was
possible to improve the estimation of where the particle traversed the
detector. Electron-hole pairs could be collected in multiple pads if an
ionizing particle passed through the detector near the edge of a pad or
at an angle that was not normal to the detector. These situations were
taken into account by attempting to merge several adjacent hits. For
Spectrometer channels, hit merging was performed in the horizontal
direction. For most Spectrometer silicon sensors (the only exception
being Type~1 sensors), pad sizes were much smaller in the horizontal
direction than in the vertical and it was not unreasonable to expect
that adjacent hits in the vertical direction were due to more than one
particle. Adjacent Spectrometer channels that had recorded more than
15\% of the energy a \ac{MIP} would deposit were used to find the merged
hit. The locations of up to eight such adjacent hits were averaged, each
weighted by the energy of the hit. The averaged position was taken as
the location of the hit, and the sum of the deposited energy of the
individual hits was taken as the energy of the merged hit. Merged hits
were required to have a total deposited energy greater than  half of a
\ac{MIP}. An example of merged hits on a Spectrometer sensor is shown in
\fig{calib:fig:specHitMerge}.

%---------------------------------------------------------------
\section{Zero-Degree Calorimeter Energy Calibration}
\label{calib:zdc}
%---------------------------------------------------------------

The {\phob} \ac{ZDC} detectors were used to measure the total energy of
particles. Raw signals measured by the detectors were corrected for
an overall offset and a calibration scale-factor.

%---------------------------------------------------------------
\subsection{{\cheren} Detectors}
\label{calib:zdc:ceren}
%---------------------------------------------------------------

Some materials have refractive properties that slow the progression of
light through the material. While light is slowed in the material, a
fast moving charged particle will not be. Thus it is possible for a
charged particle to pass through the material at speeds greater than
the speed at which light can travel through the material. When this
happens, an electromagnetic shock wave occurs. This shock wave is known
as {\cheren} radiation. The goal of the {\phob} \ac{ZDC} detectors was to
stop neutrons and measure the {\cheren} radiation generated by charged
particles in the resulting particle shower. The total {\cheren}
radiation collected could then be related to the energy of the original
neutron.

In an infinite radiating medium, all {\cheren} radiation is emitted in a
single coherent wave front, shaped like a cone. For this ideal case, the
{\cheren} light is emitted at a single angle relative to the charged
particle's trajectory. In a finite medium, there is a set of discrete
angles at which light is emitted. Nevertheless, the majority of light is
emitted at the same angle as for the ideal case, so it is of interest to
determine this angle. Because the index of refraction of the radiating
material, $n$, is known,\footnote{Note that the index of refraction may
depend on the wavelength of light traveling through the material.} it is
not difficult to calculate the angle at which light is emitted from the
{\cheren} cone to be $\theta_{C} = \arccos\prn{\frac{c}{v n}}$, where
$v$~is the speed of the charged particle and $c$~is the speed of light
in a vacuum~\cite{Leo94:cheren}. Thus, the speed of the particle must be
greater than the speed of light in the material ($c/n$), or the angle is
undefined and no {\cheren} radiation occurs.

{\cheren} detectors are commonly built to exploit the dependence of the
emission angle on particle velocity and index of refraction in two ways.
First, the index of refraction of the radiating medium can be changed to
alter the minimum kinetic energy a particle of a certain type must have
in order to be observed. This is commonly done by using a liquid or gas
whose index of refraction can be changed by varying its pressure. Such
an energy threshold can be used to reject particles from a measurement
that would otherwise contribute to the measurement's background. Second,
the speed of the particle can be identified by measuring the angle of
the {\cheren} cone. This can be used to determine the type of particle
being observed. Note that neither of these techniques were used by the
{\phob} \acp{ZDC}.

%---------------------------------------------------------------
\subsection{Photomultipliers}
\label{calib:zdc:pmts}
%---------------------------------------------------------------

Light generated from {\cheren} radiation is collected and converted to an
electric signal by \acfp{PMT}. Photomultipliers generally consist of
several different components: a photocathode that converts photons to
electrons, an electron input system that collects the photo-electrons, a
dynode string that creates an electron cascade and finally an anode from
which the electronic signal can be read out~\cite{Leo94:pmt}.

The photocathode produces electrons via the photoelectric effect. This
quantum mechanical effect is described by the well known function, the
discovery of which won Einstein his Nobel prize, $E = h \nu - \phi$,
where $E$~is the energy of the emitted electron, $\nu$~is the frequency
of the incident photon and $\phi$~is the work function of the emitting
material. Thus there is a minimum frequency of light that can produce a
photoelectron in a given material. However, even for photons having a
frequency above this threshold, it is not certain that an electron will
be emitted and the probability of producing a photoelectron in a given
material can depend strongly on the frequency of the photon. For physics
experiments, photocathode materials are chosen such that the efficiency
of producing photoelectrons, known as the quantum efficiency, is peaked
for the photon wavelengths produced by the {\cheren} or scintillator
detector being constructed. Typically, such light is in the visible
range, having wavelengths around 400~{\nm} (which appears blue).

The photoelectrons are then collected and directed to the multiplier
section. Collection in the electron input system is typically achieved
by an electric field that directs electrons onto the first dynode of the
multiplier section. A well designed electron input system can collect
all electrons from the photocathode, independent of the position from
which they were produced on the photocathode. That is, the electron
input system should collect all electrons with the same efficiency and
direct all electrons from the cathode to the dynode string in the same
amount of time.

The dynode string amplifies the photoelectron current by using a series
of secondary emission electrodes, known as dynodes (hence the name). At
the first dynode, the photoelectrons strike the electrode and release
secondary electrons from the dynode material. The secondary electrons
are then accelerated through a static electric field to the next
electrode. This process is repeated at each step of the dynode string,
resulting in an electron cascade. A well designed electron multiplier
uses dynode material that has a high secondary electron emission factor
yet does not emit electrons due to thermal energy (which would result in
noise). It is also important that the dynode material can emit secondary
electrons in a stable manner when placed under high currents. After
being amplified by the electron multiplier, the photoelectron current is
collected at the anode and passed out of the \ac{PMT} as an electric
signal to be recorded by a \ac{DAQ} system.

%---------------------------------------------------------------
\subsection{Pedestal and Noise Correction}
\label{calib:zdc:pedcor}
%---------------------------------------------------------------

As with all detectors, the \ac{ZDC} signals had to be corrected for
various offsets. The pedestal was mainly due to electronic offsets in
the readout system and leakage currents in the phototubes. Offsets that
fluctuated from one event to the next included electronic noise as well
as statistical and thermal fluctuations in secondary electron emission
within the \acp{PMT}. Thermal emission of electrons by the photocathode
and dynodes resulted in an electric current that was independent of any
light being incident on the \ac{PMT}. Statistical fluctuations, on the
other hand, originated from the inherit randomness associated with the
photoelectron and secondary electron emission processes.

The pedestal and noise of \ac{ZDC} channels was measured by observing
the \ac{ADC} signal distribution of each channel from a collection of
events in which no signal was expected to be present. Such events could
be identified by checking that the channel's \ac{TDC} produced no signal
or by requiring that the \acs{RHIC} accelerator was not colliding
particles during that particular bunch crossing. The resulting \ac{ADC}
distribution was then fit with a Gaussian, from which the mean and width
were taken to be measures of the pedestal and noise in the channel.

%---------------------------------------------------------------
\subsection{Energy Calibration}
\label{calib:zdc:gain}
%---------------------------------------------------------------

The \ac{ZDC} detectors had the great advantage that the majority of
particles it was built to observe had a known energy. Most free
spectator neutrons had an energy equal, or very close, to 100~{\gev}.
This fact was exploited to calibrate the \ac{ZDC} \ac{ADC} signal. A
\ac{ZDC} module could detect one, two or, with decreasing probability,
some larger number of neutrons. Because each neutron deposited roughly
the same amount of energy, \emph{neutron peaks} were expected (and
observed) in the \ac{ADC} distribution of a \ac{ZDC} channel from many
collisions. The first peak could then be assumed to correspond to an
incident  energy of 100~{\gev}, the second to 200~{\gev} and so on.

The energy calibration of each \ac{ZDC} detector, i.e. the Au-side and
d-side, was done separately and in several steps. First, the
pedestal-subtracted \ac{ADC} distribution of a particular \ac{ZDC}
channel was plotted for collision events in which the channel reported a
signal above the pedestal, while the other two channels of the detector
did not. The resulting one-neutron peak of this distribution was then
fit with a Gaussian. The mean of the Gaussian divided by 100~{\gev} gave
an estimate of the energy calibration factor for the channel. This
method yielded calibration factors for the three channels of a \ac{ZDC}
detector that agreed to within about 20\%.

The calibration of a \ac{ZDC} detector was then improved by varying the
calibration factors of individual channels such that the best possible
energy resolution was achieved. First, the calibration factor of the
\ac{ZDC} module nearest the \ac{IP} was held constant, while the
calibration of the middle module was varied over $\pm50\%$. The sum of
the signals in the two channels was then plotted for each step. Each
resulting one-neutron peaks was fit with a Gaussian. The best relative
calibration factor between the front two modules, $f_1$, was taken to be
that which minimized the width of the resulting peak. This process was
then repeated by holding $f_1$ constant and varying the relative
calibration of the module furthest from the \ac{IP}. At each step, the
sum of the signals from all channels was then plotted and fit with a
Gaussian. The best relative calibration factor between the back module
and the other two modules, $f_2$, was taken to be that which minimized
the width of the resulting peak. For events in which all three modules
of the deuteron-side \ac{ZDC} showed a signal above the pedestal, this
method yielded an energy resolution of $\sigma/E\approx30\%$ for that
detector ($\sigma$ being the width of the one-neutron peak).

%---------------------------------------------------------------
\section{Proton Calorimeter Energy Calibration}
\label{calib:pcal}
%---------------------------------------------------------------

Like the \acp{ZDC}, the \ac{PCAL} detectors were also used to measure
the total energy of particles. Raw signals measured by the detectors
were corrected for an overall offset and a calibration scale-factor.

%---------------------------------------------------------------
\subsection{Scintillator Detectors}
\label{calib:pcal:scint}
%---------------------------------------------------------------

Particles passing through scintillator material collide with the atoms
and molecules in the material. The excited atoms and molecules decay to
their ground state by radiating the extra energy in the form of photons.
The detector measures the number of photons produced, and from this
information the amount of energy deposited into the detector can be
determined.

Scintillator detectors are used both for their fast time response and
for the linear proportionality of light output with deposited energy. A
good scintillator material for particle detection is one that
(a)~efficiently radiates excitation energy as fluorescent light, (b)~is
transparent to the fluorescent light, (c)~emits fluorescent light at
wavelengths that can be efficiently detected (by \acp{PMT}) and
(d)~quickly returns to its ground state after being excited.

The time evolution of light output from a scintillator material can be
expressed by a two-component exponential

\begin{equation}
   \label{calib:eq:lsci}
   N = A \exp\prn{\frac{-t}{\tau_{f}}} + B \exp\prn{\frac{-t}{\tau_{s}}}
\end{equation}

\noindent%
where $\tau_{f}$ and $\tau_{s}$ are the fast and slow time-decay
constants, respectively. Note that while the rise time of light output
is finite, it is typically so much faster than the decay time that it
has been taken to be zero in \eq{calib:eq:lsci}. For some materials, the
prompt and delayed components of the decay time can depend significantly
on the type of particle that excites the scintillator. In such cases, it
is possible to identify the type of particle by examining the resulting
pulse shape (this technique was not employed for the {\phob} \acp{PCAL}).
The scintillating material used in the \ac{PCAL} detectors had a prompt
decay time of 3.2~{\ns} and no significant delayed component. Light
emitted by scintillator material is detected using photomultiplier tubes
(see \sect{calib:zdc:pmts}).

%---------------------------------------------------------------
\subsection{Pedestal and Noise Correction}
\label{calib:pcal:pedcor}
%---------------------------------------------------------------

\begin{figure}[t!]
   \begin{center}
      \includegraphics[width=0.8\linewidth]{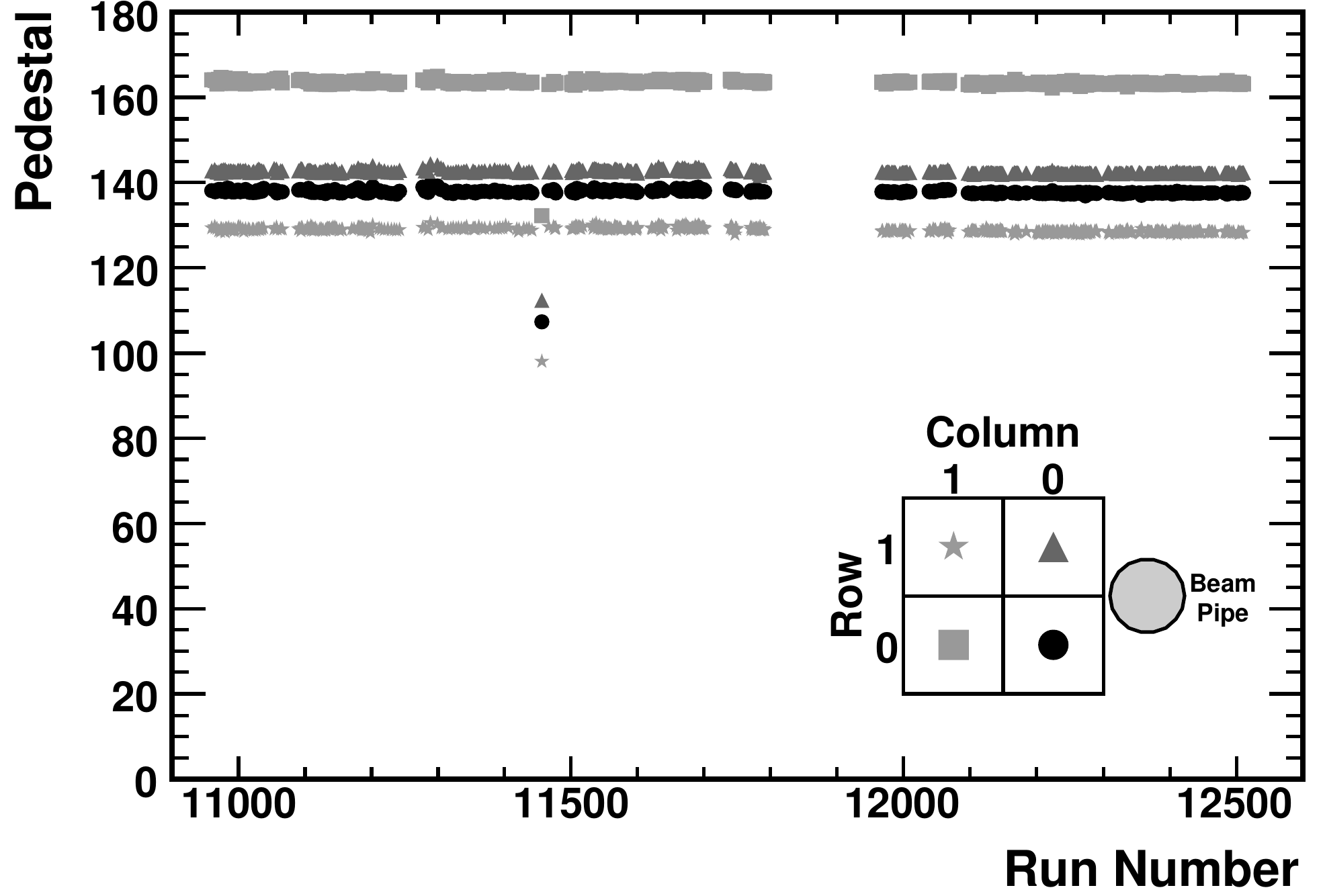}
   \end{center}
   \caption{   \label{calib:fig:pcalPzPedsdA}
      The pedestal values for each channel in the \protect\ac{d-PCAL}
      versus the {\phob} \protect\ac{DAQ} run number, which is increasing
      with time. The large gap separates the {\dAu} physics run (lower
      numbers) from the {\pp} physics run (higher numbers). Also shown is
      the row and column numbering convention for \protect\ac{PCAL}
      detectors.}
\end{figure}

The main sources of signal offsets in the \ac{PCAL} detectors were
similar to those of the \acp{ZDC}; namely electronic offsets in the
readout system and various \ac{PMT} effects. The pedestal and noise were
determined by examining the \ac{ADC} distribution of each channel over
at least six hundred \emph{heartbeat} events. Heartbeat events were not
triggered by any physical collision property, but rather were periodic
readouts of the entire detector system by the {\phob} \ac{DAQ}. Thus, it
was very improbable that any physical particle had deposited energy into
a \ac{PCAL} channel during the 100~{\ns} window in which the detector's
signals were recorded. The mean and \ac{RMS} of each distribution was
then directly calculated and taken to be the channel's pedestal and
noise, respectively. The pedestals of the \ac{PCAL} channels were found
to be quite stable over time, as shown in \fig{calib:fig:pcalPzPedsdA}.

\begin{figure}[t!]
   \centering
   \subfigure[A Pedestal Distribution]{
      \label{calib:fig:exPedDist}
      \includegraphics[width=0.4\linewidth]{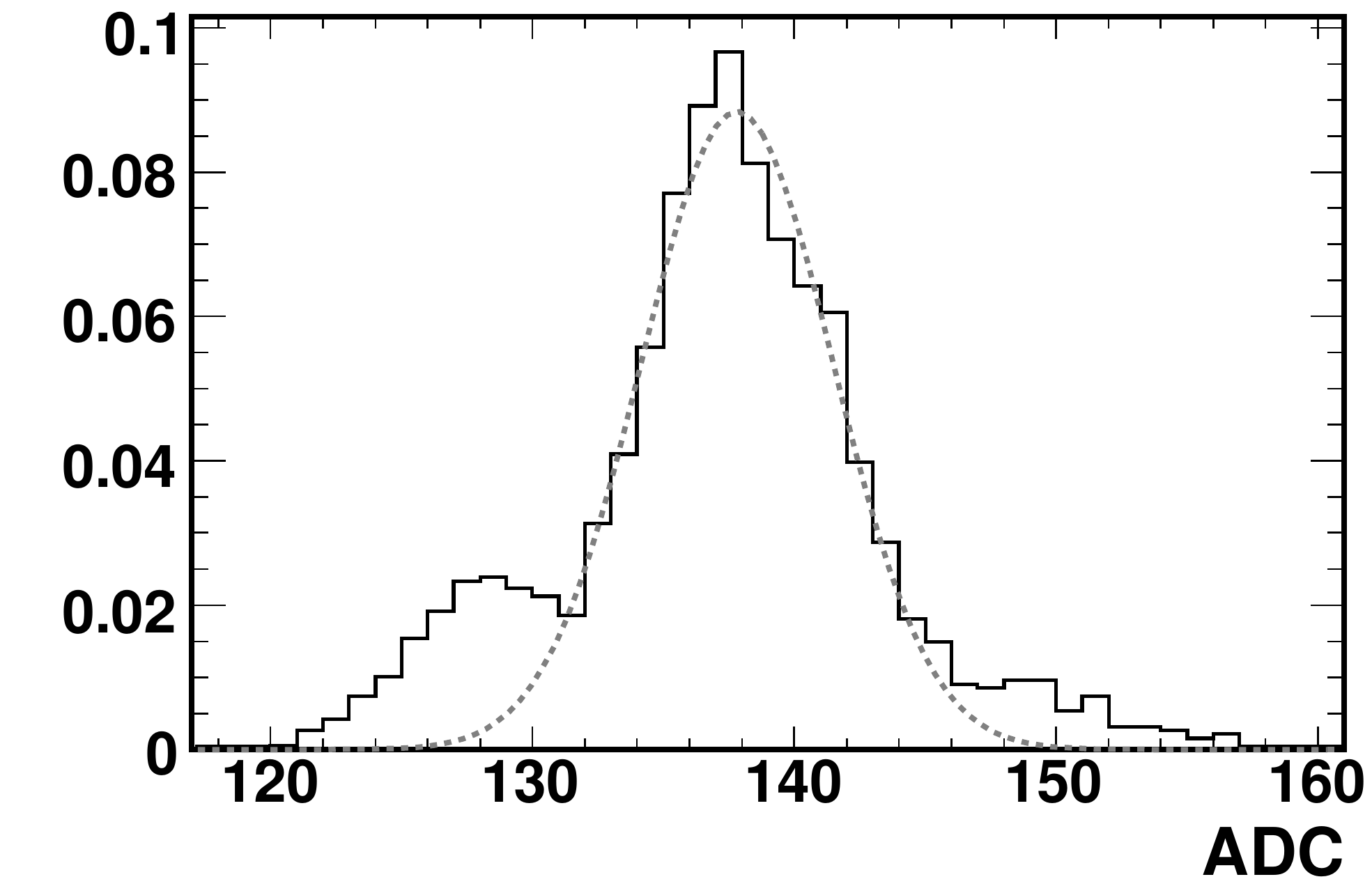}
   }
   \subfigure[ADC vs Time]{
      \label{calib:fig:pcalADCvsEvt}
      \includegraphics[width=0.4\linewidth]{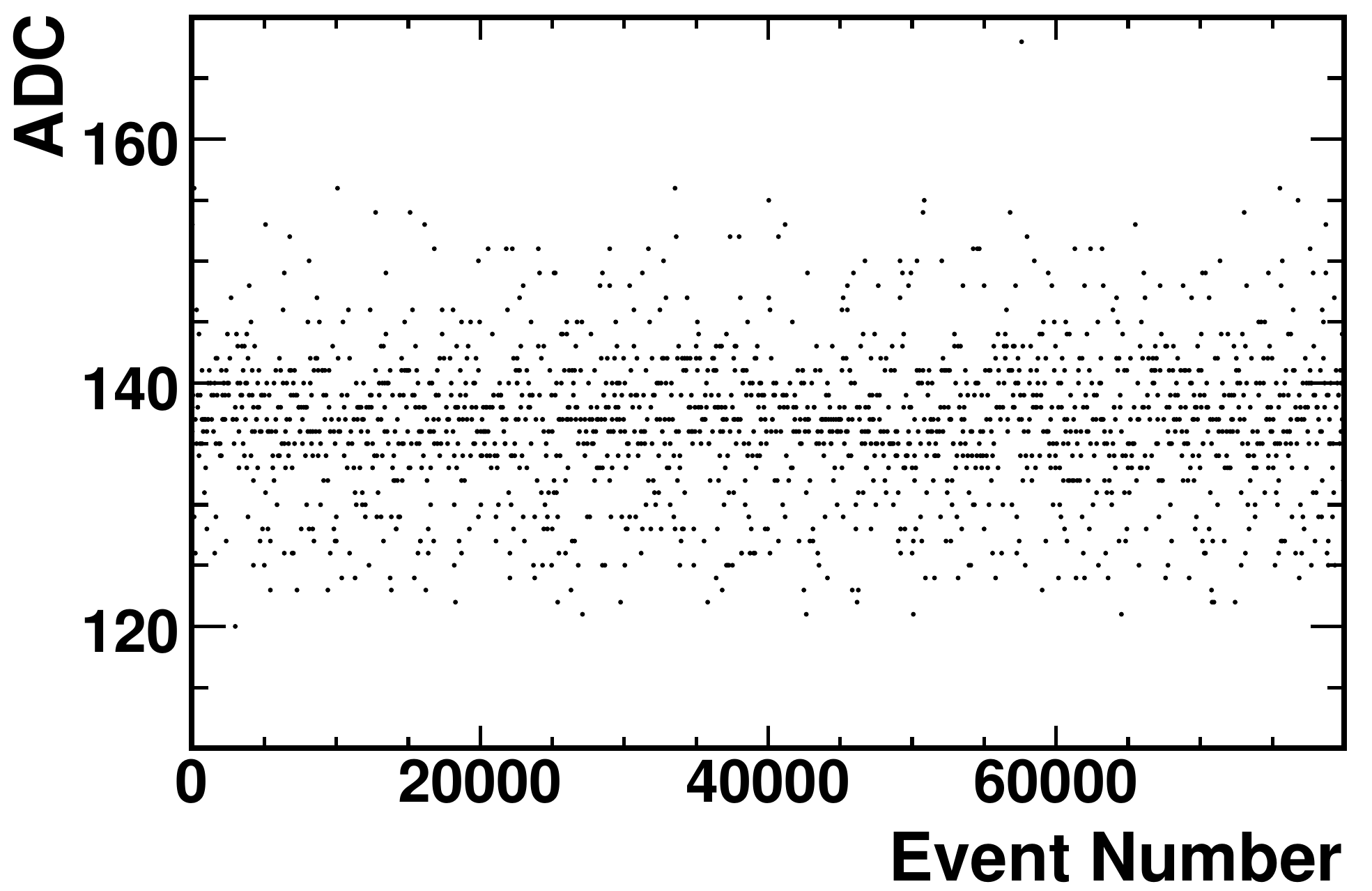}
   }
   \subfigure[Channel Correlation]{
      \label{calib:fig:pcalCMN}
      \includegraphics[width=0.4\linewidth]{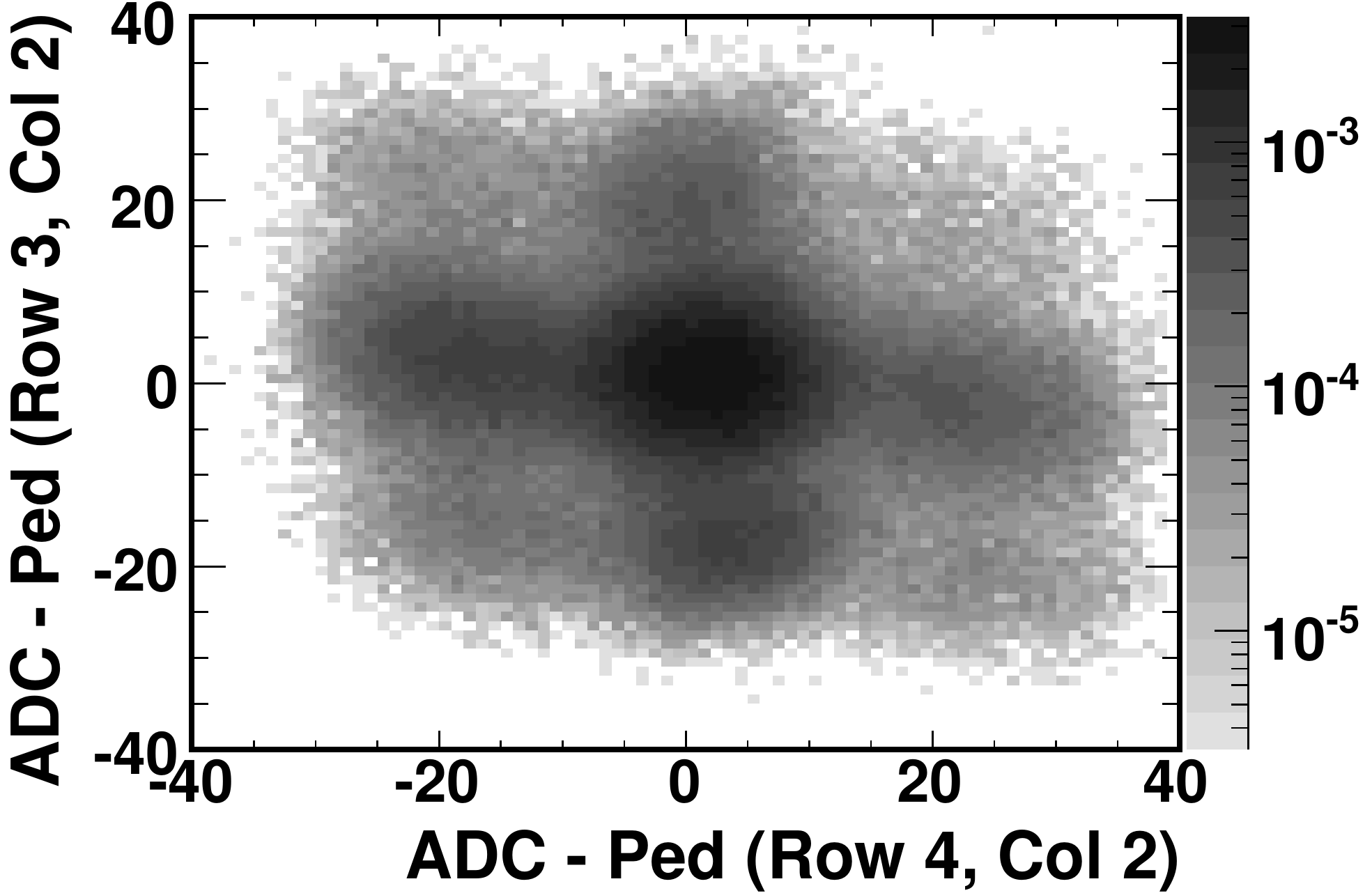}
   }
   \caption{   \label{calib:fig:pcalpeds}
      Structure in the pedestal \protect\acs{PCAL} \protect\acs{ADC}
      distribution. \subref{calib:fig:exPedDist}~The \protect\acs{ADC}
      distribution of signals in a \protect\acs{PCAL} channel from
      heartbeat events. A Gaussian fit to the central peak is shown by
      the grey, dashed line to guide the eye. Note the secondary peak at
      the low end of the pedestal and the non-Gaussian tail at the high
      end. \subref{calib:fig:pcalADCvsEvt}~Signals of a \protect\acs{PCAL}
      channel from heartbeat events versus time. No baseline drift is
      observed. \subref{calib:fig:pcalCMN}~Pedestal-subtracted
      \protect\acs{ADC} signals  from heartbeat events of one
      \protect\acs{PCAL} channel versus an adjacent channel. No
      correlation is present that would suggest a \protect\acs{CMN}
      effect.}
\end{figure}

While the pedestals were stable with time in the \ac{PCAL} detectors,
the shape of the pedestal peak was not simply a Gaussian, as would be
expected given a stable, constant offset and some random noise. In
addition to a central peak, some \ac{PCAL} channels exhibited large
tails and/or smaller, displaced peaks. An example is shown in
\fig{calib:fig:exPedDist}. While an offset that shifts or jumps with
time in a systematic, rather than random, way could explain such an
\ac{ADC} distribution, such behavior was not seen.
\Fig{calib:fig:pcalADCvsEvt} shows that the \ac{ADC} signals reported by
a single \ac{PCAL} channel were not correlated with time in a systematic
way. In addition, it was found that the effect was due to the detector,
most likely the \acp{PMT}, and not the readout system, as the \ac{ADC}
distributions were not changed by connecting the detector to an
alternate readout system. Finally, it was thought that the effect may be
due to common-mode noise. However, no correlation of signals between
channels was observed in heartbeat events, as shown in
\fig{calib:fig:pcalCMN}. Thus, since the effect could not be corrected
for, it was intentionally taken into account by estimating the pedestal
and noise from a direct calculation of the mean and \acs{RMS} of the
full distribution, rather than from a Gaussian fit to the central peak.

%---------------------------------------------------------------
\subsection{Energy Calibration}
\label{calib:pcal:gain}
%---------------------------------------------------------------

%---------------------------------------------------------------
\subsubsection{Au-PCAL Calibration}
\label{calib:pcal:aupcalgain}
%---------------------------------------------------------------

\begin{figure}[t!]
   \begin{center}
      \includegraphics[width=0.5\linewidth]{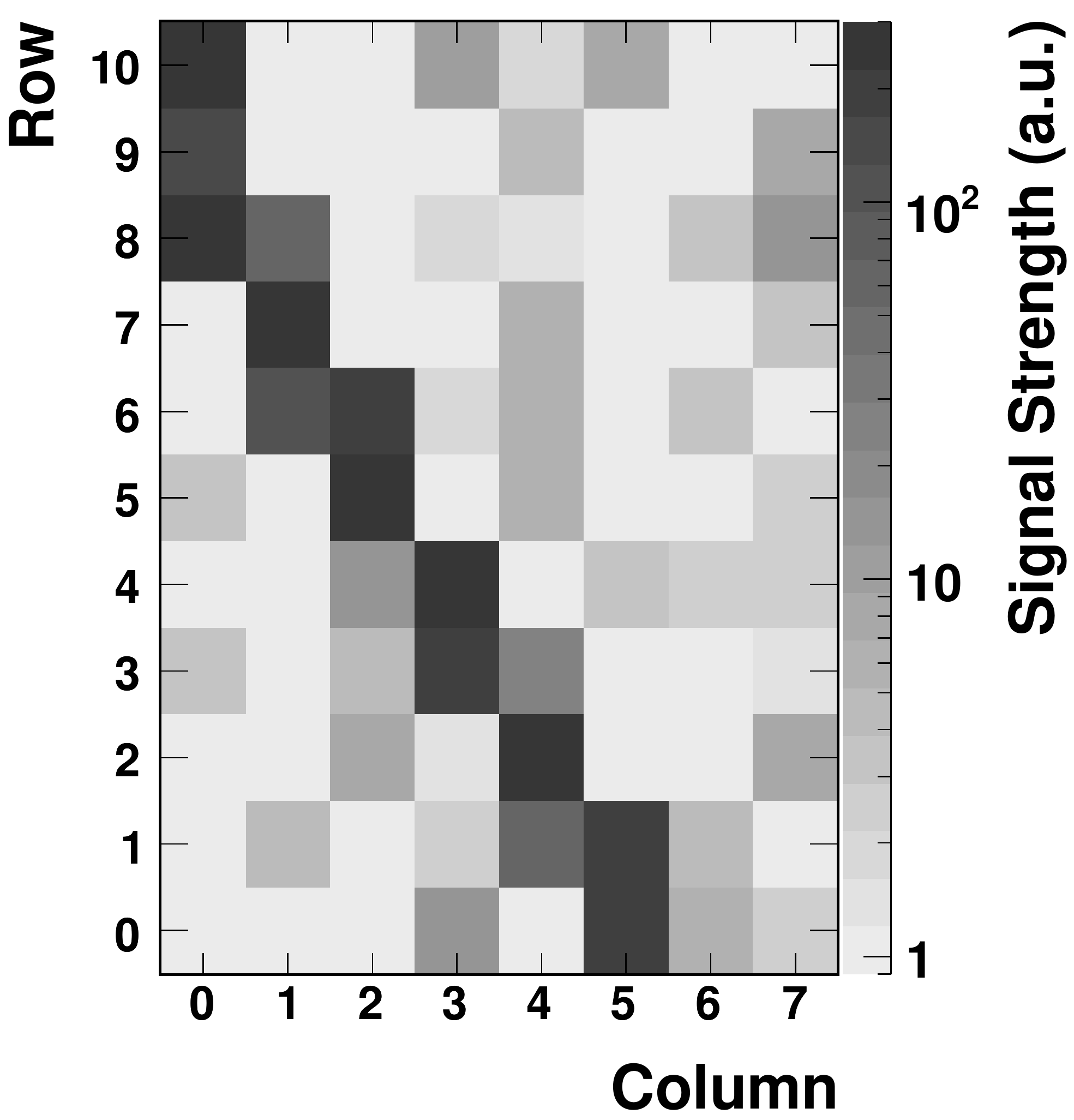}
   \end{center}
   \caption{   \label{calib:fig:pcalCosmEvtDisp}
      A display of the energy deposited into the \protect\acs{Au-PCAL}
      by a cosmic ray. Modules are represented by squares; the dark grey
      diagonal ``line'' shows the path of the cosmic ray. Light grey
      squares simply show noise in the \protect\acs{Au-PCAL} read-out
      and detector electronics.}
\end{figure}

The energy deposited into the \ac{Au-PCAL} was determined by using
cosmic rays to calibrate the detector. This was necessary because the
\ac{Au-PCAL}, unlike the \ac{ZDC} and the \ac{d-PCAL}, did not observe a
single-hadron peak in the {\dAu} data. A sample display of a cosmic ray
detected by the \ac{Au-PCAL} is presented in
\fig{calib:fig:pcalCosmEvtDisp}. To calibrate the detector with cosmic
rays, it was necessary to estimate the amount of energy that cosmic ray
muons were expected to deposit in a \ac{PCAL} module. This was done
using the Bethe-Bloch formula (see \eq{calib:eq:bb}) and assuming that,
on average, a cosmic muon traveled about 10.8~{\cm} through a
$10\times10~\cm$ calorimeter block, of which 1.9~{\cm} was through
scintillator material and 8.9~{\cm} was through lead. From this, it was
estimated that only muons having roughly 1.5~{\gev} kinetic energy were
able to traverse the entire calorimeter and deposit energy into cosmic
ray trigger modules both above and below the \ac{Au-PCAL}. The average
kinetic energy of muons incident on the calorimeter was expected to be
around 4~{\gev}. Finally, it was estimated that on average, cosmic ray
muons would deposit around 145~{\mev} of energy into a \ac{PCAL} module
(lead and scintillator combined).

This information was used to calibrate the detector by assuming that
145~{\mev} of energy deposited by a muon should produce the same amount
of scintillator light as 145~{\mev} of energy deposited by a proton
(\eq{calib:eq:esame}). However, the calibration was further complicated
by the fact that the voltages used to power the calorimeter's \acp{PMT}
during cosmic ray data taking were different from those used to take
{\dAu} physics data. The differing voltages were required since spectator
protons from a high-energy {\dAu} collision were expected to deposit much
more energy than a cosmic ray muon. Thus, it was necessary to measure
how the overall amplification of the \acp{PMT} changed with voltage.
This was done for a collection of tubes using a \acf{LED} as a light
source, and it was found that the gain of the tubes was described by

\begin{equation}
   \label{calib:eq:pmtgain}
A = k V^{8.93}
\end{equation}

\noindent%
where $A$ is the signal reported by the tube, $V$ is the voltage and $k$
is some constant of proportionality.

The energy deposited by a cosmic ray muon was

\begin{equation}
   \label{calib:eq:cmuon}
E_{C} = G_{C} A_{C}
\end{equation}

\noindent%
where $E_{C}$ is the energy deposited by the muon, $G_{C}$ is the
energy calibration factor at the cosmic ray voltage and $A_{C}$ is the
signal generated by the muon. From \eq{calib:eq:pmtgain}, it was then
possible to find the energy calibration factor for {\dAu} collisions,
$G_{D}$, to be

\begin{align}
E_{D} &= E_{C}\label{calib:eq:esame}\\
G_{D} A_{D} &= G_{C} A_{C}\notag\\
G_{D} k V_{D}^{8.93} &= G_{C} k V_{C}^{8.93}\notag\\
G_{D} &= G_{C} 
          \prn{\frac{V_{C}}{V_{D}}}^{8.93}\label{calib:eq:dgain}
\end{align}

The value of $G_{C}$ was calculated given that $E_{C} = 145~\mev$ for
the average cosmic ray muon and that the signal reported by such a muon
was just the mean of the muon-peak in the cosmic ray data, $A_{CP}$.
Thus the energy deposited by a proton from a {\dAu} collision, $E_{D}$,
as a function of the signal reported by a module, $A_{D}$, was

\begin{align}
E_{D}(A_{D}) &= G_{D} A_{D}\notag\\
      &= G_{C} \prn{\frac{V_{C}}{V_{D}}}^{8.93} A_{D}\notag\\
      &= \frac{145~\mev}{A_{CP}} 
          \prn{\frac{V_{C}}{V_{D}}}^{8.93} A_{D}\notag\\
E_{D}(A_{D}) &= 145~\mev \frac{A_{D}}{A_{CP}} 
          \prn{\frac{V_{C}}{V_{D}}}^{8.93}\label{calib:eq:prote}
\end{align}

\begin{figure}[t]
   \begin{center}
      \includegraphics[width=0.5\linewidth]{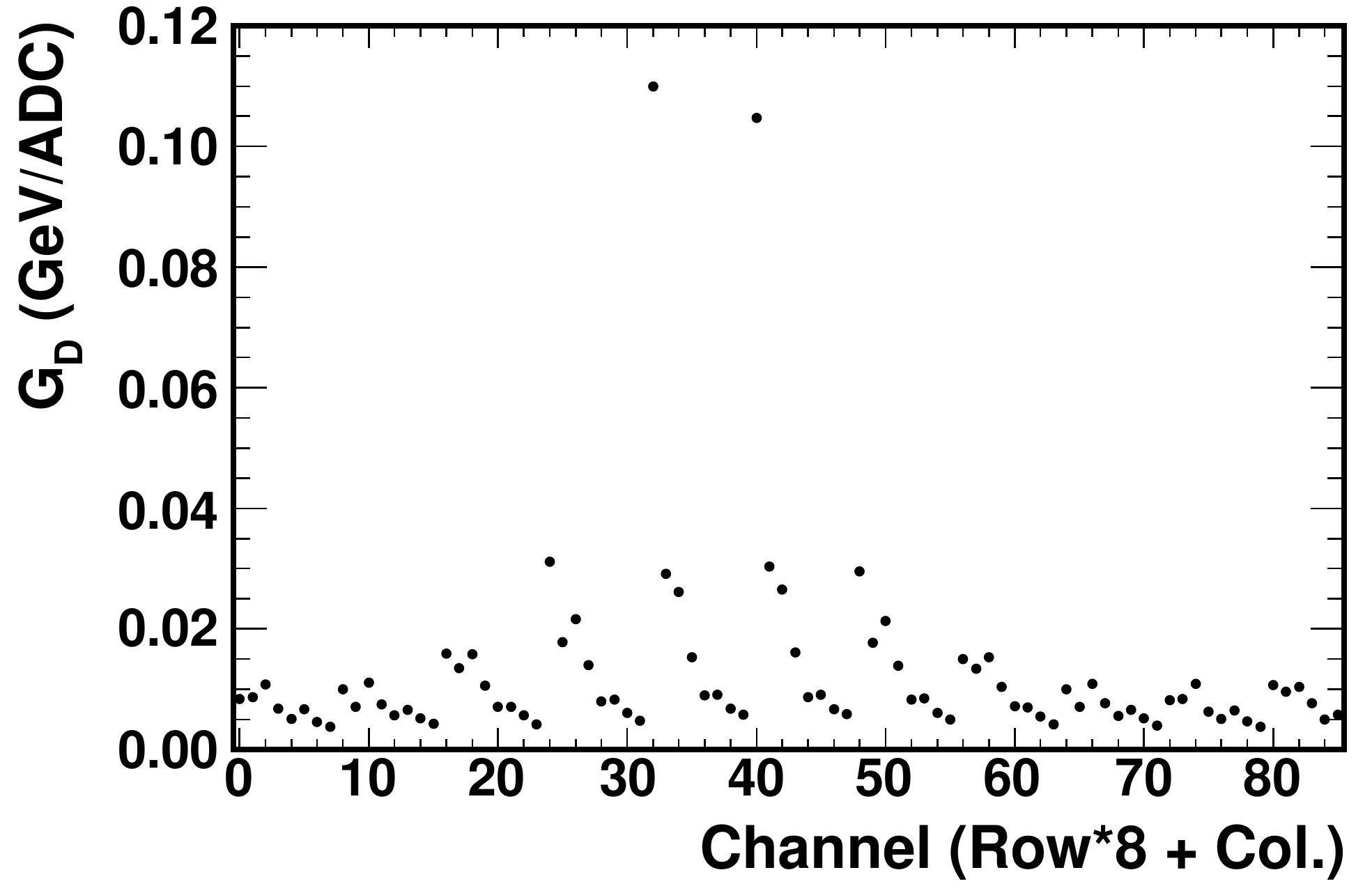}
   \end{center}
   \caption{   \label{calib:fig:pcalnzGains}
      Values of $G_{D}$ used during {\dAu} physics running for channels
      in the \protect\acs{Au-PCAL}.}
\end{figure}

Typical values of $V_{C}$ were \mbox{$\sim1450~\volts$}, while typical
values of $V_{D}$ were \mbox{$\sim1150~\volts$}. The values of $G_{D}$
used to process the {\dAu} data are presented in
\fig{calib:fig:pcalnzGains}. The spread in gain values is a result of
the variation of $V_{D}$ from module to module. Voltages used during
{\dAu} physics running were adjusted for each module such that a
majority of the dynamic range of \ac{ADC} signals was used. Note in
particular the two channels with very large gain. These channels were
nearest the beam pipe, in column zero, rows four and five (see
\fig{calib:fig:pcalCosmEvtDisp}). Much more energy was deposited into
these two channels during {\dAu} physics running than into any of the
other channels. To compensate for this, filters were installed into the
\acp{PMT} of these two channels that absorbed 80\% of the scintillator
light. This allowed these \acp{PMT} to be operated at voltages that were
closer to, but still somewhat lower than, the voltages of \acp{PMT} on
other channels. This lower voltage was the cause of the large gain
values for these two modules.

%---------------------------------------------------------------
\subsubsection{PCAL Simulations}
\label{calib:pcal:sims}
%---------------------------------------------------------------

\begin{figure}[t]
   \begin{center}
      \includegraphics[width=0.5\linewidth]{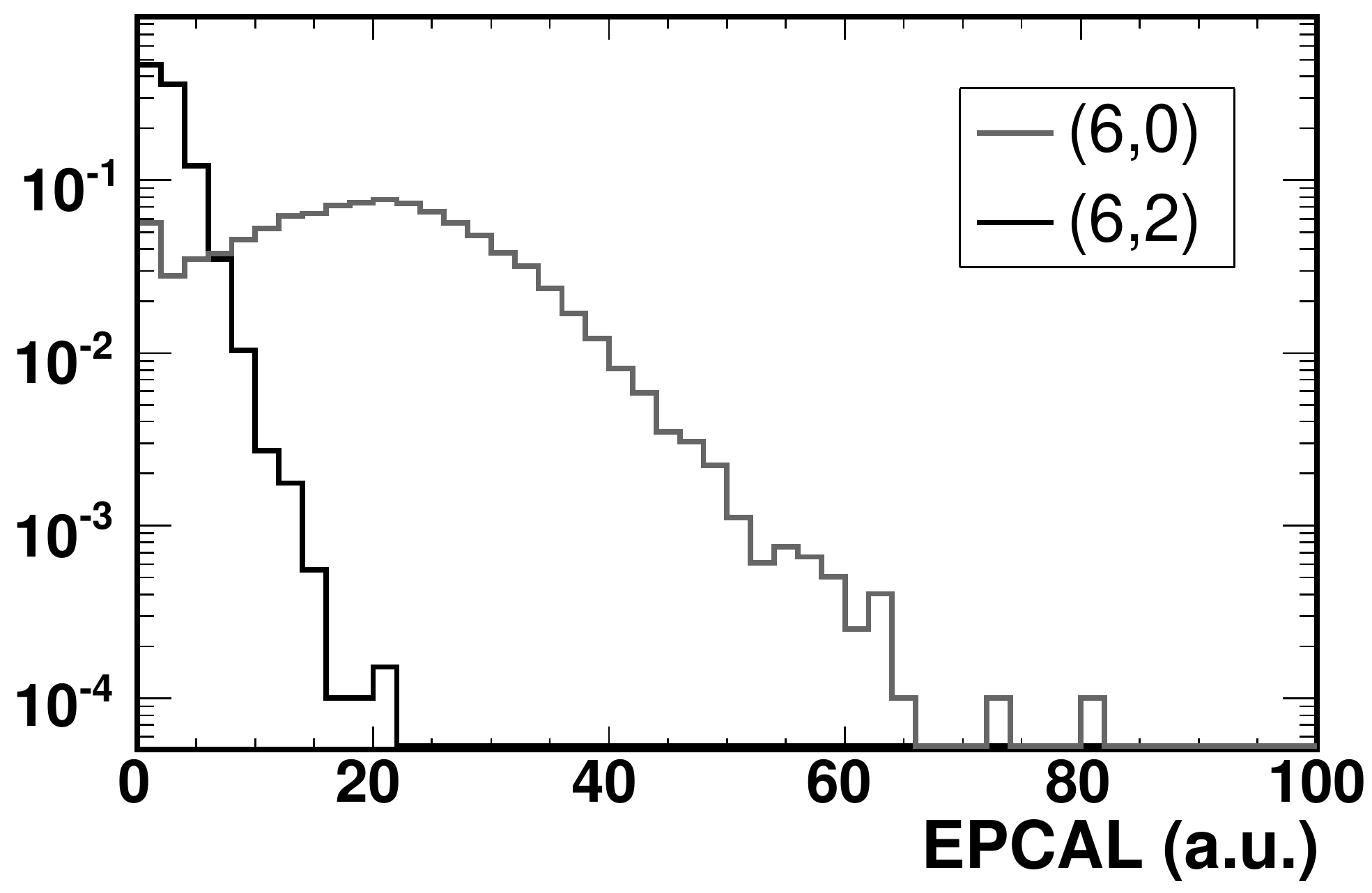}
   \end{center}
   \caption{   \label{calib:fig:pcalCol0}
      Signal distribution of two \protect\acs{Au-PCAL} modules in the
      {\dAu} data. See \fig{calib:fig:pcalCosmEvtDisp} for a diagram of
      the module positions; the beam line was closest to modules (4,0)
      and (5,0), located between rows 4 and 5 in the vertical direction.}
\end{figure}

After energy calibration, it was found that the distribution of signals
from modules closest to the beam line, those in column zero, exhibited a
different shape compared to those of other modules, as shown in
\fig{calib:fig:pcalCol0}. The modules nearest the beam line showed a
peak in the signal distribution. To investigate the origin of this
excess signal, simulations of the calorimeters were performed. First, an
extensive accounting of the \ac{RHIC} accelerator material, including
the DX-magnet structure and beam pipes, was obtained to ensure that the
simulation geometry was accurate. Then, four types of nucleons were
simulated, (a)~uninteracted deuteron nucleons, (b)~elastically scattered
deuteron nucleons, (c)~thermally radiated gold nucleons and (d)~gold
nucleons emitted from a nuclear fragmentation cascade.

All four types were simulated using very simple models. The uninteracted
deuteron nucleons were simulated using the beam momentum, including the
\mbox{$\sim1~\mrad$} angular offset of the beam and the spread in
initial vertex positions due to the width of the beam. Taken from
measurements of the spread of {\dAu} interaction vertices, the beam was
described as having a Gaussian shape in the transverse directions, with
an \ac{RMS} of 0.122~{\cm} in the horizontal and 0.097~{\cm} in the
vertical direction. The elastically scattered deuteron nucleons were
simulated by randomly sampling the transverse momentum from the relation

\begin{equation}
   \label{calib:eq:delas}
\frac{1}{\pt} \frac{d\!N}{d\!\pt} = \exp\prn{\frac{-\pt^2}{T_e}}
\end{equation}

\noindent%
where \mbox{$T_e\sim100~\mev$}, a parametrization inspired by previous
studies, see for example~\cite{Akimov:1975rm}. The longitudinal momentum
was then calculated such that the total momentum of the nucleon was
equal to the beam momentum. The beam angle and width were then added
to the initial origin and momentum of the nucleon. The gold nucleons
were simulated by randomly choosing the kinetic energy of the nucleon in
the gold nucleus rest frame by sampling the relation

\begin{equation}
   \label{calib:eq:aucascthmen}
\frac{d\!N}{d\!E} = \sqrt{E - B} \exp\prn{\frac{-(E - B)}{T}}
\end{equation}

\noindent%
where \mbox{$B\sim10~\mev$} represents a Coulomb
barrier~\cite{Hauger:1998jg} and $T$ was taken to be 50~{\mev} for the
cascade nucleons~\cite{Hauger:1998jg} and 8~{\mev} for the thermal
nucleons~\cite{Lefort:2001pa}. The angular orientation of the nucleons
in the gold nucleus rest frame was then chosen randomly, according to
the angular distributions given in~\cite{Cherry:1994ai}

\begin{equation}
   \label{calib:eq:blackgreyangl}
\frac{d\!N}{d\!\cos\theta} \propto \prn{\frac{4}{\sqrt{\pi}} \chi_0
\cos\theta}
\end{equation}

\noindent%
where $\chi_0$ is a parameter related to the velocity of the fragmenting
nucleus and the nucleon. Its value is taken to be 0.45 for cascade
nucleons and 0.11 for thermal nucleons (see~\cite{Cherry:1994ai}). The
nucleons were then boosted by the beam rapidity into the lab frame.

\begin{figure}[t]
   \centering
   \subfigure[Z-X Distribution]{
      \label{calib:fig:pcalThmSims0}
      \includegraphics[width=0.4\linewidth]{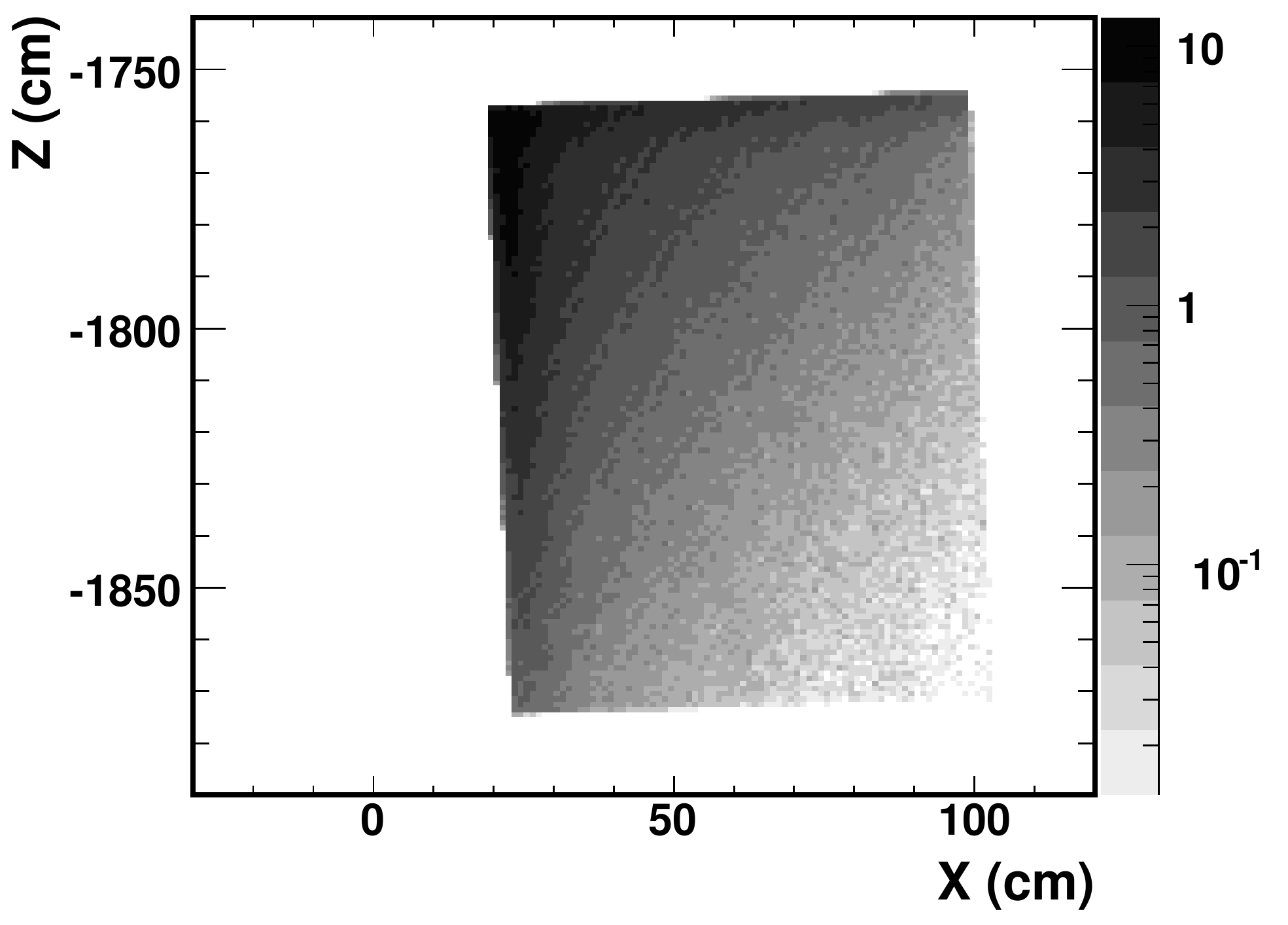}
   }
   \subfigure[Y-Z Distribution]{
      \label{calib:fig:pcalThmSims1}
      \includegraphics[width=0.4\linewidth]{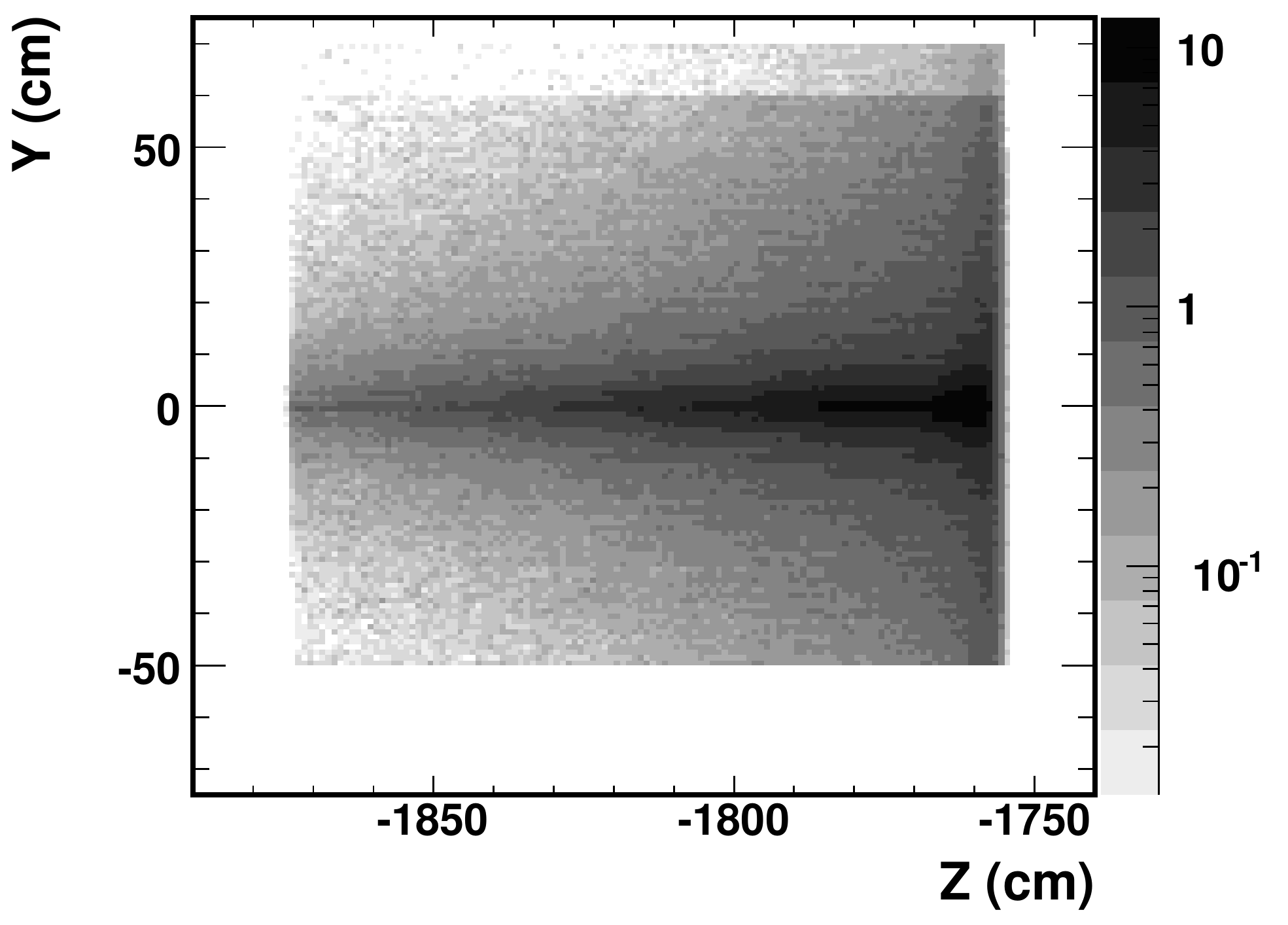}
   }
   \subfigure[Y-X Distribution]{
      \label{calib:fig:pcalThmSims2}
      \includegraphics[width=0.4\linewidth]{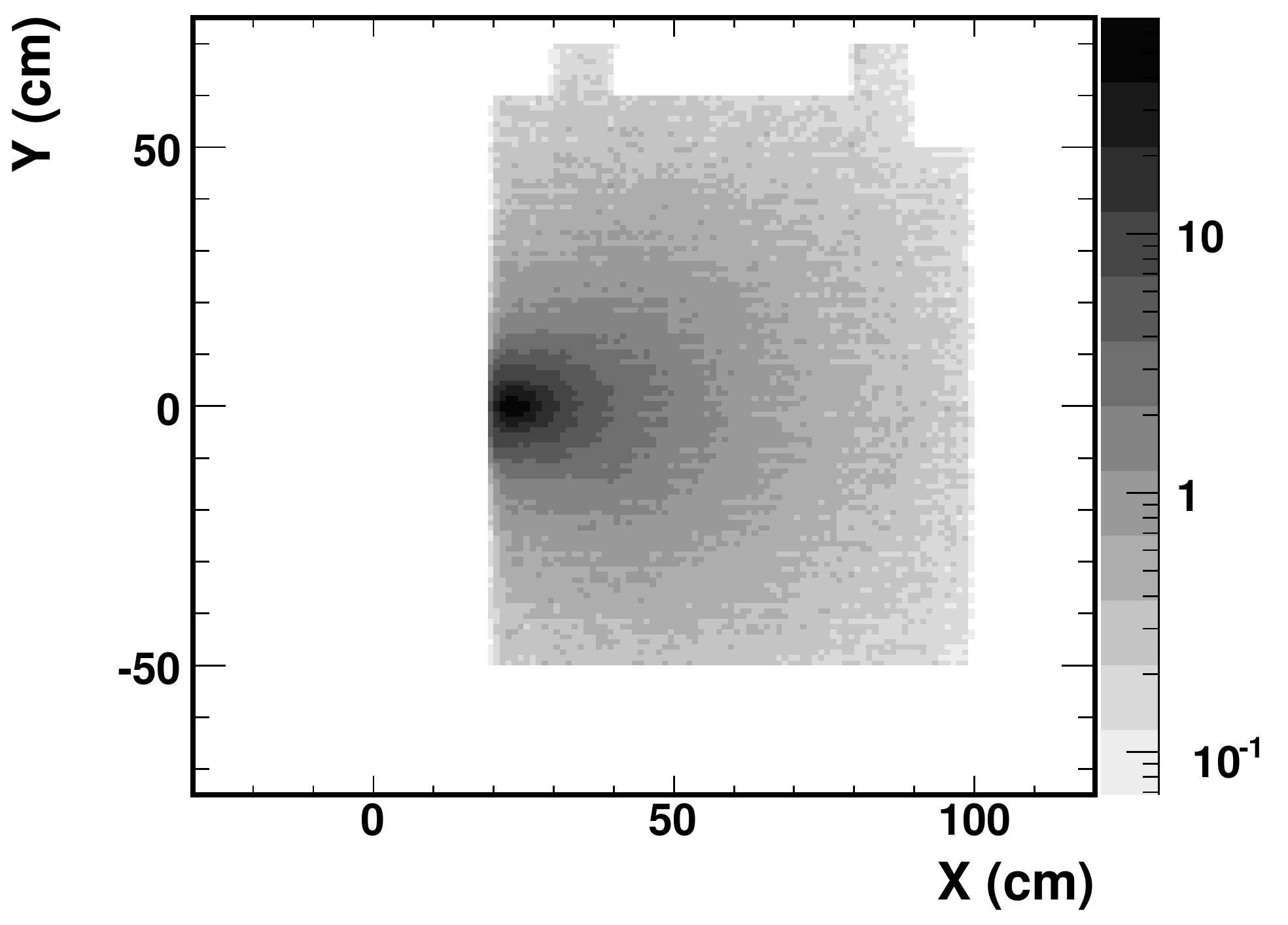}
   }
   \subfigure[Modules]{
      \label{calib:fig:pcalThmSims3}
      \includegraphics[width=0.4\linewidth]{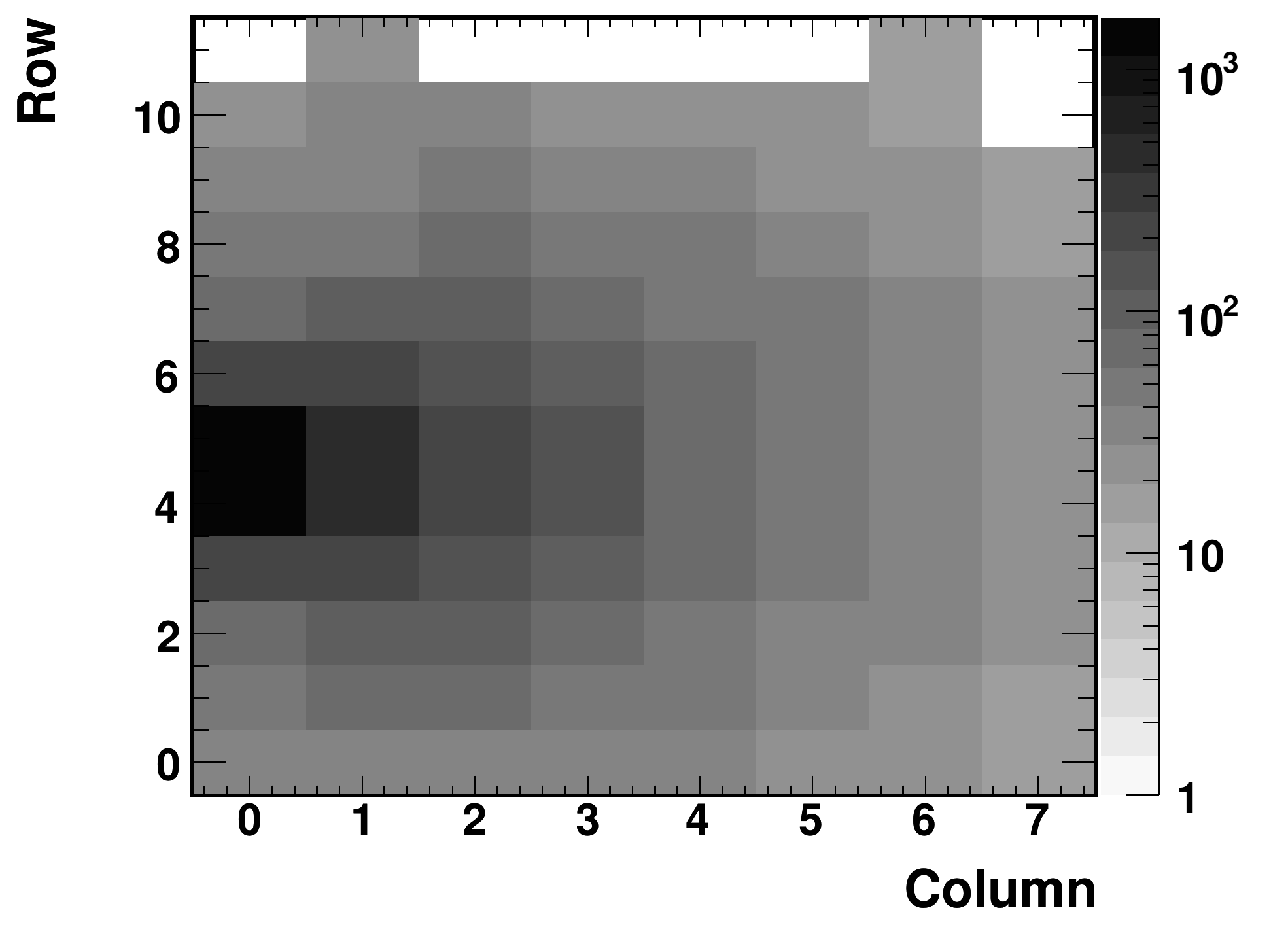}
   }
   \caption{   \label{calib:fig:thmsims}
      Energy deposited into the \protect\acs{Au-PCAL} by thermally
      radiated nucleons in simulations.}
\end{figure}

The nucleons were then swum from the \ac{IP} and through the magnetic
field of the DX-magnet using the Geant simulation package~\cite{Geant}.
Both protons and neutrons were simulated for each possible type of
signal nucleon. An example that shows the location and amount of energy
deposited into the \ac{Au-PCAL} by thermally radiated nucleons is shown
in \fig{calib:fig:thmsims}. It can be seen that the majority of the
energy from the nucleons was deposited into the two modules nearest
the beam line.

These simulations revealed that protons traversed a significant amount
of magnet iron as they passed through the DX-magnet, causing the protons
to lose energy prior to entering the calorimeter. Thus it was expected
that this detector would have a lower efficiency than the \ac{ZDC}; see
\sect{recon:nuctag:NAcent}. However, they showed that the signal from
such protons was not blocked completely, and the distribution of energy
deposited into the \acp{PCAL} in the simulations was found to agree
reasonably well with the data. Further, these simulations suggested that
a substantial fraction of the excess signal seen in the \ac{Au-PCAL}
column nearest to the beam line was due to neutron shower leakage from
the \ac{ZDC}. Attempts to correct for this effect by subtracting from
the \ac{Au-PCAL} signal an amount of energy proportional to the signal
seen in the \ac{Au-ZDC} did not improve the energy resolution. It was
assumed that fluctuations in the calorimeter signals were too large for
this procedure to work. In the end, only energy deposited into modules
from rows four and five, as well as modules in columns four through
seven, was included in the definition of the \ac{EPCAL} variable (see
\sect{recon:cent:pcal}). Signals measured by \emph{all} modules in rows
four and five were included in this variable because the energy
deposited by spectator protons was expected to be much larger than any
deposited energy due to shower leakage from the \ac{ZDC}, as seen in
\fig{calib:fig:pcalThmSims3}. Such an assumption was not made for the
signals reported by modules near the beam line in other rows.

While the signal measured by the \ac{d-PCAL} should have been influenced
by the \ac{d-ZDC} in a similar manner, it was not a significant effect
due to the smaller amount of energy that was deposited into the
\ac{d-PCAL} during a {\dAu} collision. As can be seen in
\pfig{recon:fig:dPcalVsdZDCRegs}, the effect did not contaminate the set
of nucleon-nucleus tagged data; see \sect{recon:nuctag:NA}.

%---------------------------------------------------------------
\subsubsection{d-PCAL Calibration}
\label{calib:pcal:dpcalgain}
%---------------------------------------------------------------

\begin{figure}[t!]
   \centering
   \subfigure[Varying Relative Gain]{
      \label{calib:fig:pcalDRG}
      \includegraphics[width=0.4\linewidth]{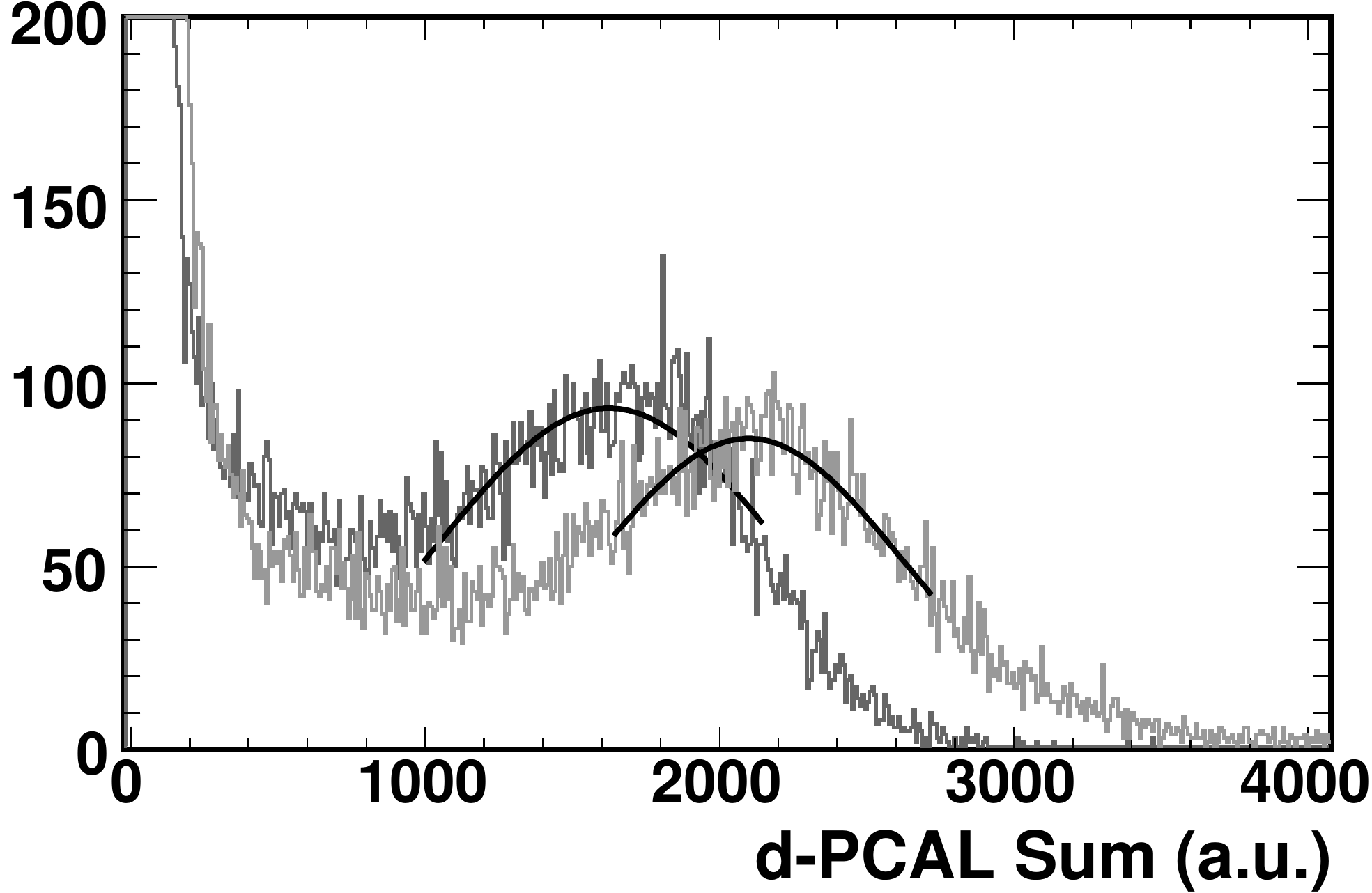}
   }
   \subfigure[Minimizing Peak Width]{
      \label{calib:fig:pcalRGW}
      \includegraphics[width=0.4\linewidth]{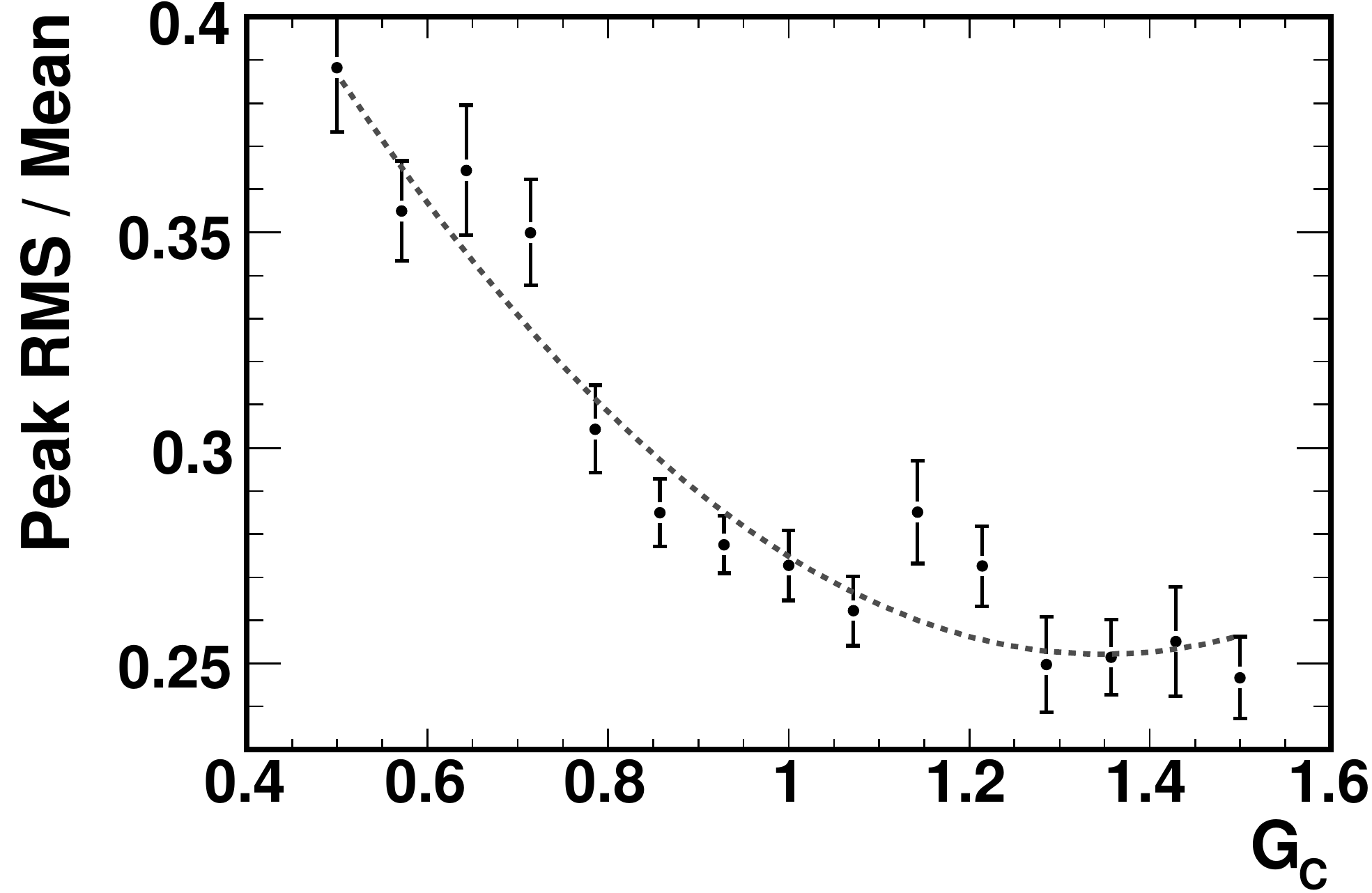}
   }
   \caption{   \label{calib:fig:relgains}
      Steps used to find the relative calibration factors for
      \protect\acs{d-PCAL} channels. \subref{calib:fig:pcalDRG}~Example
      distributions of the full \protect\acs{d-PCAL} signal using two
      different relative gain factors between the columns. The black
      lines show Gaussian fits to the peaks.
      \subref{calib:fig:pcalRGW}~The width of the peak, as measured by
      the \protect\acs{RMS}/mean of the Gaussian fit, versus relative
      gain factor. The dashed grey line shows a parabolic fit. The
      minimum of the fit was taken to be the best estimate of the
      relative calibration.}
\end{figure}

Modules in the \ac{d-PCAL} were calibrated to be consistent relative to
each other, but not to any absolute energy scale. A relative calibration
of the \ac{d-PCAL} modules was sufficient since this detector was only
used to discriminate between collision types, not to measure the energy
of the signal. Since the distribution of signals over many events from
channels in the \ac{d-PCAL} contained a peak, it was possible to
calibrate the detector using a method similar to that used for the
\acp{ZDC}. The modules were calibrated relative to each other in several
steps. First, the channel at $(\text{row,column})=(0,0)$ (see
\fig{calib:fig:pcalPzPedsdA} for the numbering convention) was chosen as
the reference channel. Next, vertically adjacent channels were
gain-matched relative to each other. This was done by varying the
relative gain, $G_{0}$, between channels (0,0) and (1,0) and plotting
the sum $A_{0,0} + G_{0}A_{1,0}$, where $A_{1,0}$ is the signal in
channel (1,0), over many events. The relative gain factor that minimized
the width of the peak in the resulting distribution was then taken to be
the best estimate of the relative calibration. The same process was
performed for the column further from the beam pipe, to find the best
relative gain factor $G_{1}$. Finally, the two columns of the detector
were calibrated relative to each other. This was done by varying the
relative gain, $G_{C}$, between column~0 and column~1 and plotting the
sum $A_{0,0} + G_{0}A_{1,0} + G_{C}(A_{0,1} + G_{1}A_{1,1})$. Examples
of this are shown in \fig{calib:fig:pcalDRG}. The relative gain factor
that minimized the width of the peak in the resulting distribution was
taken to be the best estimate of the relative gain factor between the
two columns, as shown in \fig{calib:fig:pcalRGW}.

%% file: srcs/reconChapter.tex
% $Id: reconChapter.tex,v 1.51 2006/09/05 01:05:18 cjreed Exp $
%

%---------------------------------------------------------------
\chapter{Collision Reconstruction}
\label{recon}
%---------------------------------------------------------------

\myupdate{$*$Id: reconChapter.tex,v 1.51 2006/09/05 01:05:18 cjreed Exp $*$}%
After the amount of energy deposited into each detector channel had been
determined, it was possible to deduce various physical properties of the
collision. Information about the location, magnitude and, in some cases,
time of each hit in the {\phob} detector was used to ascertain the
collision's usefulness for physics analysis, its geometry and its
location.

%---------------------------------------------------------------
\section{Collision Triggering}
\label{recon:trig}
%---------------------------------------------------------------

During the {\dAu} physics run, {\phob} made use of several different
algorithms to determine when a desirable collision had occurred. Each of
the algorithms imposed certain criteria on the responses of the fastest
detectors to trigger the read-out of the entire experiment. Three
different trigger types were used by the analysis presented in this
thesis: \acs{dAuMinBias}, \acs{dAuVertex} and \acs{dAuPeriph}.

%---------------------------------------------------------------
\subsection{The dAuMinBias Trigger}
\label{recon:trig:minbias}
%---------------------------------------------------------------

The \acf{dAuMinBias} trigger was used to build a statistical sample set
of events that most accurately represented the full {\dAu} collision
population. The bias in the sample set was minimized by placing loose
restrictions on the number of particles produced as well as the regions
of phase space filled by produced particles. This was done using the
Paddle detectors (see~\sect{exp:phobdet:trig:paddle}), which had a
large geometrical acceptance.

\begin{figure}[t]
   \begin{center}
      \includegraphics[width=\linewidth]{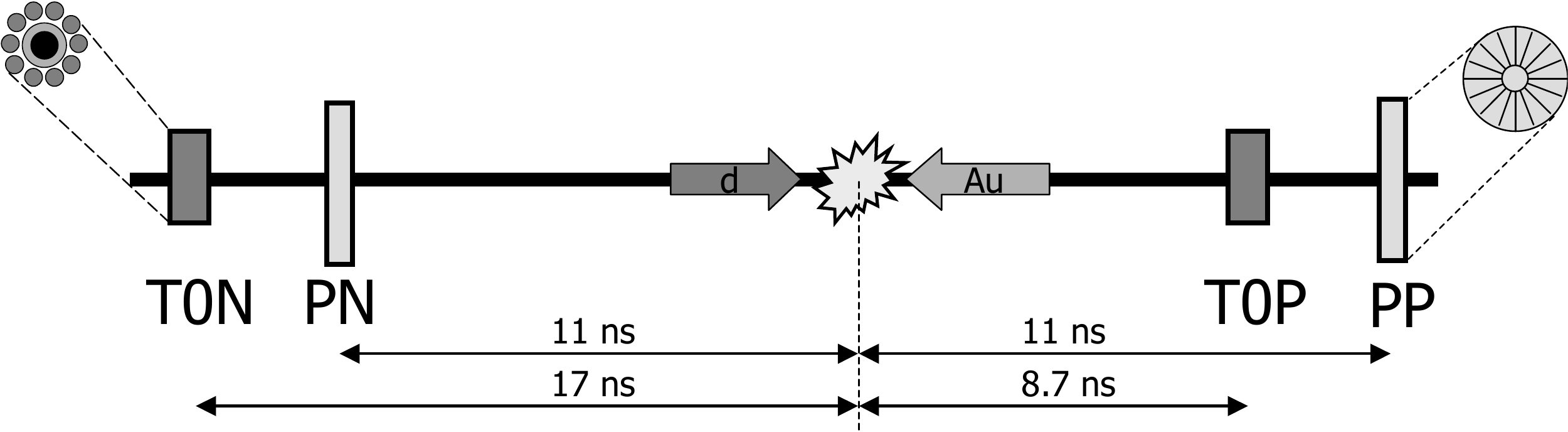}
   \end{center}
   \caption{\label{recon:fig:trgTimeDiag}
      The location of trigger detectors during the {\dAu} physics run.
      Distances are shown by the amount of time it took light to travel
      from the \protect\acs{IP} to each detector. `T0N' and `T0P' refer
      to the Au-side and d-side \protect\acsp{T0}, respectively, while
      `PN' and `PP' refer to the Paddle detectors~\cite{Bickley:2003}.}
\end{figure}

\begin{figure}[t]
   \begin{center}
      \includegraphics[width=0.6\linewidth]{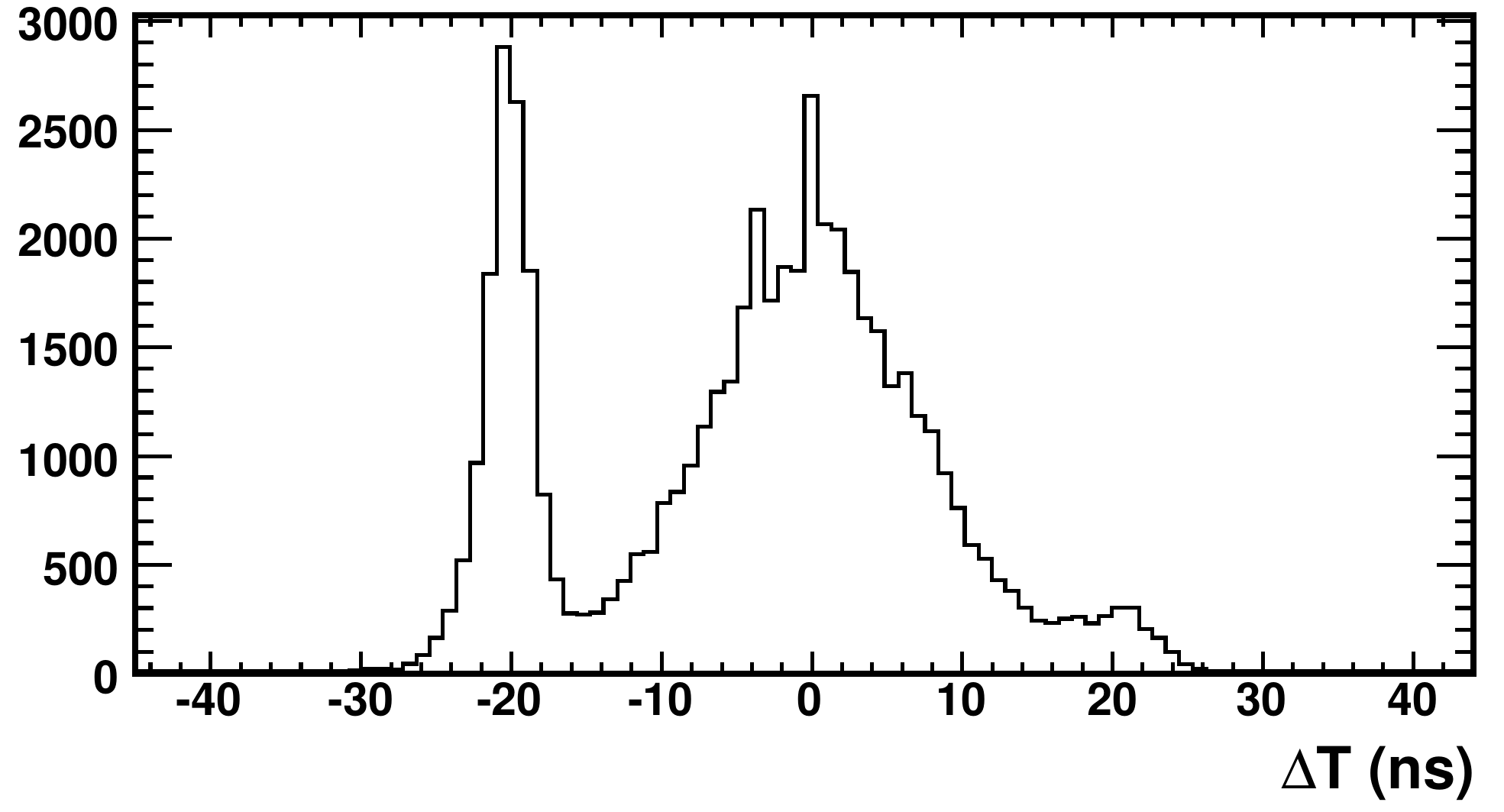}
   \end{center}
   \caption{\label{recon:fig:pdlTdiffwide}
      The paddle time difference distribution from roughly
      54k~\protect\acs{dAuMinBias} triggered events. The peaks at
      $\pm22~\ns$ can be attributed to beam-gas collisions.}
\end{figure}

The \acs{dAuMinBias} trigger required that at least one channel in each
Paddle detector report a hit during a specified time window. This
provided the first clue that some sort of collision had taken place. An
even more unbiased trigger could be imagined, in which only a single
Paddle channel needs to be hit. However, a single-arm trigger would fire
on many events that would be unusable for analysis, such as collisions
between the beam and gas inside the beam pipe. Such beam-gas collisions
can, of course, occur at any position along the beam direction. The
\acs{dAuMinBias} trigger reduced the number of beam-gas events measured
by the experiment by requiring that the two Paddle hits occurred within
38~{\ns} of each other. As shown in
\stolenfig{recon:fig:trgTimeDiag}{Bickley:2003}, the Paddles were only
22~{\ns} apart. Thus, the largest time difference between hits in the
two Paddles caused by particles produced in the same collision would
be 22~{\ns} plus some measurement error in the timing. Therefore, the
large timing window of 38~{\ns} served to select all collisions that
produced hits in both Paddles. This loose requirement was necessary to
construct a minimally biased collision sample, but it also allowed many
beam-gas collisions to be triggered, as shown in
\fig{recon:fig:pdlTdiffwide}.

%---------------------------------------------------------------
\subsection{The dAuVertex Trigger}
\label{recon:trig:davtx}
%---------------------------------------------------------------

The \acf{dAuVertex} trigger was used to record collisions that were more
likely to produce particles in the Spectrometer acceptance. This was
achieved by selecting collisions that occurred within 50~{\cm} of the
\ac{IP}. A timing coincidence between the \ac{T0} detectors was used to
enforce this vertex requirement. The \acsp{T0} had a much better timing
resolution than the Paddles (roughly an order of magnitude better),
resulting in a more precise vertex resolution, but they had a smaller
acceptance than the Paddles. This smaller acceptance introduced a
non-negligible centrality bias in the \acs{dAuVertex} trigger. Central
{\dAu} collisions yielded more particles than peripheral collisions,
making them more likely to produce a hit in each of the \ac{T0}
detectors. Thus, for a central and a peripheral collision both occurring
exactly at the \acs{IP}, the peripheral collision would be less likely
to produce particles which hit the \acsp{T0} and thereby fire the
\acs{dAuVertex} trigger. To facilitate a study of this centrality bias,
when the {\phob} experiment would take data, it would not use the
\acs{dAuVertex} trigger exclusively. Instead, when taking primarily
\acs{dAuVertex} triggered data, a small fraction of the data recorded by
the experiment would be \acs{dAuMinBias} triggered collisions.

%---------------------------------------------------------------
\subsection{The dAuPeriph Trigger}
\label{recon:trig:periph}
%---------------------------------------------------------------

The \acf{dAuPeriph} trigger was used to specifically select collisions
with large impact parameters. The goal of this trigger was to build a
set of data that complemented the centrality-biased \acs{dAuVertex}
triggered data by being biased toward peripheral collisions. This was
achieved by first keeping the vertex requirement from the
\acs{dAuVertex} trigger and then adding a further requirement that no
more than eight channels in the \mbox{Au-side} Paddle detector be hit.
The low-multiplicity requirement served to reduce the amount of central
collision data taken, thereby increasing the relative number of
peripheral collisions in the data. As done for the \acs{dAuVertex}
triggered data, a small fraction of the data recorded while running
primarily with the \acf{dAuPeriph} trigger consisted of \acs{dAuMinBias}
triggered collisions.

%---------------------------------------------------------------
\section{Vertex Reconstruction}
\label{recon:vertex}
%---------------------------------------------------------------

While the Paddle and \acs{T0} detectors were used to quickly estimate
the position of a collision during data-taking, a more precise
measurement was obtained offline, using silicon detectors. In {\AuAu}
collisions, the large number of particles released allowed the
collision location to be determined by simply finding a common origin
of tracks in the Spectrometer and Vertex detectors. The relatively low
multiplicity produced by {\dAu} collisions, on the other hand, made it
necessary for a different measurement technique to be adopted. The
location of a collision along the beam axis (the $z$~direction) was
obtained using the Octagon. The transverse vertex position was
determined using the Spectrometer and the Vertex detectors.

Since the Octagon had only a single layer of silicon, no particle tracks
could be reconstructed using the detector. However, the amount of energy
deposited into an Octagon pad could, in fact, be used to gain
information about how a particle might have traversed the silicon. Large
energy depositions were generally due either to a slow particle (see
\fig{calib:fig:betabloch}) or a particle which passed through the
silicon at a shallow angle. Most slow particles were produced not by the
{\dAu} collision itself, but rather by a secondary collision in which a
product of the {\dAu} collision interacted with some material. Such
secondaries would not necessarily be produced anywhere near the original
collision. Therefore, a hit due to a secondary particle would always
have a large energy deposition, even after attempting to correct for the
increased path length of a particle coming from the collision vertex.

\begin{figure}[t]
   \begin{center}
      \includegraphics[width=0.3\linewidth]{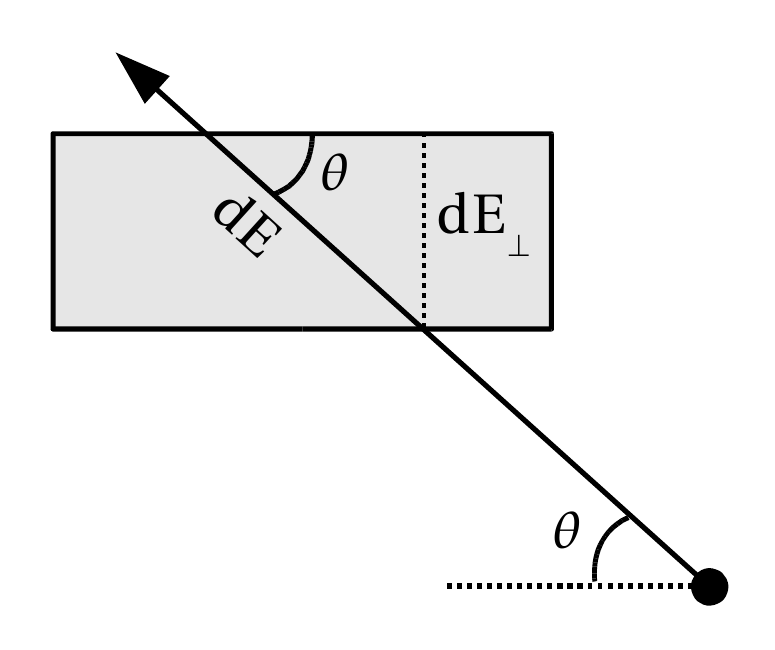}
   \end{center}
   \caption{\label{recon:fig:deAngleCor}
      Correcting the deposited energy ($\mathit{dE}$) for the incident
      angle ($\theta$) of a particle coming from the vertex.}
\end{figure}

The \ac{OctDe} vertexing algorithm assumed that each Octagon hit was due
to a primary particle passing through the silicon at an oblique angle.
First, the energy from three adjacent pads along the beam direction was
combined. The angle of the resulting merged hit was determined by the
collision position and center of the first merged silicon pad (furthest
along the beam line toward the Au-side).\footnote{There is no obvious
reason why the first pad was chosen over the center pad. This behavior
may have been unintentional.} Note that the weighted average method
described in \sect{calib:si:corrections} was \emph{not} used to find the
merged hit position. Then for some chosen test vertex, the deposited
energy of each Octagon hit was corrected for the path length of
particles through the silicon, as shown in \fig{recon:fig:deAngleCor},

\begin{equation}
   \label{recon:eq:anglecor}
\mathit{dE}_{\perp} = \mathit{dE} \sin(\theta) = \frac{\mathit{dE}}{\cosh{\eta}}
\end{equation}

\begin{equation}
   \label{recon:eq:prap}
{\nonumber}\eta = -\ln\prn{\tan\frac{\theta}{2}}
\end{equation}

\noindent%
where $\mathit{dE}$ is the total deposited energy of the hit, $\theta$
is the angle of the particle coming from the collision vertex,
$\mathit{dE}_{\perp}$ is the amount of energy a perpendicularly incident
particle would deposit and $\eta$ is the {\prap} of the particle. The
best estimate of the vertex was then taken as the one which resulted in
the fewest hits due to secondaries -- or equivalently, the largest
number of \acp{MIP}. After the deposited energy was angle-corrected, any
hit that had between 70~{\kev} and 110~{\kev} of deposited energy was
identified as being due to a \ac{MIP}. The longitudinal vertex position
resolution of the \ac{OctDe} algorithm ranged from 0.7~{\cm} for central
collisions to 1.3~{\cm} for peripheral collisions~\cite{Back:2003ff}.

\begin{figure}[t!]
   \centering
   \subfigure[Horizontal Beam Orbit]{
      \label{recon:fig:beamOrbitX}
      \includegraphics[width=0.4\linewidth]{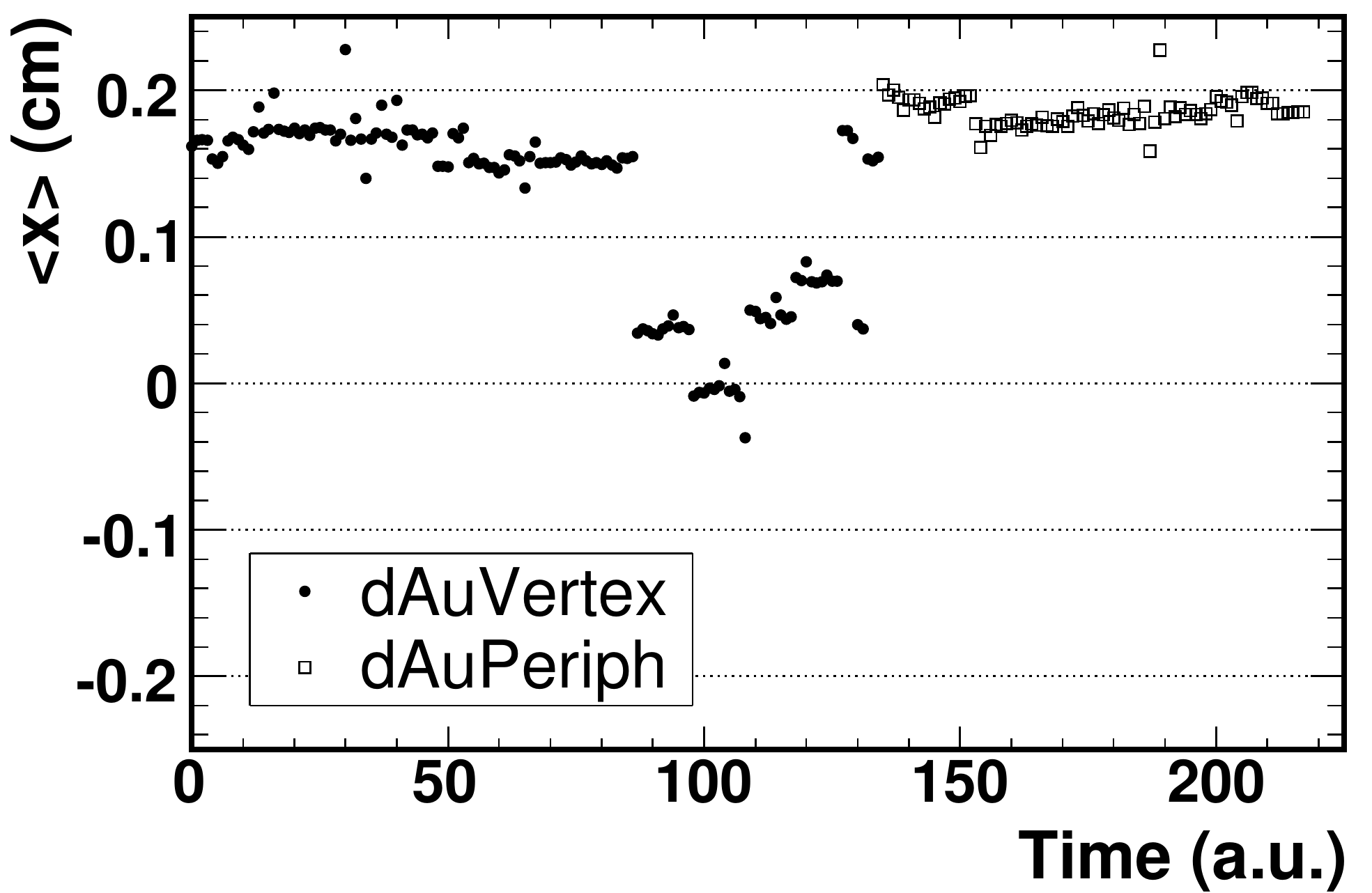}
   }
   \subfigure[Vertical Beam Orbit]{
      \label{recon:fig:beamOrbitY}
      \includegraphics[width=0.4\linewidth]{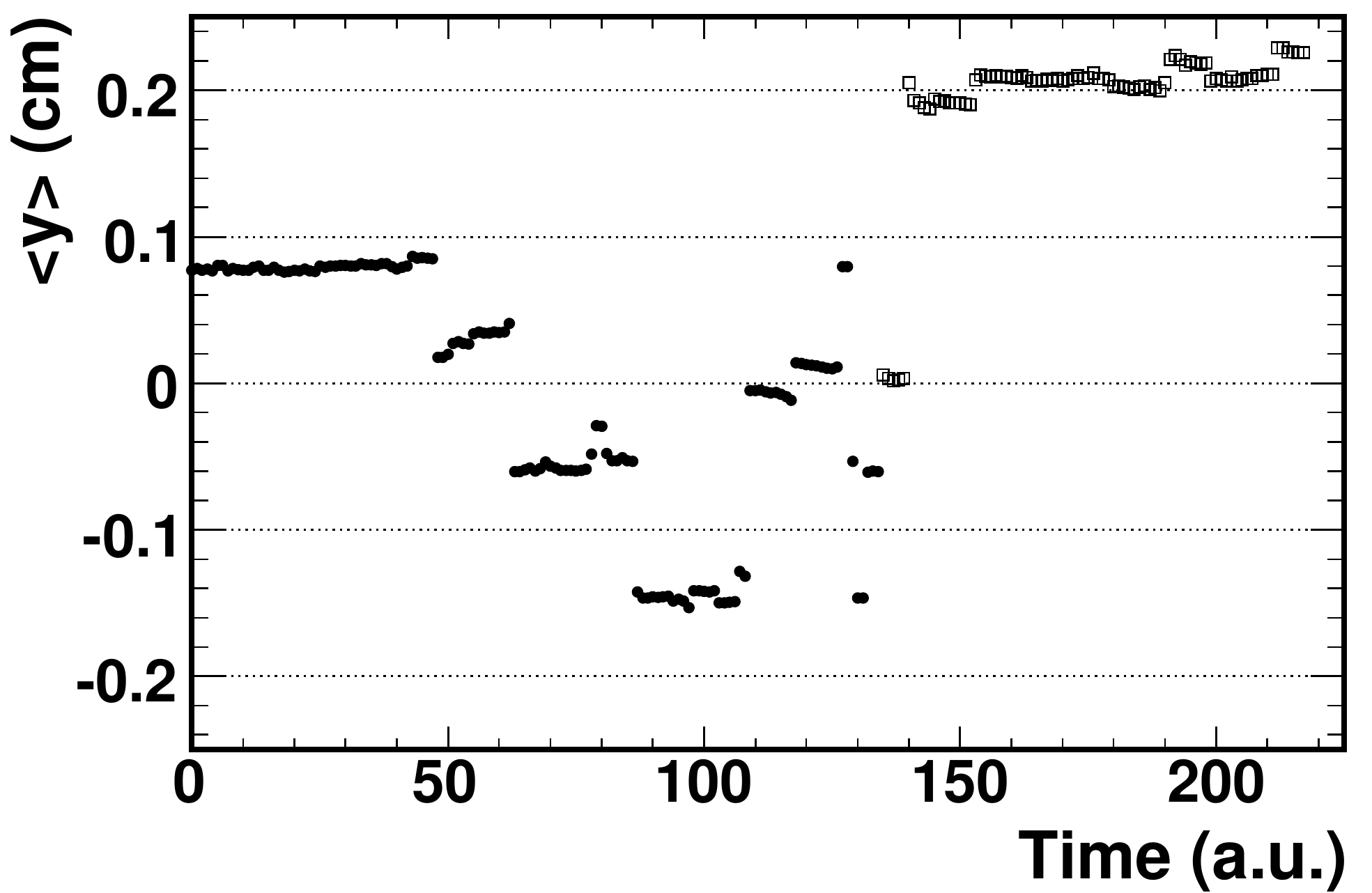}
   }
   \caption{   \label{recon:fig:beamorbits}
      The beam orbit as a function of time during the {\dAu} collision
      data taking. \subref{recon:fig:beamOrbitX}~The average horizontal,
      $x$, position of the beam. \subref{recon:fig:beamOrbitY}~The
      average vertical, $y$, position of the beam.}
\end{figure}

Tracks in the Spectrometer and Vertex detectors were used to determine
the average transverse position of collisions. Due to the low number of
tracks produced in a {\dAu} interaction, it was not possible to measure
the transverse position of each individual collision. However, since the
\acs{RHIC} beams were tightly bunched in the transverse plane, the
position of a collision was determined primarily by the position of the
beams. This allowed the average horizontal, $x$, and vertical, $y$,
positions of the \emph{beam orbit} to be determined for a given set of
collisions~\cite{abbysThesis}. The origin of straight tracks in the
magnetic field-free region of the Spectrometer and of tracks in the
highly segmented Vertex detector were determined for each event by
finding the average intersection point of the tracks in the event. The
average (over several collision events) of the $x$ and $y$ positions of
these track origins were taken to be the beam orbit for that collection
of collisions. The beam orbit showed a relatively small spread in
transverse vertex position (0.4~{\mm}) compared to the resolution
obtained by simply using the origin of single tracks~\cite{Back:2003ff}.
The beam orbit $x$ and $y$ positions are shown in
\fig{recon:fig:beamorbits} as a function of time, for the full data set
analyzed in this thesis.

%---------------------------------------------------------------
\section{Centrality Determination}
\label{recon:cent}
%---------------------------------------------------------------

The collision between a nucleus and a nucleon, or between two nuclei,
is a many-body interaction, the outcome of which can depend intimately
upon its geometry. One of the most fundamental parameters used to
describe collision geometry is the impact parameter. While the impact
parameter of a \acf{AA} collision is not directly measurable in an
experiment, certain observables can be used to classify the centrality
of collisions on a statistical basis.

\begin{figure}[t]
   \centering
   \subfigure[Front View: $\Npart$]{
      \label{recon:fig:npart}
      \includegraphics[width=0.4\linewidth]{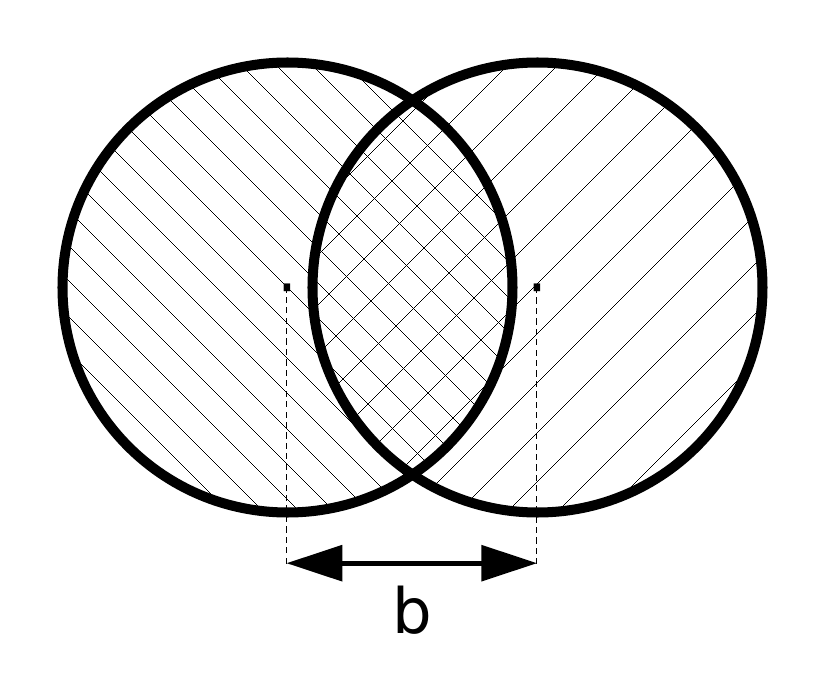}
   }
   \subfigure[Side View: $\Ncoll$]{
      \label{recon:fig:ncoll}
      \includegraphics[height=0.4\linewidth]{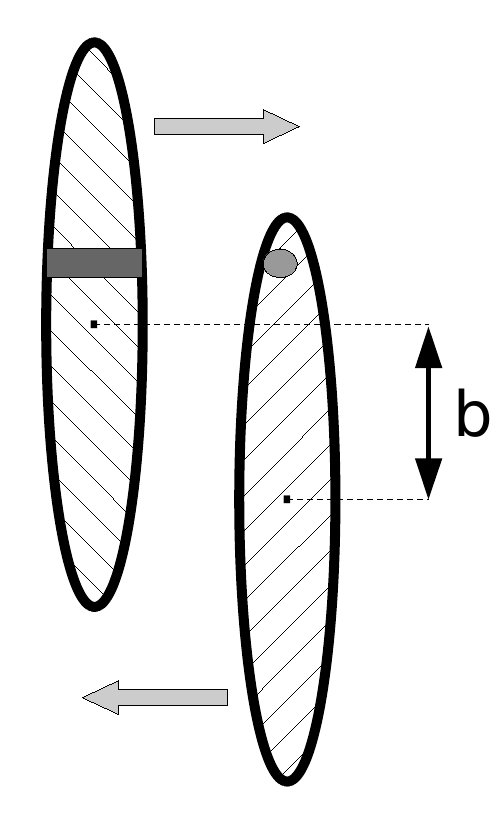}
   }
   \caption{   \label{recon:fig:npartncoll}
      Cartoons of collision geometry. The impact parameter is shown by
      'b.' \subref{recon:fig:npart}~Front view of two symmetric nuclei
      colliding. The number of nucleons in the overlap region is known
      as $\Npart$. \subref{recon:fig:ncoll}~Side view of the
      collision. The thickness of the nuclei is parametrized by the
      total number of binary collisions between nucleons, $\Ncoll$.
      The light-grey circle represents a nucleon, while the dark-grey
      box shows the effective thickness of the nucleus seen by the
      nucleon.}      
\end{figure}

Two variables that have been used to parametrize the centrality of
\acl{AA} collisions are \emph{$\Npart$} and \emph{$\Ncoll$}. These
parameters, while also not directly observable in an experiment,
correlate well with impact parameter and have simple physical
interpretations. Nucleons from one nucleus that collide (inelastically)
with nucleons from the opposing nucleus are known as participants. The
total number of participant nucleons in a collision is $\Npart$. As
shown in \fig{recon:fig:npart}, the overlap region in a collision is
defined by the participating nucleons. Nucleons outside this region,
that do not directly take part in the collision, are known as
spectators. The number of binary collisions that the participating
nucleons suffer is known as $\Ncoll$. The number of binary collisions
parametrizes the thickness of a nucleus seen by a nucleon in the
opposing nucleus, as shown in \fig{recon:fig:ncoll}. $\Ncoll$ is
commonly used to compare the results of \acs{AA} collisions to those of
\ac{NN} collisions since, in the absence of any nuclear effects, one
would expect \ac{AA} collisions to be simply a superposition of \ac{NN}
collisions. The details of determining the average $\Npart$ and $\Ncoll$
for a set of collisions will be discussed in \sect{recon:cent:pars}.

\begin{figure}[t]
   \begin{center}
      \includegraphics[width=0.6\linewidth]{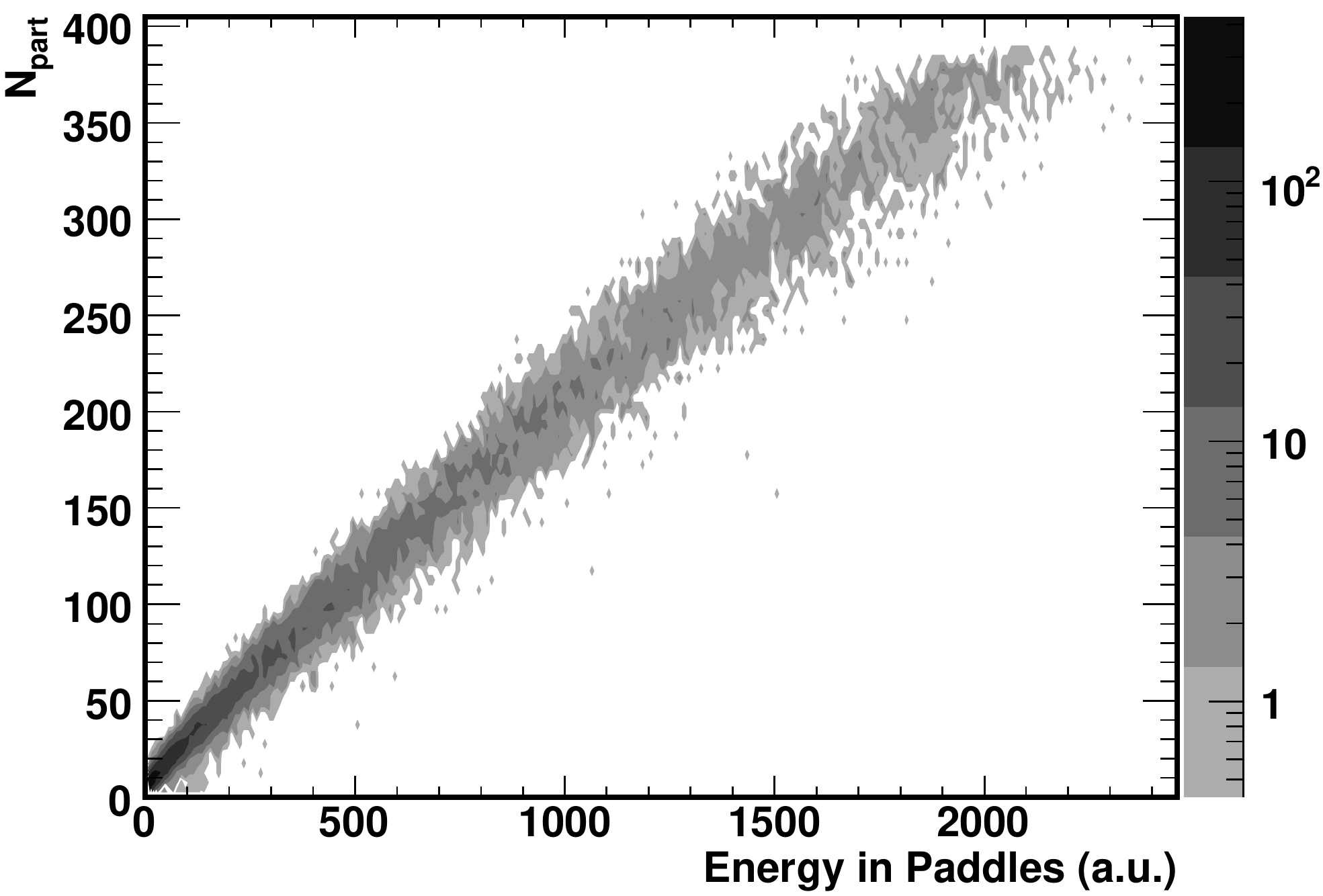}
   \end{center}
   \caption{\label{recon:fig:NpartVsPdlMeanHij}
      The amount of energy deposited into the Paddle detectors from
      {\AuAu} collisions correlated very well with $\Npart$ in the
      \protect\acs{HIJING}~\cite{Gyulassy:1994ew} model.}
\end{figure}

The average value of these parameters -- $\Npart$ and $\Ncoll$ -- can
be estimated for different classes of collisions using experimentally
measured quantities. Intuitively, one would expect that a parameter
like $\Npart$ (for example) should scale monotonically with the number
of particles produced by a collision. Such scaling was found to hold
in collision simulations, as shown in
\fig{recon:fig:NpartVsPdlMeanHij}, where the multiplicity was measured
using the Paddle detectors. It is possible to exploit this monotonic
relationship by (a)~grouping together collisions in which the measured
quantity, e.g.~the energy in the Paddles, is within some range and
(b)~using simulations to find the average of the desired parameter,
e.g.~$\Npart$, of the collisions in each group. In order to facilitate
comparison of results between different experiments, and between
experiment and theory, the groups of collisions are typically chosen
to be specific fractions of the total inelastic collision cross
section.  See~\cite{richardsThesis} for a detailed discussion of
centrality in heavy ion collisions.

%---------------------------------------------------------------
\subsection{Centrality Cuts}
\label{recon:cent:cuts}
%---------------------------------------------------------------

The values of a measured quantity that are used to divide the data into
groups of centrality classes are known as centrality cuts. These values
are not known a priori; centrality cuts must be determined using
collision data and the estimated efficiency of the detector for
recording collisions usable in an analysis. Collisions that are usable
in an analysis are those that not only pass the trigger, but also pass a
user applied \emph{event selection}. For example, a collision may occur
11~{\cm} away from the \acs{IP} and fire the \acs{dAuVertex} trigger, but
the same collision may not pass the event selection of an analysis that
requires the collision vertex to be within 8~{\cm} of the \acs{IP}. Event
selections used in the analysis presented in this thesis are described
in \sect{ana:evtsel}.

With a perfectly efficient detector and event selection, centrality cuts
could be determined in the most straight-forward way. A collision that
produces, for example, a multiplicity greater than 90\% of all other
collisions would be known to have had a centrality greater than 90\% of
all collisions. Such a collision could therefore be placed into the
\emph{top 10\% fractional cross section} centrality class (i.e.~the
\mbox{0-10\%} bin), since the chances of such a collision occuring is
less than 10\% of the total collision cross section. This reasoning was
used to define the centrality cuts for a chosen measured quantity.
One would simply divide the full distribution of the measured signals
into centrality classes that directly correspond to the desired
fractional cross section groups. For example, if one had recorded the
energy in the Paddles in 200 events and wanted to find two centrality
classes, \mbox{0-60\%} and \mbox{60-100\%} of the total cross section,
the following procedure could be used. First, the events would be
ordered by the recorded energy in the Paddles, from highest to lowest.
Then the events would be counted, starting with the event that had the
highest recorded energy in the Paddles, until 60\% of all events had
been counted. The value of the energy recorded in the Paddles of the
last event counted, number 120, would be the desired centrality cut.
That is, events that had more energy in the Paddles than this cut would
be in the \mbox{0-60\%} group, while events that had less would fall
into the \mbox{60-100\%} group.

\begin{table}[t]
   \begin{center}
      \begin{tabularx}{\linewidth}{|XXXp{2in}|}
\hline
Variable &  Type  &  $\eta$ Coverage  &  Description \\
\hline
\protect\acs{ERing} & Multiplicity & $\pm(3.0,5.4)$ & %
   Energy in the Rings \\
\protect\acs{EOct} & Multiplicity & $(-3.2,3.2)$ & %
   Energy in the Octagon \\
\protect\acs{EPCAL} & Spectators & Beam & %
   Energy of Proton Spectators \\
\hline
      \end{tabularx}
   \end{center}
   \caption{   \label{recon:tab:centvars}
      Description of centrality variables used in this analysis.}
\end{table}

An analogous procedure was adapted for finding centrality cuts in
{\phob}. Unfortunately, the {\phob} detector was not a perfectly
efficient detector. Fortunately, simulations of the detector could be
used to estimate the efficiency of an event selection as a function of
the desired \emph{centrality variable}. The centrality variable refers
to the measured quantity used to generate the centrality cuts; in the
previous example, the centrality variable was the energy recorded in
the Paddles. The centrality variables used in the analysis presented in
this thesis were \ac{ERing}, \ac{EOct} and \ac{EPCAL}. These
variables are described in \tab{recon:tab:centvars}. The centrality
cuts used in the analysis presented in this thesis were determined
using several steps.

%---------------------------------------------------------------
\subsubsection{MC Scaling}
\label{recon:cent:cuts:scaling}
%---------------------------------------------------------------

The first step in finding centrality cuts was ensuring that the
distribution of centrality variable signals in the \ac{MC} simulations
corresponded to that of the recorded data. For such a correspondence to
be expected, the event selection used in the analysis of data had to be
equivalent to that placed upon the \ac{MC}. See \sect{ana:evtsel} for
details on the event selections used in this analysis. With equivalent
event selections placed on the data and \ac{MC}, the overall
\emph{shape} of the distributions of the centrality variable signals in
\ac{MC} and data were expected to be the same. However, the absolute
scale of signals in the \ac{MC} and data were not necessarily
equivalent. For example, a signal reported by the Paddle detector for a
particular collision may be 5\% lower in the simulations than the signal
in the data would have been for the same event. As long as such a
scaling factor is independent of the centrality of collisions, it does
not affect the centrality cuts, since the cuts are based on
\emph{fractions} of the distribution.

While the absolute scale of the \ac{MC} signals was not critical for
finding the centrality cuts themselves, it was critical for finding the
efficiency. This requirement stems from the fact that it was necessary
to know the efficiency of the event selection for all values of the
centrality variable in the \emph{data}. Thus, before the efficiency
could be determined, the centrality variable distribution in the
\ac{MC} had to be scaled such that it matched the same distribution in
the data.

\begin{figure}[t]
   \centering
   \subfigure[EOct No Scaling]{
      \label{recon:fig:eoctDataHij}
      \includegraphics[width=0.4\linewidth]{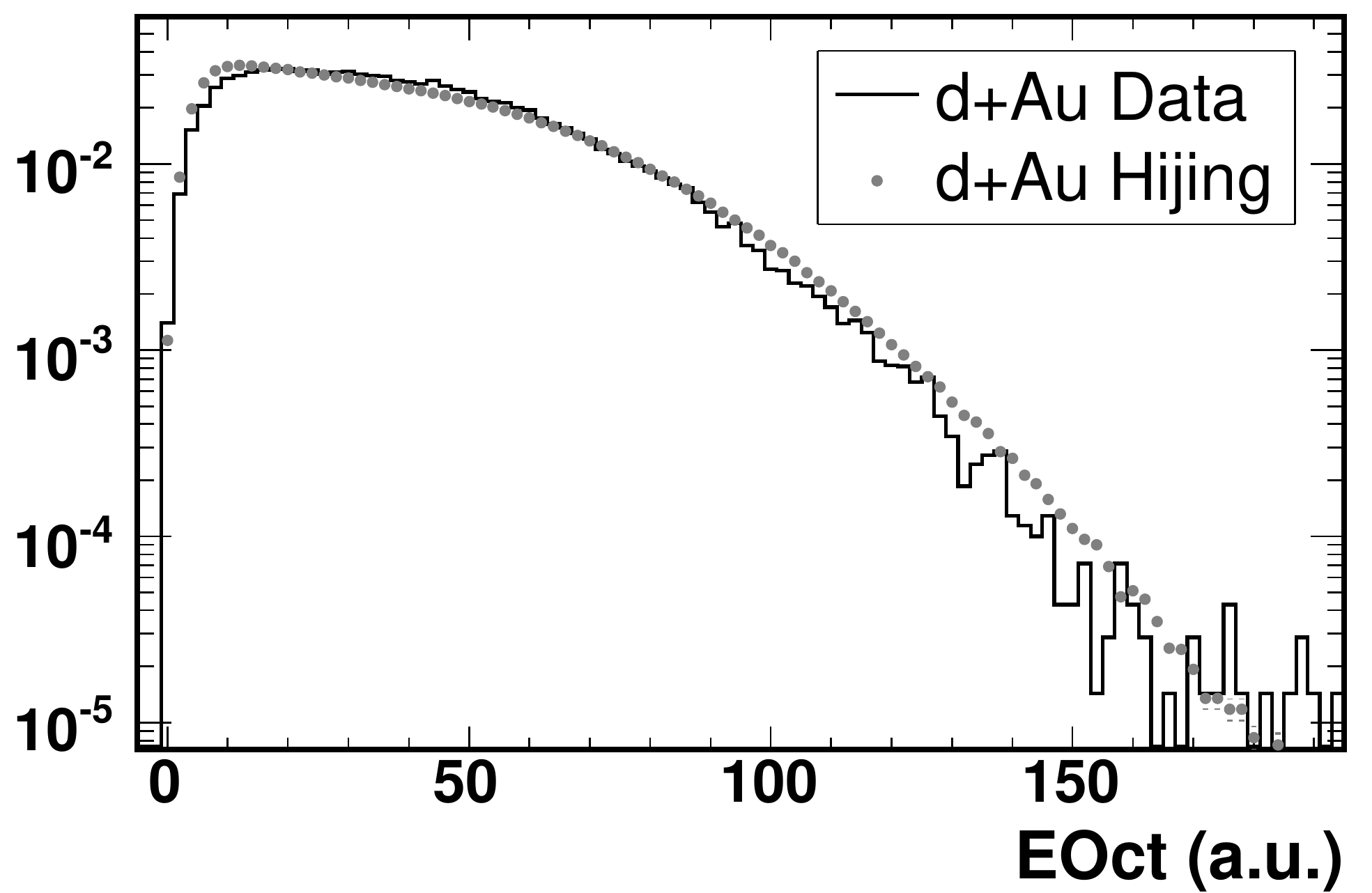}
   }
   \subfigure[EOct Scaled]{
      \label{recon:fig:eoctDataHijScaled}
      \includegraphics[width=0.4\linewidth]{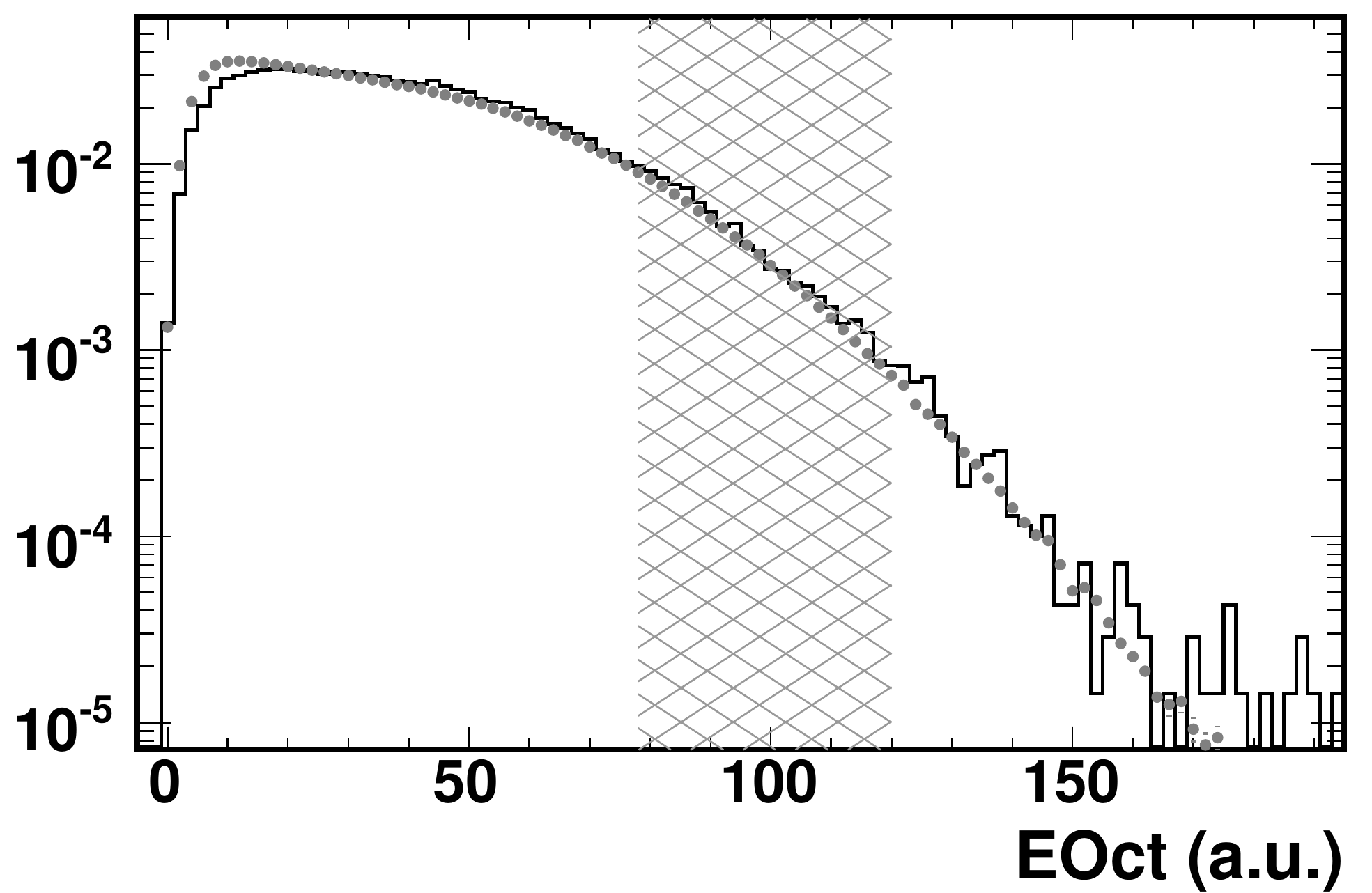}
   }
   \caption{   \label{recon:fig:eoctScale}
      Example of the scaling of \protect\acs{EOct} in \protect\acs{MC}.
      \subref{recon:fig:eoctDataHij}~Comparison of the normalized
      \protect\acs{EOct} distribution in the data and \protect\acs{MC}
      with no scaling. \subref{recon:fig:eoctDataHijScaled}~Comparison
      of \protect\acs{EOct} in the data and \protect\acs{MC}, with
      \protect\acs{EOct} signals scaled down by 5.2\% in the
      \protect\acs{MC}. The grey band shows the range of
      \protect\acs{EOct} used to determine the scaling factor.}
\end{figure}

It was assumed that for some region of sufficiently central events, the
efficiency of the event selection should not change with centrality. In
this region, then, the distribution of centrality variable signals in
the \ac{MC} could be scaled to match that of the data. This region was
different for each centrality variable. The \ac{MC} scale factor was
determined in the following way. First, a test scale factor was chosen
and the distribution of scaled signals from the \ac{MC} was generated.
This distribution and the data distribution were then independently
normalized. A $\chi^{2}$ comparison between the scaled \ac{MC} and the
data was performed for signals in the chosen region. The region used for
matching was a subset of the constant efficiency region, such that bins
with low statistics were explicitly ignored. However, the same scaling
factor was obtained when the full constant efficiency region was used to
do the matching, as expected. This process was repeated for 50 different
scaling values ranging from 90\% to 120\% (although a smaller range
could have been used). The final scaling factor was chosen to be the one
that yielded the best match according to the $\chi^{2}$ tests. An
example of the scaling of \ac{MC}, using \ac{EOct} from
\acf{HIJING}~\cite{Gyulassy:1994ew} simulations, can be seen in
\fig{recon:fig:eoctScale}.

%---------------------------------------------------------------
\subsubsection{Efficiency}
\label{recon:cent:cuts:eff}
%---------------------------------------------------------------

After the signal in the \ac{MC} simulations had been scaled to match
the signals seen in the data, the efficiency of the event selection
could be determined. The desired efficiency would be a function of the
centrality variable, and could be used to estimate the distribution of
signals that would have been measured by a perfect detector, according
to the relation

\begin{equation}
   \label{recon:eq:eff}
N_{ideal}(C) = \frac{N_{meas}(C)}{\epsilon(C)}
\end{equation}

\noindent%
where $N_{ideal}(C)$ is the number of events recorded by an ideal
detector in which the value of the centrality variable is $C$,
$N_{meas}$ is the number of events measured in the actual data and
$\epsilon$ is the estimated efficiency.

The efficiency of an event selection was estimated using the \ac{MC}
simulations. The basic procedure was simply to take the centrality
variable distribution in the \ac{MC} \emph{with} the event selection
and divide it by the same distribution in the \ac{MC} \emph{without}
the event selection. One complication to this procedure arose due to
the vertex requirement of an event selection. For example, an event
selection that requires collisions to be within 2~{\cm} of the \acs{IP}
might be 100\% efficient for events satisfying this vertex requirement.
However, if the \ac{MC} simulations generated collisions that were
randomly distributed within 4~{\cm} of the \acs{IP}, then the efficiency
would be estimated as 50\%, since half of the collisions generated by
the \acs{MC} would lie outside the vertex requirement of the event
selection. This is clearly not the desired efficiency, since any
efficiency value could be produced simply by changing the range of
vertices in which collisions are simulated!

\begin{figure}[t]
   \centering
   \subfigure[With \& Without Event Selection]{
      \label{recon:fig:eoctEffCmp}
      \includegraphics[width=0.4\linewidth]{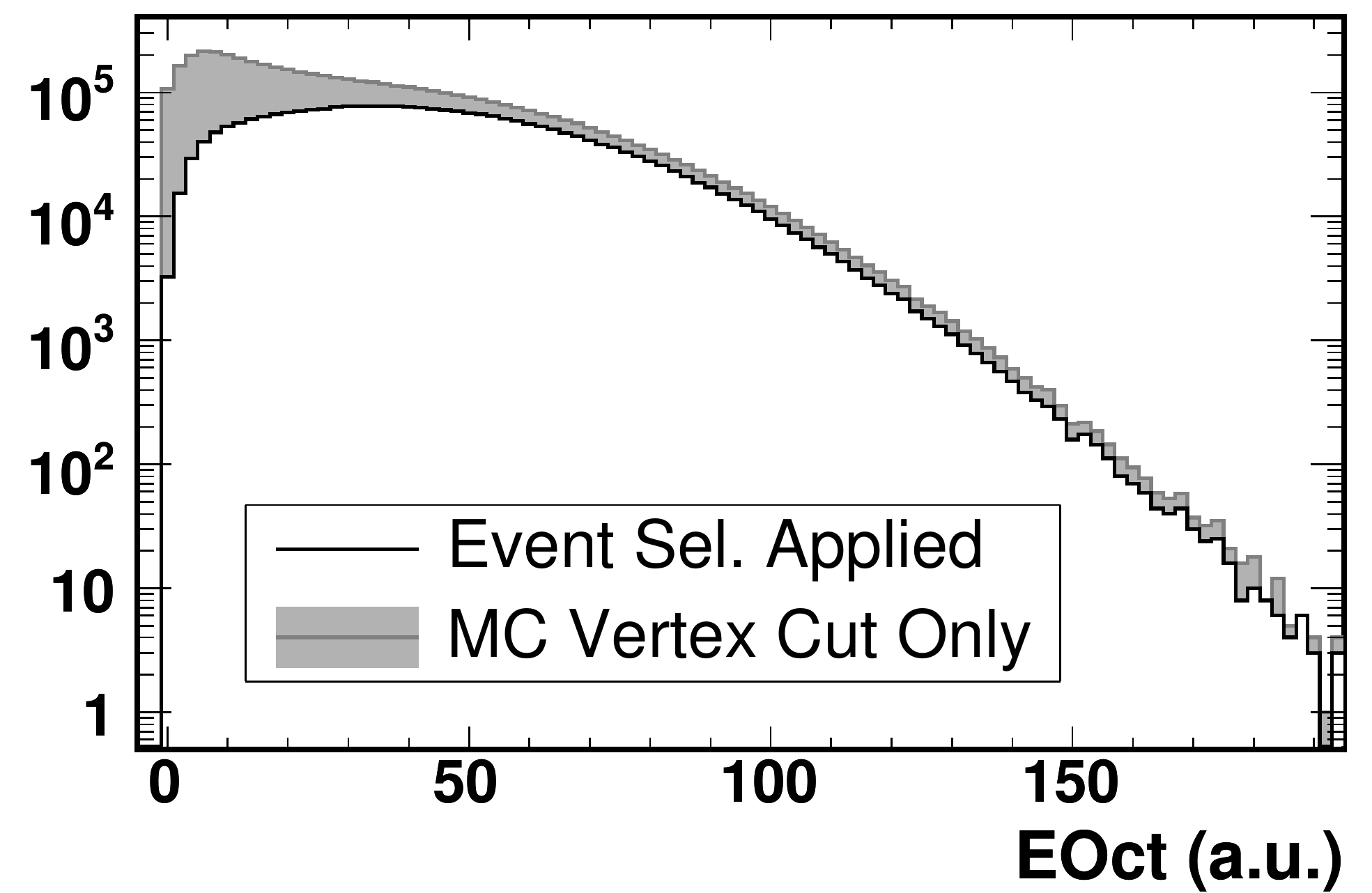}
   }
   \subfigure[EOct Efficiency]{
      \label{recon:fig:eoctEffFit}
      \includegraphics[width=0.4\linewidth]{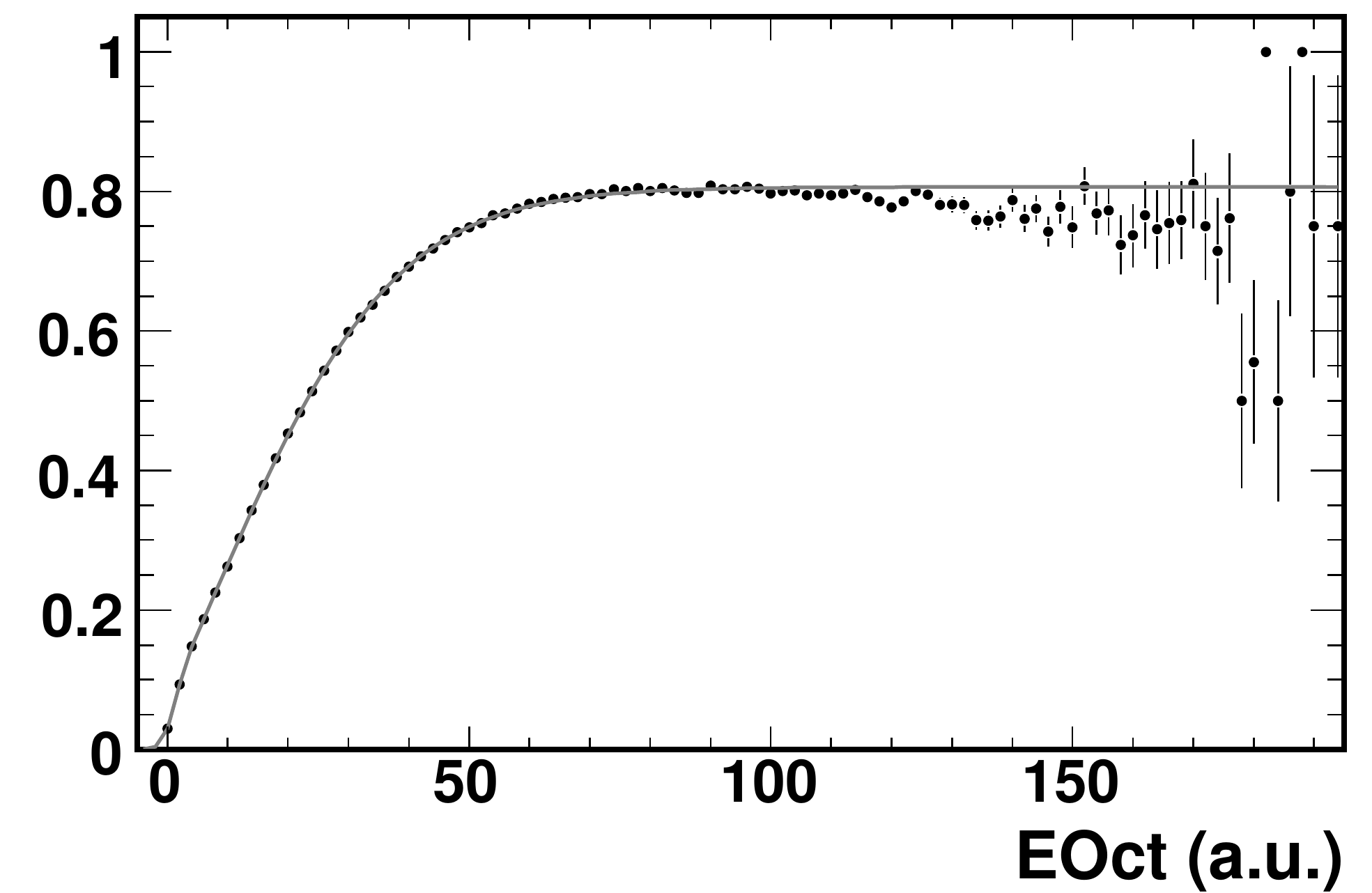}
   }
   \caption{   \label{recon:fig:eoctEff}
      The efficiency of \protect\acs{EOct} from \protect\acs{HIJING}
      using the \protect\acs{dAuSpectra} event selection (see
      \sect{ana:evtsel}). \subref{recon:fig:eoctEffCmp}~Scaled
      \protect\acs{EOct} distributions in the \protect\acs{MC} with the
      full event selection (black line) and with only a true
      \protect\acs{MC} vertex cut (grey shaded region).
      \subref{recon:fig:eoctEffFit}~The \protect\acs{EOct} efficiency
      profile is shown by the black points, while the fit is shown by
      the grey line.}
\end{figure}

The procedure used to estimate the efficiency for the analysis
presented in this thesis was as follows. First, the scaled centrality
variable distribution was produced for events in the \ac{MC} that were
generated within the vertex range required by the event selection. That
is, rather than applying no event selection, a cut was placed only on
the \emph{true} \acs{MC} vertex. This distribution was then divided by
the scaled centrality variable distribution for events in the \ac{MC}
that passed the full event selection. The result of this ratio provided
a profile of the efficiency, which was then fit with a function of
the following form:

\begin{equation}
   \label{recon:eq:efffit}
\frac{\alpha}{1 + \exp\prn{\beta_{0} - \gamma_{0} C} + %
\exp\prn{\beta_{1} - \gamma_{1} C} + \exp\prn{\beta_{2} - \gamma_{2} C}}
\end{equation}

\noindent%

where $\alpha$, $\beta_{i}$ and $\gamma_{i}$ are fit parameters. Note
that there was no physics motivation for this function; it simply
provided a good fit to each efficiency profile without requiring any
tweaking of initial parameter values or of parameter limits. The
estimated efficiency function of \acs{EOct} from \acs{HIJING} using the
\acs{dAuSpectra} event selection (see \sect{ana:evtsel}) is shown in
\fig{recon:fig:eoctEff}. It can be seen from this figure that
\mbox{$\sim20\%$} of {\dAu} collisions are lost, even those having
relatively large values of \ac{EOct}. This inefficiency is due to the
requirement that both \ac{T0} detectors be hit. It was traced to a
combination of two effects: the small acceptance of the \acp{T0} and the
relatively low number of particles produced at high {\prap} (on the
deuteron side of the interaction).

%---------------------------------------------------------------
\subsubsection{Fractional Cross Section Cuts}
\label{recon:cent:cuts:crsscncuts}
%---------------------------------------------------------------

The efficiency function was used to obtain an estimate of the
centrality variable distribution in the absence of experimental biases.
In other words, what would have been measured by an ideal detector.
Using this distribution, it was possible to determine the fractional
cross section cuts using the same method described in
\sect{recon:cent:cuts}. The procedure used in the analysis presented in
this thesis was as follows.

First, the unbiased centrality variable distribution was produced using
the estimated efficiency. This was done by filling a histogram with the
centrality variable signal from each collision event in the data that
passed the event selection. Each entry in this histogram was weighted
according to \eq{recon:eq:eff} as not one event, but rather as
$1/\epsilon(C)$ events. The integral of this histogram gave the
total number of events that would have been measured by an ideal
detector, $N^{tot}_{ideal}$.

Then, fractions of $N^{tot}_{ideal}$ corresponding to the desired
fractional cross section classes were computed. For example, the
fractional cross section bins used in this analysis were, from most
central to most peripheral, \mbox{0-20\%}, \mbox{20-40\%},
\mbox{40-70\%} and \mbox{70-100\%}. Since the widths of these bins were
20\%, 20\%, 30\% and 30\%, respectively, the most central bin contained
$N^{central}_{ideal} = 0.2{\times}N^{tot}_{ideal}$ events. More precisely, it
should contain the $N^{central}_{ideal}$ \emph{most central} events.

\begin{figure}[t]
   \centering
   \subfigure[EOct Unbiased Distribution]{
      \label{recon:fig:eoctUnbiasCuts}
      \includegraphics[width=0.4\linewidth]{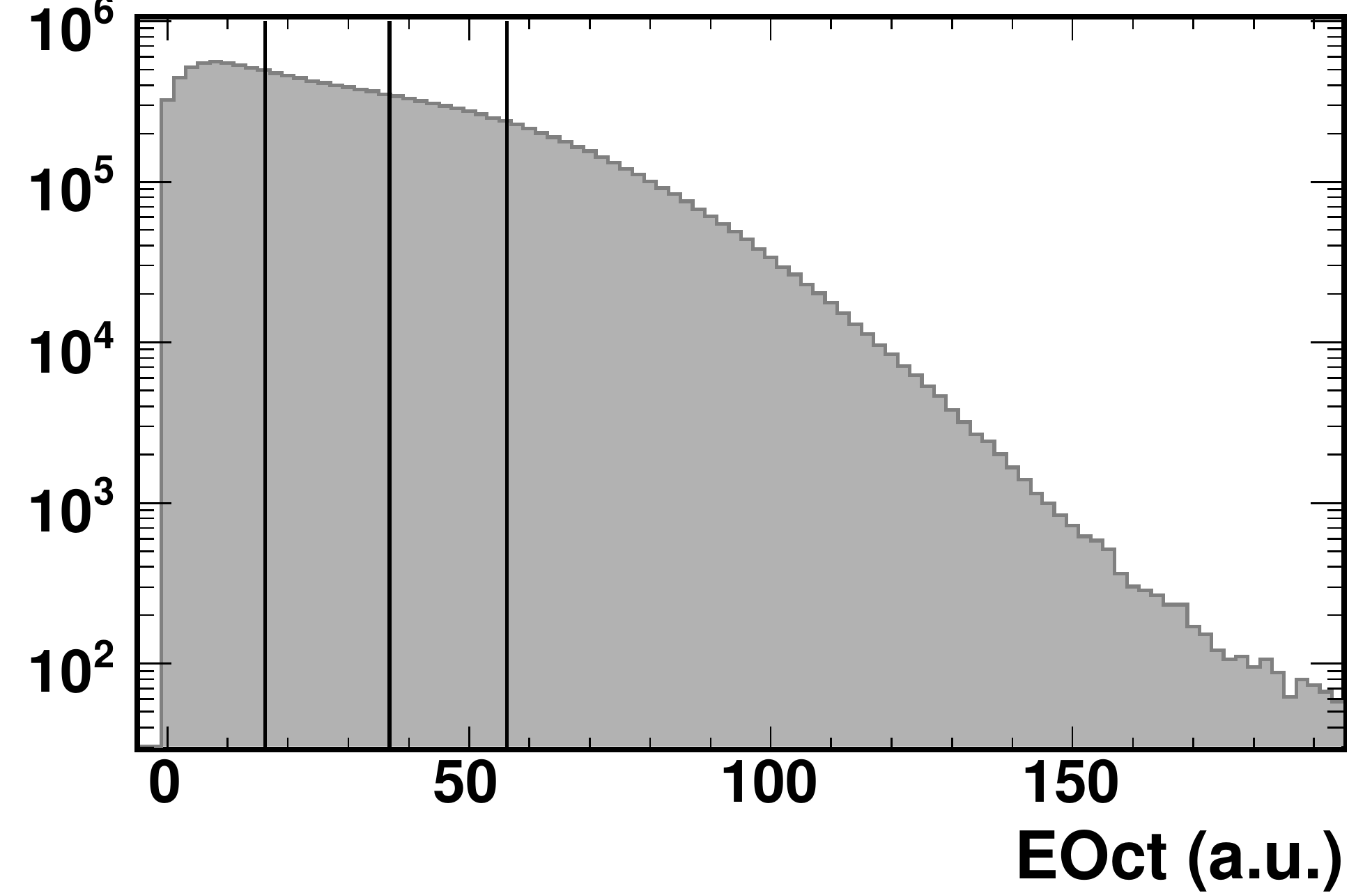}
   }
   \subfigure[EOct Centrality Bins]{
      \label{recon:fig:eoctBiasedCuts}
      \includegraphics[width=0.4\linewidth]{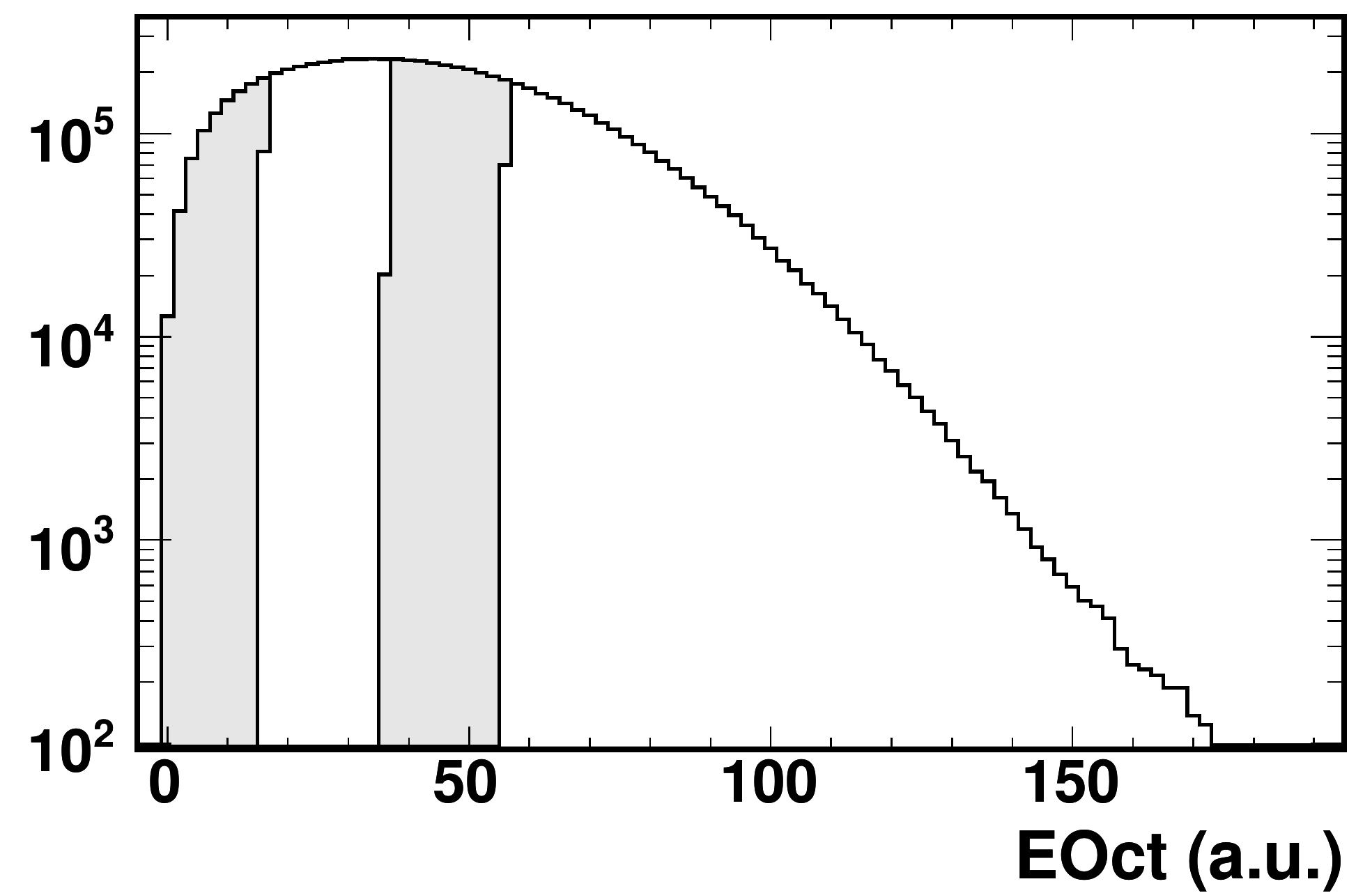}
   }
   \caption{   \label{recon:fig:eoctCuts}
      The centrality cuts for \protect\acs{EOct} from
      \protect\acs{HIJING} using the \protect\acs{dAuSpectra} event
      selection. \subref{recon:fig:eoctUnbiasCuts}~The efficiency
      corrected \protect\acs{EOct} distribution. The black lines show
      the centrality cut values. \subref{recon:fig:eoctBiasedCuts}~The
      distribution of \protect\acs{EOct} in each fractional cross
      section bin. Alternating centrality bins are shaded to guide the
      eye. A step in the shading indicates the fraction of collisions in
      the \protect\acs{EOct} bin that belong to each centrality class.}
\end{figure}

Once the number of events in each centrality class was determined, the
centrality cuts could be found. This was done by computing a cumulative
sum of the entries in each bin of the unbiased histogram, starting from
the highest bin. The bin which caused the sum to exceed
$N^{central}_{ideal}$ was known to contain the first centrality cut
value. For example, if $N^{central}_{ideal}=4$ events, the top bin of
the unbiased histogram contained 2 events having a centrality signal
between 90 and 100, and the next bin contained 5 events having a
centrality signal between 80 and 90, then the centrality cut would have
been between 80 and 90. In order to find the value of the centrality
cut, it was assumed that the shape of the distribution was flat over
the (small) width of the histogram bin:

\begin{equation}
   \label{recon:eq:cutval}
K(G) = L_{i} + W_{i} \prn{
\frac{\sum_{j=top}^{j=i}\prn{E_{j}} - N_{ideal}(G)}{E_{i}}}
\end{equation}

\noindent%
where $K(G)$ is the cut value for the centrality group, $i$ is the
histogram bin containing the centrality cut, $L_{i}$ is the lower edge
of that bin, $W_{i}$ is the width of that bin, $E_{j}$ is the number of
events contained in the $j^{th}$ histogram bin and $N_{ideal}(G)$
is the number of events that should be contained in the centrality
group (i.e.~$N^{central}_{ideal}$). Thus, for the previous example,
\eq{recon:eq:cutval} gives

\begin{equation}
   \label{recon:eq:cutvalexample}
{\nonumber}K({\rm central}) = 80 + 10 \prn{\frac{(2+5) - 4}{5}} = 86
\end{equation}

\noindent%
This procedure was then repeated for each successively more peripheral
centrality class. The centrality cuts generated in this way for
\ac{EOct} from \ac{HIJING} using the \acs{dAuSpectra} event selection
(defined in \sect{ana:evtsel:dAuSpectra}) are shown in
\fig{recon:fig:eoctCuts}.

%---------------------------------------------------------------
\subsection{Centrality Parameters}
\label{recon:cent:pars}
%---------------------------------------------------------------
   
\begin{figure}[t]
   \centering
   \subfigure[Density Profiles]{
      \label{recon:fig:nucDens}
      \includegraphics[width=0.4\linewidth]{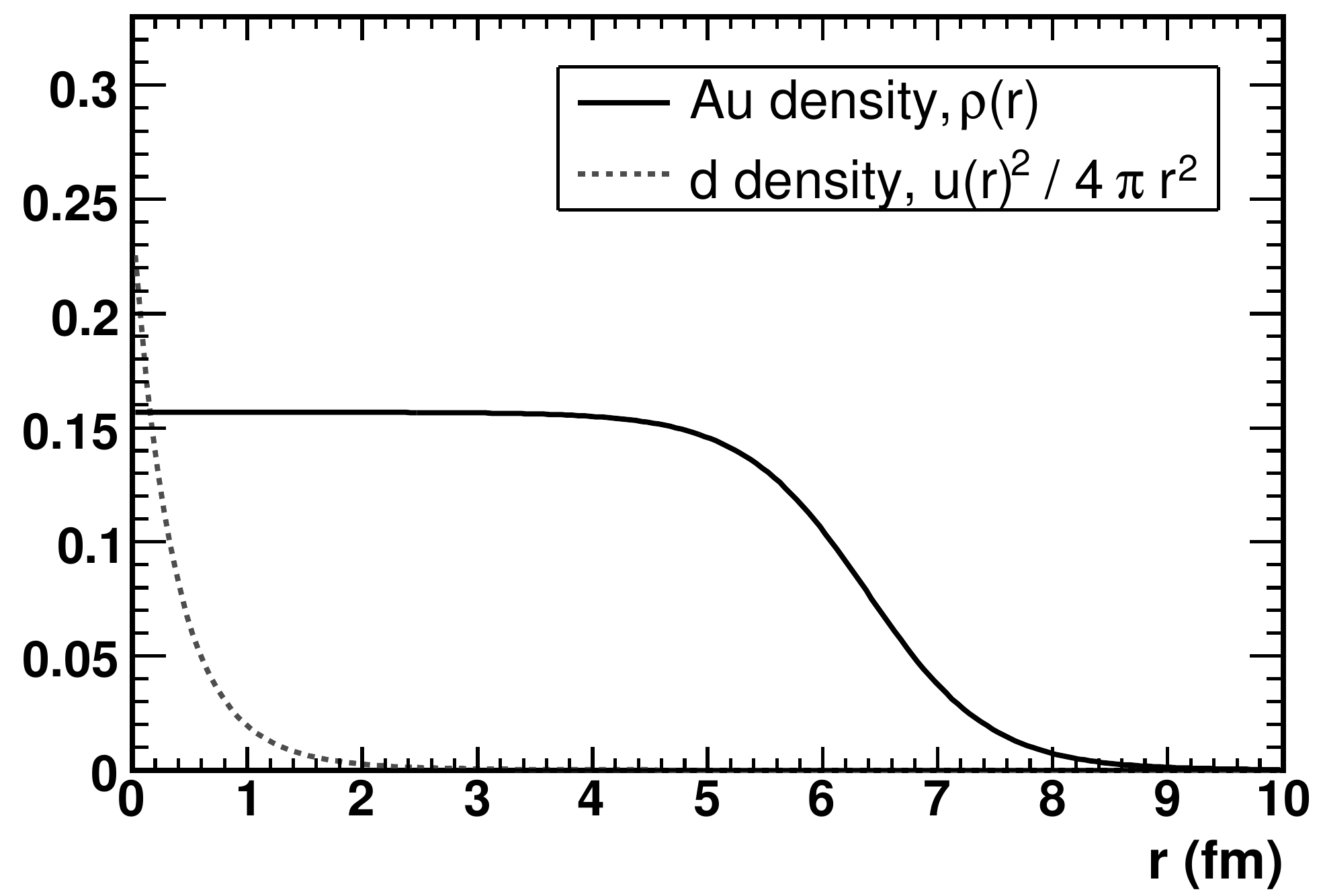}
   }
   \subfigure[Probability Profiles]{
      \label{recon:fig:nucProbs}
      \includegraphics[width=0.4\linewidth]{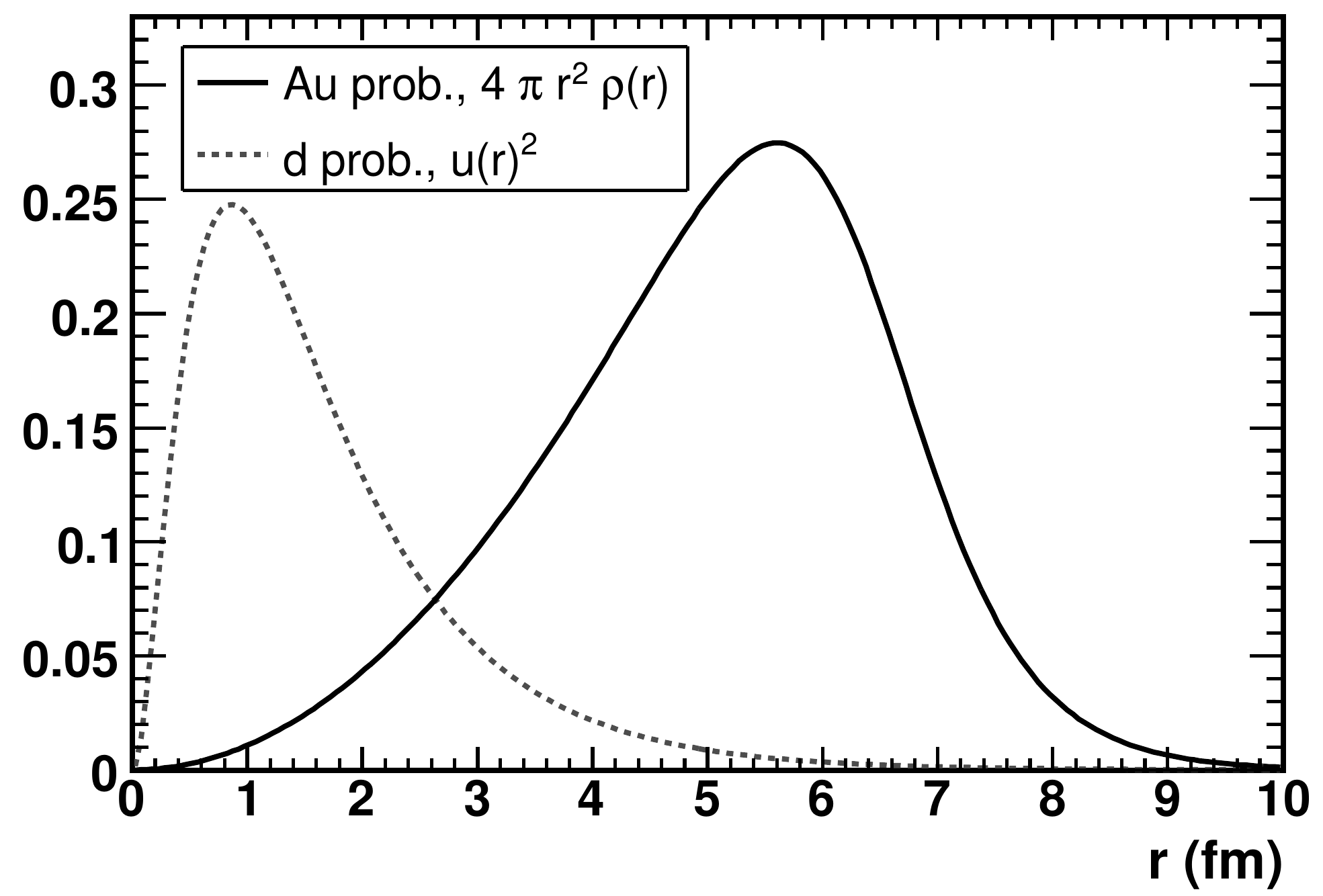}
   }
   \caption{\label{recon:fig:nucProfs}
      \subref{recon:fig:nucDens}~The nuclear density profiles used for
      gold (solid black line) and deuteron (dashed grey line) nuclei in
      the Glauber model. The variable $r$ shows the distance of each
      nucleon from the center of the nucleus.
      \subref{recon:fig:nucProbs}~The corresponding radial probability
      distributions for each nucleus.}
\end{figure}

\begin{figure}[t]
   \centering
   \subfigure[$\Npart$ vs EOct]{
      \label{recon:fig:npartVsEOct}
      \includegraphics[width=0.4\linewidth]{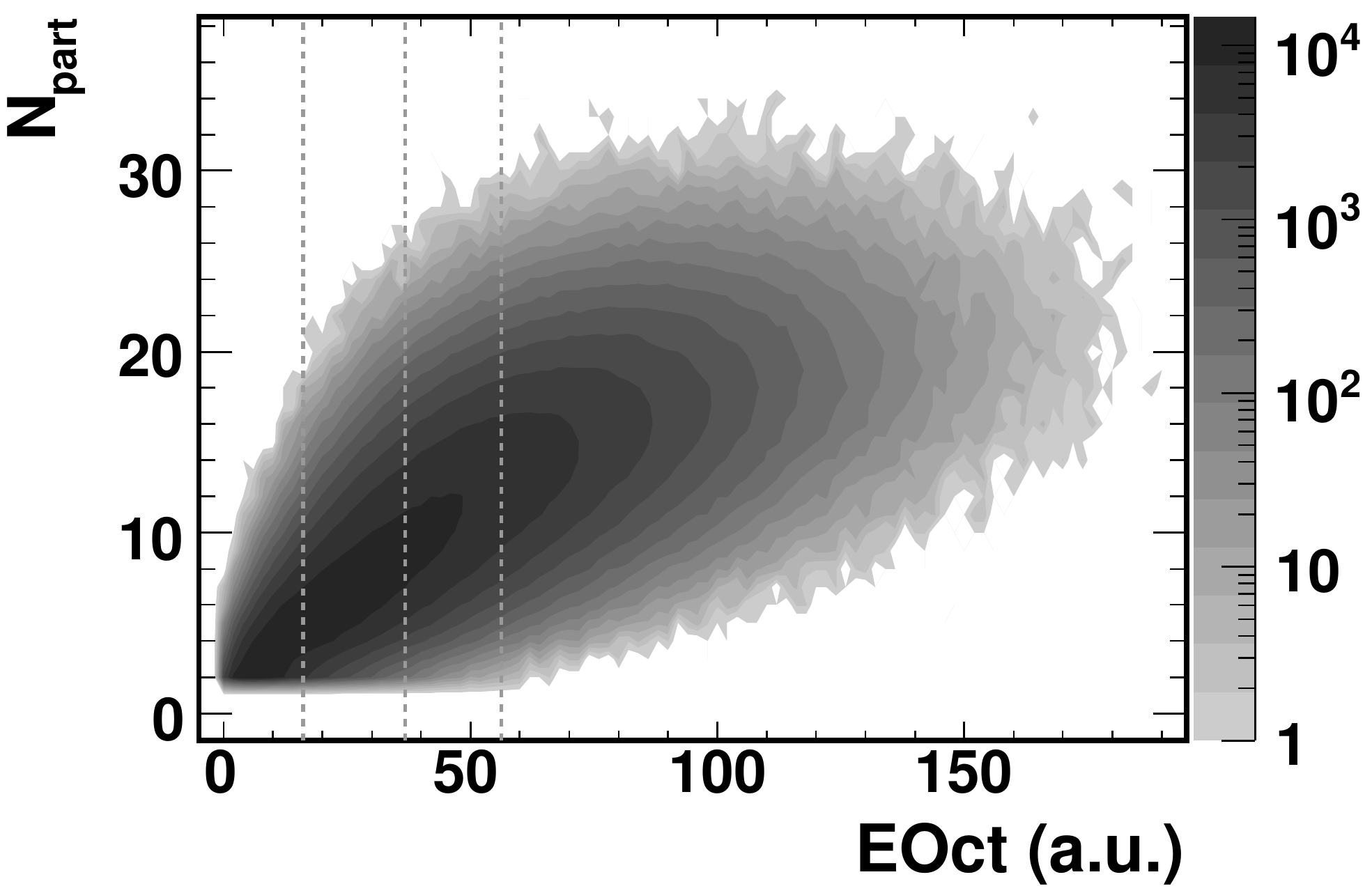}
   }
   \subfigure[$\Npart$ in Centrality Bins]{
      \label{recon:fig:npartEOct1D}
      \includegraphics[width=0.4\linewidth]{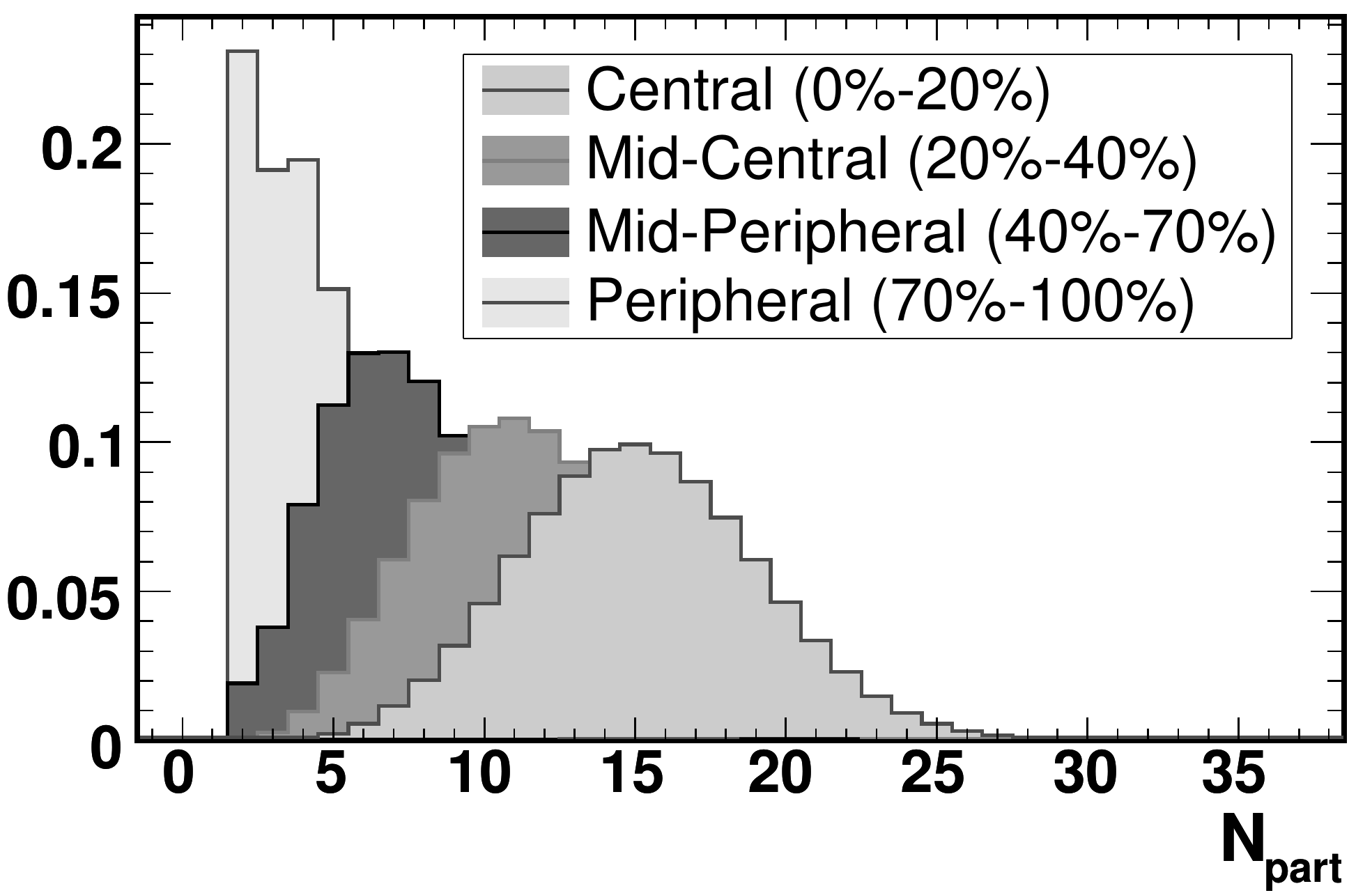}
   }
   \caption{   \label{recon:fig:npartfromeoct}
      The $\Npart$ distributions for the \protect\acs{EOct} centrality
      cuts. \subref{recon:fig:npartVsEOct}~$\Npart$ versus
      \protect\acs{EOct} in {\dAu} \protect\acs{HIJING}. The dashed grey
      lines show the centrality cut values.
      \subref{recon:fig:npartEOct1D}~The $\Npart$ distributions for each
      centrality class. Each distribution is (independently)
      normalized.}
\end{figure}

Once centrality cuts were obtained, they were applied to the \ac{MC}
simulations in order to study the properties of each centrality class.
The distribution of parameters such as $\Npart$ and $\Ncoll$ in each
centrality class could be directly examined in the \ac{MC}. Note that
the distributions of such parameters were model dependent. In both the
\ac{HIJING} and \ac{AMPT}~\cite{Zhang:1999bd} collision generators, a
Glauber model was used to estimate the number of participant nucleons
and binary collisions. In this model, the nucleons of each nucleus were
distributed according to a density function. The nucleons were then
assumed to travel in a straight line, undeflected by collisions. The
probability that nucleons would interact was taken from the total
inelastic {\pp} cross section of 41~{\mb} (which is appropriate for
$\snn=200~\gev$)~\cite{Back:2003ns}. The number of participants and
binary collisions were then directly counted. The simulations performed
for this analysis used \acs{HIJING} version~1.383, which modeled the
structure of the deuteron using the Hulthen wave
function~\cite{Lin:2003ah}

\begin{equation}
   \label{recon:eq:hulthen}
u(r) = C e^{-\alpha (2 r)} \prn{1 - e^{-\mu (2 r)}}
\end{equation}

\noindent%
where $r$ is half the distance between the nucleons in the deuteron (the
radius of the nucleus), $C$ is a constant that normalizes the
probability distribution $u(r)^2$, $\alpha=(4.38~\fm)^{-1}$ and
$\mu=(1.05~\fm)^{-1}$. The structure of the gold nucleus was modeled by
a Woods-Saxon distribution,

\begin{equation}
   \label{recon:eq:woodssaxon}
\rho(r) = \frac{\rho_{0}}{1 + \exp\prn{\frac{r - R}{a}}}
\end{equation}

\noindent%
where $r$ is the distance of a nucleon from the center of the
nucleus,$R=6.38~\fm$ describes the radius of the nucleus, $a=0.535~\fm$
describes the diffuseness of the nuclear edge and $\rho_{0}$ normalizes
the distribution. See \fig{recon:fig:nucProfs} for a diagram of these
probability densities.

For any given centrality class, the distribution of a parameter, such as
$\Npart$, could be analyzed to find the average value of that parameter.
This was done by constructing a histogram for each parameter and each
centrality class. The histogram was then filled with the value of the
parameter in every \ac{MC} collision that belonged to that centrality
class. The mean and \ac{RMS} of the parameter could then be directly
calculated from the histogram. The $\Npart$ distributions for the
\ac{EOct} centrality cuts, using \ac{HIJING} and the \acs{dAuSpectra}
event selection, are shown in \fig{recon:fig:npartfromeoct}. If the
event selection criteria were placed on the \acs{MC} when generating the
$\Npart$ distribution, then the average of the distribution would be a
\emph{biased} $\Npart$.

The bias introduced by the event selection could be undone either by
(a)~not applying the event selection or (b)~using the efficiency
(obtained according to \sect{recon:cent:cuts:eff}) to correct for the
bias. The latter method proceeded as follows. First, the efficiency of
the event was determined by evaluating the efficiency function at the
value of the centrality measure (i.e.~\ac{EOct}) in the event. Then, the
event was weighted by the inverse of this efficiency. For example, an
event with an efficiency of 50\% would be counted as two collisions in
the centrality parameter distribution. In this way, an unbiased
distribution of the centrality parameter (i.e.~$\Npart$) was
constructed. Imposing the event selection on the \ac{MC} and then using
the efficiency to ``remove'' it may seem unnecessary for the
simulations, when one could simply not impose any event selection.
However, this type of efficiency weighting could be performed on the
data (see \sect{ana:spec:centeff}), where an event selection was
required. Therefore, this was the preferred method for finding an
\emph{unbiased} average $\Npart$ for a particular centrality cut bin.

%---------------------------------------------------------------
\subsection{PCAL Centrality}
\label{recon:cent:pcal}
%---------------------------------------------------------------

Centrality cuts derived from \ac{EPCAL} signals were determined using a
related, but modified procedure. This was necessary because a reliable
model of the breakup of the gold nucleus was not implemented, so the
\ac{Au-PCAL} was not simulated in the \ac{HIJING} or \ac{AMPT}
\acl{MC}. The procedure that was developed exploited the monotonic
correlation in the {\dAu} data between the \ac{EPCAL} signal, denoted
$\spcal$, and the signal of another detector, denoted $\scorl$
(\ac{EOct}, for example).

\begin{figure}[t]
   \centering
   \subfigure[EPCAL vs EOct]{
      \label{recon:fig:pcalEoctCorl}
      \includegraphics[width=0.4\linewidth]{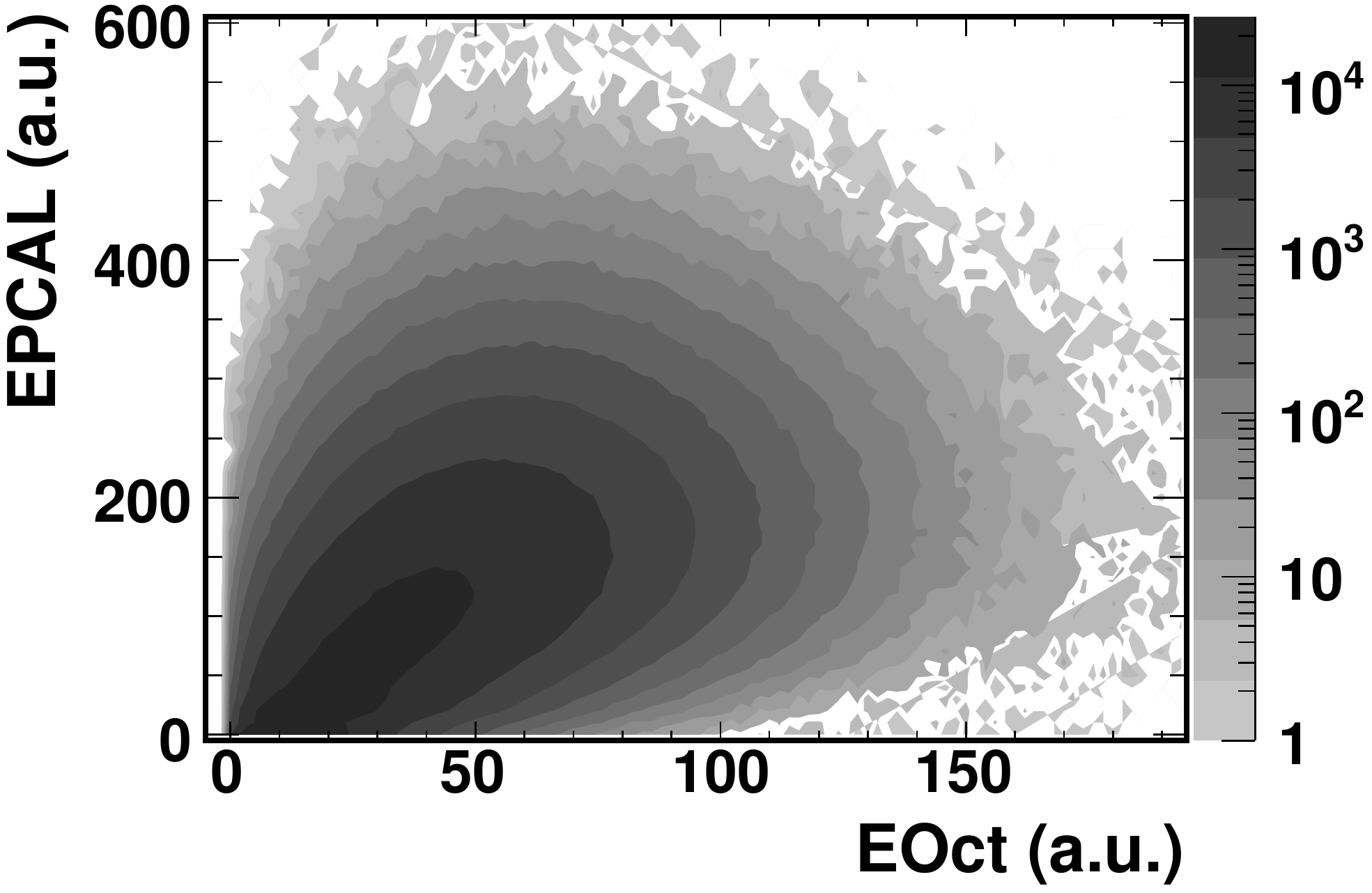}
   }
   \subfigure[EPCAL Efficiency]{
      \label{recon:fig:pcalEoctEff}
      \includegraphics[width=0.4\linewidth]{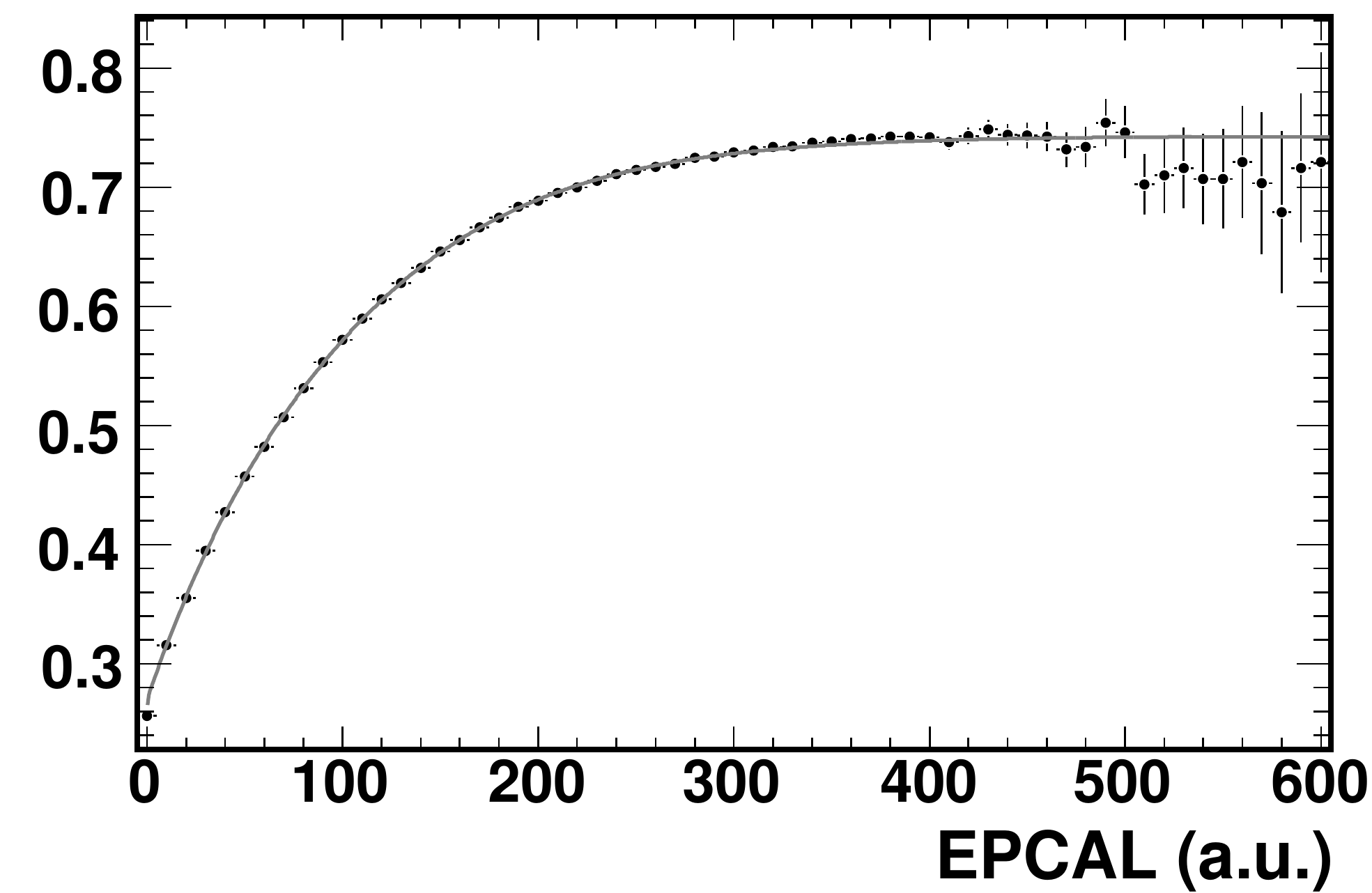}
   }
   \caption{   \label{recon:fig:pcalfromeoct}
      The \protect\acs{EPCAL} efficiency obtained from
      \protect\acs{EOct}. \subref{recon:fig:pcalEoctCorl}~The
      correlation between \protect\acs{EPCAL} and \protect\acs{EOct}
      in the {\dAu} data using the \protect\acs{dAuSpectra} event
      selection. \subref{recon:fig:pcalEoctEff}~The
      \protect\acs{EPCAL} efficiency obtained using the known
      \protect\acs{EOct} efficiency.}
\end{figure}

The method relied on being able to derive an estimate for the
efficiency of $\spcal$ using the known efficiency of $\scorl$. This
efficiency could then be used with the data to determine the centrality
cuts. The centrality cuts were found in several steps. First, the
distribution of $\spcal$ was generated for all events in the data that
passed the chosen event selection. Next, a second $\spcal$ distribution
was generated, for the same events, but in this distribution all events
were not weighted equally. Rather, an event was weighted, according to
\eq{recon:eq:eff}, by $1/\epsilon_{corl}(\scorl)$. For example, if the
correlated variable was \ac{EOct}, then in each event the histogram
would be filled with the \ac{EPCAL} signal, weighted by a certain
value. The value of the weight would be determined by evaluating the
\ac{EOct} efficiency function at the value of the \ac{EOct} signal in
the given event. The efficiency profile of $\spcal$ was then obtained
by dividing the $\spcal$ distribution by the weighted distribution.
This profile was then fit to obtain the efficiency function of $\spcal$.
An example using \ac{EOct} as the correlated signal is shown in
\fig{recon:fig:pcalfromeoct}.

\begin{figure}[t]
   \centering
   \subfigure[EPCAL Cuts]{
      \label{recon:fig:pcalEoctCuts}
      \includegraphics[width=0.4\linewidth]{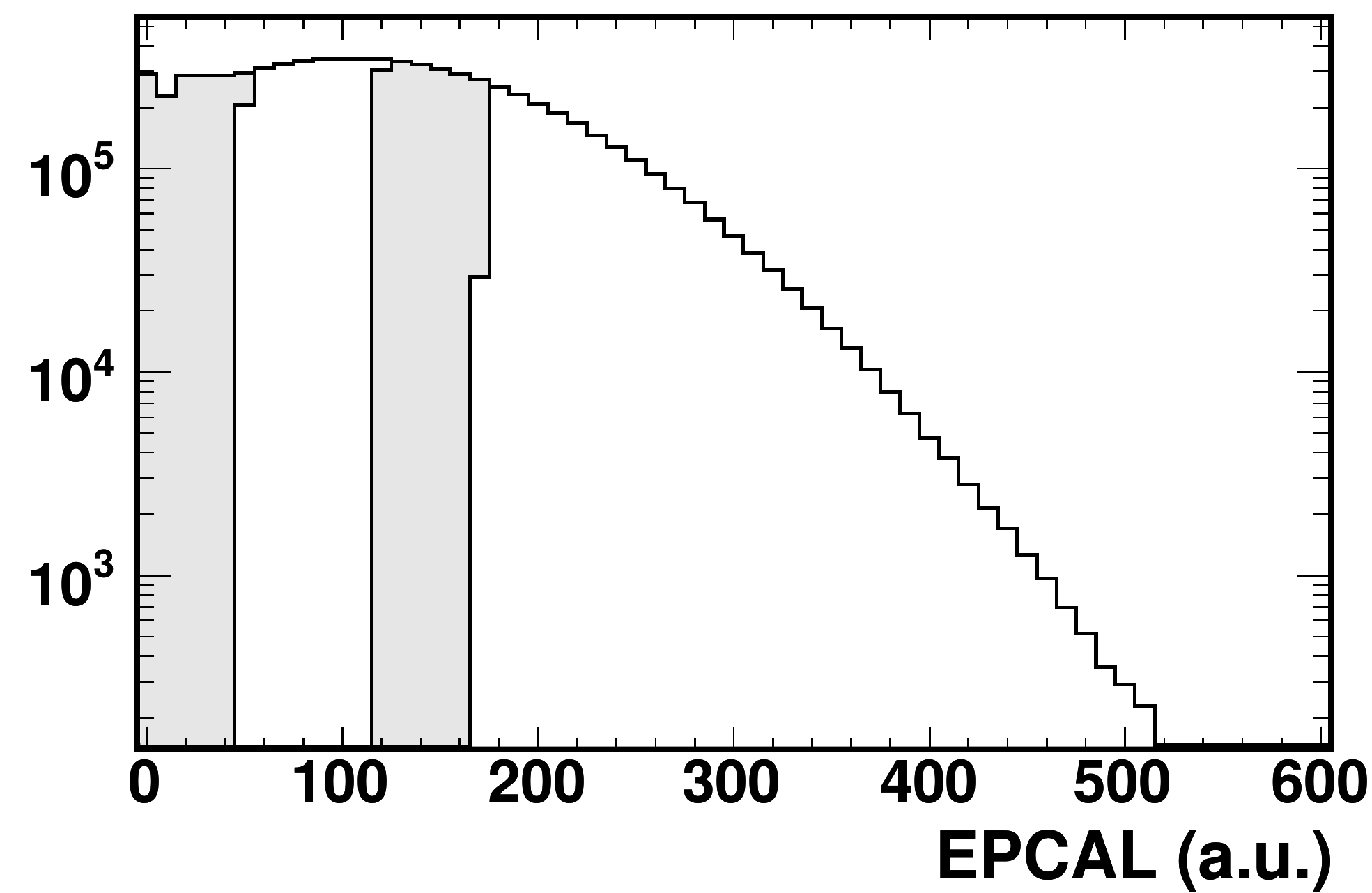}
   }
   \subfigure[EOct in Central Bin]{
      \label{recon:fig:eoctInPcalBin}
      \includegraphics[width=0.4\linewidth]{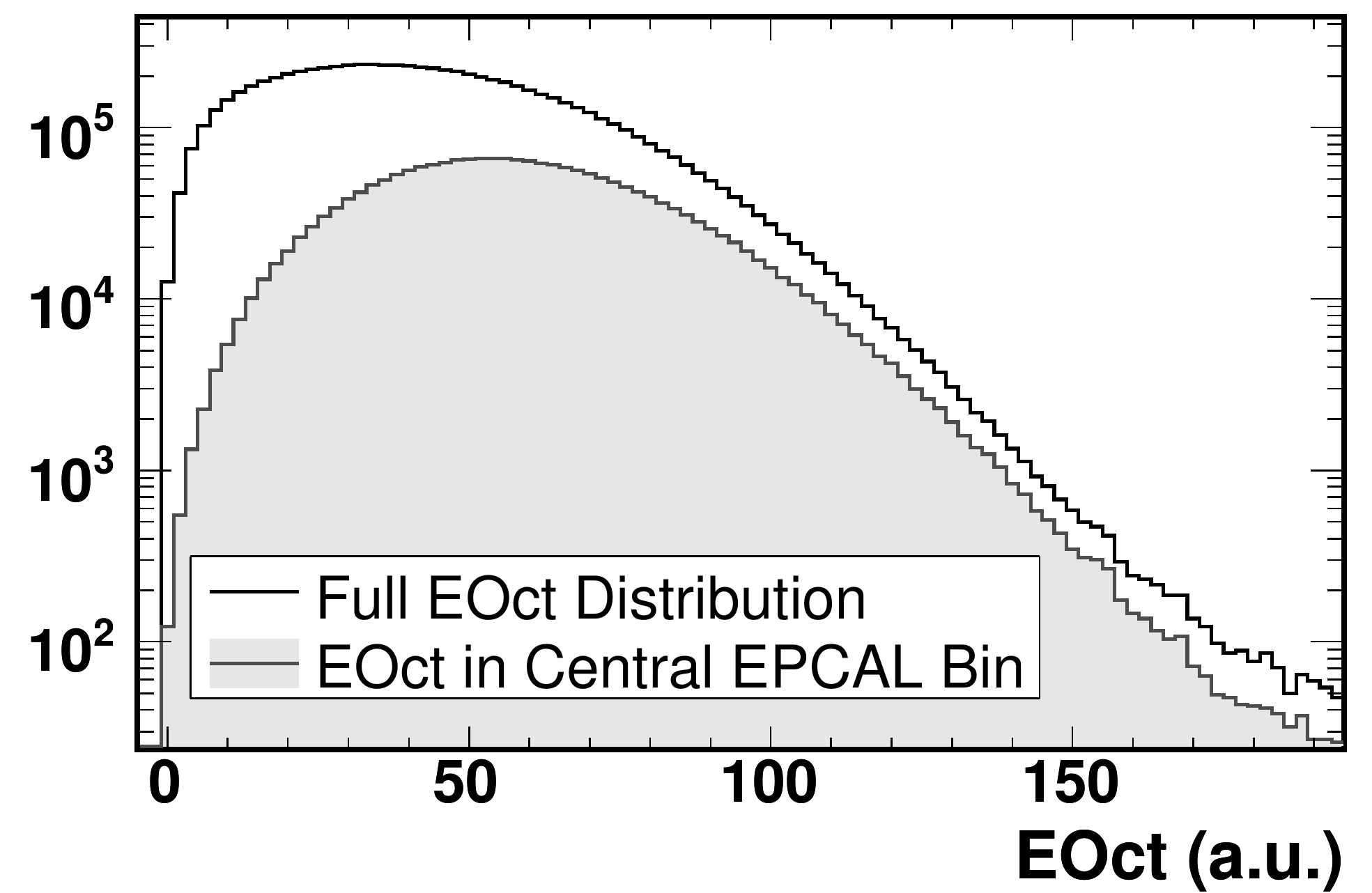}
   }
   \subfigure[$\Npart$ vs EOct Fit]{
      \label{recon:fig:npartVsEOctProf}
      \includegraphics[width=0.4\linewidth]{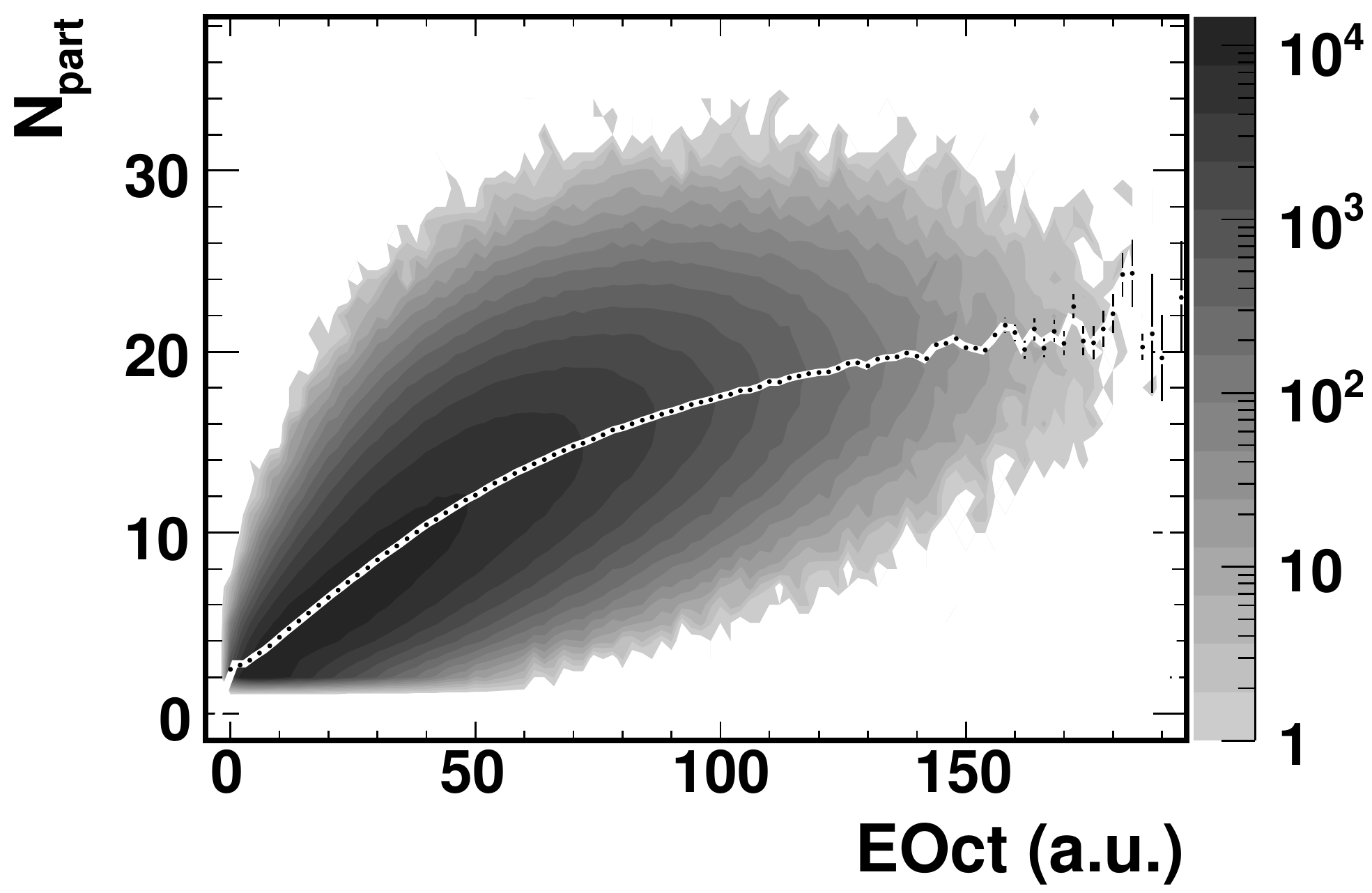}
   }
   \subfigure[$\Npart$ in Central Bin]{
      \label{recon:fig:npartFromEoctFit}
      \includegraphics[width=0.4\linewidth]{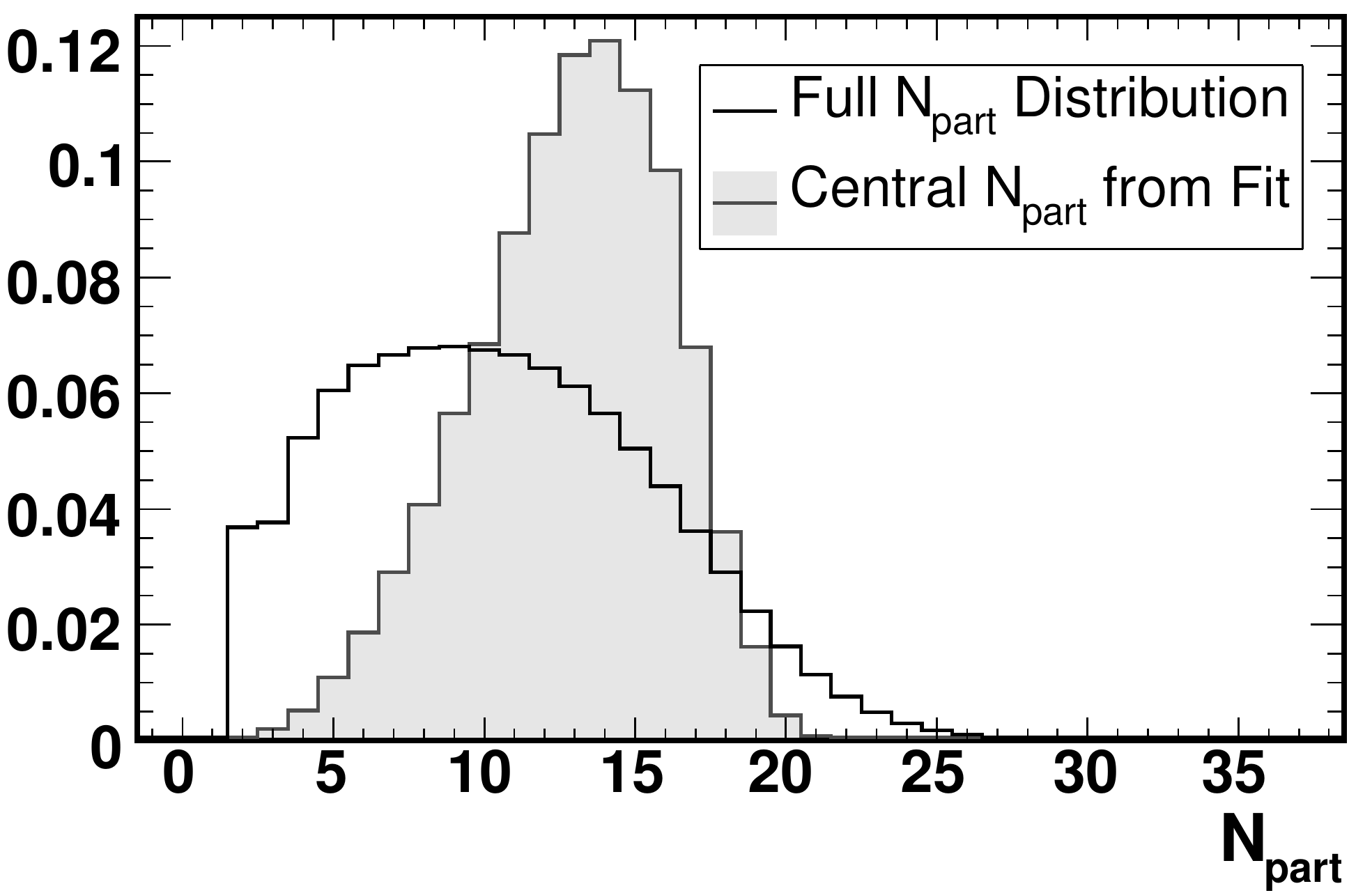}
   }
   \caption{   \label{recon:fig:pcaleoctfitnpart}
      $\Npart$ in \protect\acs{EPCAL} centrality bins obtained by the
      fit method. \subref{recon:fig:pcalEoctCuts}~The
      \protect\acs{EPCAL} centrality cuts from \protect\acs{EOct}
      with the \protect\acs{dAuSpectra} event selection.
      \subref{recon:fig:eoctInPcalBin}~The \protect\acs{EOct}
      distribution in the central \protect\acs{EPCAL} bin (grey
      histogram) compared to the full \protect\acs{EOct} distribution
      (black line). \subref{recon:fig:npartVsEOctProf}~The white line
      shows the fit to the $\Npart$ dependence on \protect\acs{EOct} in
      the \protect\acs{MC}. \subref{recon:fig:npartFromEoctFit}~The
      $\Npart$ distribution in the most central \protect\acs{EPCAL}
      bin (grey histogram) found using the fit method. Each
      distribution is (independently) normalized.}
\end{figure}

The efficiency of $\spcal$ was used to find \ac{EPCAL} centrality cuts
in the normal way (see \sect{recon:cent:cuts:crsscncuts}).
\Fig{recon:fig:pcalEoctCuts} shows the \ac{EPCAL} centrality cuts
obtained by using $\scorl=$\acs{EOct} with the \protect\acs{dAuSpectra}
event selection. The estimation of the average value of a collision
parameter, $\cpar$ (such as $\Npart$), required a new procedure due to
the lack of $\spcal$ in the simulations. Collision parameters were
determined by exploiting both the correlation of $\spcal$ with $\scorl$
and of $\scorl$ with $\cpar$. Two different procedures were developed to
find the average values of collision parameters in \ac{EPCAL}
centrality cut bins. Since \acs{EPCAL} centrality cuts could only be
applied to {\dAu} data, and not to \acs{MC}, both methods attempted to
estimate the average $\cpar$ of collisions in a centrality bin using the
$\scorl$ distribution of {\dAu} \emph{data} collisions in that bin.

First, a {\naive} approach was taken by simply fitting the dependence of
$\cpar$ on $\scorl$. The resulting function was then used to obtain an
estimate of $\cpar$ in a {\dAu} \emph{data} event given the value of
$\scorl$ in the event. The function was obtained by first generating a
two-dimensional histogram of $\cpar$ versus $\scorl$, such as the one
shown in \fig{recon:fig:npartVsEOct}. From this, a one-dimensional
\emph{profile} of the $\cpar$ dependence on $\scorl$ was obtained. The
position and error of each point in the profile distribution was
determined by the mean and \acs{RMS} of $\scorl$ bins in the
two-dimensional histogram. The profile was then interpolated by a
$3^{rd}$~degree polynomial spline with 100~knots.\footnote{That is, a
collection of $3^{rd}$~degree polynomials, each of which are fit in one
of the $100-1$ subintervals and are matched to be continuous and smooth.
See~\cite{NucRec:interp}.} \Fig{recon:fig:npartVsEOctProf} shows the
interpolated function obtained for $\scorl=$\acs{EOct} and
$\cpar=\Npart$ with the \protect\acs{dAuSpectra} event selection. From
the interpolated function, it was possible to estimate the value of
$\cpar$ given the value of $\scorl$ in a {\dAu} collision. Thus, $\cpar$
distributions could be obtained for each \acs{EPCAL} centrality bin. The
result of this method for the most central \acs{EPCAL} bin is shown in
\fig{recon:fig:npartFromEoctFit}.

\begin{figure}[t!]
   \centering
   \subfigure[EOct in Central Bin]{
      \label{recon:fig:eoctInPcalBinWt}
      \includegraphics[width=0.4\linewidth]{eoctInPcalBin}
   }
   \subfigure[EOct Weights in Central Bin]{
      \label{recon:fig:eoctWtsInPcalBin}
      \includegraphics[width=0.4\linewidth]{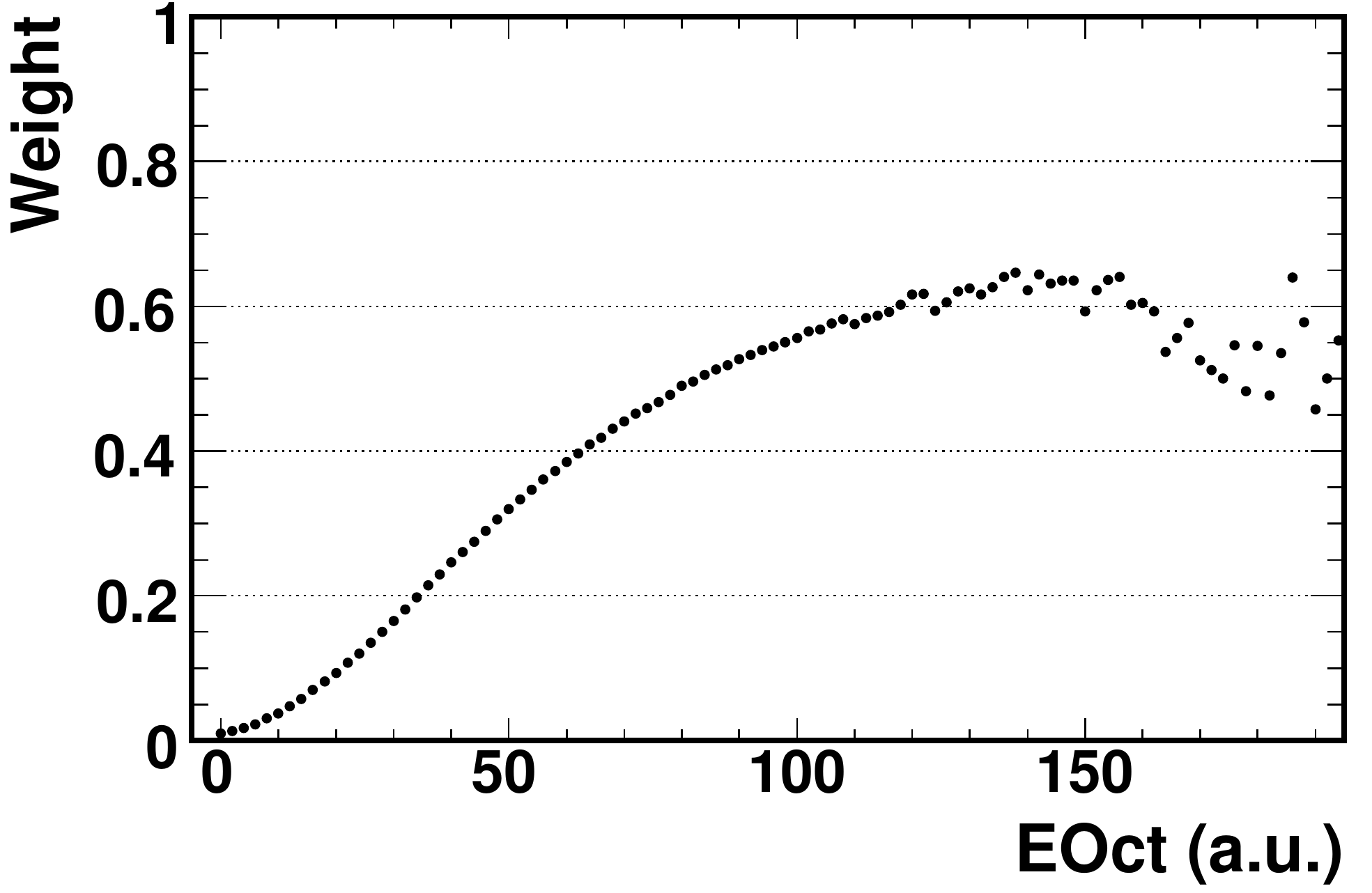}
   }
   \subfigure[$\Npart$ vs Weighted EOct]{
      \label{recon:fig:npartVsEOctWts}
      \includegraphics[width=0.4\linewidth]{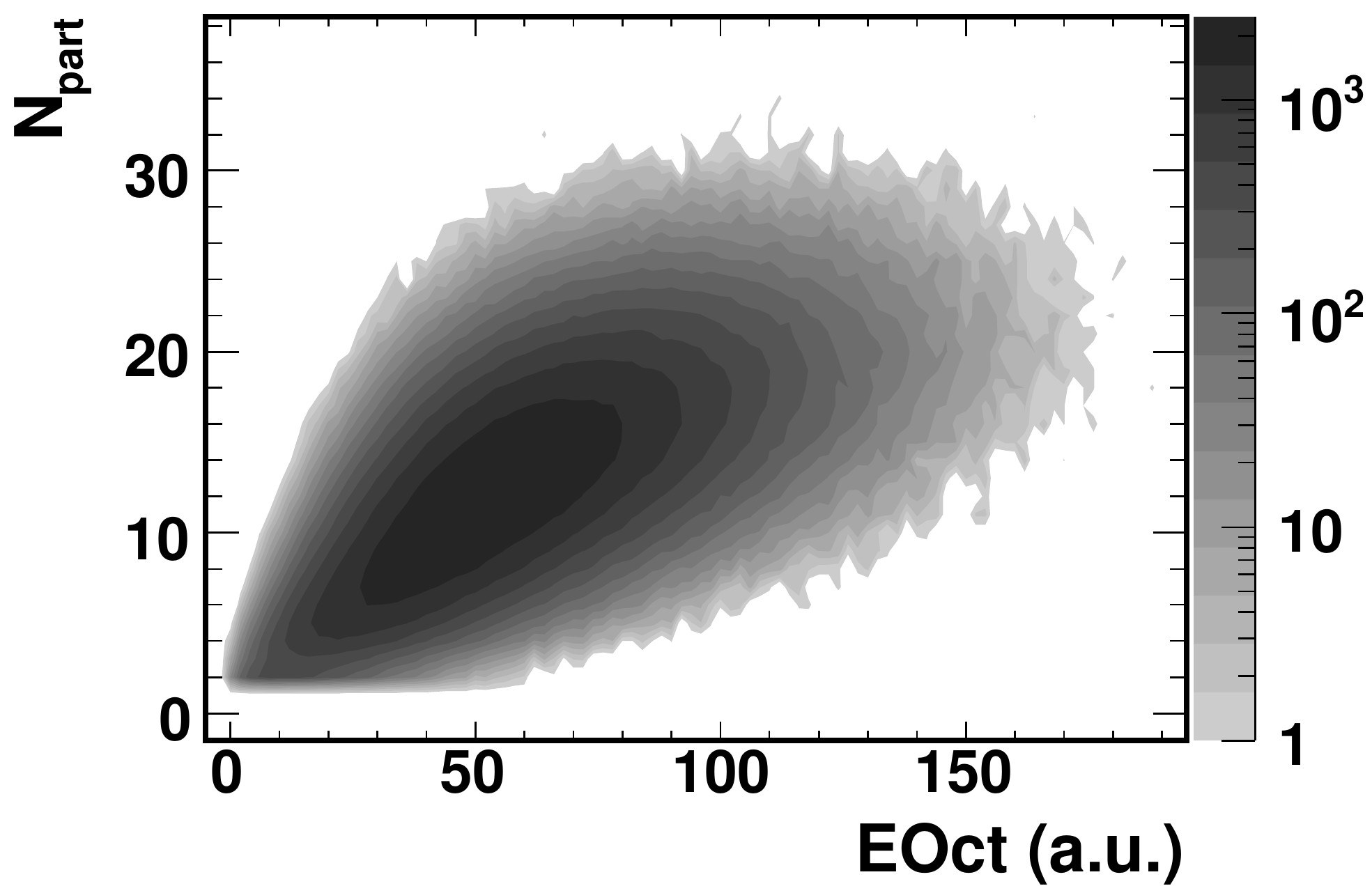}
   }
   \subfigure[$\Npart$ in Central Bin]{
      \label{recon:fig:npartEOctWts1D}
      \includegraphics[width=0.4\linewidth]{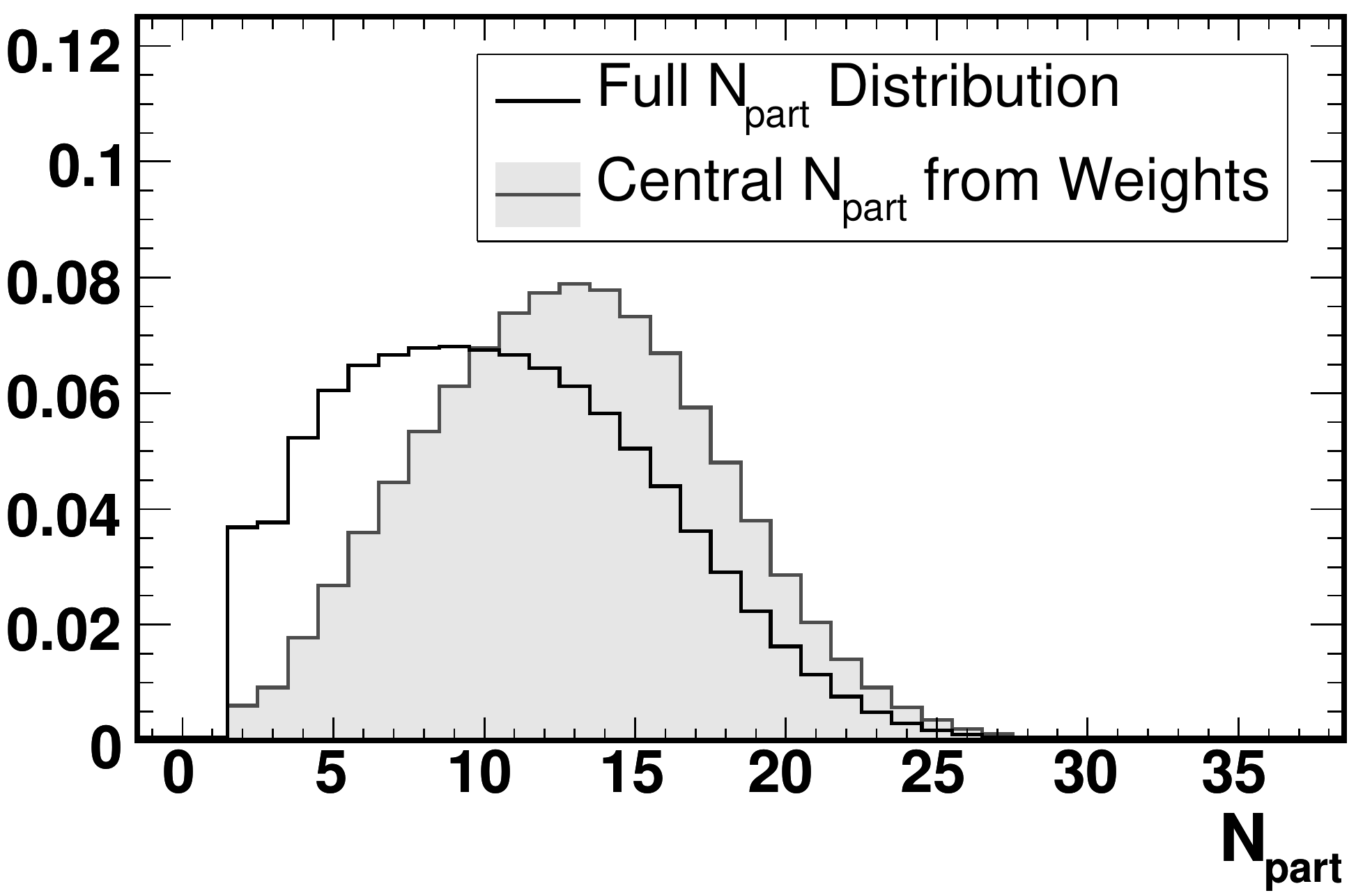}
   }
   \caption{   \label{recon:fig:pcalFromEOctWts}
      $\Npart$ in the most central \protect\acs{EPCAL} bin obtained
      by the weighting method. \subref{recon:fig:eoctInPcalBinWt}~The
      \protect\acs{EOct} distribution in the central
      \protect\acs{EPCAL} bin (grey histogram) compared to the full
      \protect\acs{EOct} distribution (black line).
      \subref{recon:fig:eoctWtsInPcalBin}~\protect\acs{EOct} weights in
      the central bin, obtained by taking the ratio of
      \protect\acs{EOct} in the central bin to the full
      \protect\acs{EOct} distribution.
      \subref{recon:fig:npartVsEOctWts}~The $\Npart$ dependence on
      \protect\acs{EOct} with \protect\acs{EOct} values weighted by
      their probability of occurring in the most central
      \protect\acs{EPCAL} bin. \subref{recon:fig:npartEOctWts1D}~The
      $\Npart$ distribution in the most central \protect\acs{EPCAL}
      bin (grey histogram) found using the weighting method. Each
      distribution is (independently) normalized. Compare to
      \fig{recon:fig:npartFromEoctFit}.}
\end{figure}

The second approach was to estimate the correlation between $\cpar$ and
$\scorl$ in an \acs{EPCAL} centrality bin. This was done by weighting
the $\scorl$ distribution in the \acs{MC}. Two-dimensional histograms of
$\cpar$ versus $\scorl$ for all \acs{EPCAL} centrality bins were
constructed. Each histogram was then filled using every \acs{MC} event, 
but all collisions were not weighted equally. Instead, a \acs{MC} event
was weighted by the likelihood that a {\dAu} collision, in the chosen
\acs{EPCAL} centrality bin, would be found having the same value of
$\scorl$. The weighted correlation between \acs{EOct} and $\Npart$ in
the most central \acs{EPCAL} bin is shown in
\fig{recon:fig:npartVsEOctWts}. The likelihood weights were obtained in
a simple manner. For each \acs{EPCAL} centrality bin, the distribution
of $\scorl$ in that bin was divided by the full $\scorl$ distribution
(see \figs{recon:fig:eoctInPcalBinWt}{recon:fig:eoctWtsInPcalBin}). The
average value of $\cpar$ in a \acs{EPCAL} centrality bin was then
estimated by projecting the two-dimensional histogram onto the $\cpar$
axis and calculating the mean, as shown in
\fig{recon:fig:npartEOctWts1D}.

%---------------------------------------------------------------
\section{Deuteron-Nucleon Tagging}
\label{recon:nuctag}
%---------------------------------------------------------------

%---------------------------------------------------------------
\subsection{The Deuteron}
\label{recon:nuctag:deuteron}
%---------------------------------------------------------------

The mass of a nucleus is always less than the sum of the masses of the
individual nucleons which make up the nucleus. This can be understood
using the famous formula $E = m c^2$. For a nucleus or a nucleon at
rest, its mass \emph{is} its total energy, and in order for the nucleus
to exist, it must be energetically favorable for the nucleons to bind
together. Thus, the total energy (i.e.~mass) of a nucleus at rest must
be less than the sum of the energy (mass) of each individual nucleon.
This difference in mass is known as the binding energy of the nucleus,

\begin{equation}
   \label{recon:eq:bindenergy}
B(A,Z) = c^2 \prn{Z M_{p} + (A-Z) M_{n} - M(A,Z)}
\end{equation}

\noindent%
where $A$ is the number of nucleons in the nucleus, $Z$ is the number
of protons, $M_{p}$ is the mass of a proton, $M_{n}$ is the mass of a
neutron, $M(A,Z)$ is the mass of the nucleus and $B(A,Z)$ is the
binding energy of the nucleus.

The deuteron is a very weakly bound nucleus. While all nuclei heavier
than Neon ($A=10$) have a binding energy above 7.4~{\mev} per
nucleon~\cite{StrucNuc:bind}, the deuteron has only

\begin{align}
   \label{recon:eq:deutbind}
B(A,Z) / A &= c^2 \prn{M_{p} + M_{n} - M_{d}} / 2\\
\notag &= \prn{938.27~{\mev} + 939.57~{\mev} - 1875.61~{\mev}} / 2\\
\notag &= 1.11~{\mev}
\end{align}

\noindent%
This weak binding energy has two consequences that are relevant to the
analysis presented in this thesis. First, as can be seen in
\fig{recon:fig:nucProfs}, it is quite possible for the nucleons of a
deuteron to be found relatively far apart. If the deuteron is in such a
state when it collides with a gold nucleus, then it is possible that
only one nucleon will actually participate in the collision. The second
consequence is that the remaining spectator nucleon can emerge from the
collision relatively unperturbed. Thus, such a collision is nearly
equivalent to the collision between a single nucleon and a gold
nucleus.

These types of {\pAu} and {\nAu} collisions were identified in the
{\dAu} data by actively looking for spectator nucleons from the
deuteron. If, in a {\dAu} collision, a spectator neutron was observed
and no spectator proton was observed, then it was inferred that the
proton interacted in a {\pAu} collision. Similarly, if the only
deuteron spectator observed was a proton, then the neutron must have
interacted in a {\nAu} collision.

%---------------------------------------------------------------
\subsection{Identifying Nucleon-Nucleus Collisions}
\label{recon:nuctag:NA}
%---------------------------------------------------------------

The deuteron spectators were measured in {\phob} using the
\ac{d-PCAL} and \ac{d-ZDC} detectors. Qualitatively, a
collision in which the \acs{d-PCAL} recorded a hit and the \acs{d-ZDC}
did not was labeled a {\nAu} interaction (and vice-versa). A hit in each
detector was defined as a signal that was within certain limits. These
limits could have been chosen tightly, so that only signals underneath
the neutron or proton peaks were used, but tight cuts would have
rejected a large number of collisions that were otherwise acceptable for
analysis. However, if the limits were too loose, then the fraction of
mis-identified nucleon-nucleus collisions would increase. The goal,
then, was to chose limits that would reject the fewest number of
collisions while still maintaining as pure a signal as possible.

Because the calorimeters were not simulated in the {\dAu} \ac{MC}, the
purity of different signal cuts could not be directly studied. Instead,
the shapes of certain centrality variable distributions, such as
\acs{ERing}, were studied in the {\dAu} data. This study was motivated
by the bias that tagging a deuteron spectator introduces on centrality:
a {\dAu} collision in which only one nucleon from the deuteron
interacts is likely to have a larger impact parameter than an average
{\dAu} collision. Thus, the \acs{ERing} distribution in a
nucleon-nucleus collision should be biased toward more peripheral
collisions when compared to the \acs{ERing} distribution of {\dAu}.

\begin{figure}[t!]
   \centering
   \subfigure[d-ZDC Regions]{
      \label{recon:fig:dZDCRegions}
      \includegraphics[width=0.4\linewidth]{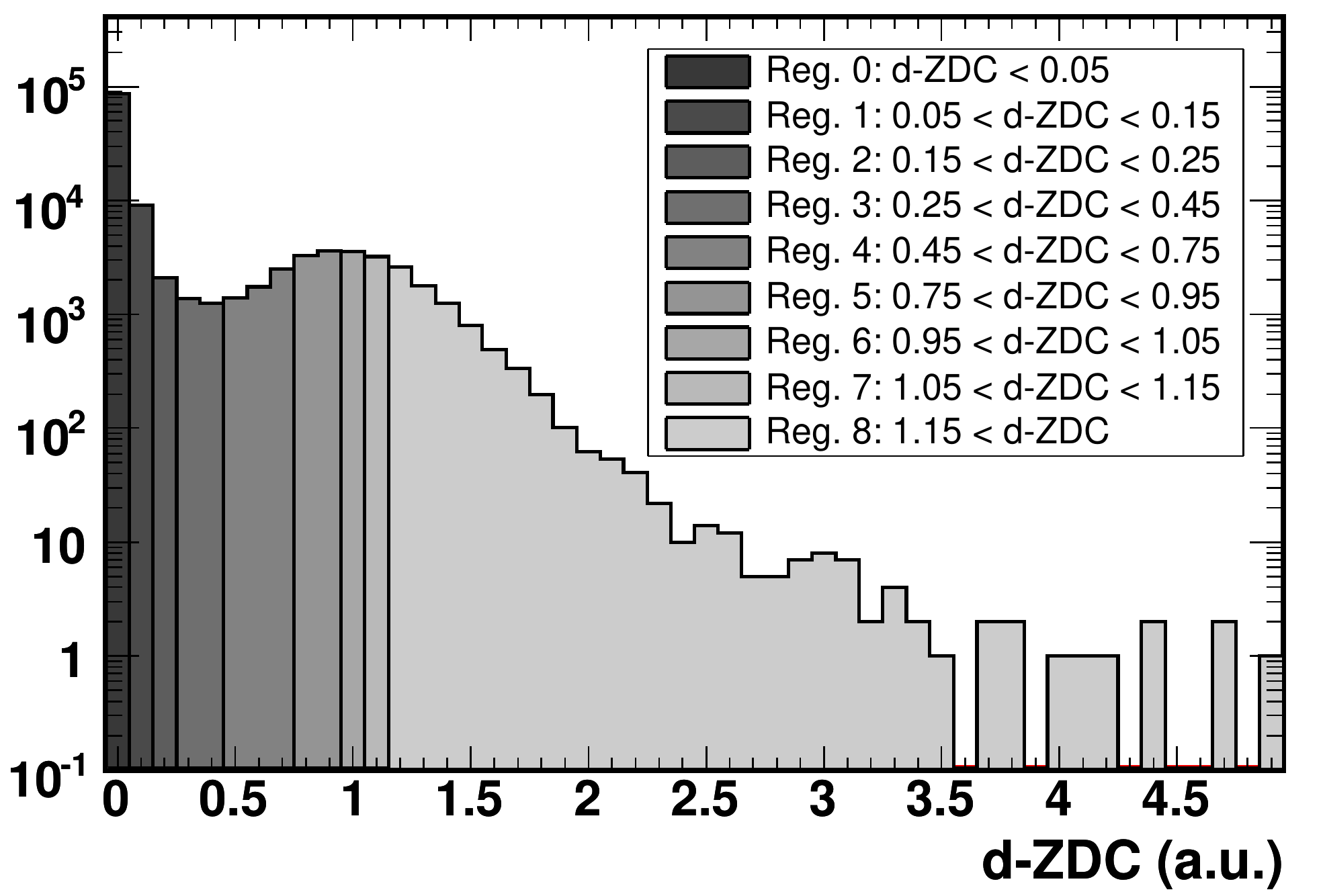}
   }
   \mbox{
      \subfigure[ERing in d-ZDC Regions]{
         \label{recon:fig:dZDCERing}
         \includegraphics[width=0.4\linewidth]{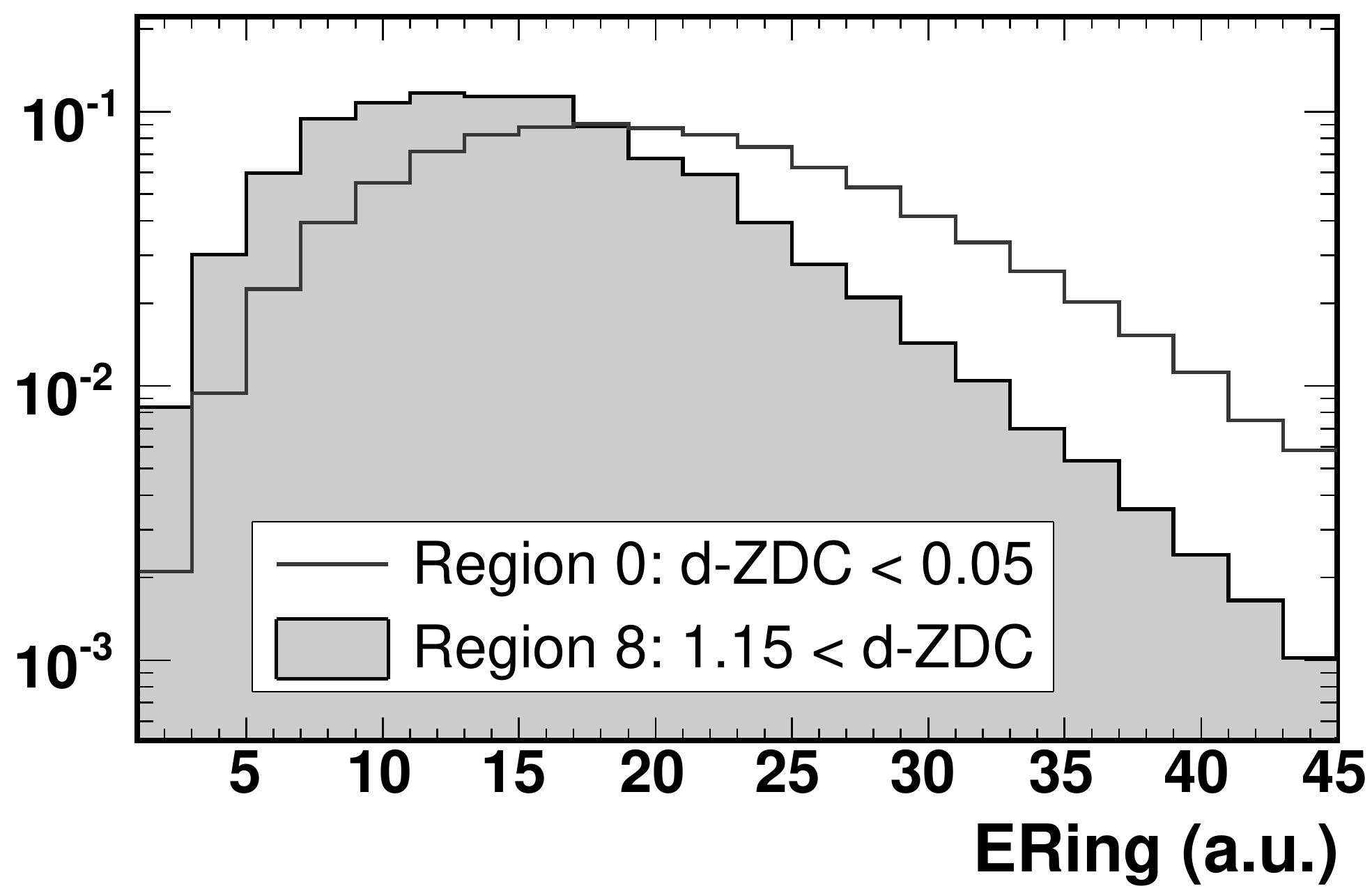}
      }
      \subfigure[Ratio: ERing in Each Reg. to Reg.~5]{
         \label{recon:fig:dZDCRatios}
         \includegraphics[width=0.4\linewidth]{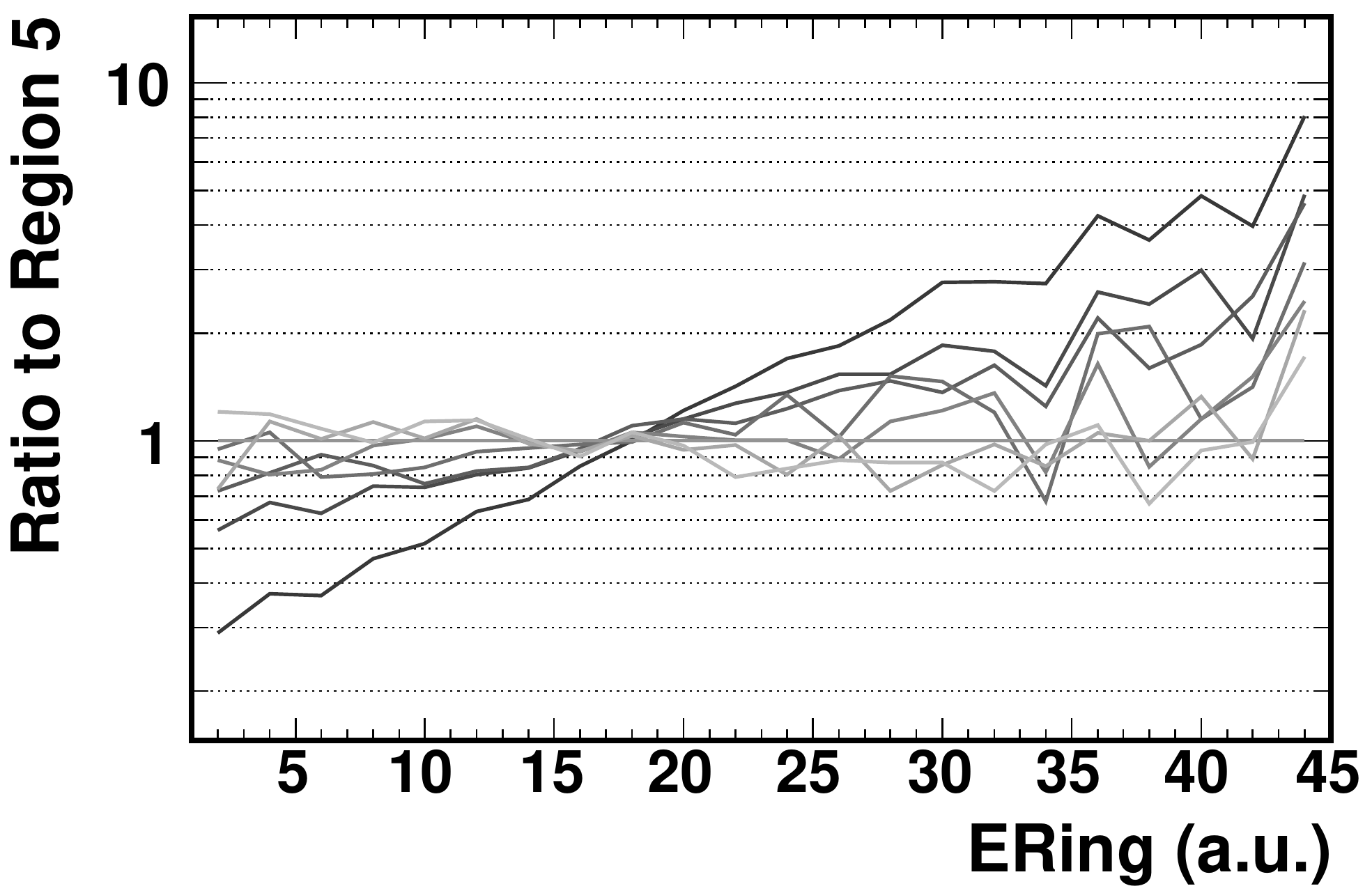}
      }
   }
   \caption{   \label{recon:fig:dZDCRegs}
      Studying the impact of \protect\acs{d-ZDC} cuts on the shape of a
      centrality variable. \subref{recon:fig:dZDCRegions}~The regions of
      \protect\acs{d-ZDC} used in the study.
      \subref{recon:fig:dZDCERing}~The normalized \protect\acs{ERing}
      distributions when no neutron is present (grey histogram) and when
      a neutron is clearly present (dark grey line).
      \subref{recon:fig:dZDCRatios}~The ratio of \protect\acs{ERing} in
      each region to that of region~5, which is under the neutron peak.
      The color of the lines follow the same color scheme as used in
      \subref{recon:fig:dZDCRegions}. Note that region~0, shown by the
      darkest grey line, in which no neutron is observed, has a
      significantly different shape from all other regions.}
\end{figure}

By studying the shape of centrality distributions for different regions
of \acs{d-ZDC} and \acs{d-PCAL} signals, it was possible to measure the
bias introduced by selecting nucleon-nucleus collisions.
\Fig{recon:fig:dZDCRegs} shows such a study of \acs{ERing}
distributions, for different regions of \acs{d-ZDC} signals. As can be
seen in \fig{recon:fig:dZDCERing}, the \acs{ERing} distribution from
collisions with the largest signals in the \acs{d-ZDC} detector is
indeed biased toward peripheral collisions when compared to the
\acs{ERing} distribution from collisions with no signal in the
\acs{d-ZDC}. The ratio of the latter distribution, \acs{ERing} in
region~0, to the distribution of \acs{ERing} from collisions with
\acs{d-ZDC} signals that are under the neutron peak (region~5) is
clearly visible in \fig{recon:fig:dZDCRatios}. This shows that
collisions with a \acs{d-ZDC} signal below 0.05~units can not be
identified as {\pAu} interactions. It also shows that the centrality
bias of collisions in regions~3 and above are roughly the same.

\begin{figure}[t!]
   \centering
   \subfigure[d-PCAL Regions]{
      \label{recon:fig:dPcalRegions}
      \includegraphics[width=0.4\linewidth]{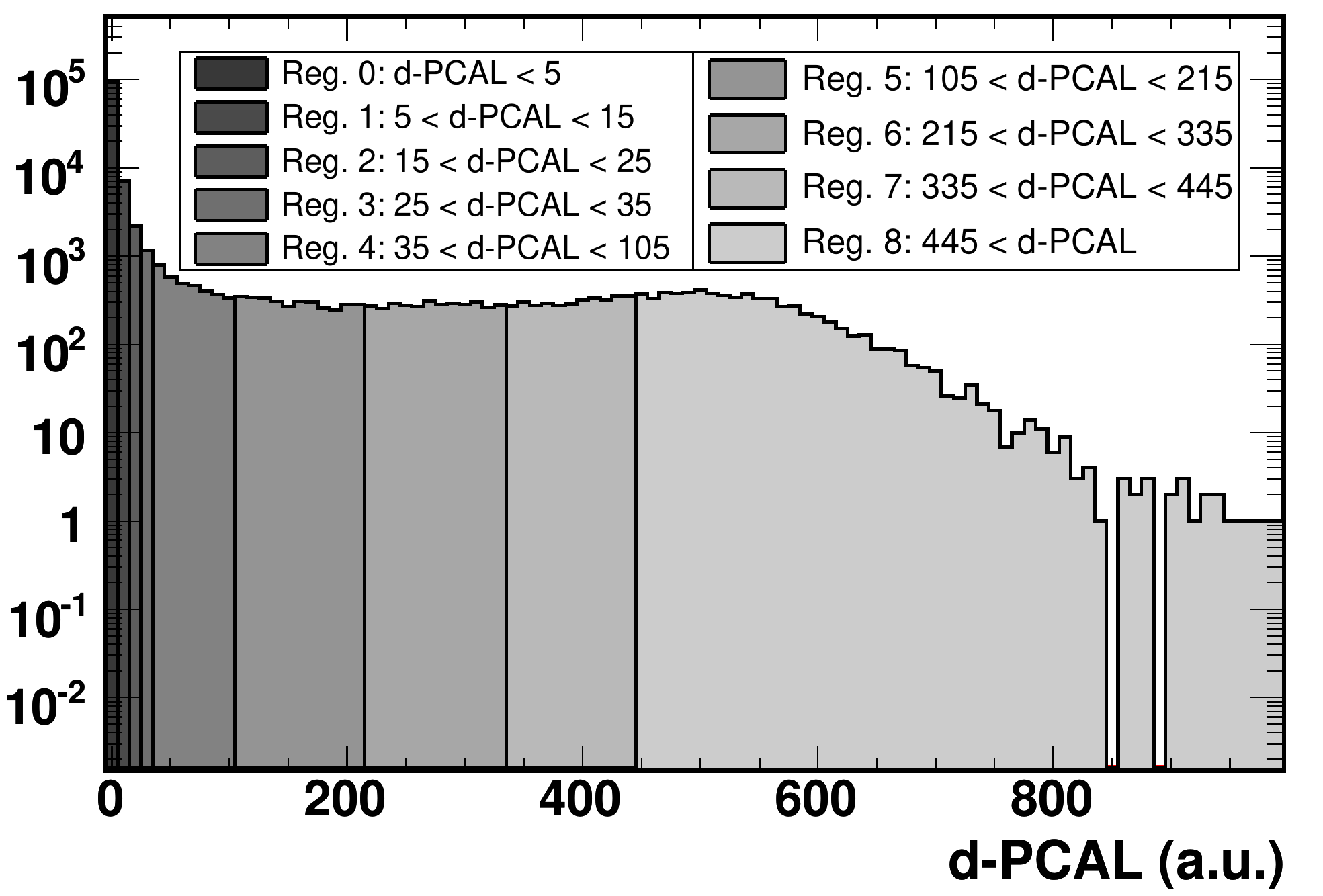}
   }
   \subfigure[Ratio: ERing in Each Reg. to Reg.~7]{
      \label{recon:fig:dPcalRatios}
      \includegraphics[width=0.4\linewidth]{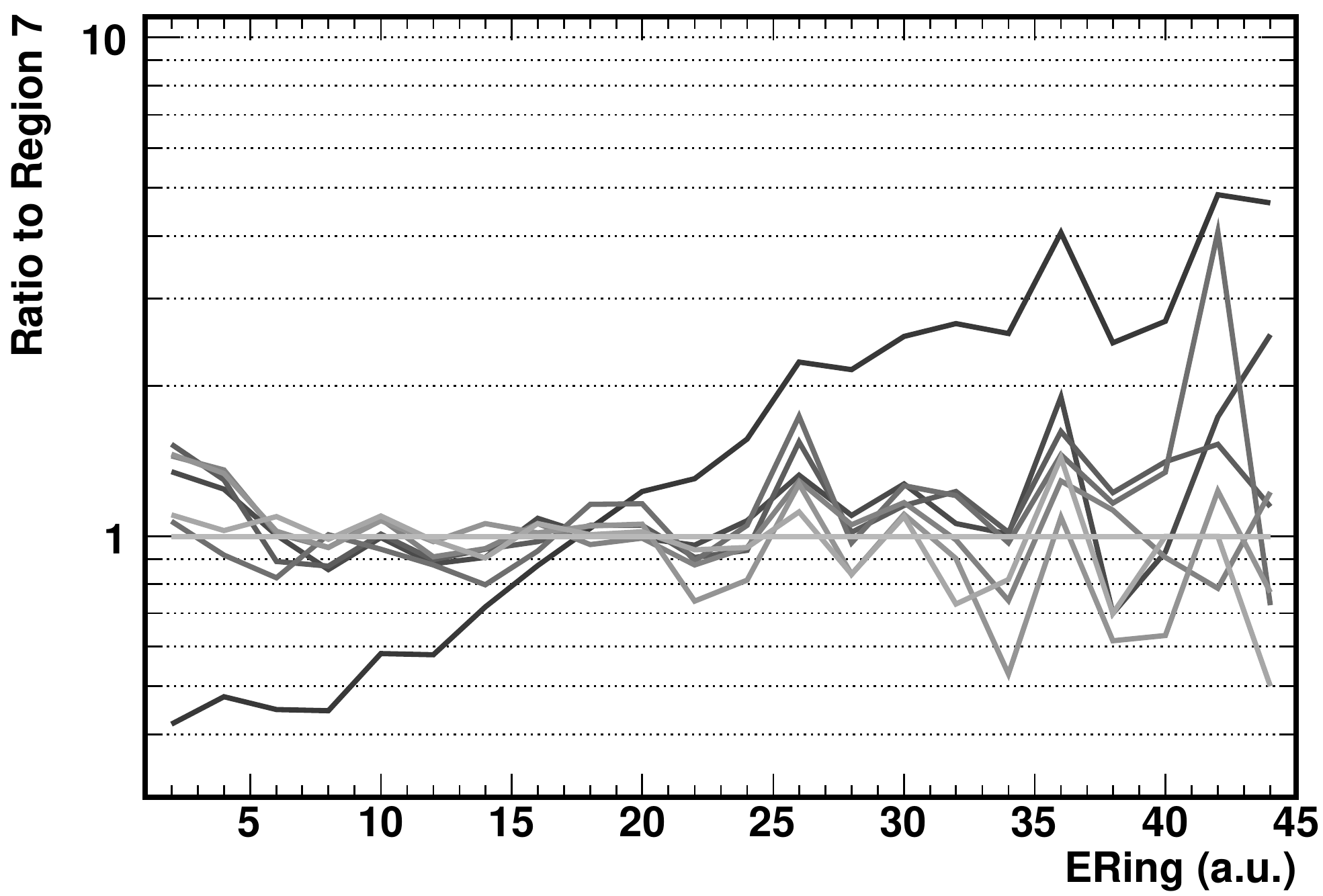}
   }
   \caption{   \label{recon:fig:dPcalRegs}
      Studying the impact of \protect\acs{d-PCAL} cuts on the shape of a
      centrality variable. \subref{recon:fig:dPcalRegions}~The regions
      of \protect\acs{d-PCAL} used in the study.
      \subref{recon:fig:dPcalRatios}~The ratio of \protect\acs{ERing} in
      each region to that of region~7. The color of the lines follow the
      same color scheme as used in \subref{recon:fig:dPcalRegions}. Note
      that region~0, in which no proton is observed, has a significantly
      different shape from all other regions.}
\end{figure}

\begin{figure}[t!]
   \centering
   \subfigure[d-PCAL vs d-ZDC]{
      \label{recon:fig:dPcalVsdZDC3D}
      \includegraphics[width=0.4\linewidth]{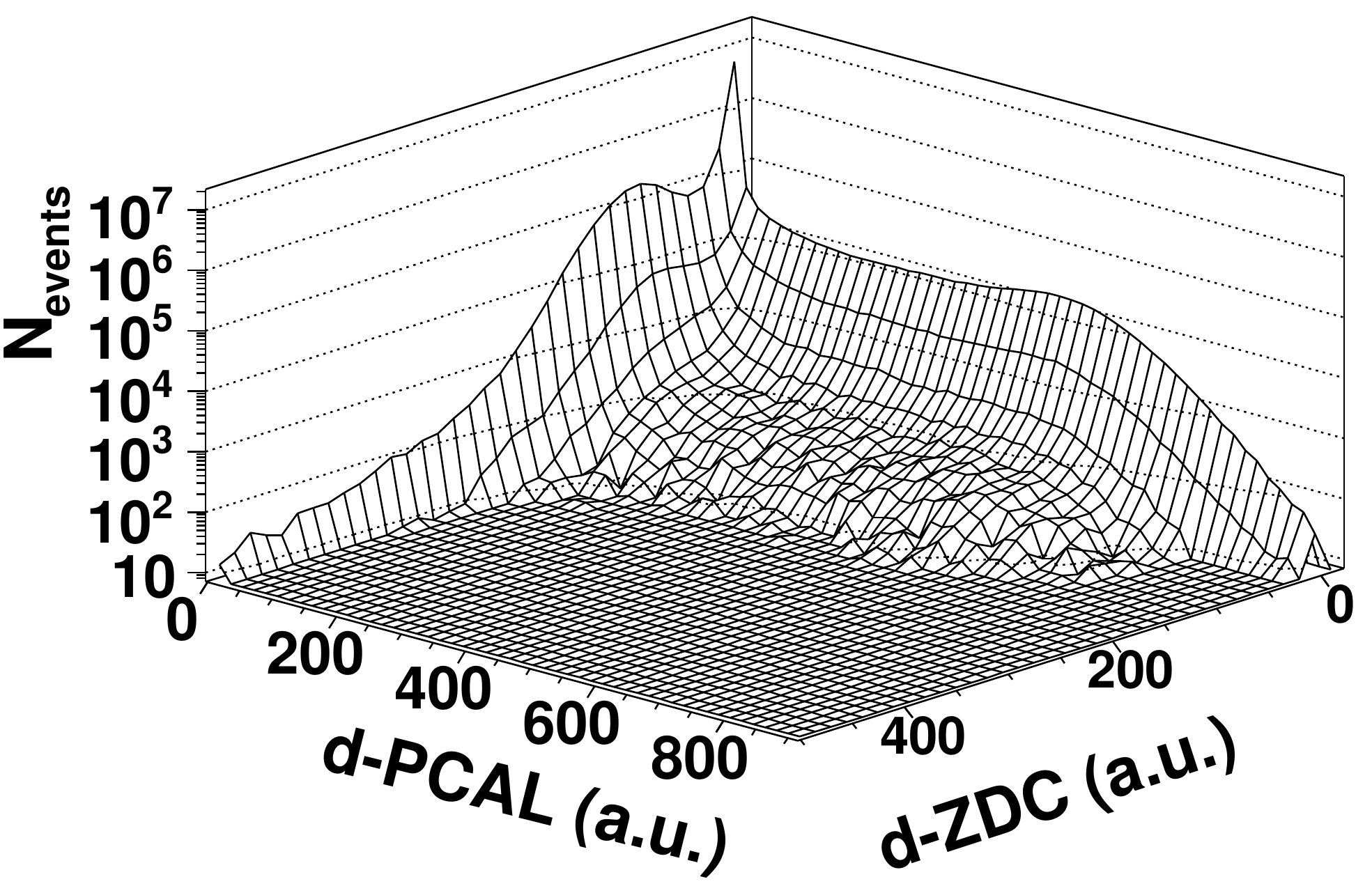}
   }
   \subfigure[d-PCAL vs d-ZDC with Tag Cuts]{
      \label{recon:fig:dPcalVsdZDC}
      \includegraphics[width=0.4\linewidth]{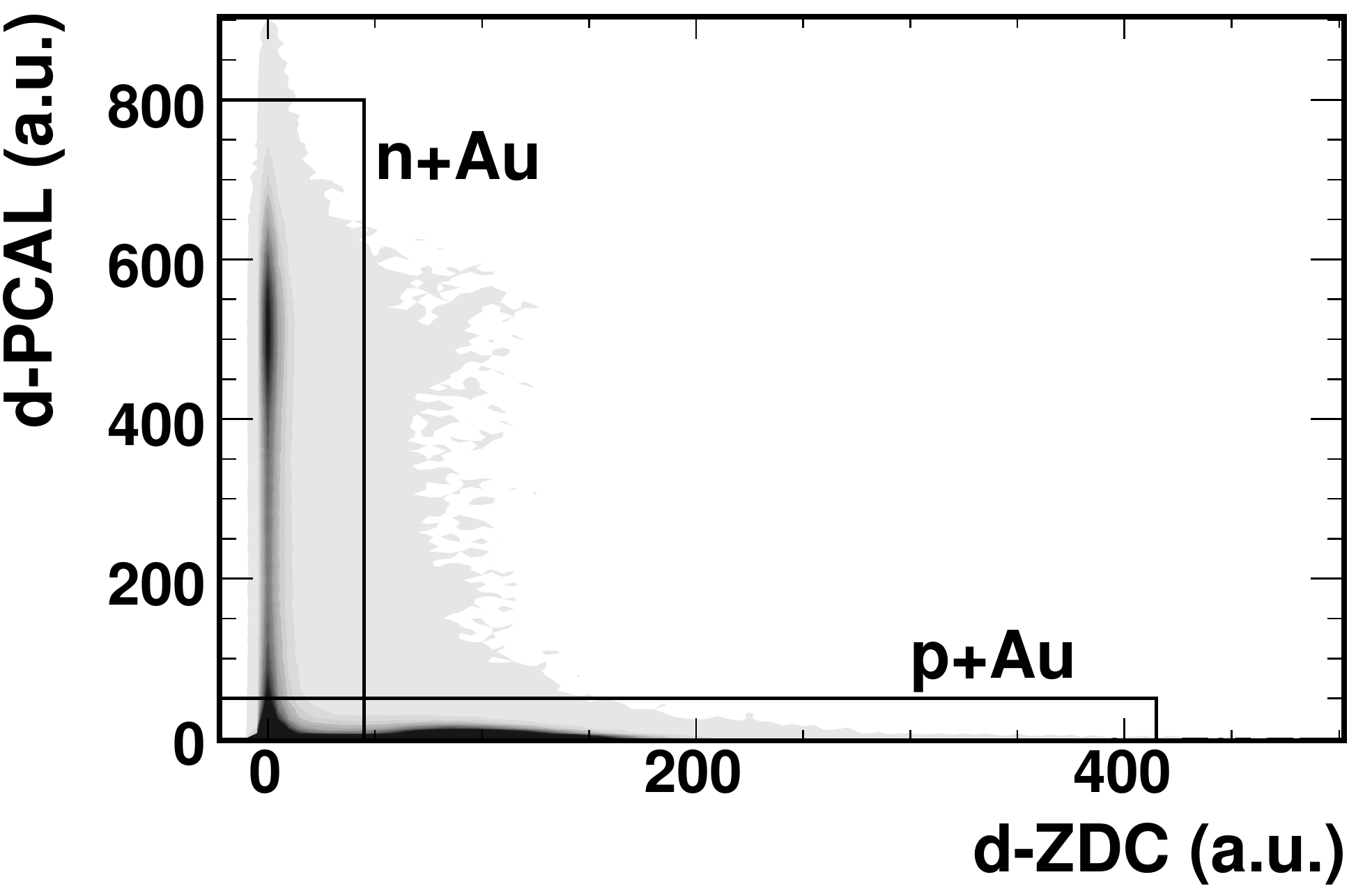}
   }
   \caption{   \label{recon:fig:dPcalVsdZDCRegs}
      The \protect\acs{d-PCAL} signal versus the \protect\acs{d-ZDC}
      signal. Note that in both plots, only bins that contain more than
      6 collisions are shown. \subref{recon:fig:dPcalVsdZDC3D}~A surface
      representation that shows the shape of each distribution.
      \subref{recon:fig:dPcalVsdZDC}~A contour representation that shows
      the lack of correlation between the two calorimeters. The boxes
      (at high \protect\acs{d-PCAL}, low \protect\acs{d-ZDC} and
      vice-versa) show the regions in which collisions were identified
      as a {\pAu} or {\nAu} collision.}
\end{figure}

A similar study for regions of \acs{d-PCAL} signals is shown in
\fig{recon:fig:dPcalRegs}. For the \acs{d-PCAL}, it is even more apparent
that regions showing any amount of signal in the calorimeter have a
similar centrality bias -- suggesting that this signal is indeed due to
a spectator from the deuteron. Therefore, the cuts were placed at the
high edge of the pedestal peak, more than $6~\sigma$ above the pedestal
(taking the average noise of a channel as $\sigma$). The final
nucleon-nucleus tagging cuts are shown in
\fig{recon:fig:dPcalVsdZDCRegs}.

%---------------------------------------------------------------
\subsection{Centrality of Nucleon-Nucleus Collisions}
\label{recon:nuctag:NAcent}
%---------------------------------------------------------------

Due to the lack of simulations of the calorimeters, it was not possible
to obtain the efficiency of an event selection that implemented
\acs{d-PCAL} or \acs{d-ZDC} signal cuts. Thus, no centrality cuts
specific to the tagged {\pAu} or {\nAu} data sets were generated.
Instead, the same cuts used for the untagged {\dAu} data were applied to
the {\pAu} and {\nAu} collision data. Of course, a centrality cut that
selected the top 20\% of the {\dAu} cross section would select a
different percentage of the {\pAu} (or {\nAu}) cross section.
Furthermore, without knowing the efficiency of the tagging procedures,
it was not possible to determine the percentage cross section of
centrality bins in the tagged {\pAu} and {\nAu} data\footnote{Such a
fractional cross section would probably not be very meaningful anyway,
since a fractional cross section bin of \emph{tagged} {\pAu} collisions
would not necessarily be the same as a fractional cross section bin of
true {\pAu} interactions.}. However, it \emph{was} possible to obtain
the average of \acs{MC} parameters, such as $\Npart$, in {\pAu} and
{\nAu} centrality bins.

The average value of \acs{MC} parameters in {\pAu} and {\nAu} centrality
bins were found using all simulated {\pAu} and {\nAu} collisions.
Ideally, one would not use all {\pAu} and {\nAu} collisions, but only
those that generated signals in the calorimeters which would pass the
nucleon tagging cuts. Since the calorimeters were not simulated, this
was not possible. Instead, the true \acs{MC} information about which
particles were in fact spectators was used. This procedure assumed that
the distribution of a \acs{MC} parameter, such as $\Npart$, was the same
in tagged {\pAu} ({\nAu}) collisions  as it was in true {\pAu} ({\nAu})
collisions. However, these distributions could be different if some
events that passed the tagging cuts were not really {\pAu} ({\nAu})
collisions or if the chance that a spectator would deposit energy in a
calorimeter depended on the centrality of the collision. Note that the
chance of \emph{observing} a spectator is distinct from the chance of
\emph{producing} a spectator, which certainly depends on centrality.
Thus, the validity of using true \acs{MC} information to select {\pAu}
and {\nAu} collisions in the simulations rested on three assumptions.
First, that it was not possible for a nucleon of the deuteron to both
interact with the gold nucleus and to deposit a measurable amount of
energy into one of the calorimeters. Second, that if such a nucleon did
not interact, it would be observed by the calorimeters with some
efficiency, but that this efficiency was \emph{independent} of the
centrality of the {\dAu} collision. Finally, that the \acs{d-PCAL} and
\acs{d-ZDC} detected only deuteron spectators.

\begin{figure}[t!]
   \centering
   \subfigure[Proton Energy Cut]{
      \label{recon:fig:mctagPenergy}
      \includegraphics[width=0.4\linewidth]{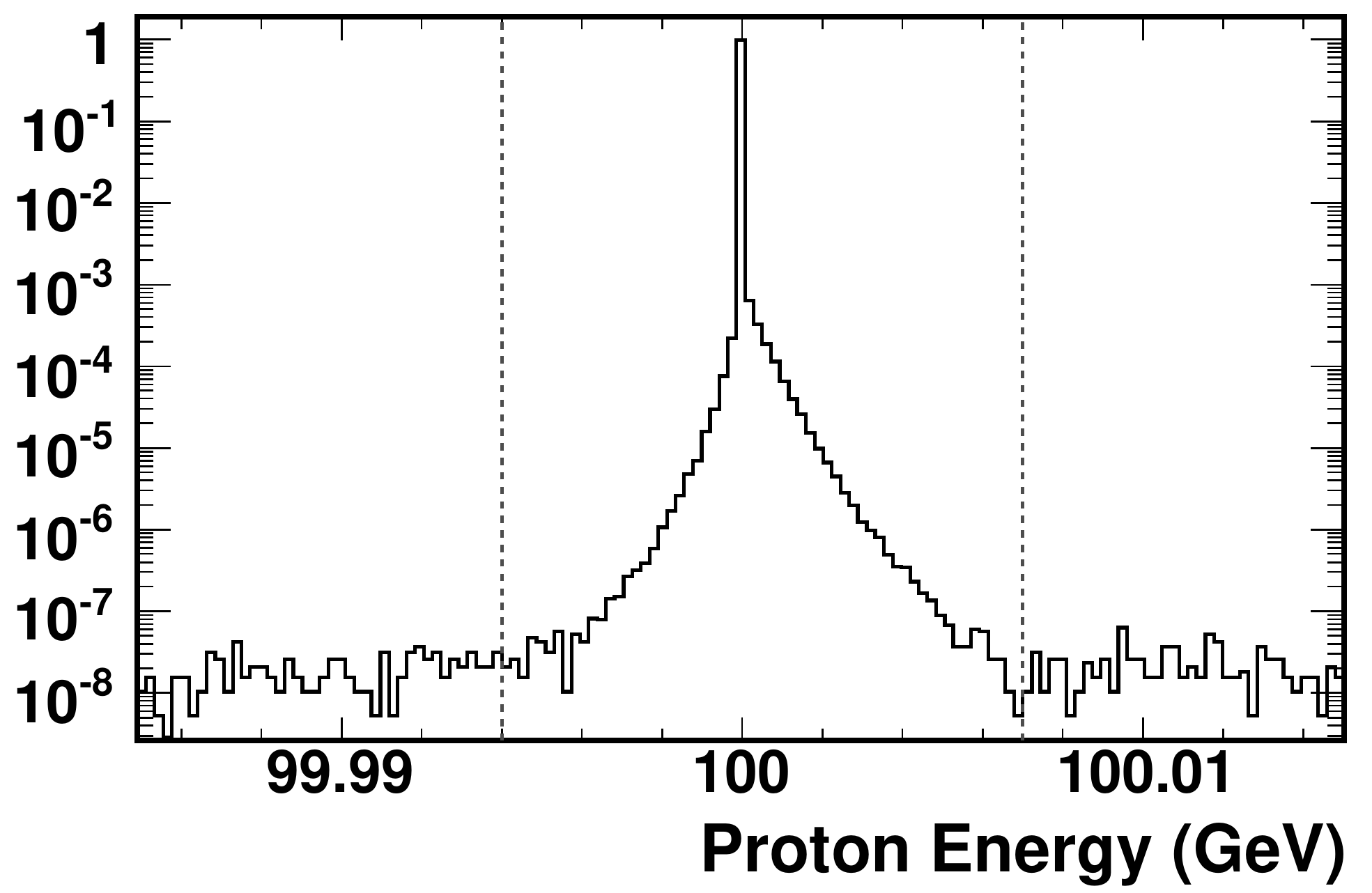}
   }
   \subfigure[Neutron Energy Cut]{
      \label{recon:fig:mctagNenergy}
      \includegraphics[width=0.4\linewidth]{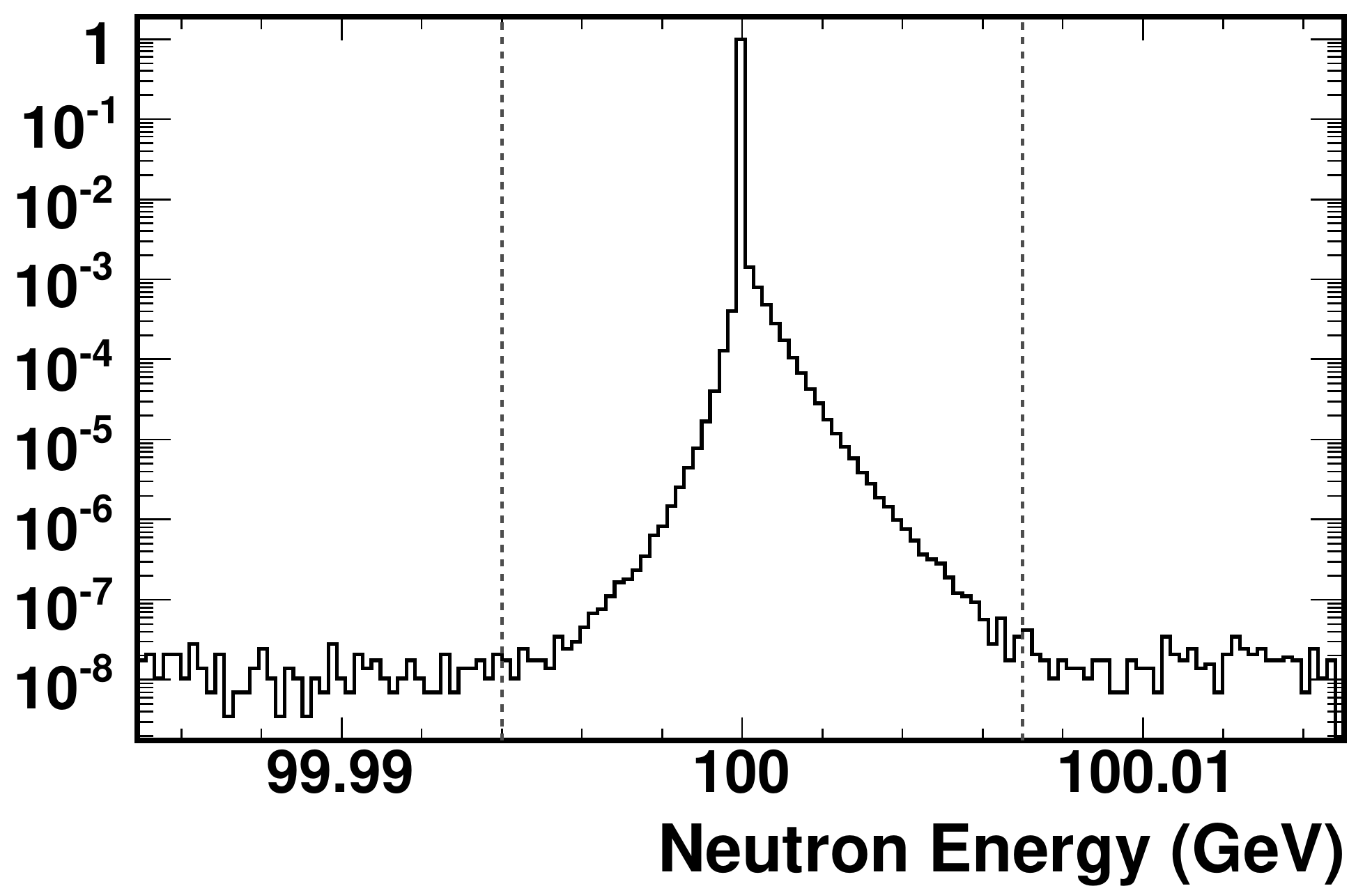}
   }
   \caption{   \label{recon:fig:mctagenergy}
      The energy cuts used to find spectators in simulated {\dAu}
      collisions are shown by the dashed grey lines.
      \subref{recon:fig:mctagPenergy}~The energy of protons in the
      \protect\acs{HIJING} \protect\acs{MC}. 
      \subref{recon:fig:mctagNenergy}~The energy of neutrons in the
      \protect\acs{HIJING} \protect\acs{MC}.}
\end{figure}

Given these assumptions, a simple method was developed to find
spectators in simulated collisions. The procedure was to use a
restrictive energy cut: any proton or neutron having between
99.994~{\gev} and 100.007~{\gev} of energy was assumed to be a
spectator. Spectators were identified as coming from a particular
nucleus by their momentum in the beam direction; positive $p_{z}$
implied a deuteron-spectator and negative $p_{z}$ implied a
gold-spectator. The distributions of proton and neutron energy in
\acs{HIJING} simulations are shown in \fig{recon:fig:mctagenergy}. For
\ac{HIJING}, which reported the number of participants of each nucleus
(but not separately for neutrons and protons), this energy cut almost
always yielded the correct number of deuteron spectators. A value
different from that reported by \ac{HIJING} occurred only once in every
500,000 collisions.

The same procedure was used for the \ac{AMPT} simulations. In the
original \ac{AMPT} model, dated August~18, 2003, the number of deuteron
and gold participants reported did not come directly from the Glauber
model. Instead, some kinematic cuts were used to identify the
spectators, thereby determining the number of participants. In order to
obtain consistent results between \ac{AMPT} and \ac{HIJING}, the
\ac{AMPT} model used in the analysis presented in this thesis was
altered. In this altered \ac{AMPT}, the number of participants of each
nucleus was taken directly from the Glauber model (as run by the
\ac{HIJING} portion of the full \ac{AMPT} model). However, the energy
cut could not be used to reproduce the number of participants reported
by the Glauber model. This was due to the hadron transport
model~\cite{Li:1995pr} incorporated in \ac{AMPT}, which was used to
model the interactions between hadrons formed after a {\dAu} collision.
In this model, it was possible for hadrons that did not interact in the
initial collision between the nuclei (i.e.~spectators in the Glauber
model) to later participate in the final stages of the collision. Note
that the energy cuts could also not reproduce the number of participants
reported by the original \ac{AMPT} model. This was due to the fact that
the original \ac{AMPT} counted the number of participants prior to
running the hadron transport model. Nevertheless, the discrepancy
between the number of deuteron participants reported by the Glauber
model and the number that could be identified using the energy cuts was
a feature of the \ac{AMPT} model. It did not invalidate the procedure
used to identify the deuteron participant nucleons.

\begin{figure}[t]
   \centering
   \subfigure[\protect\acs{ERing} Distributions]{
      \label{recon:fig:dataTagERingDists}
      \includegraphics[width=0.4\linewidth]{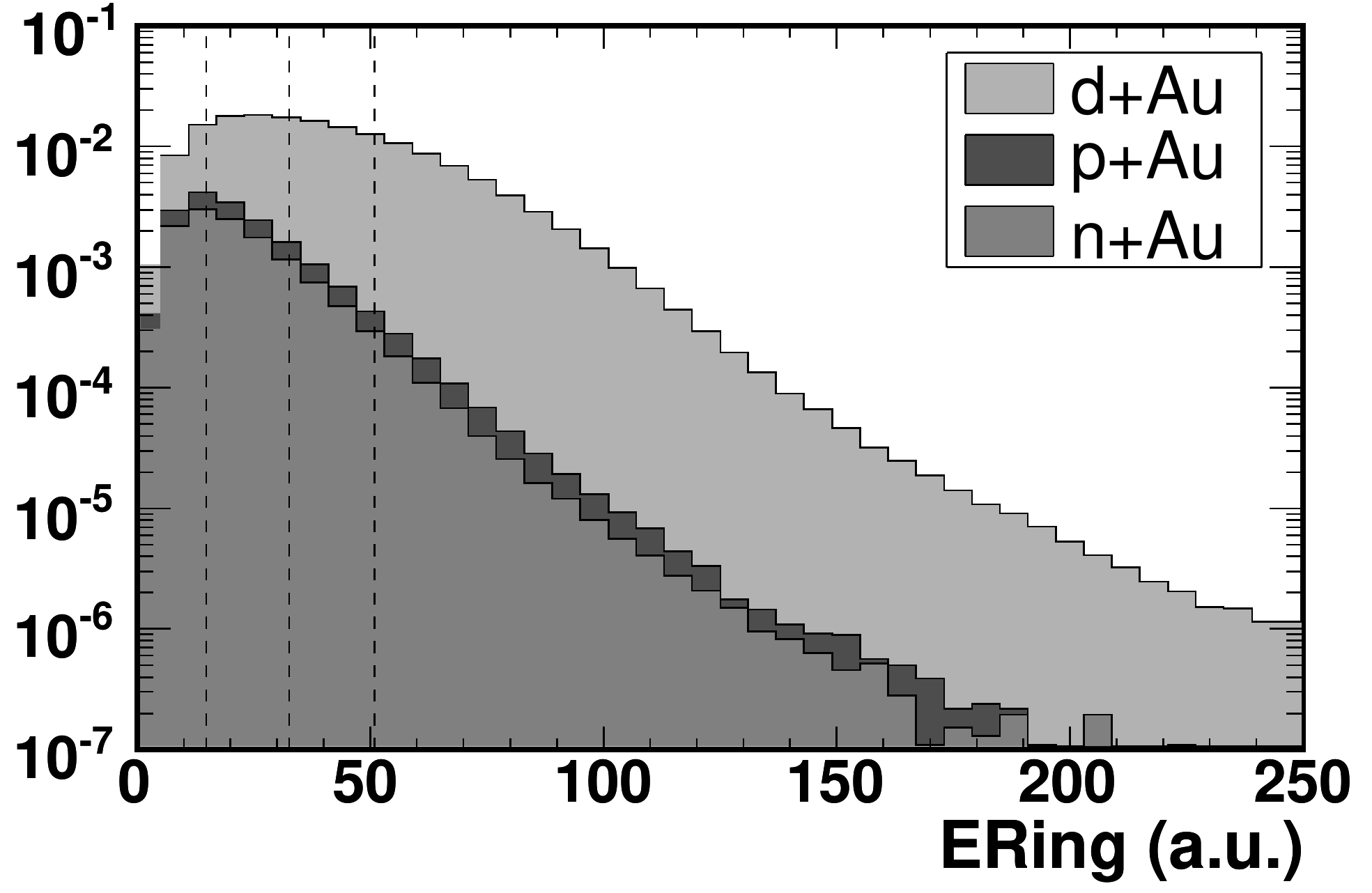}
   }
   \subfigure[{\nAu} / {\dAu} in Data and \protect\acs{MC}]{
      \label{recon:fig:dataMCTagERingRatio}
      \includegraphics[width=0.4\linewidth]{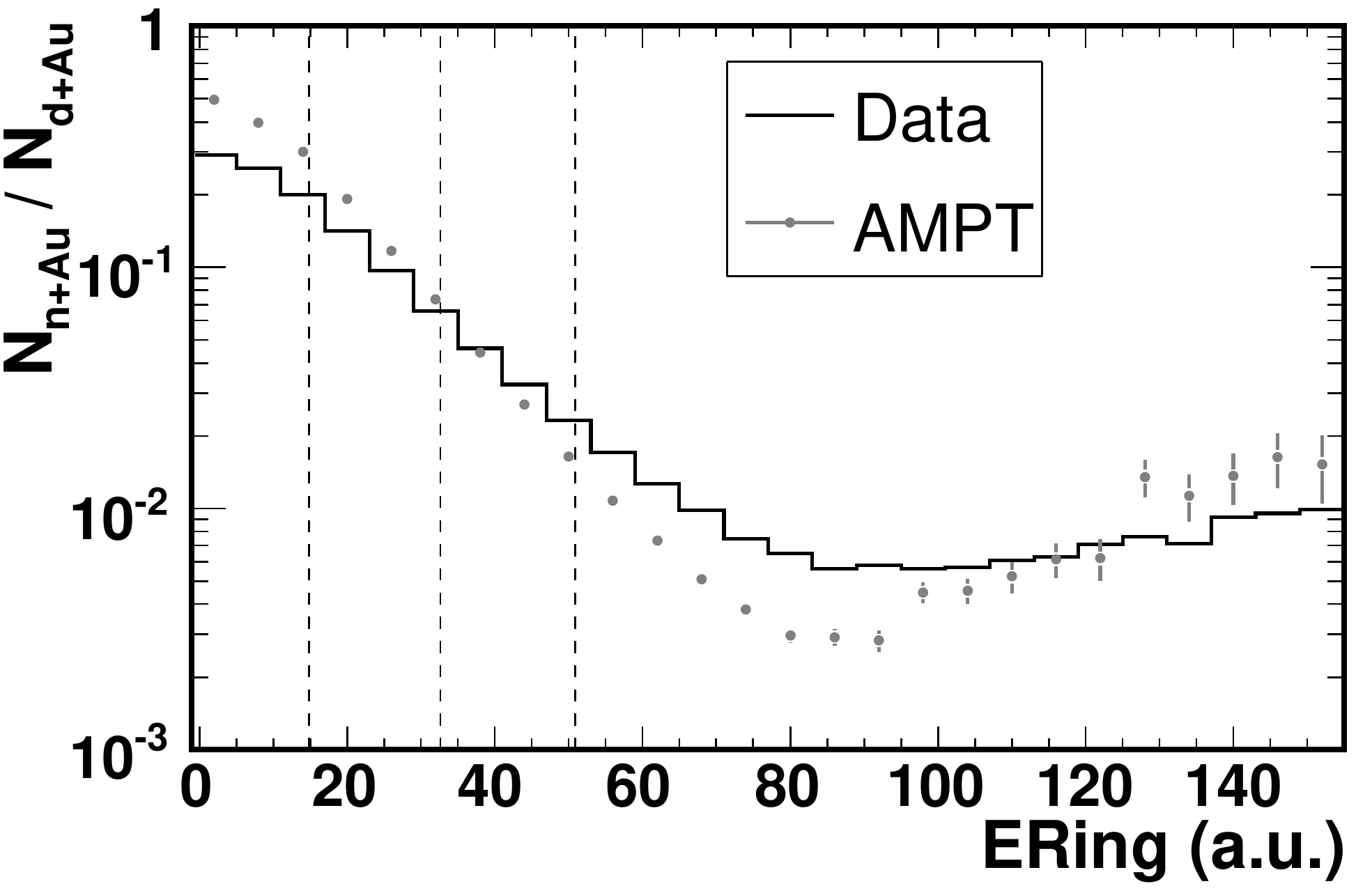}
   }
   \caption{   \label{recon:fig:tageringdists}
      \subref{recon:fig:dataTagERingDists}~The \protect\acs{ERing}
      distribution of {\dAu}, {\nAu} and {\pAu} interactions as observed
      in the data. \subref{recon:fig:dataMCTagERingRatio}~The ratio of
      the \protect\acs{ERing} distribution in {\nAu} to that of {\dAu}
      for both the data (black line) and the \protect\acs{AMPT}
      simulations (grey points).}
\end{figure}

The \ac{ERing} distribution observed in the data is presented in
\fig{recon:fig:dataTagERingDists} for {\dAu}, {\pAu} and {\nAu}
collisions. The dashed vertical lines show the centrality cut
positions.  The \mbox{$\sim35\%$} difference between the number of
{\pAu} and {\nAu} collisions observed, which was found to be independent
of centrality, reflects the different efficiencies of the two
calorimeters.  It can be seen that the tagged nucleon-nucleus
interactions were biased toward more peripheral collisions. This was
studied in more detail by looking at the ratio of the \ac{ERing}
distribution in {\nAu} to the \ac{ERing} distribution in {\dAu}, as
shown in \fig{recon:fig:dataMCTagERingRatio}. That this ratio is not
constant as a function of \ac{ERing} is a consequence of the peripheral
bias of tagged {\nAu} collisions. Such a bias was expected, as the
chance of one of the nucleons of the deuteron avoiding a collision with
the gold nucleus should decrease with decreasing impact parameter. To
test whether this bias was the same in the simulations as it was in the
data, the analogous ratio was determined for {\nAu} and {\dAu}
interactions in \ac{AMPT}, also shown in
\fig{recon:fig:dataMCTagERingRatio}. This comparison revealed that the
bias was indeed qualitatively similar in the mid-peripheral and
mid-central bins. However, the relative shape of the \ac{ERing}
distribution of {\nAu} and {\dAu} interactions in \ac{AMPT} differed
from that of the data. Thus, the centrality of ``central'' {\nAu}
collisions was qualitatively different in \ac{AMPT} as compared to the
data. Therefore, the average number of participants of a \emph{central}
nucleon-nucleus interaction may not have been estimated accurately by
the simulations. Some possible implications of this discrepancy will be
discussed in \sect{rslt:rda:ncollpAdA}.

The simulations were also used to obtain a rough estimate of the tagging
efficiency. Due to the discrepancy between data and simulation, these
efficiencies were not used in the analysis presented in this thesis. As
long as these efficiencies are independent of centrality, the would not
impact the measurements performed in this thesis. Nevertheless, they
were estimated by comparing fraction of tagged collisions observed in
the data  to the fraction of nucleon-nucleus collisions that occurred
(and passed the event selection) in the simulations. This comparison
gave an estimate of the average efficiency with which a nucleon-nucleus
collision would be successfully tagged. It was found that
\mbox{$\sim63\%$} of {\pAu} interactions and \mbox{$\sim46\%$} of {\nAu}
interactions would be tagged using the procedure described in
\sect{recon:nuctag:NA}.

%% file: srcs/trackChap.tex
% $Id: trackChap.tex,v 1.17 2006/09/05 01:18:31 cjreed Exp $
%

%---------------------------------------------------------------
\chapter{Particle Reconstruction}
\label{track}
%---------------------------------------------------------------

\begin{figure}[t]
   \begin{center}
      \includegraphics[width=0.95\linewidth]{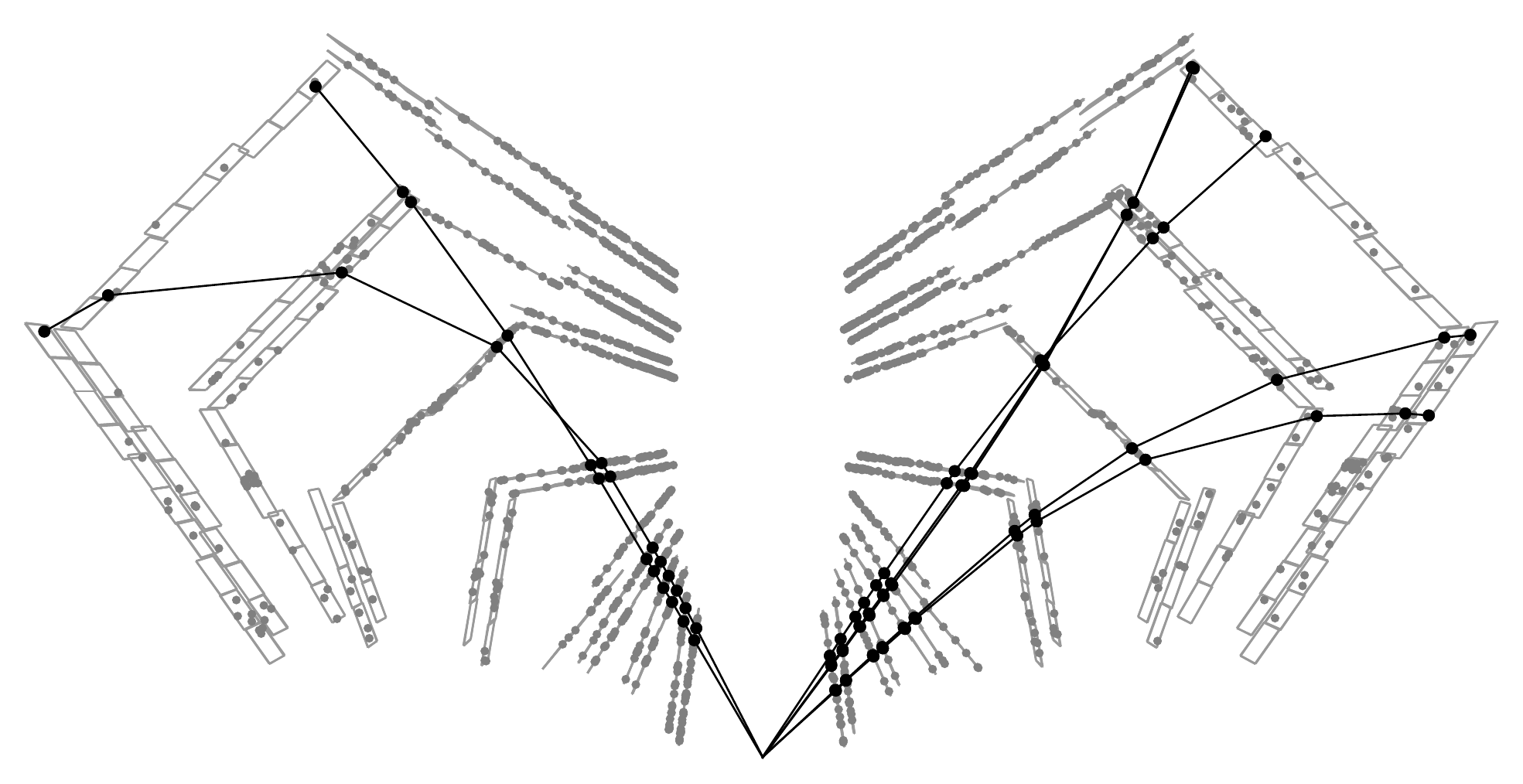}
   \end{center}
   \caption{\label{track:fig:AuAuTracks}
      Tracked particles in a central {\AuAu} collision. The grey dots
      show all merged hits in the Spectrometer. The black dots show the
      hits assigned to a track. The black lines are straight line
      segments that connect the hits on a given track, to guide the
      eye.}
\end{figure}

\myupdate{$*$Id: trackChap.tex,v 1.17 2006/09/05 01:18:31 cjreed Exp $*$}%
Calibrated signals in the Spectrometer detector were used to measure the
properties of individual (charged) particles. The charge and momentum of
a particle could be determined by the curvature of its trajectory
through the magnetic field. The full trajectory of a particle consisted
of two distinct sections, as each particle followed a straight path
before entering the magnetic field and a curved path while traveling
through the field. Particles were measured by first finding the straight
path followed by the particle, then \emph{tracking} its movement through
the magnetic field and finally fitting the two sections together. All
tracking procedures were performed separately for the two Spectrometer
arms. An example of tracked particles in a {\AuAu} collision is shown in
\fig{track:fig:AuAuTracks}.

%---------------------------------------------------------------
\section{Straight Track Finding}
\label{track:straight}
%---------------------------------------------------------------

The first step in reconstructing the trajectories of particles in a
{\dAu} collision was to locate all straight tracks. A straight track was
the path taken by a particle in the first six layers of a Spectrometer
arm. As seen in \fig{exp:fig:magfield}, the strength of the magnetic
field in this region was negligible, so it did not significantly affect
the motion of charged particles. Due to the small acceptance of the
{\phob} Spectrometer and the low multiplicity of {\dAu} collisions, only
one straight track was observed in an average \acs{dAuVertex} triggered
collision event.

Straight tracks were found in a relatively simple manner. First, all
combinations of hits on the first and fourth layers of each arm were
taken as candidates for straight tracks. Next, each track candidate was
extrapolated back to $x=0$, the nominal horizontal position (transverse
to the beam direction) of collisions in the detector. This was done so
that the distance of the track origin from the beam orbit could be
computed. The beam orbit was observed to have no slope in either the
horizontal or vertical plane. Since tracks in the Spectrometer
acceptance were also very nearly horizontal, only the vertical distance
between the track origin and the beam orbit was considered. If the
height of the track (at $x=0$) was more than 2.5~{\cm} above or below
the beam orbit position, then the track was rejected as not having come
from the {\dAu} collision.

Then, for each candidate, an attempt was made to find hits on the
remaining layers that could be associated with the track. For a given
layer, hits within 2~{\mm} of the track candidate were examined, and the
hit closest to the track was then associated with that track. For hits
on the fifth and sixth layers, this constraint was relaxed in the
vertical direction to account for the height of the pad, which was
larger than 2~{\mm}. Of course, hits on each layer could not be found
for every two-hit track candidate, since every candidate was not
necessarily representative of a physical particle. To increase the
probability that a straight track was in fact a reconstruction of the
trajectory of a physical particle, it was required that each straight
track have a hit on at least four layers of the Spectrometer.

This procedure was repeated for a different set of track candidates:
those formed by the combination of hits on the third and fourth layers.
For these candidates, the first two layers of the spectrometer were
ignored completely. The only hits to be associated with these tracks
were those found on the fifth and sixth layers of the Spectrometer.
However, the requirements that each straight track extrapolate back to
within 2.5~{\cm} of the beam height and have hits on at least four
layers of the Spectrometer were maintained.

The direction of a straight track was found by performing a
least-squares fit of a straight line to the hits associated with the
track. Two linear fits were performed: one in the $x-z$ (horizontal)
plane and one in the $y-z$ (vertical in the beam direction) plane. The
horizontal fit used only those hits in the first four layers, to reduce
effects of the magnetic field. The vertical fit used all hits associated
with the track. The fit procedure was performed each time a new hit was
associated with the track. This allowed the best estimate of the
trajectory to be used when searching for hits close to the track in
successive layers.

%---------------------------------------------------------------
\section{Curved Track Finding}
\label{track:curved}
%---------------------------------------------------------------

The curved trajectory of a particle was reconstructed by chaining
together track segments that shared common physical properties.
Combinations of two hits on successive layers of a Spectrometer arm were
used to form the track segments. A Hough transform
procedure~\cite{Hough:361242} was used to determine the polar angle,
$\theta$, and the inverse of the momentum, $1/p$, of a particle with a
trajectory containing the track segment. The polar angle was defined as
the angle between the beam line and the trajectory of the particle at
its point of origin (the collision). Because the track segments were
short, straight and used for performing a Hough transform, they were
referred to as Hough-sticks.

\begin{figure}[t]
   \begin{center}
      \includegraphics[width=0.6\linewidth]{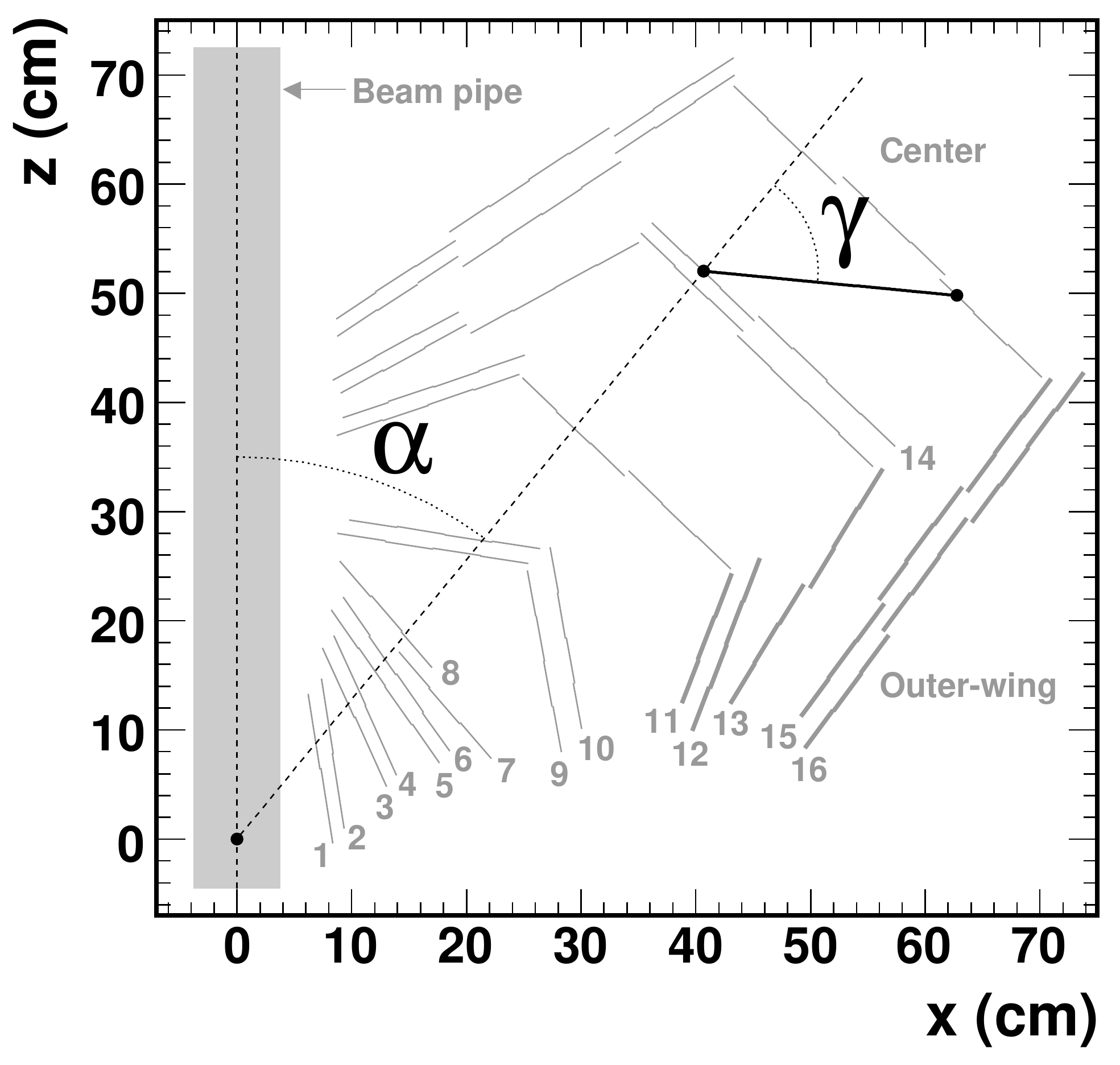}
   \end{center}
   \caption{\label{track:fig:houghAngles}
      The angles describing a Hough-stick, in this case in the central
      region between layers 14 and 15, shown as the solid black line.
      The bold sensors represent the outer-wing. The corresponding
      angled sensors near the beam form the inner-wing (not labeled).}
\end{figure}

\begin{table}[t]
   \begin{center}
      \begin{tabular}{|ccc|}
\hline
      & Central Region  & Outer-Wing Region\\
Stick & Layers Paired   & Layers Paired\\
\hline
A     &  9 - 10         &  9 - 10\\
B     & 10 - 11         & 10 - 11\\
C     & 11 - 13         & 11 - 13\\
D     & 13 - 14         & 13 - 15\\
E     & 14 - 15         & 15 - 16\\
\hline
      \end{tabular}
   \end{center}
   \caption{   \label{track:tab:houghsticks}
      Layers used when pairing hits to form Hough-sticks. The
      Spectrometer geometry, shown in \fig{track:fig:houghAngles},
      motivated the layers that were (and that were not) used in the
      Hough tracking.}
\end{table}

The first step in chaining together Hough-sticks was to generate the
sticks themselves. This was done by pairing a hit on a specific layer of
the Spectrometer with another hit on a successive layer. In total, six
hits were used to form the five sticks of a Hough-chain. Two separate
methods of pairing hits were used: one that could find sticks only in
the \emph{central} part of the spectrometer, and another that could find
sticks in the \emph{outer-wing} (away from the beam).
\Fig{track:fig:houghAngles} shows the distinction between the center and
outer-wing sections of a Spectrometer arm. \Tab{track:tab:houghsticks}
shows the layer pairs used to form Hough-sticks in the central and
outer-wing regions of the Spectrometer. All combination of hits on these
layers were used to form as many Hough-sticks as possible.

Next, the physical properties of each Hough-stick were determined. This
was done by calculating two angles: $\alpha$, the polar angle of the
\emph{hit} closest to the beam, and $\gamma$, the relative angle between
the two hits~\cite{krisThesis}. These angles are shown in
\fig{track:fig:houghAngles} for a Hough-stick connecting layers 14 and
15 of the central region of the Spectrometer. 

The vertex point used to define $\alpha$ did not come from the
\ac{OctDe} vertexing algorithm, due to the poor resolution of this
vertexing procedure. Instead, for {\dAu} collisions, the vertex was
taken to be the intersection of a straight track and the beam orbit.
That is, for each straight track candidate, the vertex was assumed to be
the interaction of the track and the beam orbit. The full curved
tracking was then run with this assumption. This process was then
repeated for each straight track. Finally, any duplicate tracks
resulting from running the tracking multiple times were explicitly
removed, as described in \sect{track:momfit:dups}.

Note that the angles $\alpha$ and $\gamma$ simply described the location
of the two hits at the ends of the Hough-stick; they did not describe
properties of the particle track. That is, $\alpha$ was the polar angle
of the first hit in the stick, while $\theta$ was the polar angle of the
track (at its origin). The physical parameters, $\theta$ and $1/p$, of a
particle whose trajectory would contain the hits at the ends of the
Hough-stick were obtained using a simple lookup table, called a
Hough-table.

Hough-tables were stored as histograms with \mbox{$20~\alpha$-bins} and
\mbox{$20~\gamma$-bins}. There was one Hough-table for each combination
of: (a)~the five pairs of layers used to form sticks, (b)~the two
regions of the Spectrometer used to find sticks (center and outer-wing),
(c)~the two electric charges of charged particles, (d)~the 50~vertex
regions of 0.5~{\cm} width between $z=-15~\cm$ and $z=+10~\cm$, and
(e)~the two physical parameters that were to be looked up. The z-axis of
a Hough-table stored the value of the physical parameter in question
(such as~$\theta$) that a track with a given $\alpha$ and $\gamma$ would
have. Since the histograms were discretely binned, the value of the
physical parameters were directly accessible only for values of $\alpha$
and $\gamma$ at the center of a bin. Polynomial fits were used to
extrapolate between bin centers. For the $\theta$ lookup tables, the
polynomial fits were performed along the $\alpha$ direction, but the
lookup was still discrete in the $\gamma$ direction. For the $1/p$
lookup tables, the polynomial fits were performed along the $\gamma$
direction, but the lookup was discrete in the $\alpha$ direction.

The Hough-tables were filled using \ac{MC} simulations of single tracks.
For this purpose, only charged pions entering one Spectrometer arm were
simulated. In addition, only one orientation of the magnetic field was
used. These simplifications were implemented under the valid assumptions
that (a)~the vast majority of particles produced in nucleus-nucleus
collisions at \acs{RHIC} were pions, (b)~one Spectrometer arm was a
mirror image of the other (about the beam line), and (c)~the trajectory
of a positive pion in one magnetic field orientation was the same as
that of a negative pion (with equivalent momentum) in the opposite field
orientation. The range of the $\alpha$ and $\gamma$ axes in each
Hough-table were determined dynamically by the range of values observed
in the simulations. For a given particle, the appropriate cell of a
Hough-table was filled with either $1/p$ or $\theta$. Each cell of the
Hough-table was then normalized by the number of particles that had been
used to fill that cell. Thus, the value of $\theta$ or $1/p$ stored in a
Hough-table was ultimately the \emph{average} parameter for particles in
that cell. In a similar manner, tables were constructed to store the
average relative \emph{error} of $1/p$ and $\theta$. For a given
simulated particle, the relative error was calculated as the difference
between the true value of the parameter of the particle and the average
value of the parameter for the cell in the table, normalized by the true
value.

Once the physical parameters $\theta$ and $1/p$ had been determined for
each Hough-stick, it was possible to form a Hough-chain. First,
Hough-sticks that shared hits -- that is, the outer hit of one stick was
the same as the inner hit of another stick -- were considered for
chaining. Since a single particle would have had definite values of
$\theta$ and $1/p$, all Hough-sticks formed by the hits of a single
particle should have had the same physical parameters. Thus, if $\theta$
and $1/p$ for each of the two sticks being considered were similar, then
they were chained together. A physical parameter of one stick was said
to be similar to that of a second stick if the difference between them
was not more than six times the error on that parameter. The error on
the parameter was calculated by summing quadratically the error of the
parameter on each stick.

Finally, a series of checks were performed on the Hough-chain to ensure
its quality. First, only complete chains were kept; any chain with less
than five sticks was rejected. Then, requirements were placed on the
vertical displacement of the hits on a chain to reject chains that did
not follow a smooth trajectory in the vertical direction. Next, chains
formed by hits that had excessive differences in deposited energy were
rejected. Finally, a $\chi^2$-statistic (see \eq{ana:eq:chi2}) was
calculated for the Hough-chain, taking into account the errors on each
Hough-stick as well as fluctuations in the positions of hits away from
the expected particle trajectory in the $x-z$ plane. Hough-chains with
large values of $\chi^2$ were rejected as having a low probability of
representing a physical particle.

%---------------------------------------------------------------
\section{Momentum Determination}
\label{track:momfit}
%---------------------------------------------------------------

The first step in finalizing the trajectory of an observed particle, and
measuring its momentum, was to match the curved section of the
trajectory with the straight section. Then, all hits on the full track
were fit numerically to obtain the best measurement of the trajectory
and physical properties of the particle that generated the hits.
Finally, a check was performed to clean up duplicate tracks that shared
a large number of hits and were most likely reconstructions of the same
physical particle.

%---------------------------------------------------------------
\subsection{Full Track Construction}
\label{track:momfit:join}
%---------------------------------------------------------------

A full collection of hits that could have been caused by a single
particle was constructed by matching straight tracks with Hough-chains.
The first step in this procedure was to find every straight track which
could be fit to a straight line with a \mbox{$\chi^2$-probability}
greater than 1\%. Next, each of these straight tracks were compared to
all the Hough-chains in the collision event. Four requirements had to be
met for a straight track to be joined with a Hough-chain. The first
requirement was that the polar angles of the straight track and the
Hough-chain did not differ by more than 15~{\mrad}. The second
requirement was a very loose one. It ensured that the difference between
the mean deposited energy (per~{\cm} of silicon) of hits in the straight
track and the mean $\mathit{dE}/\mathit{dx}$ of hits in the Hough-chain
be less than 80\% of the average $\mathit{dE}/\mathit{dx}$ of all hits
on the track. The next two requirements imposed constraints on the
distances of hits from the expected trajectory of a particle. The
$\chi^2$-statistic described in \sect{track:curved} was again checked,
to further reject chains that had a low probability of representing a
particle. The final requirement ensured that the full track would follow
a straight line in the vertical direction. First, the vertical slope of
the straight track was computed. Then, the vertical deviation of hits in
the Hough-chain from this straight line were used to compute a $\chi^2$
sum, labeled $\chi^{2}_{y}$. The $\chi^{2}_{y}$ per degree of freedom
(in this case, the six hits on the chain) was required to be less than
5. If, and only if, all of these requirements were met, then the
straight track and Hough-chain were combined together into a full track,
known as a SpecTrack. The constraints enforced during track matching
were designed to be loose requirements, to produce as many reasonable
track candidates as possible.

%---------------------------------------------------------------
\subsection{Trajectory Fitting}
\label{track:momfit:fit}
%---------------------------------------------------------------

The true trajectory and momentum of a particle was estimated using the
hits on the SpecTrack. The time-honored method for doing this is to fit
a simple, analytical form of the trajectory to the hits. However, due to
the complexity of the magnetic field shape in the region of the
Spectrometer arms, this procedure was not used in {\phob}. Instead, the
trajectory of a particle was estimated numerically using the measured
strength of the magnetic field at a large number of points (this map is
shown in \fig{exp:fig:magfield}). The charge, momentum and point of
origin of the SpecTrack obtained from the track matching was used to
simulate the trajectory of a pion through the magnetic field. This
trajectory simulation proceeded by calculating the location and momentum
vector of the particle at each step along its path. For SpecTracks in
the central region of the Spectrometer, the distance between steps along
the trajectory was fixed at 10~{\cm}. Since the simulation of particles
through the magnetic field was computationally intensive, large step
sizes were chosen in this region to reduce computing time. However, when
the particle would pass through a region in which the magnetic field was
changing, the step size was reduced by half to improve the accuracy of
the simulation. Since SpecTracks in the outer-wing of the Spectrometer
could spend a significant amount of time in such a region, a variable
step size was implemented for these tracks. The step size of tracks in
the outer-wing depended on the momentum of the particle as 
\mbox{$2.5~\cm + k \cdot p$}, where $p$ is the momentum of the particle
and the constant $k$ was taken to be simply $1~\cm/\gev$.

The resulting trajectory could then be used to calculate a $\chi^2$
statistic that accounted for the deviation of hits on the SpecTrack from
the expected particle position. These deviations, known as residuals,
were used to reject SpecTracks whose hits were located too far from the
expected particle trajectory. However, the trajectory resulting from the
numerical estimation was that of an \emph{ideal} particle. This was due
to the relative simplicity of the numerical simulation. Effects such as
energy loss inside the silicon and deflections of a particle passing
through some material, known as multiple scattering, were not
considered. Thus, it was \emph{expected} that the trajectory of a
physical particle could differ from that of an ideal particle with the
same charge, origin and momentum. These differences would obviously
affect the $\chi^2$ computed for a SpecTrack and had to be taken into
account. For SpecTracks that passed the residual cut, this was done by
looking up the expected deviations in a covariance, or error, matrix.
Diagonal elements of this matrix stored the deviation of physical
particles from ideal particles at a particular layer of silicon.
Off-diagonal elements stored the correlation between deviations in
different layers of silicon.

\begin{table}[t]
   \begin{center}
      \begin{tabular}{|cccc|}
\hline
Parameter   & Minimum            & Maximum            & Number of Bins\\
\hline
Charge      & -1~$e$             & +1~$e$             & 2\\
$1/p$       & 0.1~$(\mom)^{-1}$  & 10~$(\mom)^{-1}$   & 40\\
$\theta$    & 0.25~{\rad}        & 1.75~{\rad}        & 30\\
$z_0$       & -20~{\cm}          & 10~{\cm}           & 60\\
\hline
      \end{tabular}
   \end{center}
   \caption{   \label{track:tab:covar}
      Bins of physical parameters used to construct a covariance matrix.
      `$z_0$' refers to the particle's point of origin along the
      beam direction.}
\end{table}

The covariance matrix was constructed prior to data-taking using
simulations of physical particles. First, the physical parameters of a
pion were chosen: charge, point of origin, total momentum and polar
angle. These parameters were then fed into the numerical trajectory
estimating procedure described above, to obtain the path of an ideal
particle. Next, a pion with the same physical parameters was fully
simulated, taking effects such as energy loss and multiple scattering
into account. From these, the deviations between the ideal and physical
particle trajectories were computed. A different covariance matrix was
constructed for each combination of the different physical parameters
used to simulate particles. Similar values of physical parameters were
binned together to generate a covariance matrix, as described in
\tab{track:tab:covar}. In each bin, 5000~sets of physical parameters
were chosen randomly, and each set was then fed into the numerical and
full particle simulations.

Using the simulated ideal particle trajectory together with the
appropriate covariance matrix, the $\chi^2$ of a particular SpecTrack
and momentum hypothesis was computed. This process was repeated
iteratively to find the momentum vector that minimized the $\chi^2$ of a
SpecTrack. More precisely, the physical parameters varied during the
fitting routine were (a)~the inverse of the total momentum $1/p$,
(b)~the polar angle $\theta$, (c)~the azimuthal angle $\phi$, (d)~the
point of origin along the beam $z_0$, and (e)~the point of origin in the
vertical direction $y_0$. In order to minimize $\chi^2$ while varying
all five parameters, a simplex minimization technique was adopted. This
technique chose 6 points in the 5-dimensional parameter space at which
$\chi^2$ was calculated. The shape created by drawing connecting lines
between the points is known as a simplex. The points were randomly
chosen around the initial physical parameters determined by track
merging. The point with the largest $\chi^2$, and therefore worst fit,
was moved through the opposite face of the simplex in such a way as to
reduce the volume of the simplex. This procedure was then repeated,
allowing the simplex to fall into the region of lowest $\chi^2$.
Minimization was stopped when the $\chi^2$ calculated at the points of
the simplex did not deviate by more than 0.001 units.
See~\cite{NucRec:simplex} for more information on simplex minimization
techniques.

%---------------------------------------------------------------
\subsection{Duplicate Track Rejection}
\label{track:momfit:dups}
%---------------------------------------------------------------

The final step in reconstructing particles in the Spectrometer was to
find and clean up duplicate tracks. Duplicate tracks were groups of
SpecTracks that were thought to have a high probability of all
describing the same physical particle. First, all tracks with less than
eleven hits were rejected. Then, to find duplicate tracks, each pair of
the remaining SpecTracks was compared. If a pair of tracks shared more
than two hits, then they were assumed to represent the same particle. In
addition, if there were less than five Spectrometer layers in which
(a)~both tracks had a hit and (b)~the hits were on different pads, then
the tracks were assumed to be duplicates. For a pair of duplicate
tracks, the SpecTrack with the highest fit probability, calculated using
the $\chi^2$ of the track, was kept while the other was rejected.

%% file: srcs/anaChap.tex
% $Id: anaChap.tex,v 1.46 2006/09/05 02:20:50 cjreed Exp $
%

%---------------------------------------------------------------
\chapter{Obtaining Hadron Spectra}
\label{ana}
%---------------------------------------------------------------

\myupdate{$*$Id: anaChap.tex,v 1.46 2006/09/05 02:20:50 cjreed Exp $*$}%
Particles observed in the Spectrometer were used to measure the
transverse momentum, $\pt$, spectra in {\dAu} collisions. The spectra
were measured by the \emph{invariant yield} of particles in a certain
range of $\pt$. This yield represents the number of particles produced
in an average collision that have a transverse momentum in the given
range. It is defined as

\begin{align}
   \label{ana:eq:invyield1}
E \frac{d^{3}\!N}{d^{3}\!\vec{p}} &= E
   \frac{d^{3}\!N}{\mathit{dp_x}\mathit{dp_y}\mathit{dp_z}}\\
\notag &= \frac{E}{\pt} \frac{d^{3}\!N}{\mathit{d\pt}\mathit{d\phi}
   \mathit{dp_z}}
\end{align}

\noindent%
where $N$ is the number of charged hadrons observed having a momentum in
the range \mbox{($\vec{p}$, $\vec{p} + d^{3}\!\vec{p}$)} and an energy
\mbox{$E=\sqrt{m^2 + \vec{p}^2}$}. Note that these equations are given
in natural units, with \mbox{$c=1$}. Since the distribution of produced
particles should be azimuthally symmetric on average, it is possible to
average the yield over the azimuthal angle $\phi$,

\begin{align}
   \label{ana:eq:invyield2}
E \frac{d^{3}\!N}{d^{3}\!\vec{p}} &= 
   \frac{E}{2 \pi \pt } \frac{d^{2}\!N}{\mathit{d\pt}\mathit{dp_z}}\\
\notag &= \frac{E}{2 \pi \pt } \frac{d^{2}\!N}{\mathit{d\pt}
   (m_T \cosh(y) \mathit{dy})}
\end{align}

\noindent%
where the identity \mbox{$p_z = m_T \sinh(y)$} was used to find
\mbox{$\mathit{dp_z} = m_T \cosh(y) \mathit{dy}$}. Using the fact
that \mbox{$E = m_t \cosh(y)$},

\begin{equation}
   \label{ana:eq:invyield}
E \frac{d^{3}\!N}{d^{3}\!\vec{p}} =
   \frac{1}{2 \pi \pt } \frac{d^{2}\!N}{\mathit{d\pt} \mathit{dy}}
\end{equation}

To see that this quantity does not change under Lorentz transformations,
note that for a boost in the beam direction,

\begin{align}
\notag p'_x &= p_x\\
\notag p'_y &= p_y\\
\notag p'_z &= \gamma \prn{p_z - \beta E}\\
\notag E'   &= \gamma \prn{E - \beta p_z}
\end{align}

\noindent%
So that the differential \mbox{$d^{3}\!\vec{p}'$} can be expressed

\begin{align}
\notag d^{3}\!\vec{p}' &= \mathit{dp'_x}\mathit{dp'_y}\mathit{dp'_z}\\
\notag &= \mathit{dp_x}\mathit{dp_y} \gamma \prn{\mathit{dp_z} - \beta
   \mathit{dE}}
\end{align}

\noindent%
From the definition of energy,

\begin{align}
\notag E^2 &= m^2 + p_x^2 + p_y^2 + p_z^2\\
\notag 2 E \mathit{dE} &= 2 p_z \mathit{dp_z}
\end{align}

\noindent%
Thus

\begin{align}
\notag d^{3}\!\vec{p}' &= 
   \mathit{dp_x}\mathit{dp_y} \gamma \prn{\mathit{dp_z} - \beta
   \frac{p_z}{E}\mathit{dp_z}}\\
\notag &= \mathit{dp_x}\mathit{dp_y} \frac{1}{E} \gamma
   \prn{E - \beta p_z} \mathit{dp_z}\\
\notag &= \mathit{dp_x}\mathit{dp_y} \frac{E'}{E} \mathit{dp_z}\\
\Rightarrow \frac{E'}{d^{3}\!\vec{p}'} &= 
   \frac{E}{d^{3}\!\vec{p}} \label{ana:eq:invariant}
\end{align}

\noindent%
and since the number of particles produced in a collision is clearly
independent of the observer's reference frame, \eq{ana:eq:invariant}
shows that the invariant yield (\eq{ana:eq:invyield}) is indeed
invariant.

For this analysis, the mass of an observed particle was not
determined. Thus it was not possible to measure the rapidity of
particles in the Spectrometer. Instead, the {\prap}, $\eta$, was
measured, since $\eta$ is determined by the polar angle of a particle
(see \eq{recon:eq:prap}). Thus, the quantity measured in this analysis
was approximately equal to the invariant yield,

\begin{equation}
   \label{ana:eq:invyielmeas}
E \frac{d^{3}\!N}{d^{3}\!\vec{p}} \approx
   \frac{1}{2 \pi \pt } \frac{d^{2}\!N}{\mathit{d\pt} \mathit{d\eta}}
\end{equation}

This is a good approximation for particles moving at relativistic
speeds, since

\begin{align}
y &= \frac{1}{2} \ln \prn{\frac{E + p_z}{E - p_z}} \label{ana:eq:rap}\\
\notag &\approx \frac{1}{2} \ln \prn{\frac{p + p_z}{p - p_z}}
   \qquad \text{for}\ m^2 \ll E^2\\
\notag &= \frac{1}{2} \ln 
   \prn{\frac{p + p \cos(\theta)}{p - p \cos(\theta)}}\\
\notag &= \frac{1}{2} \ln 
   \prn{\frac{1 + \cos(\theta)}{\sin(\theta)} \ 
      \frac{\sin(\theta)}{1 - \cos(\theta)}}\\
\notag &= \frac{1}{2} \ln 
   \prn{\frac{1}{\tan(\theta / 2)} \ 
      \frac{1}{\tan(\theta / 2)}}\\
y &\approx - \ln \prn{\tan(\theta / 2)}\ \equiv\ \eta
   \label{ana:eq:prapapprox}
\end{align}

\noindent%
Note that the fifth step was reached using the half-angle formulas

\begin{equation*}
\tan\prn{\frac{\theta}{2}} = \frac{\sin(\theta)}{1 + \cos(\theta)}
 = \frac{1 - \cos(\theta)}{\sin(\theta)}
\end{equation*}

%---------------------------------------------------------------
\section{Event Selection}
\label{ana:evtsel}
%---------------------------------------------------------------

The first step in measuring the yield of charged hadrons was to impose
certain requirements on the collision events used for the analysis. The
requirements were used to ensure that only {\dAu} collisions, as opposed
to beam-gas collisions, were analyzed. Further, they were used to select
collisions that produced particles in the acceptance of the detector.
Ideally, when comparing data to simulation, the same event selection
would be used on both the data and simulation. However, this was not
possible, mainly because the arrival time of a signal in a fast
detector, like the Paddles or \acsp{T0}, was not determined in the
simulation. Therefore, the restrictions placed on the simulations were
chosen to correspond as closely as possible to those placed on the data.

%---------------------------------------------------------------
\subsection{Minimum Bias Selection}
\label{ana:evtsel:minbias}
%---------------------------------------------------------------

A minimum bias event selection was used when matching the distribution
of a centrality variable in a simulation to the same distribution in the
data. The matching procedure is described in
\sect{recon:cent:cuts:scaling}. The goal of this event selection was
similar to that of the \acs{dAuMinBias} \emph{trigger}, described in
\sect{recon:trig:minbias}. That is, to build a sample of {\dAu}
collisions that was not biased toward any particular type of {\dAu}
collision. This was necessary when matching a distribution of signals in
data and \acs{MC}, since any bias introduced by an event selection could
affect the distribution differently in the data or the simulation.

%---------------------------------------------------------------
\subsubsection{Data Selection}
\label{ana:evtsel:minbias:data}
%---------------------------------------------------------------

\begin{table}[t]
   \begin{center}
      \begin{tabularx}{\linewidth}{|llX|}
\hline
Variable & Condition & Summary\\
\hline
PdlDouble & True & \protect\acs{dAuMinBias} triggered\\
\protect\acs{OctDe} Vertex ($v_z$) & \mbox{$-10~{\cm} < v_z < +10~{\cm}$} &
   Collision in Spectrometer acceptance\\
\protect\acs{OctDe} Vertex & Valid & Vertex reconstruction succeeded\\
NotPrePileUp & True & No signals from the previous collision\\
NotPostPileUp & True & No signals from the following collision\\
d-side Paddle & Hit & One or more signals on d-side Paddle\\
Au-side Paddle & Hit & One or more signals on Au-side Paddle\\
\protect\acs{IsCol} & True & Remove beam-gas collisions\\
\hline
      \end{tabularx}
   \end{center}
   \caption{   \label{ana:tab:minbias:data}
      Minimum bias event selection for {\dAu} data.}
\end{table}

The conditions required by the minimum bias event selection on {\dAu}
collisions are shown in \tab{ana:tab:minbias:data}. First, collisions
that were recorded by the \acs{dAuMinBias} trigger were selected. Such
collisions were always present in the data, even when the experiment was
running primarily with the \acs{dAuVertex} trigger. Then, collisions
that occurred within 10~{\cm} of the \acs{IP} were selected. This
condition was imposed to ensure that particles produced by the
collisions were within the acceptance of the Silicon detectors, and to
correspond to the requirement imposed by the \acs{dAuSpectra} event
selection (see \sect{ana:evtsel:dAuSpectra}). The vertex of each
collision was required to have been successfully reconstructed by the
\acs{OctDe} vertexing algorithm, described in \sect{recon:vertex}. The
longitudinal position of the \acs{OctDe} vertex was then required to be
within 10~{\cm} of $z=0$.

Next, a check was performed to ensure that the event had not recorded
collision \emph{pile-up}. Due to the high collision rate during the
{\dAu} physics run at \acs{RHIC}, it was possible for more than one
collision to occur while the {\phob} detector was being read-out. This
could cause signals from more than one collision to be recorded in a
single data event, an effect known as pile-up. To determine whether a
collision contained pile-up, the time at which the collision was
triggered was compared to that of the previous and following events. If
the previous collision occurred less than 5~{\us} before the collision in
question, then signals in the Silicon detectors would not have had
enough time to decay away. Thus, if the {\phob} \ac{DAQ} received a
trigger signal less than 5~{\us} after it had received the previous
trigger signal, then the current collision event would be marked, and
later (during the analysis) rejected. On the other hand, if two
collisions occurred within 500~{\ns}, then particles from the later
collision could create signals in the Silicon while the former collision
was still being read-out. Thus, collisions triggered less than 500~{\ns}
before the next event were also marked and rejected.

Finally, steps were taken to reduce the number of beam-gas collisions
that could enter into the analysis. This was done by requiring at
least one hit on both the \mbox{d-side} and \mbox{Au-side} Paddle
detectors.  In addition, background collisions were further rejected,
without placing a bias on the data, by the requirement known as
\ac{IsCol}. This was done using two types of selections: a single-arm
timing cut and a double-arm cut. The single arm timing cut required
that the first signal reported by each side of a detector occurred at
a reasonable time. For these timing cuts, the arrival time of a signal
was determined relative to the collider's crossing clock
(see~\sect{exp:RHIC}). A reasonable signal timing was determined by
averaging the signal's arrival time over a group of events. Signals in
the \acs{T0} detector were required to occur within 3~{\ns} of this
average, while signals in the {\cheren} were required to occur within
5~{\ns} of the average. For the Paddle detectors, signals were
required to occur within one standard deviation of the mean
time. Double-arm cuts were performed only if both sides of the
detector had recorded a hit, to avoid introducing a bias on the
data. If both sides of the detector had been hit, then the timing of
the signals on each side were required to be reasonably close. The
time difference, between the first signal on the Au-side and the first
signal on the d-side, was required to be less than 5~{\ns} for the
\acs{T0}, 7~{\ns} for the {\cheren} and 8~{\ns} for the Paddle
detectors.

%---------------------------------------------------------------
\subsubsection{MC Selection}
\label{ana:evtsel:minbias:MC}
%---------------------------------------------------------------

\begin{table}[t]
   \begin{center}
      \begin{tabular}{|lll|}
\hline
Variable & Condition & Summary\\
\hline
\protect\acs{OctDe} Vertex ($v_z$) & \mbox{$-10~{\cm} < v_z < +10~{\cm}$} &
   Collision in Spectrometer acceptance\\
\protect\acs{OctDe} Vertex & Valid & Vertex reconstruction succeeded\\
d-side Paddle & Hit & One or more signals on d-side Paddle\\
Au-side Paddle & Hit & One or more signals on Au-side Paddle\\
\hline
      \end{tabular}
   \end{center}
   \caption{   \label{ana:tab:minbias:mc}
      Minimum bias event selection for {\dAu} \protect\acs{MC}.}
\end{table}

The conditions required by the minimum bias event selection on simulated
{\dAu} collisions are shown in \tab{ana:tab:minbias:mc}. For \acs{MC},
the event selection was simplified because background collisions were
not simulated, nor was any triggering necessary. Thus, only cuts on the
Paddle multiplicity and \acs{OctDe} vertex were imposed. These cuts were
the same as those used for the data, namely that each Paddle detector
had a hit and that the \acs{OctDe} vertex was successfully reconstructed
to be within 10~{\cm} of the \acs{IP}.

%---------------------------------------------------------------
\subsection{{\dAu} Spectra Selection}
\label{ana:evtsel:dAuSpectra}
%---------------------------------------------------------------

For the {\dAu} charged hadron $\pt$ spectra analysis, a different event
selection was used. The goal of this event selection was to produce a
collection of {\dAu} collisions with as little background (i.e.~beam-gas
interactions) as possible. In addition, the \acf{dAuSpectra} event
selection threw out collisions that had a low probability of producing
particles in the acceptance of the Spectrometer.

%---------------------------------------------------------------
\subsubsection{Data Selection}
\label{ana:evtsel:dAuSpectra:data}
%---------------------------------------------------------------

\begin{table}[t]
   \begin{center}
      \begin{tabularx}{\linewidth}{|llX|}
\hline
Variable & Condition & Summary\\
\hline
\protect\acs{OctDe} Vertex ($v_z$) & \mbox{$-10~{\cm} < v_z < +10~{\cm}$} &
   Collision in Spectrometer acceptance\\
NotPrePileUp & True & No signals from the previous collision\\
NotPostPileUp & True & No signals from the following collision\\
d-side Paddle & Hit & One or more signals on d-side Paddle\\
Au-side Paddle & Hit & One or more signals on Au-side Paddle\\
AllT0Diagonal & True & \protect\acs{dAuVertex} triggered 
   and reasonable \protect\acs{T0} timing\\
OctDeT0 & True & \protect\acs{OctDe} vertex and \protect\acs{T0} vertex agree\\
\hline
      \end{tabularx}
   \end{center}
   \caption{   \label{ana:tab:dAuSpectra:data}
      \protect\acs{dAuSpectra} event selection for {\dAu} data.}
\end{table}

The conditions imposed by the \ac{dAuSpectra} event selection are shown
in \tab{ana:tab:dAuSpectra:data}. Like the minimum bias event selection,
events with pile-up were filtered out and background collisions were
rejected by requiring a hit on each of the Paddle detectors. To further
reduce the number of background collisions that would be analyzed,
several more cuts were used. First was the ``AllT0Diagonal'' condition,
which required that each event was recorded by either the
\acs{dAuVertex} or the \acs{dAuPeriph} trigger. The AllT0Diagonal
condition also removed collisions that occurred between discordant
bunches\footnote{Bunches were timed to collide at the \protect\acs{IP}
by design. If, for example, bunches $A$ and $B$ were properly timed,
then a collision between bunch $A$ and any bunch \emph{other} than $B$
would be improperly timed.}. This was done by (a)~requiring that both
\acs{T0} detectors were hit, (b)~imposing single arm timing cuts on the
\acs{T0} detectors and (c)~imposing double arm timing cuts on the
\acs{T0} detectors. The timing cuts are described in
\sect{ana:tab:minbias:data}. Note that in the minimum bias event
selection, the timing cuts were only imposed if each \acs{T0} was hit.
For the \ac{dAuSpectra} event selection, these cuts were imposed on all
events. This produced a more pure {\dAu} collision sample, but that
sample was biased toward higher multiplicity collisions.

\begin{figure}[t!]
   \centering
   \mbox{
      \subfigure[h$+$, B$+$]{
         \label{ana:fig:trkZ_PCPP}
         \includegraphics[width=0.4\linewidth]{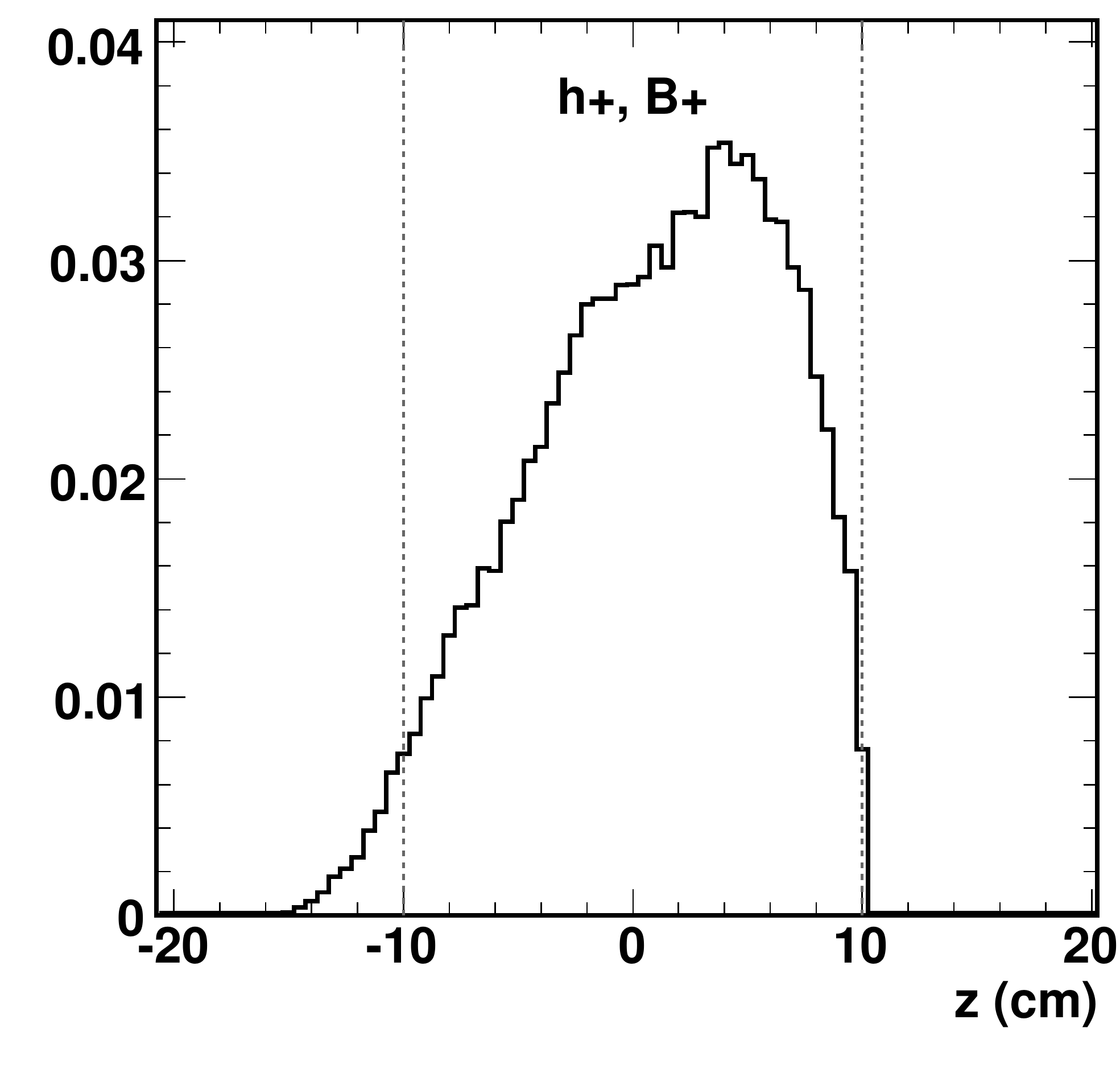}
      }
      \subfigure[h$-$, B$+$]{
         \label{ana:fig:trkZ_NCPP}
         \includegraphics[width=0.4\linewidth]{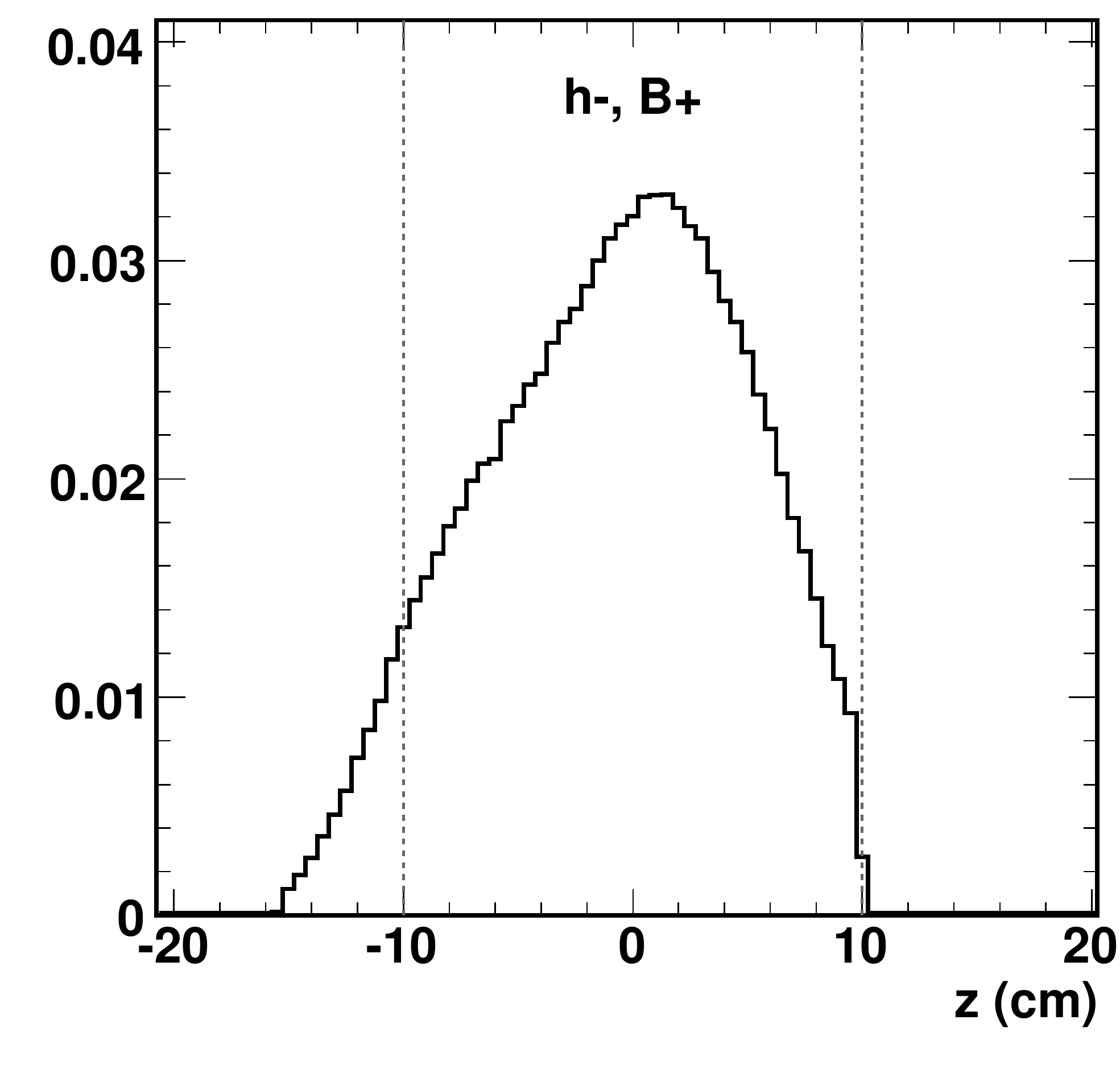}
      }
   }
   \mbox{
      \subfigure[h$-$, B$-$]{
         \label{ana:fig:trkZ_NCNP}
         \includegraphics[width=0.4\linewidth]{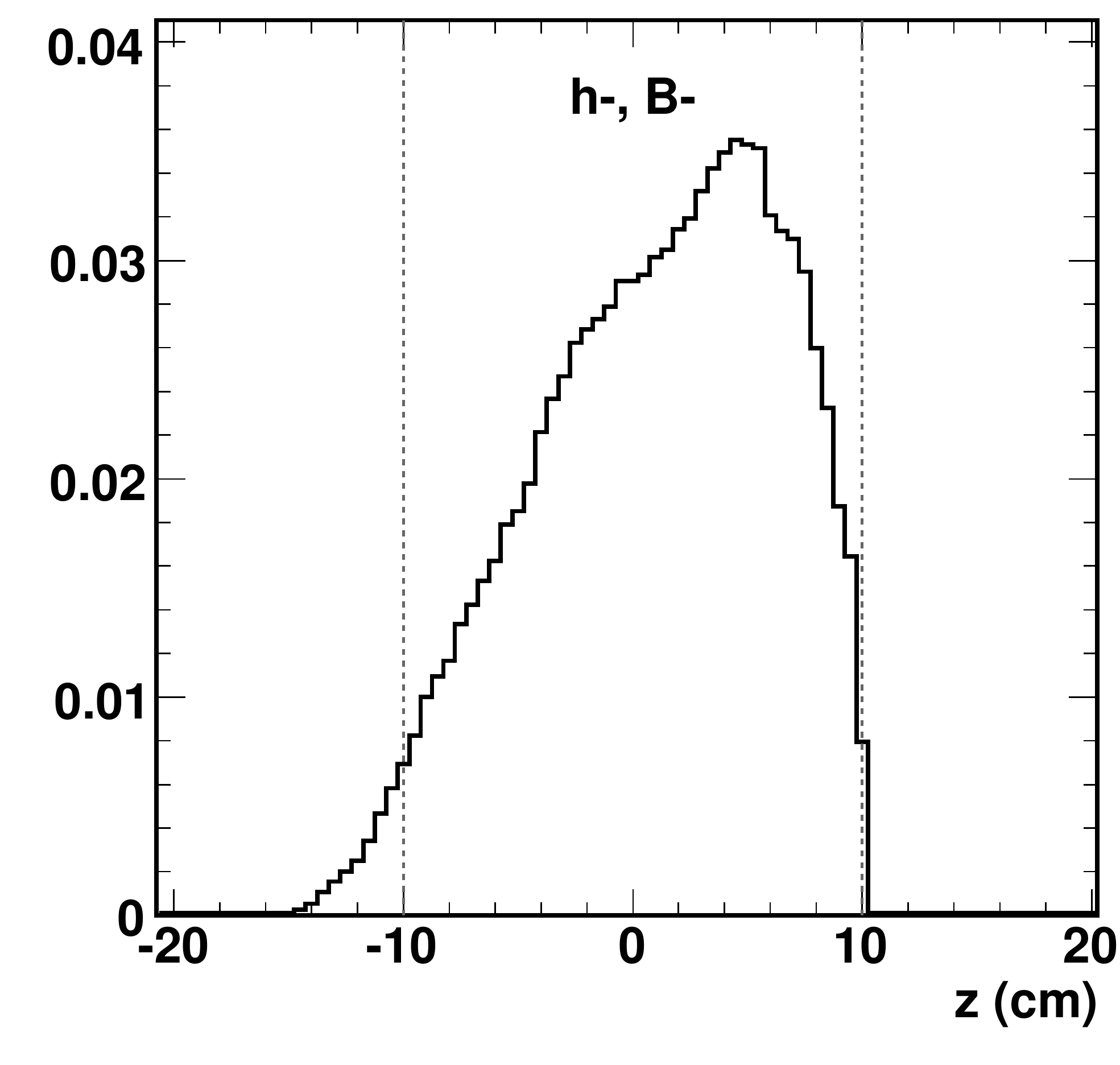}
      }
      \subfigure[h$+$, B$-$]{
         \label{ana:fig:trkZ_PCNP}
         \includegraphics[width=0.4\linewidth]{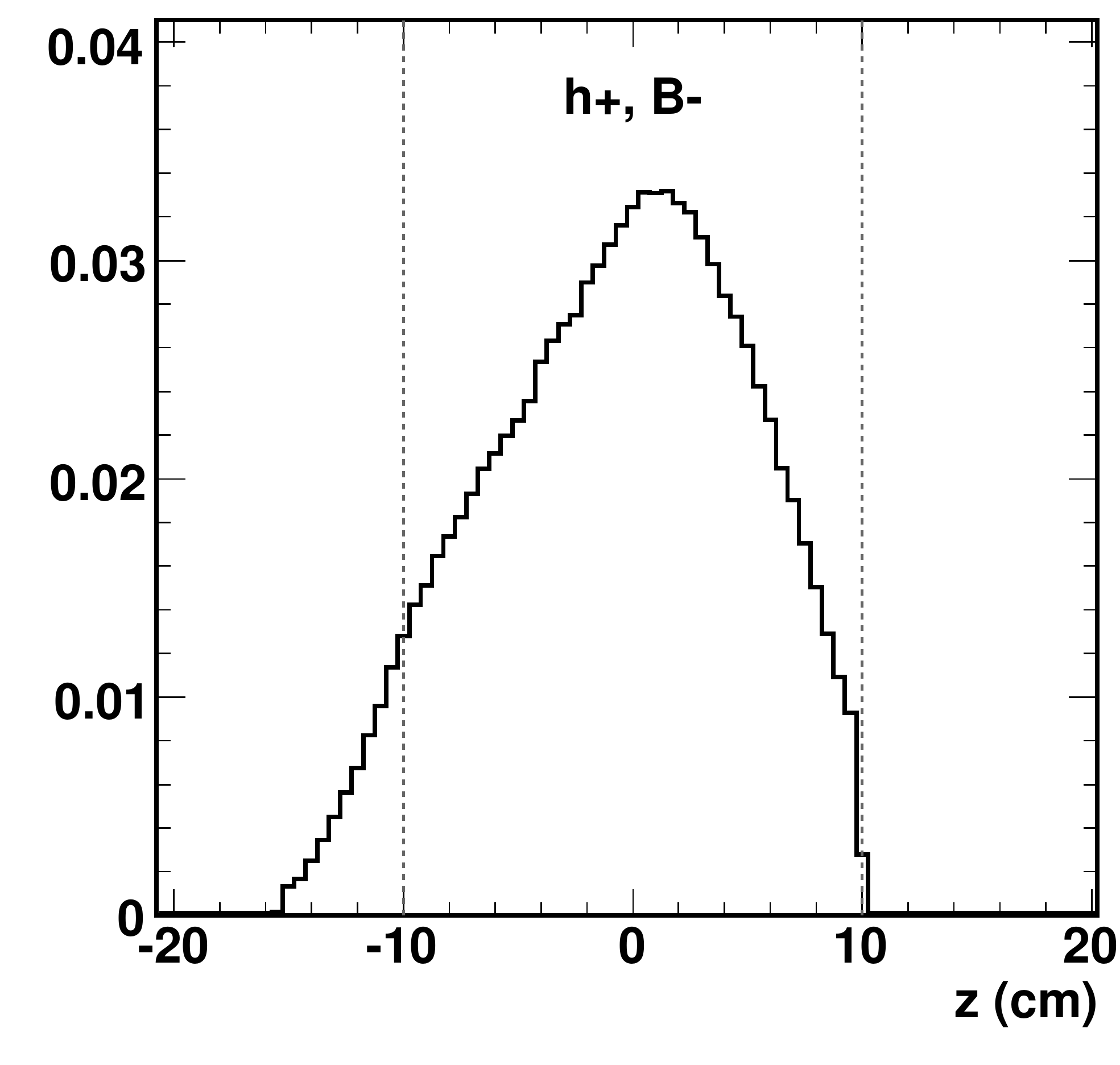}
      }
   }
   \caption{   \label{ana:fig:trkZ}
      The longitudinal track origin, as determined by the track
      reconstruction, for each combination of electric charge and magnet
      polarity (see~\sect{exp:phobdet:spec:mag}.
      \subref{ana:fig:trkZ_PCPP}~Positive hadrons, positive magnetic
      field. \subref{ana:fig:trkZ_NCPP}~Negative hadrons, positive
      magnetic field. \subref{ana:fig:trkZ_NCNP}~Negative hadrons,
      negative magnetic field. \subref{ana:fig:trkZ_PCNP}~Positive
      hadrons, negative magnetic field.}
\end{figure}

To reduce the number of events that were unlikely to produce particles
in the Spectrometer acceptance, a vertex cut was imposed. As shown in
\fig{ana:fig:trkZ}, relatively few particles that originated further
than 10~{\cm} from the \acs{IP} could be reconstructed. While some
charged particles having a longitudinal origin 10~to~15~{\cm} from the
\acs{IP} in the Au-direction could be reconstructed, the Spectrometer
had a complicated acceptance in that region (as a function of transverse
momentum). Because of this, only collisions that occurred within 10~{\cm}
of $z=0$, as determined by the \acs{OctDe} vertexing algorithm, were
used in the analysis. \acs{T0} information was used to reject events for
which the \acs{OctDe} vertexing algorithm yielded an unreasonable
vertex. The ``OctDeT0'' condition required that the \acs{OctDe} vertex
agree with the \acs{T0} vertex to better than 25~{\cm}. This cut also
served to further reduce background collisions between discordant
bunches in which the \acs{T0} timing appeared good, but a corresponding
collision vertex could not be found.

%---------------------------------------------------------------
\subsubsection{MC Selection}
\label{ana:evtsel:dAuSpectra:MC}
%---------------------------------------------------------------

\begin{table}[t]
   \begin{center}
      \begin{tabular}{|lll|}
\hline
Variable & Condition & Summary\\
\hline
\protect\acs{OctDe} Vertex ($v_z$) & \mbox{$-10~{\cm} < v_z < +10~{\cm}$} &
   Collision in Spectrometer acceptance\\
d-side Paddle & Hit & One or more signals on d-side Paddle\\
Au-side Paddle & Hit & One or more signals on Au-side Paddle\\
d-side \protect\acs{T0} & Hit & 
   One or more signals on d-side \protect\acs{T0}\\
Au-side \protect\acs{T0} & Hit & 
   One or more signals on Au-side \protect\acs{T0}\\
\hline
      \end{tabular}
   \end{center}
   \caption{   \label{ana:tab:dAuSpectra:mc}
      \protect\acs{dAuSpectra} event selection for {\dAu} \protect\acs{MC}.}
\end{table}

The conditions required by the \ac{dAuSpectra} event selection on
simulated {\dAu} collisions are shown in \tab{ana:tab:dAuSpectra:mc}.
The \acs{OctDe} vertex requirement was kept the same for \acs{MC} as for
data, as were the requirements that both Paddle and \acs{T0} detectors
be hit. However, due to the lack of timing information in the
simulations, no timing cuts were applied to the \acsp{T0}.

%---------------------------------------------------------------
\section{Track Selection}
\label{ana:trksel}
%---------------------------------------------------------------

Once an event had passed the event selection, certain cuts were imposed
on the reconstructed particles. The curved tracking itself took steps to
remove duplicate tracks and to discard tracks that had large residuals,
as discussed in \chap{track}. However, further steps were taken to
remove particle tracks that were not reliably reconstructed.

%---------------------------------------------------------------
\subsection{Fit Probability Cut}
\label{ana:trksel:fitprobcut}
%---------------------------------------------------------------

First, tracks with poor momentum fits were removed. This was done by
making a cut on the fit probability, calculated from the $\chi^2$
statistic described in \sect{track:curved}. Given the $\chi^2$ value of
the track, $\chi^2_t$, and the number of hits on the track, $N_h$, the
fit probability was calculated as

\begin{equation}
   \label{ana:eq:fitprob}
P\prn{\frac{N_h}{2},\frac{\chi^2_t}{2}} = 
   \frac{1}{\Gamma\prn{\frac{N_h}{2}}} 
   {\int_{{\chi^2_t}/2}^{+\infty}} e^{-t} t^{(N_h/2)-1} \mathit{dt}
\end{equation}

\noindent%
where $P$ represents the probability that, given the same physical
particle, the tracking procedure would yield a worse fit purely by
chance. The probability is normalized by the so-called Gamma function,

\begin{equation}
   \label{ana:eq:gamma}
\Gamma\prn{\frac{N_h}{2}} = \int_0^{+\infty} e^{-u} u^{(N_h/2) - 1} \mathit{du}
\end{equation}

\begin{figure}[t!]
   \centering
   \subfigure[$\chi^2$ Distribution]{
      \label{ana:fig:chi2dist}
      \includegraphics[width=0.4\linewidth]{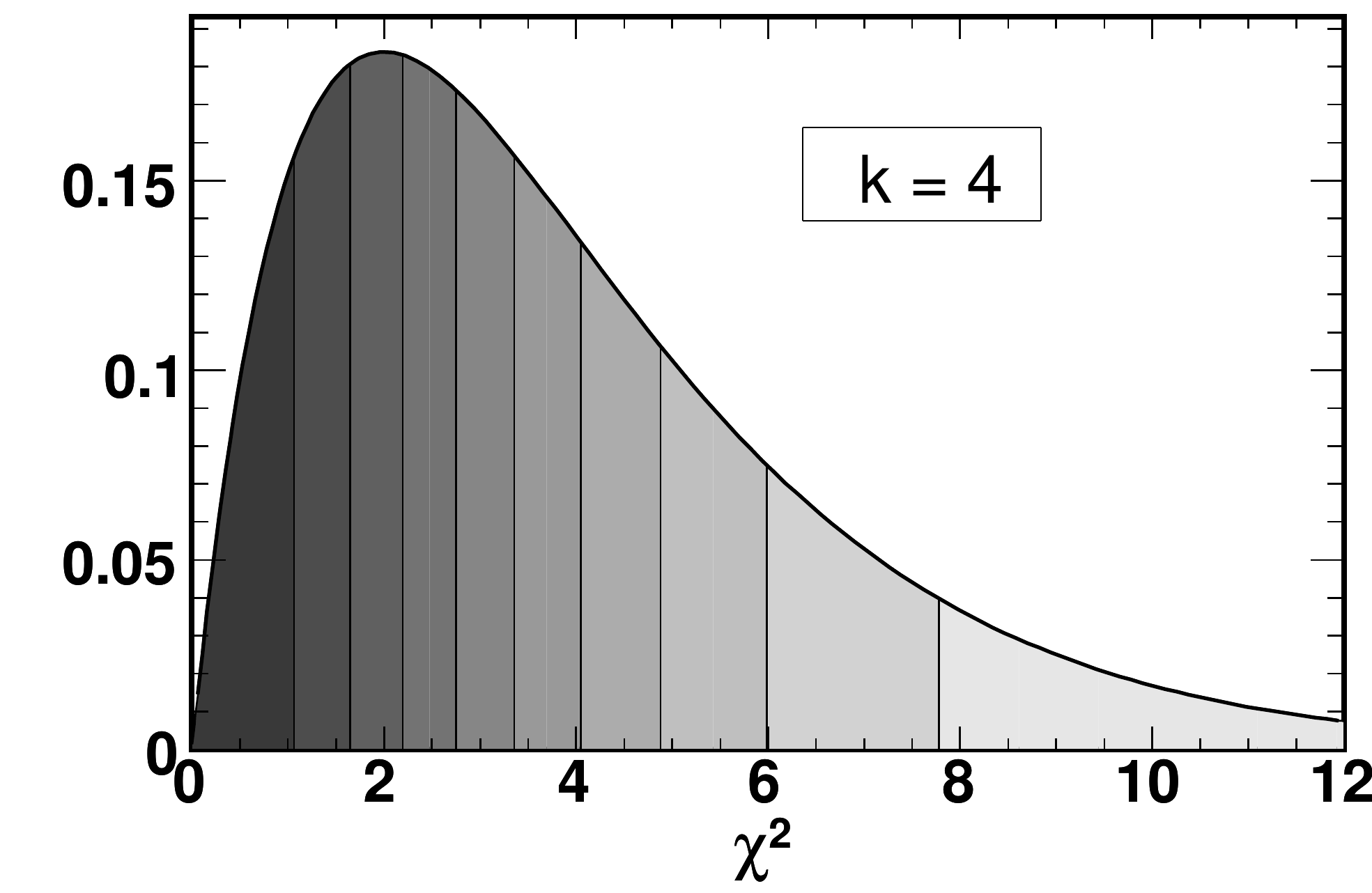}
   }
   \subfigure[$\chi^2$ Probability]{
      \label{ana:fig:chi2prob}
      \includegraphics[width=0.4\linewidth]{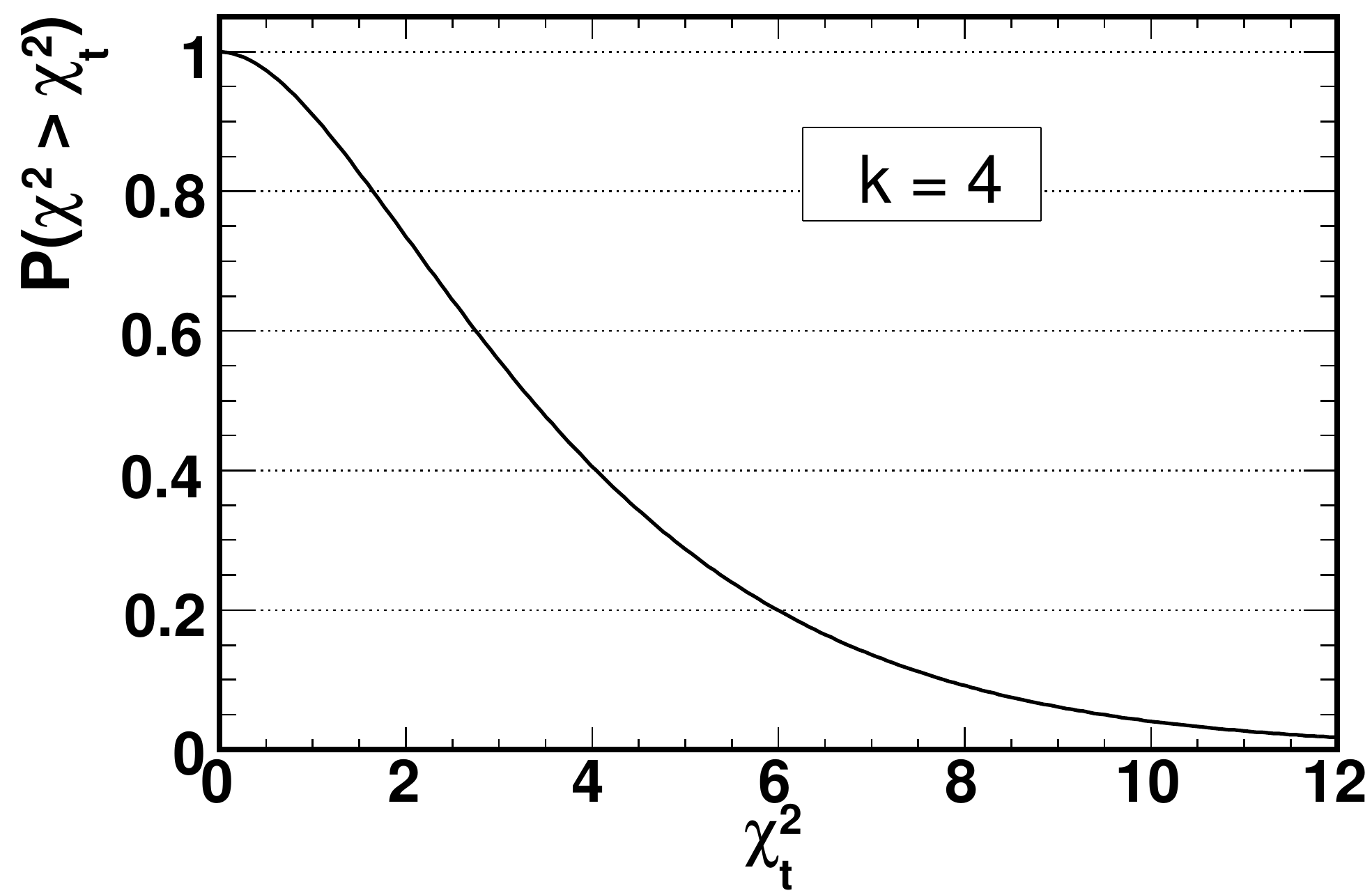}
   }
   \caption{   \label{ana:fig:chi2}
      \subref{ana:fig:chi2dist}~The $\chi^2$ (probability density)
      distribution with four degrees of freedom. See text for an
      explanation of the shaded regions. \subref{ana:fig:chi2prob}~The
      corresponding $\chi^2$ complementary cumulative distribution
      function, which gives the probability that an observed $\chi^2$
      could be greater than some $\chi^2_t$ purely by chance.}
\end{figure}

The distribution of fit probabilities obtained for all the tracks was
expected to be flat. To see this, suppose one performed measurements of
a set of random variables $y_i$. Suppose further that each individual
measurement was a sampling of a Gaussian distribution with mean $\mu_i$
(thus the expected value) and standard deviation $\sigma_i$ (thus the
measurement error). Then the sum

\begin{equation}
   \label{ana:eq:chi2}
\chi^2 = \sum_{i=1}^{N} \prn{\frac{y_i - \mu_i}{\sigma_i}}^2
\end{equation}

\noindent%
would follow a $\chi^2$ distribution (by definition), with \mbox{$k = N
- \nu$} degrees of freedom, where $\nu$ is the number of parameters used
to determine each $\mu_i$. For example, the $\chi^2$ of a track could be
calculated by taking each $y_i$ to be the position of a hit on the
track, each $\mu_i$ to be the expected hit position and each $\sigma_i$
to be the relevant measurement error. Then the values obtained by
calculating the sum in \eq{ana:eq:chi2} for each track would be
distributed according to a $\chi^2$ distribution. The $\chi^2$
distribution is shown in \fig{ana:fig:chi2dist} (for $k=4$) and is
defined by

\begin{equation}
   \label{ana:eq:chi2pdf}
\mathscr{P}\prn{\chi^2} = \frac{(1/2)^{k/2}}{\Gamma(k/2)}
   {\prn{\chi^2}}^{k/2 - 1} e^{-\chi^{2}/2}
\end{equation}

The probability calculated by \eq{ana:eq:fitprob} is simply the integral
of this function over values above the measured $\chi^2_t$, and is shown
in \fig{ana:fig:chi2prob} for $k=4$. The vertical lines in
\fig{ana:fig:chi2dist} show the $\chi^2$ values corresponding to 90\%
fit probability \mbox{$\prn{\chi^2 = 1.06}$}, 80\% probability
\mbox{$\prn{\chi^2 = 1.64}$}, and so on. Notice that the area of the
shaded regions between these lines are all equal -- as they have to be,
since they all represent a 10\% wide probability ``bin.'' Thus, any
measured $\chi^2_t$ value is equally likely to fall into any one of the
10\% wide probability bins.

\begin{figure}[t]
   \centering
   \subfigure[Fit Probability Cut]{
      \label{ana:fig:trkPrbCut}
      \includegraphics[width=0.4\linewidth]{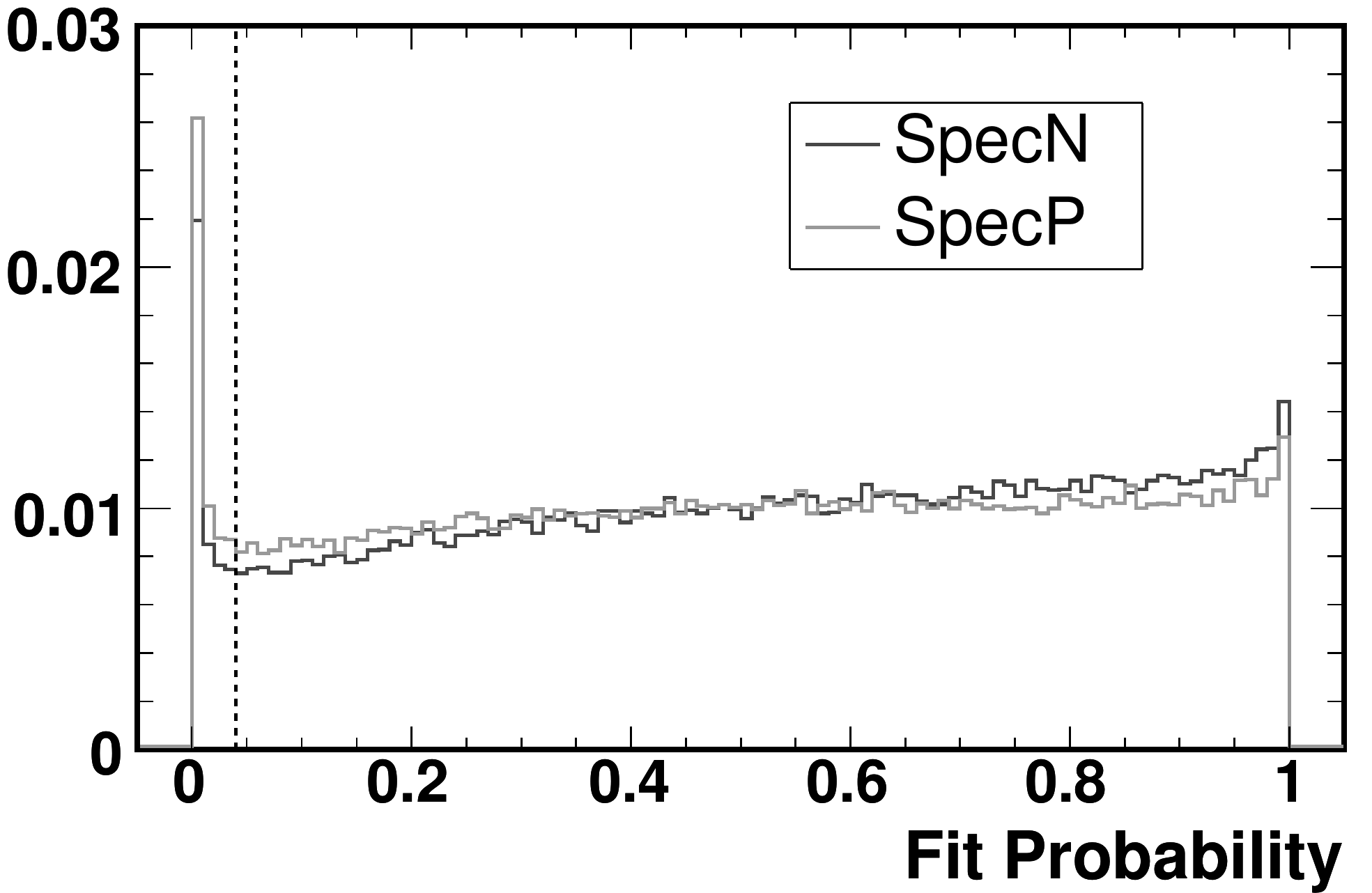}
   }
   \subfigure[{\Prap} Cut]{
      \label{ana:fig:trkEtaCut}
      \includegraphics[width=0.4\linewidth]{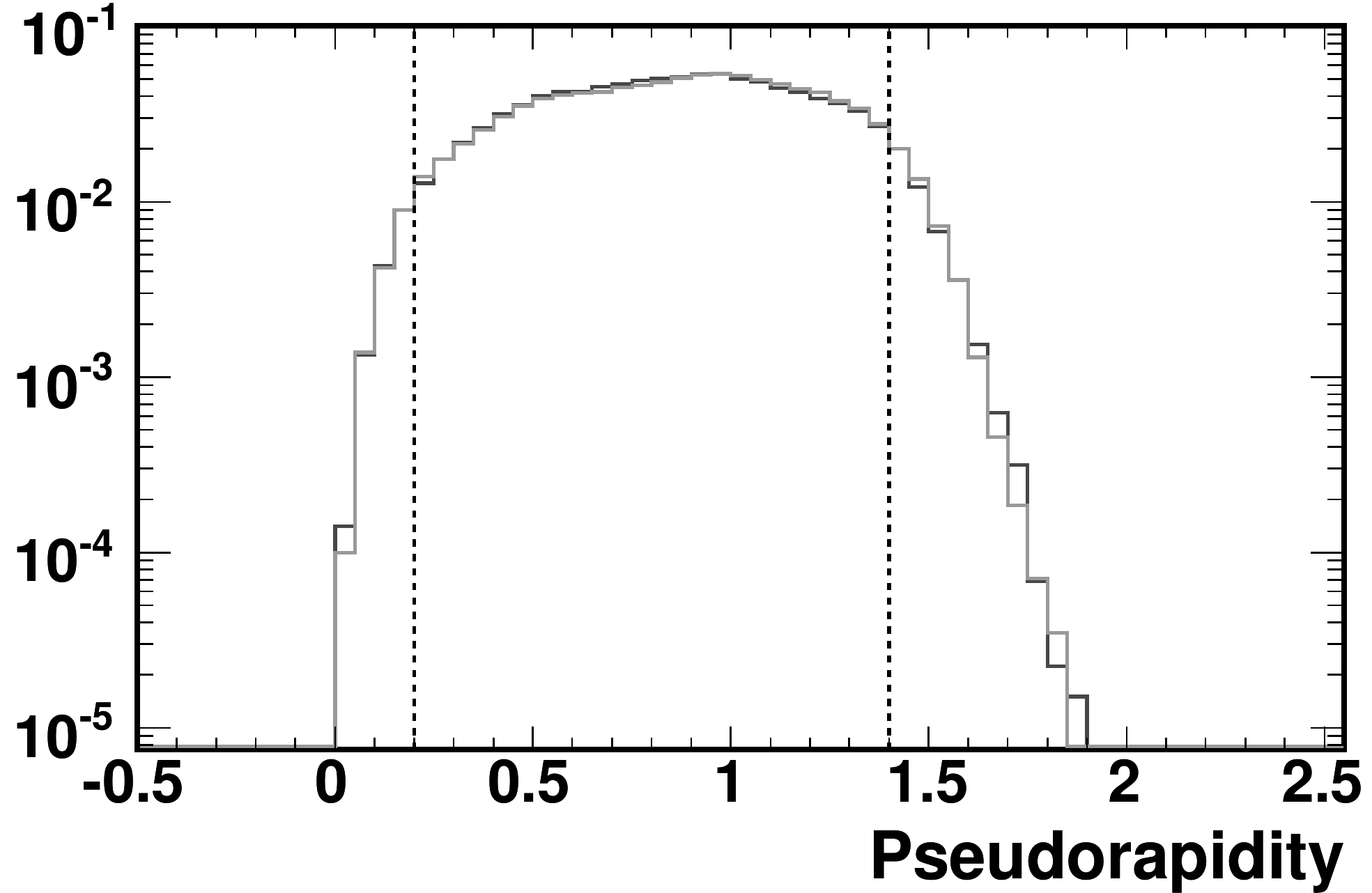}
   }
   \caption{\label{ana:fig:otherTrkCuts}
      Cuts made to select tracks for the analysis. Black dashed lines
      show the cuts, solid dark grey lines show the distribution for
      \protect\acs{SpecN} and solid light grey lines show the
      distribution for \protect\acs{SpecP}.
      \subref{ana:fig:trkPrbCut}~Tracks with fit probability below 4\%
      were not used in the analysis. \subref{ana:fig:trkEtaCut}~Tracks
      outside the Spectrometer's acceptance, \mbox{$0.2 < \eta < 1.4$},
      were cut.}
\end{figure}

It follows that, if the values of $\chi^2_t$ calculated for each
reconstructed particle track were actually distributed according to
\eq{ana:eq:chi2pdf}, then the distribution of fit probabilities seen in
the data would be flat. Any deviation from flatness would indicate that
the errors were not properly calculated. As seen in
\fig{ana:fig:trkPrbCut}, there was an excess of tracks with a fit
probability below 4\%. It was inferred that the errors used for these
tracks (taken from the covariance matrices) did not properly describe
the deviation of hits from the reconstructed particle trajectory. Thus,
it was unlikely that the reconstructed track actually described a
physical particle that produced the observed distribution of hits.
Therefore, these tracks were not used in the analysis.

%---------------------------------------------------------------
\subsection{Spectrometer Acceptance Cut}
\label{ana:trksel:etacut}
%---------------------------------------------------------------

The next track quality cut ensured that only tracks that were within the
{\prap} acceptance of the Spectrometer were used. This cut was performed
mainly to simplify the {\prap} normalization (the ``$\mathit{d\eta}$''
in \eq{ana:eq:invyielmeas}). The {\prap} distribution of tracks in the
{\dAu} data is shown in \fig{ana:fig:trkEtaCut}. Note that the
distribution drops rapidly outside of the cuts. Only tracks with a
{\prap} value between 0.2~and~1.4 units were used in the analysis.

%---------------------------------------------------------------
\subsection{Distance to Beam Cut}
\label{ana:trksel:dcacut}
%---------------------------------------------------------------

Finally, a cut was implemented to remove particles known as secondaries.
That is, particles that were not directly produced by the {\dAu}
collision. For a {\AuAu} collision, this would have been done by
requiring that a reconstructed particle be produced a short distance
from the collision vertex. However, because the vertexing resolution in
{\dAu} was roughly two orders of magnitude larger (in the beam
direction), a different procedure was used for the {\dAu} analysis.

\begin{figure}[t!]
   \centering
   \subfigure[DCA Cut]{
      \label{ana:fig:trkDCACut}
      \includegraphics[width=0.4\linewidth]{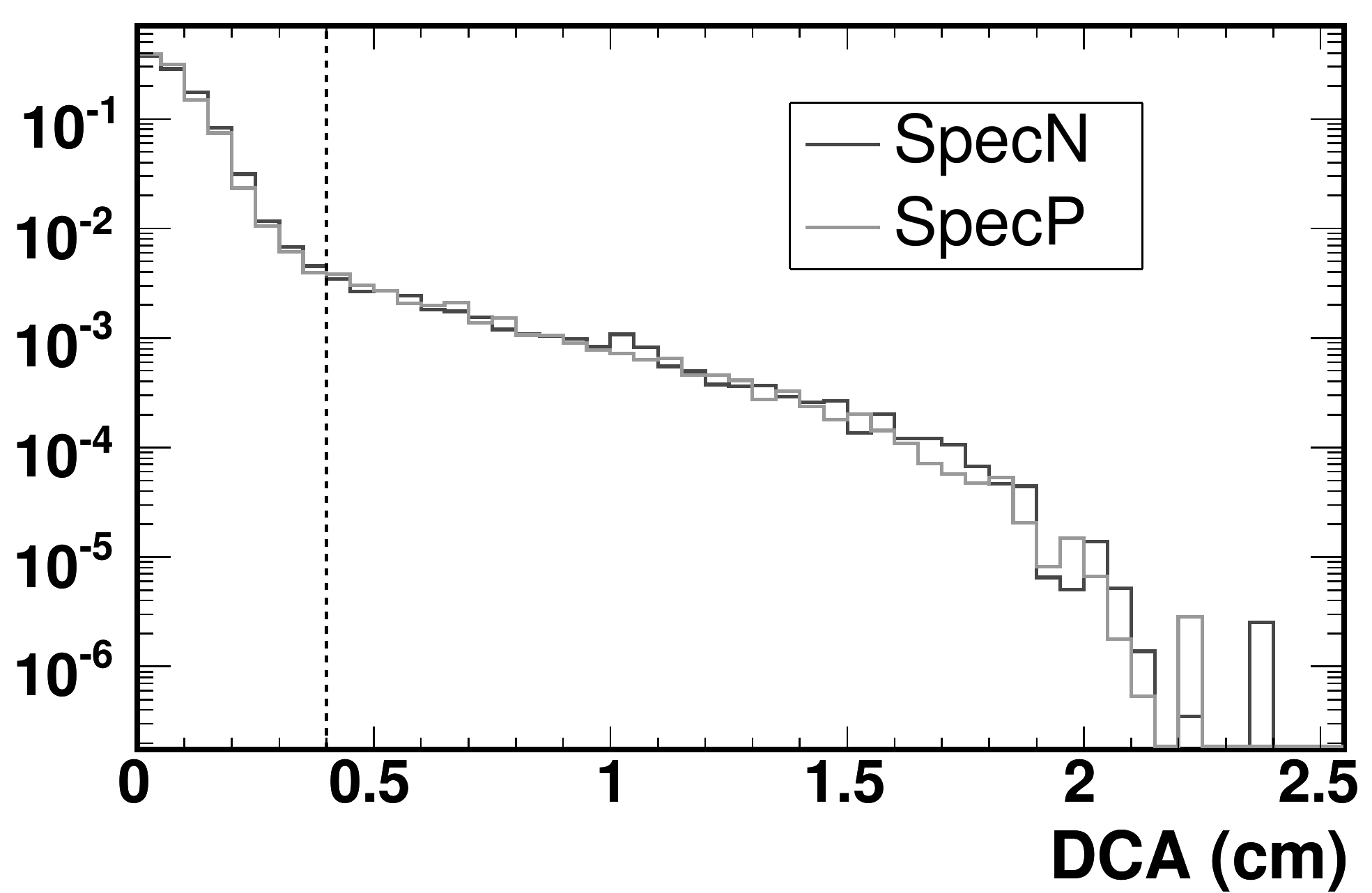}
   }
   \mbox{
      \subfigure[HIJING DCA]{
         \label{ana:fig:secDCAHijing}
         \includegraphics[width=0.4\linewidth]{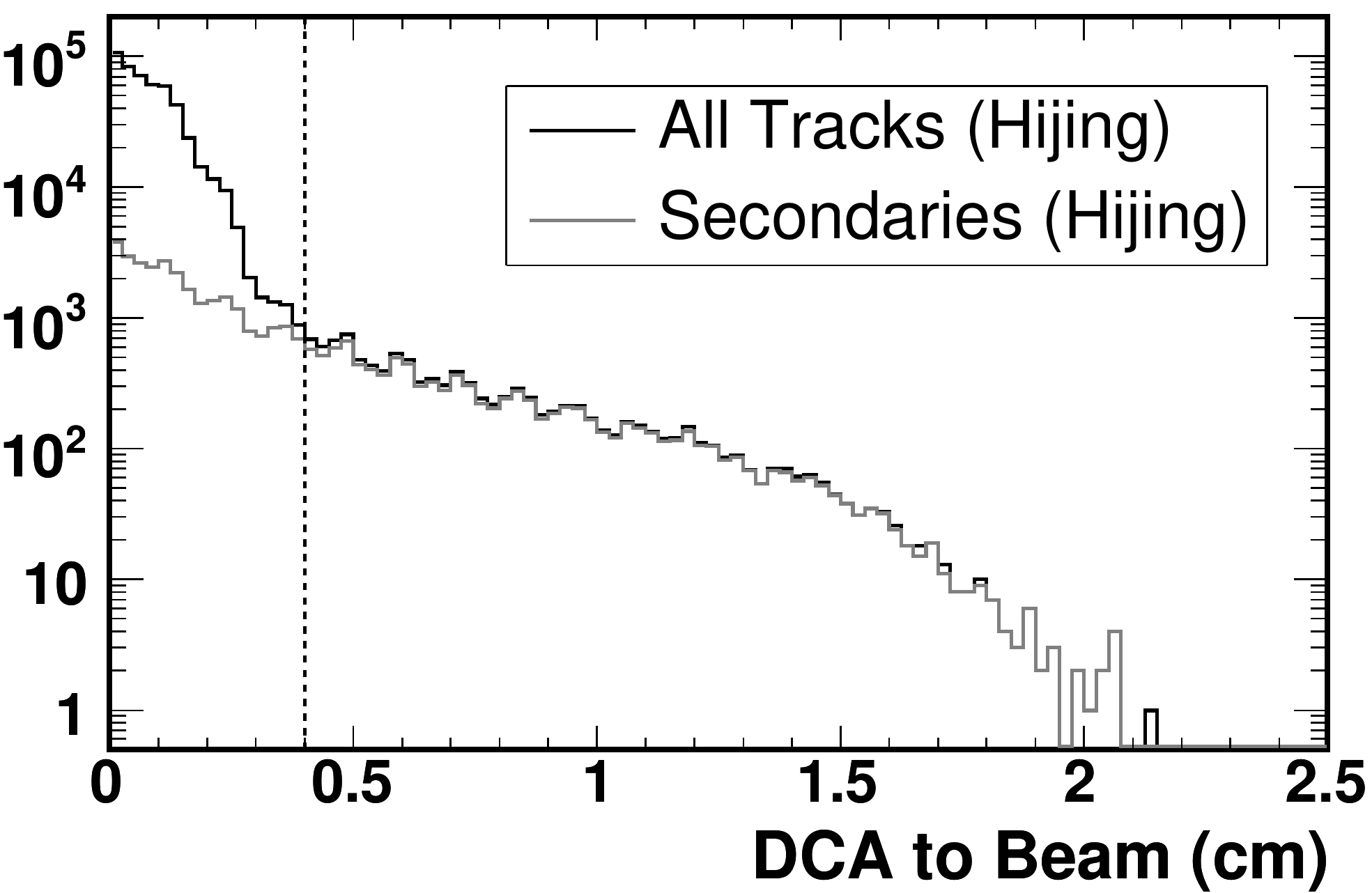}
      }
      \subfigure[Rejected Primaries (HIJING)]{
         \label{ana:fig:ratioDCAHijing}
         \includegraphics[width=0.4\linewidth]{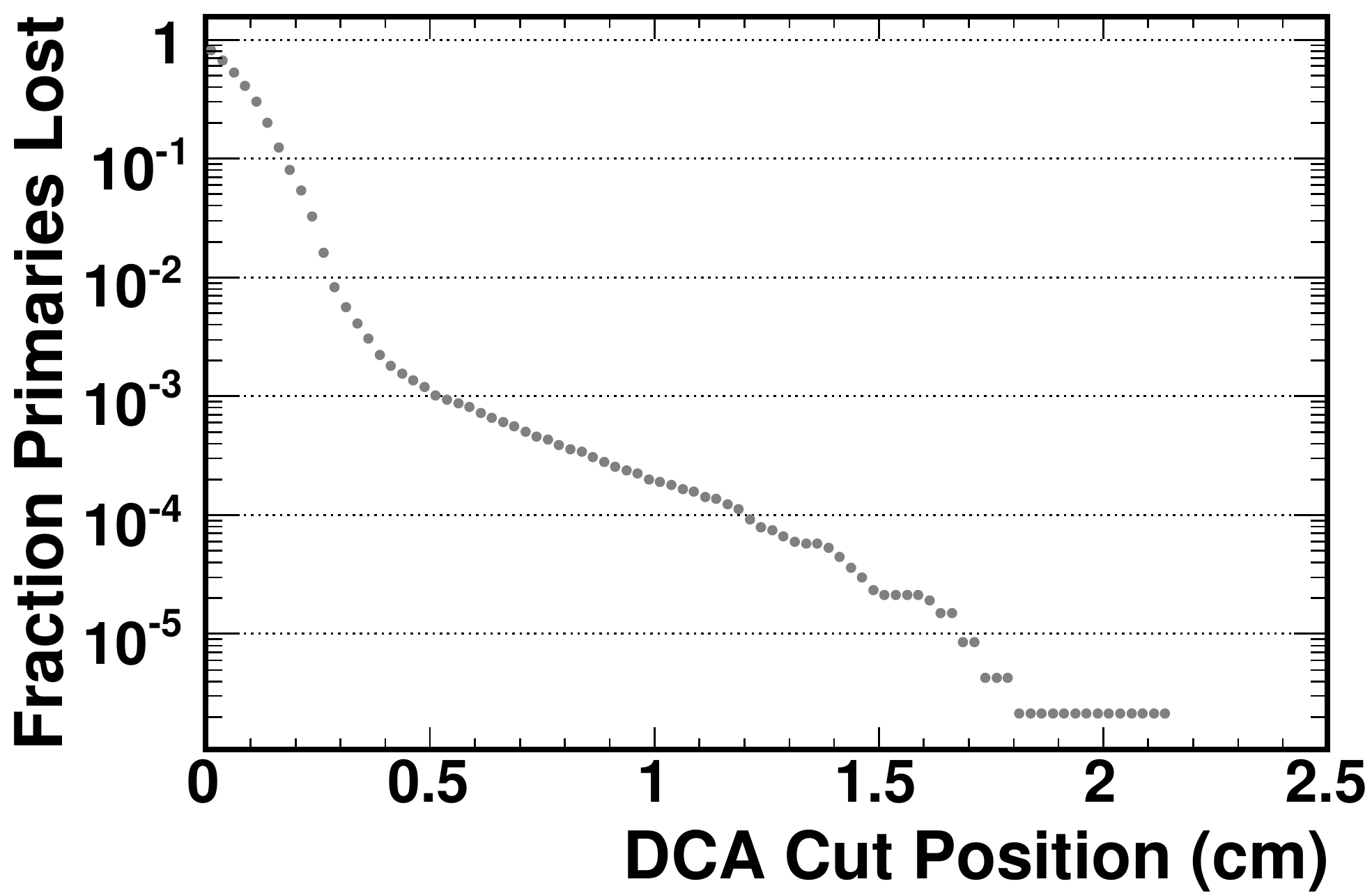}
      }
   }
   \caption{   \label{ana:fig:dcaTrkCuts}
      Tracks that did not originate within 4~{\mm} of the beam orbit
      were cut. \subref{ana:fig:trkDCACut}~The $\dcab$ distribution
      observed in the {\d+Au} data. \subref{ana:fig:secDCAHijing}~The
      $\dcab$ distribution seen in the \protect\acs{HIJING} simulations.
      \subref{ana:fig:ratioDCAHijing}~The fraction of primaries in the
      \protect\acs{HIJING} simulations that would be rejected by cutting
      at a given $\dcab$.}
\end{figure}

In this procedure, the trajectory of the particle was extrapolated back
to the beam line, using the particle's initial momentum vector. Then,
the distance of closest approach, $\dcab$, between the three-dimensional
line of the particle's trajectory and the the three-dimensional line of
the beam orbit was calculated. The distribution of this distance for
tracks in the {\dAu} data is shown in \fig{ana:fig:trkDCACut}. The peak
below distances of 4~{\mm} was due mainly to primary particles (those
produced directly by the {\dAu} collision), while the tail of the
distribution was due to secondaries. This could be seen easily in
\ac{HIJING} simulations, as shown in \fig{ana:fig:secDCAHijing}. Thus,
only particles that came within 4~{\mm} of the beam orbit were used in
the analysis. This cut rejected less than 0.2\% of all primaries, as can
be seen in \fig{ana:fig:ratioDCAHijing}, which shows the fraction of
primaries rejected as a function of the $\dcab$ cut. The accuracy with
which the \ac{HIJING} simulations reproduced the track distributions
seen in the {\dAu} data is discussed in \sect{ana:spec:ghostsec}.

%---------------------------------------------------------------
\section{Measuring Hadron Spectra}
\label{ana:spec}
%---------------------------------------------------------------

With an ideal experiment, it would be possible to observe every
collision, no matter how peripheral, and to measure every particle
produced by the collision, with no contribution from any background
(such as secondaries). In that case, measuring the transverse momentum
spectra of hadrons would be easy. The experimenter would simply count
the number of charged hadrons observed in bins of $\pt$ and {\prap},
weighting each particle by \mbox{$(2 \pi \pt)^{-1}$}. Each bin would
then be divided by the number of collisions, and normalized by the bin
widths.

In a real experiment, it is not possible to obtain such a direct
measurement of the charged hadron spectra. While the fundamental
procedure for finding the spectra remains the same, various corrections
are needed to account for experimental effects that cause the number of
\emph{observed} charged particles to differ from the number of
\emph{produced} charged particles. These corrections were determined by
estimating the effects of experimental imperfections using simulations. 

%---------------------------------------------------------------
\subsection{Acceptance and Efficiency}
\label{ana:spec:acceff}
%---------------------------------------------------------------

\begin{table}[t]
   \begin{center}
      \begin{tabular}{|ccc|}
\hline
Charge & Polarity & Bending Direction\\
\hline
$\rm{h}^+$ & B$+$ & Toward the beam line\\
$\rm{h}^+$ & B$-$ & Away from the beam line\\
$\rm{h}^-$ & B$+$ & Away from the beam line\\
$\rm{h}^-$ & B$-$ & Toward the beam line\\
\hline
      \end{tabular}
   \end{center}
   \caption{   \label{ana:tab:benddir}
      Definition of the bending direction of a charged hadron in the
      {\phob} magnetic field.}
\end{table}

\begin{figure}[t]
   \begin{center}
      \includegraphics[width=0.6\linewidth]{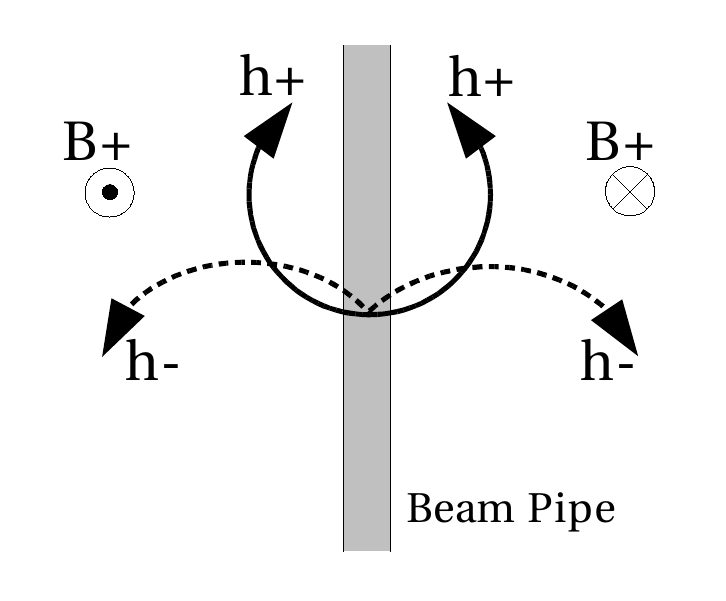}
   \end{center}
   \caption{\label{ana:fig:benddir}
      The bending direction of positive (solid arrow) and negative
      (dashed arrow) hadrons in a positively oriented magnetic field.}
\end{figure}

The largest correction made to the spectra accounted for the limited
acceptance of the {\phob} Spectrometer and for the efficiency of the
tracking procedure. The correction depended on the transverse momentum
of the particle, the electric charge of the particle, the longitudinal
collision position, the polarity of the magnetic field and on the
Spectrometer arm used to reconstruct the particle. The dependence on
charge and magnetic field were related, however. A positively charged
pion with momentum $p$ traveling through a magnetic field of positive
polarity (B$+$) follows the same trajectory as a negatively charged pion
with momentum $p$ traveling through a magnetic field with opposite
(B$-$) polarity. Thus, the correction depended on the so-called bending
direction of the particle, described in \tab{ana:tab:benddir}, rather
than on the four combinations of charge and polarity. Note that a
magnetic field of one polarity, say B$+$, would be oriented up-wards on
one side of the beam line and down-wards on the other. The result was
that particles of the same charge would bend in the same direction,
either away from or toward the beam line, no matter which Spectrometer
arm they traveled through, as shown in \fig{ana:fig:benddir}.

\begin{table}[t]
   \begin{center}
      \begin{tabular}{|ccc|}
\hline
Parameter & Minimum & Maximum\\
\hline
$z_0$ & $-$10~{\cm} & $+$10~{\cm}\\
$\pt$ & 0.15~{\mom} & 7.0~{\mom}\\
$y$ & 0 & 1.5\\
$\phi$ & $-$0.2~{\rad} & $+$0.2~{\rad}\\
\hline
      \end{tabular}
   \end{center}
   \caption{   \label{ana:tab:acceffTrkThrow}
      The physical parameters used to simulate pions for the
      Spectrometer acceptance and tracking efficiency corrections.}
\end{table}

The size of the correction was determined by simulating the response of
the Spectrometer and tracking procedure to individual particles. Ten
million charged pions were simulated and separately reconstructed in
each of the eight combinations of: $\pi^+$ or $\pi^-$, B$+$ or B$-$, and
reconstruction in \ac{SpecN} or \ac{SpecP}. The simulated pions were
generated using four parameters to describe the particle's trajectory:
longitudinal origin, $z_0$, transverse momentum, $\pt$, rapidity, $y$,
and azimuthal angle, $\phi$. For each particle, random values of these
parameters were chosen between the limits shown in
\tab{ana:tab:acceffTrkThrow}. To account for the collision position
dependence, the corrections were determined in four bins of $z_0$, each
5~{\cm} wide. Particles were simulated over a range of azimuthal angle
that was sufficient to cover the acceptance of one Spectrometer arm, but
was smaller than $\pi$. Therefore, an extra factor of \mbox{$\pi /
(0.4~\rad) \approx 7.85$} was required to correct the measured spectra
up to the azimuthally averaged yield described in
\eq{ana:eq:invyielmeas}.

The simulation of each particle generated ``hits'' in the Spectrometer.
This allowed the tracking procedure to be run on a simulated event
containing only one particle. The tracking would then either fail to
find any track, or it would produce a single reconstructed particle
track. Very rarely, multiple tracks would be found; these events were
explicitly discarded, as this effect of the tracking was taken into
account by the ghost correction discussed in \sect{ana:spec:ghostsec}. 

To find the correction, the number of particles simulated was recorded
in bins of $\pt$, $z_0$, Spectrometer arm and bending direction. Only
particles with \mbox{$0.2 < \eta < 1.4$} were used to find the
correction (to correspond with the {\prap} range used in the analysis).
Then the number of successfully reconstructed tracks in each bin was
recorded. A track was successfully reconstructed if (a)~the tracking
procedure produced a track and (b)~the reconstructed particle passed the
track selection cuts described in \sect{ana:trksel}. Note that only the
true simulation parameters of a particle were used to choose in which
bin the track should be counted. For example, suppose two $\pt$ bins
were used, where bin $A$ counted tracks with \mbox{$\pt < 1~\mom$} and
bin $B$ counted tracks with \mbox{$\pt > 1~\mom$.} If a particle was
simulated with \mbox{$\pt = 0.9~\mom$} but reconstructed as having
\mbox{$\pt = 1.1~\mom$,} then both the number of simulated and the
number of reconstructed particles in bin $A$ would be increased. This
was done so that the momentum resolution of the tracking (see
\sect{ana:spec:momres}) could be accounted for separately from the
efficiency of the tracking. The true value of $z_0$ was also used, to
prevent vertexing efficiency from entering into this correction.

\begin{figure}[t!]
   \centering
   \subfigure[$-10~\cm < z_0 < -5~\cm$]{
      \label{ana:fig:acceffvb1}
      \includegraphics[width=0.4\linewidth]{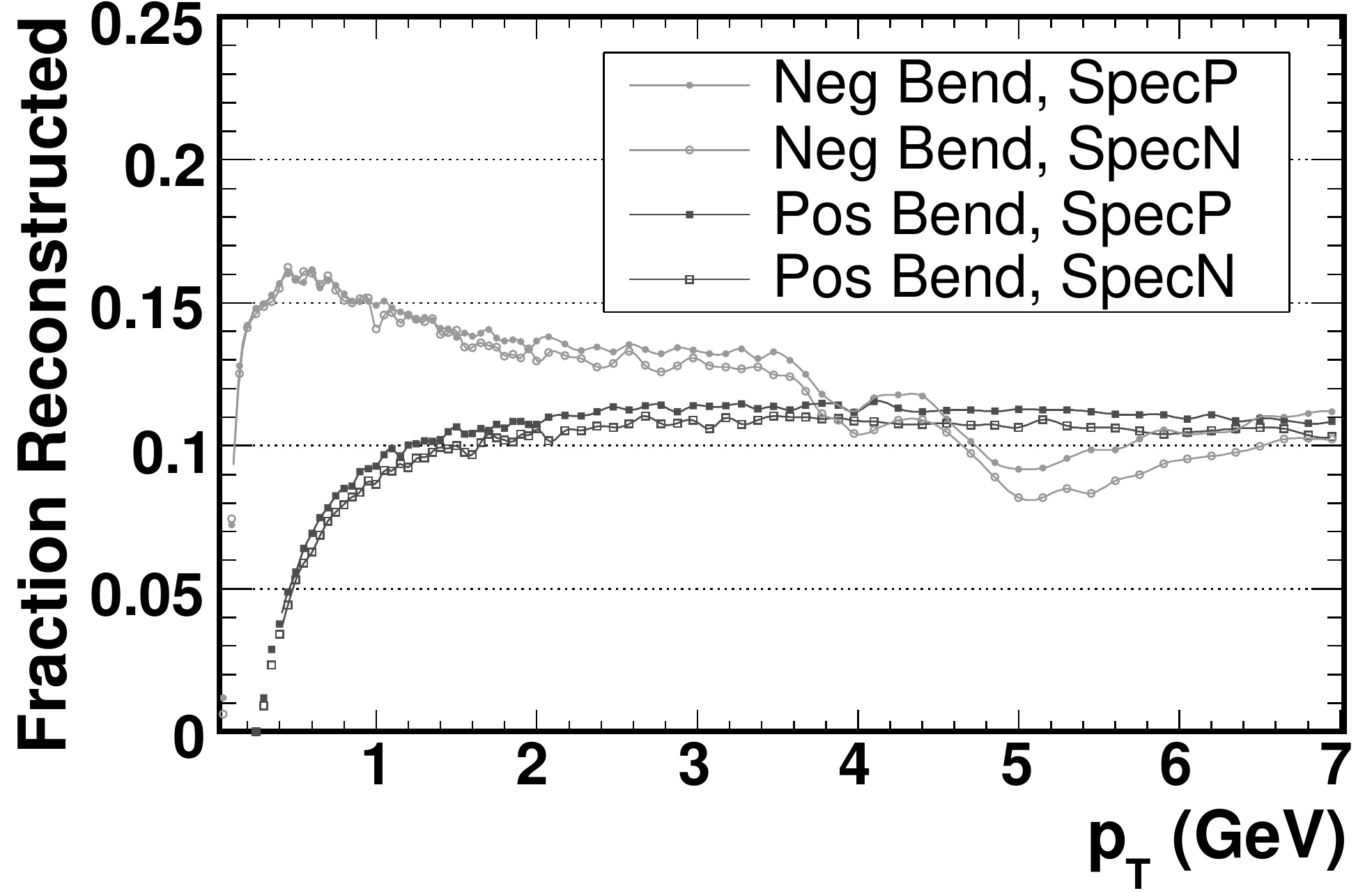}
   }
   \subfigure[$-5~\cm < z_0 < 0~\cm$]{
      \label{ana:fig:acceffvb2}
      \includegraphics[width=0.4\linewidth]{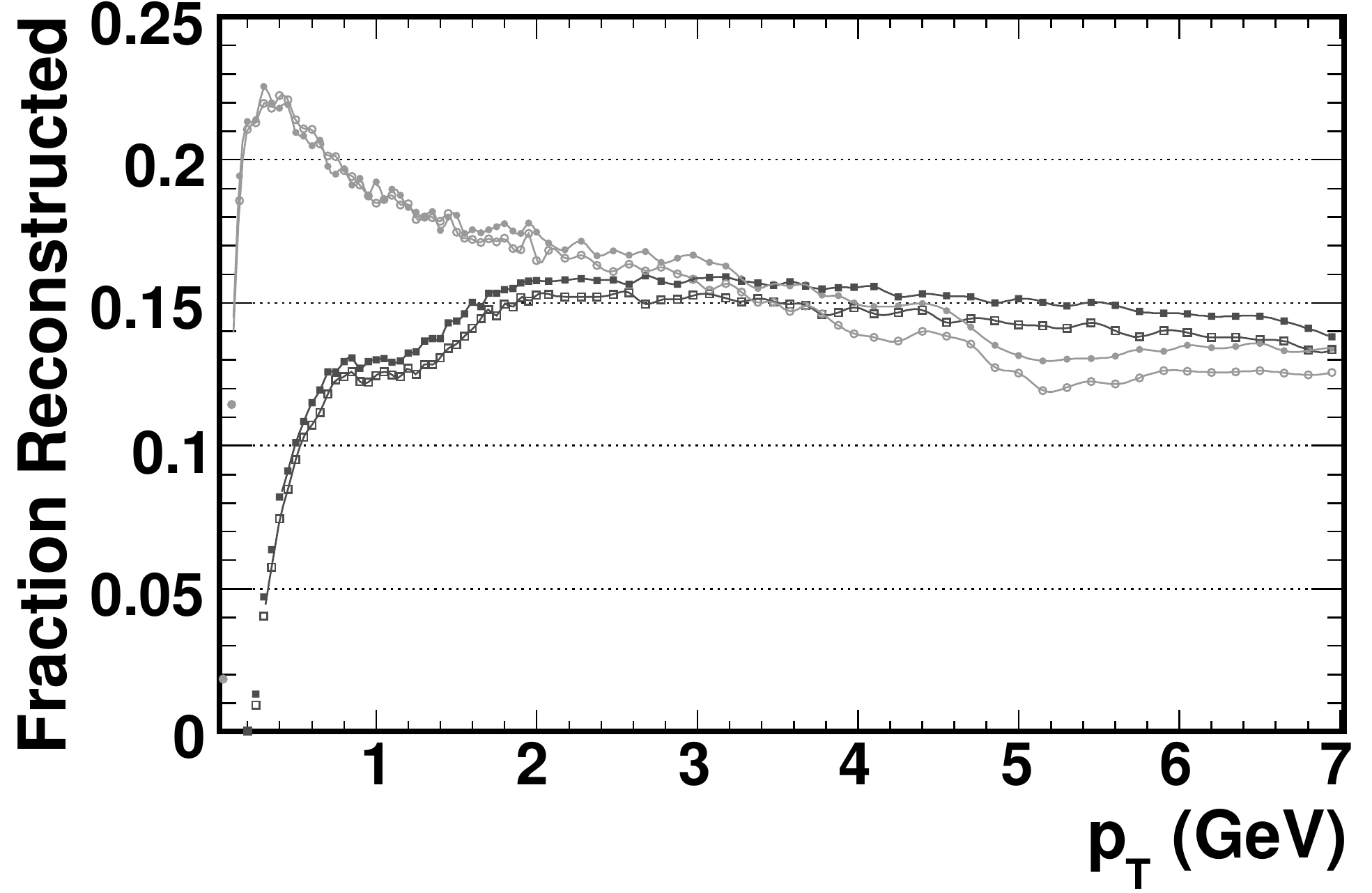}
   }
   \subfigure[$0~\cm < z_0 < +5~\cm$]{
      \label{ana:fig:acceffvb3}
      \includegraphics[width=0.4\linewidth]{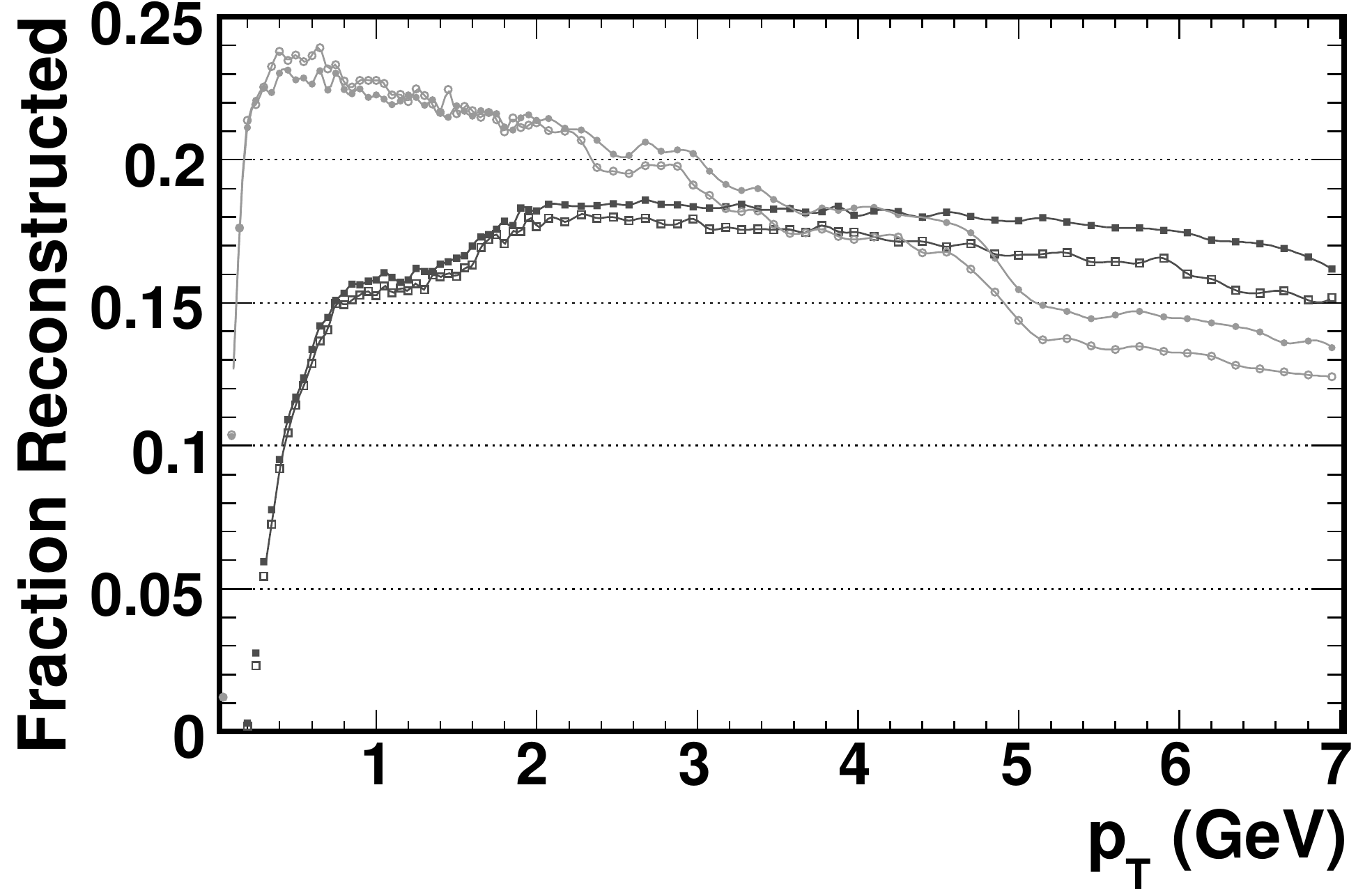}
   }
   \subfigure[$+5~\cm < z_0 < +10~\cm$]{
      \label{ana:fig:acceffvb4}
      \includegraphics[width=0.4\linewidth]{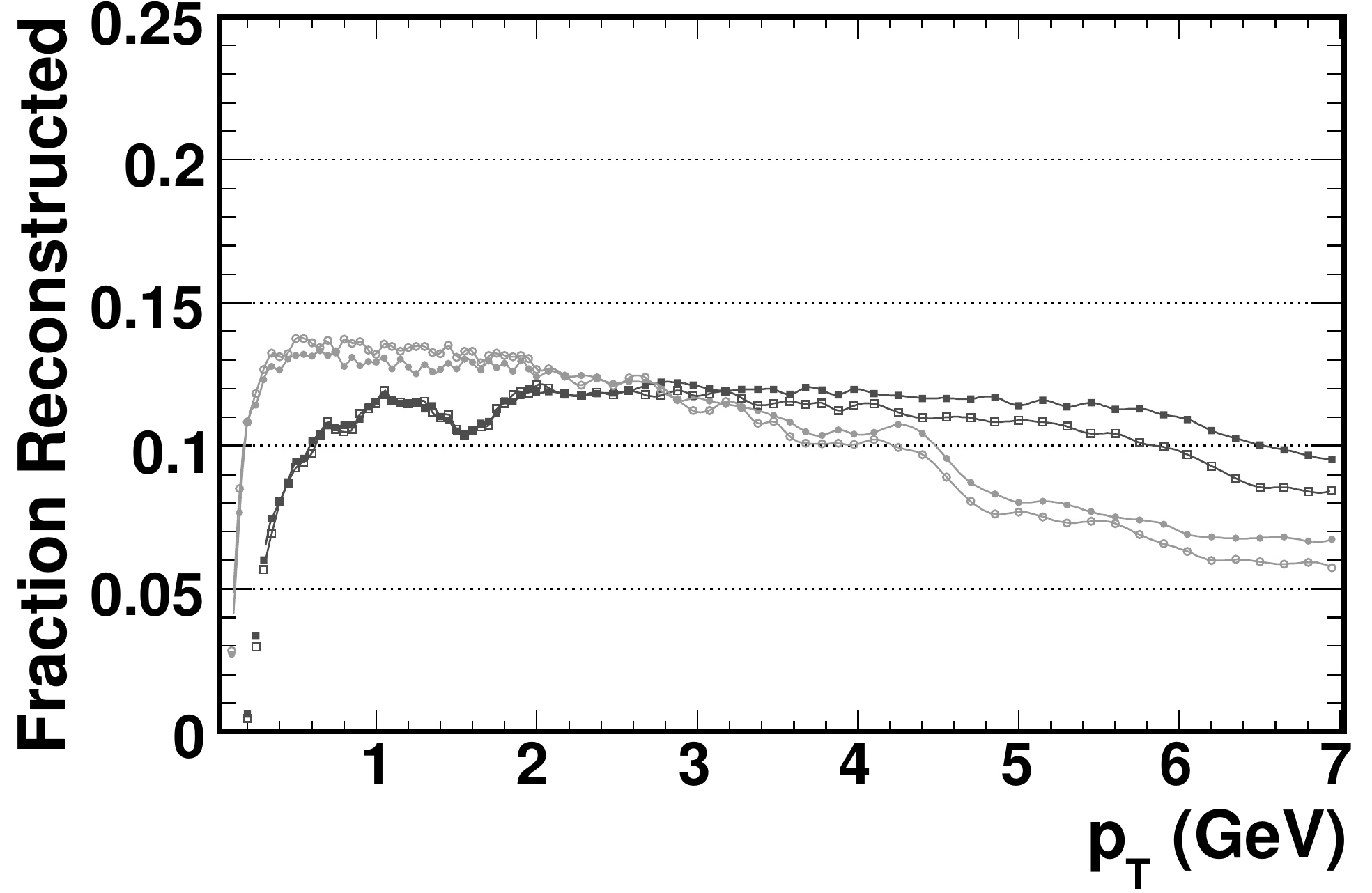}
   }
   \caption{   \label{ana:fig:accEffCorrs}
      The Spectrometer acceptance and tracking efficiency corrections as
      determined by the fraction of successfully reconstructed pions in
      bins of $\pt$, track origin ($z_0$), Spectrometer arm and bending
      direction.}
\end{figure}

\begin{table}[t]
   \begin{center}
      \begin{tabular}{|cccccc|}
\hline
 &  & \multicolumn{4}{c|}{Minimum $\pt$ (\mom)} \\
Bending Dir. & Spec. Arm & $z_0 = (-10~\cm,~-5~\cm)$
 & $(-5,~0)$ & $(0,~5)$ & $(5,~10)$\\
\hline
Toward & \protect\acs{SpecP} & 0.39 & 0.29 & 0.29 & 0.29\\
Toward & \protect\acs{SpecN} & 0.39 & 0.29 & 0.29 & 0.29\\
Away & \protect\acs{SpecP} & 0.09 & 0.09 & 0.09 & 0.09\\
Away & \protect\acs{SpecN} & 0.09 & 0.09 & 0.09 & 0.09\\
\hline
      \end{tabular}
   \end{center}
   \caption{   \label{ana:tab:minPt}
      The minimum $\pt$ used to measure particles in each track origin,
      Spectrometer arm and bending direction bin.}
\end{table}

The value of the correction in a bin was the fraction of simulated
particles that were successfully reconstructed. The corrections were
stored as histograms, with separate histograms for each combination of
track origin bin, Spectrometer arm and bending direction. Each histogram
contained bins of $\pt$, with $0.05~\mom$ wide bins up to $2~\mom$,
$0.1~\mom$ wide bins up to $4~\mom$ and $0.15~\mom$ wide bins above
$4~\mom$. The histograms were then fit with $3^{rd}$~degree polynomial
splines. The fits allowed interpolation between the $\pt$ bin centers
and were used to estimate the corrections for particles having any
transverse momentum up to $7~\mom$. The minimum $\pt$ used to fit each
spline was not the lowest point in the histogram. Instead, a minimum was
chosen to correspond to the point at which the tracking efficiency
dropped to roughly 30\% of the maximum. This was done to avoid
systematic errors resulting from (a)~interpolating between points in a
rapidly changing function and (b)~measuring particles in a region of
$\pt$ where the efficiency may not be well determined. The minimum $\pt$
values used in the spline fits, and therefore also the minimum $\pt$
used in the spectra measurements, are show in \tab{ana:tab:minPt}. The
resulting corrections are shown in \fig{ana:fig:accEffCorrs}.

%---------------------------------------------------------------
\subsection{Ghost and Secondary Particles}
\label{ana:spec:ghostsec}
%---------------------------------------------------------------

While the efficiency of the tracking procedure could be understood by
simulating single particles, further studies were necessary to estimate
how often a successfully reconstructed track was \emph{not} actually a
measurement of a primary particle. Such tracks introduced errors on the
number of primary particles per {\dAu} collision counted by the
analysis. These errors were corrected for by investigating two sources
of such mis-identified tracks. One source was reconstructed tracks that
did not actually correspond to any physical particle, but were
constructed out of hits from different particles (or detector noise).
These tracks were known as ghosts. The other source was successfully
reconstructed secondaries.

Full {\dAu} collision simulations were used to estimate the number of
ghosts and secondaries that would be observed in the data. This was done
by first running the tracking procedure on each simulated event. The
reconstructed tracks were then matched to the underlying simulated
particles by finding shared hits. That is, for a given Silicon pad that
was hit by a reconstructed track, all simulated particles that deposited
energy into the same pad were said to share a hit with the track. The
simulated particle that shared the most hits with a reconstructed track
was the best match for that track.

\begin{figure}[t!]
   \centering
   \subfigure[Ghosts]{
      \label{ana:fig:ghostfrac}
      \includegraphics[width=0.4\linewidth]{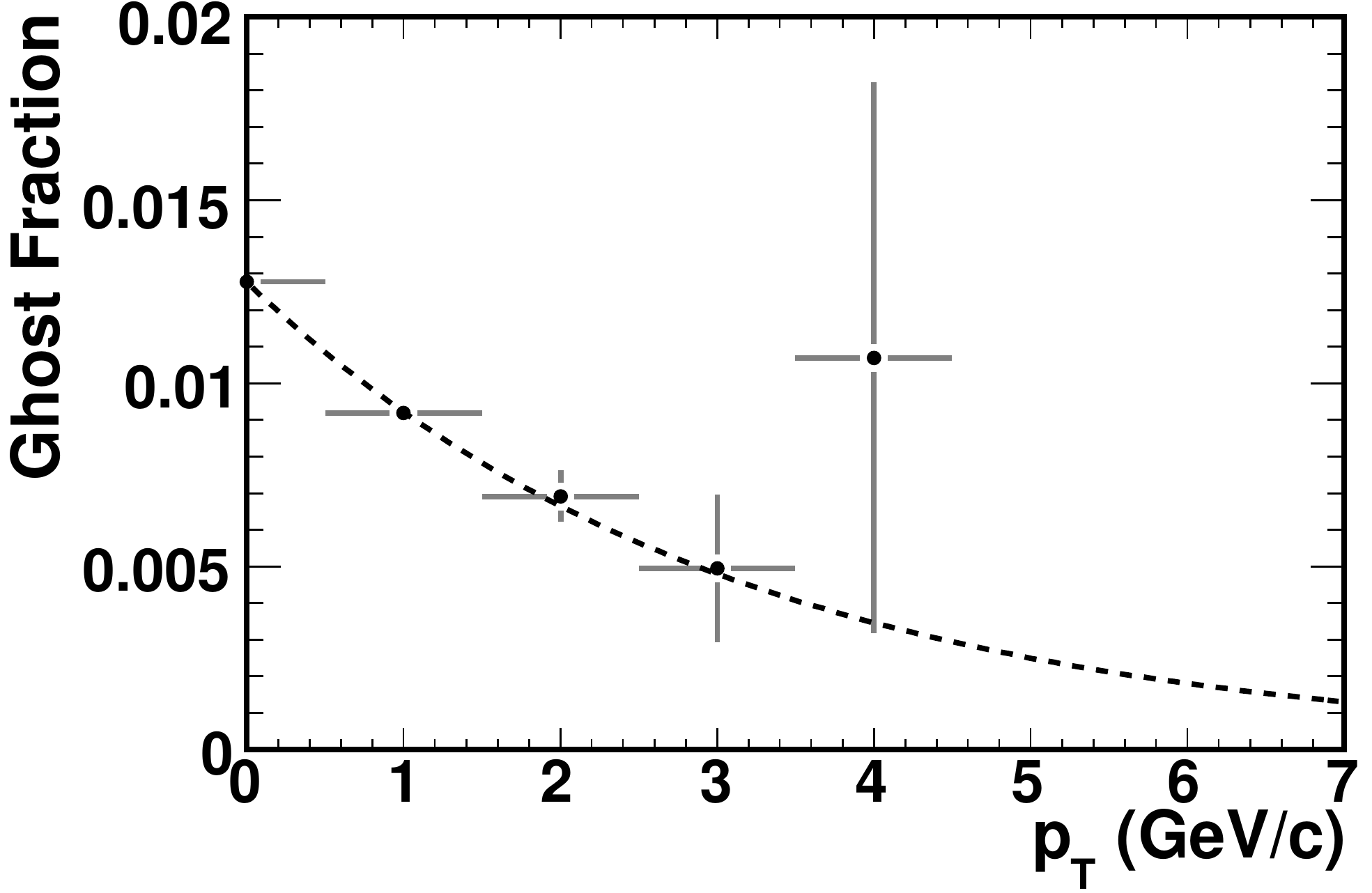}
   }
   \subfigure[Secondaries]{
      \label{ana:fig:secondaryfrac}
      \includegraphics[width=0.4\linewidth]{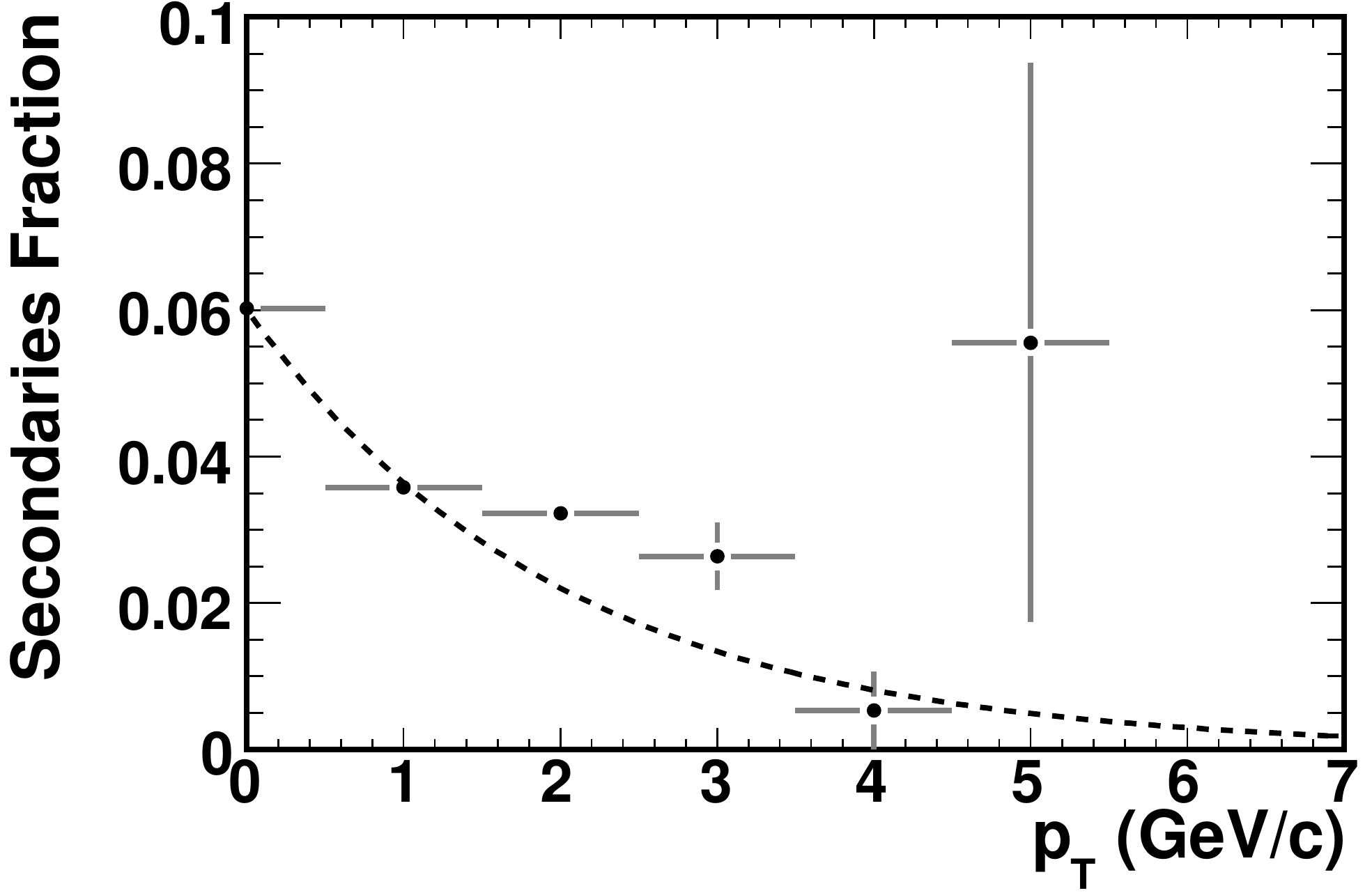}
   }
   \caption{   \label{ana:fig:ghostsecondcor}
      Ghost and secondary tracks. The dashed black lines show
      the fits used in the analysis.
      \subref{ana:fig:ghostfrac}~The fraction of successfully
      reconstructed tracks due to ghosts.
      \subref{ana:fig:secondaryfrac}~The fraction of successfully
      reconstructed tracks due to secondaries.}
\end{figure}

Tracks due to secondaries and ghosts were then easy to identify.
Secondary tracks were successfully reconstructed tracks (i.e.~those that
passed the track selection described in \sect{ana:trksel}) whose best
matching simulated particle was not a primary. Ghost tracks were
successfully reconstructed tracks that did not share ten or more hits
with any simulated particle. The ghost (secondary) correction was then
simply the ratio of the number of reconstructed ghost (secondary) tracks
to the total number of successfully reconstructed tracks, in bins of
reconstructed transverse momentum. The corrections were originally
calculated in each centrality bin, however it was found that neither the
secondary nor the ghost correction depended on centrality.
\Fig{ana:fig:ghostsecondcor} shows the fractions of ghost and secondary
tracks expected to be present in the data.

\begin{figure}[t!]
   \centering
   \subfigure[Fit Probability]{
      \label{ana:fig:fitPrbComp}
      \includegraphics[width=0.4\linewidth]{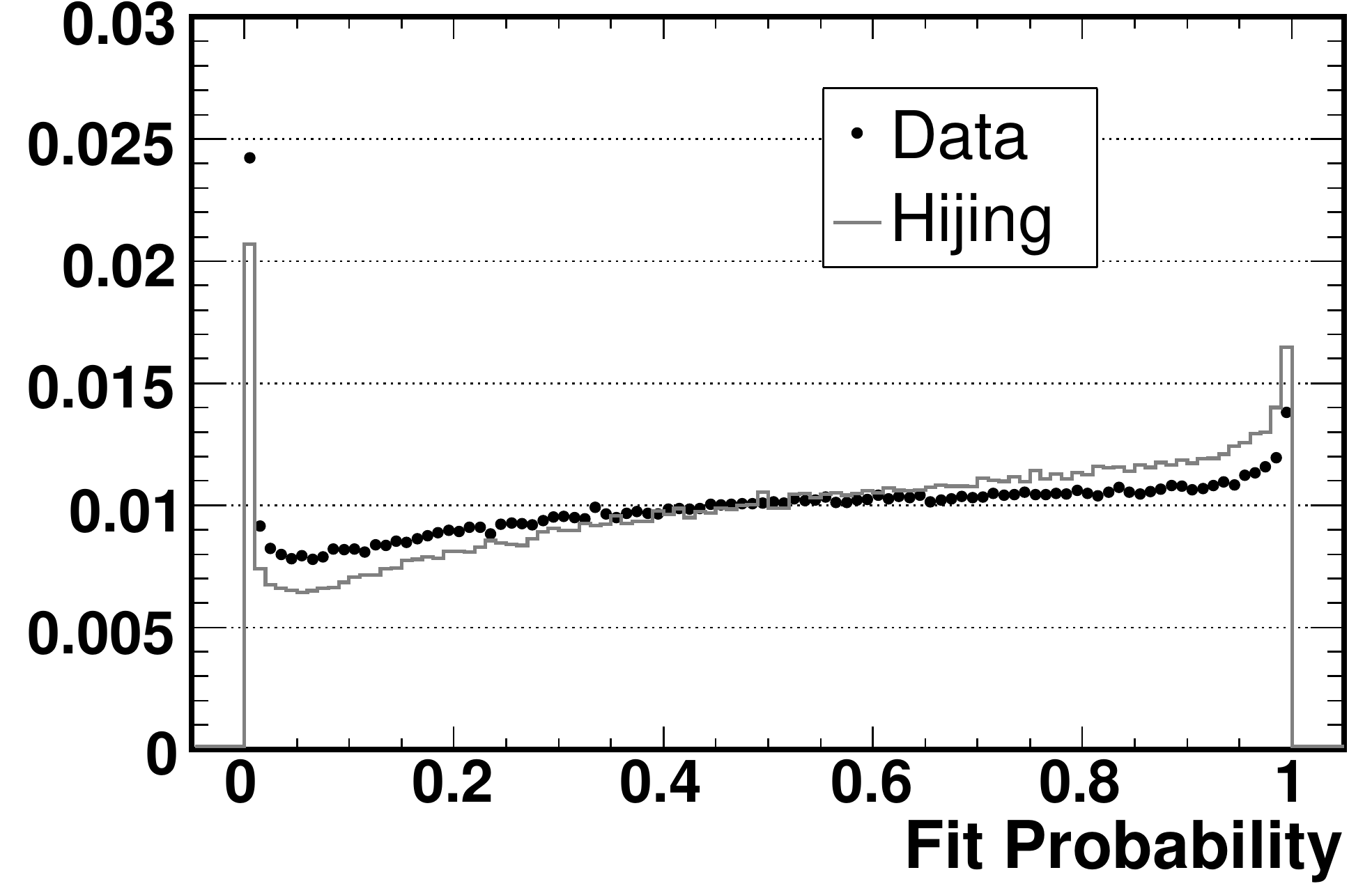}
   }
   \subfigure[DCA Beam]{
      \label{ana:fig:dcaComp}
      \includegraphics[width=0.4\linewidth]{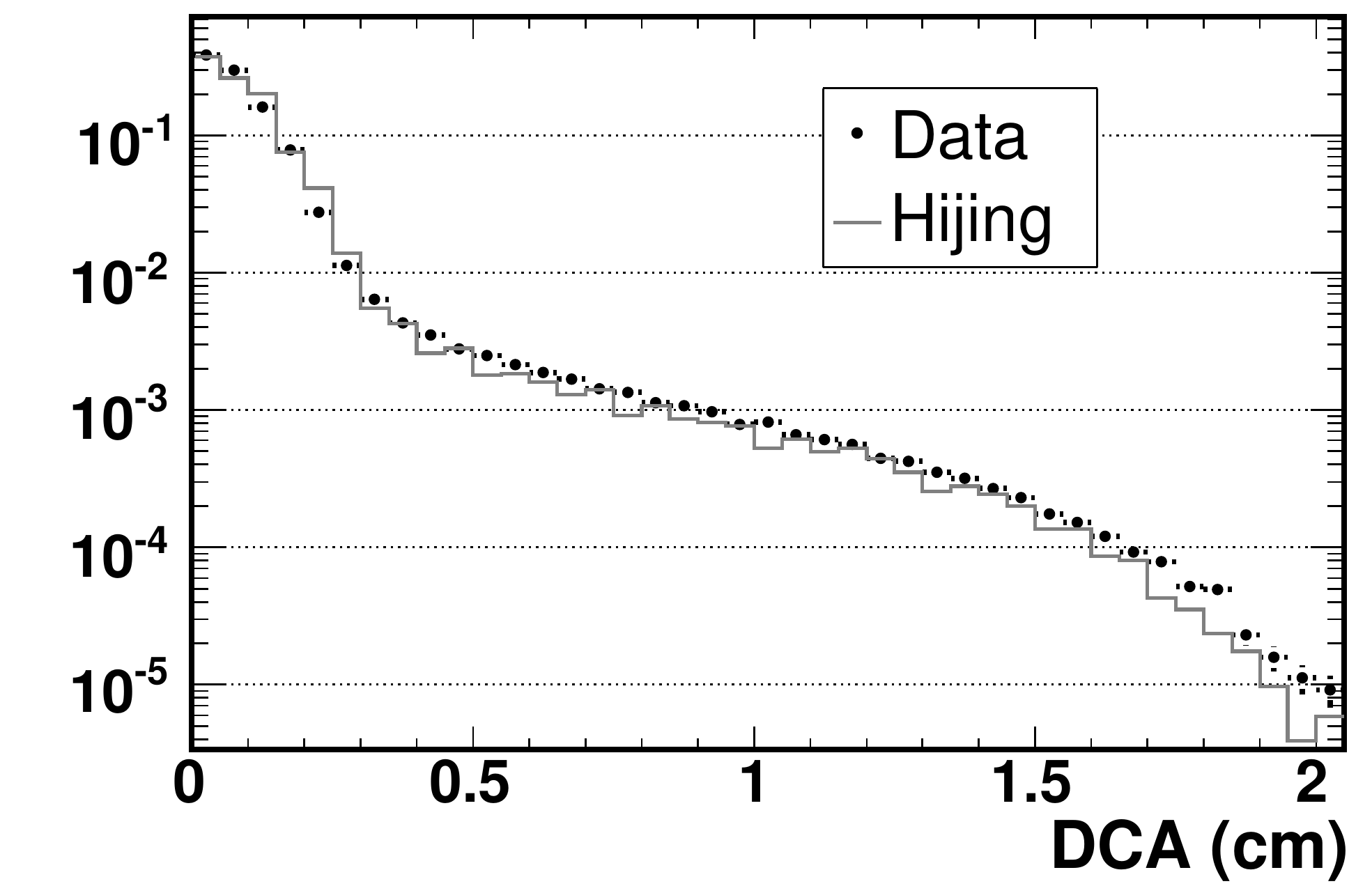}
   }
   \caption{   \label{ana:fig:datamccomp}
      Comparison of reconstructed track distributions from
      \protect\acs{HIJING} (grey line) and {\dAu} data (black points).
      \subref{ana:fig:fitPrbComp}~The fit probability distributions.
      \subref{ana:fig:dcaComp}~The distance of closest approach to the
      beam orbit distributions.}
\end{figure}

\begin{table}[t]
   \begin{center}
      \begin{tabular}{|cc|}
\hline
Collisions & Tracks with Fit Prob. $<$ 4\%\\
\hline
{\dAu} Data & 4.96\%\\
AMPT & 4.08\%\\
Hijing & 4.14\%\\
\hline
      \end{tabular}
   \end{center}
   \caption{   \label{ana:tab:fitprbcut}
      The fraction of tracks rejected by the fit probability cut.}
\end{table}

Since these corrections relied on \ac{MC} simulations, it was necessary
to ensure that the simulations provided a reasonably accurate
description of reality. The $\dcab$ and the fit probability
distributions from \ac{HIJING} simulations are compared to the
distributions from the {\dAu} data in \fig{ana:fig:datamccomp}. The
corresponding distributions from \ac{AMPT} were nearly identical. This
allowed the results of both \ac{AMPT} and \ac{HIJING} simulations to be
combined to find the ghost and secondary track corrections with better
statistics. While the fit probability distributions differed slightly
between data and simulation, the fraction of tracks rejected by the fit
probability cut was essentially the same in the data as it was in the
simulations, as shown in \tab{ana:tab:fitprbcut}. Thus, the simulations
provided an accurate description of the observed tracking behavior and
detector response. 

%---------------------------------------------------------------
\subsection{Dead and Hot Spectrometer Pads}
\label{ana:spec:deadchan}
%---------------------------------------------------------------

Silicon channels that did not function properly introduced another
source of error into the analysis. There were two types of problematic
channels: those that reported a signal without being hit (a hot channel)
and those that reported no signal when hit (a dead channel). Hot
channels were identified as those that reported either many more hits
than the average channel and/or much more energy per hit than the
average channel. Dead channels were identified as those that reported
many fewer hits than the average channel and/or much less energy per hit
than the average channel.

During the {\dAu} spectra analysis, signals from hot and dead channels
were used when reconstructing tracks. The effects of including these
channels was studied using both {\dAu} data and \ac{MC} simulations. It
was assumed that hot channels would increase the number of ghost tracks
in the data, since the tracking could (wrongly) associate a false hit on
a hot channel with a track. On the other hand, dead channels were
assumed to decrease the number of tracks that could be reconstructed,
since hits from particles passing through these channels could not be
observed.

\begin{table}[t]
   \begin{center}
      \begin{tabular}{|cccc|}
\hline
 & Data & Single Particle Simulation & \\
Spectrometer Arm & Masked / Unmasked & Masked / Unmasked & Correction\\
\hline
\protect\acs{SpecP} & 0.8410 & 0.8157 & 1.031\\
\protect\acs{SpecN} & 0.7640 & 0.6840 & 1.117\\
\hline
      \end{tabular}
   \end{center}
   \caption{   \label{ana:tab:deadchan}
      The hot and dead channel correction.}
\end{table}

To estimate the effect of hot channels, one million {\dAu} collisions
were reprocessed with hot and dead channels masked out. That is, hits on
those channels were not used to reconstruct tracks. This reduced the
number of tracks that were successfully reconstructed in these
collisions by about 16\% in \ac{SpecP} and 24\% in \ac{SpecN}, as shown
in \tab{ana:tab:deadchan}. The magnitude of this difference quantified
the relative number of ghost tracks created by hot channels. Note that
these ghost tracks were not accounted for in the correction described in
\sect{ana:spec:ghostsec} since there were no hot channels in the
simulations. 

The reduced efficiency of the tracking due to dead channels was
estimated using single-track simulations, similar to those described in
\sect{ana:spec:acceff}. One million single tracks were simulated in each
combination of $\pi^+$ or $\pi^-$, B$+$ or B$-$, and reconstruction in
\ac{SpecN} or \ac{SpecP}. For these simulations, track reconstruction
was run with the hot and dead channels masked out. This reduced the
single track reconstruction efficiency, as compared to the efficiency
described in \sect{ana:spec:acceff}, by about 18\% in \ac{SpecP} and
32\% in \ac{SpecN}, as shown in \tab{ana:tab:deadchan}. The reason
masked out channels had such a significant impact on the tracking
efficiency was due to the large number of hits required on a track, as
discussed in \chap{track}.

\begin{table}[t]
   \begin{center}
      \begin{tabular}{|ccc|}
\hline
Spectrometer Arm & Hot Channels & Dead Channels\\
\hline
\protect\acs{SpecP} & 0.98\% & 2.07\%\\
\protect\acs{SpecN} & 1.39\% & 3.55\%\\
\hline
      \end{tabular}
   \end{center}
   \caption{   \label{ana:tab:deadhot}
      The fraction of hot and dead channels in each Spectrometer arm.}
\end{table}

The final correction to the number of tracks observed in the data
without masking hot or dead channels was then

\begin{equation}
   \label{ana:eq:dcmcorr}
C_{HD} = \prn{\frac{\text{Data Masked}}{\text{Data Unmasked}}} \bigg /
   \prn{\frac{\text{Single Track Masked}}{\text{Single Track Unmasked}}}
\end{equation}

\noindent%
where $C_{HD}$ is the value of the hot and dead channel correction for a
Spectrometer arm. The first term of \eq{ana:eq:dcmcorr} reduces the
number reconstructed tracks by the fraction of ghosts expected to
originate from hot channels. The second term reduces the efficiency of
the tracking that was estimated in simulations of a Spectrometer with no
dead channels. It accounts for the reduced number of Spectrometer pads
available due to dead channels. Since the raw number of tracks will be
divided by the tracking efficiency (see \eq{ana:eq:trkwt}), this term is
in the denominator.

The values used in this analysis to correct for hot and dead channels
are shown in \tab{ana:tab:deadchan}. The fraction of hot and dead
channels in each Spectrometer arm was not large, as can be seen in
\tab{ana:tab:deadhot}. However, the correction for \ac{SpecN} was much
larger than the correction for \ac{SpecP}. This was due mainly to the
fact that three out of the four sensors on the fifth plane of \ac{SpecN}
had many problematic channels.

%---------------------------------------------------------------
\subsection{Event Selection Efficiency}
\label{ana:spec:centeff}
%---------------------------------------------------------------

As discussed in \sect{recon:cent:cuts:eff}, the efficiency of the
detector to observe collisions that satisfy the chosen event selection
was estimated using \ac{MC} simulations. This efficiency estimate can be
used to correct the observed hadron spectra, in order to measure the
number of charged particles produced per collision, rather than the
number of charged particles per \emph{observed} collision. These
quantities would be equal if the efficiency of measuring a collision did
not depend on the properties of that collision. For example, suppose one
could detect half of all collisions, irrespective of the properties of
those collisions. In that case, on average one would also detect half of
the tracks produced by all the collisions that actually occurred. Then
the ratio of the number of tracks observed to the number of collisions
observed is of course the same as ratio of the number of tracks actually
produced to the number of collisions that actually occurred. However, as
seen in \sect{recon:cent:cuts:eff}, the efficiency of measuring a
collision did in fact depend on the centrality of that collision.

\begin{figure}[t!]
   \centering
   \subfigure[EOct Efficiency]{
      \label{ana:fig:EOctEfficiencies}
      \includegraphics[width=0.4\linewidth]{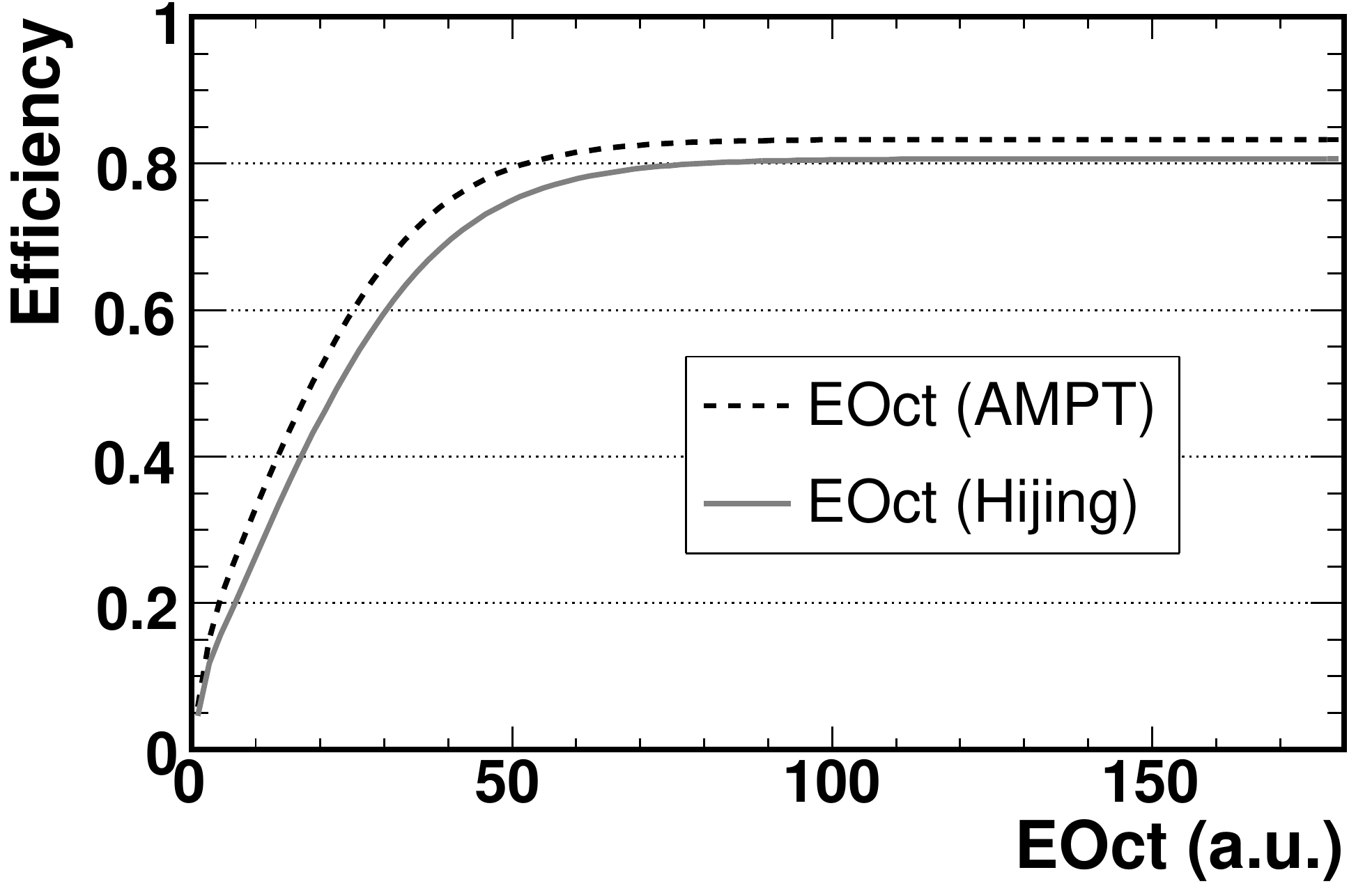}
   }
   \subfigure[ERing Efficiency]{
      \label{ana:fig:ERingEfficiencies}
      \includegraphics[width=0.4\linewidth]{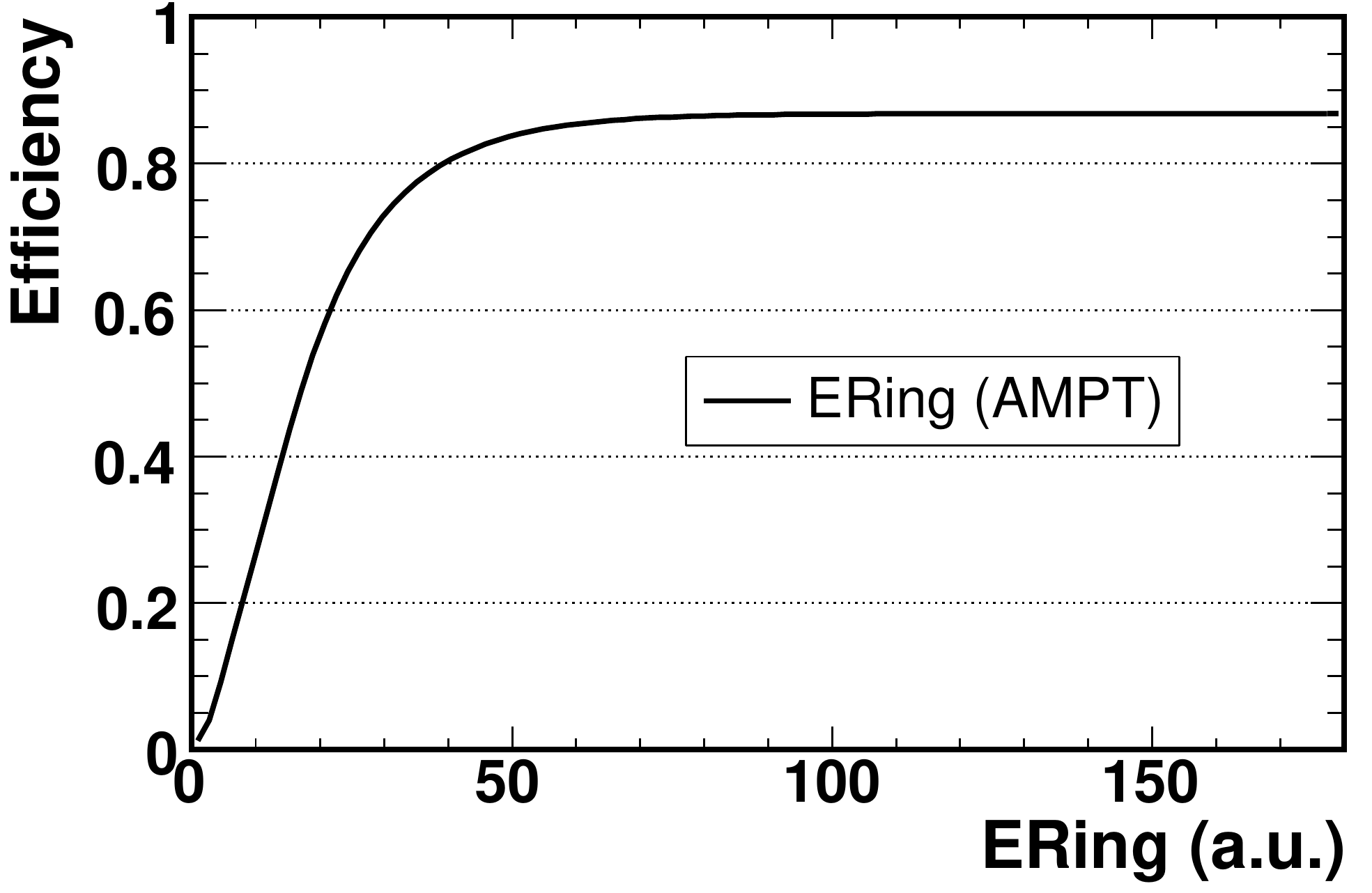}
   }
   \subfigure[Au-PCAL Efficiency]{
      \label{ana:fig:PCALEfficiencies}
      \includegraphics[width=0.4\linewidth]{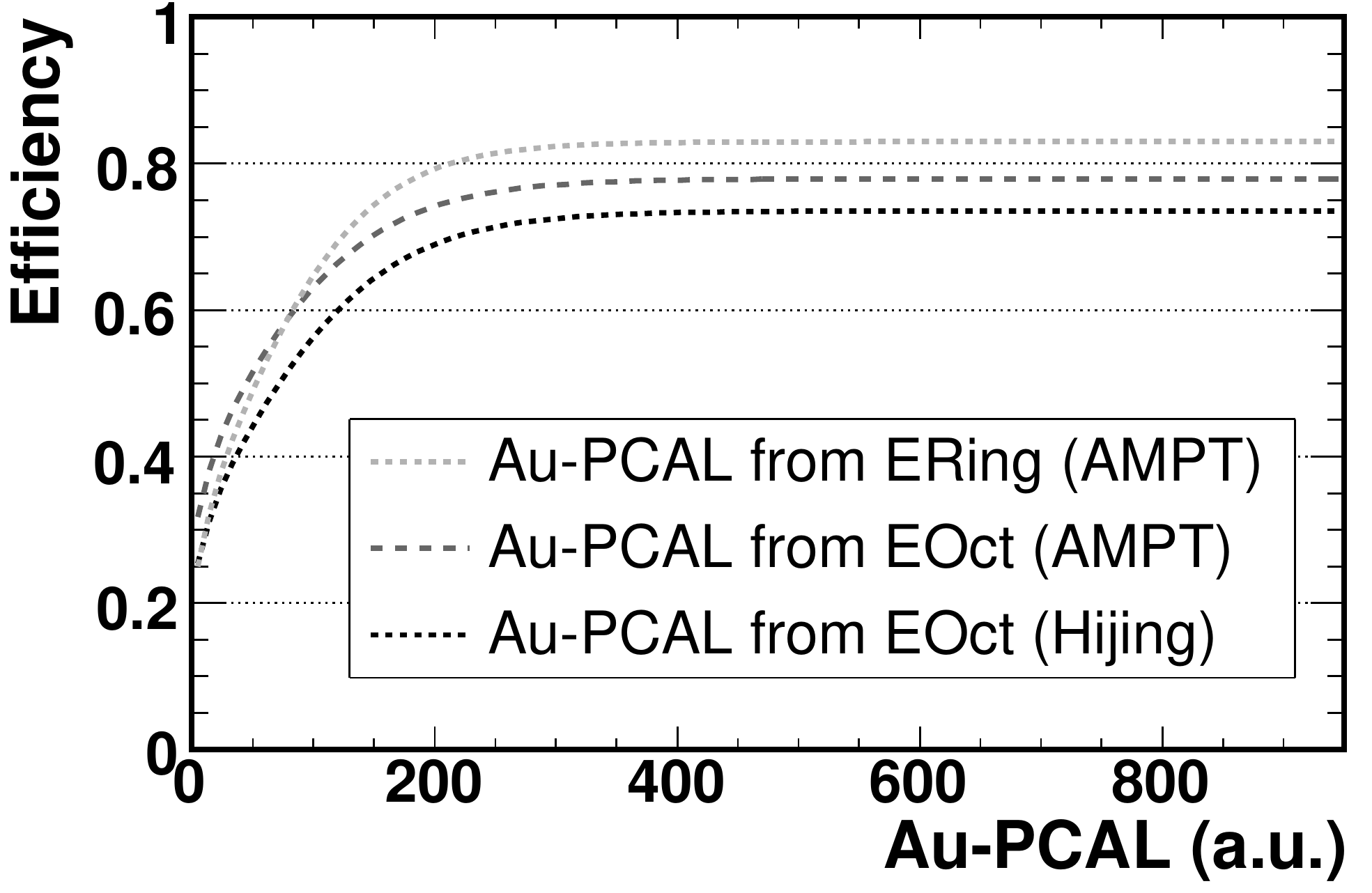}
   }
   \caption{   \label{ana:fig:efficiencies}
      The \protect\acs{dAuSpectra} event selection efficiency as a
      function of the different centrality measures.}
\end{figure}

The centrality dependence of the event selection efficiency could be
accounted for in different ways. First, a model (i.e.~Glauber) could be
used to parametrize the bias introduced by the varying efficiency. This
would be done by (a)~applying the event selection to a group of
simulated collisions, (b)~using a model to calculate a centrality
parameter like {\Npart} for each selected collision and then (c)~finding
the average of the centrality parameter for that group of collisions.
This procedure would give, for example, a biased number of participants
({\Nptbs}) for each centrality bin. For the efficiency shown in
\fig{ana:fig:ERingEfficiencies}, {\Nptbs} of the most central bin would
be equal to {\Npart} in that bin, since the efficiency is flat in the
central bin. However, {\Nptbs} of the most peripheral bin would be
significantly larger than the average number of participants for all
collisions in the peripheral bin. In this way, the measured spectra would
be the number of charged hadrons produced by collisions having an
average number of participants equal to {\Nptbs}.

In this analysis, a different procedure was used; namely, the estimated
efficiency was used to correct the spectra. This provided a measurement
of the average number of particles produced by collisions in a certain
fractional cross section bin. Applying this correction was
straight-forward. For any given collision, the efficiency was determined
using the chosen centrality measure (i.e.~\acs{EOct}). Then, both the
number of tracks and the number of collisions were weighted by this
efficiency. For example, if collision $A$ had an efficiency of 100\% and
one track was observed, then the total number of tracks and collisions
would be increased by one. Whereas if collision $B$ had an efficiency of
only 50\% and two tracks were observed, then the total number of tracks
would be increased by four and the total number of collisions would be
increased by two. The \acs{dAuSpectra} event selection efficiency
corrections for each centrality measure used in the analysis is shown in
\fig{ana:fig:efficiencies}.

%---------------------------------------------------------------
\subsection{Event Normalization}
\label{ana:spec:evtnorm}
%---------------------------------------------------------------

Taking all the corrections into account, the average charged hadron
yield per collision, $\ave{Y}$, in a particular bin (of $\pt$,
fractional cross section and electric charge) was simply the total
corrected number of tracks in that bin, $t_{\rm tot}$, divided by the total
corrected number of events in that bin, $n_{\rm tot}$,

\begin{equation}
   \label{ana:eq:aveyield}
\ave{Y(\pt,~\text{crs.~scn.},~\text{charge})} = 
   \frac{t_{\rm tot}}{n_{\rm tot}}
   = \frac{\sum^{N}_{e=1} \prn{\frac{1}{\epsilon_{e}}
      \sum^{T_e}_{i=1} \prn{w_i}}}
   {\sum^{N}_{e=1} \prn{\frac{1}{\epsilon_{e}}}}
\end{equation}

\noindent%
where $\epsilon_{e}$ is the event selection efficiency for the
collision, $w_i$ is the weight given to each track by the corrections,
$N$ is the (uncorrected) number of measured collisions and $T_e$ is the
(uncorrected) number of reconstructed tracks in the event. The full
track weighting was calculated by

\begin{equation}
   \label{ana:eq:trkwt}
w = \frac{\prn{1 - G(\pt)}~\prn{1 - S(\pt)}~C_{HD}(\text{arm})~\prn{7.85 /
   A(\pt, z_0, \text{arm}, \text{bend})}}
   {2 \pi \pt~\Delta\eta}
\end{equation}

\noindent%
where $\pt$ is the transverse momentum of the particle, $z_0$ is the
longitudinal collision vertex, `arm' is the Spectrometer arm in which
the track was reconstructed, `bend' is the bending direction of the
track, $G$ is the fraction of ghost tracks, $S$ is the fraction of
secondaries, $C_{HD}$ is the hot and dead channel correction, $A$ is the
acceptance and efficiency correction and \mbox{$\Delta\eta = 1.2~{\rm
units}$} is the range of {\prap} over which particles were counted.

The event normalization was complicated, however, by the dependence of
the acceptance on bending direction and collision vertex, shown in
\tab{ana:tab:minPt}. For example, a positive hadron having
$\pt=250~\mmom$ could be measured with the {\phob} magnet in negative
polarity, but not with the magnet in the positive polarity. During the
analysis, the number of particles were counted separately for each
combination of charge, magnet polarity and spectrometer arm. This was
necessary both to apply the acceptance and efficiency corrections (see
\sect{ana:spec:acceff}) and to allow the separate charged spectra to be
measured. The number of collisions observed was counted separately for
each magnet polarity.

It was then necessary to properly sum the corrected number of tracks in
a given $\pt$ bin and to sum the corresponding corrected number of
events. However, the example above suggests that in certain ranges of
$\pt$, only collisions from one magnet polarity should be counted when
measuring particles of a particular charge. Therefore, the number of
collisions used to normalize a particular hadron spectra was dependent
on $\pt$.

\begin{table}[t]
   \begin{center}
      \begin{tabular}{|cccc|}
\hline
Collision Num. & Magnet Polarity & $\pt$ ({\mom}) & Observed?\\
\hline
1 & B$-$ & 0.21 & Yes \\
2 & B$+$ & 0.25 & No \\
3 & B$-$ & 0.42 & Yes \\
4 & B$+$ & 0.35 & Yes \\
5 & B$-$ & 0.55 & Yes \\
\hline
      \end{tabular}
   \end{center}
   \caption{   \label{ana:tab:evtNormExMeas}
      Measurements made by the simple example detector described in
      the text.}
\end{table}

\begin{figure}[t]
   \centering
   \subfigure[Raw Tracks]{
      \label{ana:fig:evtNormExRawTrks}
      \includegraphics[width=0.4\linewidth]{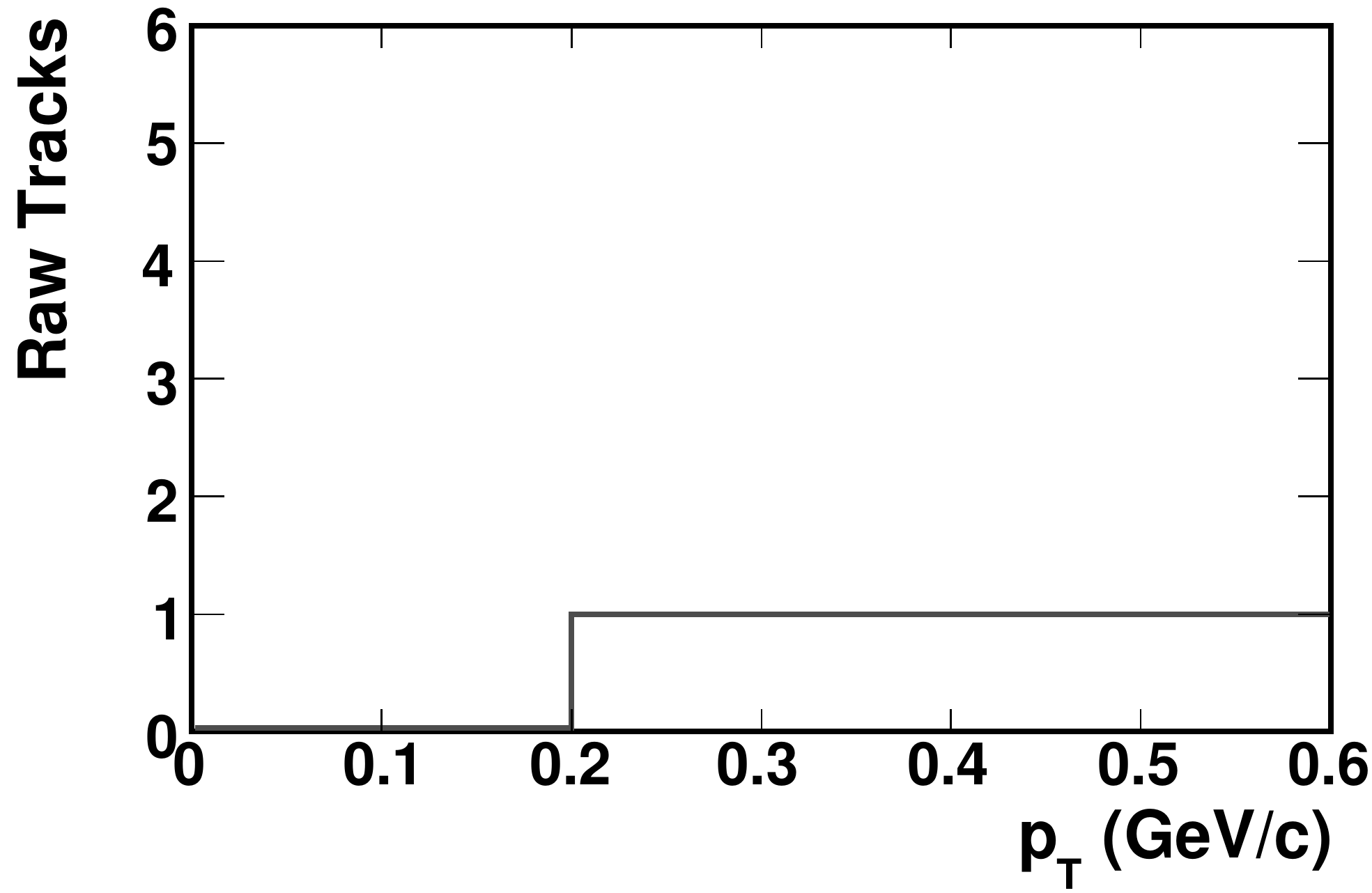}
   }
   \subfigure[Summed Collisions]{
      \label{ana:fig:evtNormExEvts}
      \includegraphics[width=0.4\linewidth]{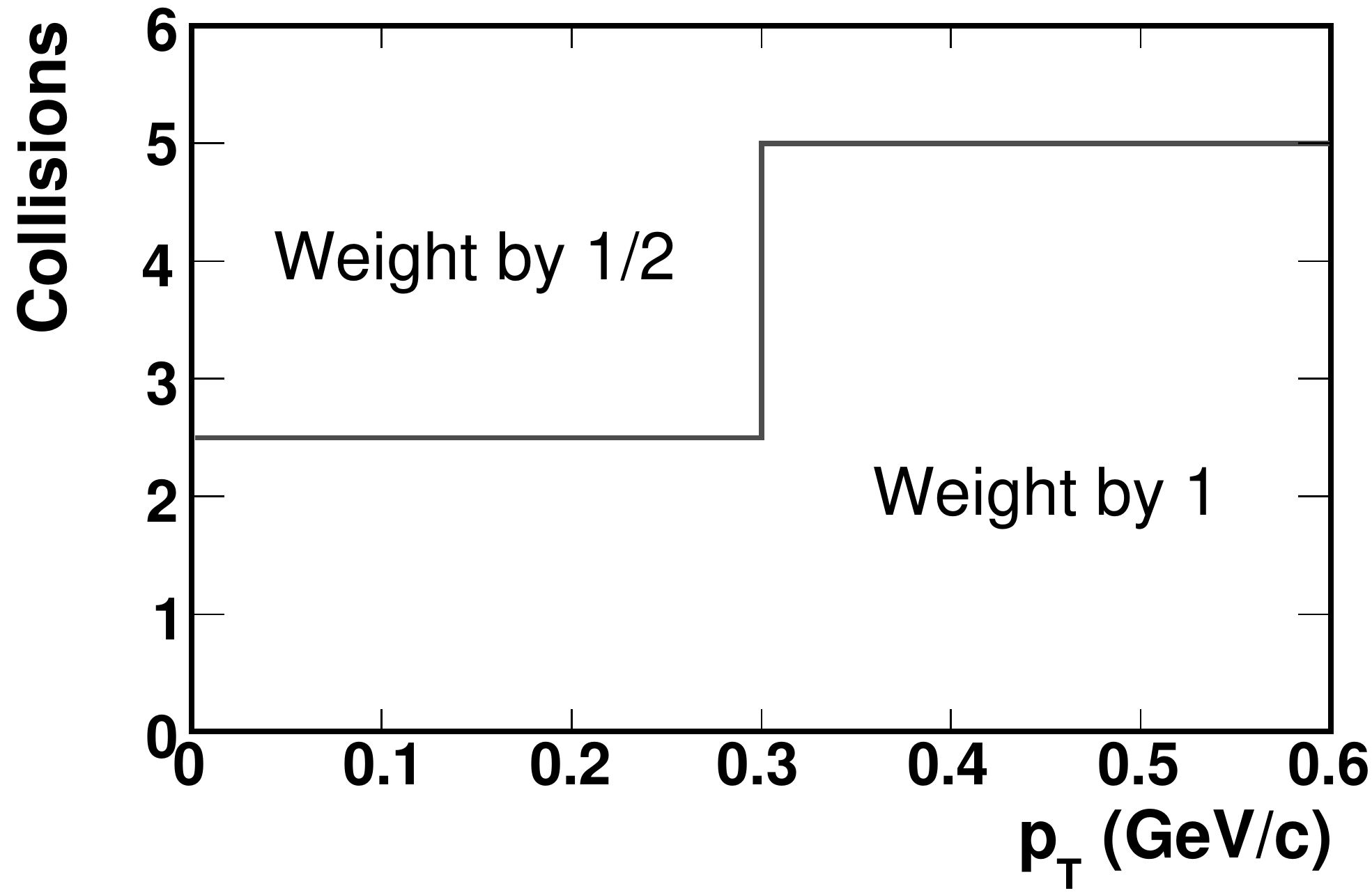}
   }
   \caption{   \label{ana:fig:evtNormEx}
      A simple example of the event normalization procedure for a
      detector which measures particles having $\pt<0.3~\mom$ in only
      one of two possible magnet polarities.}
\end{figure}

A very simple example can illustrate this subtle point. Imagine a nearly
perfect detector that has 100\% efficiency inside its acceptance, but
can not observe particles below $\pt=0.3~\mom$ when its magnet polarity
is positive. This detector is then used to measure the yield of hadrons
from five interactions. These raw measurements are shown in
\tab{ana:tab:evtNormExMeas}. Thus, the raw $\pt$ spectrum of these
collisions is presented in \fig{ana:fig:evtNormExRawTrks} for $\pt$ bins
of 0.1~{\mom}. It is clear from \tab{ana:tab:evtNormExMeas} that the
final yield of positive hadrons should be 2/5 in the $0.2\to0.3$ bin,
and 1/5 in the higher bins. However, this result is not obtained by
simply dividing the raw $\pt$ spectrum shown in
\fig{ana:fig:evtNormExRawTrks} by the number of collisions (5). The
reason is that for $\pt<0.3~\mom$, only one of the two possible magnet
polarities will allow a particle to be detected. Thus, when normalizing
the raw spectrum for $\pt<0.3~\mom$, the number of collisions should be
weighted by one half, as shown in \fig{ana:fig:evtNormExEvts}. Dividing
the raw $\pt$ spectrum shown in \fig{ana:fig:evtNormExRawTrks} by the
number of collisions versus $\pt$ shown in \fig{ana:fig:evtNormExEvts}
then produces the correct yield. While this is a simple example, an
analogous procedure was used to normalize the hadron yields measured in
this thesis.

%---------------------------------------------------------------
\subsection{Statistical Errors}
\label{ana:spec:staterr}
%---------------------------------------------------------------

To find the statistical error of the average yield given by
\eq{ana:eq:aveyield}, it was necessary to treat each measured collision
as a sampling of two random (but correlated) variables: the corrected
number of tracks per event, $t$, and the corrected number of collisions
per event, $n$. Note that the average of $t/n$ is equal to $t_{\rm
tot}/n_{\rm tot}$. The error on the mean of the ratio $t/n$ in bin $B$
was then calculated in the usual way,

\begin{equation}
   \label{ana:eq:staterr}
\sigma\prn{\ave{Y_B}} = \frac{\ave{Y_B}}{\sqrt{N}}
   \sqrt{\frac{\sigma^2_t}{\ave{t}^2} + \frac{\sigma^2_n}{\ave{n}^2}
      - 2 \frac{\cov(t, n)}{\ave{t} \ave{n}}}
\end{equation}

\noindent%
where $N$ is the (uncorrected) number of measured collisions. The mean
and standard deviation of $n$ and $t$ were calculated by

\begin{align}
\ave{n} &= \frac{1}{N} \sum_{e=1}^{N}\prn{n_e}
   & \sigma^2_n &= \frac{1}{N} \sum_{e=1}^{N}\prn{n_e - \ave{n}}^2
   & \text{where} && n_e &= \frac{1}{\epsilon_{e}} \label{ana:eq:nerr}\\
\ave{t} &= \frac{1}{N} \sum_{e=1}^{N}\prn{t_e}
   & \sigma^2_t &= \frac{1}{N} \sum_{e=1}^{N}\prn{t_e - \ave{t}}^2
   & \text{where} && t_e &= \frac{1}{\epsilon_{e}}
      \sum^{T_e}_{j=1} \prn{\frac{1}{w_j}} \label{ana:eq:terr}
\end{align}

\noindent%
The covariance was calculated by

\begin{equation}
\cov(t, n) = \frac{1}{N} \sum_{e=1}^{N} \prn{n_e - \ave{n}}
      \prn{t_e - \ave{t}} \label{ana:eq:covtn}
\end{equation}

\noindent%
Calculation of these variances was simplified by exploiting the fact that

\begin{align}
\sigma^2_x &= \frac{1}{N} \sum_{e=1}^{N}\prn{x_e - \ave{x}}^2
 = \frac{1}{N} \sum_{e=1}^{N}\prn{x_e^2 - 2 x_e \ave{x} + \ave{x}^2}\notag\\
 &= \frac{1}{N} \sum_{e=1}^{N}\prn{x_e^2} - 2 \ave{x}
   \prn{\frac{1}{N} \sum_{e=1}^{N} x_e} + \ave{x}^2
 = \ave{\smash[t]{x^2}} - 2 \ave{x}^2 + \ave{x}^2\notag\\
\sigma^2_x &= \ave{\smash[t]{x^2}} - \ave{x}^2 \label{ana:eq:sig}\\
\intertext{and}
\cov(t, n) &= \frac{1}{N} \sum_{e=1}^{N} \prn{n_e - \ave{n}}
   \prn{t_e - \ave{t}}
 = \frac{1}{N} \sum_{e=1}^{N} \prn{n_e t_e - t_e \ave{n} - \ave{t} n_e + 
   \ave{t} \ave{n}}\notag\\
 &= \ave{n t} - \ave{n} \frac{1}{N} \sum_{e=1}^{N} \prn{t_e}
   - \ave{t} \frac{1}{N} \sum_{e=1}^{N} \prn{n_e} + \ave{t} \ave{n}\notag\\
\cov(t, n)  &= \ave{n t} - \ave{n} \ave{t} \label{ana:eq:cov}
\end{align}

\noindent%
where $x$ can be either $t$ or $n$. \Eqs{ana:eq:sig}{ana:eq:cov} allowed
the variances to be computed by summing different quantities in a single
processing of the data. Without these identities, it would have been
necessary to analyze the data once to calculate the mean values
$\ave{n}$ and $\ave{t}$, and then to analyze the data a second time to
calculate the standard deviations from these means.

Note that for the special case in which the event selection efficiency
is constant and the number of particles produced follows a Poisson
distribution, \eq{ana:eq:staterr} reduces to the standard form
\mbox{$\sigma\prn{\ave{Y_B}} = \sqrt{\ave{Y_B} / N}$}. To see this, take
\mbox{$\epsilon_{e} \rightarrow 1$}. Then

\begin{align*}
n_e = \frac{1}{\epsilon_{e}} &= 1 & \sigma^2_n 
   &= 0 & \cov(t, n) &= 0\\
\ave{n} = \frac{1}{N} \sum_{e=1}^{N}\prn{n_e} &= 1
   & \ave{Y_B} &= \ave{\frac{t}{n}} = \ave{t}  & \sigma^2_t &= \ave{t}\\
\end{align*}

\noindent%
so

\begin{align*}
\sigma\prn{\ave{Y_B}} &= \frac{\ave{Y_B}}{\sqrt{N}}
   \sqrt{\frac{\sigma^2_t}{\ave{t}^2} + \frac{\sigma^2_n}{\ave{n}^2}
      - 2 \frac{\cov(t, n)}{\ave{t} \ave{n}}}\\
   &= \frac{\ave{Y_B}}{\sqrt{N}}
      \sqrt{\frac{\ave{t}}{\ave{t}^2}}
   = \sqrt{\frac{\ave{Y_B}^2}{N \ave{t}}}\\
\sigma\prn{\ave{Y_B}} &= \sqrt{\frac{\ave{Y_B}}{N}}
   \quad \text{for} \quad \epsilon_{e} = 1
\end{align*}

\noindent%
which is the error on the number of particles per collision in a
particular bin, when each collision is weighted equally and when the
number of particles per collision in the bin follows a Poisson
distribution.

%---------------------------------------------------------------
\subsection{Momentum Resolution and Binning}
\label{ana:spec:momres}
%---------------------------------------------------------------

Once the spectra had been produced, a final correction was necessary.
This correction accounted for the momentum resolution of the tracking
algorithm and for the varying $\pt$ bin sizes. Because the invariant
yield of charged hadrons followed a steeply falling curve, errors in the
reconstructed momentum of a particle led to an excess of particles at
higher $\pt$. To see this, consider an example in which measurements are
made of an extremely steeply falling slope. Suppose that the yield were
measured in only two bins, with bin $A$ containing particles with
$\pt<1~\mom$ and bin $B$ containing particles with $\pt>1~\mom$. Suppose
further that after measuring 50 collisions, bin $A$ contained 400
particles, giving $Y_A=8$ and bin $B$ contained 10, giving $Y_B=0.2$. If
an additional particle were produced with a transverse momentum of
$0.9~\mom$, then with an ideal detector, it would be counted in bin $A$,
resulting in $Y_A=8.02$; a difference of only 0.25\%. However, if the
same particle were measured inaccurately as having $\pt=1.1~\mom$, then
the particle would be counted in bin $B$, resulting in $Y_B=0.22$; a
difference of 10\%!

\begin{figure}[t]
   \centering
   \subfigure[Reconstructed Momentum]{
      \label{ana:fig:momreso}
      \includegraphics[width=0.6\linewidth]{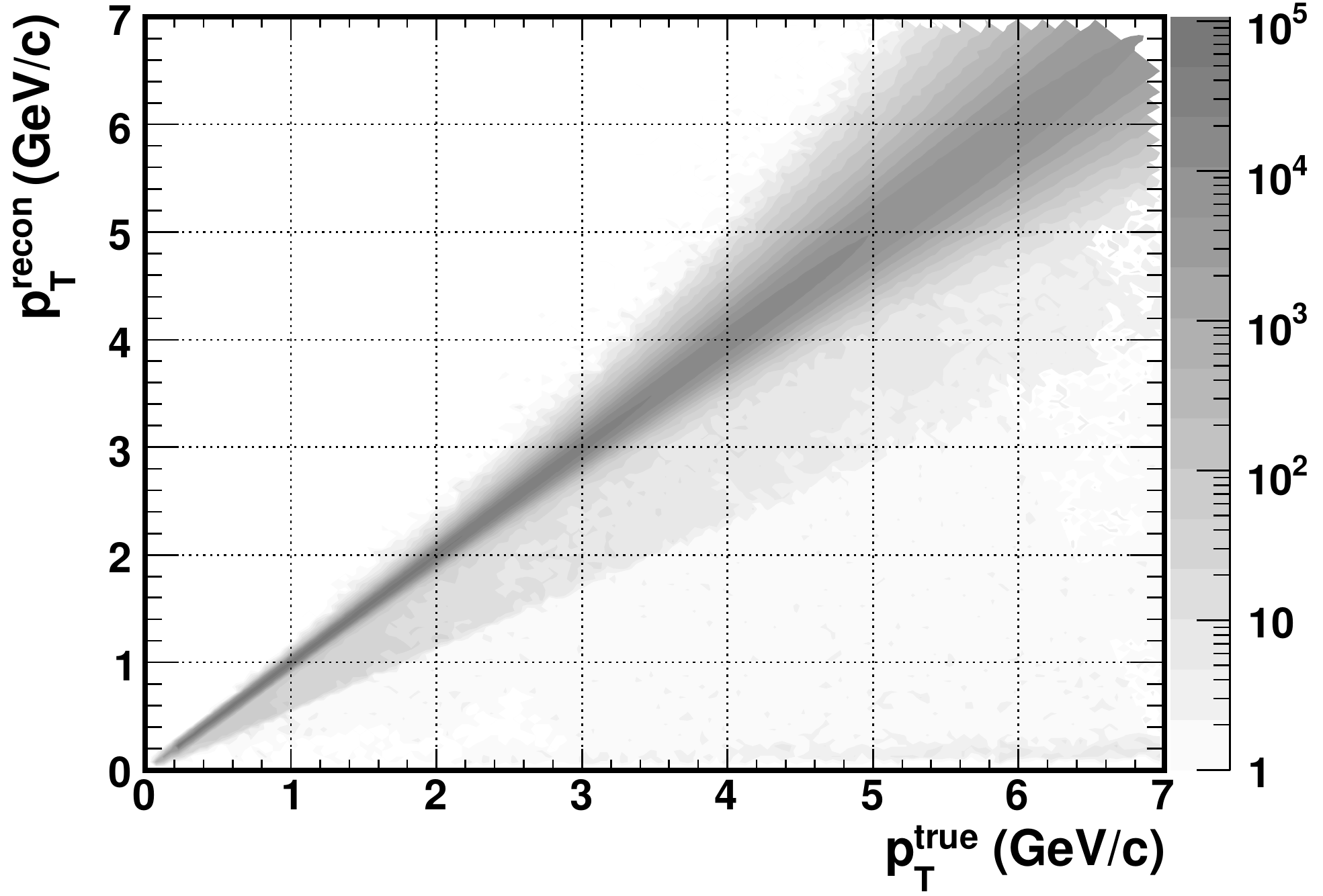}
   }
   \mbox{
      \subfigure[$\pt\approx1~\mom$]{
         \label{ana:fig:momresExDist0}
         \includegraphics[width=0.4\linewidth]{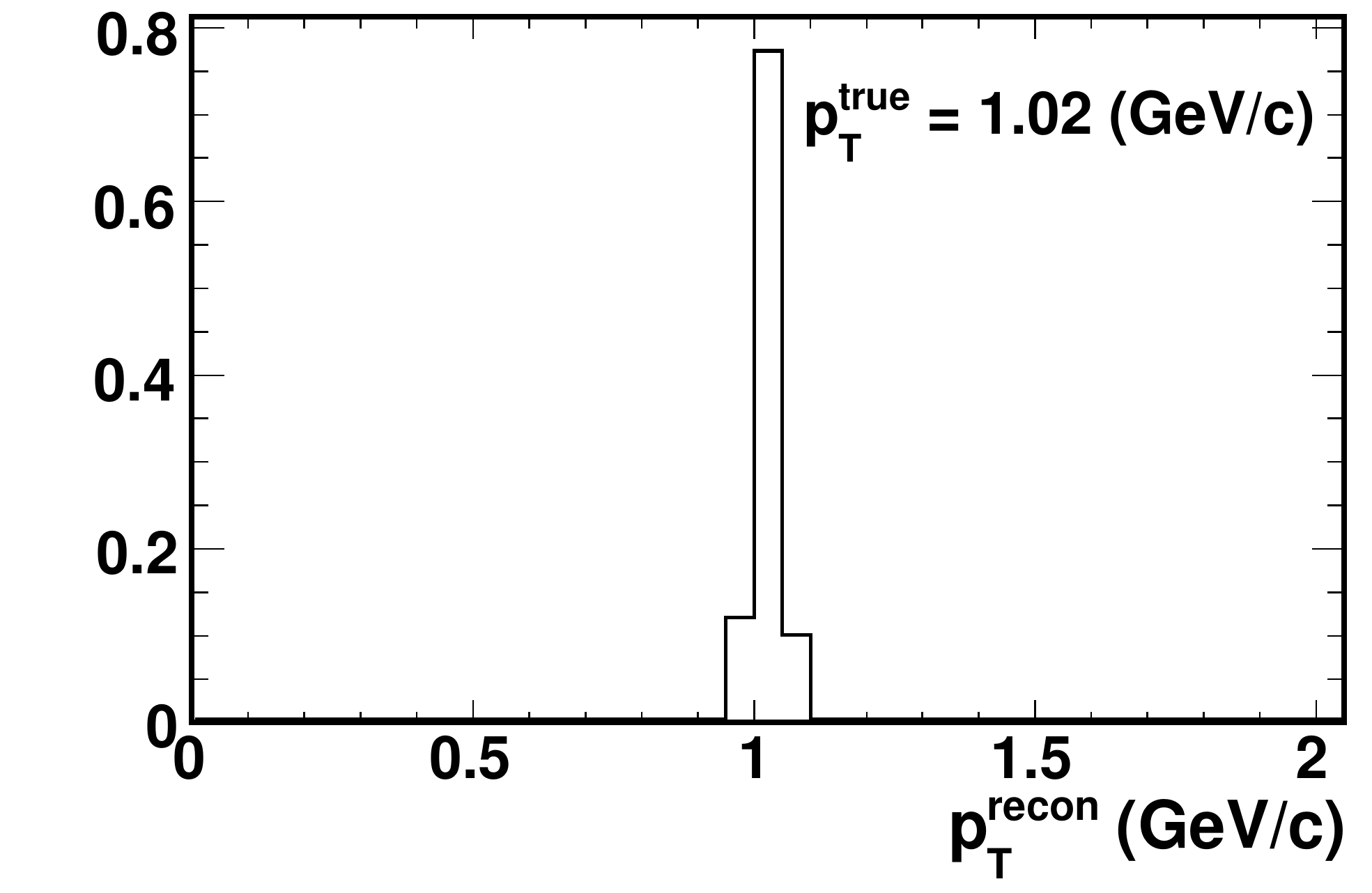}
      }
      \subfigure[$\pt\approx3~\mom$]{
         \label{ana:fig:momresExDist1}
         \includegraphics[width=0.4\linewidth]{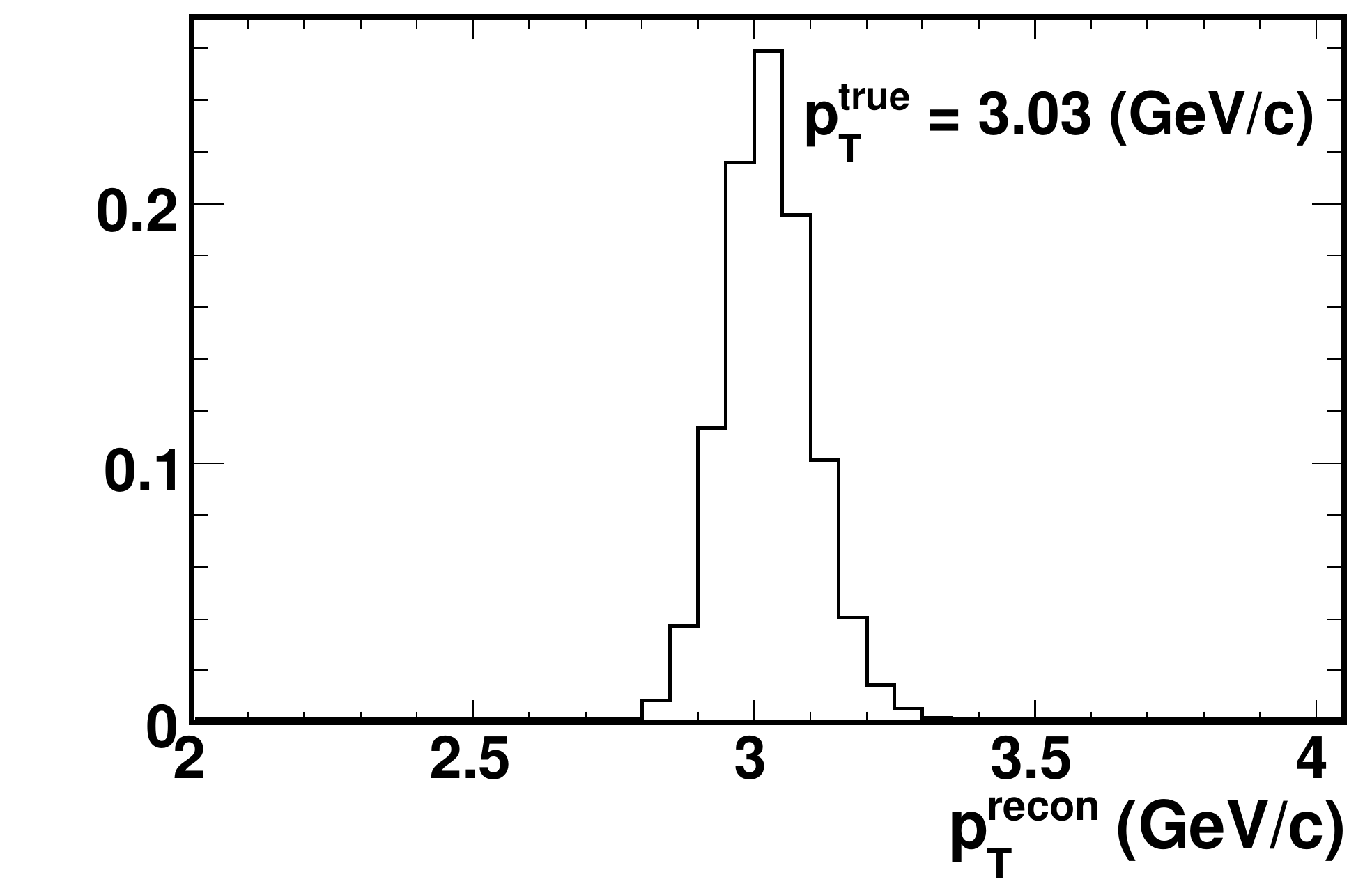}
      }
   }
   \caption{   \label{ana:fig:momresoAndEx}
      \subref{ana:fig:momreso}~The reconstructed transverse momentum of
      simulated pions as a function of the true $\pt$ of the particle.
      \subref{ana:fig:momresExDist0}~The reconstructed momentum
      distribution for particles having $\pt\approx1~\mom$.
      \subref{ana:fig:momresExDist1}~The reconstructed momentum
      distribution for particles having $\pt\approx3~\mom$.}
\end{figure}

To correct for this effect, the momentum resolution of the tracking
procedure was estimated using the single track simulations described in
\sect{ana:spec:acceff}. From these simulations, the reconstructed
momentum could be compared to the true momentum of the particle, as
shown in \fig{ana:fig:momreso}. Note that the histogram in this figure
was very finely binned, and that bins could be combined evenly to
produce the same $\pt$ binning as used in the spectra analysis.

\begin{figure}[t!]
   \centering
   \subfigure[Fit to Yield]{
      \label{ana:fig:MRfityield}
      \includegraphics[width=0.4\linewidth]{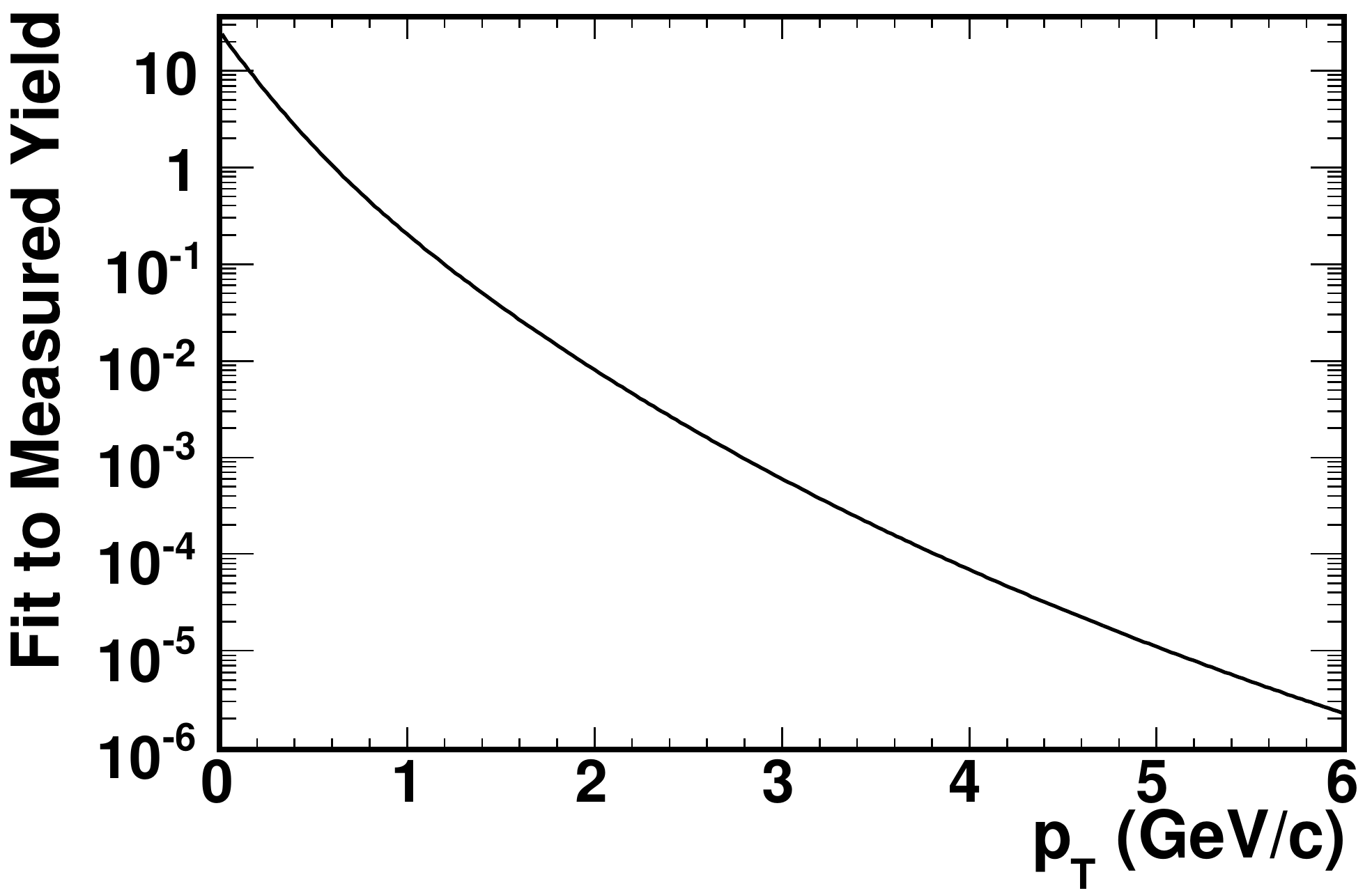}
   }
   \subfigure[Smeared Yield]{
      \label{ana:fig:MRsmeared}
      \includegraphics[width=0.4\linewidth]{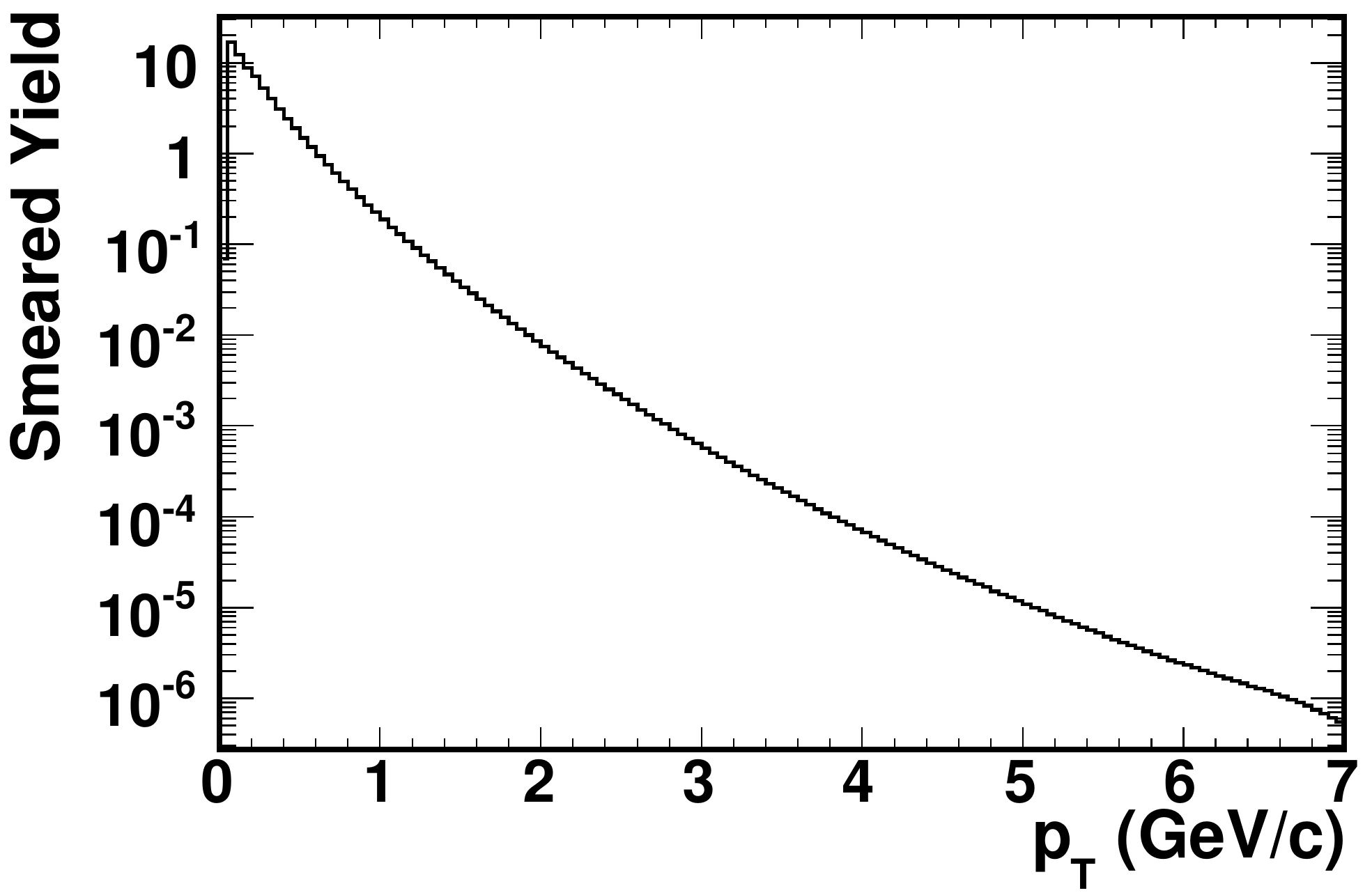}
   }
   \subfigure[Rebinned, Smeared Yield]{
      \label{ana:fig:MRrebinned}
      \includegraphics[width=0.4\linewidth]{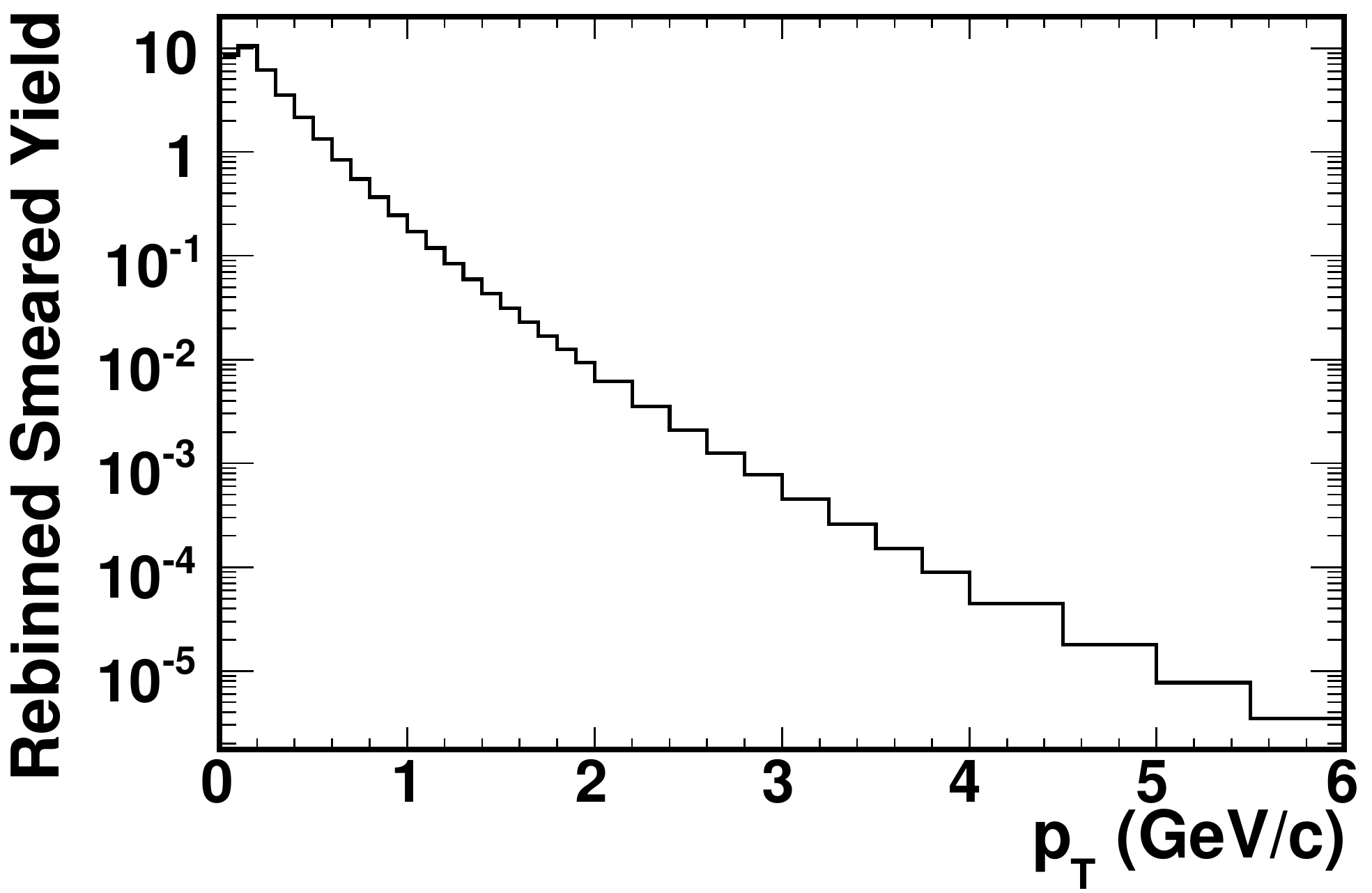}
   }
   \subfigure[Correction]{
      \label{ana:fig:MRcorr}
      \includegraphics[width=0.4\linewidth]{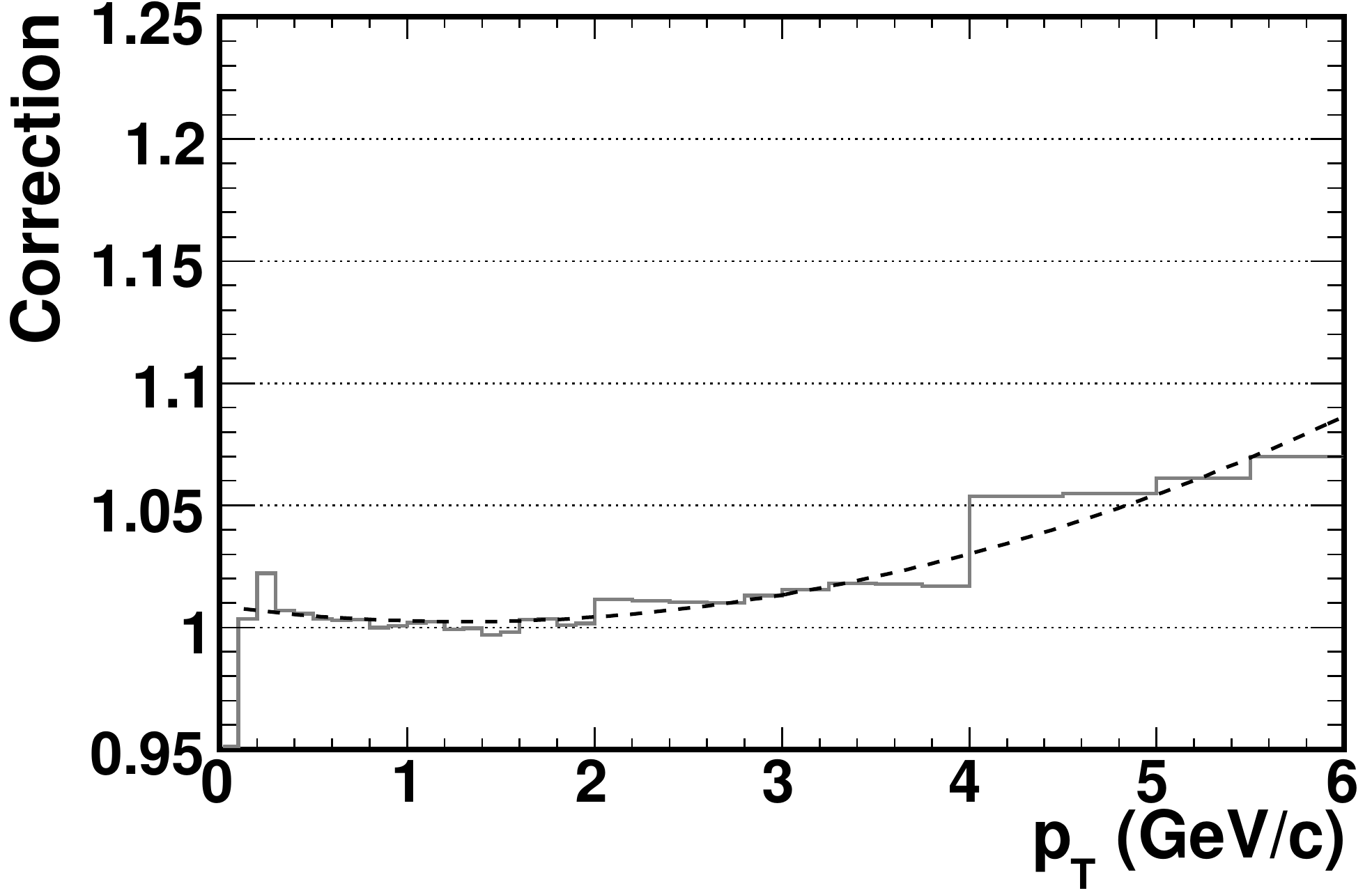}
   }
   \caption{   \label{ana:fig:momrescor}
      Example of the steps necessary to determine the momentum
      resolution and binning correction. \subref{ana:fig:MRfityield}~A
      sample fit to some measured yield (data points not shown).
      \subref{ana:fig:MRsmeared}~The smeared yield obtained by
      normalizing the momentum resolution to the measured yield.
      \subref{ana:fig:MRrebinned}~The smeared yield rebinned to match
      the binning of the data. \subref{ana:fig:MRcorr}~The correction,
      obtained by dividing
      \subref{ana:fig:MRrebinned}~by~\subref{ana:fig:MRfityield}.}
\end{figure}

The correction was determined in several steps. First, the measured
$\pt$ spectra was fit with a smooth function, an example of which can be
seen in \fig{ana:fig:MRfityield}. Next, the reconstructed $\pt$
distribution in each $\pttrue$ bin of \fig{ana:fig:momreso} was
independently normalized. The distribution in each $\pttrue$ bin was
then scaled up by the value of the yield at the center of the bin, as
determined by the spectra fit. Next, this weighted two-dimensional
histogram was projected onto the $\ptrecon$ axis. This resulted in a
hadron spectra distribution that was smeared by the momentum resolution,
as seen in \fig{ana:fig:MRsmeared}. Bins in this distribution were then
combined in such a way as to recreate the same binning used in the
analysis, as shown in \fig{ana:fig:MRrebinned}. Finally, this smeared
and re-binned distribution was then divided by the original fit to the
spectra, as shown in \fig{ana:fig:MRcorr}. This histogram was then fit
with a quadratic function to remove statistical fluctuations due to the
number of tracks simulated. As can be seen in \fig{ana:fig:MRcorr}, this
fit was smooth, and thus removed some of the correction that accounted
for abrupt changes in bin size (note the jump at $\pt=4~\mom$). However,
the discrepancy was less than 3\%, and the systematic error on this
correction was larger than that, as will be discussed in the next
section.

%---------------------------------------------------------------
\subsection{Systematic Errors}
\label{data:syserr}
%---------------------------------------------------------------

%---------------------------------------------------------------
\subsubsection{Systematics of the Yield Measurements}
\label{data:syserr:spectra}
%---------------------------------------------------------------

Uncertainties in the precision of the spectra measurement originated
from the number of observed collisions and particles, and were estimated
by the statistical errors discussed in \sect{ana:spec:staterr}.
Uncertainties in the \emph{accuracy} of the measurement originated from
the methods used to determine the various corrections. The largest such
systematic uncertainty was associated with the largest correction: the
acceptance and efficiency functions discussed in \sect{ana:spec:acceff}.
To estimate this systematic error, the yield of {\dAu} collisions was
reconstructed using several different subsets of the data. With perfect
corrections, the yield measured in each subset of the data would be
identical. Thus, the amount by which the yield varies from subset to
subset gives an estimate of the systematic error associated with the
correction.

\begin{figure}[p]
   \centering
   \subfigure[Acceptance/Efficiency]{
      \label{data:fig:acceffSysErr}
      \includegraphics[width=0.4\linewidth]{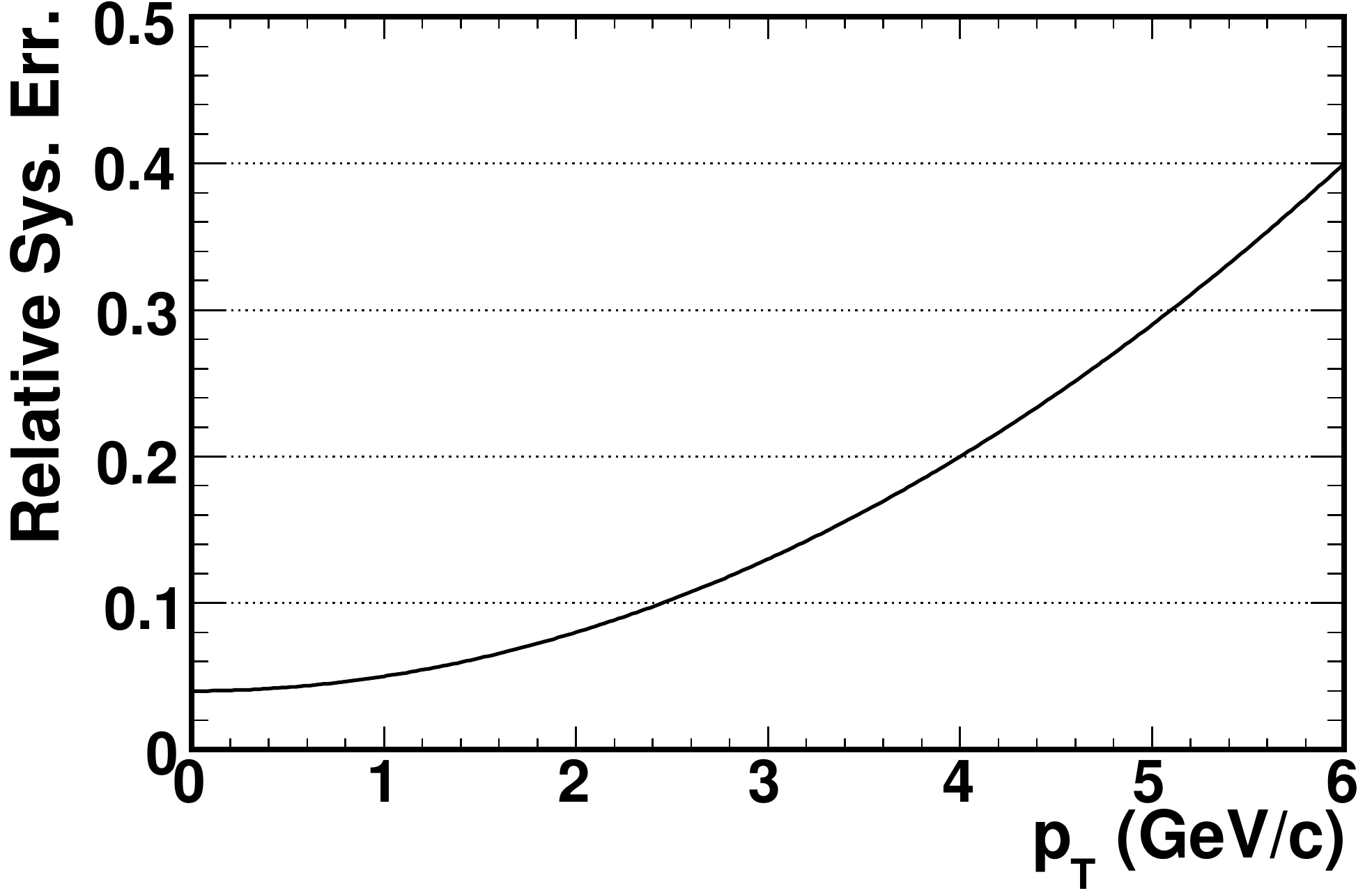}
   }
   \subfigure[Momentum Resolution]{
      \label{data:fig:momresSysErr}
      \includegraphics[width=0.4\linewidth]{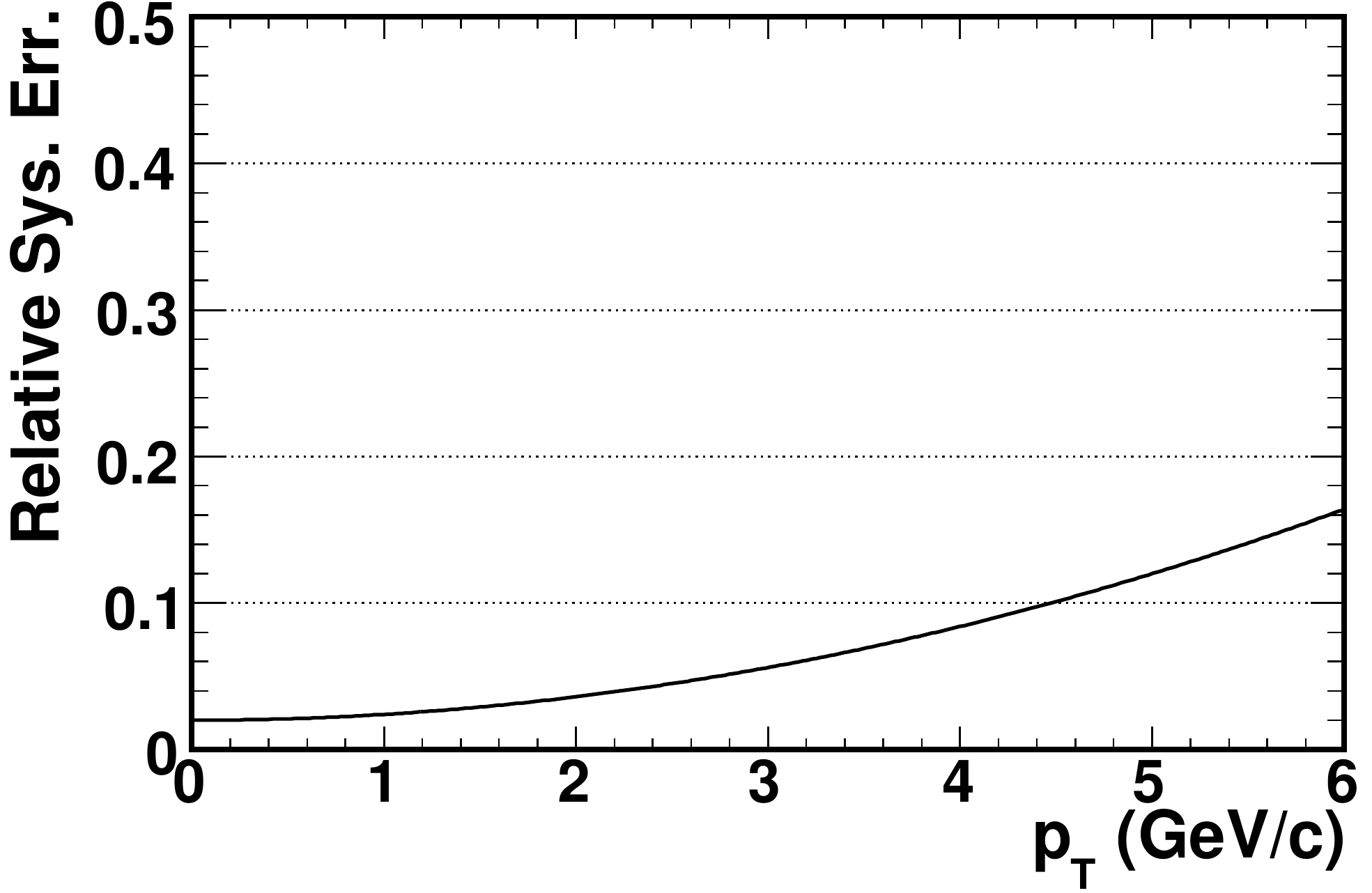}
   }
   \subfigure[Secondaries]{
      \label{data:fig:secondSysErr}
      \includegraphics[width=0.4\linewidth]{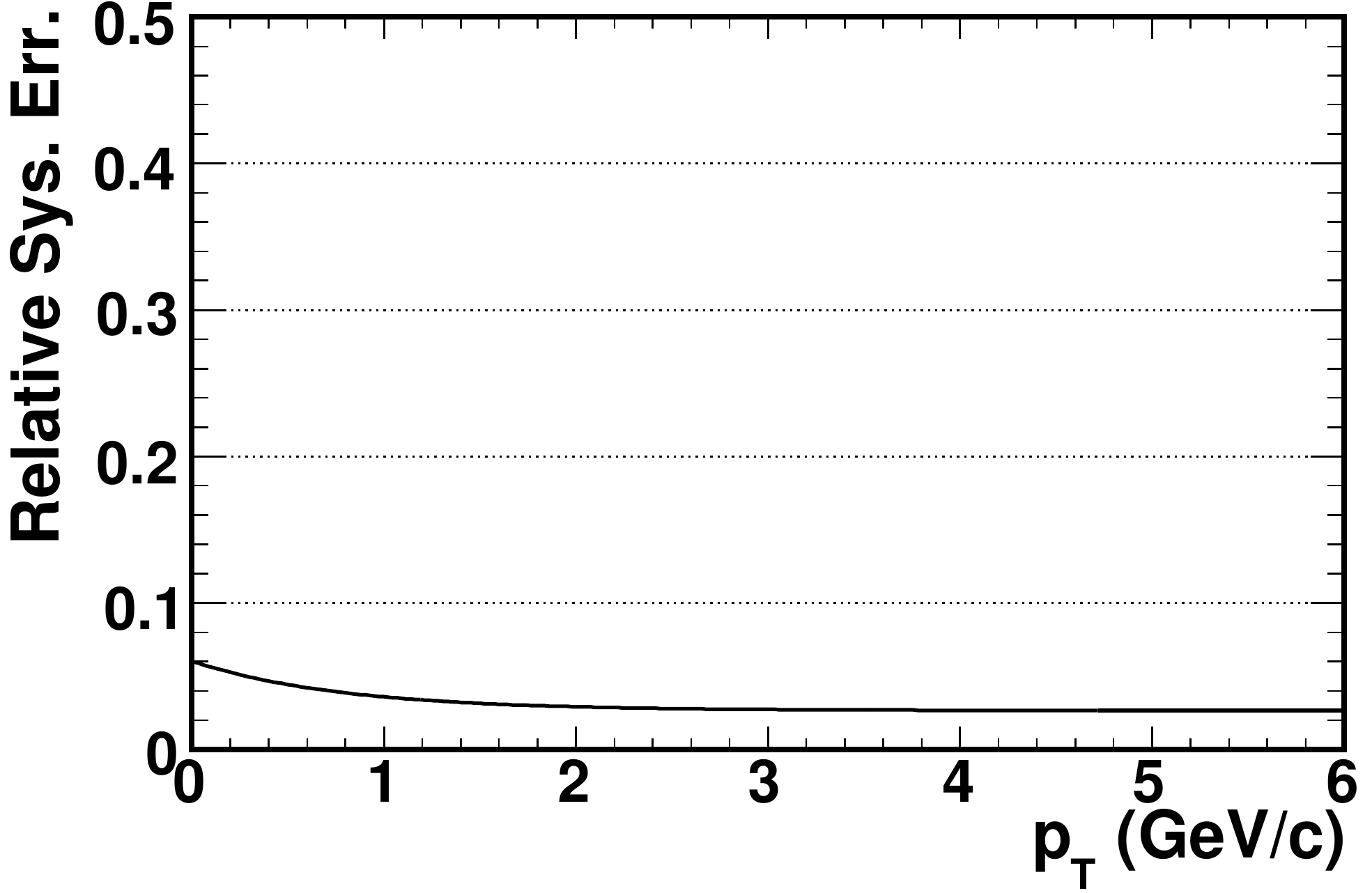}
   }
   \subfigure[Dead/Hot Channels]{
      \label{data:fig:deadhotSysErr}
      \includegraphics[width=0.4\linewidth]{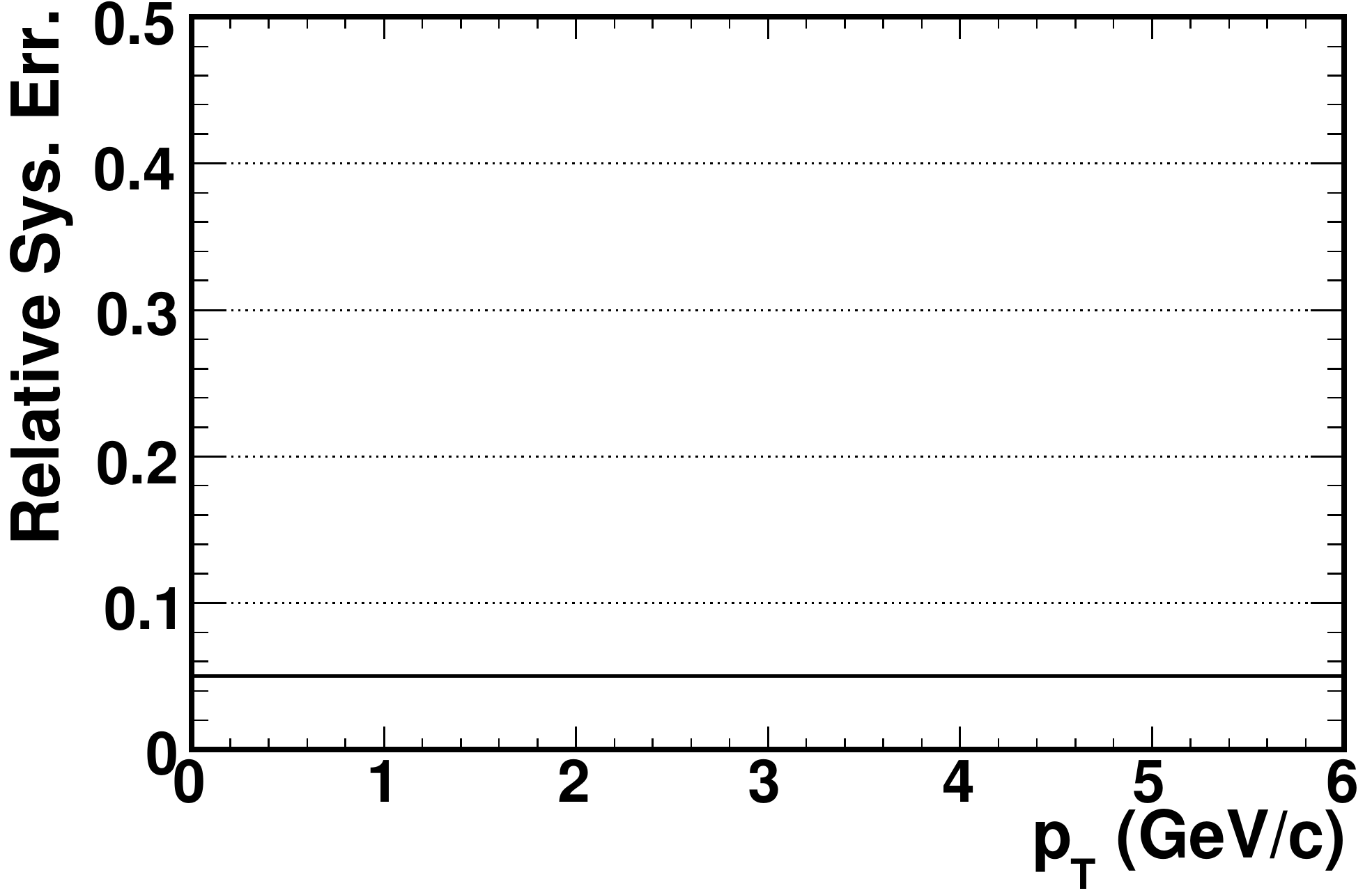}
   }
   \subfigure[Ghosts]{
      \label{data:fig:ghostSysErr}
      \includegraphics[width=0.4\linewidth]{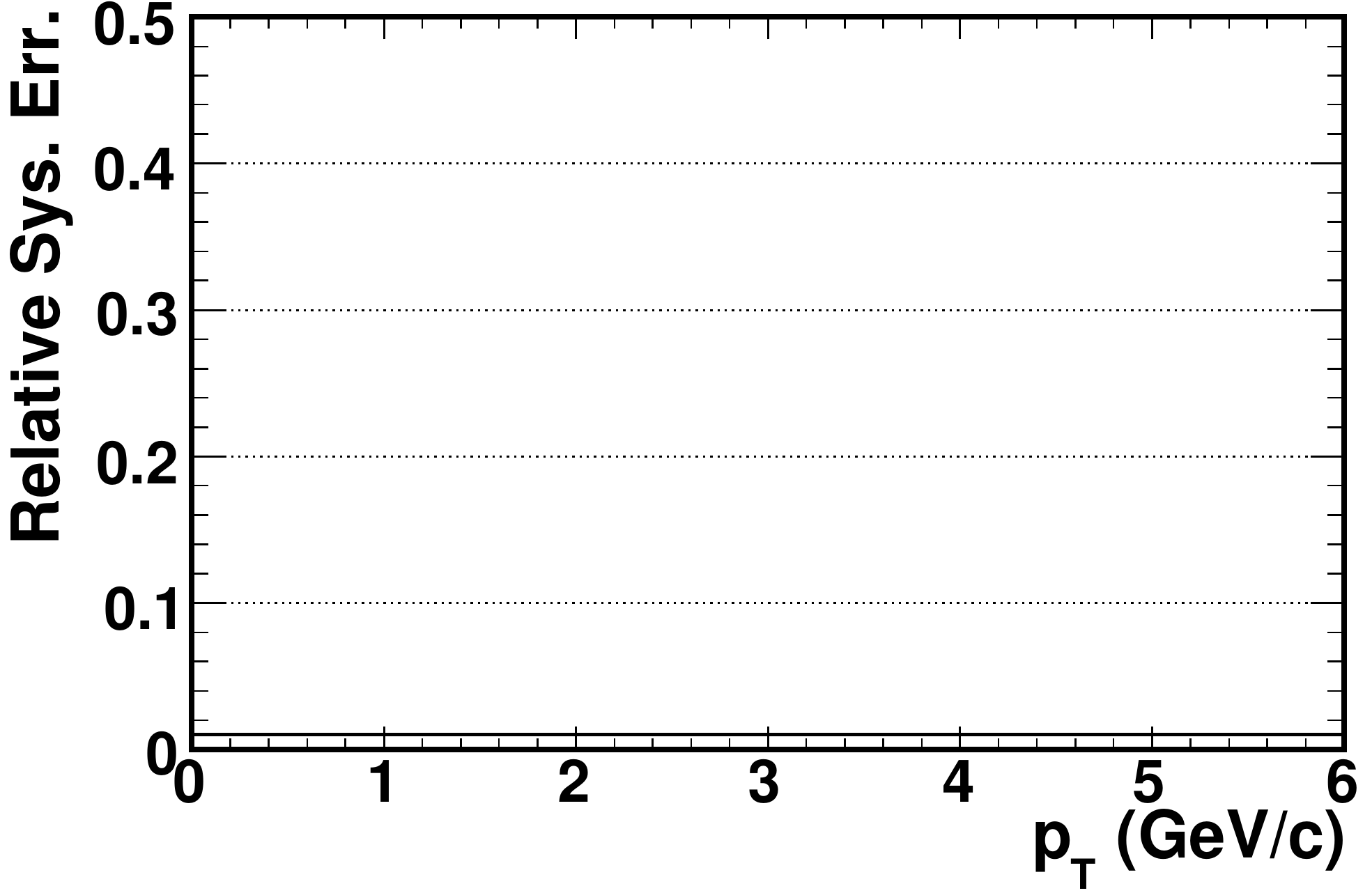}
   }
   \subfigure[Deuteron Tagging]{
      \label{data:fig:tagSpecSysErr}
      \includegraphics[width=0.4\linewidth]{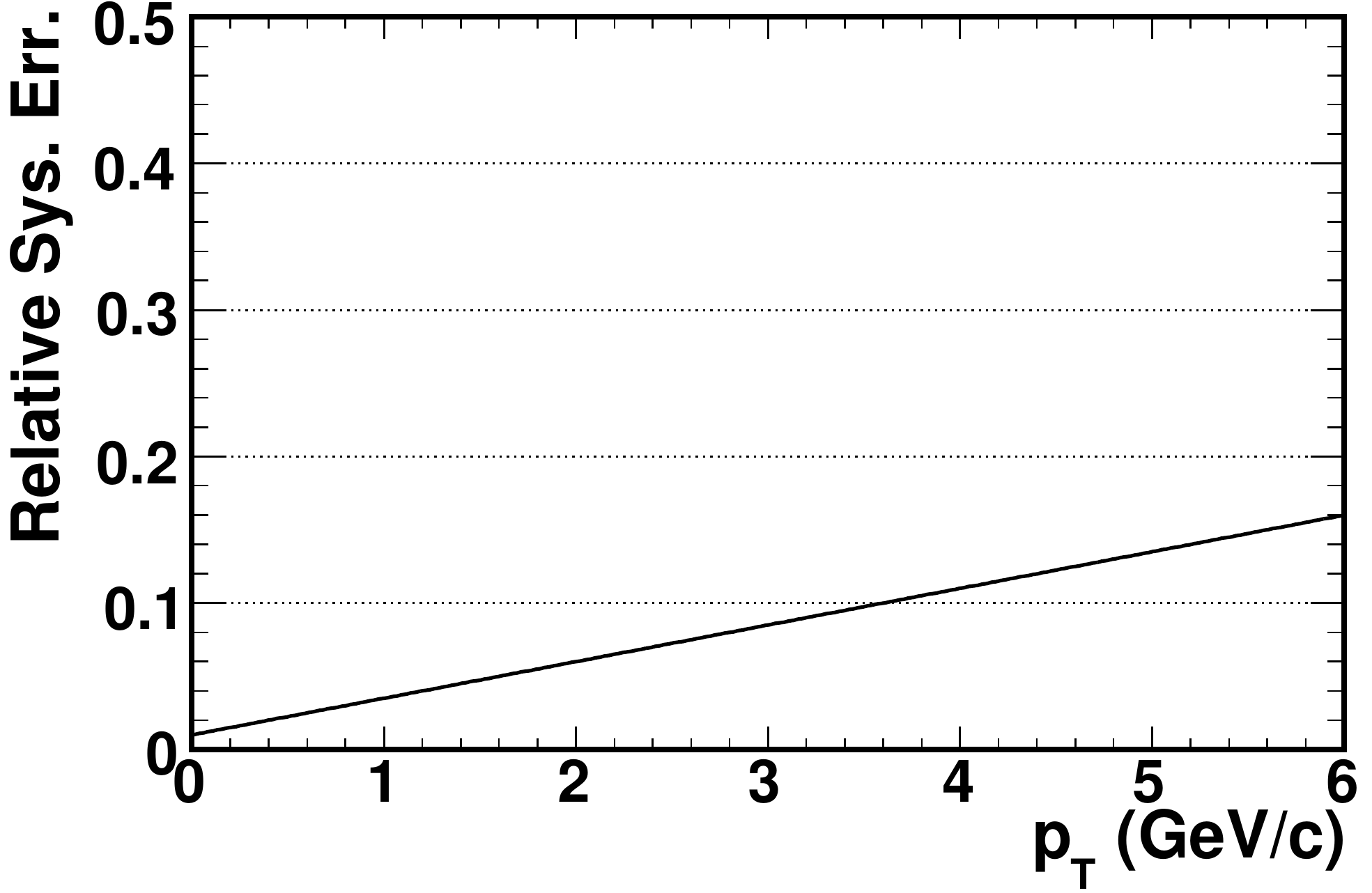}
   }
   \subfigure[Total d+Au Sys. Err.]{
      \label{data:fig:totalSysErr}
      \includegraphics[width=0.4\linewidth]{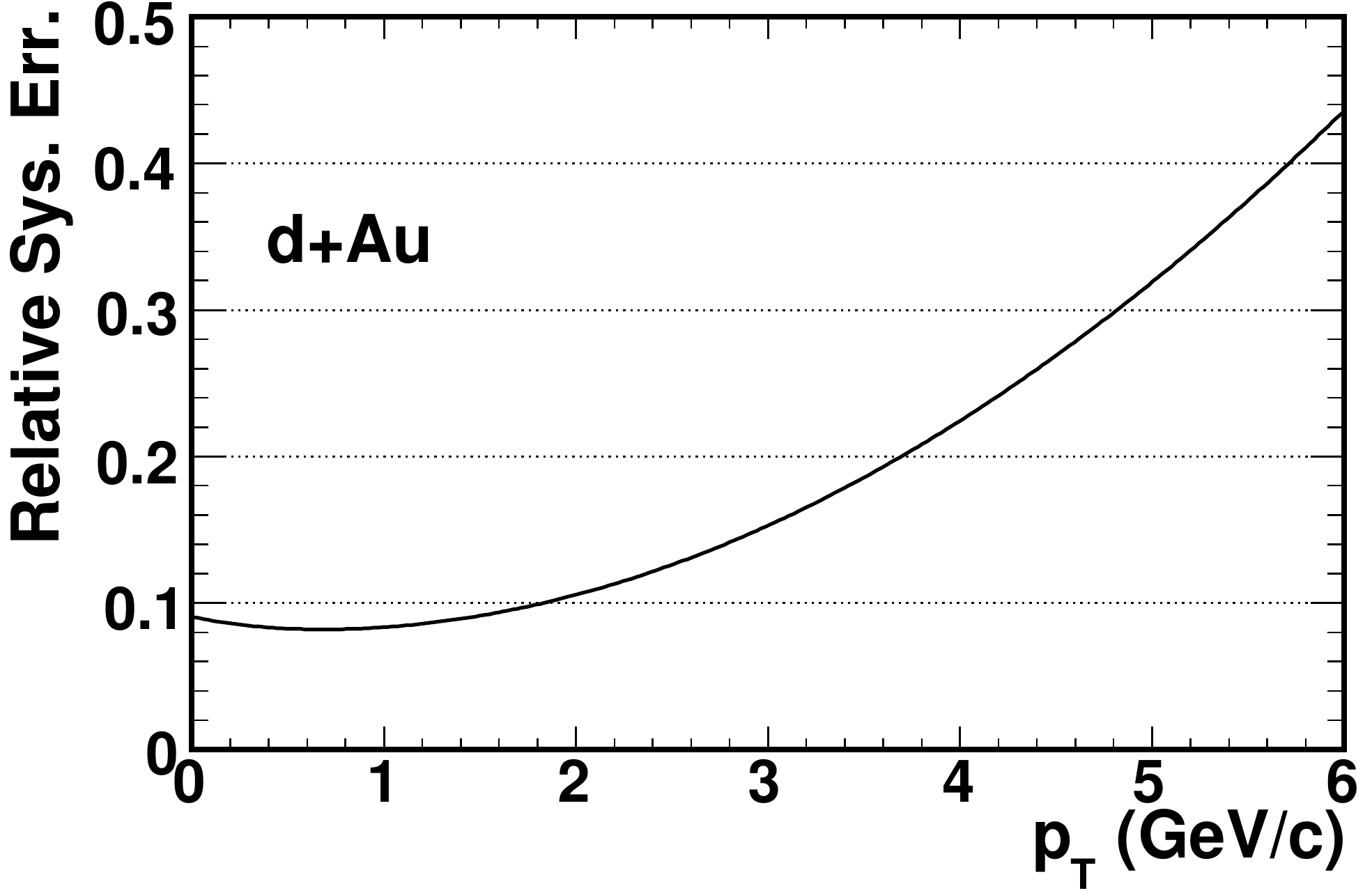}
   }
   \subfigure[Total N+Au Sys. Err.]{
      \label{data:fig:totalTagSysErr}
      \includegraphics[width=0.4\linewidth]{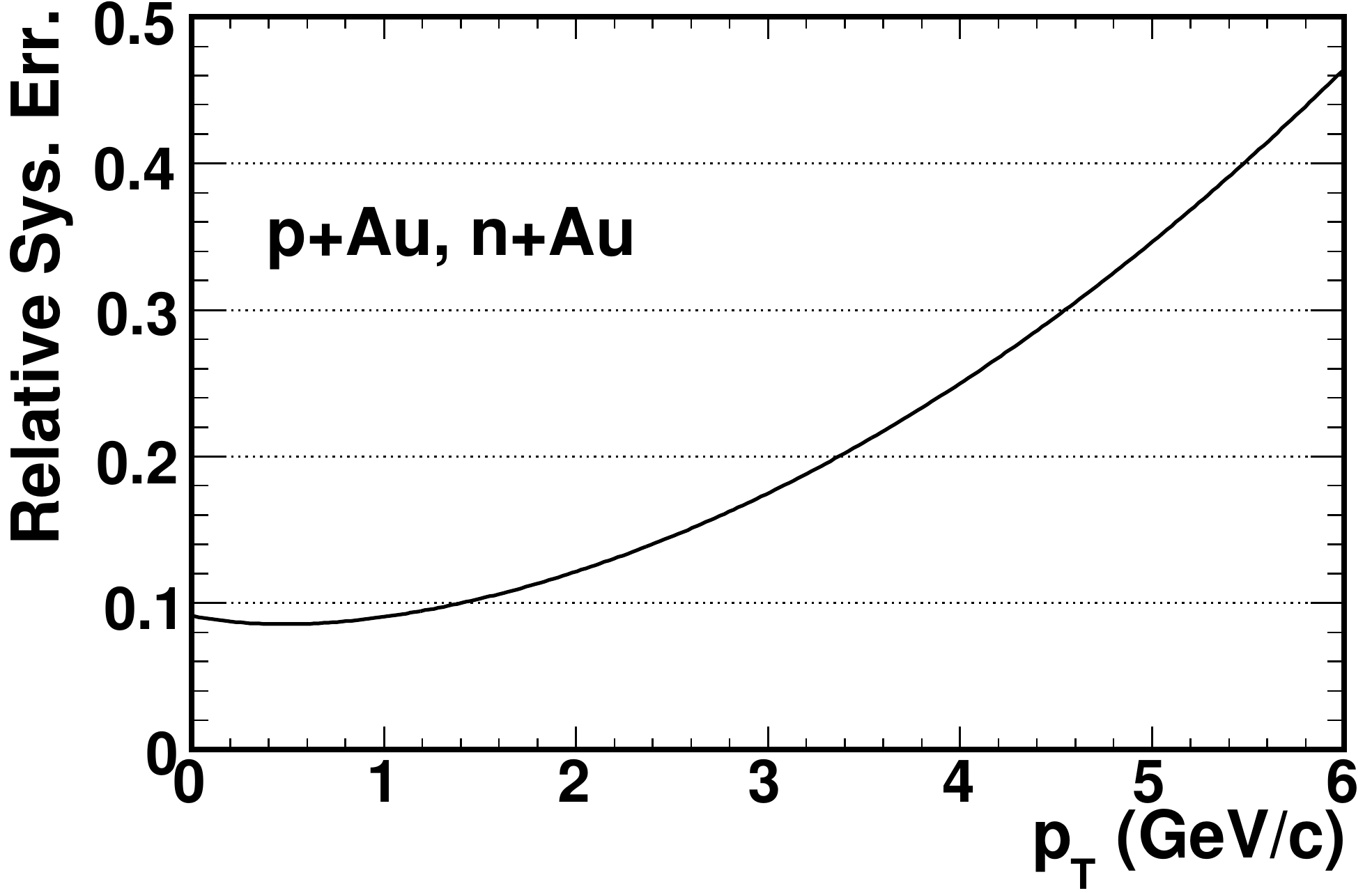}
   }
   \caption{   \label{data:fig:syserrs}
      Contributions to the systematic error associated with uncertainty
      in the various corrections. All errors are relative and represent
      a 90\%~confidence level. The total systematic error is the
      quadrature sum of the individual contributions.}
\end{figure}

For example, the yield of positive hadrons measured using only data
taken with the {\phob} magnet at positive polarity (B$+$) was compared
to the same yield measured with B$-$ data. Since the bending direction
of particles in these two measurements were different, the acceptance
and efficiency corrections would be different. Therefore, variations in
the positive hadron yield measured by these two subsets of the data
could be associated with systematic uncertainties in the acceptance and
efficiency correction. A host of such studies were done, and the
resulting systematic error, similar to that used in~\cite{jaysThesis},
is shown in \fig{data:fig:acceffSysErr} as a function of $\pt$.

The systematic uncertainty in the dead and hot channel correction was
estimated in a similar way, by comparing the yields measured separately
in each Spectrometer arm. In {\phob} previous studies, discrepancies
between the yield measured in each arm were not understood,
see~\cite{jaysThesis}. These effects have since been traced to a
significant difference in the number of malfunctioning channels in the
two arms, as discussed in \sect{ana:spec:deadchan}. With the corrections
applied separately to the two arms, discrepancies in the yields were
reduced from \mbox{$\sim10\%$} to \mbox{$\lesssim3\%$}. The systematic
error shown in \fig{data:fig:deadhotSysErr} represents both
uncertainties associated with the correction and with the (small)
differences in the yield measured separately with each Spectrometer arm.

For some corrections, such studies could not easily be performed. In
each of these cases, the uncertainty was taken to be of the same
magnitude as the correction itself. This was true for the momentum
resolution and ghost corrections, see
\figs{data:fig:momresSysErr}{data:fig:ghostSysErr} respectively. For the
secondary correction, the systematic error was larger than the
correction at higher values of $\pt$ to take into account the low
statistics used to estimate the correction in this region, as shown
in \fig{data:fig:secondSysErr}.

Finally, an uncertainty in the yield measured for nucleon-nucleus
interactions was estimated. This was done by shifting the \acs{d-PCAL}
and \acs{d-ZDC} energy cuts used to tag {\pAu} and {\nAu} collisions
(see~\sect{recon:nuctag}). If the minimum energy used to identify a
proton or a neutron in one of the calorimeters was increased, it should
affect only the number of interactions (and therefore particles) used to
make the measurement. It should not have an impact on the yield. Thus,
changing the tagging cut values provided an estimate of this additional
systematic uncertainty in the yield of nucleon-nucleus interactions, and
is presented in \fig{data:fig:tagSpecSysErr}.

The total systematic error of the measured hadron yield was determined
by taking the quadrature sum of the various contributions. Thus, the
total systematic error on the yield of {\dAu} collisions, shown in
\fig{data:fig:totalSysErr}, was not the same as the systematic error on
the yield of nucleon-nucleus interactions, shown in
\fig{data:fig:totalTagSysErr}. All of these systematic errors are
thought to be conservative, and represent a 90\% confidence level.

%---------------------------------------------------------------
\subsubsection{Systematics of the Centrality Measurements}
\label{data:syserr:centrality}
%---------------------------------------------------------------

In depth studies of the accuracy with which centrality parameters like
$\Npart$ could be estimated were performed prior to this analysis, as
described in~\cite{richardsThesis}. First, the dependence on the
simulations was studied by varying the centrality efficiency and
observing the effects on the estimated centrality parameter. The degree
to which the centrality efficiency could vary was estimated by dividing
the simulated data up into vertex bins and comparing the efficiency
determined in each bin. Finally, the dependence on the deuteron wave
function was studied by comparing the results of simulations performed
using the Hulthen wave function shown in \eq{recon:eq:hulthen} to the
results of simulations done using a Woods-Saxon distribution.

\begin{table}[t]
   \begin{center}
      \begin{tabular}{|cccccc|}
\hline
 & \multicolumn{5}{c|}{Relative Systematic Error} \\
Centrality Bin & $\Ncoll$ & $\Npart$ & $\NparA$ & $\Npard$ & $\nu$ \\
\hline
0-20\%   & 6.1\%  & 7.1\%  & 7.4\%  & 5.5\%  & 6.1\%  \\
20-40\%  & 7.1\%  & 8.3\%  & 7.9\%  & 5.3\%  & 7.1\%  \\
40-70\%  & 12.0\% & 11.1\% & 13.0\% & 12.0\% & 12.0\% \\
70-100\% & 30.3\% & 22.4\% & 26.0\% & 16.7\% & 30.3\% \\
\hline
      \end{tabular}
   \end{center}
   \caption{   \label{ana:tab:centsyserr}
      Systematic errors (90\%~C.L.) on the centrality parameters found
      with \protect\acs{ERing} and \protect\acs{EOct} cuts.}
\end{table}

In addition, the dependence of the estimated centrality parameters on
the specific collision model was studied. This was done by making a
simple Glauber \ac{MC} to find the distribution of each centrality
parameter. This distribution was then scaled to match the data, the
centrality cuts used on the data were then applied and the centrality
parameter was extracted directly. This could then be compared to the
result of the full centrality technique. To make the distribution
obtained by the simple \ac{MC} more realistic, further studies were done
in which some Gaussian smearing was applied to the \ac{MC}
distributions. In addition, several different techniques were used to
scale the \ac{MC} distribution to match the data. Next, effects due to
efficiency were taken into account by (a) scaling the \ac{MC}
distribution, (b)~applying the efficiency to the distribution,
(c)~smearing the distribution, (d)~correcting for the efficiency,
(e)~applying the centrality cuts and (f)~extracting the average
centrality parameter in each bin. Finally, other collision simulations
were used to perform the full centrality procedure. The results of these
studies are shown in \tab{ana:tab:centsyserr}.

These studies were done to estimate the systematic error on centrality
parameters that were determined using similar techniques to those
presented in this thesis. In fact, it is believed that the centrality
technique described in \sect{recon:cent} should be more accurate.
Therefore, the systematic uncertainties shown in
\tab{ana:tab:centsyserr} were thought to provide a conservative estimate
of the errors relevant to this analysis.

\begin{table}[t]
   \begin{center}
      \begin{tabular}{|rccccc|}
\hline
 & \multicolumn{5}{c|}{Relative Systematic Error} \\
Centrality Bin & $\Ncoll$ & $\Npart$ & $\NparA$ & $\Npard$ & $\nu$ \\
\hline
0-20\% {\dAu} & 13.5\% & 10.7\% & 13.3\% & 5.9\% & 11.7\% \\
       {\pAu} & 27.7\% & 23.1\% & 28.0\% & 5.5\% & 28.7\% \\
       {\nAu} & 25.7\% & 23.1\% & 26.1\% & 5.5\% & 25.7\% \\
\hdashline[1pt/4pt]
20-40\% {\dAu} & 11.5\% & 11.5\% & 12.0\% & 6.6\% & 9.3\% \\
       {\pAu} & 30.8\% & 27.3\% & 31.0\% & 5.3\% & 30.8\% \\
       {\nAu} & 28.9\% & 27.3\% & 30.1\% & 5.3\% & 28.9\% \\
\hdashline[1pt/4pt]
40-70\% {\dAu} & 14.4\% & 13.7\% & 15.8\% & 13.4\% & 13.9\% \\
       {\pAu} & 29.5\% & 24.6\% & 30.9\% & 12.0\% & 29.5\% \\
       {\nAu} & 29.5\% & 24.6\% & 30.9\% & 12.0\% & 29.5\% \\
\hdashline[1pt/4pt]
70-100\% {\dAu} & 31.0\% & 23.0\% & 26.7\% & 16.7\% & 30.1\% \\
       {\pAu} & 30.0\% & 22.4\% & 26.0\% & 16.7\% & 30.0\% \\
       {\nAu} & 30.1\% & 22.5\% & 26.1\% & 16.7\% & 30.1\% \\
\hline
      \end{tabular}
   \end{center}
   \caption{   \label{ana:tab:pcalsyserr}
      Total systematic uncertainties (90\%~C.L.) associated with
      \protect\acs{EPCAL} based centrality cuts.}
\end{table}

However, additional systematic errors were needed to estimate
uncertainties in the new centrality procedure used to find parameters
(like $\Npart$) in \ac{EPCAL} bins described in \sect{recon:cent:pcal}.
These uncertainties were estimated by applying the procedure to a
variable for which the centrality parameters could be determined in the
more traditional manner. For example, the $\Npart$ estimated using
\ac{EOct} centrality cuts could be compared to the $\Npart$ estimated
using the \ac{EOct} cut values, but employing an ``\ac{EOct} from
\ac{ERing}'' method to determine the average $\Npart$ value in the
simulations. That is, centrality parameters were determined by the
method described in \sect{recon:cent:pcal}, in which the observed
correlation between \ac{EOct} and \ac{ERing} in the {\dAu} data was
applied to the simulations. Discrepancies between the centrality
parameters determined using the two methods gave an estimate of
systematic biases introduced by the new technique.

In principal, these estimates could be used to correct for biases
introduced onto the centrality parameters determined using \ac{EPCAL}
cuts. However, similar studies performed using different centrality
measures produced different estimates of the systematic biases. This was
also true for variations of existing centrality measures which were
constructed to more closely approximate the \ac{EPCAL} resolution. Since
there was no specific centrality measure that was thought to provide a
particularly accurate estimate (that is, a better estimate than any
other measure) of the \ac{EPCAL} systematics, no correction could be
performed. Instead, these studies were used to determine the additional
systematic uncertainty in the centrality parameters estimated by
\ac{EPCAL} centrality cuts. The total systematic uncertainties for
\ac{EPCAL} cuts are shown in \tab{ana:tab:pcalsyserr}.

%% file: srcs/data.tex
% $Id: data.tex,v 1.17 2006/08/18 01:05:31 cjreed Exp $
%

%---------------------------------------------------------------
\chapter{Spectra Measurements}
\label{data}
%---------------------------------------------------------------

\myupdate{$*$Id: data.tex,v 1.17 2006/08/18 01:05:31 cjreed Exp $*$}%
After the raw transverse momentum spectra had been corrected, it was
possible to measure the invariant charged hadron yield, as a function of
transverse momentum, at an average {\prap} of 0.8 units (slightly
forward in the deuteron direction). The spectra were measured in three
different collision systems, {\dAu}, {\pAu} and {\nAu}, by the
identification of nucleon-nucleus collisions described in
\sect{recon:nuctag}. For each system, the $\pt$ spectrum of positive
hadrons, $\hpos$, negative hadrons, $\hneg$, and of the average,
$\have$, was measured. Each spectrum was also measured in four bins of
centrality. To study the effects of the centrality classification
technique, six different variables were used as a centrality measure.
Thus, a total of 216 spectra were measured.

Note that data presented for the first time in this thesis have not been
reviewed by the {\phob} collaboration.

%---------------------------------------------------------------
\section{Invariant Yield Data}
\label{data:invyield}
%---------------------------------------------------------------

\begin{figure}[t]
   \begin{center}
      \includegraphics[width=\linewidth]{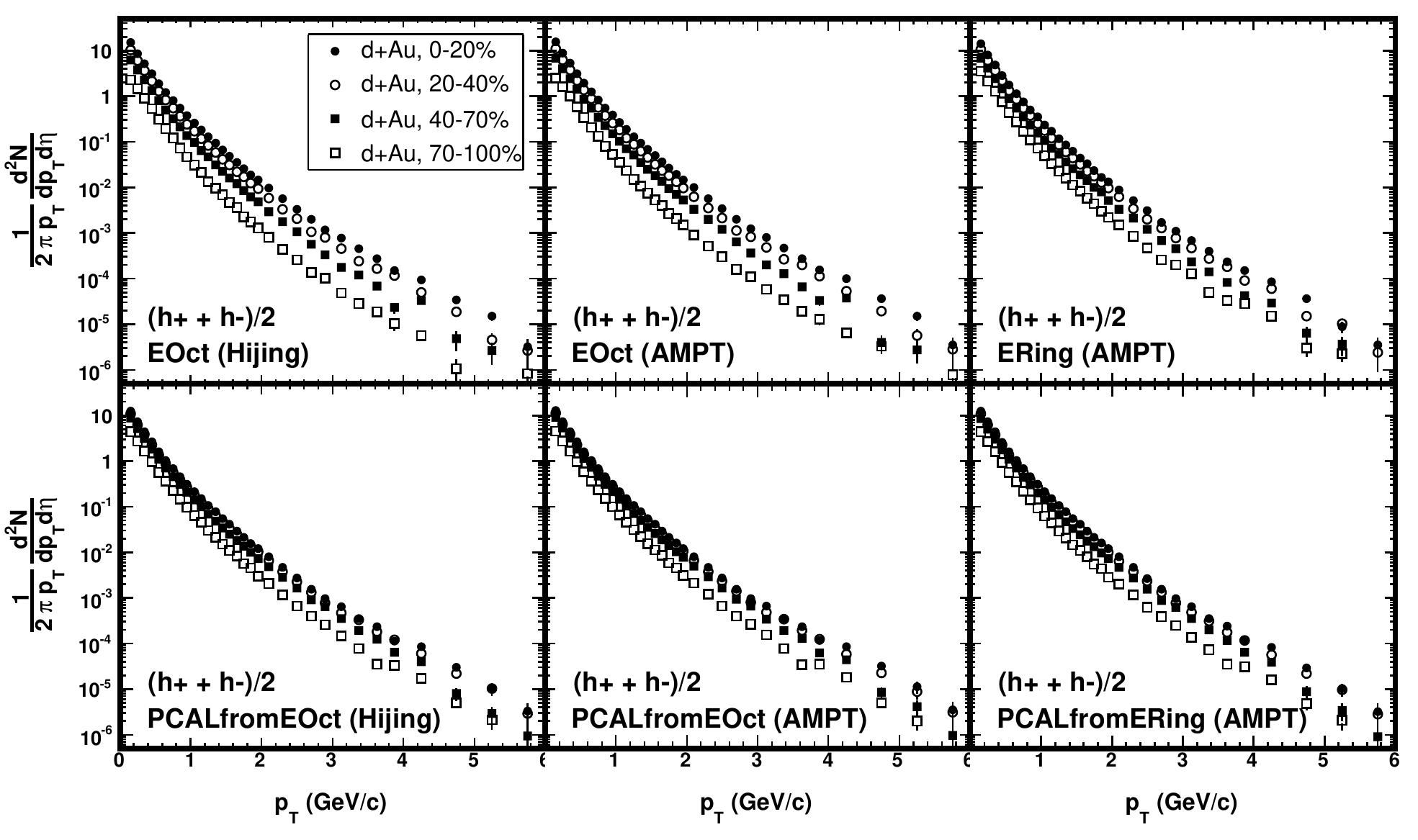}
   \end{center}
   \caption{\label{data:fig:ntSpectraNoTagAve}
      The invariant yield of charged hadrons emitted by {\dAu}
      collisions in four centrality bins. The variable used as the
      centrality measure is indicated in each plot. Only statistical
      errors are shown.}
\end{figure}

The average hadron yield in {\dAu}, measured in four centrality bins by
six different centrality measures, is shown in
\fig{data:fig:ntSpectraNoTagAve}. Note that previous {\phob}
results~\cite{Back:2003ns} are different (by a few percent) from those
presented in this thesis. The differences, while not big, are due to
changes in both the data used and in the analysis procedure itself. A
different data set, which included collisions recorded using a
high-$\pt$ trigger, was used in the former analysis. In addition, an
antiquated and less efficient version of the {\phob} Hough tracking
procedure was used in that analysis.

The analysis procedure presented in this thesis further differed from
the analysis of~\cite{Back:2003ns} in the way corrections were applied.
In the former analysis, the acceptance and efficiency corrections were
taken as smooth fits to the single track simulations, resulting in the
removal of some of the features seen in \pfig{ana:fig:accEffCorrs}. In
addition, corrections were not applied separately for each spectrometer
arm; including both the acceptance and efficiency correction, as well as
the dead and hot channel correction (which was in fact quite different
for the two arms, see \sect{ana:spec:deadchan}). Further, a subtle
correction was needed in the former analysis to account for the
vertex resolution. This correction was necessary because tracks were
counted only between \mbox{$\abs{z}<10~\cm$} based on the origin of the
\emph{track}, while collisions were counted in the same range, but based
on the reconstructed vertex of the \emph{collision}. This correction was
avoided entirely in the analysis presented in this thesis by making both
cuts using the vertex of the collision.

Finally, the method of centrality determination was different in the
analysis presented in~\cite{Back:2003ns}. In that work, \ac{ERing} cuts
were determined using \ac{HIJING}, rather than \ac{AMPT}. Because the
\ac{ERing} distribution in \ac{HIJING} did not have the same shape as in
the data, it was not possible to determine centrality using the method
described in \sect{recon:cent}. Instead, \ac{HIJING} was used to
determine the overall efficiency of the event selection and the average
efficiency in each cross section bin. These average efficiencies were
then assumed to be the same in the data, and were used to determine the
appropriate centrality cuts. Since this method was unable to produce an
event selection efficiency as a function of \ac{ERing}, it was not
possible for the former analysis to correct for that efficiency (see
\sect{ana:spec:centeff}).

\begin{figure}[t!]
   \begin{center}
      \includegraphics[width=\linewidth]{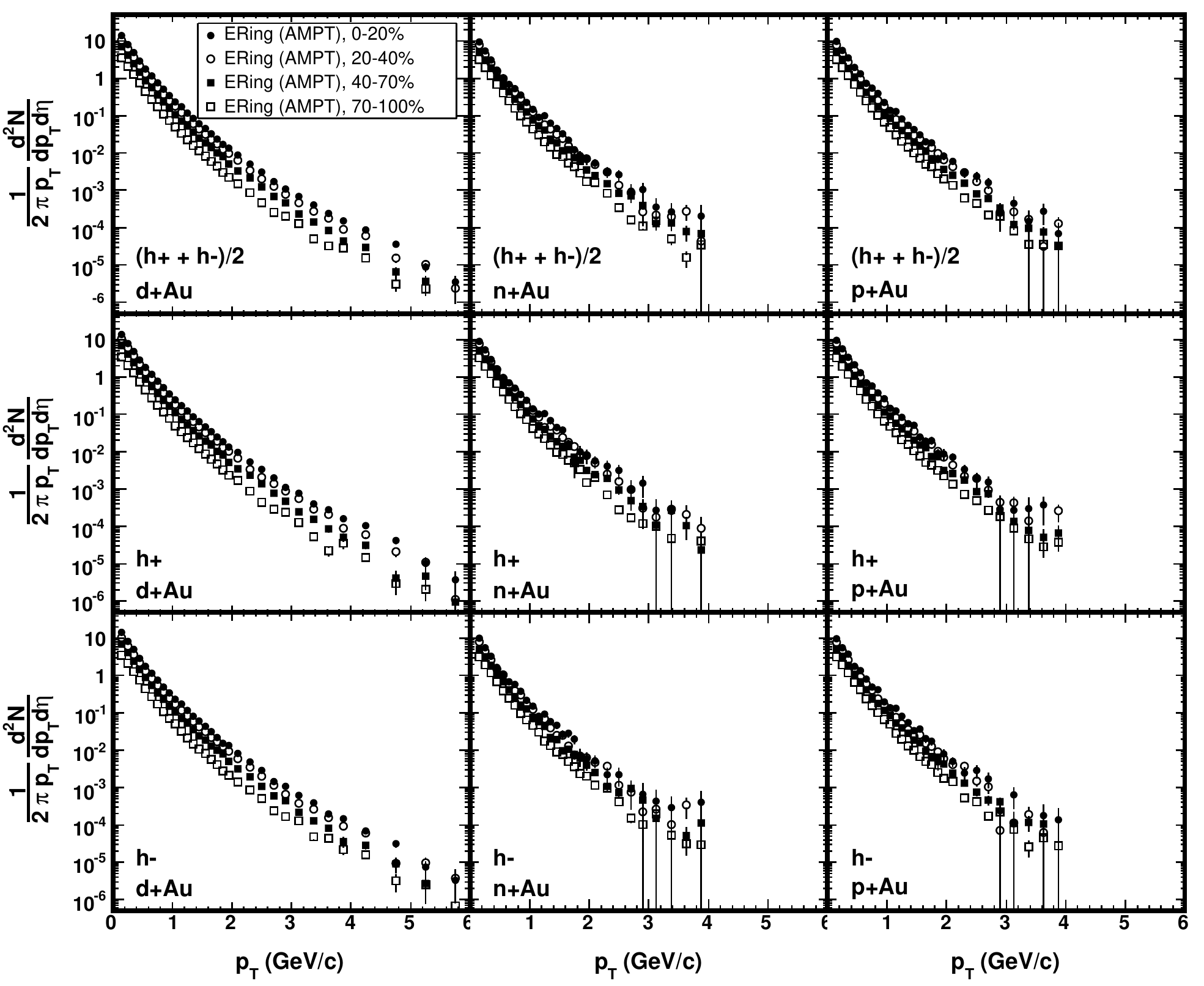}
   \end{center}
   \caption{\label{data:fig:ntSpecERingAMPT}
      The invariant yield of $\have$, $\hpos$ and $\hneg$ in four
      centrality bins determined using the \protect\acs{ERing}
      centrality variable. The spectra for {\dAu}, {\nAu} and {\pAu} are
      shown in separate columns. Only statistical errors are shown.}
\end{figure}

The positive, negative and average charged hadron spectra are shown in
\fig{data:fig:ntSpecERingAMPT} for {\dAu}, {\pAu} and {\nAu}. Only the
\ac{ERing} centrality measure was used for these plots. Note that the
difference in the $\pt$ range between {\dAu} and the nucleon-nucleus
spectra is due to statistics; fewer {\pAu} and {\nAu} collisions were
collected.

%---------------------------------------------------------------
\section{Fitting the Spectra}
\label{data:fitspec}
%---------------------------------------------------------------

The charged hadron spectra were fit using the sum of an exponential in
transverse mass, \mbox{$\mt = \sqrt{m^2 + \pt^2}$}, and a power law in
transverse momentum. These choices were motivated only loosely by
physics. That is, the essential functional form of the fit was borrowed
from previous physical arguments, but the fit parameters were not
treated as physically meaningful. The only goal was to obtain a good fit
to the data. The exponential term was inspired by theoretical arguments
for the production of soft (low $\pt$) hadrons, both in nucleon-nucleon
collisions~\cite{Gatoff:1992cv} and in nucleus-nucleus
collisions~\cite{Schaffner-Bielich:2002rt}. Since the unidentified
hadron spectra were composed primarily of pions~\cite{Veres:2004kc}, the
charged pion mass, \mbox{$\mpi=0.13957~\gev$}, was used to calculate
$\mt$.\footnote{An effective mass could have been used, based on the
relative abundance of particle species produced as a function of $\pt$.
However, such a complication was not necessary to fit the data.} The
power law term of the fit was taken from leading-order perturbative
\ac{QCD} calculations of jet production in heavy ion
collisions~\cite{Fries:2002kt}. This makes it primarily relevant for the
hard part of the spectrum (above a few {\mom}). However, previous
empirical observations have shown that such a power law fit well to
nucleon-nucleon data over a wide range of transverse
momentum~\cite{Bocquet:1995jq}. The full form of the function used to
fit the spectra was as follows.

\begin{equation}
   \label{data:eq:rawfit}
\frac{1}{2 \pi \pt} \frac{d^{2}\!N}{\mathit{d\pt} \mathit{d\eta}} 
   = A \prn{1 + \frac{\pt}{p_0}}^{-n} + 
     B \exp\prn{\frac{-\sqrt{\pt^2 + \mpi^2}}{T}}
\end{equation}

An observable used extensively in the analysis presented in this thesis
was the integral of the spectra over $\pt$; that is, the average
$\dndeta$ at $\eta=0.8$. This integral was obtained directly from the
fit to the data. Because \eq{data:eq:rawfit} could be integrated
analytically, it was possible to redefine one of the fit parameters,
namely $A$, in terms of the integral of the function (and the other fit
parameters).

\begin{align}
\frac{d\!N}{\mathit{d\eta}} 
   &= \int_0^{+\infty} 2 \pi \pt \sqb{A \prn{1 + \frac{\pt}{p_0}}^{-n} + 
     B \exp\prn{\frac{-\sqrt{\pt^2 + \mpi^2}}{T}}} d\!\pt
     \label{data:eq:fitinteg} \\
   &\equiv I_{\mathrm{pl}} + I_{\mathrm{exp}} \notag
\end{align}

\noindent%
Integrating the power-law term first:

\begin{align}
I_{\mathrm{pl}}
   &= \int_0^{+\infty} 
      \prn{2 \pi \pt} A \prn{1 + \frac{\pt}{p_0}}^{-n} d\!\pt \notag \\
   &= 2 \pi A \int_1^{+\infty} p_0^2 (u-1) u^{-n} d\!u \qquad
      \text{where} \quad u = 1 + \frac{\pt}{p_0} \notag \\
   &= 2 \pi A p_0^2 \sqb{ \sqb{\frac{1}{1-n} u^{1-n} (u-1)}_{1}^{+\infty}
      - \frac{1}{1-n} \int_1^{+\infty} u^{1-n} d\!u } \qquad \text{by parts}
         \notag \\
   &= 2 \pi A p_0^2 \sqb{ 0 + \frac{1}{n-1}
      \sqb{\frac{1}{2-n} u^{2-n}}_1^{+\infty}} \notag \\
I_{\mathrm{pl}}
   &= 2 \pi A p_0^2 \frac{1}{(n-1) (n-2)} \label{data:eq:integratePL}
\end{align}

\noindent%
where it is assumed that $n>2$. Similarly, the exponential term can be
integrated.

\begin{align}
I_{\mathrm{exp}}
   &= \int_0^{+\infty} 
      \prn{2 \pi \pt} B \exp\prn{\frac{-\sqrt{\pt^2 + \mpi^2}}{T}} d\!\pt
      \notag \\
   &= 2 \pi B \int_{\mpi}^{+\infty} \prn{\mt e^{-\mt / T}} d\!\mt \notag \\
I_{\mathrm{exp}}
   &= 2 \pt B T (\mpi+T) e^{-\mpi / T} \qquad \text{by parts}
   \label{data:eq:integrateEXP}
\end{align}

Using \eqs{data:eq:integratePL}{data:eq:integrateEXP}, the fit parameter
$A$ of \eq{data:eq:fitinteg} can be defined in terms of the physical
variable $\dndeta$.

\begin{align*}
\frac{d\!N}{\mathit{d\eta}} &= I_{\mathrm{pl}} + I_{\mathrm{exp}} \\
   &= \frac{2 \pi A p_0^2}{(n-1) (n-2)} + 2 \pi B T (\mpi+T) e^{-\mpi / T}\\
\Longleftrightarrow
A &= \frac{(n-1) (n-2)}{2 \pi p_0^2} \sqb{\frac{d\!N}{\mathit{d\eta}} - 
      2 \pi B T (\mpi+T) e^{-\mpi / T}}
\end{align*}

\noindent%
This allows \eq{data:eq:fitinteg} to be re-written as

\begin{multline}
\frac{1}{2 \pi \pt} \frac{d^{2}\!N}{\mathit{d\pt} \mathit{d\eta}} =
   \frac{(n-1) (n-2)}{2 \pi p_0^2} \sqb{\frac{d\!N}{\mathit{d\eta}} - 
      2 \pi B T (\mpi+T) e^{-\mpi / T}} \prn{1 + \frac{\pt}{p_0}}^{-n} \\
   + B \exp\prn{\frac{-\sqrt{\pt^2 + \mpi^2}}{T}}
   \label{data:eq:fitwithinteg}
\end{multline}

\begin{table}[t]
   \begin{center}
      \begin{tabular}{|lll|}
\hline
Parameter & Minimum & Maximum\\
\hline
$\dndeta$ & 0 & 92\\
$T$ & 0~{\gev} & 0.500~{\gev}\\
$B$ & 0 & 40 \\
$p_0$ & 0~{\mom} & 4~{\mom}\\
$n$ & 6 & 17\\
\hline
      \end{tabular}
   \end{center}
   \caption{   \label{data:tab:fitparranges}
      The ranges over which parameters were varied while fitting the
      spectra.}
\end{table}

The advantage of \eq{data:eq:fitwithinteg} is that it contains a
physically meaningful fit parameter, namely $\dndeta$. This was the
function fit to the transverse momentum spectra. The ranges over which
the fit parameters were varied are shown in \tab{data:tab:fitparranges}.
The multiplicity and the statistical error on the multiplicity
measurement were then extracted directly from the fit.

As a cross-check on the $\dndeta$ value obtained from the fit, an
estimate of $\dndeta$ was calculated from the data. This was done by
finding the integral of the data in the most straight-forward way.
First, the content of each $\pt$ bin was multiplied by \mbox{$2 \pi \pt
W_b$}, where $W_b$ is the width of the $\pt$ bin. Then the sum, $S_b$,
of these terms for each $\pt$ bin was obtained. Finally, the spectra was
extrapolated to \mbox{$\pt = 0~\mom$} by calculating a straight line from
the two lowest $\pt$ data points. The integral of this line (below the
lowest $\pt$ bin) was then added to the sum $S_b$. The resulting total
provided a rough estimate of $\dndeta$. Typically, this value differed
from the $\dndeta$ obtained from the fit by about 1\%. The two estimates
always agreed to better than 5\%.

The systematic uncertainty on the integrated yield was determined using
the following method. First, all points of a given measured spectrum
were shifted up to the maximum of their systematic error. Then, this new
$\pt$ spectrum was fit and a maximum $\dndeta$ value was obtained. Next,
all points on the spectrum were shifted down to the minimum of their
systematic error. Again, this new spectra was fit and a minimum
$\dndeta$ value was extracted. Using this method, it was found that the
minimum and maximum integrated yields were consistently within about 9\%
of the measured integrated yield. Thus, a relative systematic
uncertainty (90\%~C.L.) on the integrated yield of 9\% was determined.
This error was found to be independent of centrality bin, centrality
measure, the specific hadron spectra (i.e.~average or charged hadrons)
and the collision system.

\begin{figure}[t]
   \centering
   \subfigure[$\hneg$ Fit]{
      \label{data:fig:exampleFitGood}
      \includegraphics[width=0.4\linewidth]{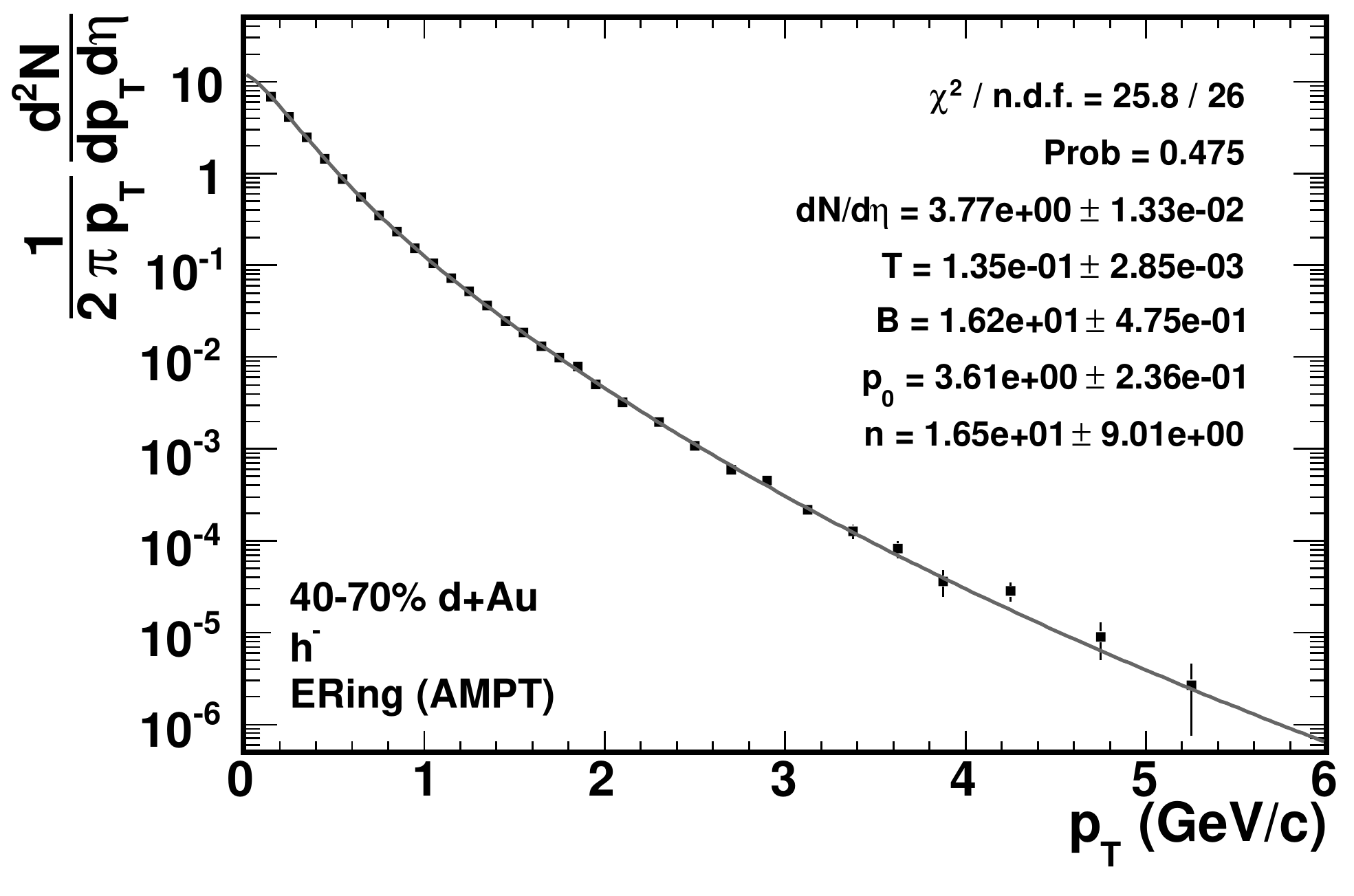}
   }
   \subfigure[$\hneg$ Fit (Linear)]{
      \label{data:fig:exampleFitGoodLinear}
      \includegraphics[width=0.4\linewidth]{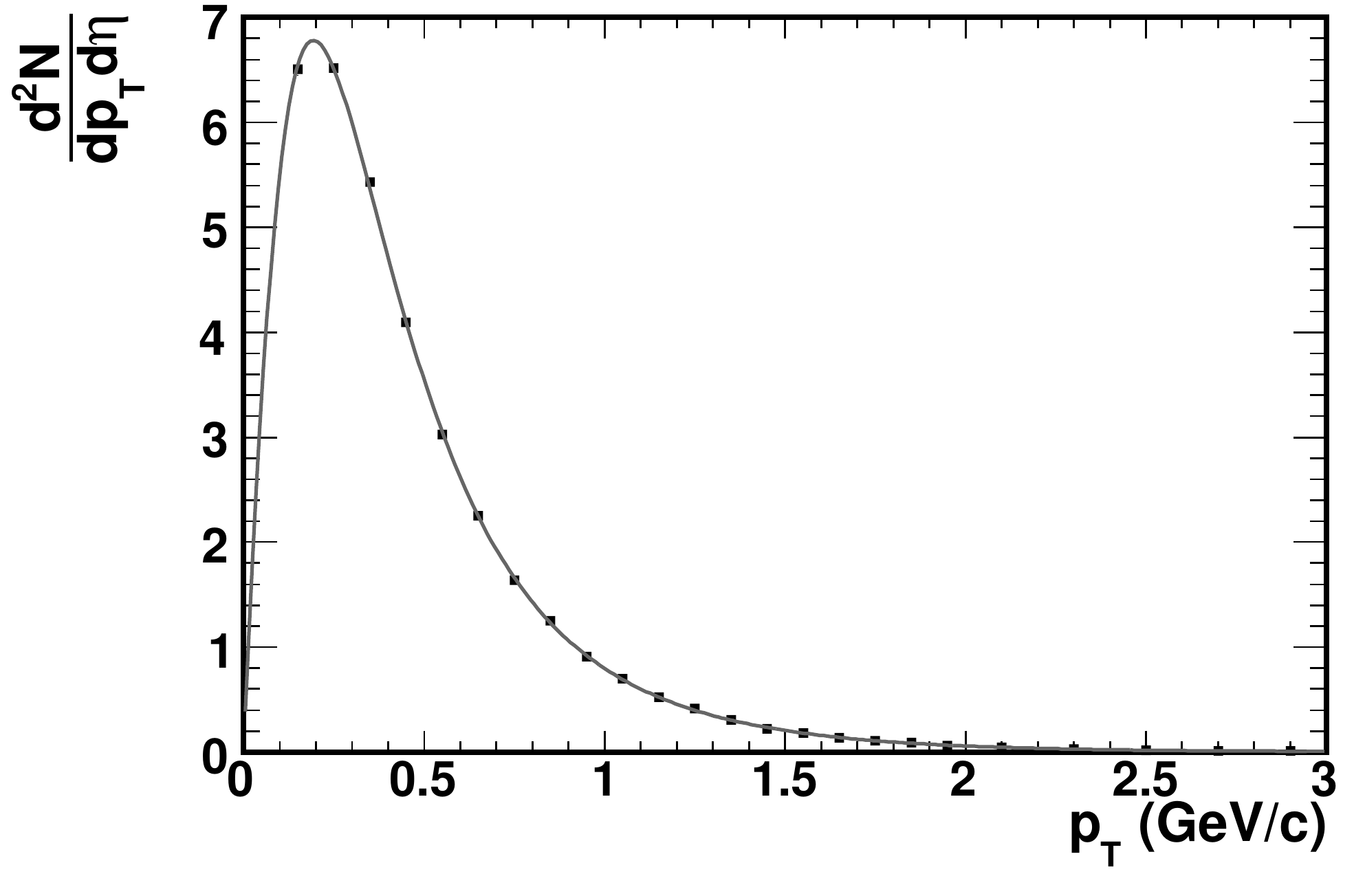}
   }
   \subfigure[Data / Fit Ratio]{
      \label{data:fig:exampleFitRatioGood}
      \includegraphics[width=0.4\linewidth]{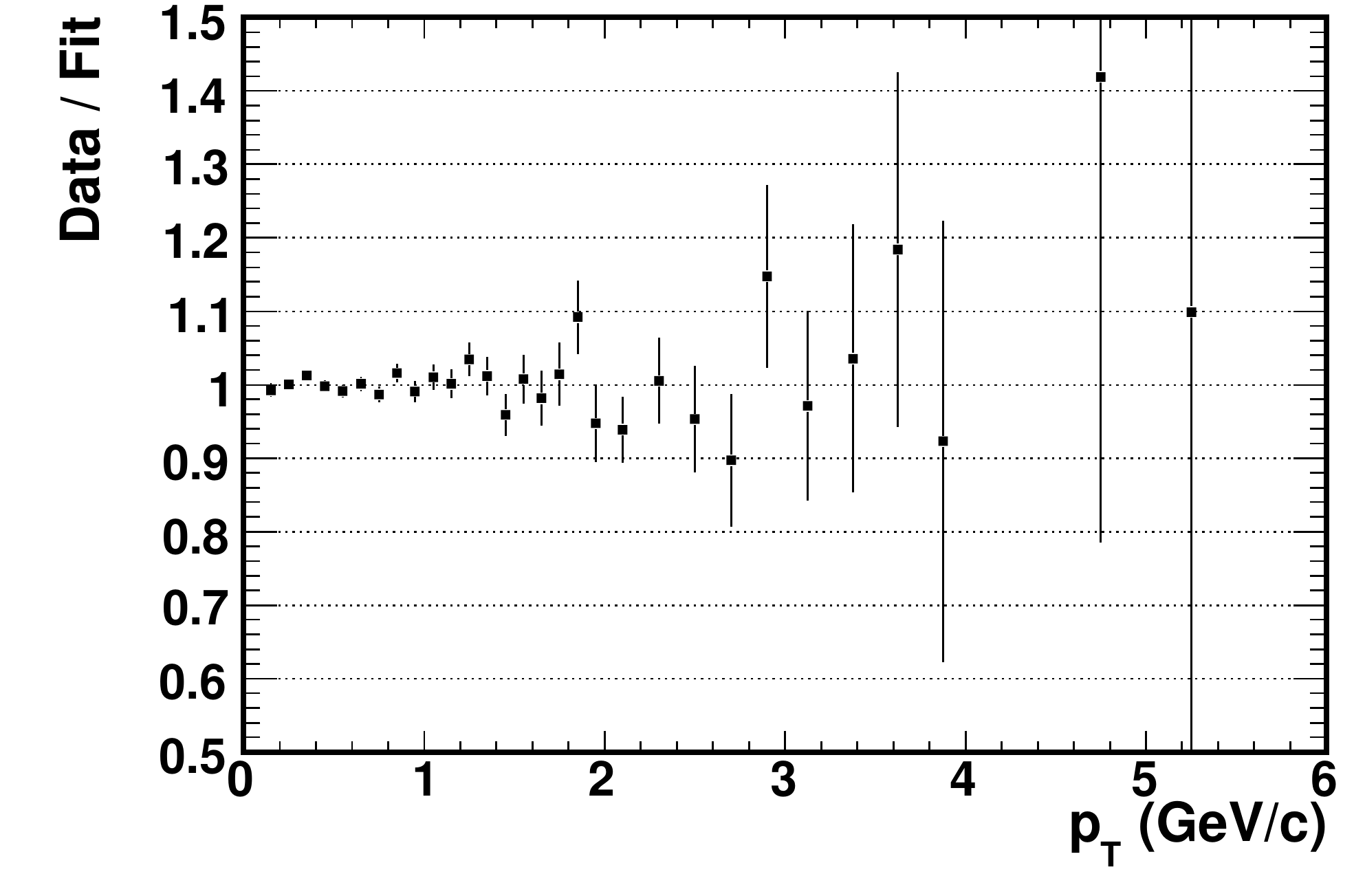}
   }
   \caption{   \label{data:fig:exFitGood}
      An example of a good fit to spectra that was generated with a
      large number of tracks. \subref{data:fig:exampleFitGood}~The fit
      of the $\hneg$ spectra in the 40-70\% most central {\dAu} (as
      found using the \protect\acs{ERing} centrality variable).
      \subref{data:fig:exampleFitGoodLinear}~The transverse momentum
      distribution (not invariant yield) and fit on a linear scale. Note
      the smaller $\pt$ range displayed.
      \subref{data:fig:exampleFitRatioGood}~Ratio of the data to the
      fit.}
\end{figure}

\begin{figure}[t!]
   \centering
   \subfigure[$n$ vs $p_0$]{
      \label{data:fig:contour_p0_n}
      \includegraphics[width=0.4\linewidth]{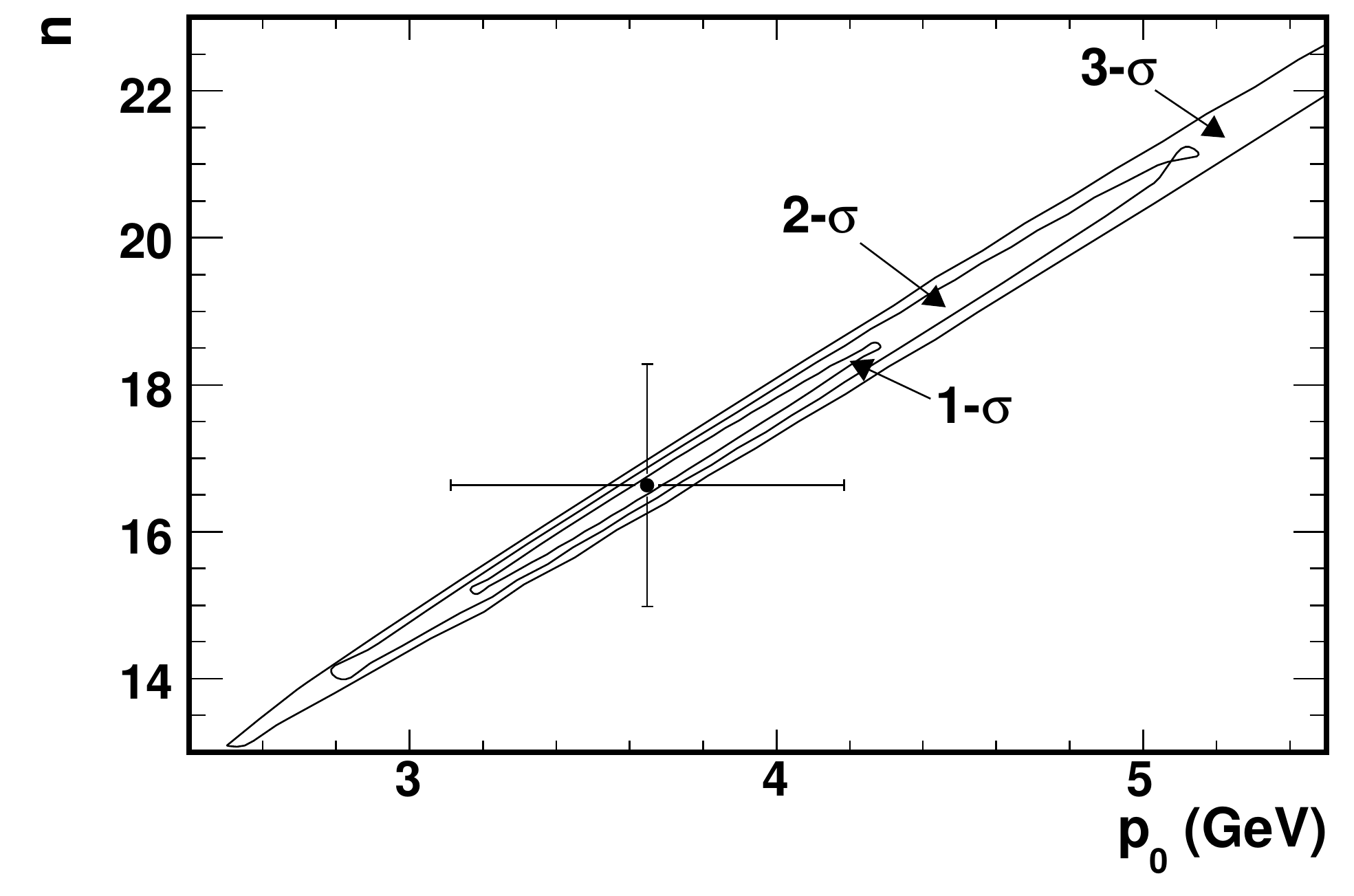}
   }
   \subfigure[$T$ vs $\dndeta$]{
      \label{data:fig:contour_dndy_T}
      \includegraphics[width=0.4\linewidth]{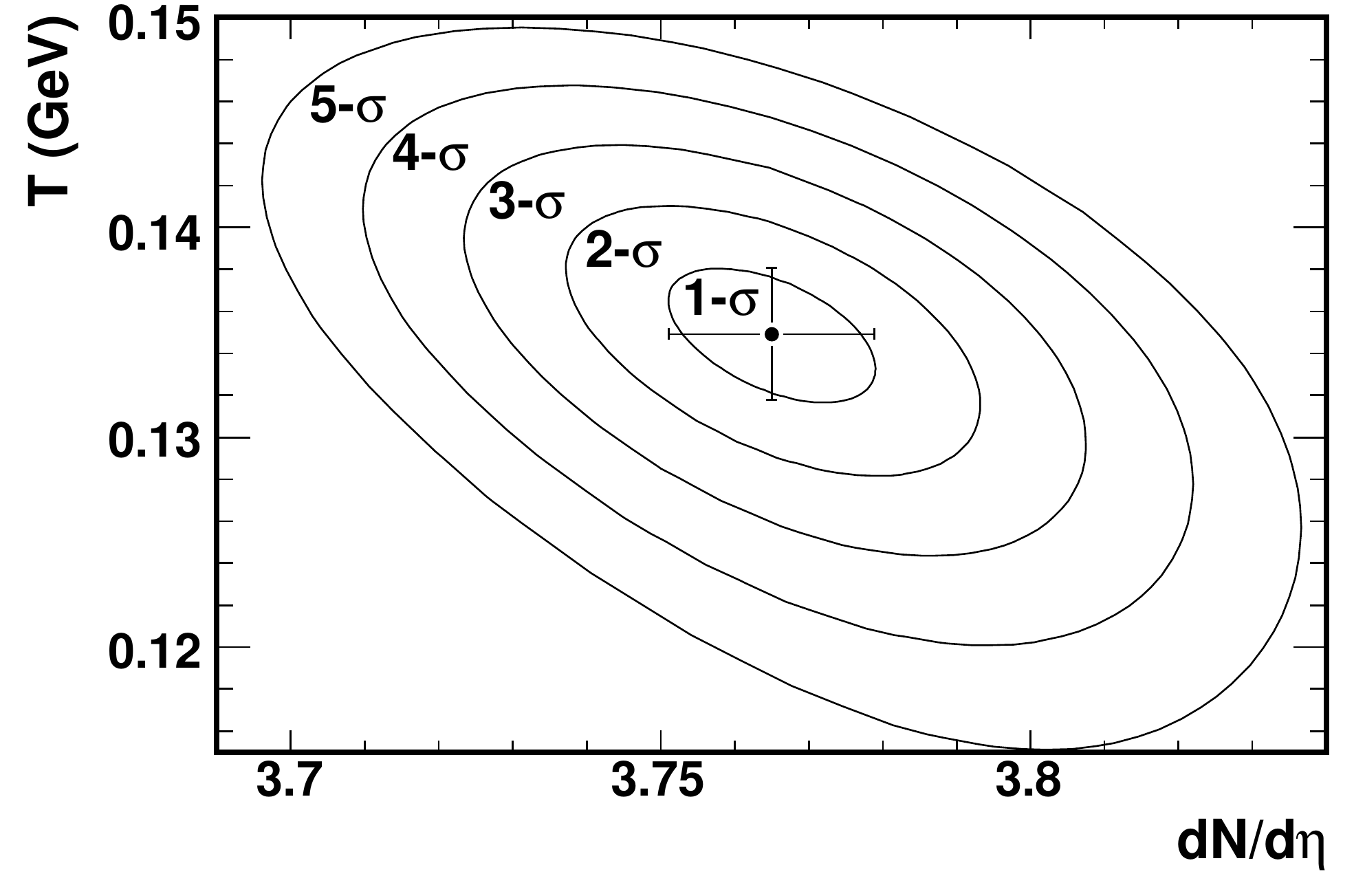}
   }
   \caption{   \label{data:fig:fitcontours}
      The $\chi^{2}$ contours for the fit to the 40-70\% most central
      $\hneg$ spectra in {\dAu}. The sigma contours are labeled, and the
      points show the minima found by the fit.
      \subref{data:fig:contour_p0_n}~The $n$ and $p_0$ parameter
      space. The correlation between these two parameters is evident.
      \subref{data:fig:contour_dndy_T}~The $T$ and $\dndeta$
      parameter space.}
\end{figure}

An example of a fit to the measured spectra can be seen in
\fig{data:fig:exFitGood}. The spectra shown in \fig{data:fig:exFitGood}
has good statistics; that is, it was produced using a large number of
tracks over much of the $\pt$ range (around 142,000 total). The
transverse momentum distribution of particles in the center of mass
frame is shown in \fig{data:fig:exampleFitGoodLinear}. This spectra was
fit well by \eq{data:eq:fitwithinteg}, as seen by both the fit
probability (see \fig{data:fig:exampleFitGood}) and the ratio of the
data to the fit (see \fig{data:fig:exampleFitRatioGood}). As noted in
the UA1 paper~\cite{Bocquet:1995jq}, the parameters $p_0$ and $n$ are
highly correlated. This can be seen in \fig{data:fig:contour_p0_n},
which shows the first three $\chi^2$ contours for these variables, as
determined for the fit shown in \fig{data:fig:exFitGood}. However, the
$\dndeta$ parameter does not show a strong correlation with the other
parameters; see \fig{data:fig:contour_dndy_T} for the $\chi^2$ contours
in the \mbox{$T-\dndeta$} plane. Thus, it was assumed that the errors
reported by the fit for $\dndeta$ were reasonable.

\begin{figure}[t!]
   \centering
   \subfigure[$\hneg$ Fit]{
      \label{data:fig:exampleFitBad}
      \includegraphics[width=0.4\linewidth]{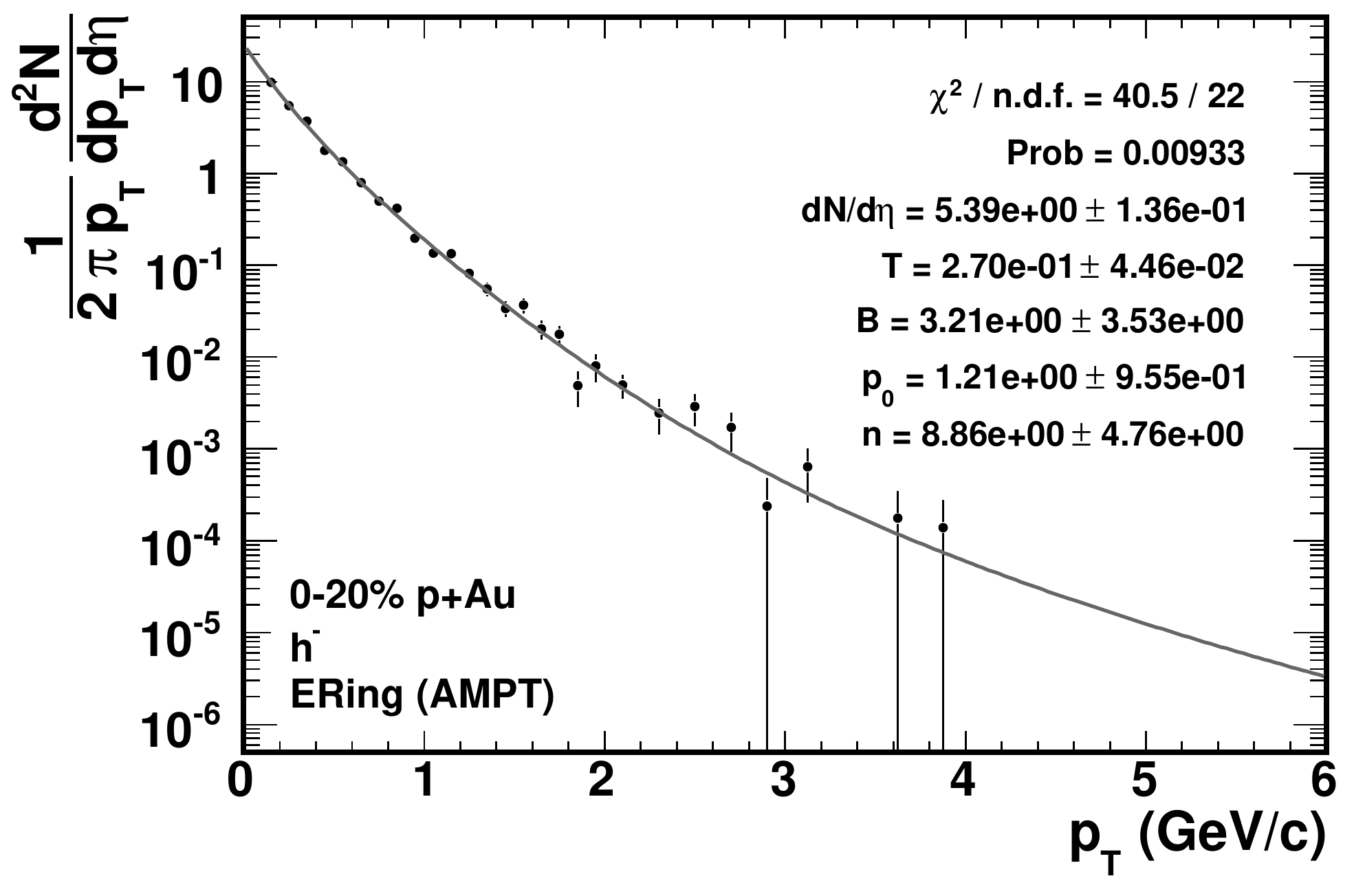}
   }
   \subfigure[Data / Fit Ratio]{
      \label{data:fig:exampleFitRatioBad}
      \includegraphics[width=0.4\linewidth]{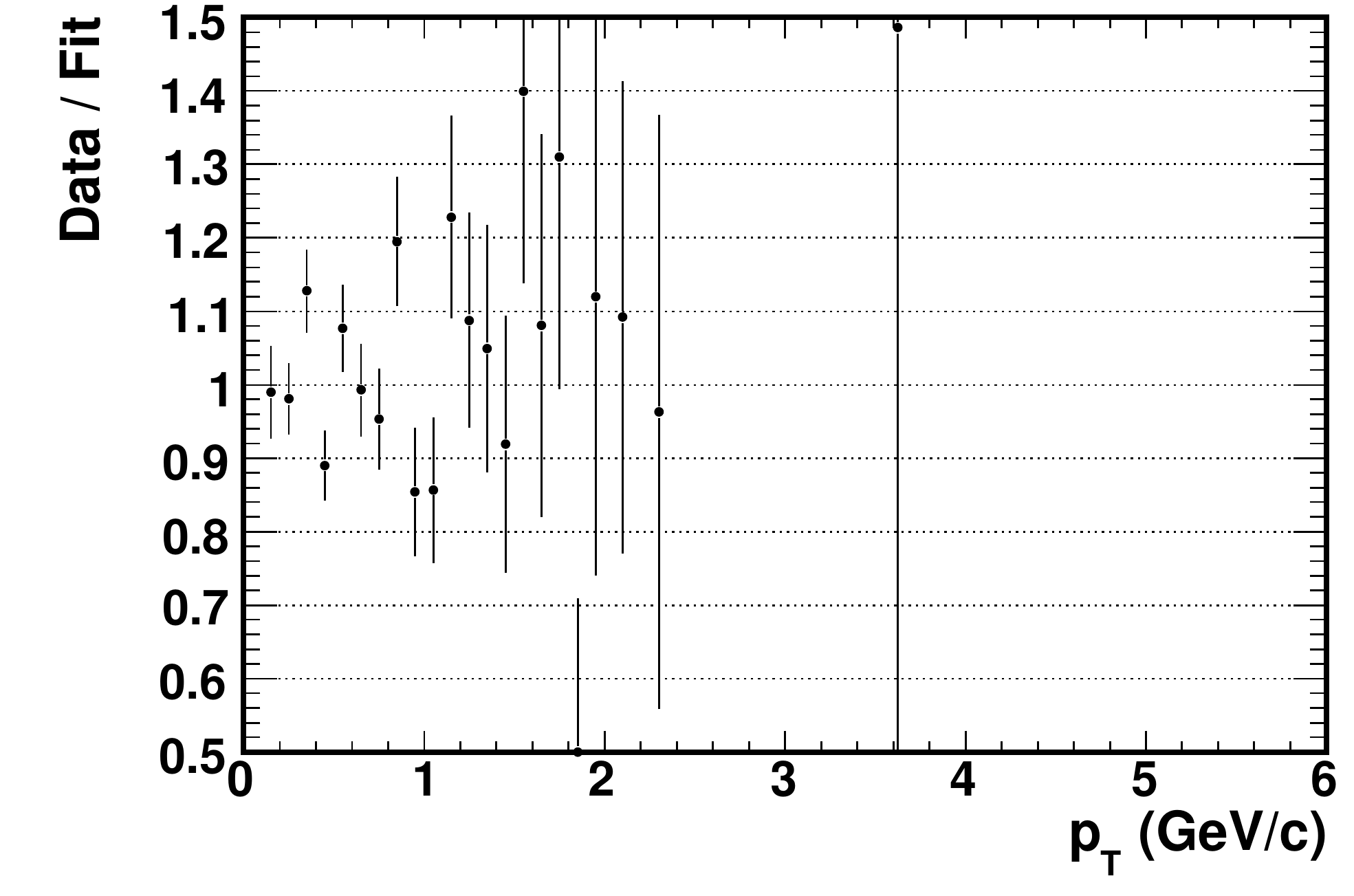}
   }
   \caption{   \label{data:fig:exFitBad}
      The fit of the $\hneg$ spectra in the 0-20\% most central {\pAu}
      (as found using the \protect\acs{ERing} centrality variable). This
      is an example of a somewhat poor fit to spectra that was generated
      with a relatively small number of tracks.}
\end{figure}

An example of a fit to spectra that was produced with fewer tracks
(around 3,400) is shown in \fig{data:fig:exFitBad}. Despite the low
statistics, fits to such spectra were not necessarily poor (as
determined by $\chi^2$). However, the fit shown in
\fig{data:fig:exFitBad} was intended to serve not only as an example of
a fit to spectra with low statistics, but also as an example of an
atypical (poor) fit.

%---------------------------------------------------------------
\section{Centrality Results}
\label{data:cent}
%---------------------------------------------------------------

Six different centrality measures were used to find centrality cuts for
the {\dAu} data. These measures were built from combinations of three
variables (see \ptab{recon:tab:centvars}) and two \ac{MC} simulations,
\ac{HIJING} and \ac{AMPT}. The combinations are shown in
\tab{data:tab:centvars}. As described in \sect{recon:cent}, the \ac{MC}
simulations were used to estimate the efficiency of the trigger and
event selection as a function of the centrality measure. To obtain this
estimation, it was required that the shape of the centrality measure
distribution in the simulations matched (possibly with some constant
scaling factor) the same distribution observed in the collision data.
Because such a matching could not be obtained for the \ac{ERing}
distribution in \ac{HIJING}, \ac{ERing} cuts using the \ac{HIJING}
simulations were not determined for this analysis.

\begin{table}[t]
   \begin{center}
      \begin{tabular}{|lll|}
\hline
Cut Variable &  \protect\acs{MC} Simulation & Description \\
\hline
\protect\acs{ERing} & \protect\acs{AMPT} (\small{8-18-03}) & %
   Energy in the Rings \\
\protect\acs{EOct} & \protect\acs{AMPT} (\small{8-18-03}) & %
   Energy in the Octagon \\
\protect\acs{EOct} & \protect\acs{HIJING} (\small{1.383}) & %
   Energy in the Octagon \\
\protect\acs{EPCAL} & \protect\acs{AMPT} (\small{8-18-03}) & %
   Energy in the \protect\acs{Au-PCAL}, correlation with
   \protect\acs{ERing} used \\
\protect\acs{EPCAL} & \protect\acs{AMPT} (\small{8-18-03}) & %
   Energy in the \protect\acs{Au-PCAL}, correlation with
   \protect\acs{EOct} used \\
\protect\acs{EPCAL} & \protect\acs{HIJING} (\small{1.383}) & %
   Energy in the \protect\acs{Au-PCAL}, correlation with
   \protect\acs{EOct} used \\
\hline
      \end{tabular}
   \end{center}
   \caption{   \label{data:tab:centvars}
      Description of centrality measures used in this analysis. See
      \sect{recon:cent}.}
\end{table}

Each centrality measure was used to find centrality cuts corresponding
to fractional cross section bins containing the 0-20\%, 20-40\%, 40-70\%
and 70-100\% most central {\dAu} collisions. The cuts determined for the
\ac{EOct} centrality measure, using \ac{HIJING} simulations, are shown
in \pfig{recon:fig:eoctCuts}. These same cuts were used for the
nucleon-nucleus collision systems {\pAu} and {\nAu}. However, the
fraction of the nucleon-nucleus cross section to which the bins
corresponded was left undetermined (see \sect{recon:nuctag:NAcent}). The
average unbiased $\Npart$ estimated for each centrality bin of each
centrality measure is shown in \tab{data:tab:localNpartdAu}.

\input{srcs/centables/localNpart_dAu_table.tex}

The widths of the {\dAu} centrality bins used in this analysis were the
same as those used in the previous {\phob} {\dAu} $\pt$ spectra
analysis~\cite{Back:2003ns}. However, neither the procedure to determine
the centrality bins nor the \ac{MC} simulations used were the same
between the two analyses. In addition, all centrality parameters (such
as $\Npart$) used in the analysis presented in this thesis were
\emph{unbiased}. That is, the estimated trigger and event selection
efficiency was used to correct for biases introduced by the centrality
dependence of the efficiency, as described in \sect{recon:cent:pars}.
Thus the average $\Npart$ values estimated in the two analyses were not
expected to be identical (although they do agree within systematic
uncertainties). A comparison of the biased and unbiased $\Npart$ values
estimated for the \ac{ERing} centrality cuts found using \ac{AMPT}
simulations is shown in \tab{data:tab:eringBiasUnbias}. Note that the
average $\Npart$ in the central bins are essentially the same, which is
reasonable since the efficiency was independent of centrality in this
region (see \pfig{ana:fig:efficiencies}). However, the efficiency was
found to decrease with decreasing centrality, and the resulting bias can
be seen in the average $\Npart$ values estimated for the two peripheral
bins.

\begin{table}[t]
   \begin{center}
      \begin{tabular}{|llllll|}
\hline
\multicolumn{6}{|c|}{d+Au $\Npart$ with \protect\acs{dAuSpectra} %
event selection} \\
\hline
 & & \multicolumn{4}{c|}{$\ave{\Npart} \pm \text{Sys. Err.}$} \\
Variable & Type & 70-100\% & 40-70\% & 20-40\% & 0-20\% \\
\hline
\protect\acs{ERing} & Unbiased & $3.12\pm0.70$& $6.31\pm0.70$& $10.62\pm0.88$&
$15.42\pm1.09$\\
\protect\acs{ERing} & Biased & $3.43\pm0.77$& $6.50\pm0.72$& $10.66\pm0.88$&
$15.44\pm1.10$\\
\hline
      \end{tabular}
   \end{center}
   \caption{   \label{data:tab:eringBiasUnbias}
      Biased and unbiased $\Npart$ values obtained using
      \protect\acs{ERing}, \protect\acs{AMPT} and the
      \protect\acs{dAuSpectra} event selection.}
\end{table}

In addition to the two centrality parameters shown in
\pfig{recon:fig:npartncoll}, $\Npart$ and $\Ncoll$, three other
parameters were studied in the \ac{MC}. Two were simple components of
$\Npart$. Both the average number of deuteron participants ($\Npard$)
and the average number of participating gold nucleons ($\NparA$) were
estimated for each centrality bin. The fifth centrality parameter was
particularly useful for describing the geometry of {\dAu} collisions. It
was defined as the number of collisions per deuteron participant,
\mbox{$\nu \equiv \Ncoll / \Npard$}~\cite{Back:2003ff}. The average
unbiased $\nu$ estimated for each centrality bin of each centrality
measure is shown in \tab{data:tab:localNudAu}. Note that $\ave{\nu}$ was
estimated by finding the average of the ratio \mbox{$\Ncoll / \Npard$}
in the simulated collisions, rather than taking the ratio of
\mbox{$\ave{\smash[tb]{\Ncoll}} / \ave{\smash[tb]{\Npard}}$}. The
average of each centrality parameter in each {\dAu} centrality bin as
found using \ac{ERing} and \ac{AMPT} is shown in
\tab{data:tab:eringAllCpars}.

\begin{table}[t]
   \begin{center}
      \begin{tabular}{|lllll|}
\hline
\multicolumn{5}{|c|}{d+Au centrality parameters found with %
\protect\acs{ERing} cuts} \\
\hline
 & \multicolumn{4}{c|}{$\ave{\text{Parameter}} \pm \text{Sys. Err.}$} \\
Parameter & 70-100\% & 40-70\% & 20-40\% & 0-20\% \\
\hline
$\avencl$ & $2.00\pm0.60$& $4.99\pm0.60$& $9.43\pm0.67$& $14.49\pm0.88$\\
$\avenpt$ & $3.12\pm0.70$& $6.31\pm0.70$& $10.62\pm0.88$& $15.42\pm1.09$\\
$\ave{\smash[tb]{\NparA}}$ & $1.92\pm0.50$& $4.64\pm0.60$& $8.70\pm0.69$& $13.43\pm0.99$\\
$\ave{\smash[tb]{\Npard}}$ & $1.20\pm0.20$& $1.67\pm0.20$& $1.93\pm0.10$& $1.99\pm0.11$\\
$\ave{\nu}$ & $1.73\pm0.52$& $3.27\pm0.39$& $5.20\pm0.37$& $7.55\pm0.46$\\
\hline
      \end{tabular}
   \end{center}
   \caption{   \label{data:tab:eringAllCpars}
      Unbiased centrality parameters obtained using
      \protect\acs{ERing}, \protect\acs{AMPT} and the
      \protect\acs{dAuSpectra} event selection.}
\end{table}

\input{srcs/centables/localNu_dAu_table.tex}

%% file: srcs/centables/localNpart_dAu_table.tex
\cpartab{d+Au unbiased}{\Npart}{\protect\acs{dAuSpectra}}{
\protect\acs{EOct} & \protect\acs{HIJING} & $3.73\pm0.84$& $7.48\pm0.83$& $11.32\pm0.94$& $15.15\pm1.08$\\ 
\protect\acs{EOct} & \protect\acs{AMPT} & $3.75\pm0.84$& $7.52\pm0.83$& $11.28\pm0.94$& $15.02\pm1.07$\\ 
\protect\acs{ERing} & \protect\acs{AMPT} & $3.12\pm0.70$& $6.31\pm0.70$& $10.62\pm0.88$& $15.42\pm1.09$\\ 
\protect\acs{EPCAL} (\protect\acs{EOct})  & \protect\acs{HIJING}& $5.31\pm1.22$& $8.25\pm1.13$& $10.49\pm1.21$& $12.19\pm1.30$\\ 
\protect\acs{EPCAL} (\protect\acs{EOct})  & \protect\acs{AMPT}& $5.26\pm1.21$& $8.30\pm1.14$& $10.46\pm1.21$& $12.07\pm1.29$\\ 
\protect\acs{EPCAL} (\protect\acs{ERing})  & \protect\acs{AMPT}& $4.12\pm0.94$& $7.36\pm1.01$& $10.43\pm1.20$& $12.82\pm1.37$\\ 
}{localNpartdAu}{t!}

%% file: srcs/centables/localNu_dAu_table.tex
\cpartab{d+Au unbiased}{\nu}{\protect\acs{dAuSpectra}}{
\protect\acs{EOct} & \protect\acs{HIJING} & $2.06\pm0.62$& $3.84\pm0.46$& $5.56\pm0.39$& $7.47\pm0.46$\\ 
\protect\acs{EOct} & \protect\acs{AMPT} & $2.06\pm0.62$& $3.80\pm0.46$& $5.51\pm0.39$& $7.37\pm0.45$\\ 
\protect\acs{ERing} & \protect\acs{AMPT} & $1.73\pm0.52$& $3.27\pm0.39$& $5.20\pm0.37$& $7.55\pm0.46$\\ 
\protect\acs{EPCAL} (\protect\acs{EOct})  & \protect\acs{HIJING}& $2.80\pm0.84$& $4.18\pm0.58$& $5.23\pm0.49$& $6.04\pm0.71$\\ 
\protect\acs{EPCAL} (\protect\acs{EOct})  & \protect\acs{AMPT}& $2.75\pm0.83$& $4.17\pm0.58$& $5.19\pm0.48$& $5.95\pm0.70$\\ 
\protect\acs{EPCAL} (\protect\acs{ERing})  & \protect\acs{AMPT}& $2.20\pm0.66$& $3.73\pm0.52$& $5.17\pm0.48$& $6.30\pm0.74$\\ 
}{localNudAu}{t!}

%% file: srcs/results.tex
\chapter{Studies of d+Au, p+Au and n+Au Spectra}
\label{rslt}
%---------------------------------------------------------------

\myupdate{$*$Id: results.tex,v 1.38 2006/09/04 20:41:22 cjreed Exp $*$}%
With the measured spectra fully corrected, it was possible to study
particle production in {\pAu}, {\nAu} and {\dAu} collisions at a center
of mass energy of 200~{\gev} per nucleon pair. The extreme asymmetry of
these collision systems, in addition to the small transverse area of the
interaction region (as compared to {\AuAu}), necessitated a clear
understanding of the centrality of the collisions. To this end, the
impact of the choice of centrality measure on the spectra measurement
was investigated. Further studies of the transverse momentum spectra
were then conducted. First, the assumption that {\dAu} collisions rather
than nucleon-nucleus collisions could be used as a control experiment
for {\AuAu}, in the sense of a collision system that includes nuclear
effects but not any extended volume medium effects, was tested. Then, a
systematic study of the modification of hadron production in {\dAu} with
respect to {\pbarp} was performed. This study examined the change in
shape of the spectrum, both as a function of the amount of initial
nuclear material probed by the deuteron and as a function of the
outgoing particle density. Finally, charged hadron production in {\pAu}
was compared to production in {\nAu}.

Note that data presented for the first time in this thesis have not been
reviewed by the {\phob} collaboration.

%---------------------------------------------------------------
\section{Significance of the Chosen Centrality Technique}
\label{rslt:cent}
%---------------------------------------------------------------

It has been generally accepted that the technique used to determine and
to parameterize the centrality of collisions can impact the desired
measurement; see for example~\cite{Antinori:2001qn}
and~\cite{Olszewski:2002ie}. Precisely how measurements of particle
production in {\dAu} and nucleon-nucleus collisions were affected by the
techniques discussed in \sect{recon:cent} was studied.

%---------------------------------------------------------------
\subsection{Fractional Cross Section}
\label{rslt:cent:crsscn}
%---------------------------------------------------------------

\begin{figure}[t]
   \begin{center}
      \includegraphics[width=0.8\linewidth]{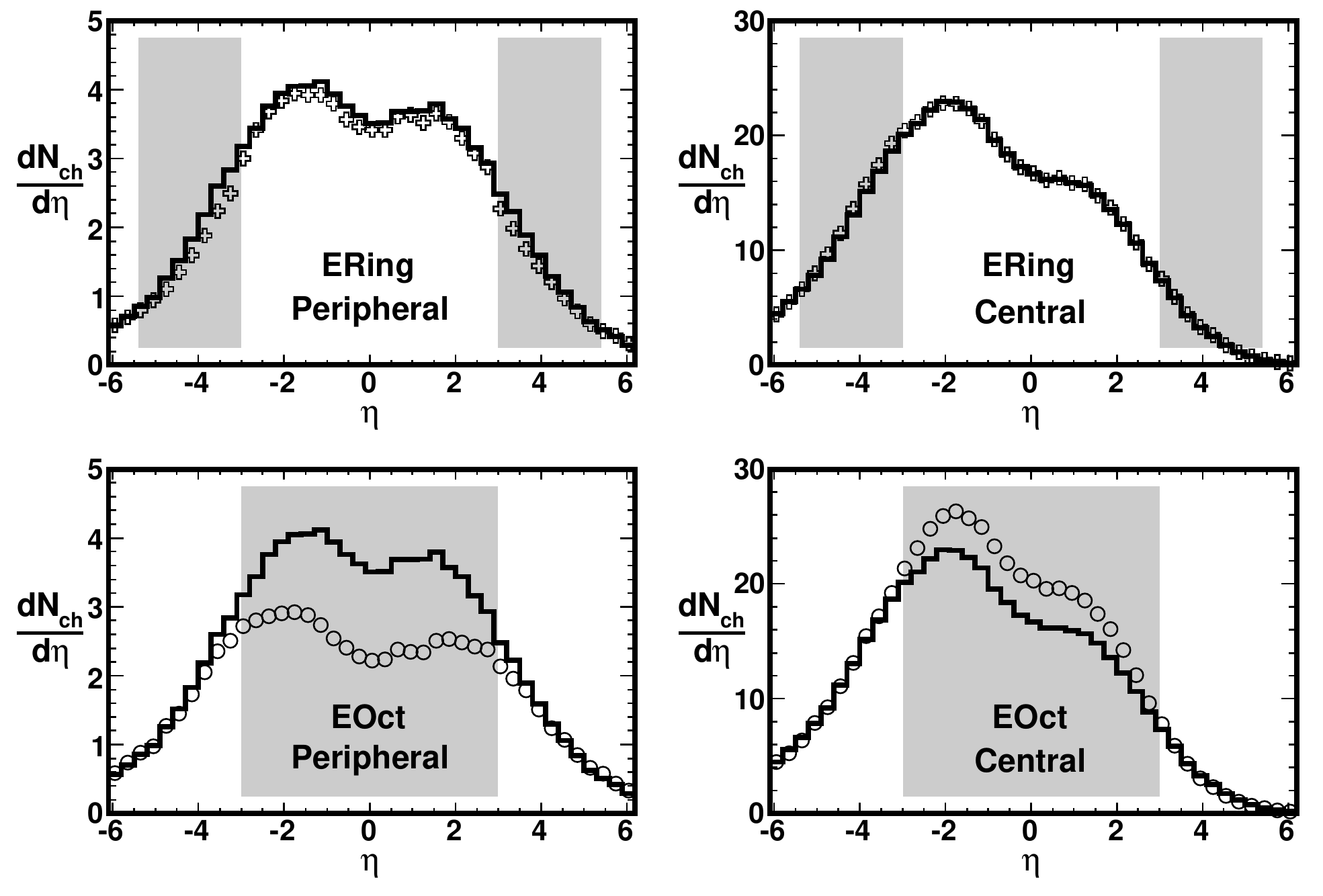}
   \end{center}
   \caption{\label{rslt:fig:eoctERingBias}
      The bias introduced by the choice of centrality measure.
      Reconstructed {\dAu} multiplicity distributions in
      \protect\acs{HIJING} are shown by the open symbols. The unbiased,
      true multiplicity distribution is shown by the solid line. Shaded
      areas indicate the {\prap} region over which the centrality
      variable is measured~\cite{phobWhitePaper}.}
\end{figure}

The {\phob} experiment used a multiplicity-based measurement, the amount
of energy deposited into the Paddles~\cite{Back:2001xy}, to determine
the centrality of {\AuAu} collisions. Similar multiplicity-based
measurements were used to classify the centrality of {\dAu} collisions.
However, it was found that in {\dAu}, special care had to be taken to
ensure that the choice of centrality measure did not bias the physics
analysis. As discussed in~\cite{phobWhitePaper}, different biases were
observed for centrality variables that used data from different ranges
of {\prap}. This can be seen by comparing the reconstructed multiplicity
distributions to the true distributions in \ac{HIJING} collisions, for
both \ac{EOct} and \ac{ERing} centrality cuts. As shown in
\stolenfig{rslt:fig:eoctERingBias}{phobWhitePaper}, centrality cuts
based on the \ac{ERing} variable produced little or no bias on the
multiplicity reconstructed in the central {\prap} region. On the other
hand, centrality cuts based on \ac{EOct} did bias the reconstructed
multiplicity. The average peripheral collision, as selected by
\ac{EOct}, was seen to have a lower multiplicity in the region of
{\prap} over which \ac{EOct} was measured than did the average unbiased
collision in the same percent cross-section bin. The opposite was true
for central collisions, where the average central collision selected by
\ac{EOct} had a higher multiplicity in the central region of {\prap}
than did the average unbiased central collision.

\begin{figure}[t]
   \centering
   \subfigure[$\dnchdeta$ vs Centrality Bin]{
      \label{rslt:fig:crsScnMultERingEOct}
      \includegraphics[width=0.4\linewidth]{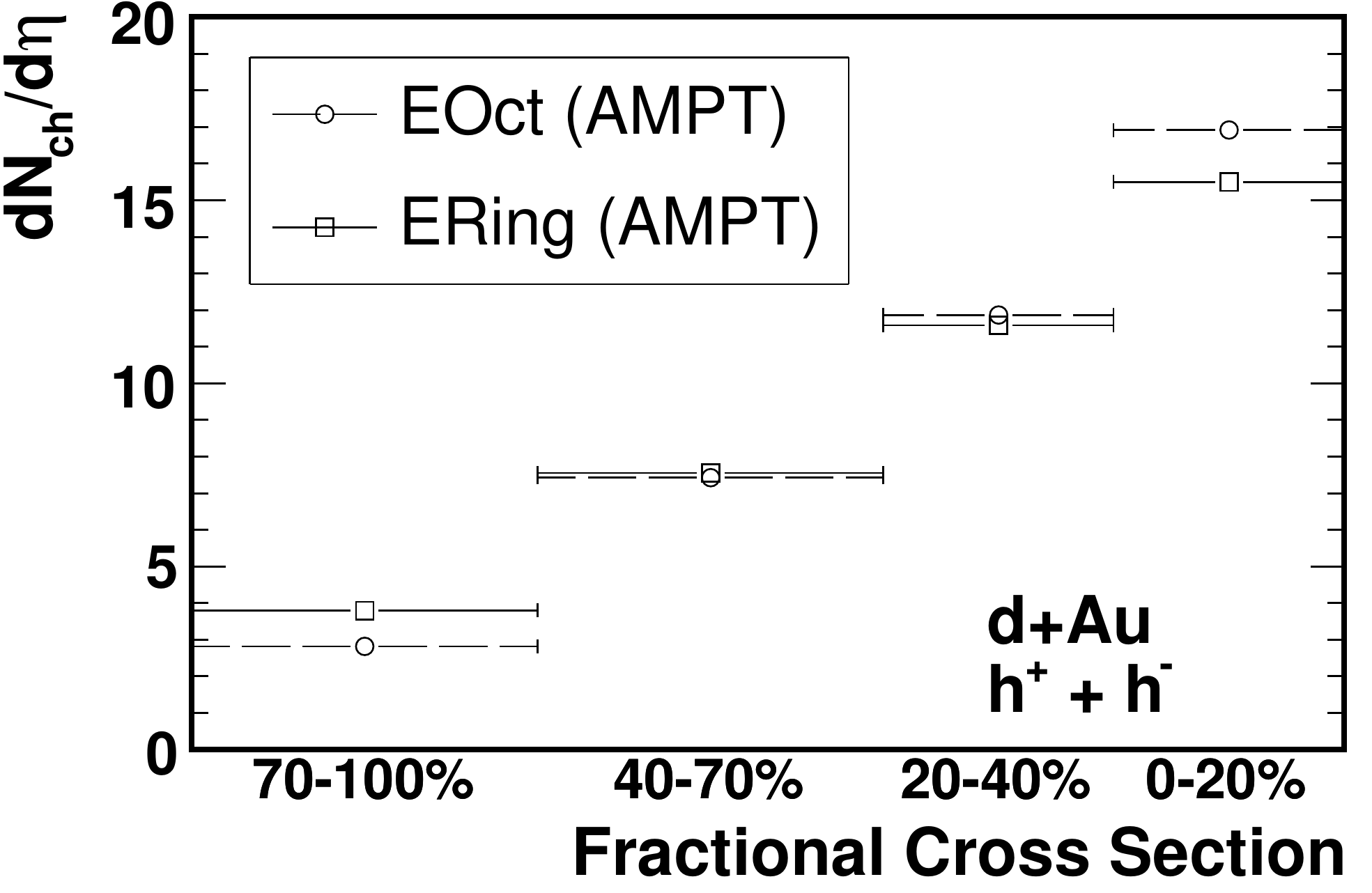}
   }
   \subfigure[$\dnchdeta$ Percent Difference]{
      \label{rslt:fig:crsScnMultPctDiffERingEOct}
      \includegraphics[width=0.4\linewidth]{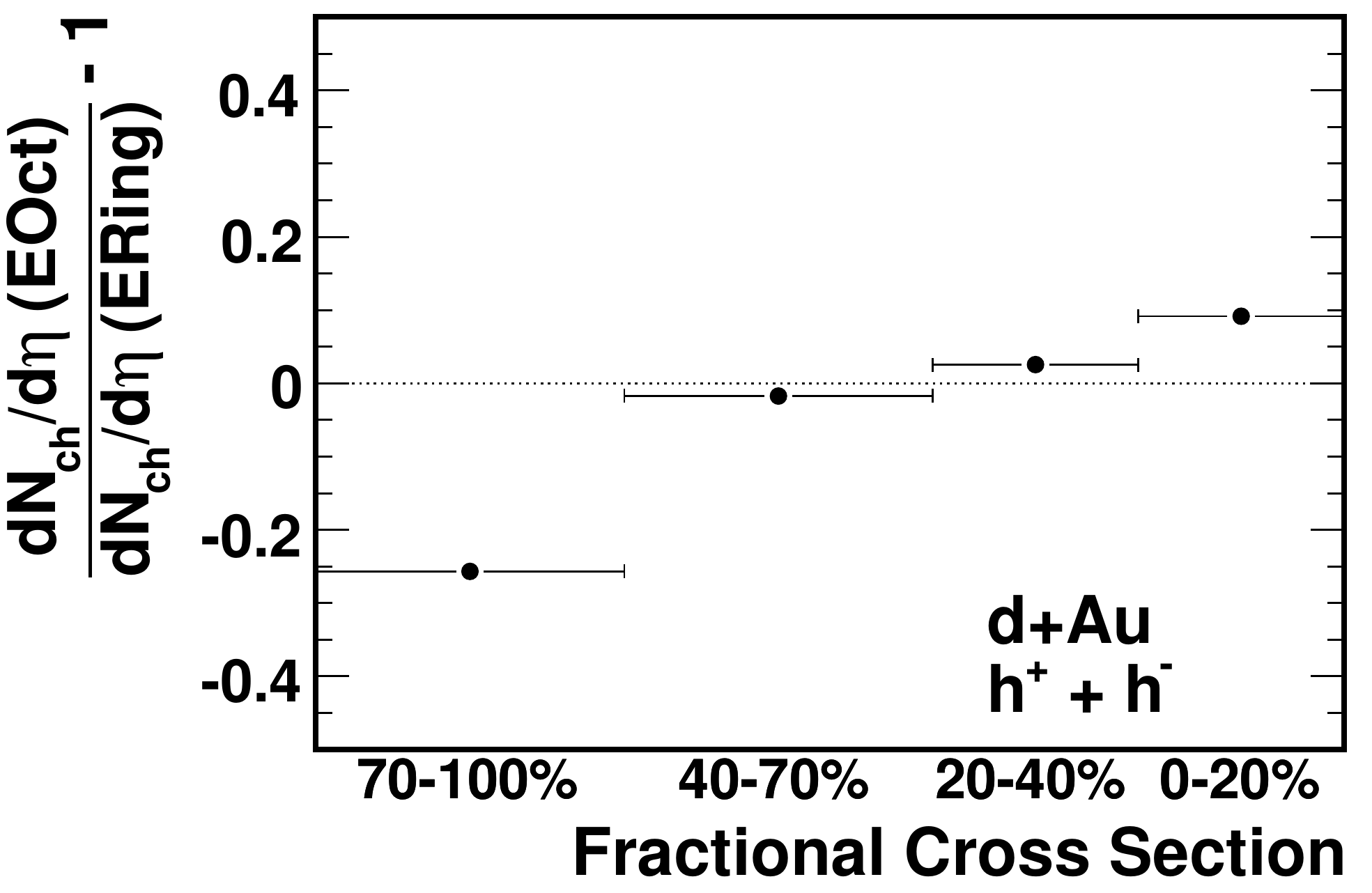}
   }
   \caption{   \label{rslt:fig:crsScnMult}
      \subref{rslt:fig:crsScnMultERingEOct}~The measured $\dnchdeta$ at
      $\aveeta=0.8$ as a function of fractional cross
      section for both \protect\acs{EOct} and \protect\acs{ERing}
      centrality cuts. Horizontal error bars represent the width of the
      bin. \subref{rslt:fig:crsScnMultPctDiffERingEOct}~The fractional
      difference between the integrated yield measured using
      \protect\acs{ERing} and \protect\acs{EOct}.}
\end{figure}

\begin{figure}[t]
   \begin{center}
      \includegraphics[width=0.4\linewidth]{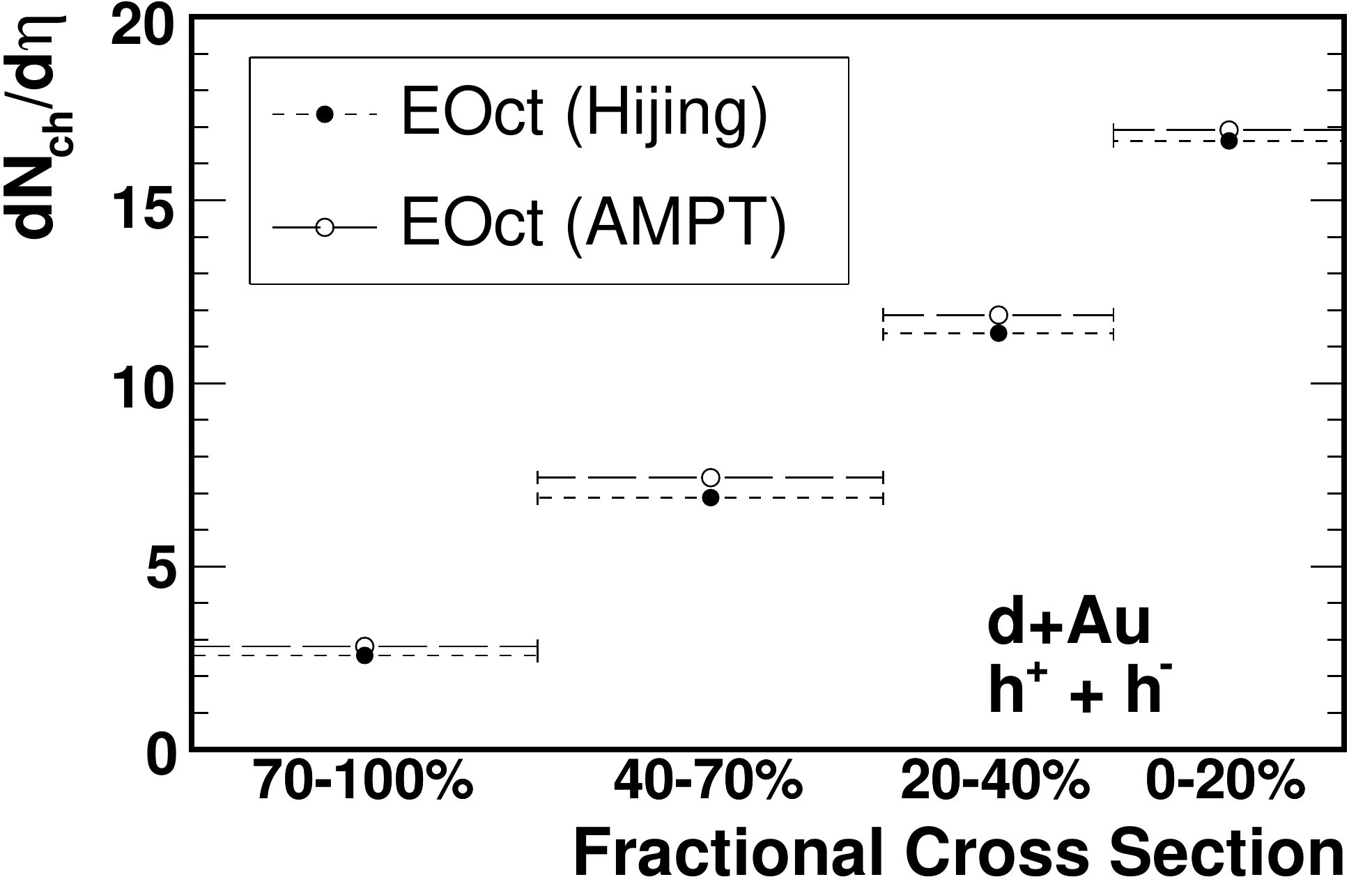}
   \end{center}
   \caption{\label{rslt:fig:crsScnMultEOctHijAMPT}
      The measured $\dnchdeta$ at $\aveeta=0.8$ in
      \protect\acs{EOct} centrality bins is the same (within 10\%)
      whether \protect\acs{HIJING} or \protect\acs{AMPT} was used to
      find the centrality cuts.}
\end{figure}

This biasing effect was also seen in the spectra data presented in this
thesis. As shown in \fig{rslt:fig:crsScnMult}, the integrated yield
measured in each centrality bin is dependent upon the choice of
centrality measure. The $\dndeta$ values obtained from the fits to the
\mbox{$(h^{+} + h^{-})/2$} data were doubled, to obtain the integrated
yield of all charged hadrons. The bias introduced by \ac{EOct}
centrality cuts on the (near {\mrap}) integrated yield can be seen in
\fig{rslt:fig:crsScnMultPctDiffERingEOct}. In peripheral events, the
integrated yield measured using \ac{EOct} cuts is lower than the
integrated yield measured using \ac{ERing} cuts, while in the most
central events, the opposite effect is seen. This is the same bias
observed in the \ac{HIJING} simulations shown in
\fig{rslt:fig:eoctERingBias}. That the bias was observed both in
\ac{HIJING} simulations as well as in the data -- with cuts made using
efficiency estimates from \ac{AMPT} -- suggests that the effect is not
somehow a product of the chosen model. Indeed, the integrated yield
measured using \ac{EOct} cuts found with \ac{HIJING} agrees (to better
than 10\% in each centrality bin) with the integrated yield measured
using cuts found with \ac{AMPT}, as shown in
\fig{rslt:fig:crsScnMultEOctHijAMPT}.

%---------------------------------------------------------------
\subsection{$\Npart$ Parametrization}
\label{rslt:cent:npt}
%---------------------------------------------------------------

\begin{figure}[t!]
   \centering
   \subfigure[$\dnchdeta$ vs $\avenpt$]{
      \label{rslt:fig:eringEoctMultVsNpart}
      \includegraphics[width=0.4\linewidth]{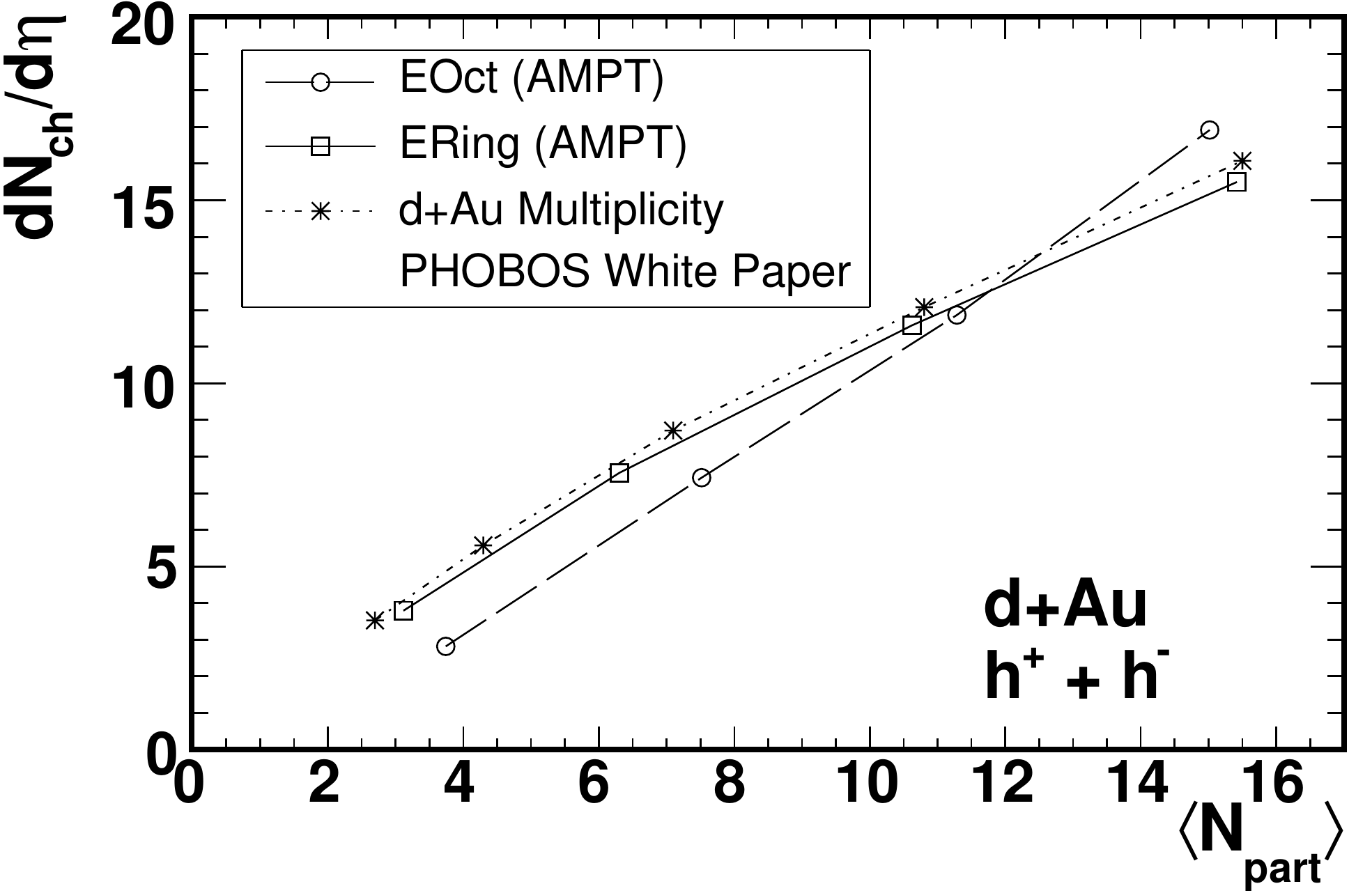}
   }
   \subfigure[$\avenpt$ Percent Difference]{
      \label{rslt:fig:npartERingEOctPctDiff}
      \includegraphics[width=0.4\linewidth]{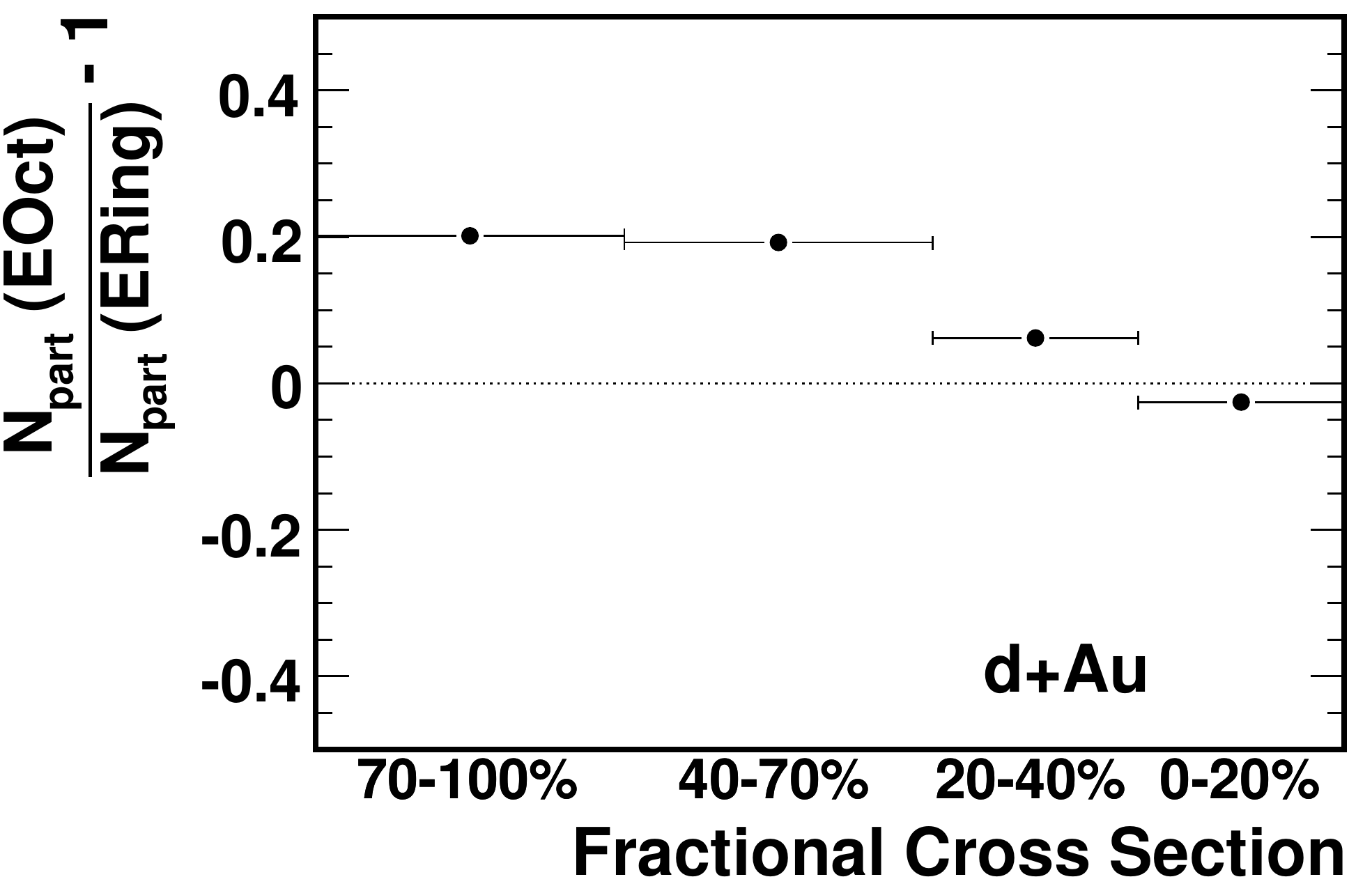}
   }
   \caption{   \label{rslt:fig:eringEoctMult}
      \subref{rslt:fig:eringEoctMultVsNpart}~The measured $\dnchdeta$ at
      $\aveeta=0.8$ as a function of $\avenpt$ for both
      \protect\acs{EOct} and \protect\acs{ERing} centrality cuts. The
      lines simply connect the data points to guide the eye. The data
      shown as asterisks are taken from~\cite{phobWhitePaper}. Five
      centrality bins determined by \protect\acs{ERing} and
      \protect\acs{HIJING} were used for that analysis.
      \subref{rslt:fig:npartERingEOctPctDiff}~The percent difference in
      $\avenpt$ as estimated using \protect\acs{EOct} and
      \protect\acs{ERing} centrality cuts.}
\end{figure}

Thus it is clear that the classification of {\dAu} interactions by
fractional cross section does not select a unique set of collisions.
However, it has been observed that the total charged particle
multiplicity scales with the number of participating
nucleons\footnote{In this thesis, a participant is a nucleon that has
undergone at least one \emph{inelastic} collision, and is equivalently
known as a ``wounded nucleon.''} in hadron-nucleus~\cite{Elias:1978ft,
Bialas:1976ed}, deuteron-nucleus~\cite{Back:2004mr, Bialas:2004su} and
heavy ion collisions~\cite{Back:2003xk}. Therefore, one may expect that
the multiplicity near {\mrap} may be more accurately parametrized by
$\Npart$, rather than by fractional cross section.

The integrated yield of charged hadrons is shown as a function of
$\avenpt$ for two different centrality measures in
\fig{rslt:fig:eringEoctMultVsNpart}. The data labeled ``{\dAu}
Multiplicity'' were taken from~\cite{phobWhitePaper}. To ensure that the
data from the multiplicity measurement were compatible with the data
from the spectra analysis, the {\prap} range of the spectra had to be
accounted for. This was done by averaging the multiplicity measurements
of $\dndeta$ at $\eta=0.7$ and $\eta=0.9$, to match the $\dndeta$ at
$\aveeta=0.8$ obtained from the spectra. The multiplicity measurement
used \ac{ERing} as the centrality measure, but used \ac{HIJING}
simulations to estimate $\Npart$. As seen in
\fig{rslt:fig:eringEoctMultVsNpart}, the multiplicity measurement
from~\cite{phobWhitePaper} agrees quite well with the $\dndeta$ obtained
from the spectra analysis using \ac{ERing} cuts. Such a comparison of
the two results provides an important cross-check on the two independent
analyses, and adds confidence in the accuracy of the results.

With centrality parametrized by $\Npart$, the bias on the integrated
yield reconstructed with \ac{EOct} centrality cuts remains. Indeed it
must, since while the integrated yield reconstructed in the most
peripheral \ac{EOct} bin was \emph{lower} than in the same \ac{ERing}
bin (see \fig{rslt:fig:crsScnMultPctDiffERingEOct}), the average number
of participants was estimated to be \emph{larger} in the peripheral
\ac{EOct} bin than in the peripheral \ac{ERing} bin, as shown in
\fig{rslt:fig:npartERingEOctPctDiff} and \ptab{data:tab:localNpartdAu}.
$\Npart$ being larger in the most peripheral \ac{EOct} bin and smaller
in the most central \ac{EOct} bin, as compared to the corresponding
\ac{ERing} bins, is a consequence of the poorer $\Npart$ resolution
predicted by the models for \ac{EOct} as compared to \ac{ERing}. That
is, collisions with the same $\Npart$ will have a wider spread in
\ac{EOct} values than in \ac{ERing} values. These fluctuations tend to
diminish the distinction between the average number of participants in
neighboring centrality bins.

%---------------------------------------------------------------
\subsection{Two-Component Parametrization}
\label{rslt:cent:ncoll}
%---------------------------------------------------------------

\begin{figure}[t]
   \centering
   \subfigure[Two-Component Parametrization]{
      \label{rslt:fig:eringEoctMultVsNcolNpt}
      \includegraphics[width=0.4\linewidth]{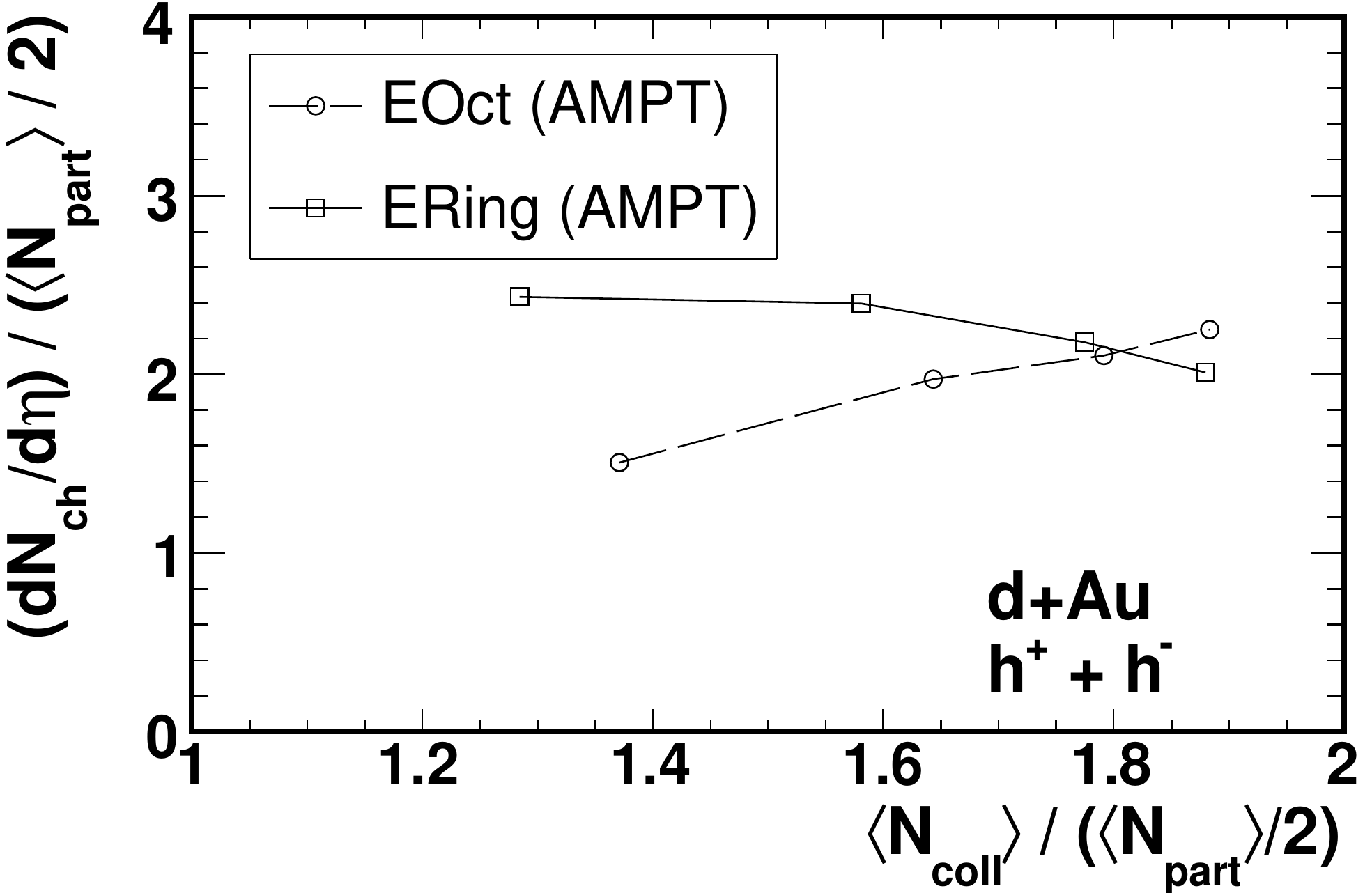}
   }
   \subfigure[$\dnchdeta$ vs $\ave{\nu}$]{
      \label{rslt:fig:eringEoctMultVsNu}
      \includegraphics[width=0.4\linewidth]{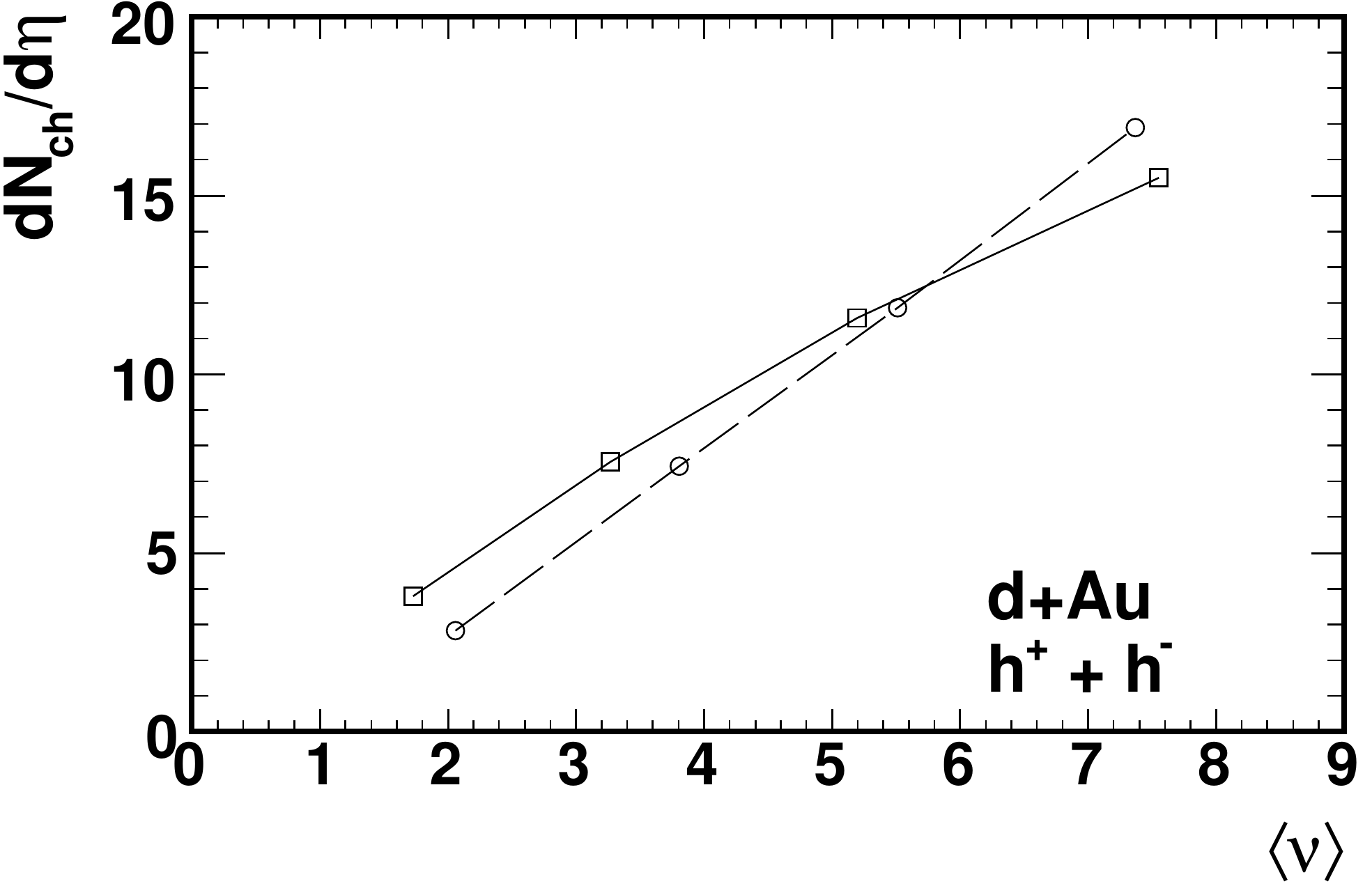}
   }
   \caption{   \label{rslt:fig:eringEOctNptNcl}
      \subref{rslt:fig:eringEoctMultVsNcolNpt}~The measured $\dnchdeta$ 
      per participant pair at $\aveeta=0.8$ as a function of
      $\avencl/(\avenpt/2)$. The integrated yield bias from
      \protect\acs{EOct}  centrality cuts is not resolved by the
      two-component parametrization.
      \subref{rslt:fig:eringEoctMultVsNu}~$\dnchdeta$ as a function of
      \ave{\nu}.}
\end{figure}

While the \emph{total} multiplicity in various collision systems has
been shown to scale with the number of participants, the {\mrap}
multiplicity in heavy ion collisions at $\snn = 200~\gev$ is better
described by a two-component model~\cite{Back:2002uc}. The motivation
for such a model came from the expectation that as the collision energy
increases, hard-scattering (i.e.~jet production) becomes more important;
however see~\cite{Back:2002uc} for a measurement of the magnitude of
this effect. Further, processes such as jet production are expected to
scale with the number of inelastic collisions in the absence of final
state effects (such as energy loss of produced particles traveling
through a dense medium). Thus, a two component parametrization of the
{\mrap} multiplicity of heavy ion collisions was
proposed~\cite{Wang:2000bf, Kharzeev:2000ph}

\begin{equation}
   \label{rslt:eq:twocomp}
\frac{d\!N_{\mathrm{ch}}}{\mathit{d\eta}} = n_{\mathrm{pp}}
   \sqb{(1-x) (\avenpt / 2) + x \avencl}
\end{equation}

\noindent%
where $n_{\mathrm{pp}}$ is the multiplicity in {\pp} and $x$ is the
fraction of charged particle production due to hard scattering
processes. This implies that the {\mrap} multiplicity per participant
pair should scale with the number of collisions per participant pair

\begin{align}
\frac{1}{\avenpt / 2} \frac{d\!N_{\mathrm{ch}}}{\mathit{d\eta}} &= 
   n_{\mathrm{pp}} \sqb{(1-x) + x \frac{\avencl}{\avenpt / 2}}
% \notag \\
%\frac{1}{\avenpt / 2} \frac{d\!N_{\mathrm{ch}}}{\mathit{d\eta}} &\propto
%   \frac{\avencl}{\avenpt / 2}
   \label{rslt:eq:twocompscale}
\end{align}

\noindent%
since both $n_{\mathrm{pp}}$ and $x$ are independent of centrality. The
integrated yield per participant pair measured in {\dAu} collisions by the
spectra analysis using both \ac{EOct} and \ac{ERing} cuts as a function
of \mbox{$\avencl / ( \avenpt / 2)$} is shown in
\fig{rslt:fig:eringEoctMultVsNcolNpt}. It is clear that this
parametrization does not account for the bias imposed on the
integrated yield by the \ac{EOct} cuts. The bias is also present when the
centrality of {\dAu} collisions is parametrized by
$\ave{\nu}\equiv\ave{\smash[tb]{\Ncoll/\Npard}}$~\cite{Back:2003ff}, as
seen in \fig{rslt:fig:eringEoctMultVsNu}.

%---------------------------------------------------------------
\subsection{Au-PCAL Centrality Cuts}
\label{rslt:cent:pcal}
%---------------------------------------------------------------

Centrality cuts based on the amount of spectator material should provide
a reliable estimate of centrality parameters, such as
$\ave{\nu}$~\cite{DeMarzo:1984eg}. This was the motivation for
constructing the \ac{Au-PCAL}; to measure the energy of spectator
protons (those that did not suffer an inelastic collision) of the gold
nucleus. However, unlike previous fixed target and emulsion experiments
that could directly observe such particles~\cite{DeMarzo:1984eg,
Brick:1989dm, Cherry:1994ai, Chemakin:1999jd, Abbott:1990ka}, the
{\phob} \ac{Au-PCAL} was, for a number of reasons, not able to count the
number of spectator protons emerging from a collision. First, no
tracking detectors were constructed to observe these particles; only the
total energy was measured. Second, this energy was seen to fluctuate
significantly for collisions of a similar centrality (see
\pfig{recon:fig:pcalEoctCorl}). Simulations of single nucleons traveling
at or near beam rapidity suggested that the fluctuations had two main
sources. One was leakage of the neutron-induced hadron shower in the
\ac{ZDC} into nearby \ac{Au-PCAL} modules. The other was due to the
passage of protons through a significant amount of iron as they were
bent out of the DX-magnet, see \pfig{exp:fig:protpaths}. Another
possible source of fluctuations was related to the break-up of the
nucleus, and was extremely difficult to simulate. As spectator protons
leave the interaction region, they may be bound inside clusters such as
deuterons or alpha particles. These clusters would have a different
charge to mass ratio than a free proton, and would not be bent by the
\ac{RHIC} magnets into the \ac{Au-PCAL} detector. Due to the difficulty
of accurately simulating such processes, the magnitude of these effects
was not determined.

\begin{figure}[t]
   \centering
   \subfigure[EPCAL Cuts]{
      \label{rslt:fig:crsScnPCAL}
      \includegraphics[width=0.4\linewidth]{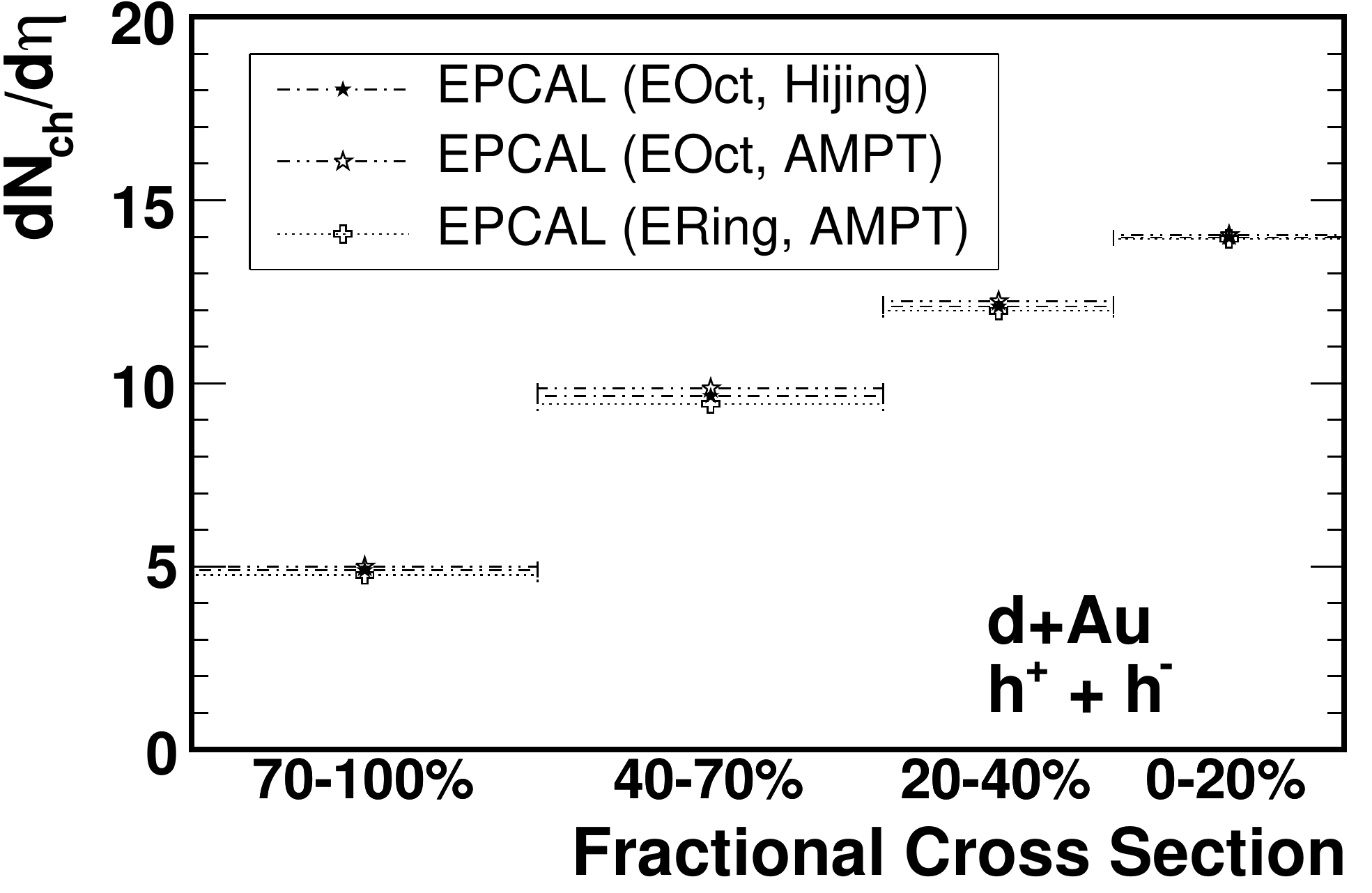}
   }
   \subfigure[EPCAL and ERing Cuts]{
      \label{rslt:fig:crsScnPCALERing}
      \includegraphics[width=0.4\linewidth]{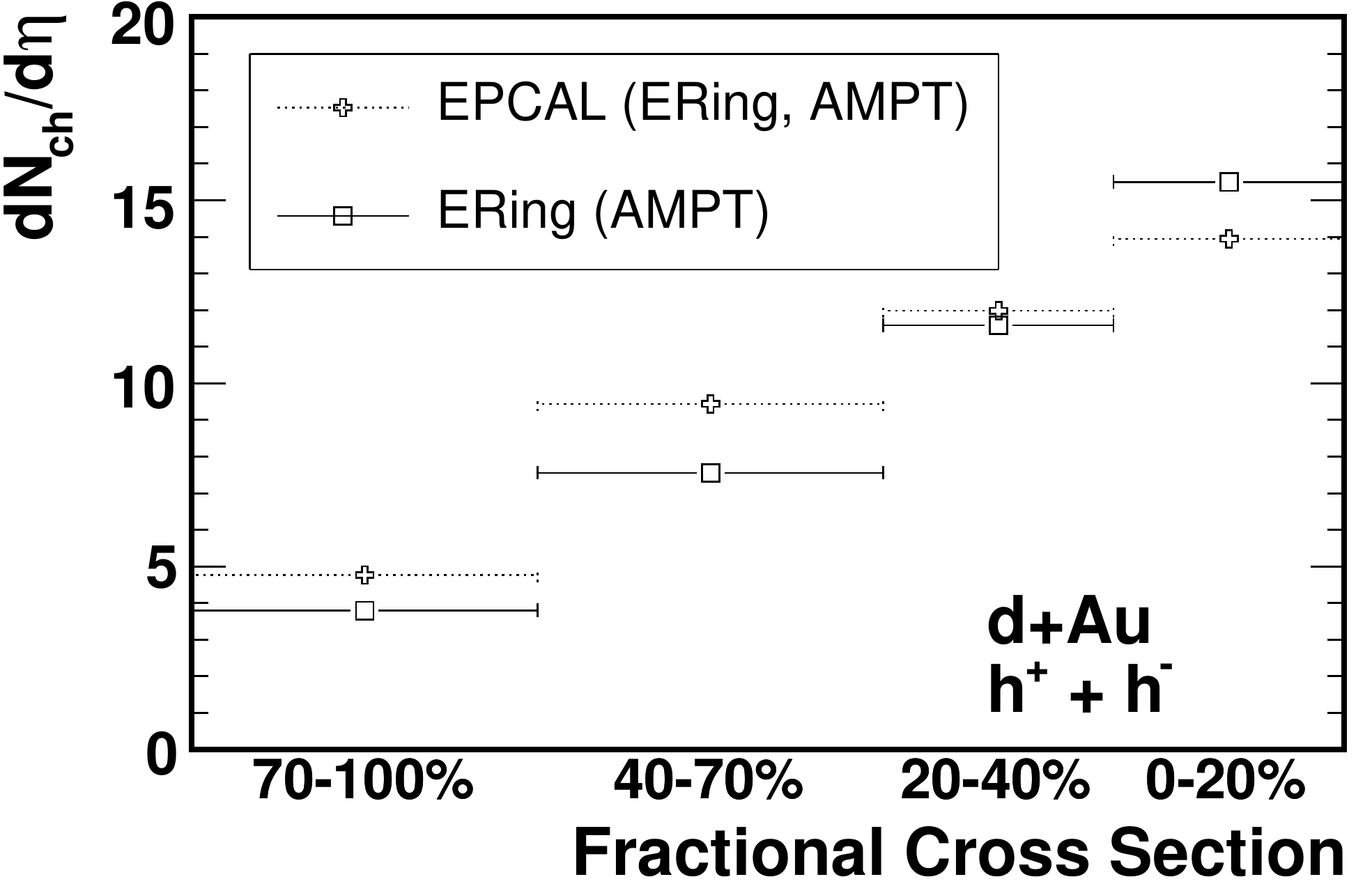}
   }
   \caption{   \label{rslt:fig:crsScnPCALcomp}
      \subref{rslt:fig:crsScnPCAL}~Comparison of the measured
      $\dnchdeta$ at $\aveeta=0.8$ obtained using \protect\acs{EPCAL}
      cuts with various correlations used to estimate efficiency.
      \subref{rslt:fig:crsScnPCALERing}~The integrated yield reconstructed
      in \protect\acs{EPCAL} centrality bins compared to
      \protect\acs{ERing} centrality bins.}
\end{figure}

Despite these complications, there was no reason to suspect that
centrality cuts based on \ac{EPCAL} would cause any of the biasing
effects seen with \ac{EOct} cuts. That is, \ac{EOct} was thought to bias
measurements of multiplicity in the {\mrap} region since \ac{EOct}
itself was a measurement of multiplicity in the {\mrap} region. However,
energy deposited in the \ac{Au-PCAL} was not due to produced particles
at all (having been shielded from such background, see
\sect{exp:phobdet:calo:pcalshield}) and therefore should not have biased
measurements of produced particles. Thus it was expected that the
multiplicity measured using \ac{EPCAL} centrality cuts should be the
same regardless of what \ac{MC} simulation and variable were used to
estimate the efficiency (see \sect{recon:cent:pcal}). This is indeed the
case, as shown in \fig{rslt:fig:crsScnPCAL}. However, since the
measurements presented in \fig{rslt:fig:crsScnPCAL} were performed using
the same data set, any differences in the measurements can only be due
to differences in the efficiency estimates.

The integrated yield measured using \ac{EPCAL} cuts, with efficiency
estimated using \ac{ERing} in \ac{AMPT} simulations, is compared to the
integrated yield measured using \ac{ERing} cuts in
\fig{rslt:fig:crsScnPCALERing}. While both the \ac{EPCAL} and \ac{ERing}
centrality measures are thought to impose little or no bias on the
integrated yield measurement, the integrated yield in the most peripheral
\ac{EPCAL} bin is higher than that of the most peripheral \ac{ERing}
bin, and the opposite behavior is seen in the most central bin. One
possible explanation for this is that the trigger and event selection
efficiency may have been incorrectly estimated. This is almost certainly
true for peripheral events, where fluctuations in the correlation
between \ac{EPCAL} and \ac{ERing} resulted in a range of finite values
of \ac{ERing} in collisions that deposited no energy in the
\ac{Au-PCAL}, which in turn led to an unrealistically large estimate of
the efficiency to detect collisions having very small values of
\ac{EPCAL}. In addition, large fluctuations in the \ac{EPCAL} signal for
collisions of similar centrality would reduce the centrality resolution
of \ac{EPCAL}, causing collisions in the peripheral bin of \ac{EPCAL} to
be more central than collisions in the peripheral bin of \ac{ERing}.

\begin{figure}[t]
   \centering
   \subfigure[$\avenpt$ Parametrization]{
      \label{rslt:fig:nptPCALERing}
      \includegraphics[width=0.4\linewidth]{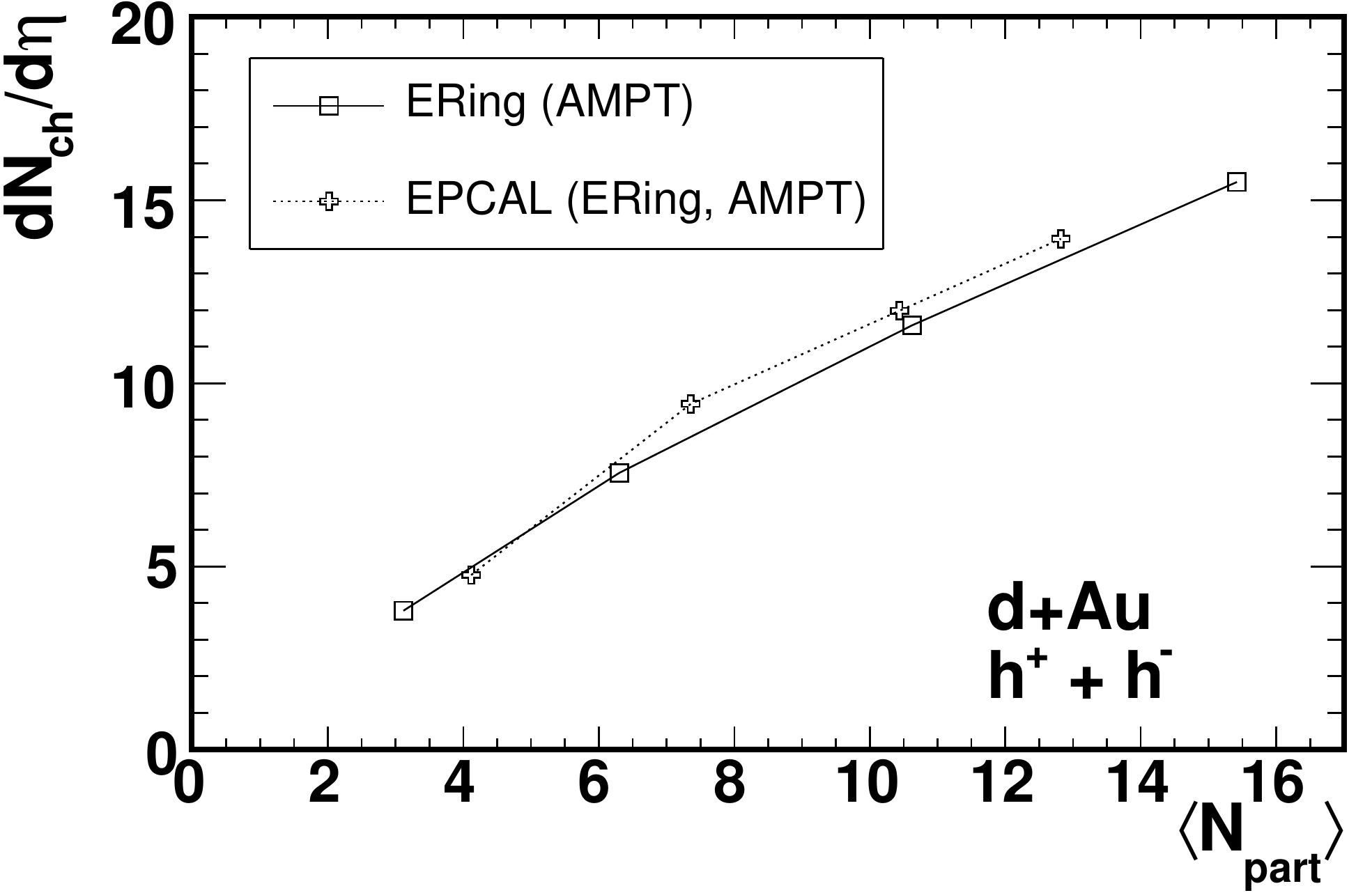}
   }
   \subfigure[$\ave{\nu}$ Parametrization]{
      \label{rslt:fig:nuPCALERing}
      \includegraphics[width=0.4\linewidth]{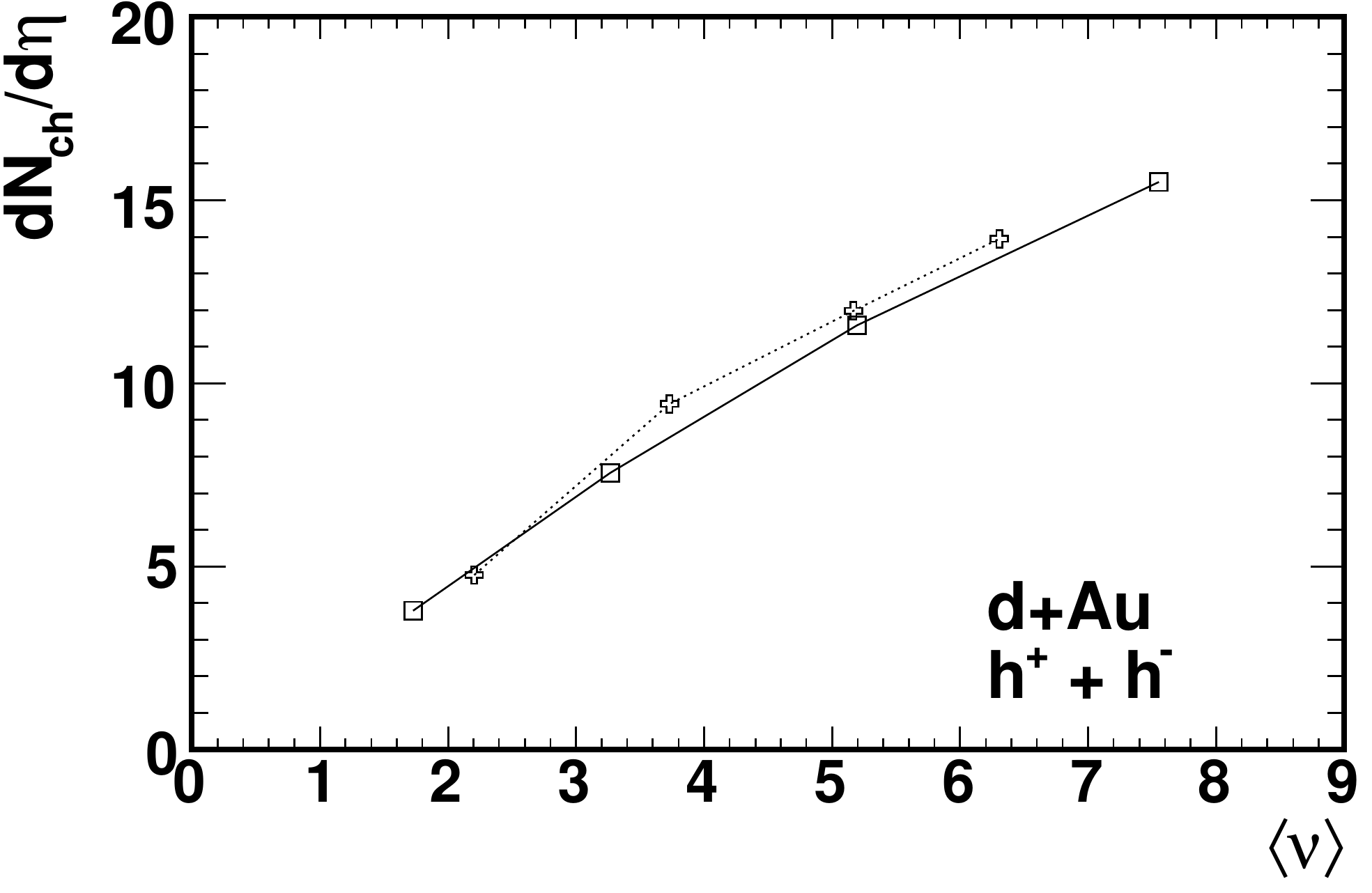}
   }
   \caption{   \label{rslt:fig:nptNuPcalERing}
      Comparison of the measured $\dnchdeta$ at $\aveeta=0.8$ obtained
      using \protect\ac{EPCAL} and \protect\ac{ERing} centrality cuts.
      \subref{rslt:fig:nptPCALERing}~Centrality is parametrized by
      $\avenpt$. \subref{rslt:fig:nuPCALERing}~Centrality is
      parametrized by $\nu$.}
\end{figure}

\begin{figure}[t]
   \centering
   \subfigure[EPCAL Cuts from EOct]{
      \label{rslt:fig:nptPCALEOct}
      \includegraphics[width=0.4\linewidth]{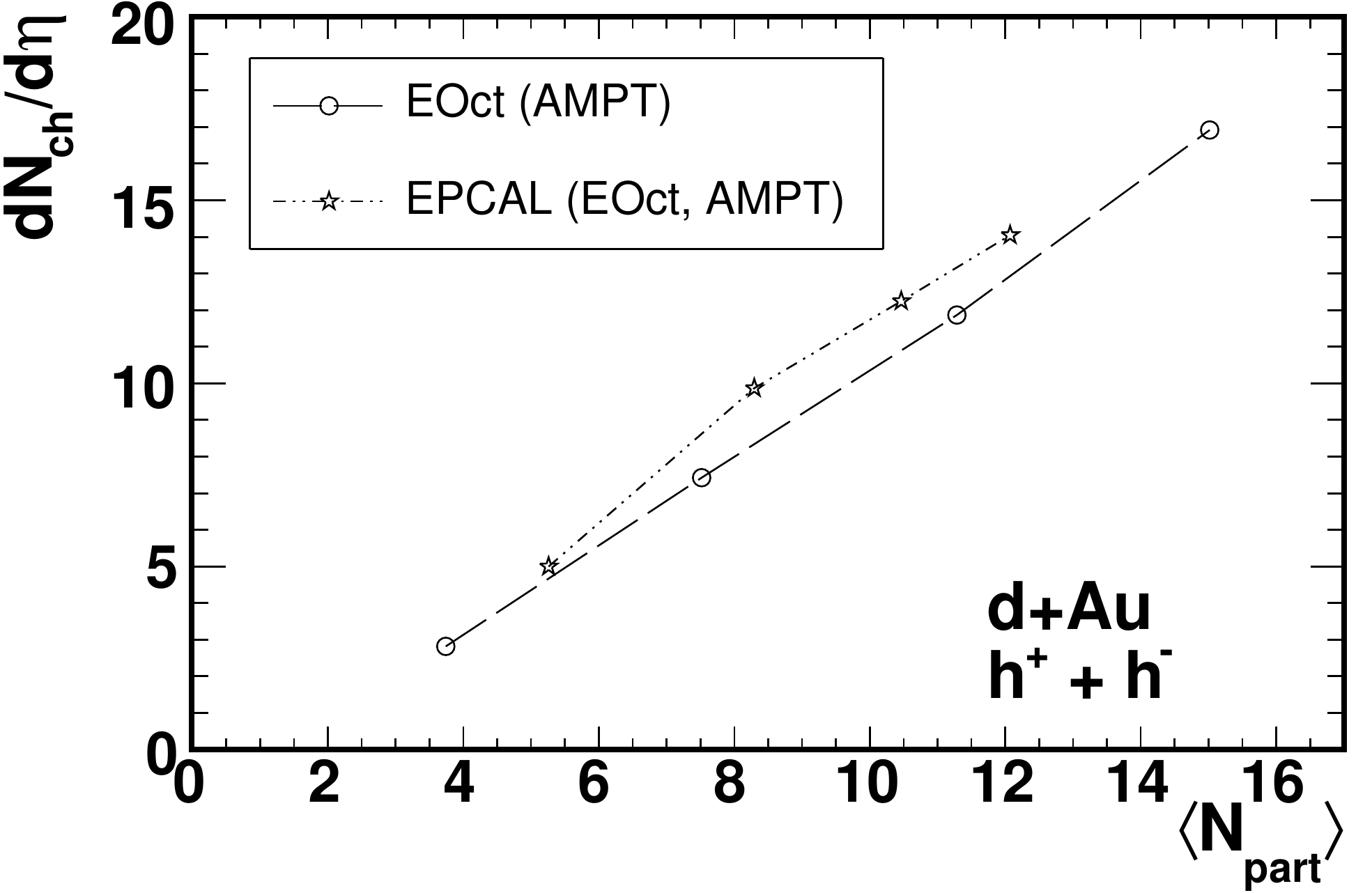}
   }
   \subfigure[Comparison of EPCAL Cuts]{
      \label{rslt:fig:nptPCALEOctERing}
      \includegraphics[width=0.4\linewidth]{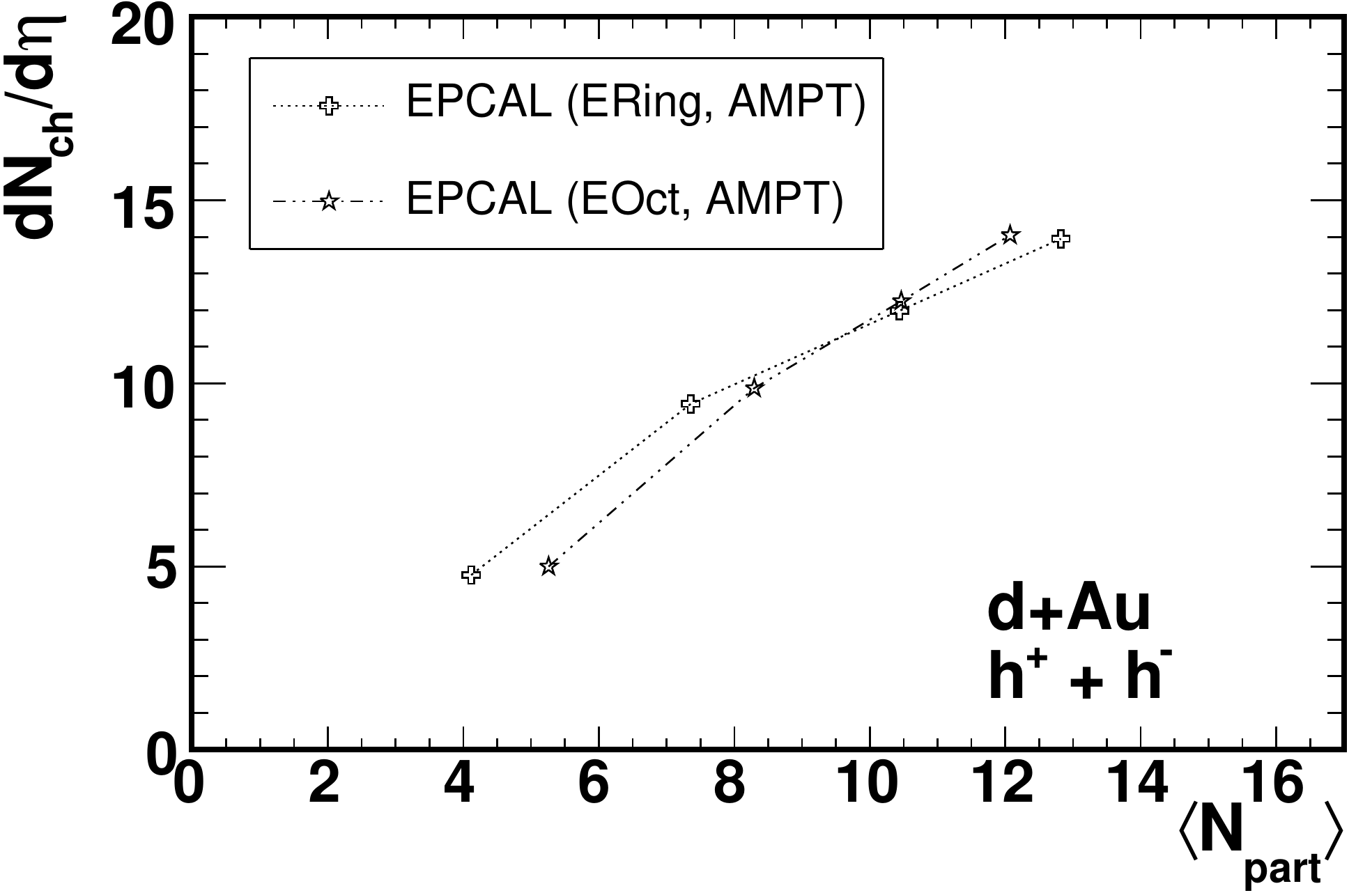}
   }
   \caption{   \label{rslt:fig:nptNuPcalEOctCmp}
      The measured $\dnchdeta$ at $\aveeta=0.8$ obtained using
      \protect\ac{EPCAL} centrality cuts.
      \subref{rslt:fig:nptPCALEOct}~The integrated yield in
      \protect\ac{EPCAL} bins determined using \protect\ac{EOct} follows
      the same trend as the integrated yield in \protect\ac{EOct} bins.
      \subref{rslt:fig:nptPCALEOctERing}~Estimates of $\avenpt$ are
      biased by the choice of simulation variable used to estimate
      $\avenpt$. Compare to \fig{rslt:fig:eringEoctMultVsNpart}.}
\end{figure}

Both of these effects lead to the conclusion that assigning fractional
cross section widths to the centrality bins of \ac{EPCAL}, found using
the method described in \sect{recon:cent:pcal}, is suspect. Since the
fractional cross section that each bin of \ac{EPCAL} represents is not
well determined, the centrality of the bins may be better described by
collision geometry parameters such as $\Npart$. This indeed seems to be
the case, as suggested by \fig{rslt:fig:nptNuPcalERing} which shows that
the integrated yield measured in bins of \ac{EPCAL} using \ac{ERing} agrees
with the integrated yield measured in bins of \ac{ERing}. However, the
integrated yield measured in bins of \ac{EPCAL} that were determined using
\ac{EOct} in \ac{AMPT} simulations, shown in \fig{rslt:fig:nptPCALEOct},
follow the same trend as the integrated yield measured in bins of \ac{EOct}.
This is not entirely surprising; it is not unreasonable to expect that
the estimation of centrality parameters such as $\avenpt$ was biased by
the variable used in the simulations. This bias can be seen by comparing
the integrated yield reconstructed in \ac{EPCAL} centrality bins as a
function of $\avenpt$, where $\avenpt$ was determined using either
\ac{ERing} or \ac{EOct} in the simulations, as shown in
\fig{rslt:fig:nptPCALEOctERing} (compare to
\fig{rslt:fig:eringEoctMultVsNpart}). While studies were performed to
try to correct for this bias on centrality parameters like $\Npart$, no
reliable method for doing so was found. 

%---------------------------------------------------------------
\subsection{Summary of Centrality Discussion}
\label{rslt:cent:summary}
%---------------------------------------------------------------

\begin{figure}[t]
   \begin{center}
      \includegraphics[width=0.6\linewidth]{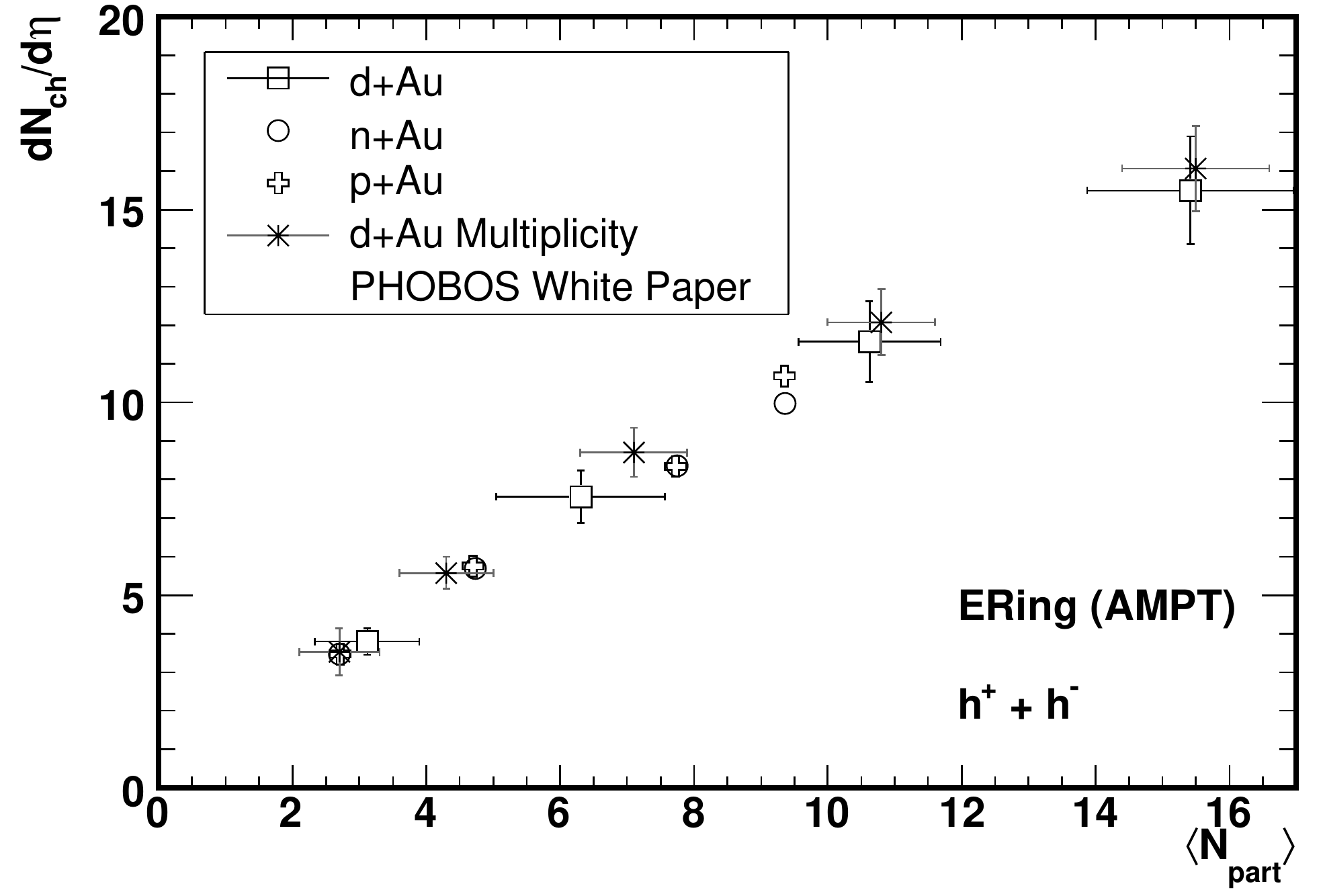}
   \end{center}
   \caption{\label{rslt:fig:eringTagMult}
      The measured $\dnchdeta$ at $\aveeta=0.8$ obtained using
      \protect\ac{ERing} centrality cuts. Systematic errors (90\%~C.L.)
      are shown for the {\dAu} measurements; statistical errors are
      negligible. Systematic errors on the nucleon-nucleus measurements
      are not shown, but are of similar order. A consistent dependence
      of the multiplicity on $\Npart$ is observed for the three
      different collision systems.}
\end{figure}

These studies make it clear that measurements of produced particles in
the {\mrap} region are biased, particularly in small systems, by
centrality cuts placed on a variable that is itself a measure of {\mrap}
multiplicity. The bias was not found to depend on the choice of
simulation model used to determine the triggering efficiency and the
parameterizations of centrality (such as $\Npart$). The least biased
centrality variable was found to be \ac{ERing}, most likely due to the
fact that it measures particles away from {\mrap}. While it is believed
that centrality cuts on \ac{EPCAL} also impose little or no bias on
measurements of produced particles near {\mrap}, it was found that these
fractional cross section cuts selected a different class of collisions
than did fractional cross section cuts based on \ac{ERing} or \ac{EOct}.
Further, attempts to describe the centrality of \ac{EPCAL} bins by
parameters like $\Npart$ were shown to preserve the bias of the
centrality variable (i.e.~\ac{EOct}) used to estimate the average of the
parameter in the \ac{MC} simulations. That is, while the integrated
yield reconstructed in \ac{EPCAL} centrality bins may not itself have
been biased, attempts to quantify the centrality of \ac{EPCAL} bins were
biased. For this reason, the majority of the discussion in this chapter
will be on results found using \ac{ERing} centrality cuts. Using these
cuts gives a consistent set of results for {\dAu}, {\pAu} and {\nAu}
collisions, as shown in \fig{rslt:fig:eringTagMult} and discussed in the
next section.

%---------------------------------------------------------------
\section{d+Au as a Control Experiment}
\label{rslt:rda}
%---------------------------------------------------------------

The yield of hadrons in {\dAu} collisions played a vital role in the
investigation of particle production in high energy {\AuAu}
collisions~\cite{dAuPressRelease}. In the absence of any nuclear or
produced medium effects, a heavy ion collision would essentially be a
collection of independent nucleon-nucleon collisions\footnote{While one
might expect that particle production in any given binary interaction would
depend on the number of collisions previously suffered by either nucleon,
such dependence is not observed~\cite{Elias:1978ft}.}. In this {\naive}
picture, the yield of an average {\AuAu} collision would be determined
simply by the yield of an average nucleon-nucleon collision multiplied
by the number of binary collisions taking place in the {\AuAu} system.
The nuclear modification factor, $R_X$ given by \eq{rslt:eq:nucmod}, is
therefore a convenient measure with which to test the assumption of
binary collision scaling.

\begin{equation}
   \label{rslt:eq:nucmod}
R_X = \frac{d^{2}\!N_X / \mathit{d\pt} \mathit{d\eta}}{%
   \avencl d^{2}\!N_{\mathrm{NN}} / \mathit{d\pt} \mathit{d\eta}}
   \qquad X = \text{\AuAu, \pAu, etc.}
\end{equation}

This ratio was studied as a function of transverse momentum in order to
separate soft processes from hard scattering processes. Previous studies
of \ac{QCD} have shown that short-range (hard processes) and long-range
(soft processes) interactions factorize in the theory for simpler
collision systems~\cite{Collins:1985ue}. Assuming that factorization
holds for high energy nucleus-nucleus collisions, binary collision
scaling should be expected for large momentum transfer collisions that
produce high-$\pt$ particles. For $\Ncoll$ scaling to hold at
high-$\pt$, the chance for a hard scattering process to occur in a
binary collision should not depend on the number of collisions either of
the two nucleons had previously suffered. It is interesting to note that
in {\AuAu} collisions at $\snn = 200~\gev$, the average number of
collisions per participant has been estimated to be around
4.7~\cite{Wang:2000bf} to 5.2~\cite{Back:2003ff}. The number of
collisions suffered by the deuteron or proton in central {\dAu} or
{\pAu} collisions, respectively, can be about 60\% higher, however, as
shown in \tab{rslt:tab:eringCentPars}. Note that the approximate scaling
of the {\mrap} multiplicity with $\avenpt$, shown in
\fig{rslt:fig:eringTagMult}, cannot be assumed to hold over a wide range
of transverse momentum, since the multiplicity is dominated by low-$\pt$
particles.

The nucleon-nucleon reference used for studies of heavy ion collisions
at $\snn = 200~\gev$ came from the UA1  data~\cite{Albajar:1989an} of
the {\pbarp} inelastic cross section. As described
in~\cite{Back:2003ns}, corrections were applied to the UA1 results to
account for (a)~the conversion from rapidity to {\prap} and (b)~the
difference between the UA1 acceptance (\mbox{$\abs{\smash[bt]{\eta}} <
2.5$}) and the {\phob} acceptance (\mbox{$0.2 < \eta < 1.4$}). The
{\pbarp} reference spectrum was fit by the function

\begin{align}
\frac{1}{2 \pi \pt} \frac{d^{2}\!N}{\mathit{d\pt} \mathit{d\eta}}
   = & \frac{50.9}{2 \pi} \prn{1 + \frac{\pt}{1.59}}^{-11.2} \notag \\
     & \times \frac{\pt}{\sqrt{\pt^2 + 0.00545}} 
      \qquad (y \rightarrow \eta) \notag \\
     & \times \biggl[ \frac{\ln \prn{\exp(1.065 + 0.004 \pt)^{\;40} + 
      \exp(0.85 + 0.07 \pt)^{\;40}}}{40} \notag \\
     & \quad - 0.12 e^{-2 \pt} +  0.04 e^{-3.66 \pt} \biggl] 
      \qquad (\text{acceptance correction})
   \label{rslt:eq:ppfit}
\end{align}

\noindent%
The shape of the {\pbarp} spectra described by \eq{rslt:eq:ppfit} can be
seen in \fig{rslt:fig:exRlYldppda}. An inelastic {\pbarp} cross section
of 41~{\mb} was used to estimate the yield of {\pbarp} collisions given
the differential cross section measurements from UA1.

\begin{figure}[t]
   \begin{center}
      \includegraphics[width=0.6\linewidth]{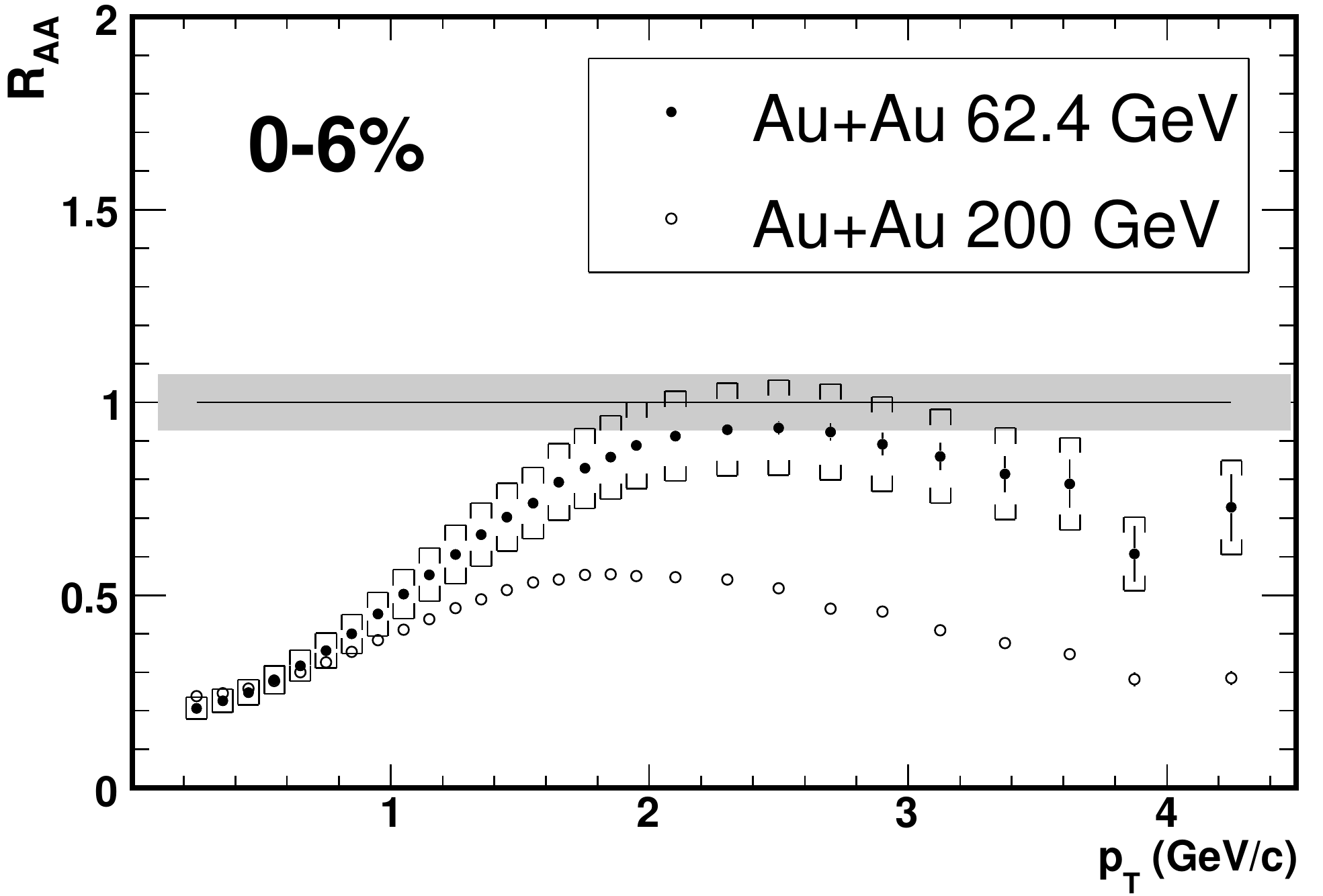}
   \end{center}
   \caption{\label{rslt:fig:RAA20062Central}
      The nuclear modification factor measured by {\phob} in the 0-6\%
      most central {\AuAu} collisions. Brackets on $R_{AA}$ show the
      systematic uncertainties. The solid line shows the expectation of
      $\Ncoll$ scaling and the grey band shows the systematic
      uncertainty on the overall scale due to
      $\Ncoll$~\cite{phobWhitePaper}.}
\end{figure}

The nuclear modification factor of nucleus-nucleus collisions at
\ac{RHIC} has been studied extensively for {\AuAu} interactions at
$\snn = 62.4~\gev$~\cite{Back:2004ra, Arsene:2006pn},
$130~\gev$~\cite{Adcox:2001jp, Adcox:2002pe, Adler:2002xw} and
$200~\gev$~\cite{Back:2003qr, Adams:2003kv, Adler:2003qi, Adler:2003au},
as well as for {\CuCu} interactions at $\snn = 62.4$ and
$200~\gev$~\cite{Alver:2005nb}. As an example of these studies, the
{\phob} measurement of the nuclear modification factor in central
{\AuAu} collisions, $R_{AA}$, at two center of mass energies is shown in
\stolenfig{rslt:fig:RAA20062Central}{phobWhitePaper}.

One of the fundamental conclusions drawn from examination of the
nuclear modification factor was that the production of charged hadrons
in central {\AuAu} collisions at $\snn = 200~\gev$ is highly
suppressed with respect to binary collision scaling. This can be seen
clearly in \stolenfig{rslt:fig:RAA20062Central}{phobWhitePaper}.
However, it was not known from the nucleus-nucleus data alone whether
the suppression was due to initial or final state effects. That is, were
in fact fewer high-$\pt$ hadrons being produced because of some
modification of the initial nuclei, such as the one described
in~\cite{Kharzeev:2002pc}? Or were the expected number of particles
being produced, but losing energy during transit through some medium and
being observed as lower-$\pt$ hadrons, as described
in~\cite{Baier:1996sk}?

The hadron production of nucleon-nucleus collisions at the same center
of mass energy could distinguish between the two possibilities. Any
initial effects that modify the {\Au} nucleus should still be present in
{\NAu} interactions\footnote{`N' here refers to `nucleon,' rather than
nitrogen.}. On the other hand, final state effects caused by a dense
medium in {\AuAu} collisions are not expected to be present in {\NAu}
interactions, as the system size should be too small to produce a medium
that high-$\pt$ particles would interact with. Thus, {\NAu} collisions
would serve as a control experiment for {\AuAu} interactions. At
\ac{RHIC}, these studies were performed using {\dAu} rather than {\NAu}
collisions~\cite{Back:2003ns, Adams:2003im, Adler:2003ii, Adler:2006xd,
Arsene:2004ux, Arsene:2003yk}. This was due to technical reasons: the
\ac{RHIC} machine required mechanical changes in order to perform {\pAu}
collisions ({\nAu} were not possible). However, the reasonable
assumption was made that, due to the small size and weak binding of the
deuteron nucleus, {\dAu} collisions would provide as good a control
experiment for {\AuAu} interactions as {\NAu} collisions would.

%---------------------------------------------------------------
\subsection{$\Ncoll$ Scaling from {\pAu} to {\dAu}}
\label{rslt:rda:ncollpAdA}
%---------------------------------------------------------------

\begin{figure}[t]
   \begin{center}
      \includegraphics[width=0.8\linewidth]{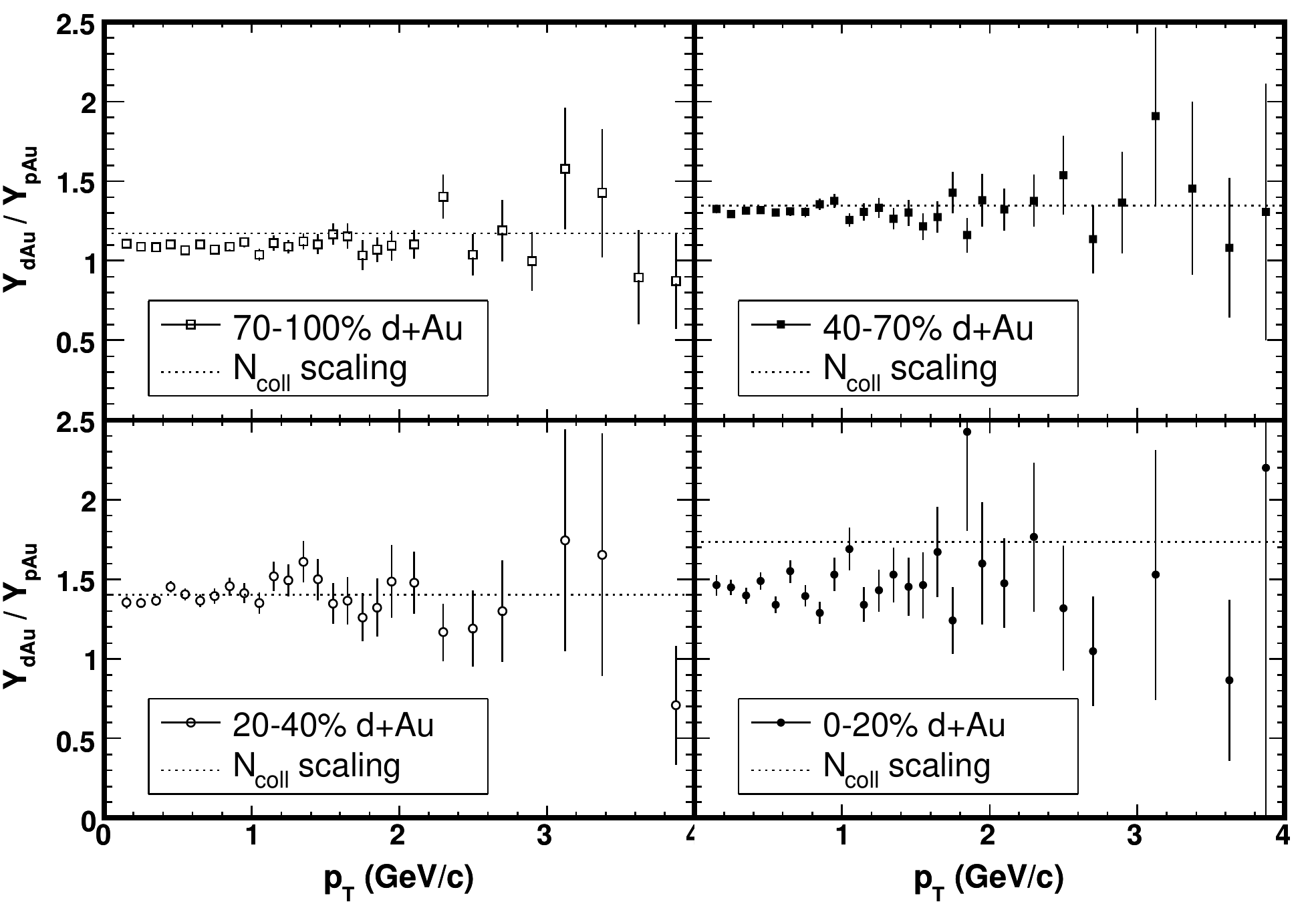}
   \end{center}
   \caption{\label{rslt:fig:dAupAuNcollScaling}
      Ratio of the invariant yield of hadrons in {\dAu} collisions to
      the yield in {\pAu} collisions in each \protect\acs{ERing}
      centrality bin. The dotted line marks the expectation of $\Ncoll$
      scaling. Fractional cross section widths refer only to the {\dAu}
      cross section. See \tab{rslt:tab:eringCentPars} for the relevant
      centrality parameters in each bin. Systematic errors not shown
      (see text).}
\end{figure}

\begin{table}[t]
   \begin{center}
      \begin{tabular}{|rlllll|}
\hline
\multicolumn{6}{|c|}{Centrality Parameters in \protect\acs{ERing} Bins} \\
Cent. Bin & $\ave{b}$ & $\avenpt$ & $\avencl$ & $\ave{\nu}$ & $\dndeta$ \\
\hline
0-20\% {\dAu} & $3.31$ & $15.42\pm1.09$ & $14.49\pm0.88$ & $7.55\pm0.46$ & $15\pm1.4$ \\
 {\pAu} & $6.09$ & $9.36\pm0.66$ & $8.36\pm0.51$ & $8.36\pm0.51$ & $11\pm0.99$ \\
\hdashline[1pt/4pt]
20-40\% {\dAu} & $4.67$ & $10.62\pm0.88$ & $9.43\pm0.67$ & $5.20\pm0.37$ & $12\pm1.1$ \\
 {\pAu} & $6.38$ & $7.72\pm0.64$ & $6.72\pm0.48$ & $6.72\pm0.48$ & $8.3\pm0.75$ \\
\hdashline[1pt/4pt]
40-70\% {\dAu} & $6.29$ & $6.31\pm0.70$ & $4.99\pm0.60$ & $3.27\pm0.39$ & $7.6\pm0.68$ \\
 {\pAu} & $7.18$ & $4.70\pm0.52$ & $3.70\pm0.44$ & $3.70\pm0.44$ & $5.8\pm0.52$ \\
\hdashline[1pt/4pt]
70-100\% {\dAu} & $7.76$ & $3.12\pm0.70$ & $2.00\pm0.60$ & $1.73\pm0.52$ & $3.8\pm0.34$ \\
 {\pAu} & $8.01$ & $2.71\pm0.61$ & $1.71\pm0.51$ & $1.71\pm0.51$ & $3.5\pm0.32$ \\
\hline
      \end{tabular}
   \end{center}
   \caption{   \label{rslt:tab:eringCentPars}
      Comparison of centrality parameters in each \protect\acs{ERing}
      bin for the {\dAu} and {\pAu} collision systems. Parameters of the
      {\nAu} collision system were found to be equivalent to those of the
      {\pAu} system. Errors represent systematic uncertainties
      (90\%~C.L.). The systematic error on impact parameter was not
      determined (the statistical uncertainty is negligible).}
\end{table}

The analysis presented in this thesis provided a set of data with which
this assumption could be tested. By using {\dAu} collisions to study the
nuclear modification factor of a system with minimal final state
effects, it was assumed implicitly that hadron production in {\dAu}
interactions scale with $\Ncoll$ relative to {\NAu}.
\Fig{rslt:fig:dAupAuNcollScaling} shows the ratio of the average charged
hadron yield in {\dAu} collisions to the yield in {\pAu} collisions. The
dotted line shows the expectation of $\Ncoll$ scaling. Systematic errors
on the spectra measurements are not shown in this figure, as they are
expected to be highly correlated between {\dAu} and {\pAu}, and are thus
unnecessary for the comparison of the systems. While it was also
expected that a strong correlation exists between systematic
uncertainties in the number of collisions in the two systems, the
magnitude of correlation was not determined. Therefore, the correct
systematic uncertainty in the ratio of $\Ncoll$ is not well understood.

The ratio of the yields was taken between spectra in a given \ac{ERing}
centrality bin. Recall that the fractional cross section included in a
centrality bin was estimated only for {\dAu} collisions, and not for
{\pAu} collisions (see \sect{recon:nuctag:NAcent}). Therefore, the
average impact parameter of {\dAu} collisions may be quite different
from that of {\pAu} collisions in the same \ac{ERing} centrality bin.
This was indeed the case, as seen in \tab{rslt:tab:eringCentPars}, which
compares centrality parameters of the two systems in each bin.

The most extreme difference in impact parameter between the two systems
occurred in the central bin. In the \ac{AMPT} simulations, the impact
parameter of the average {\pAu} collision in this bin was 80\% larger
than the average {\dAu} collision in the same bin. Since the impact
parameter measured the distance between the \emph{center} of the
deuteron and the center of the gold nucleus, it is possible that the
most central \ac{ERing} bin contained {\dAu} collisions in which the
nucleons of the deuteron were unusually far apart. The fact that, in the
simulations at least, the proton of the average central {\pAu} collision
suffered a comparable number of binary interactions as each deuteron
nucleon in the average central {\dAu} collision provided further
evidence for this situation.

While the simulations suggest a geometry of a central {\pAu} collision
that is not thought to be physically unreasonable (large impact
parameter and large deuteron ``radius''), they may not provide an
accurate simulation of the data. That is, there may be some effects,
experimental or physical, present in the analysis of tagged {\pAu}
collisions that were not present in the simulations. As discussed in
\sect{recon:nuctag:NAcent}, the fraction of {\dAu} collisions that were
tagged as {\pAu} interactions was qualitatively different in the data
and the simulations. The basic implication of this is that a larger
number of very central {\pAu} collisions, relative to the number of
{\dAu} collisions, were found in the data than in the simulations.
Consequently, the average number of collisions estimated by the
simulations for central {\pAu} interactions may have been too small, if
the simulations were biased toward less central interactions. If this
were indeed the case, then the expectation based on $\Ncoll$ scaling
shown in the central bin of \fig{rslt:fig:dAupAuNcollScaling} would be
too large.

Thus, the data presented in \fig{rslt:fig:dAupAuNcollScaling} should not
be interpreted as implying that central {\pAu} interactions were
especially efficient at producing particles. Further studies into the
reliability of the estimation of centrality parameters (like $\Ncoll$)
and of the tagging procedure in both the data and \ac{MC} are required
before such a statement can be made.

%---------------------------------------------------------------
\subsection{An Ideal $R_{AA}$ Reference}
\label{rslt:rda:RAAreference}
%---------------------------------------------------------------

While no evidence was found of $\Ncoll$ scaling violations between
{\pAu} and {\dAu} interactions, the availability of {\pAu} and {\nAu}
data allowed the construction of an ideal reference for {\AuAu}
collisions. Previous studies performed by the NA49 
collaboration~\cite{Fischer:2002qp, Rybicki:2004jd} have suggested that
hadron production of nucleus-nucleus collisions may be better understood
through careful consideration of the neutron content of the nucleus.
With this in mind, an ideal nuclear modification factor variable may be
defined as follows.

\begin{equation}
   \label{rslt:eq:RNAdef}
R_{NA} = \frac{0.40 \prn{Y_{pA} / \avencl^{pA}} + 
   0.60 \prn{Y_{nA} / \avencl^{nA}}}{Y_{p\bar{p}}}
\end{equation}

\noindent%
where $Y_{X}$ is the yield, $\dnchdeta$, in the specified collision
system. The definition of $R_{NA}$ shown in \eq{rslt:eq:RNAdef} takes
into account the fact that the gold nucleus consists of 60\% neutrons
and 40\% protons. 

\begin{figure}[t]
   \begin{center}
      \includegraphics[width=0.8\linewidth]{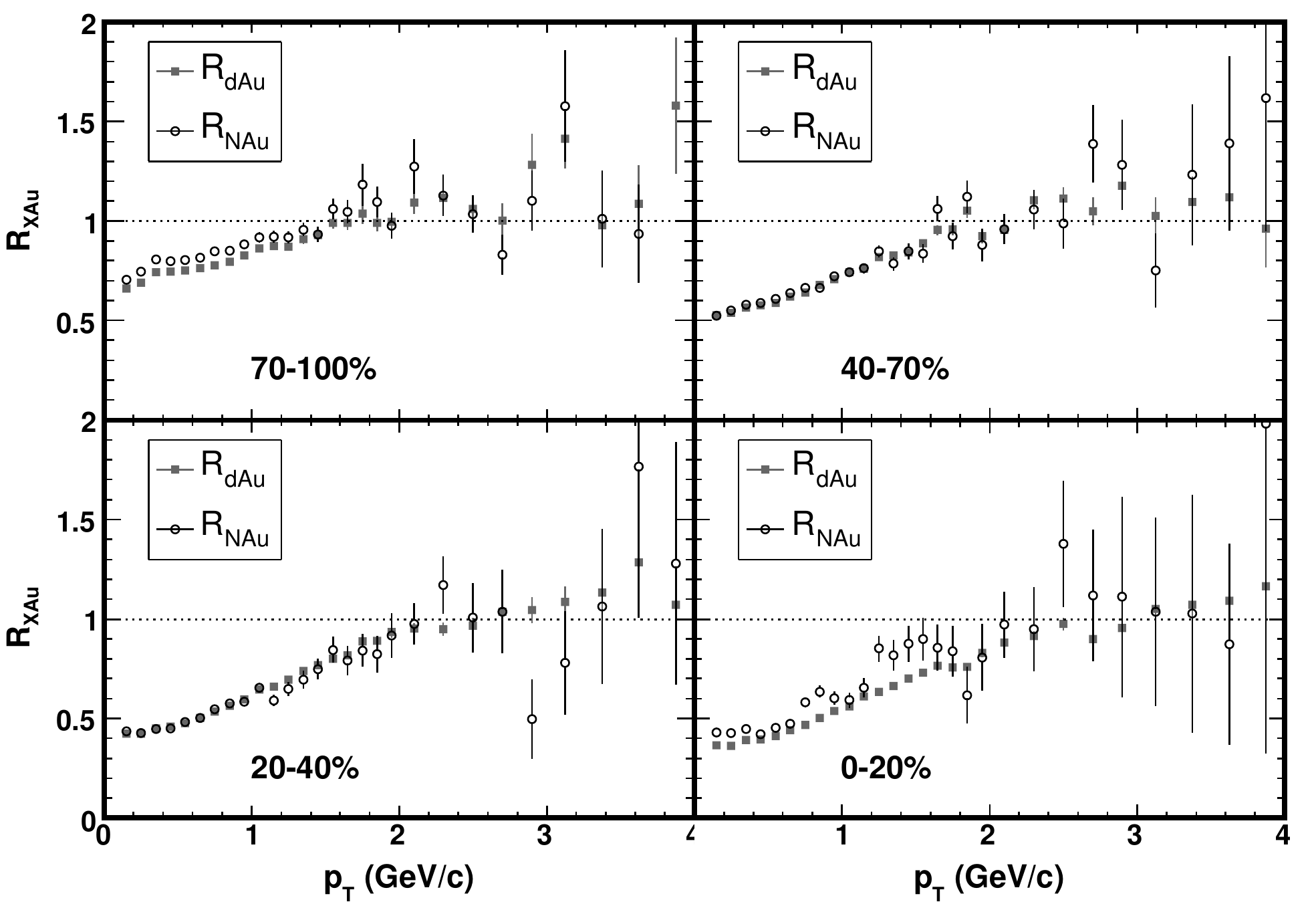}
   \end{center}
   \caption{\label{rslt:fig:RdAuRNAuCmp}
      Comparison of $R_{dA}$ and $R_{NA}$, as defined by
      \eq{rslt:eq:RNAdef}, in each \protect\acs{ERing} centrality bin.
      Systematic errors are not shown (see text).}
\end{figure}

The nuclear modification factor $R_{NA}$ is shown in
\fig{rslt:fig:RdAuRNAuCmp}. Also shown is $R_{dA}$, measured using the
\ac{ERing} centrality cuts. Systematic errors are not shown on this
figure, as all systematic effects are expected to be highly correlated
between {\dAu} and {\NAu}, and would therefore be unimportant to the
comparison of the two systems. Systematic uncertainties in $\Ncoll$
would affect the overall scale of the ratios and systematic
uncertainties in the spectra measurements would tend to shift the {\dAu}
and {\NAu} points by the same amount.

Not surprisingly, no significant difference between $R_{NA}$ and
$R_{dA}$ is observed. However, this measurement bolsters the
conclusions drawn from the nuclear modification factor measurements
from {\dAu} collisions~\cite{phobWhitePaper, Back:2003ns}; namely,
that high-$\pt$ hadron production in central {\AuAu} collisions is
significantly suppressed with respect to the expectation of binary
collision scaling of {\pbarp}, while the production of {\dAu}
collisions is not (compare
\figs{rslt:fig:RAA20062Central}{rslt:fig:RdAuRNAuCmp}). It should be
noted that no claim of binary collision scaling in {\dAu} or {\NAu}
interactions has been made. In addition to uncertainties on estimates
of the number of collisions, it has been observed that the nuclear
modification factor in {\dAu} exhibits a dependence on
{\prap}~\cite{Back:2004bq, Arsene:2003yk, Arsene:2004ux,
Arsene:2006pn}.  Thus, the apparent tendency of $R_{NA}$ and $R_{dA}$
to take the value of one at high-$\pt$ is likely a consequence of the
{\phob} {\prap} acceptance. Further, as will be discussed in the next
section, the hadron production of {\dAu} collisions is known to be
\emph{enhanced} with respect to binary collision scaling, in a certain
range of transverse momentum. Any statement that {\dAu} lacks a
suppression of high-$\pt$ hadrons is therefore contingent upon the
magnitude of this enhancement; see~\cite{Accardi:2005fu} for a
discussion.

Nevertheless, the stark discrepancy between {\NAu} and {\AuAu}
collisions at $\snn = 200~\gev$ demonstrate that final state effects
play a much stronger role in the high-$\pt$ hadron production of
central {\AuAu} collisions than do initial state effects. While the
{\prap} dependence of $R_{dA}$ may provide evidence of some initial
modification of the gold nucleus~\cite{Kharzeev:2003wz,
Jalilian-Marian:2005jf}, it is clear that interactions with some
dense, large volume medium produced only in the nucleus-nucleus system
form the dominant source of high-$\pt$ hadron suppression in {\AuAu}
collisions. The data presented in this thesis prove that this
conclusion was not biased by the the use of deuteron-nucleus rather
than nucleon-nucleus interactions as the reference for {\AuAu}.

%---------------------------------------------------------------
\section{Centrality Dependence of Spectra}
\label{rslt:shape}
%---------------------------------------------------------------

\begin{figure}[t!]
   \centering
   \mbox{
      \subfigure[Invariant Yields]{
         \label{rslt:fig:exRlYldppda}
         \includegraphics[width=0.4\linewidth]{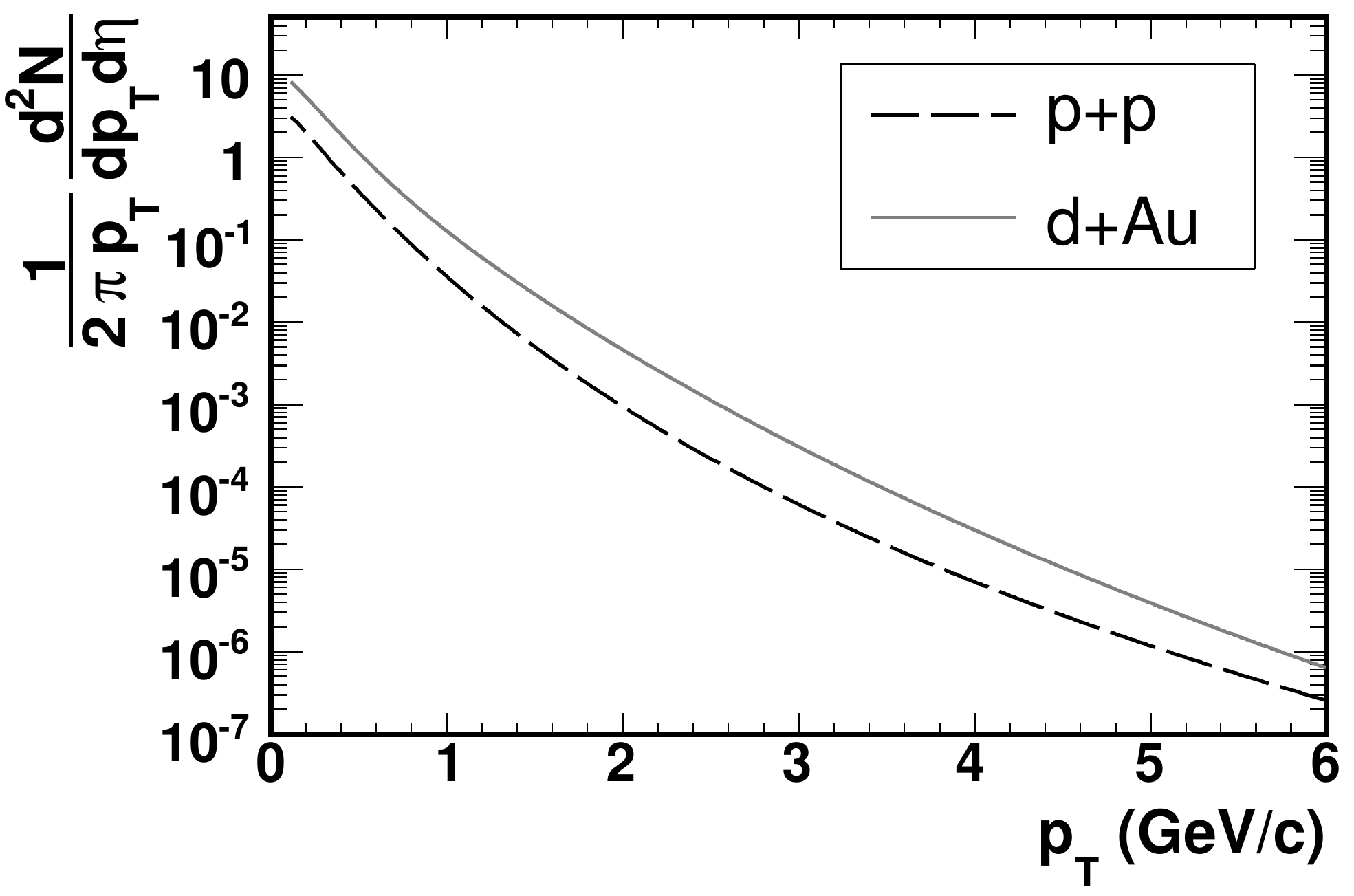}
      }
      \subfigure[Matched at Low $\pt$]{
         \label{rslt:fig:exRlYldScaled}
         \includegraphics[width=0.4\linewidth]{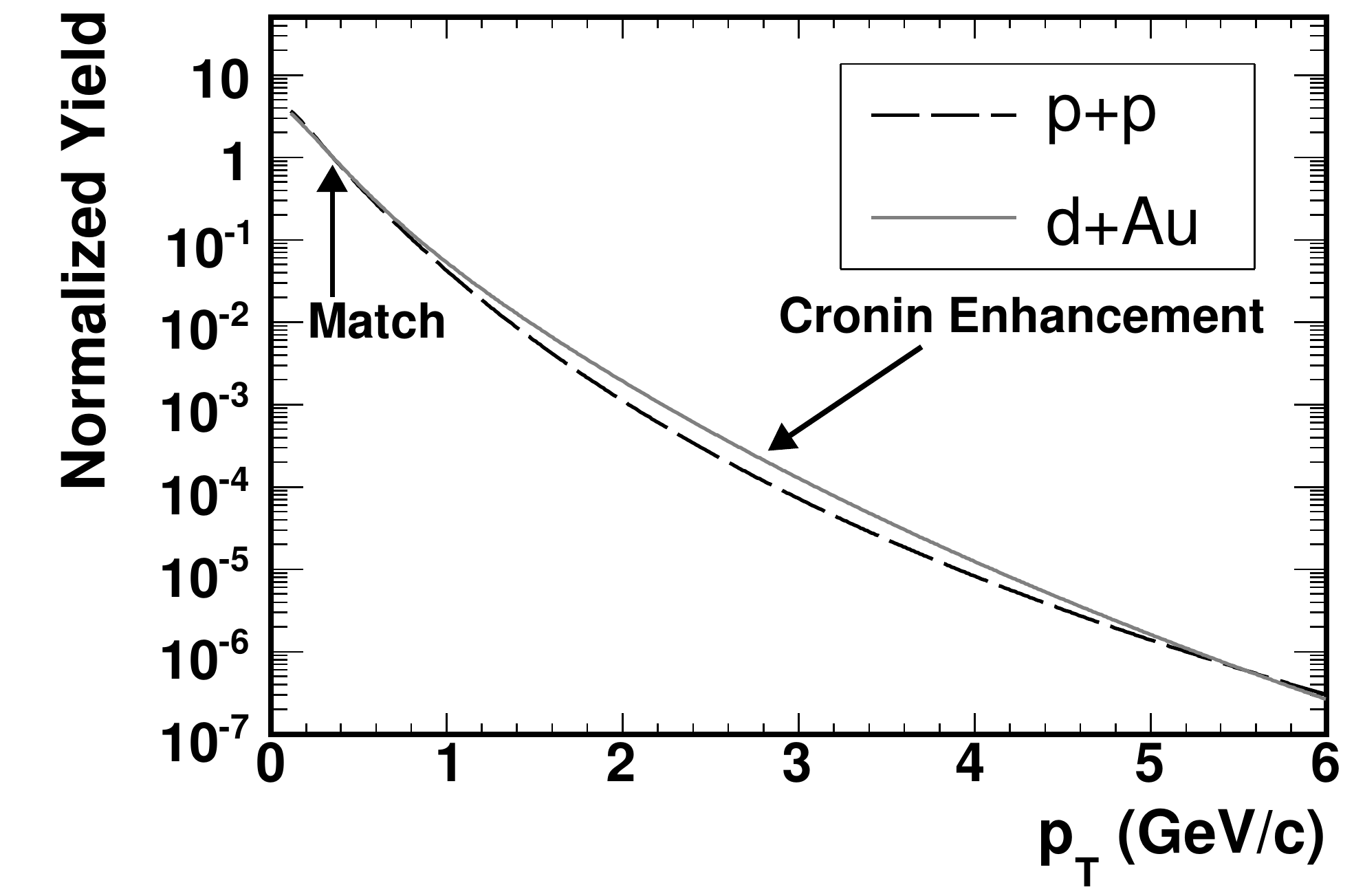}
      }
   }
   \subfigure[Relative Yield vs $\pt$]{
      \label{rslt:fig:exRlYldCronin}
      \includegraphics[width=0.4\linewidth]{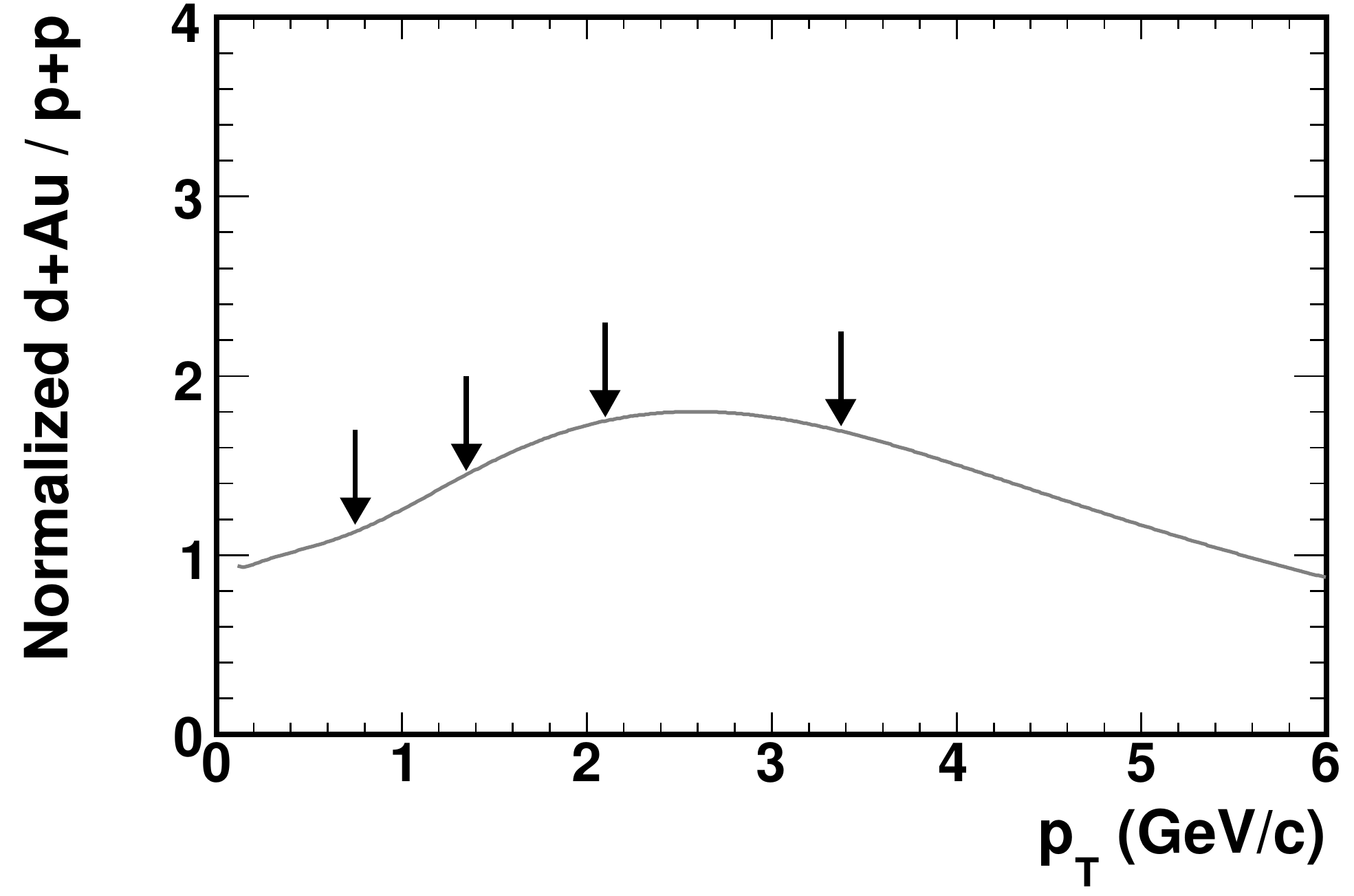}
   }
   \caption{   \label{rslt:fig:exrelyld}
      An example of the steps followed to determine the relative yield.
      \subref{rslt:fig:exRlYldppda}~The invariant yield of {\dAu}
      compared to {\pbarp}. \subref{rslt:fig:exRlYldScaled}~The yields
      are scaled to match at $\pt = 0.35~\mom$. The difference in shape
      of the two spectra is apparent.
      \subref{rslt:fig:exRlYldCronin}~The ratio of the scaled yields.
      The arrows mark the $\pt$ values at which the centrality
      dependence of the relative yield was studied.}
\end{figure}

The shape of the nuclear modification factor seen in
\figs{rslt:fig:RAA20062Central}{rslt:fig:RdAuRNAuCmp} is understood to
be related to the so-called Cronin effect. This effect refers to the
enhancement of hadron production in proton-nucleus
collisions~\cite{Cronin:1974zm} relative to {\pp} collisions scaled by
the effective thickness of the nucleus. General aspects of the
enhancement of inclusive charged hadron production (that is,
unidentified hadrons) in {\pAu} collisions can be described by models in
which partons undergo multiple scattering at the initial impact of the
{\pAu} collision~\cite{Accardi:2005fu}. However, the observed difference
in the strength of enhancement for mesons and baryons~\cite{Shao:2006qi}
is not easily explained by initial state partonic scattering models.
While other theories, such as those based on the recombination model of
hadronization~\cite{Hwa:2004zd}, may be better suited to describe the
enhancement of individual hadron species, it remains safe to say that
the Cronin effect is not a thoroughly understood phenomenon. Of
particular importance in the study of this effect is the dependence of
the enhancement on the nuclear thickness probed by the projectile
(i.e.~the deuteron in a {\dAu} collision)~\cite{Accardi:2003jh}.

The centrality dependence of the nuclear modification factor in {\dAu}
and {\AuAu} collisions at \ac{RHIC} has been studied
extensively~\cite{phobWhitePaper, Adcox:2004mh, Adams:2005dq,
Arsene:2004fa}. A particularly convenient method for exploring how the
shape of the transverse momentum spectra changes relative to {\pbarp}
was suggested in~\cite{Back:2003ns}. This involved studying the
centrality dependence of the charged hadron yield in {\dAu} collisions
relative to {\pp} at several values of $\pt$. The procedure for
determining the so-called relative yield was as follows.

\begin{figure}[t]
   \begin{center}
      \includegraphics[width=0.8\linewidth]{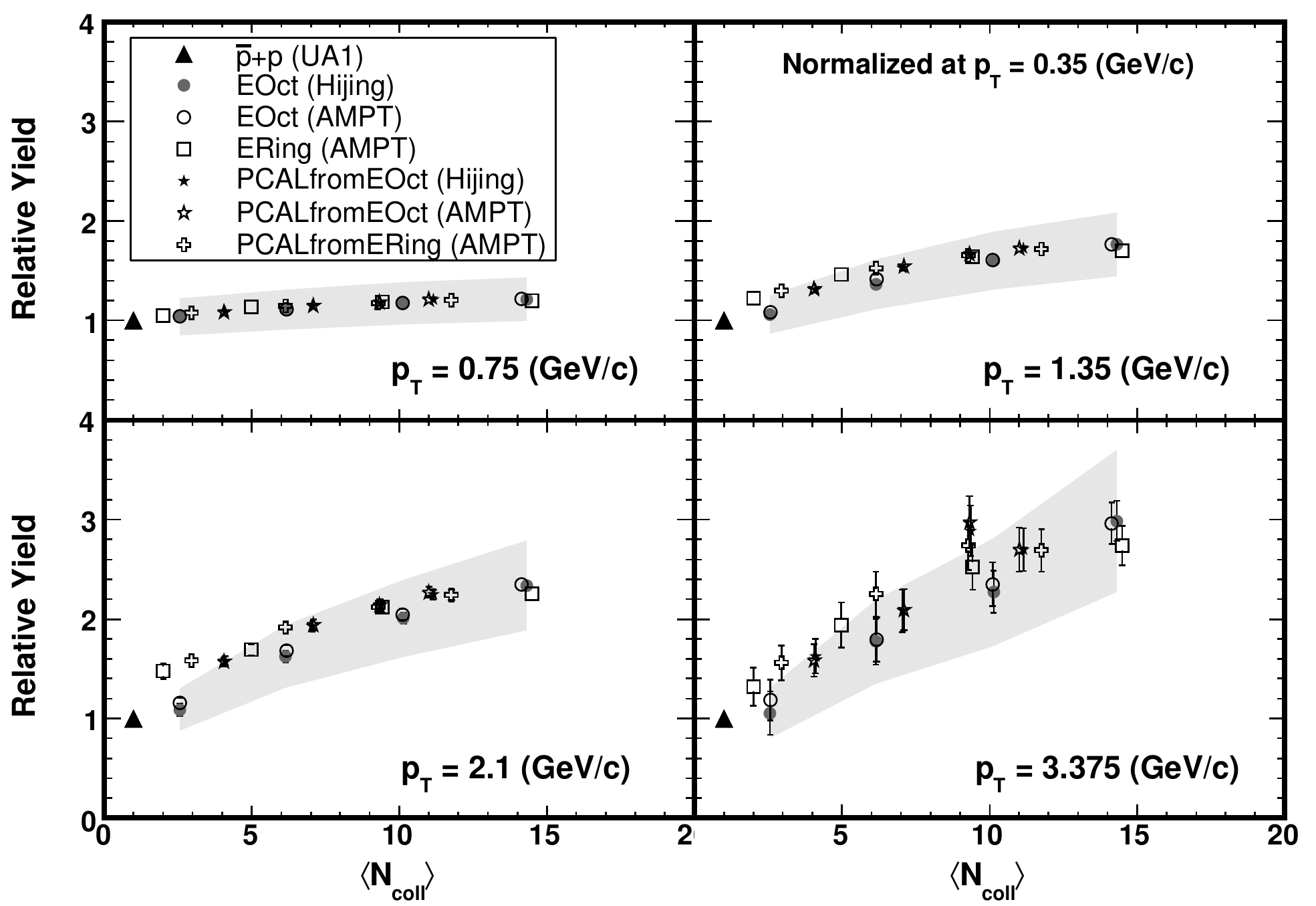}
   \end{center}
   \caption{\label{rslt:fig:relYldvsNcoll}
      The average hadron yield of {\dAu} collisions relative to {\pbarp}
      as a function of $\avencl$. Statistical errors are represented by
      bars on the points. The systematic error (90\%~C.L.) for one group
      of points, \protect\acs{EOct} (\protect\acs{HIJING}), is shown by
      the grey band. See text for a discussion of the systematic errors.
      A dependence of the relative yield on both centrality and $\pt$ is
      seen, but a bias introduced by the chosen centrality measure is
      also clearly visible.}
\end{figure}

First, the transverse momentum spectra in a particular {\dAu} centrality
bin was compared to the spectra in {\pbarp}. An example of this is shown
in \fig{rslt:fig:exRlYldppda}. To compare only the shape of the two
spectra, they were then normalized. As shown in
\fig{rslt:fig:exRlYldScaled}, the {\dAu} spectra was matched to the
{\pbarp} spectra at $\pt = 0.35~\mom$. While this specific value of
$\pt$ was arbitrary, it was intentionally chosen to be in the region of
soft hadron production. Matching the {\dAu} spectra to the {\pbarp}
spectra served to remove any trivial ``enhancement'' of hadron
production in {\dAu} that was simply due to the larger number of
nucleon-nucleon collisions occurring in that system. However, matching in
this way did not assume $\Ncoll$ scaling, nor did it have any effect on
the relative shape of the spectra. Next, the ratio of the normalized
{\dAu} spectra and the {\pbarp} spectra was determined. From this ratio,
certain transverse momentum values were chosen, as shown in
\fig{rslt:fig:exRlYldCronin}. Finally, the centrality dependence of the
normalized ratio, or \emph{relative yield}, at the chosen $\pt$ values
was studied.

\begin{figure}[t]
   \begin{center}
      \includegraphics[width=0.8\linewidth]{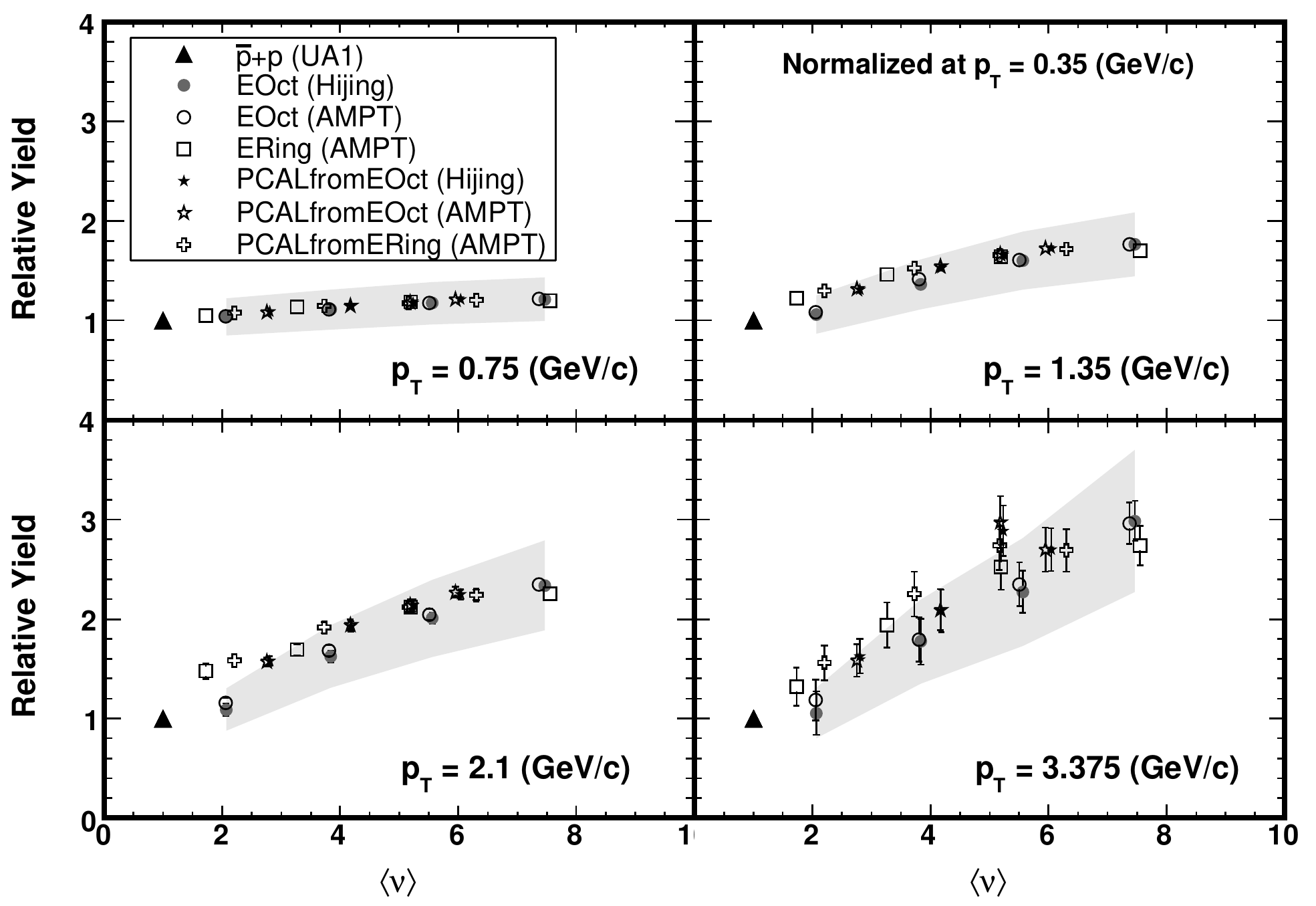}
   \end{center}
   \caption{\label{rslt:fig:relYldvsNu}
      The average hadron yield of {\dAu} collisions relative to {\pbarp}
      as a function of $\ave{\nu}$. Statistical errors are represented
      by bars on the points. A dependence of the relative yield on both
      centrality and $\pt$ is seen, but a bias introduced by the
      chosen centrality measure is also clearly visible.}
\end{figure}

The relative yield of {\dAu} collisions to {\pbarp} is shown in
\fig{rslt:fig:relYldvsNcoll} as a function of $\avencl$, for four
different values of transverse momentum. Different centrality measures
were used to determine the relative yield. Error bars on the points
represent statistical uncertainties, while the grey band shows the
systematic error associated with the relative yield of one group of
points, those found using the \acs{EOct} (\acs{HIJING}) centrality cuts.
It is expected that systematic effects on the relative yield are highly
correlated between the spectra measured with different centrality cuts.
Thus, shifts in the relative yield will tend to move all points
together. Systematic uncertainties in the number of collisions are shown
in~\appndx{app:centresults}. It is clear from
\fig{rslt:fig:relYldvsNcoll} that the centrality biases discussed in
\sect{rslt:cent} influence the measurement of the relative yield
parametrized by $\Ncoll$. Motivated by models of Cronin enhancement that
attribute the change in the shape of the {\dAu} spectrum to initial
partonic scattering, the relative yield is presented as a function of
$\ave{\nu}$ in \fig{rslt:fig:relYldvsNu}. Here again, the relative yield
shows a dependence on both centrality and transverse momentum, but the
observed dependence is still biased by the choice of centrality measure.

\begin{figure}[t]
   \begin{center}
      \includegraphics[width=0.8\linewidth]{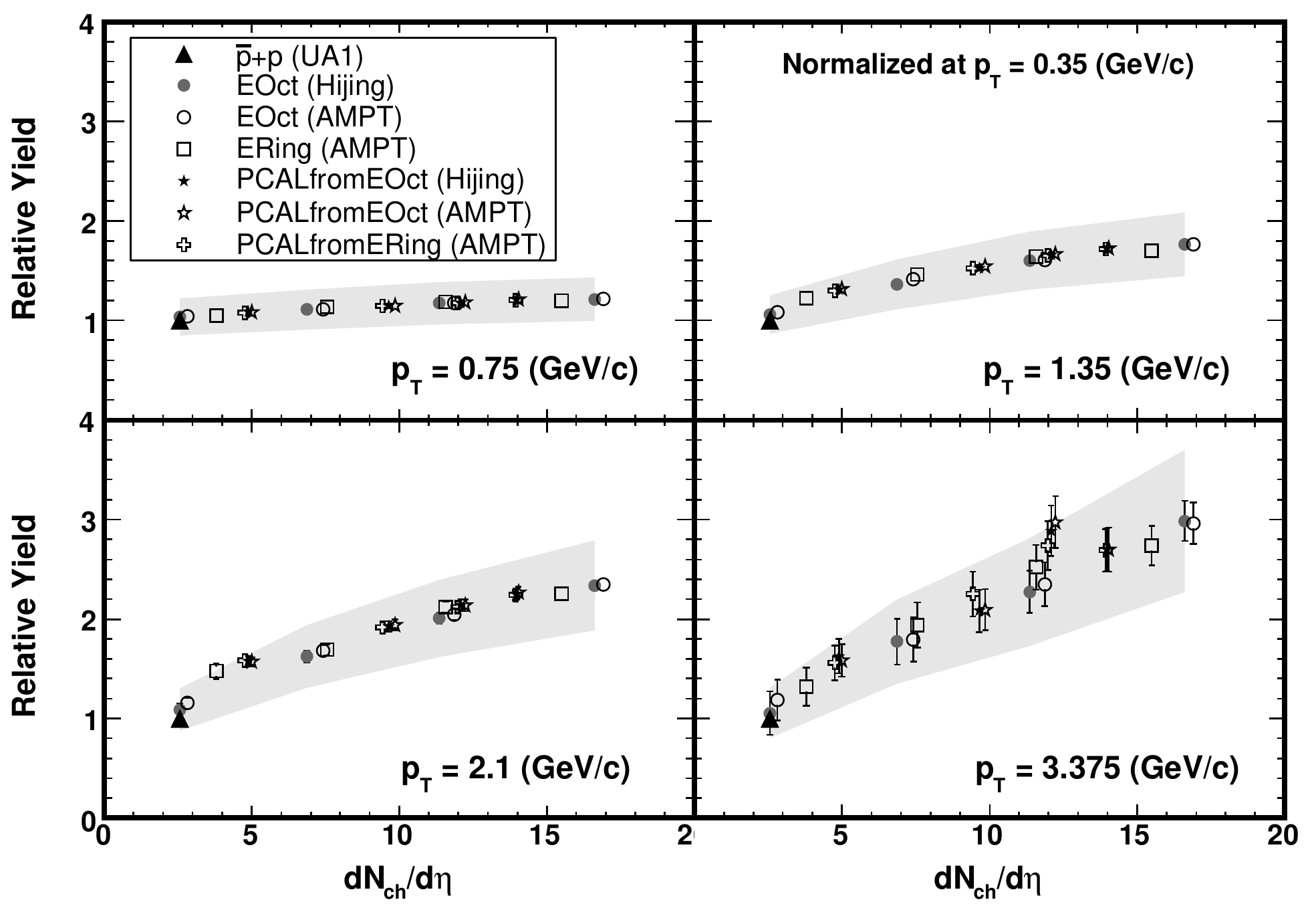}
   \end{center}
   \caption{\label{rslt:fig:relYldvsdNdeta}
      The average hadron yield of {\dAu} collisions relative to {\pbarp}
      as a function of $\dnchdeta$. Dependence of the relative yield on
      both centrality and $\pt$ is apparent and is observed to smoothly
      extrapolate back to {\pbarp}.}
\end{figure}

The relative yield as a function of $\dnchdeta$ is presented in
\fig{rslt:fig:relYldvsdNdeta}. With centrality parametrized by the
experimentally measured integrated yield, no bias or model dependence is
introduced by the choice of centrality measure. From these data, the
conclusion can be drawn that the shape of the {\dAu} spectra depends on
both centrality and $\pt$. Were the shape of the {\dAu} spectra
identical to the {\pbarp} spectra, the relative yield would be constant
at one for all values of $\pt$ and centrality. Instead, the {\dAu}
spectra show an enhancement over {\pbarp} that increases with
centrality. The strength of this enhancement is observed to increase at
higher-$\pt$. It would be interesting to study the relative yield of
much higher-$\pt$ hadrons, on the order of 10 to 100~{\mom}, to see
whether the shape of the {\pbarp} spectra is recovered in hard
scattering processes (see \fig{rslt:fig:exRlYldScaled}). However, such
particles are produced very rarely and too few were present in the
{\phob} data set to allow such a study.

\begin{figure}[t]
   \begin{center}
      \includegraphics[width=0.8\linewidth]{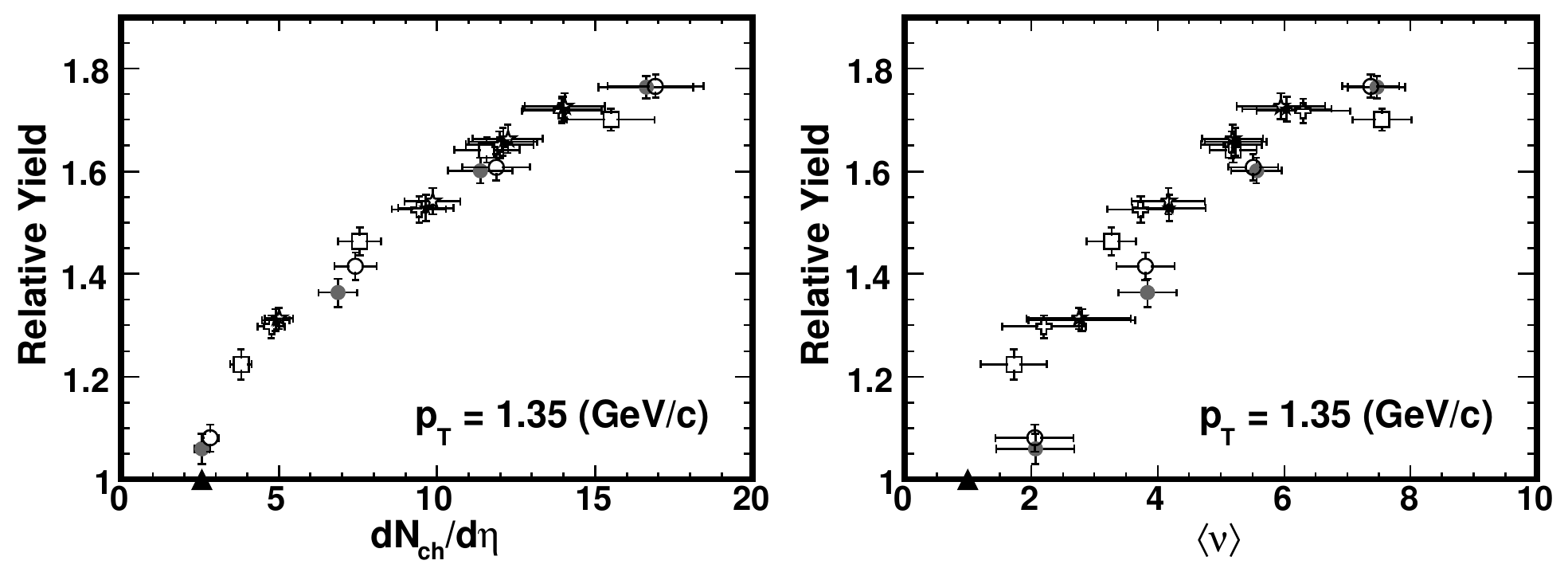}
   \end{center}
   \caption{\label{rslt:fig:relYldScaleCmp}
      The average yield of hadrons having $\pt=1.35~\mom$ from {\dAu}
      collisions relative to {\pbarp} as a function of both $\dnchdeta$
      and $\ave{\nu}$. Vertical error bars represent \emph{statistical}
      uncertainties in the relative yield. Horizontal error bars
      represent \emph{systematic} uncertainties (90\%~C.L.) in either
      $\dnchdeta$ or $\ave{\nu}$.} See \fig{rslt:fig:relYldvsdNdeta} for
      a description of the symbols.
\end{figure}

Nevertheless, the data presented in \fig{rslt:fig:relYldvsdNdeta} show
two intriguing properties. One is that the relative yield of {\dAu}
collisions is observed to extrapolate smoothly to the {\pbarp} yield as
the {\dAu} collisions become more peripheral. Thus, distortions of
{\dAu} spectra caused by nuclear effects diminish in a smooth way as the
amount of nuclear material probed by the deuteron is reduced. The other
is that this centrality dependence of the shape of the {\dAu} spectra
does not seem to depend on the choice of centrality measure, when the
centrality of the collisions are parametrized by their integrated yield.
This can be more easily seen in \fig{rslt:fig:relYldScaleCmp}, which
shows a detailed comparison of the relative yield as a function of both
$\dnchdeta$ and $\ave{\nu}$, at $\pt=1.35~\mom$. Vertical error bars
represent statistical uncertainties, and therefore points could shift
vertically independently. However, the horizontal error bars shown in
this figure represent systematic uncertainties, and are expected to be
highly correlated between different centrality measures (so the points
would shift together in the horizontal direction). From this comparison,
it is clear that the relative yield of {\dAu} extrapolates back to
{\pbarp} more smoothly when parametrized by the measured integrated
yield.

Thus, such model independent parameterizations of centrality may provide
the most unbiased, and therefore best, method with which to study the
centrality dependence of hadron production in nucleon-nucleus and
nucleus-nucleus system. Further, the common dependence of the relative
yield on the {\mrap} multiplicity among different centrality measures
suggests that the outgoing particle density may play an important role
in the Cronin effect. This observation provides additional evidence that
the enhancement may be better described by models that consider final
state effects, rather than initial partonic scattering only.

%---------------------------------------------------------------
\section{Comparison of p+Au and n+Au}
\label{rslt:pn}
%---------------------------------------------------------------

The availability of both {\pAu} and {\nAu} collision data presented a
unique opportunity to study baryon transport in nucleon-nucleus
collisions. Since a {\pAu} collision contains one more charge than an
{\nAu} collision, a search for this extra charge near the {\mrap} region
was possible. Previous measurements~\cite{Alber:1997sn} of {\pAu}
collisions at $\snn = 19.4~\gev$ found that the number of net protons (p
- \={p}) per unit of rapidity was less than one in the region of
{\mrap}. In addition, studies have shown a decrease in the {\mrap} net
proton yield with increasing center of mass energy;
see~\cite{conorsThesis} for a discussion. Further, it has been observed
that hadrons traversing nuclear material do not lose more than about
two units of rapidity~\cite{Busza:1983rj}. Thus, it was expected that any
charge asymmetry between hadrons measured at {\mrap} in {\pAu} and
{\nAu} collisions would be small.

Nevertheless, a comparison of charged hadron production in {\pAu} and
{\nAu} allowed the transport of charge from the projectile proton, and
from the projectile proton only, to be studied. Assuming that baryons
from the gold nucleus underwent transport to {\mrap} by the same process
in {\pAu} and {\nAu} collisions, excess charge at {\mrap} due to protons
in the gold nucleus would not play a role.

\begin{figure}[t]
   \centering
   \subfigure[$S_{pn}^{+}$]{
      \label{rslt:fig:asymmMultPos}
      \includegraphics[width=0.4\linewidth]{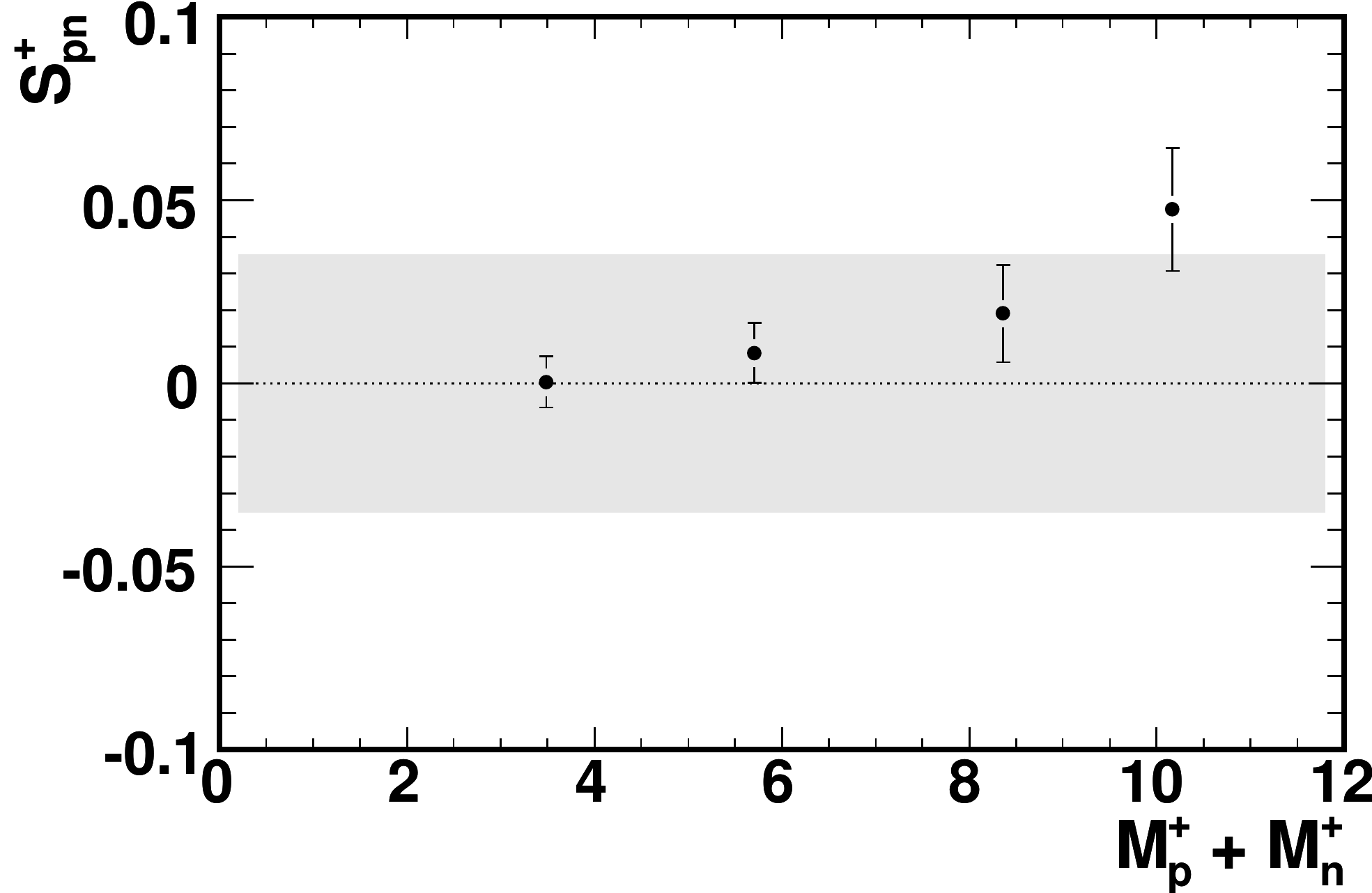}
   }
   \subfigure[$S_{pn}^{-}$]{
      \label{rslt:fig:asymmMultNeg}
      \includegraphics[width=0.4\linewidth]{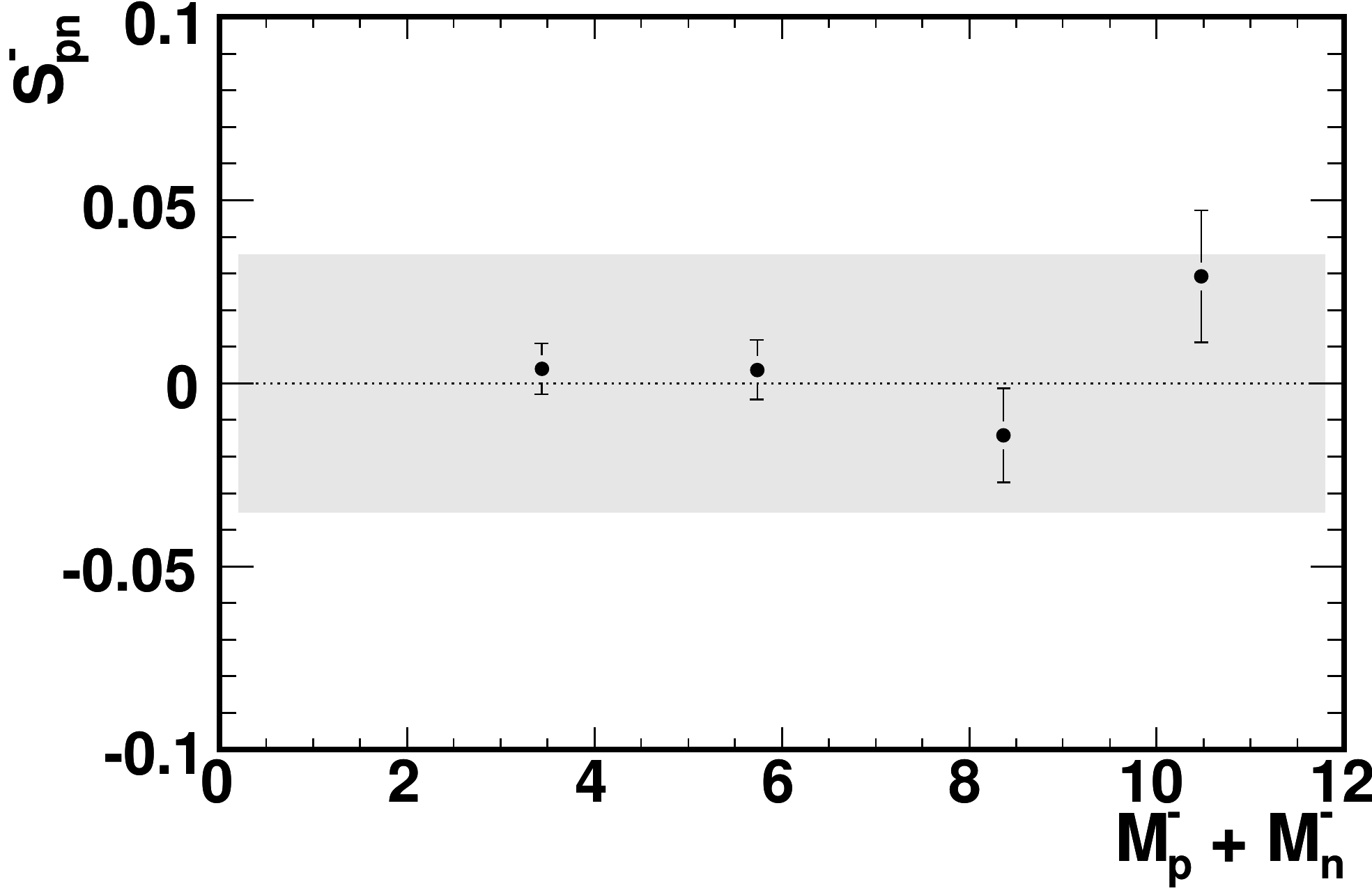}
   }
   \caption{   \label{rslt:fig:asymmMult}
      The asymmetry of charged hadrons in {\pAu} and {\nAu} collisions
      at $\aveeta=0.8$ as a function of centrality. The grey band shows
      the systematic uncertainty (90\%~C.L.) in the overal scale of the
      ratio. \subref{rslt:fig:asymmMultPos}~The relative difference of
      positive hadron production between {\pAu} and {\nAu} collisions.
      \subref{rslt:fig:asymmMultNeg}~The relative difference of negative
      hadron production between {\pAu} and {\nAu} collisions.}
\end{figure}

Simple charge conservation would imply that the \emph{total} number of
positive particles emerging from a {\pAu} collision should be greater
(by one) than the number emerging from a {\nAu} collision. Whether this
charge asymmetry would be present near {\mrap} was studied using the
following observable.

\begin{equation}
   \label{rslt:eq:spn}
S_{pn}^{+} = \frac{M_{p}^{+} - M_{n}^{+}}{M_{p}^{+} + M_{n}^{+}}
\end{equation}

\noindent%
where $M_{p}^{+}$ denotes the integrated yield, $\dndeta$, of positive
hadrons in {\pAu} collisions. The notation used for negative hadrons and
{\nAu} collisions should be obvious.

The charge asymmetry defined by \eq{rslt:eq:spn} is presented in
\fig{rslt:fig:asymmMult} for both positive and negative hadrons. The
grey band in each figure represents the systematic uncertainty in the
asymmetry ratio due to uncertainties in the measured integrated yield.
Only uncertainties specific to reconstructing the nucleon-nucleus $\pt$
spectra (see \fig{data:fig:tagSpecSysErr}) contribute to this systematic
error, as all other effects divide out in the ratio. No significant
asymmetry between {\pAu} and {\nAu} collisions is observed at $\aveeta =
0.8$, which is slightly forward on the deuteron side. However, the
possibility for a similar measurement to be done using particles
produced in more forward {\prap} regions is an exciting one. Such a
study may be possible using the {\phob} data and a particle tracking
procedure that makes use of the inner-wing of the Spectrometer (see
\sect{track:curved}).

%% file: srcs/conc.tex
% $Id: conc.tex,v 1.11 2006/09/05 01:18:31 cjreed Exp $

%---------------------------------------------------------------
\chapter{Summary}
\label{conc}
%---------------------------------------------------------------

The dynamics of strongly interacting matter under extreme conditions has
been studied experimentally by the {\phob} experiment. It was previously
reported that the yield of charged hadrons with large transverse momenta
in central {\AuAu} collisions at $\snn=200~\gev$ is highly suppressed
with respect to binary collision scaling~\cite{Back:2003qr,
Adams:2003kv, Adler:2003qi, Adler:2003au}. Two interpretations of this
observation were proposed. The first suggested that the initial state of
the nucleons in the nuclei were altered such that the
production of high-$\pt$ particles occurred less often in
nucleus-nucleus collisions than in nucleon-nucleon
interactions~\cite{Kharzeev:2002pc}. The second hypothesis held that the
\emph{production} of high-$\pt$ particles was unaltered, but that the
particles lost momentum due to subsequent interactions with a dense,
strongly interacting medium~\cite{Baier:1996sk}. These theories could be
tested using nucleon-nucleus interactions, in which any nuclear effects,
such as the modification of the initial nuclei, would be present, but a
dense medium would not be produced. Due to technical considerations, the
\ac{RHIC} facility provided {\dAu} collisions to serve as the valuable
control experiment. This decision involved two assumptions: firstly
that differences between a {\dAu} and a nucleon-nucleus interaction were
due only to the increased number of binary collisions, and secondly that
the difference in the neutron content of the ``projectile'' in {\AuAu},
{\dAu} and {\pbarp} interactions was unimportant.

The validity of these assumptions was tested by the analysis presented
in this thesis. Detectors were added to the {\phob} experiment that
measured the energy of very forward-going single protons; both on the
gold and deuteron exit side of the interaction region. The latter
detector was used in conjunction with the \ac{ZDC} detector to select
{\dAu} collisions in which only the proton or neutron of the deuteron
had interacted. Thus, samples of {\pAu} and {\nAu} collisions at
$\snn=200~\gev$ were obtained. Comparisons of the charged hadron yield
in {\dAu} and {\pAu} collisions were used to validate the assumption
of binary collision scaling from {\pAu} to {\dAu}. Further, an ideal
nuclear modification reference for {\AuAu} collisions was constructed
using a weighted combination of the yields in {\pAu} and {\nAu}
interactions. This data verified that no bias was introduced through
the use of a deuteron projectile rather than a nucleon projectile. In
addition, it supported the conclusion that the large suppression of
high-$\pt$ hadron yields seen in {\AuAu} interactions are not
observed in nucleon-nucleus collisions at $\snn=200~\gev$. Thus, this
data disfavors a hypothesis involving some modification of the initial
nuclei and proves that final state effects play a significant role in
the suppression of high-$\pt$ hadrons in nucleus-nucleus
collisions.

Further, a comparison of the yield of positively and negatively
charged hadrons in {\pAu} and {\nAu} collisions was conducted. This
comparison facilitated a rather unique study of charge transport, in
that any excess positive charge yielded by {\pAu} interactions could
be attributed to the proton from the deuteron nucleus in the initial
state. No significant asymmetry between the charged hadron yields in
{\pAu} and {\nAu} was observed at $\aveeta=0.8$.

Studies of the centrality dependence of the charged hadron yield of
{\dAu} collisions were also performed. A method of centrality
determination using the proton calorimeter on the gold exit side of the
interaction region was developed. In addition, the centrality of
collisions was ascertained using variables related to the multiplicity
in different regions of {\prap}. It was found that in small systems like
{\dAu}, the choice of centrality variable can significantly impact the
resulting measurement. Variables based on the multiplicity in the
{\mrap} region were found to bias measurements performed in the {\mrap}
region such that the average collision in the most central bin was more
central than either the fractional cross section or the number of
participants would suggest. It was found that the least biased
centrality variable used in measurements at {\mrap} was one based on the
multiplicity in a forward region of {\prap}. The \ac{PCAL} centrality
variables, based on the amount of nuclear spectator material, were not
expected to introduce any bias on measurements at {\mrap}. However, due
to the lack of reliable simulations of \ac{PCAL} signals in {\dAu}
collisions, attempts to quantify the centrality of \ac{EPCAL} bins using
multiplicity based variables were influenced by the biases of those
variables.

The multiple centrality variables were also used to study the centrality
dependence of the shape of the charged hadron spectra in {\dAu}
collisions. The modification of the $\pt$ spectrum of nucleon-nucleus
interactions as compared to a {\naive} superposition of nucleon-nucleon
collisions is known as the Cronin effect~\cite{Cronin:1974zm}. This
effect is characterized by an enhancement in the production of hadrons
having \mbox{$\pt\sim2.5~\mom$}. The centrality dependence of this
enhancement was studied in nucleus-nucleus interactions for several
values of $\pt$. One may expect that as the amount of nuclear material
the deuteron interacts with is decreased, the shape of the {\dAu}
spectrum should smoothly approach that of {\pbarp}. This behavior is
indeed seen when the centrality of {\dAu} collisions are parameterized
by the multiplicity measured near {\mrap}. Further, the same smooth
extrapolation to {\pbarp} was observed for the spectra as measured by
all centrality variables, in spite of whatever bias each centrality
variable may have introduced. On the other hand, no such smooth scaling
for all centrality variables was present when the centrality of {\dAu}
collisions was parameterized by $\Ncoll$ or $\nu$. This observation
seems to suggest that the Cronin effect may be driven by the
density of outgoing particles, in addition to the number of scatterings a
nucleon suffers during the initial collision.

%% file: srcs/appendix.tex
% $Id: appendix.tex,v 1.4 2006/08/18 01:05:32 cjreed Exp $
%
% Appendix

\input{srcs/centRes}

\input{srcs/specRes}

\input{srcs/tam}

%% file: srcs/centRes.tex
% $Id: centRes.tex,v 1.1 2006/08/17 21:42:16 cjreed Exp $

%---------------------------------------------------------------
\chapter{Centrality Results}
\label{app:centresults}
%---------------------------------------------------------------

The centrality parameters, such as $\Npart$, obtained using the various
centrality measures, such as \ac{ERing}, are presented in the following
tables.

\input{srcs/centables/Ncoll_dAu_table.tex}
\input{srcs/centables/Ncoll_pAu_table.tex}
\input{srcs/centables/Ncoll_nAu_table.tex}
\input{srcs/centables/Npart_dAu_table.tex}
\input{srcs/centables/Npart_pAu_table.tex}
\input{srcs/centables/Npart_nAu_table.tex}
\input{srcs/centables/Npart-Au_dAu_table.tex}
\input{srcs/centables/Npart-Au_pAu_table.tex}
\input{srcs/centables/Npart-Au_nAu_table.tex}
\input{srcs/centables/Npart-d_dAu_table.tex}
\input{srcs/centables/Npart-d_pAu_table.tex}
\input{srcs/centables/Npart-d_nAu_table.tex}
\input{srcs/centables/Nu_dAu_table.tex}
\input{srcs/centables/Nu_pAu_table.tex}
\input{srcs/centables/Nu_nAu_table.tex}
\input{srcs/centables/B_dAu_table.tex}
\input{srcs/centables/B_pAu_table.tex}
\input{srcs/centables/B_nAu_table.tex}

%% file: srcs/centables/Ncoll_dAu_table.tex
\cpartab{d+Au unbiased}{\Ncoll}{\protect\acs{dAuSpectra}}{
\protect\acs{EOct} & \protect\acs{HIJING} & $2.56\pm0.77$& $6.15\pm0.74$& $10.14\pm0.72$& $14.31\pm0.87$\\ 
\protect\acs{EOct} & \protect\acs{AMPT} & $2.57\pm0.77$& $6.18\pm0.74$& $10.11\pm0.72$& $14.14\pm0.86$\\ 
\protect\acs{ERing} & \protect\acs{AMPT} & $2.00\pm0.60$& $4.99\pm0.60$& $9.43\pm0.67$& $14.49\pm0.88$\\ 
\protect\acs{EPCAL} (\protect\acs{EOct})  & \protect\acs{HIJING}& $4.10\pm1.27$& $7.05\pm1.02$& $9.36\pm1.07$& $11.15\pm1.50$\\ 
\protect\acs{EPCAL} (\protect\acs{EOct})  & \protect\acs{AMPT}& $4.05\pm1.26$& $7.10\pm1.02$& $9.32\pm1.07$& $11.01\pm1.48$\\ 
\protect\acs{EPCAL} (\protect\acs{ERing})  & \protect\acs{AMPT}& $2.95\pm0.92$& $6.15\pm0.89$& $9.29\pm1.06$& $11.77\pm1.58$\\ 
}{NcolldAu}{h!}

%% file: srcs/centables/Ncoll_pAu_table.tex
\cpartab{p+Au unbiased}{\Ncoll}{\protect\acs{dAuSpectra}}{
\protect\acs{EOct} & \protect\acs{HIJING} & $2.03\pm0.61$& $4.29\pm0.52$& $6.57\pm0.47$& $8.25\pm0.50$\\ 
\protect\acs{EOct} & \protect\acs{AMPT} & $2.01\pm0.60$& $4.17\pm0.50$& $6.30\pm0.45$& $7.87\pm0.48$\\ 
\protect\acs{ERing} & \protect\acs{AMPT} & $1.71\pm0.51$& $3.70\pm0.44$& $6.72\pm0.48$& $8.36\pm0.51$\\ 
\protect\acs{EPCAL} (\protect\acs{EOct})  & \protect\acs{HIJING}& $2.42\pm0.73$& $3.18\pm0.94$& $3.82\pm1.10$& $4.41\pm1.14$\\ 
\protect\acs{EPCAL} (\protect\acs{EOct})  & \protect\acs{AMPT}& $2.34\pm0.70$& $3.06\pm0.90$& $3.65\pm1.05$& $4.18\pm1.08$\\ 
\protect\acs{EPCAL} (\protect\acs{ERing})  & \protect\acs{AMPT}& $1.98\pm0.59$& $2.87\pm0.85$& $3.83\pm1.11$& $4.83\pm1.24$\\ 
}{NcollpAu}{h!}

%% file: srcs/centables/Ncoll_nAu_table.tex
\cpartab{n+Au unbiased}{\Ncoll}{\protect\acs{dAuSpectra}}{
\protect\acs{EOct} & \protect\acs{HIJING} & $2.04\pm0.61$& $4.29\pm0.51$& $6.62\pm0.47$& $8.23\pm0.50$\\ 
\protect\acs{EOct} & \protect\acs{AMPT} & $2.02\pm0.61$& $4.18\pm0.50$& $6.33\pm0.45$& $7.86\pm0.48$\\ 
\protect\acs{ERing} & \protect\acs{AMPT} & $1.70\pm0.51$& $3.73\pm0.45$& $6.75\pm0.48$& $8.36\pm0.51$\\ 
\protect\acs{EPCAL} (\protect\acs{EOct})  & \protect\acs{HIJING}& $2.41\pm0.73$& $3.17\pm0.94$& $3.79\pm1.17$& $4.39\pm1.22$\\ 
\protect\acs{EPCAL} (\protect\acs{EOct})  & \protect\acs{AMPT}& $2.34\pm0.70$& $3.06\pm0.91$& $3.63\pm1.12$& $4.18\pm1.16$\\ 
\protect\acs{EPCAL} (\protect\acs{ERing})  & \protect\acs{AMPT}& $1.96\pm0.59$& $2.85\pm0.84$& $3.82\pm1.18$& $4.82\pm1.33$\\ 
}{NcollnAu}{h!}

%% file: srcs/centables/Npart_dAu_table.tex
\cpartab{d+Au unbiased}{\Npart}{\protect\acs{dAuSpectra}}{
\protect\acs{EOct} & \protect\acs{HIJING} & $3.73\pm0.84$& $7.48\pm0.83$& $11.32\pm0.94$& $15.15\pm1.08$\\ 
\protect\acs{EOct} & \protect\acs{AMPT} & $3.75\pm0.84$& $7.52\pm0.83$& $11.28\pm0.94$& $15.02\pm1.07$\\ 
\protect\acs{ERing} & \protect\acs{AMPT} & $3.12\pm0.70$& $6.31\pm0.70$& $10.62\pm0.88$& $15.42\pm1.09$\\ 
\protect\acs{EPCAL} (\protect\acs{EOct})  & \protect\acs{HIJING}& $5.31\pm1.22$& $8.25\pm1.13$& $10.49\pm1.21$& $12.19\pm1.30$\\ 
\protect\acs{EPCAL} (\protect\acs{EOct})  & \protect\acs{AMPT}& $5.26\pm1.21$& $8.30\pm1.14$& $10.46\pm1.21$& $12.07\pm1.29$\\ 
\protect\acs{EPCAL} (\protect\acs{ERing})  & \protect\acs{AMPT}& $4.12\pm0.94$& $7.36\pm1.01$& $10.43\pm1.20$& $12.82\pm1.37$\\ 
}{NpartdAu}{h!}

%% file: srcs/centables/Npart_pAu_table.tex
\cpartab{p+Au unbiased}{\Npart}{\protect\acs{dAuSpectra}}{
\protect\acs{EOct} & \protect\acs{HIJING} & $3.03\pm0.68$& $5.29\pm0.59$& $7.57\pm0.63$& $9.25\pm0.66$\\ 
\protect\acs{EOct} & \protect\acs{AMPT} & $3.01\pm0.67$& $5.17\pm0.57$& $7.30\pm0.61$& $8.87\pm0.63$\\ 
\protect\acs{ERing} & \protect\acs{AMPT} & $2.71\pm0.61$& $4.70\pm0.52$& $7.72\pm0.64$& $9.36\pm0.66$\\ 
\protect\acs{EPCAL} (\protect\acs{EOct})  & \protect\acs{HIJING}& $3.42\pm0.77$& $4.18\pm1.03$& $4.82\pm1.31$& $5.41\pm1.25$\\ 
\protect\acs{EPCAL} (\protect\acs{EOct})  & \protect\acs{AMPT}& $3.34\pm0.75$& $4.06\pm1.00$& $4.65\pm1.27$& $5.18\pm1.20$\\ 
\protect\acs{EPCAL} (\protect\acs{ERing})  & \protect\acs{AMPT}& $2.98\pm0.67$& $3.87\pm0.95$& $4.83\pm1.32$& $5.83\pm1.35$\\ 
}{NpartpAu}{h!}

%% file: srcs/centables/Npart_nAu_table.tex
\cpartab{n+Au unbiased}{\Npart}{\protect\acs{dAuSpectra}}{
\protect\acs{EOct} & \protect\acs{HIJING} & $3.04\pm0.68$& $5.29\pm0.59$& $7.62\pm0.63$& $9.23\pm0.66$\\ 
\protect\acs{EOct} & \protect\acs{AMPT} & $3.02\pm0.68$& $5.18\pm0.58$& $7.33\pm0.61$& $8.86\pm0.63$\\ 
\protect\acs{ERing} & \protect\acs{AMPT} & $2.70\pm0.61$& $4.73\pm0.53$& $7.75\pm0.64$& $9.36\pm0.66$\\ 
\protect\acs{EPCAL} (\protect\acs{EOct})  & \protect\acs{HIJING}& $3.41\pm0.77$& $4.17\pm1.03$& $4.79\pm1.31$& $5.39\pm1.25$\\ 
\protect\acs{EPCAL} (\protect\acs{EOct})  & \protect\acs{AMPT}& $3.34\pm0.75$& $4.06\pm1.00$& $4.63\pm1.26$& $5.18\pm1.20$\\ 
\protect\acs{EPCAL} (\protect\acs{ERing})  & \protect\acs{AMPT}& $2.96\pm0.67$& $3.85\pm0.95$& $4.82\pm1.31$& $5.82\pm1.34$\\ 
}{NpartnAu}{h!}

%% file: srcs/centables/Npart-Au_dAu_table.tex
\cpartab{d+Au unbiased}{\NparA}{\protect\acs{dAuSpectra}}{
\protect\acs{EOct} & \protect\acs{HIJING} & $2.44\pm0.63$& $5.75\pm0.75$& $9.40\pm0.74$& $13.16\pm0.97$\\ 
\protect\acs{EOct} & \protect\acs{AMPT} & $2.45\pm0.64$& $5.77\pm0.75$& $9.35\pm0.74$& $13.03\pm0.96$\\ 
\protect\acs{ERing} & \protect\acs{AMPT} & $1.92\pm0.50$& $4.64\pm0.60$& $8.70\pm0.69$& $13.43\pm0.99$\\ 
\protect\acs{EPCAL} (\protect\acs{EOct})  & \protect\acs{HIJING}& $3.86\pm1.03$& $6.56\pm1.04$& $8.67\pm1.04$& $10.29\pm1.36$\\ 
\protect\acs{EPCAL} (\protect\acs{EOct})  & \protect\acs{AMPT}& $3.81\pm1.02$& $6.60\pm1.04$& $8.64\pm1.03$& $10.17\pm1.35$\\ 
\protect\acs{EPCAL} (\protect\acs{ERing})  & \protect\acs{AMPT}& $2.79\pm0.74$& $5.71\pm0.90$& $8.60\pm1.03$& $10.90\pm1.45$\\ 
}{Npart-AudAu}{h!}

%% file: srcs/centables/Npart-Au_pAu_table.tex
\cpartab{p+Au unbiased}{\NparA}{\protect\acs{dAuSpectra}}{
\protect\acs{EOct} & \protect\acs{HIJING} & $2.03\pm0.53$& $4.29\pm0.56$& $6.57\pm0.52$& $8.25\pm0.61$\\ 
\protect\acs{EOct} & \protect\acs{AMPT} & $2.01\pm0.52$& $4.17\pm0.54$& $6.30\pm0.50$& $7.87\pm0.58$\\ 
\protect\acs{ERing} & \protect\acs{AMPT} & $1.71\pm0.44$& $3.70\pm0.48$& $6.72\pm0.53$& $8.36\pm0.62$\\ 
\protect\acs{EPCAL} (\protect\acs{EOct})  & \protect\acs{HIJING}& $2.42\pm0.63$& $3.18\pm0.98$& $3.82\pm1.15$& $4.41\pm1.15$\\ 
\protect\acs{EPCAL} (\protect\acs{EOct})  & \protect\acs{AMPT}& $2.34\pm0.61$& $3.06\pm0.94$& $3.65\pm1.10$& $4.18\pm1.09$\\ 
\protect\acs{EPCAL} (\protect\acs{ERing})  & \protect\acs{AMPT}& $1.98\pm0.51$& $2.87\pm0.89$& $3.83\pm1.15$& $4.83\pm1.26$\\ 
}{Npart-AupAu}{h!}

%% file: srcs/centables/Npart-Au_nAu_table.tex
\cpartab{n+Au unbiased}{\NparA}{\protect\acs{dAuSpectra}}{
\protect\acs{EOct} & \protect\acs{HIJING} & $2.04\pm0.53$& $4.29\pm0.56$& $6.62\pm0.52$& $8.23\pm0.61$\\ 
\protect\acs{EOct} & \protect\acs{AMPT} & $2.02\pm0.53$& $4.18\pm0.54$& $6.33\pm0.50$& $7.86\pm0.58$\\ 
\protect\acs{ERing} & \protect\acs{AMPT} & $1.70\pm0.44$& $3.73\pm0.49$& $6.75\pm0.53$& $8.36\pm0.62$\\ 
\protect\acs{EPCAL} (\protect\acs{EOct})  & \protect\acs{HIJING}& $2.41\pm0.63$& $3.17\pm0.98$& $3.79\pm1.18$& $4.39\pm1.23$\\ 
\protect\acs{EPCAL} (\protect\acs{EOct})  & \protect\acs{AMPT}& $2.34\pm0.61$& $3.06\pm0.95$& $3.63\pm1.13$& $4.18\pm1.17$\\ 
\protect\acs{EPCAL} (\protect\acs{ERing})  & \protect\acs{AMPT}& $1.96\pm0.51$& $2.85\pm0.88$& $3.82\pm1.18$& $4.82\pm1.35$\\ 
}{Npart-AunAu}{h!}

%% file: srcs/centables/Npart-d_dAu_table.tex
\cpartab{d+Au unbiased}{\Npard}{\protect\acs{dAuSpectra}}{
\protect\acs{EOct} & \protect\acs{HIJING} & $1.29\pm0.22$& $1.73\pm0.21$& $1.92\pm0.10$& $1.98\pm0.11$\\ 
\protect\acs{EOct} & \protect\acs{AMPT} & $1.30\pm0.22$& $1.74\pm0.21$& $1.93\pm0.10$& $1.99\pm0.11$\\ 
\protect\acs{ERing} & \protect\acs{AMPT} & $1.20\pm0.20$& $1.67\pm0.20$& $1.93\pm0.10$& $1.99\pm0.11$\\ 
\protect\acs{EPCAL} (\protect\acs{EOct})  & \protect\acs{HIJING}& $1.45\pm0.24$& $1.68\pm0.23$& $1.82\pm0.12$& $1.89\pm0.11$\\ 
\protect\acs{EPCAL} (\protect\acs{EOct})  & \protect\acs{AMPT}& $1.45\pm0.24$& $1.70\pm0.23$& $1.82\pm0.12$& $1.90\pm0.11$\\ 
\protect\acs{EPCAL} (\protect\acs{ERing})  & \protect\acs{AMPT}& $1.33\pm0.22$& $1.65\pm0.22$& $1.83\pm0.12$& $1.92\pm0.11$\\ 
}{Npart-ddAu}{h!}

%% file: srcs/centables/Npart-d_pAu_table.tex
\cpartab{p+Au unbiased}{\Npard}{\protect\acs{dAuSpectra}}{
\protect\acs{EOct} & \protect\acs{HIJING} & $1.00\pm0.17$& $1.00\pm0.12$& $1.00\pm0.05$& $1.00\pm0.05$\\ 
\protect\acs{EOct} & \protect\acs{AMPT} & $1.00\pm0.17$& $1.00\pm0.12$& $1.00\pm0.05$& $1.00\pm0.05$\\ 
\protect\acs{ERing} & \protect\acs{AMPT} & $1.00\pm0.17$& $1.00\pm0.12$& $1.00\pm0.05$& $1.00\pm0.05$\\ 
\protect\acs{EPCAL} (\protect\acs{EOct})  & \protect\acs{HIJING}& $1.00\pm0.17$& $1.00\pm0.12$& $1.00\pm0.05$& $1.00\pm0.05$\\ 
\protect\acs{EPCAL} (\protect\acs{EOct})  & \protect\acs{AMPT}& $1.00\pm0.17$& $1.00\pm0.12$& $1.00\pm0.05$& $1.00\pm0.05$\\ 
\protect\acs{EPCAL} (\protect\acs{ERing})  & \protect\acs{AMPT}& $1.00\pm0.17$& $1.00\pm0.12$& $1.00\pm0.05$& $1.00\pm0.05$\\ 
}{Npart-dpAu}{h!}

%% file: srcs/centables/Npart-d_nAu_table.tex
\cpartab{n+Au unbiased}{\Npard}{\protect\acs{dAuSpectra}}{
\protect\acs{EOct} & \protect\acs{HIJING} & $1.00\pm0.17$& $1.00\pm0.12$& $1.00\pm0.05$& $1.00\pm0.05$\\ 
\protect\acs{EOct} & \protect\acs{AMPT} & $1.00\pm0.17$& $1.00\pm0.12$& $1.00\pm0.05$& $1.00\pm0.05$\\ 
\protect\acs{ERing} & \protect\acs{AMPT} & $1.00\pm0.17$& $1.00\pm0.12$& $1.00\pm0.05$& $1.00\pm0.05$\\ 
\protect\acs{EPCAL} (\protect\acs{EOct})  & \protect\acs{HIJING}& $1.00\pm0.17$& $1.00\pm0.12$& $1.00\pm0.05$& $1.00\pm0.05$\\ 
\protect\acs{EPCAL} (\protect\acs{EOct})  & \protect\acs{AMPT}& $1.00\pm0.17$& $1.00\pm0.12$& $1.00\pm0.05$& $1.00\pm0.05$\\ 
\protect\acs{EPCAL} (\protect\acs{ERing})  & \protect\acs{AMPT}& $1.00\pm0.17$& $1.00\pm0.12$& $1.00\pm0.05$& $1.00\pm0.05$\\ 
}{Npart-dnAu}{h!}

%% file: srcs/centables/Nu_dAu_table.tex
\cpartab{d+Au unbiased}{\nu}{\protect\acs{dAuSpectra}}{
\protect\acs{EOct} & \protect\acs{HIJING} & $2.06\pm0.62$& $3.84\pm0.46$& $5.56\pm0.39$& $7.47\pm0.46$\\ 
\protect\acs{EOct} & \protect\acs{AMPT} & $2.06\pm0.62$& $3.80\pm0.46$& $5.51\pm0.39$& $7.37\pm0.45$\\ 
\protect\acs{ERing} & \protect\acs{AMPT} & $1.73\pm0.52$& $3.27\pm0.39$& $5.20\pm0.37$& $7.55\pm0.46$\\ 
\protect\acs{EPCAL} (\protect\acs{EOct})  & \protect\acs{HIJING}& $2.80\pm0.84$& $4.18\pm0.58$& $5.23\pm0.49$& $6.04\pm0.71$\\ 
\protect\acs{EPCAL} (\protect\acs{EOct})  & \protect\acs{AMPT}& $2.75\pm0.83$& $4.17\pm0.58$& $5.19\pm0.48$& $5.95\pm0.70$\\ 
\protect\acs{EPCAL} (\protect\acs{ERing})  & \protect\acs{AMPT}& $2.20\pm0.66$& $3.73\pm0.52$& $5.17\pm0.48$& $6.30\pm0.74$\\ 
}{NudAu}{h!}

%% file: srcs/centables/Nu_pAu_table.tex
\cpartab{p+Au unbiased}{\nu}{\protect\acs{dAuSpectra}}{
\protect\acs{EOct} & \protect\acs{HIJING} & $2.03\pm0.61$& $4.29\pm0.52$& $6.57\pm0.47$& $8.25\pm0.50$\\ 
\protect\acs{EOct} & \protect\acs{AMPT} & $2.01\pm0.60$& $4.17\pm0.50$& $6.30\pm0.45$& $7.87\pm0.48$\\ 
\protect\acs{ERing} & \protect\acs{AMPT} & $1.71\pm0.51$& $3.70\pm0.44$& $6.72\pm0.48$& $8.36\pm0.51$\\ 
\protect\acs{EPCAL} (\protect\acs{EOct})  & \protect\acs{HIJING}& $2.42\pm0.73$& $3.18\pm0.94$& $3.82\pm1.10$& $4.41\pm1.14$\\ 
\protect\acs{EPCAL} (\protect\acs{EOct})  & \protect\acs{AMPT}& $2.34\pm0.70$& $3.06\pm0.90$& $3.65\pm1.05$& $4.18\pm1.08$\\ 
\protect\acs{EPCAL} (\protect\acs{ERing})  & \protect\acs{AMPT}& $1.98\pm0.59$& $2.87\pm0.85$& $3.83\pm1.11$& $4.83\pm1.24$\\ 
}{NupAu}{h!}

%% file: srcs/centables/Nu_nAu_table.tex
\cpartab{n+Au unbiased}{\nu}{\protect\acs{dAuSpectra}}{
\protect\acs{EOct} & \protect\acs{HIJING} & $2.04\pm0.61$& $4.29\pm0.51$& $6.62\pm0.47$& $8.23\pm0.50$\\ 
\protect\acs{EOct} & \protect\acs{AMPT} & $2.02\pm0.61$& $4.18\pm0.50$& $6.33\pm0.45$& $7.86\pm0.48$\\ 
\protect\acs{ERing} & \protect\acs{AMPT} & $1.70\pm0.51$& $3.73\pm0.45$& $6.75\pm0.48$& $8.36\pm0.51$\\ 
\protect\acs{EPCAL} (\protect\acs{EOct})  & \protect\acs{HIJING}& $2.41\pm0.73$& $3.17\pm0.94$& $3.79\pm1.17$& $4.39\pm1.26$\\ 
\protect\acs{EPCAL} (\protect\acs{EOct})  & \protect\acs{AMPT}& $2.34\pm0.71$& $3.06\pm0.91$& $3.63\pm1.12$& $4.18\pm1.20$\\ 
\protect\acs{EPCAL} (\protect\acs{ERing})  & \protect\acs{AMPT}& $1.96\pm0.59$& $2.85\pm0.84$& $3.82\pm1.18$& $4.82\pm1.38$\\ 
}{NunAu}{h!}

%% file: srcs/centables/B_dAu_table.tex
\cpartab{d+Au unbiased}{b}{\protect\acs{dAuSpectra}}{
\protect\acs{EOct} & \protect\acs{HIJING} & $7.56$& $5.89$& $4.50$& $3.45$\\ 
\protect\acs{EOct} & \protect\acs{AMPT} & $7.43$& $5.81$& $4.49$& $3.47$\\ 
\protect\acs{ERing} & \protect\acs{AMPT} & $7.76$& $6.29$& $4.67$& $3.31$\\ 
\protect\acs{EPCAL} (\protect\acs{EOct})  & \protect\acs{HIJING}& $6.89$& $5.74$& $4.93$& $4.37$\\ 
\protect\acs{EPCAL} (\protect\acs{EOct})  & \protect\acs{AMPT}& $6.82$& $5.65$& $4.90$& $4.37$\\ 
\protect\acs{EPCAL} (\protect\acs{ERing})  & \protect\acs{AMPT}& $7.31$& $5.99$& $4.88$& $4.12$\\ 
}{BdAu}{h!}

%% file: srcs/centables/B_pAu_table.tex
\cpartab{p+Au unbiased}{b}{\protect\acs{dAuSpectra}}{
\protect\acs{EOct} & \protect\acs{HIJING} & $7.97$& $7.13$& $6.54$& $6.24$\\ 
\protect\acs{EOct} & \protect\acs{AMPT} & $7.85$& $7.06$& $6.53$& $6.21$\\ 
\protect\acs{ERing} & \protect\acs{AMPT} & $8.01$& $7.18$& $6.38$& $6.09$\\ 
\protect\acs{EPCAL} (\protect\acs{EOct})  & \protect\acs{HIJING}& $7.83$& $7.56$& $7.35$& $7.17$\\ 
\protect\acs{EPCAL} (\protect\acs{EOct})  & \protect\acs{AMPT}& $7.74$& $7.48$& $7.29$& $7.12$\\ 
\protect\acs{EPCAL} (\protect\acs{ERing})  & \protect\acs{AMPT}& $7.90$& $7.55$& $7.23$& $6.93$\\ 
}{BpAu}{h!}

%% file: srcs/centables/B_nAu_table.tex
\cpartab{n+Au unbiased}{b}{\protect\acs{dAuSpectra}}{
\protect\acs{EOct} & \protect\acs{HIJING} & $7.96$& $7.14$& $6.54$& $6.24$\\ 
\protect\acs{EOct} & \protect\acs{AMPT} & $7.84$& $7.05$& $6.51$& $6.24$\\ 
\protect\acs{ERing} & \protect\acs{AMPT} & $7.99$& $7.17$& $6.36$& $6.09$\\ 
\protect\acs{EPCAL} (\protect\acs{EOct})  & \protect\acs{HIJING}& $7.83$& $7.57$& $7.36$& $7.18$\\ 
\protect\acs{EPCAL} (\protect\acs{EOct})  & \protect\acs{AMPT}& $7.73$& $7.47$& $7.29$& $7.12$\\ 
\protect\acs{EPCAL} (\protect\acs{ERing})  & \protect\acs{AMPT}& $7.89$& $7.55$& $7.23$& $6.93$\\ 
}{BnAu}{h!}

%% file: srcs/specRes.tex
% $Id: specRes.tex,v 1.1 2006/08/18 01:05:46 cjreed Exp $

%---------------------------------------------------------------
\chapter{Spectra Results}
\label{app:spectraresults}
%---------------------------------------------------------------

The invariant yield of positive, negative and average charged hadrons
are presented in the following figures as a function of centrality, in
four centrality bins as determined by six different centrality measures,
for {\dAu}, {\pAu} and {\nAu} interactions.

\begin{figure}[h]
   \begin{center}
      \includegraphics[width=\linewidth]{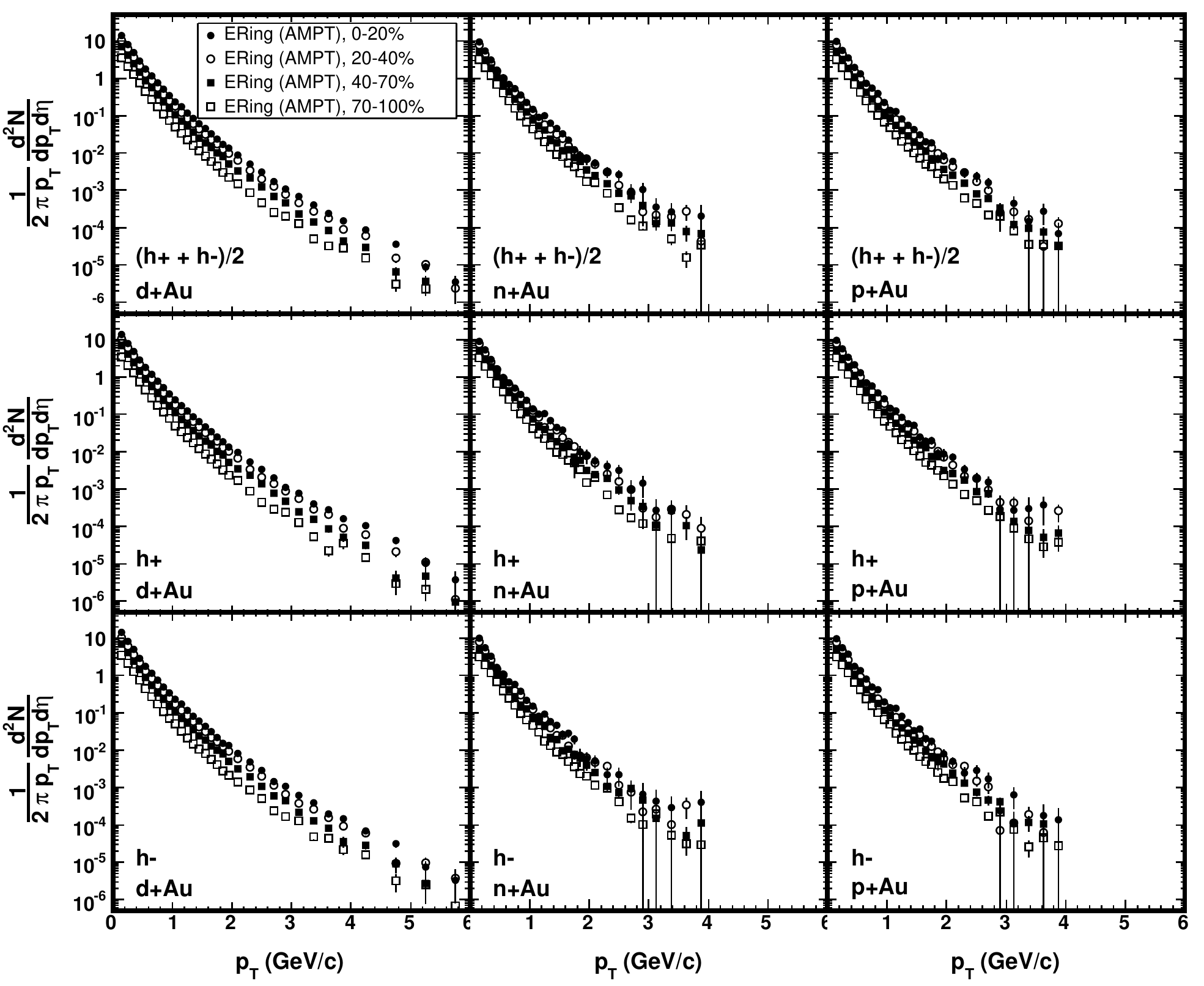}
   \end{center}
   \caption{\label{app:fig:ntSpecERingAMPT}
      The invariant yield of $\have$, $\hpos$ and $\hneg$ in four
      centrality bins determined using \protect\acs{AMPT} and the
      \protect\acs{ERing} centrality variable. The spectra for {\dAu},
      {\nAu} and {\pAu} are shown in separate columns. Only statistical
      errors are shown.}
\end{figure}

\begin{figure}[h]
   \begin{center}
      \includegraphics[width=\linewidth]{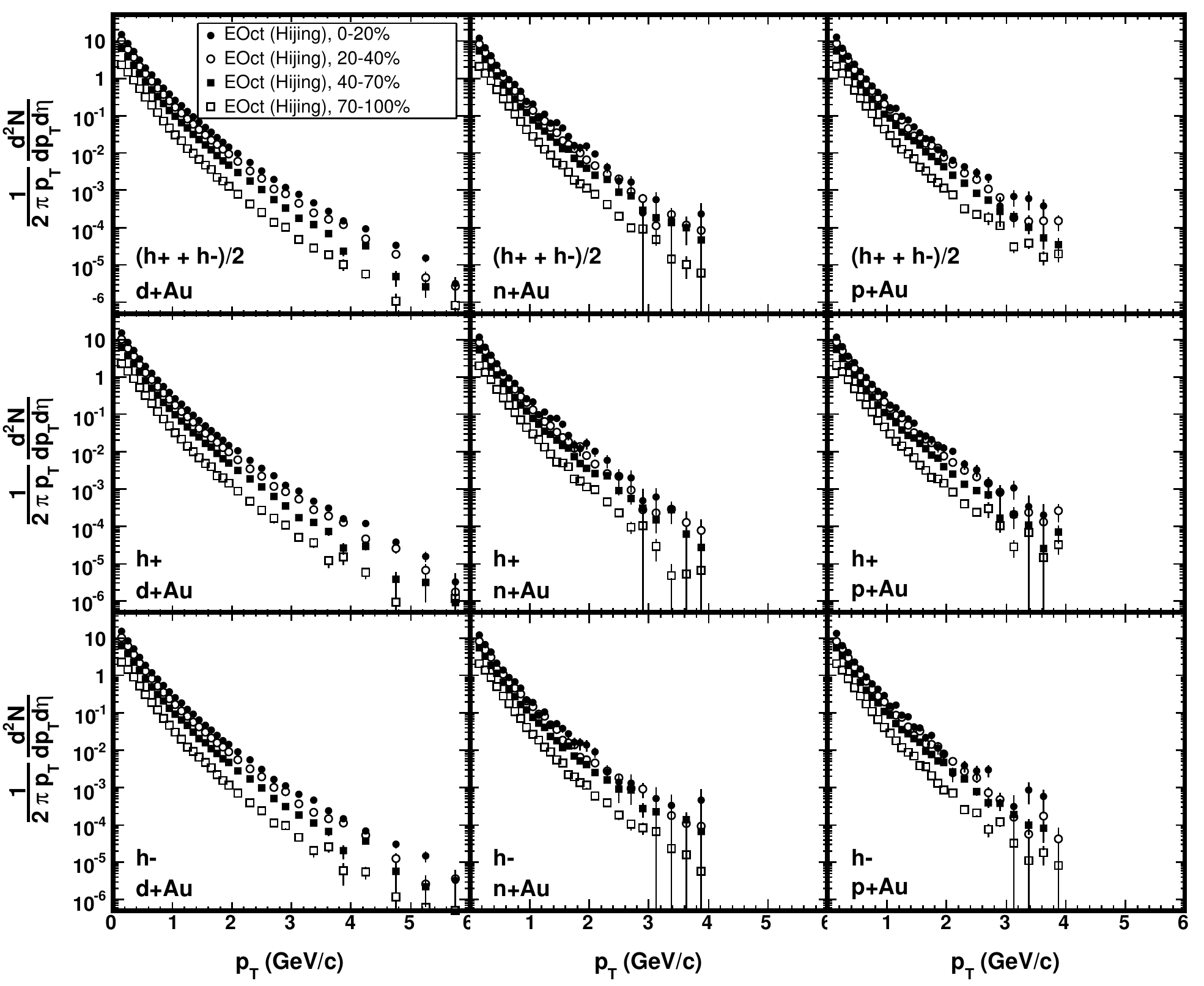}
   \end{center}
   \caption{\label{app:fig:ntSpecEOctHijing}
      The invariant yield of $\have$, $\hpos$ and $\hneg$ in four
      centrality bins determined using \protect\acs{HIJING} and the
      \protect\acs{EOct} centrality variable. The spectra for {\dAu},
      {\nAu} and {\pAu} are shown in separate columns. Only statistical
      errors are shown.}
\end{figure}

\begin{figure}[h]
   \begin{center}
      \includegraphics[width=\linewidth]{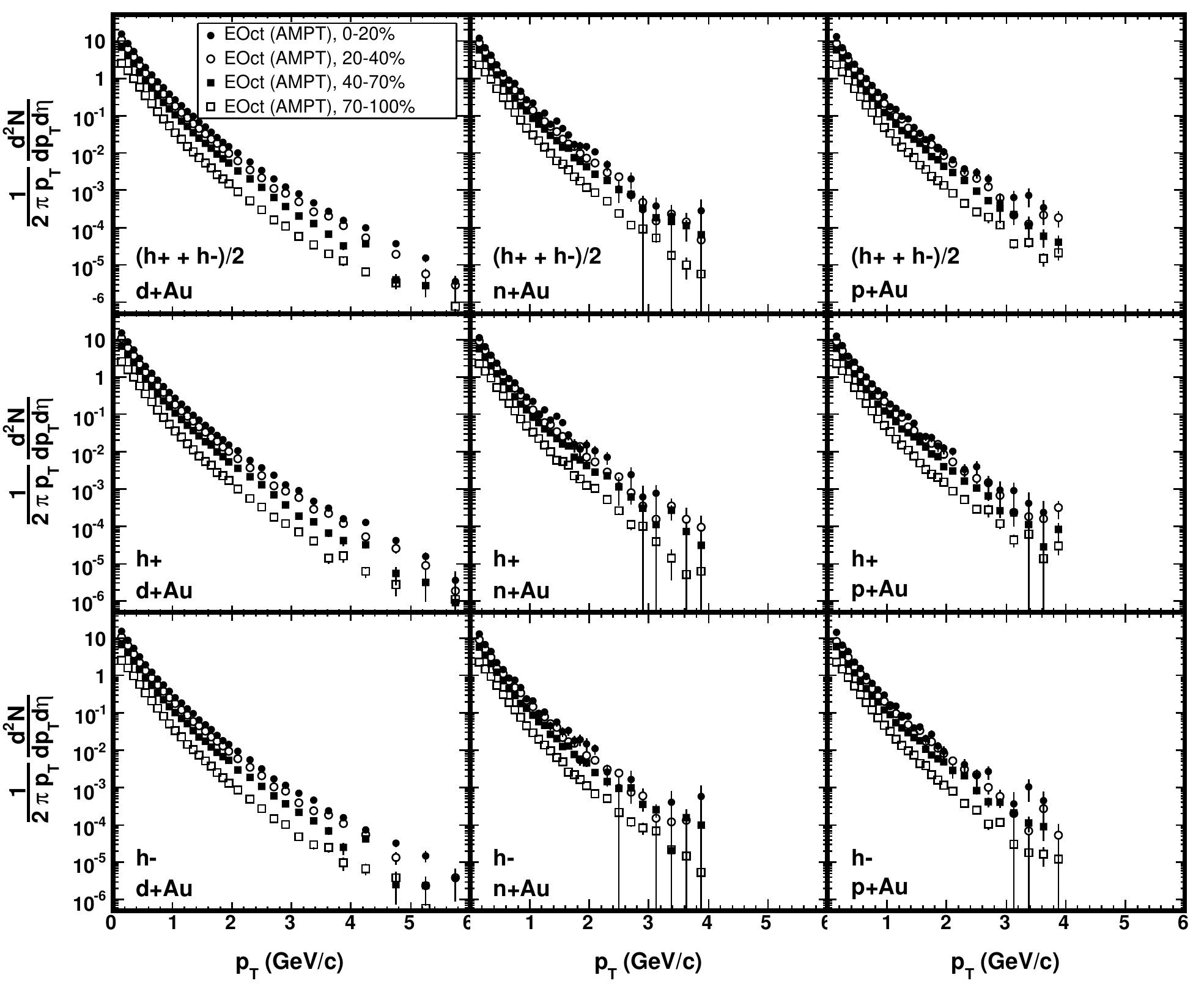}
   \end{center}
   \caption{\label{app:fig:ntSpecEOctAMPT}
      The invariant yield of $\have$, $\hpos$ and $\hneg$ in four
      centrality bins determined using \protect\acs{AMPT} and the
      \protect\acs{EOct} centrality variable. The spectra for {\dAu},
      {\nAu} and {\pAu} are shown in separate columns. Only statistical
      errors are shown.}
\end{figure}

\begin{figure}[h]
   \begin{center}
      \includegraphics[width=\linewidth]{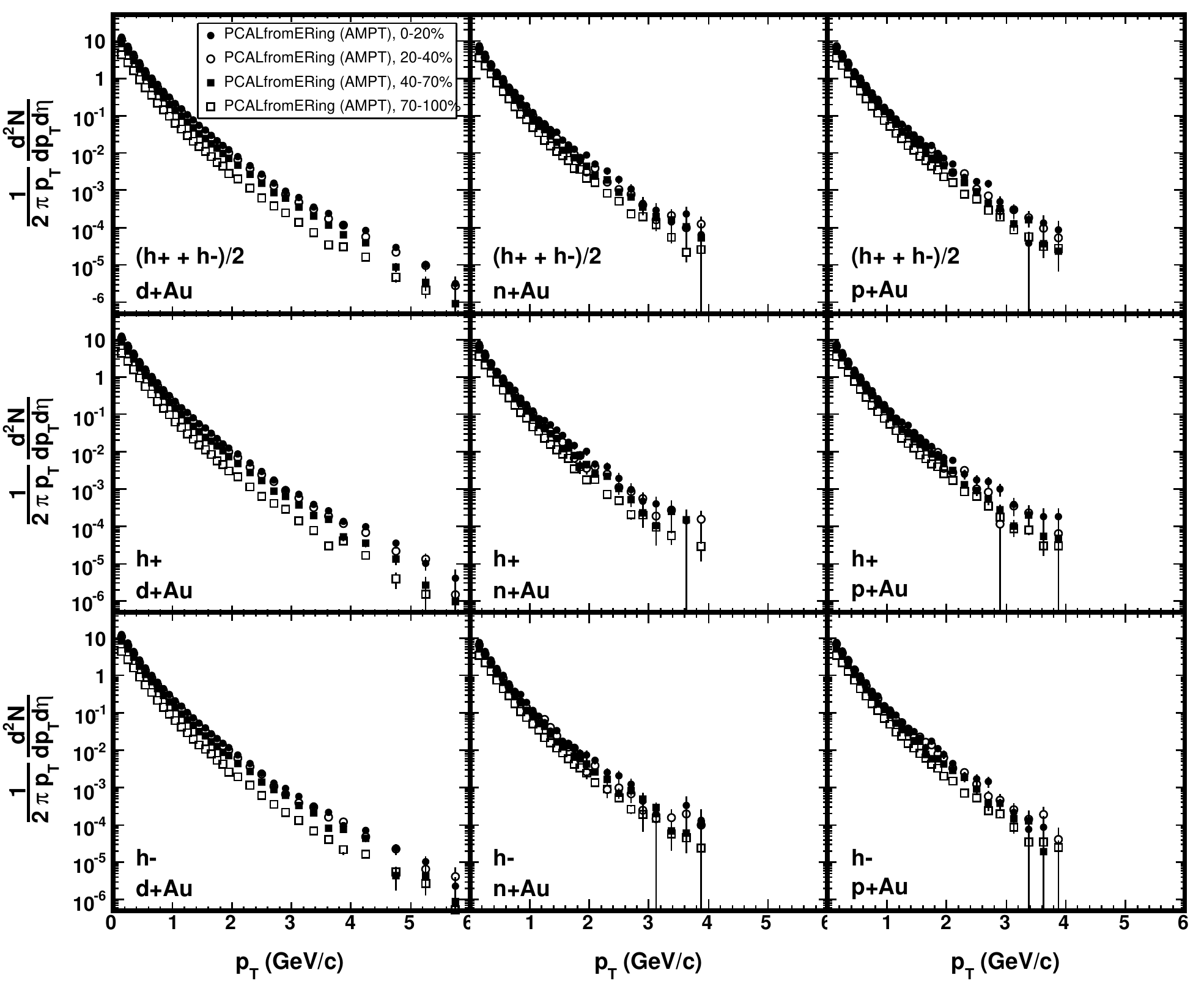}
   \end{center}
   \caption{\label{app:fig:ntSpecPCALfromERingAMPT}
      The invariant yield of $\have$, $\hpos$ and $\hneg$ in four
      centrality bins determined using \protect\acs{AMPT} and the
      \protect\acs{EPCAL} centrality variable. The trigger efficiency as
      a function of \protect\acs{EPCAL} was determined using
      \protect\acs{ERing}. The spectra for {\dAu}, {\nAu} and {\pAu} are
      shown in separate columns. Only statistical errors are shown.}
\end{figure}

\begin{figure}[h]
   \begin{center}
      \includegraphics[width=\linewidth]{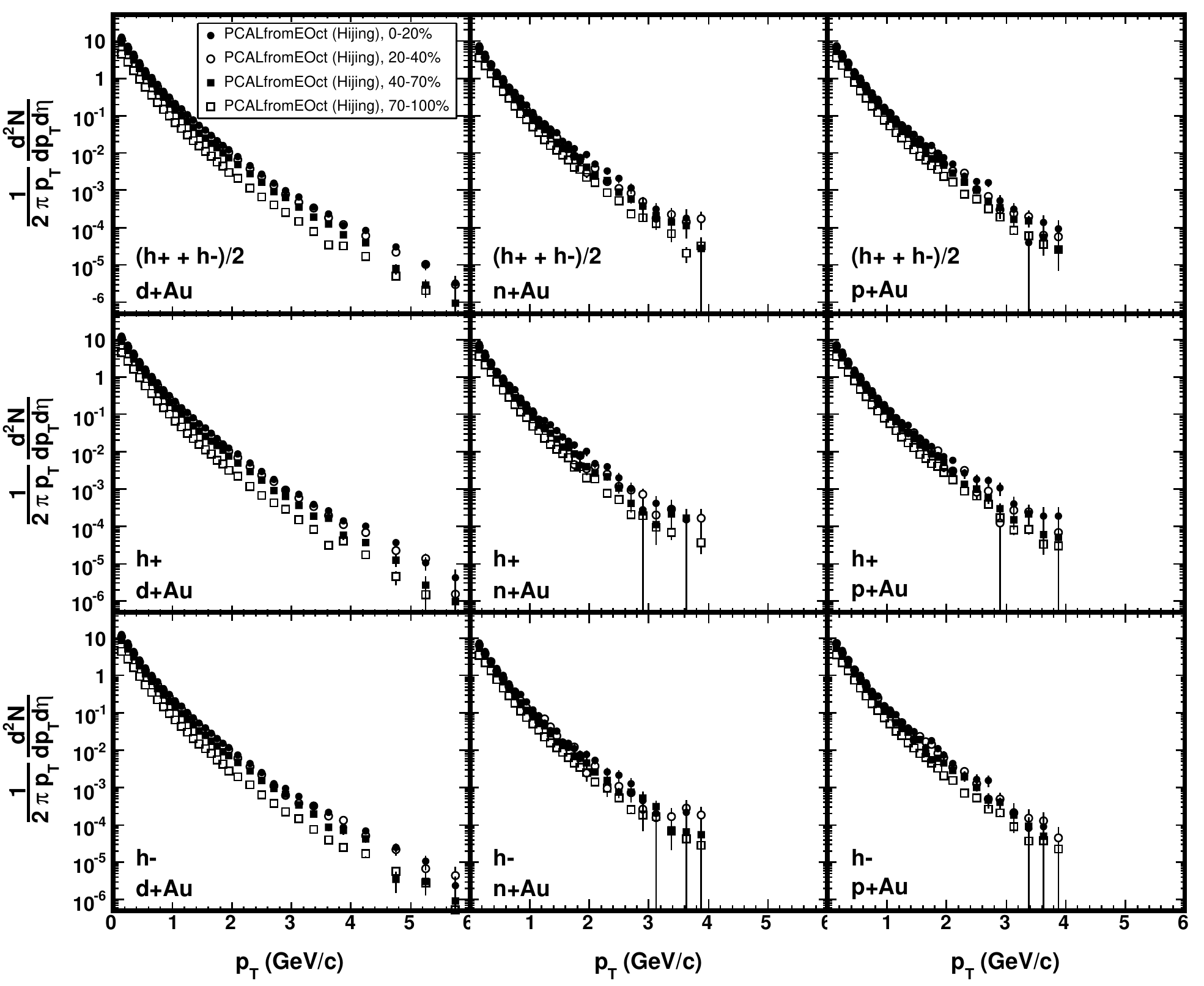}
   \end{center}
   \caption{\label{app:fig:ntSpecPCALfromEOctHijing}
      The invariant yield of $\have$, $\hpos$ and $\hneg$ in four
      centrality bins determined using \protect\acs{HIJING} and the
      \protect\acs{EPCAL} centrality variable. The trigger efficiency as
      a function of \protect\acs{EPCAL} was determined using
      \protect\acs{EOct}. The spectra for {\dAu},
      {\nAu} and {\pAu} are shown in separate columns. Only statistical
      errors are shown.}
\end{figure}

\begin{figure}[h]
   \begin{center}
      \includegraphics[width=\linewidth]{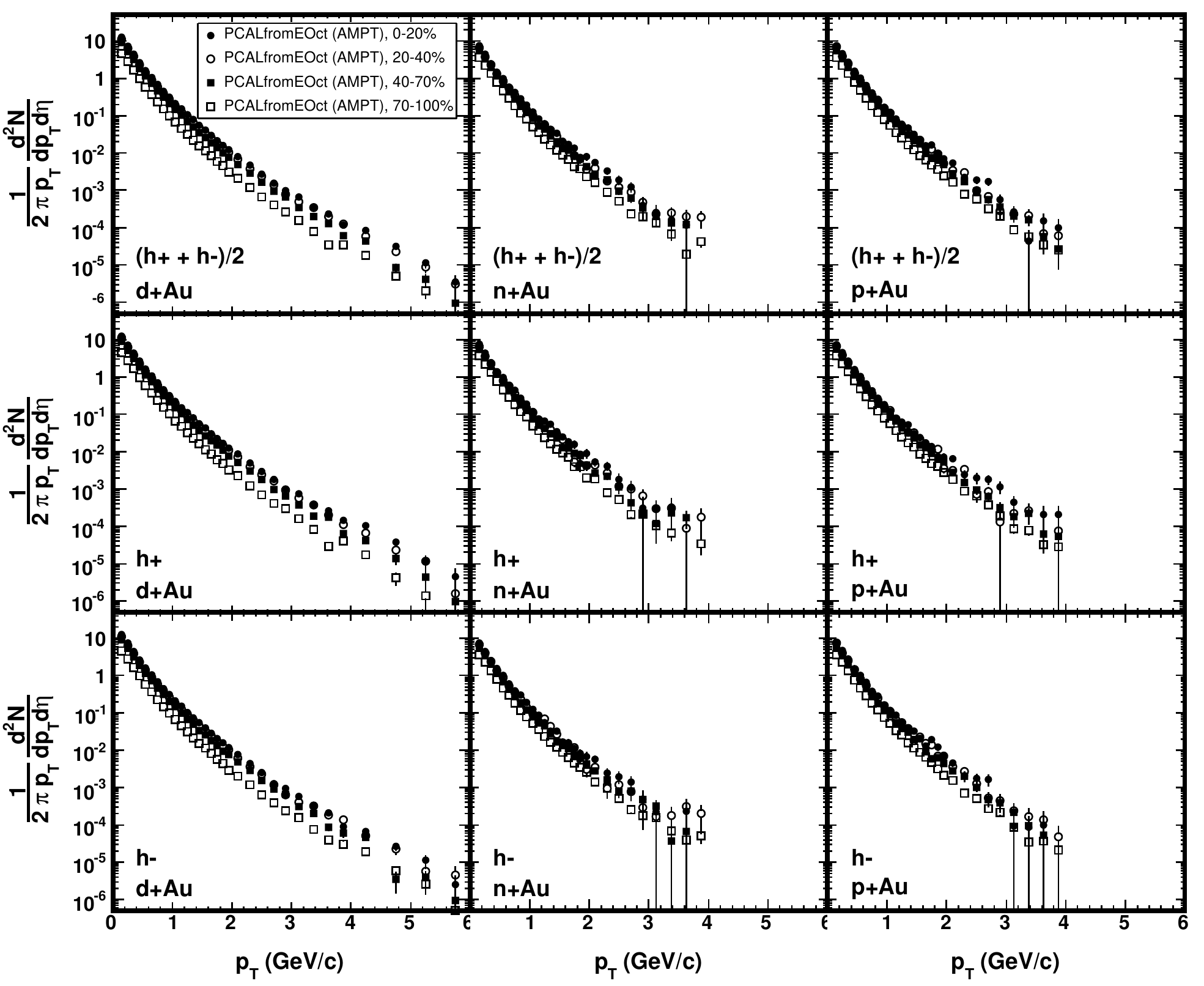}
   \end{center}
   \caption{\label{app:fig:ntSpecPCALfromEOctAMPT}
      The invariant yield of $\have$, $\hpos$ and $\hneg$ in four
      centrality bins determined using \protect\acs{AMPT} and the
      \protect\acs{EPCAL} centrality variable. The trigger efficiency as
      a function of \protect\acs{EPCAL} was determined using
      \protect\acs{EOct}. The spectra for {\dAu},
      {\nAu} and {\pAu} are shown in separate columns. Only statistical
      errors are shown.}
\end{figure}

%% file: srcs/tam.tex
% $Id: tam.tex,v 1.19 2006/09/05 02:20:50 cjreed Exp $
%

%---------------------------------------------------------------
\chapter{Computing at {\phob}}
\label{tam}
%---------------------------------------------------------------

\myupdate{$*$Id: tam.tex,v 1.19 2006/09/05 02:20:50 cjreed Exp $*$}%
Major upgrades to the {\phob} \ac{DAQ} system prior to the {\dAu}
physics run, and then again prior to the subsequent {\AuAu} run, greatly
increased the amount of data that could be taken by the experiment. All
told, the {\phob} experiment had recorded around a billion collisions by
the time the detector was decommissioned in June, 2005. The first data
storage format developed by {\phob}, and used for all its publications
prior to 2006, was not suitable for efficient storage and access to such
a large volume of data. In this format, data from an average
reconstructed {\AuAu} collision (consisting of calibrated detector
signals, reconstructed vertices and Spectrometer tracks) took up over
100~{\kB} of storage space. Due to the size of collisions in this
format, it was not possible to store all of the data in a location that
was easily accessible by physicists. The need to store this data on the
{\phob} computer farm at \ac{RCF}, where it could be repeatedly
analyzed, led to two major software projects. The first was the
development of a new, highly efficient data storage structure known as
an \ac{AnT}. The second was a new framework in which the {\ttree} objects
of \ac{ROOT}~\cite{Brun:1997pa} (as well as \acsp{AnT}) could be
analyzed.

%---------------------------------------------------------------
\section{Analysis Trees in {\phob}}
\label{tam:ant}
%---------------------------------------------------------------

The \ac{AnT} format reduced the size of the average collision data by
more than a factor of two, while also increasing the efficiency with
which it could be processed. This was achieved by using the \ac{ROOT}
package's {\ttree} class as the backbone of the format. The {\ttree}
class was designed to provide improvements in data storage and
processing efficiency over the standard \acs{ROOT} file \ac{I/O}.

\begin{figure}[t]
   \begin{center}
      \includegraphics[width=\linewidth]{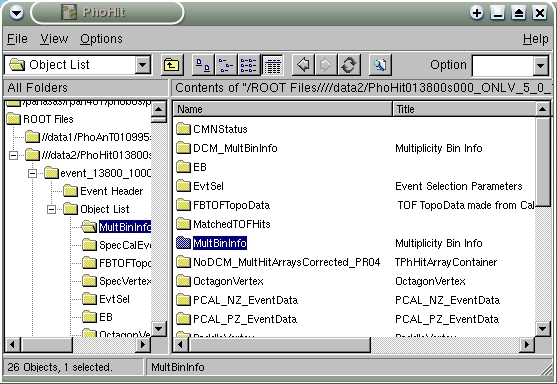}
   \end{center}
   \caption{\label{tam:fig:phoHitBrowser}
      A display of the contents of a {\phob} data file prior to
      \protect\acs{AnT}, using {\tbrow}. The selected item is the
      MultBinInfo object of event number~1000, which contains the
      \protect\acs{ERing} value for the event.}
\end{figure}

The first {\phob} data format consisted of various objects, such as hits
and tracks, all contained in a single collision event container. Each
event was independently written to a \acs{ROOT} file, as illustrated in
\fig{tam:fig:phoHitBrowser}. The major disadvantage of this structure
was the inefficiency of data access. For example, to generate an
\acs{ERing} distribution, it was necessary to (a)~read an entire event
from the file into memory, (b)~locate the ``MultBinInfo'' object that
stored \acs{ERing}, (c)~fill a histogram with the \acs{ERing} value and
(d)~delete the event object. These four steps would then need to be
repeated for each collision.

\begin{figure}[t]
   \begin{center}
      \includegraphics[width=\linewidth]{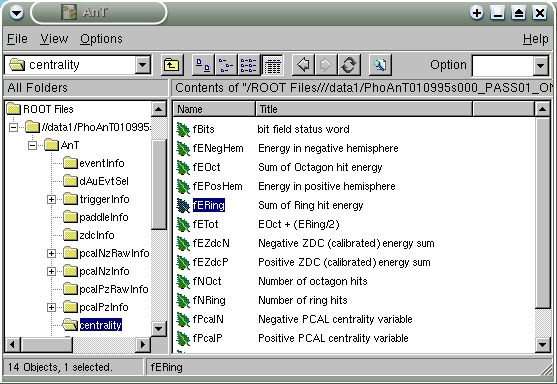}
   \end{center}
   \caption{\label{tam:fig:antBrowser}
      A display of the contents of an \protect\acs{AnT} file.  The
      selected item is the \protect\acs{ERing} leaf of the centrality
      branch, which contains the \protect\acs{ERing} value of each
      event.}
\end{figure}

The \ac{AnT} format improved this situation by storing objects in the
branches of a {\ttree}. A {\ttree} can be pictured as a large table, the
columns of which are buffers (branches) filled with data (leaves) and
the rows of which correspond to entries in the tree. In the case of
\acp{AnT}, each entry was a single collision event. This structure,
illustrated in \fig{tam:fig:antBrowser}, had two main advantages. First,
branches in the tree could be read independently. Thus, continuing the
example above, to access the \ac{OctDe} vertex information, only the
branch containing vertices would need to be read (and not the entire
event). This provided a major increase in processing speed; analyses
that did not need to process raw Silicon hits ran an order of magnitude
faster with the \ac{AnT} data format. The second advantage was that data
could be compressed far beyond that of a standard \ac{ROOT} file.
Because each entry in any particular branch stored data of the same
type, there was no need to write out the same descriptive header
information for every entry. With the standard \acs{ROOT} \acs{I/O}, the
full object information (header included) would be written to the file
for every object in every event. For more information on the {\ttree}
class, see~\cite{RootUserGuide}.

An \ac{AnT} was a \ac{ROOT} file that contained one {\ttree}. Each
branch of this tree contained a part of the collision event data: raw
Silicon hits, merged hits, reconstructed tracks, vertices, etc. New
classes were created to store this data in the most efficient way
possible. All data members of such a class were made \code{public} to
increase the ease of access. The type of a data member was chosen so as
to use the minimum number of bytes possible. For example, a merged hit
stored the number of the Silicon layer in which it was located. Since
this number could not be more than~255, as there were 16 Spectrometer
layers, only one byte was used to store this information.

Disk space and processing efficiency was further improved through the
use of {\tclon}~\cite{RootUserGuide}. A {\tclon} was a special container
that stored multiple objects of the same class. Since each object in the
container was of the same type, the memory for these objects was
allocated only once. Thus, if a {\tclon} was used to store hits in an
event, the first hit in every event would be read into the same memory
space. On the other hand, with a standard collection class such as
{\tobjar}, memory would be allocated (and de-allocated) each time a new
event was read. The \ac{AnT} format used {\tclon} objects extensively to
store groups of related objects, such as the list of merged Spectrometer
hits and the list of reconstructed tracks.

%---------------------------------------------------------------
\section{Parallel ROOT Facility}
\label{tam:proof}
%---------------------------------------------------------------

The \acf{PROOF} package was designed to allow the analysis of a large
set of data in an efficient and interactive manner. This was achieved by
parallelizing the reading and processing of data on a computing cluster
with distributed storage. An analysis run under \ac{PROOF} began at the
user's computer, referred to as the client. There, the set of data to be
processed was specified, as were the cluster \acp{CPU}, referred to as
slaves, that would be used for processing. At the start of the analysis,
a \ac{PROOF} master was initialized. The master node coordinated the
data processing by fulfilling each slave's request for work. That is,
idle slaves would ask the master for instruction on which data to
process, and the master would send a packet of work in return. This
ensured that faster slaves processed more data than slower slaves, and
optimized disk \ac{I/O} by allowing a slave to process local data
whenever possible. Finally, after the slaves had completed processing,
the results from each slave were collected and merged into a single
result, which was then passed back to the client. Thus, \ac{PROOF}
allowed the user to exploit the advantages of parallel processing,
without sacrificing the simplicity of a local, interactive \ac{ROOT}
analysis. For more information on \ac{PROOF}, see
\cite{Ballintijn:2003yt}.

%---------------------------------------------------------------
\section{Tree-Analysis Modules}
\label{tam:tam}
%---------------------------------------------------------------

To analyze data stored in the \ac{AnT} format, a software package known
as \acf{TAM} was developed. \ac{TAM} provided the infrastructure for the
processing of data stored in \ac{ROOT} trees using \emph{modules}. A
module was a user-created object structured like a {\tsel}
(see~\cite{RootUserGuide}), but which had certain advantages over the
\ac{ROOT} selector. As with a {\tsel}, a \acs{TAM} module had the
benefit of automated tree interaction (that is, the user did not have to
explicitly call \code{TBranch::SetAddress} nor loop over entries) and it
interfaced well with the \ac{PROOF} package. However, modules were not
subject to some of the drawbacks of selectors. Selectors were typically
large monolithic macros, generated by \code{TTree:MakeSelector}, making
them difficult to share with other users in a collaboration. These large
macros also tended to result in users copying blocks of code in order to
gain certain functionality. Both of these issues could be avoided by the
use of \ac{TAM}, which allowed users to separate parts of an analysis
into different modules and allowed them to run other users' modules as
part of their own analysis. In addition, one of the main goals of
\ac{TAM} was to make running an analysis with \ac{PROOF} transparent for
the user; that is, one could process data with or without \ac{PROOF} and
not need to make any changes to a \ac{TAM} module.

\begin{table}[t]
   \begin{center}
      \begin{tabularx}{\linewidth}{|l X|}
\hline
Class & Description\\
\hline
{\tamm} & The base class of all user-created modules.\\
{\tams} & The manager. Scheduled the execution of modules and the interaction
with the tree.\\
{\tamo} & An output container. Handled the merging of objects under
\protect\acs{PROOF}.\\
\hline
      \end{tabularx}
   \end{center}
   \caption{   \label{tam:tab:classes}
      The classes in the \protect\acs{TAM} package.}
\end{table}

The \ac{TAM} package was designed around three main goals. First, to
provide a very general, modular framework for analyzing data in
\ac{ROOT} trees. Second, to hide, as much as possible, all interaction
with the tree itself from the user. Third, to ensure compatibility with
\ac{PROOF} and to make the use of \ac{PROOF} transparent to the user.
This led to the development of three classes: {\tamm}, the basis of all
user-created modules, {\tams}, which managed the modules and interaction
with the tree, and {\tamo}, which managed the output of modules. These
classes are summarized in \tab{tam:tab:classes}.

%---------------------------------------------------------------
\subsection{Modules}
\label{tam:tam:mod}
%---------------------------------------------------------------

An analysis in \ac{TAM} was performed using a hierarchy of modules. That
is, a module could have any number of submodules. This capability was
provided by the base class of {\tamm}, \code{TTask}
(see~\cite{RootUserGuide}). This feature allowed a module to control
the processing of submodules, either through error handling (described
in \sect{tam:tam:mod:err}) or by directly accessing a submodule. It also
allowed a user to package an analysis into a ``supermodule,'' a concept
borrowed from the earlier {\phob} analysis framework. A supermodule was
a collection of modules that had been organized to produce some
particular output, which could then be easily utilized by any user. For
example, a supermodule could be constructed to determine the reaction
plane\footnote{The plane formed by the beam and impact parameter
vectors.} of a collision. This supermodule would then be run by all
users doing analyses that require knowledge of the reaction plane.

\begin{table}[t]
   \begin{center}
      \begin{tabularx}{\linewidth}{|l l X|}
\hline
Function & Executed On & Description\\
\hline
\code{Begin} & Client & Startup code. Typically not used.\\
\code{SlaveBegin} & Slave & Startup code. Output (i.e.~histograms) was
initialized and branches were requested here.\\
\code{Notify} & Slave & Called when a new file was opened.
Typically not used.\\
\code{Process} & Slave & Called during the event loop. Data was loaded and
histograms were filled here.\\
\code{SlaveTerminate} & Slave & Finishing code. Typically not used.\\
\code{Terminate} & Client & Finishing code.\\
\hline
      \end{tabularx}
   \end{center}
   \caption{   \label{tam:tab:modfcns}
      Functions \protect\acs{TAM} package.}
\end{table}

At its most basic, a module would perform its analysis by (a)~making
some histograms, (b)~loading some data and (c)~filling the histograms.
This flow of execution was broken up into six functions, as shown in
\tab{tam:tab:modfcns}. The \code{Begin} function was called (by {\tams})
first, and was used to run initializing code on the client computer.
This function was not usually needed by a module. Next, the
\code{SlaveBegin} function was called on the \ac{PROOF} slave. This
function was used to initialize any output, such as a histogram, and to
request the branches that would be needed (see \sect{tam:tam:mod:data}).
Then the event loop would begin, as each entry of the tree was processed
by \ac{TAM}. The \code{Notify} function would be called on the slave
whenever a new file had been opened. This function was not typically
used by a module, but could in principal have been useful for loading
calibrations or some other file-dependent information. The
\code{Process} function was called on the slave for each entry of the
tree. This was typically the most important function of a module, in
which the data was loaded, some event selection was applied, analysis
calculations were performed and histograms were filled. After all
entries of the tree had been processed, the \code{SlaveTerminate}
function was called on each slave, to perform any post-event-loop tasks
that needed to be performed prior to the merging of output from each
slave. This function was not typically used by a module. Finally, after
the output of each \ac{PROOF} slave had been merged (see
\sect{tam:tam:out:merg}), the \code{Terminate} function was called. This
function was used to finish the analysis; for example fitting a function
to the final distribution could have been done here.

%---------------------------------------------------------------
\subsubsection{Requesting Data}
\label{tam:tam:mod:data}
%---------------------------------------------------------------

Modules did not directly retrieve data from the tree; rather, they
instructed \ac{TAM} to do so. This was done using two functions,
\code{RequestBranch} and \code{LoadBranch}. The former function, called
during \code{SlaveBegin}, was used to inform the {\tams} that the module
may use data from the specified branch. \code{LoadBranch}, called during
\code{Process}, initiated the actual loading of data out of the tree.
This structure was used to ensure efficient reading of the data. By
forcing a user to load entries from the tree branch-by-branch, the user
was encouraged to load only the minimal amount of data necessary for the
current processing. That is, one would generally load some sort of event
selection branch first, perform the selection, and then, if the current
event passed the selection, load the remaining data.

The \code{RequestBranch} function was a templated function that took
(a)~the name of the branch being requested and (b)~the pointer used by
the module to access the data. The function was templated to allow
\ac{TAM} to check that the pointer type was appropriate for the data
stored in the branch, as described in \sect{tam:tam:sel:typechk}. The
\code{LoadBranch} function took only the name of the branch. The actual
loading of the data by {\tams} is described in \sect{tam:tam:sel:data}.

A typical {\phob} module would load a branch that contained general
information about an event, such as the event number in the file. A
simple example of such a module follows. First, the module would
declare the pointer it would use to access the data.

\begin{CodeBox}
class TExampleTAM : public TAModule {
private:
   TPhAnTEventInfo*  \textbf@fEvtInfo#; // event info from the AnT
\end{CodeBox}

\noindent%
While the pointer would eventually be made to point to the data by
\ac{TAM}, as will be described in \sect{tam:tam:sel:data}, it would
first be initialized to zero in the constructor.

\begin{CodeBox}
TExampleTAM::TExampleTAM(const Char_t* name,
                         const Char_t* title) :
   TAModule(name, title),
   \textbf@fEvtInfo#(0) {
}
\end{CodeBox}

\noindent%
The appropriate branch, named ``eventInfo'' in this example, would then
be requested in \code{SlaveBegin}.

\begin{CodeBox}
void TExampleTAM::SlaveBegin() {
   ReqBranch("eventInfo",\textbf@fEvtInfo#);
}
\end{CodeBox}

\noindent%
Finally, in \code{Process}, the data could be loaded and used.

\begin{CodeBox}
void TExampleTAM::Process() {
   LoadBranch("eventInfo");
   Info("Process","This is event number [%d]",
        \textbf@fEvtInfo#->fEventNum);
}
\end{CodeBox}

%---------------------------------------------------------------
\subsubsection{Error Handling}
\label{tam:tam:mod:err}
%---------------------------------------------------------------

\begin{table}[t]
   \begin{center}
      \begin{tabularx}{\linewidth}{|l X|}
\hline
Action Name & Description\\
\hline
kSuccess & Print a warning and continue processing.\\
kAbortModule & Print an error message and stop this module (and its
submodules) from processing the current event.\\
kAbortEvent & Print an error message and stop all modules from
processing the current event.\\
kAbortAnalysis & Print a break message and stop all modules from any
further processing.\\
\hline
      \end{tabularx}
   \end{center}
   \caption{   \label{tam:tab:errhand}
      The flow control options of \code{SendError}.}
\end{table}

\ac{TAM} provided some basic functionality for error handling via the
\code{SendError} function. This function was similar to the
\code{TObject::Error} and \code{Warning} functions, except that it
granted modules the ability to control the flow of processing. There
were four levels of flow control, summarized in \tab{tam:tab:errhand}.

The \code{kSuccess} option merely printed a warning. The flow of
processing was not altered. The \code{kAbortModule} option printed an
error message and stopped this module, and its submodules, from further
processing of the current event. The module would return to its normal
state at the next event. That is, if called during \code{Process}, the
module would not be active for the current entry, but would return to
normal when processing the next entry of the tree. If called during
\code{SlaveBegin}, for example, then the module would return to normal
for the first \code{Notify} or \code{Process} call (whichever came
next). The \code{kAbortEvent} option printed an error message and
stopped all modules from further processing of the current event. The
modules would return to normal before processing the next event.
Finally, the \code{kAbortAnalysis} option printed a break message and
stopped \emph{all} further processing of all modules. An example showing
the usage of \code{SendError} follows.

\begin{CodeBox}
void TExampleTAM::Process() {
   LoadBranch("eventInfo");
   const Int_t num = \textbf@fEvtInfo#->fEventNum;
   if (num > 100) {
      SendError(kAbortModule,"Process",
                "Event number [%d] too big.",
                \textbf@fEvtInfo#->fEventNum);
      return;
   }
}
\end{CodeBox}

While not useful for error handling, one other method of flow control
available to modules was the \code{SkipEvent} function. This function
controlled the flow of processing at the \code{kAbortModule} level, but
did not print any error message. This was useful for allowing a module
to implement an event selection and prevent submodules from processing
rejected events.

%---------------------------------------------------------------
\subsubsection{Interacting With Other Modules}
\label{tam:tam:mod:other}
%---------------------------------------------------------------

Often during an analysis, it would be useful for modules to access data
that was not stored in the tree. Two distinct situations were typical:
(a)~data objects that were relevant only for the current entry in the
tree and (b)~more static data objects that were relevant throughout the
analysis.

Objects relevant to the current entry of the tree could be made
available to any module and would be properly disposed of before the
processing of the next entry. This functionality was provided by the
\code{AddObjThisEvt} function, which stored the object in
\code{THashList} for fast-lookup (see~\cite{CompProg:hash} for a
discussion of hash tables). Any module could then access the object
using the \code{FindObjThisEvt} function. Such functionality would be
useful, for example, if module $A$ produced some tracks from hits stored
in the tree and module $B$ used those tracks to generate some momentum
spectra. All objects passed to the \code{AddObjThisEvt} function would
be automatically deleted by \ac{TAM} before the processing of the next
tree entry. If a module wanted to prevent this deletion, and deny other
modules access to the object, the object could be removed from the event
via the \code{RemoveObjThisEvt} function. Objects added to the event
were required to have a unique name, such that the name could be hashed.
For an object that did not inherit from \code{TNamed}, \ac{TAM} would
store the object under the class name.

For the more static data objects, a different interface was used. The
\code{PublishObj} function was used to make an object available to all
modules, all throughout the analysis. The \code{FindPublicObj} function
could be used to access the public object. This functionality would be
useful for supplying modules with some calibration objects, for example.
Like objects added to the event, public objects were also required to
have a unique name. This restriction was due to the fact that the
\ac{PROOF} input list was used to send such objects to the slave
computers, and \ac{PROOF} lists were name-based. Unlike objects added to
the event, it was the user's responsibility to clean up public objects.
In order to prevent dangling pointer issues, it was important for the
user to call the \code{RetractObj} function before deleting the object.
This function would remove the object from the list of public objects.

%---------------------------------------------------------------
\subsubsection{Using PROOF}
\label{tam:tam:mod:proof}
%---------------------------------------------------------------

\ac{TAM} was designed to make running an analysis with \ac{PROOF} as
simple as possible. One of the (somewhat superficial) ways in which 
this was accomplished was by making the syntax similar for analyses run
with and without \ac{PROOF}. While no change at all needed to be made to
modules when running with \ac{PROOF}, it was of course necessary to
start the analysis in a different way.

An example analysis may be scripted as follows. First, the module
hierarchy would be built.

\begin{CodeBox}
TMyTAMMod* myMod    = new TMyTAMMod;
TMySubMod* mySubMod = new TMySubMod;
myMod->Add(mySubMod);
\end{CodeBox}

Then, for an analysis \emph{without} \ac{PROOF}, the modules would be
added directly to the {\tams}.

\begin{CodeBox}
TAMSelector* mySel = new TAMSelector;
mySel\textbf@->AddInput(myMod);#
\textbf@tree->Process(mySel);#
TList* output = mySel->GetModOutput();
\end{CodeBox}

For an analysis \emph{using} \ac{PROOF}, the modules would be added to
the \code{TDset} object and the output would be obtained from
\ac{PROOF}, rather than from the selector. However, the syntax was
similar, as emphasized by the bold text.

\begin{CodeBox}
dset\textbf@->AddInput(myMod);#
\textbf@dset->Process("TAMSelector");#
TList* output = gProof->GetOutputList();
\end{CodeBox}

This is essentially all that is required to run an analysis under
\ac{TAM} with \ac{PROOF}. Other issues, such as porting the module
hierarchy to the slaves and extracting their output (see
\sect{tam:tam:out:extract}), are handled by \ac{TAM}.

%---------------------------------------------------------------
\subsection{The Selector}
\label{tam:tam:sel}
%---------------------------------------------------------------

While {\tamm} enforced the structure of a module, most of the features
of the \ac{TAM} package were implemented in the {\tams} class. The
module hierarchy was maintained by the selector, using its own {\tamm}
to store other modules. This top-most module did no processing and was
merely a (hidden) container for the user's modules. All of the module
processing calls, such as \code{SlaveBegin} or \code{Process}, were
called by the selector. Care was taken by the selector to ensure that
the list of objects associated with the event was properly cleaned at
the end of each process call, and that each module's pointers to data
were reset to zero before it could analyze the next entry of the tree
(to prevent dangling pointers). Most of the interaction with the tree
was handled by the nested class \code{TAMSelector::TAMBranchInfo}. This
class was used to load the data of a requested class, set the modules'
pointer and to ensure data integrity. As described in
\sect{tam:tam:future}, this nested class evolved into an extension to
\ac{TAM} that opened the door for users to control the loading of data
from the tree (i.e.~to allow for event mixing type analyses).

%---------------------------------------------------------------
\subsubsection{Loading Data}
\label{tam:tam:sel:data}
%---------------------------------------------------------------

\begin{table}[t]
   \begin{center}
      \begin{tabularx}{\linewidth}{|l l X|}
\hline
Type & Variable & Description\\
\hline
\code{Bool\_t} & \code{fIsLoaded} & True if \code{LoadBranch} had already been
called by some module for this branch.\\
\code{Bool\_t} & \code{fIsClass} & True if this branch in the tree stored an
object (as opposed to a list of numbers).\\
\code{Bool\_t} & \code{fLeafSizeConst} & True if this branch stored a list
of numbers that each used the same number of bytes.\\
\code{TBranch*} & \code{fBranch} & The branch object from the current tree.\\
\code{void*} & \code{fBAddr} & The memory location into which data from
the tree would be read.\\
\code{vector<BranchPtr\_t*>} & \code{fUsrAddresses} & List of pointers used
by each module to access data from this branch. \code{BranchPtr\_t} was
templated to preserve the type of pointer used by each module.\\
\hline
      \end{tabularx}
   \end{center}
   \caption{   \label{tam:tab:branchinfo}
      The information about each branch stored by \newline
      \code{TAMSelector::TAMBranchInfo}.}
\end{table}

The loading of data from the tree was managed by {\tams}. The selector
stored a (hash) table of \code{TAMSelector::TAMBranchInfo} objects, each
of which held information about a branch that had been requested by a
module. The information stored about each branch is summarized in
\tab{tam:tab:branchinfo}. This table was built up by the modules'
\code{ReqBranch} calls. Each time \code{ReqBranch} was called, a new
entry in the table would be made if necessary, and the module's pointer
would be added to the list of pointers for that branch,
\code{fUsrAddresses}. The name of each \code{TAMBranchInfo} object was
simply the name of the branch in the tree, and this name was used for
the hash lookups. Thus, whenever a module would call \code{LoadBranch},
{\tams} would retrieve the \code{TAMBranchInfo} object for the specified
branch. Then, if \code{fIsLoaded} was false, the data for the branch
would be loaded through a simple \code{fBranch->GetEntry} call. Finally,
the pointer for each module would be set to point to the data in memory
and \code{fIsLoaded} would be set to true.

The memory address into which the branch was read depended on the type
of data stored in the tree. For branches that stored (a)~objects or
(b)~lists of numbers, each of which were the same size,\footnote{For
example, a list of floats (4~bytes) and integers (4~bytes), but not a
list of floats and shorts (2~bytes).} a simple call to
\code{TTree::SetBranchAddress} during \code{Notify} was made. However, a
more complicated situation arose when the branch stored a list of
numbers that were not all the same size. As discussed in the \ac{ROOT}
User's Guide~\cite{RootUserGuide}, the data would be read into memory as
a simple array of numbers. Thus, for the branch

\begin{CodeBox}
*Br 0 :MyParticle : PID/S:Momentum[3]/F *
\end{CodeBox}

14~bytes of data would be read from the tree. The first two bytes would
be the particle identity, stored as a \code{Short\_t}, and the next
12~bytes would be the three components of the momentum vector, each
stored as a \code{Float\_t}. The structure used to access this data by a
modules would take the following form.

\begin{CodeBox}
struct MyParticle_t {
   Short_t PID;
   Float_t Momentum[3];
};
\end{CodeBox}

Thus, in order to properly access the data, one would need to be sure
that the start of the momentum array was only two bytes after the
address of \code{PID}. However, C++ makes no such guarantee. A compiler
could buffer the structure such that each variable would be evenly
spaced in memory -- that is, each variable would be 4~bytes apart. To
prevent errors that would result in such a situation, \ac{TAM}
individually set the address of each \emph{leaf} of the branch. This was
done by exploiting the \code{TDataMember} class of \ac{ROOT}, which gave
the offset of each variable in the structure (that is, the number of
bytes between a variable and the start of the structure). For
\code{TDataMember} to have access to the make-up of the structure, it was
necessary for the user to add the structure to the \ac{ROOT} dictionary,
in the usual way.

\begin{CodeBox}
\#pragma link C++ class MyParticle_t+;
\end{CodeBox}

%---------------------------------------------------------------
\subsubsection{Type Checking}
\label{tam:tam:sel:typechk}
%---------------------------------------------------------------

To further ensure that the data being read from the tree was properly
accessed by the modules, \ac{TAM} implemented type checking. Type
checking was run during \code{Notify}, prior to any data being loaded
from the tree. For branches that stored an object, the type checking was
simple. A loop over the list of module pointers, \code{fUsrAddresses},
was performed. The type of each pointer, as reported by
\code{type\_info}, was required to be the same as the type of the class
stored in the branch.

For a branch that stored a list of numbers, the type checking was a bit
more thorough. The list of leaves was obtained from the branch.
Concurrently, the list of variables in the structure was obtained using
\code{TClass::GetListOfDataMembers}. These lists were then looped over
together. The type name of the leaf in the tree was required to be the
same as the name of the variable type, as reported by
\code{TDataMember::GetTypeName}. While this check was being performed,
the size of each variable in the structure was also checked, such that
the \code{TAMBranchInfo::fLeafSizeConst} variable could be set
appropriately.

%---------------------------------------------------------------
\subsection{Output}
\label{tam:tam:out}
%---------------------------------------------------------------

The {\tamo} class was used to store the output of a module. This class
was explicitly designed to make running \ac{TAM} under \ac{PROOF}
transparent. It handled the merging of objects as they were returned
from the \ac{PROOF} slaves. It also attempted to associate a module's
pointers to its output objects with the merged objects in \ac{PROOF}'s
output list. This issue was important, since a module would create a
histogram, for example, during \code{SlaveBegin}. Thus, the histogram
would exist only on the slave computer. When the slaves finished
processing, the histogram would be stored in the output list of
\ac{PROOF} and the module's pointer to the histogram would be zero on
the client. {\tamo} would automatically set the module's pointer to
point to the merged histogram.

\begin{figure}[t]
   \begin{center}
      \includegraphics[width=\linewidth]{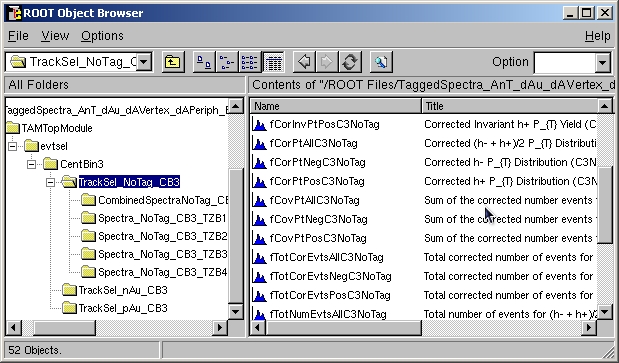}
   \end{center}
   \caption{\label{tam:fig:tamoutbrowser}
      Browsing the hierarchy of module output.}
\end{figure}

In addition, the {\tamo} class was browsable, as shown in 
\fig{tam:fig:tamoutbrowser}. The browser allowed the user to visually
navigate the module hierarchy. This was especially useful since output
objects were retrieved using the \code{FindOutput} function, which
required the user to know which module had produced the output. Thus,
the browser was useful for locating output objects in a large module
hierarchy.

%---------------------------------------------------------------
\subsubsection{Merging}
\label{tam:tam:out:merg}
%---------------------------------------------------------------

\ac{TAM} stored in the output list of \ac{PROOF} the {\tamo} object of
the top-most, hidden module. Thus, when the slave computers finished
processing, \ac{PROOF} would automatically call \code{Merge} on these
{\tamo} objects (one for each slave). The module output objects
contained both (a)~a list of the output objects for its module and (b)~a
list of the {\tamo} objects associated with each submodule. This way,
the module hierarchy was preserved in the output.

The first task performed by \code{TAMOutput::Merge} was to merge the
output objects of the current module. This was done by looping over the
corresponding {\tamo} objects from each slave. A list of objects to
merge was then assembled. That is, the first output object from each
slave was put into a list. Then, the objects in this list were merged.
For example, the first output object from the first slave would be
combined with the first output object from the second slave using the
\code{TObject::Merge} function. Then, this combined object would be
merged with the first output object from the third slave; and so on. The
same would be repeated for the second output object from each slave,
then the third, and so on. The final result would be a single {\tamo}
class for the module whose list of output objects is filled with the
combined results of each slave.

The next task was to properly merge the output of submodules, while
preserving the module hierarchy. This was done by looping through the
list of slaves and generating a list of the first submodule's {\tamo}
from each slave. This list of first submodules was then sent into
\code{TAMOutput::Merge}. The procedure was then repeated for the second
submodule from each slave, and so on. Note that because this loop over
submodule output objects was contained in the \code{TAMOutput::Merge}
function, the recursive behavior required to merge, for example,
sub-submodule output, was obtained automatically. For example, the
output of the first submodule (of the first module) would be merged
before the output of the second module would be merged.

%---------------------------------------------------------------
\subsubsection{Extraction from PROOF Output List}
\label{tam:tam:out:extract}
%---------------------------------------------------------------

A module stored an output object, such as a histogram, using the
\code{AddOutput} function. Because modules created their output objects
during \code{SlaveBegin}, there was an issue with modules running under
\ac{PROOF} trying to access these object directly. For example,
suppose a module did the following. In \code{SlaveBegin}, some output
object, a histogram, was created.

\begin{CodeBox}
void TExampleTAM::SlaveBegin() {
   ReqBranch("eventInfo",fEvtInfo);
   fEvtNum = new TH1F("hEvtNum","Event Numbers",100,0,100);
}
\end{CodeBox}

\noindent%
Then, in \code{Process}, the histogram was filled.

\begin{CodeBox}
void TExampleTAM::Process() {
   LoadBranch("eventInfo");
   fEvtNum->Fill(fEvtInfo->fEventNum);
}
\end{CodeBox}

\noindent%
Finally, in \code{Terminate}, the histogram was displayed.

\begin{CodeBox}
void TExampleTAM::Terminate() {
   \textbf@fEvtNum#->Draw();
}
\end{CodeBox}

\noindent%
This code would work if the module were run without \ac{PROOF}. However,
under \ac{PROOF}, the module would crash during \code{Terminate}. The
reason is simple: \code{fEvtNum} was created on the slave computer. On
the client computer, \code{fEvtNum} would still be zero (shown in bold
in the example above). The merged histogram would be sitting in the
module's list of output objects, and it would be up to the user to write
special code when running under \ac{PROOF} in order to extract the
object from the output list.

\ac{TAM} avoided this situation, and made running under \ac{PROOF}
transparent, by doing the work for the user, starting with
\code{AddOutput}. This function stored both the output object as well as
information about how the module accessed the output object. First, the
output object itself was added to the list of output objects in the
{\tamo}. Then, the address \emph{of} the pointer (not the address the
pointer pointed to) that was sent to \code{AddOutput} was checked
against the address of each of the module's member variables. If a match
was found, then information about this variable was stored in a hash
table. Because \code{AddOutput} was called during \code{SlaveBegin} on
the slave computers, it was not necessary to store the address of the
pointer (since this address would be different on the client computer).
Instead, both the name of the member variable and the name of the output
object were stored in the table. The name of the output object was used
as the hash key for fast-lookup.

Finally, back on the client computer, each module's member variables
could be made to point to the appropriate (merged) output object. This
was done by looping through the list of output objects. For a given
output object, an attempt was made to find an entry in the member
variable table that stored the name of the output object in question. If
one could be found, then the name of the module's member variable was
retrieved from that entry of the table. Using the variable name, the
\code{TClass::GetDataMember} function was called to access the
appropriate member variable. This member variable was then explicitly
set to point to the corresponding output object. Note that this method
could only work when the variable passed to \code{AddOutput} was (a)~a
member variable of the module and (b)~a pointer to an object -- and not,
say, the address of an instance. The former requirement was enforced
implicitly, since entries in the lookup table could only be made for
member variables. The latter requirement was explicitly enforced. The
\code{GetFullTypeName} function of \code{TDataMember} was used to get
the type of the module's member variable as a string. This string was
required to have the form \code{Class*}.

%---------------------------------------------------------------
\subsection{Data Loader Plug-ins}
\label{tam:tam:future}
%---------------------------------------------------------------

While the version of \ac{TAM} described in this thesis provided a
feature-rich environment in which users could process \ac{ROOT} trees,
it was not easily adapted to event-mixing type analyses. That is, this
version of \ac{TAM} assumed that the user was processing a single tree
and that each entry could be processed independently. To accommodate
analyses for which these assumptions are not valid, the concept of a
data loading plug-in was developed. The goal was for users to have the
ability to control the way in which data would be loaded. An analysis
would still be driven by a single tree, but the user could control how
the data would be read from the tree. For example, in an event mixing
analysis, one might use certain properties of the current collision,
such as the vertex location, to generate a mixed event from collisions
in a \emph{second} tree that have a similar vertex. Such a procedure
could be coded by a user into the plug-in. \ac{TAM} would then access
data via this plug-in, rather than by the traditional
\code{TAMSelector::TAMBranchInfo} class. Thus, the same module would be
used to process a mixed event or a collision event; only the plug-in
used to load the data would change. As of this writing, development of
the plug-in extension to \ac{TAM} is maturing rapidly, and preliminary
versions are being tested.

%% file: srcs/acronyms.tex
% $Id: acronyms.tex,v 1.40 2006/09/04 20:25:49 cjreed Exp $
%
% Acronyms

\chapter{List of Acronyms}

\noindent%
Facilities:

\begin{acronym}[dAuMinBias]
\acro{AGS}           {Alternating Gradient Syn\-chro\-tron\acroextra{
                      (\hrefurl{http://www.bnl.gov/bnlweb/facilities/AGS.asp})}}
\acro{BNL}           {Brookhaven National Laboratory\acroextra{
                        (\hrefurl{http://www.bnl.gov/})}}
\acro{RCF}           {RHIC Computing Facility\acroextra{
                        (\hrefurl{http://www.rhic.bnl.gov/RCF/})}}
\acro{RHIC}          {Relativistic Heavy Ion Collider\acroextra{ 
                        (\hrefurl{http://www.bnl.gov/RHIC/})}}
\end{acronym}

\noindent%
\phob\ Hardware:

\begin{acronym}[dAuMinBias]
\acro{ADC}           {Analog-to-Digital Converter}
\acro{Au-PCAL}       {Proton Calorimeter on the Au-exit side}
\acro{Au-ZDC}        {Zero-Degree Calorimeter on the Au-exit side}
\acro{d-PCAL}        {Proton Calorimeter on the d-exit side}
\acro{d-ZDC}         {Zero-Degree Calorimeter on the d-exit side}
\acro{DAC}           {Digital-to-Analog Converter}
\acro{DAQ}           {Data Acquisition}
\acro{DRAM}          {Dynamic Random Access Memory}
\acro{FEC}           {Front-End Con\-trol\-ler}
\acro{FPDP}          {Front Panel Data Port\acroextra{
                        (\hrefurl{http://www.fpdp.com/})}}
\acro{HPSS}          {High Performance Storage System\acroextra{
                 (\hrefurl{http://www.hpss-collaboration.org/hpss/index.jsp})}}
\acro{LED}           {Light-Emitting Diode}
\acro{PCAL}          {Proton Calorimeter}
\acro{PMT}           {Photomultiplier Tube}
\acro{SpecTrig}      {Spectrometer Trigger}
\acro{SRAM}          {Static Random Access Memory}
\acro{T0}            {Time-Zero Counter}
\acro{TDC}           {Time-to-Digital Converter}
\acro{TOF}           {Time-of-Flight}
\acro{VME}           {VERSAmodule Eurocard}
\acro{ZDC}           {Zero-Degree Calorimeter}
\end{acronym}

\noindent%
Experiment Terminology:

\begin{acronym}[dAuMinBias]
\acro{AA}            {nucleus-nucleus}
\acro{AMPT}          {A Multi-Phase Transport}
\acro{AnT}           {Analysis Tree}
\acro{CMN}           {Common-Mode Noise}
\acro{ART}           {A Relativistic Transport}
\acro{CPU}           {Central Processing Unit}
\acro{dAuMinBias}    {\dAu\ Minimum Bias}
\acro{dAuPeriph}     {\dAu\ Peripheral}
\acro{dAuSpectra}    {\dAu\ Spectra}
\acro{dAuVertex}     {\dAu\ Vertex}
\acro{EOct}          {Energy in Octagon}
\acro{EPCAL}         {Energy in Au-PCAL}
\acro{ERing}         {Energy in Rings}
\acro{HIJING}        {Heavy Ion Jet Interaction Generator}
\acro{I/O}           {Input/Output}
\acro{IP}            {Nominal Interaction Point}
\acro{IsCol}         {Is Collision}
\acro{L0}            {Level 0}
\acro{L1}            {Level 1}
\acro{L2}            {Level 2}
\acro{MC}            {Monte Carlo}
\acro{NN}            {nucleon-nucleon}
\acro{OctDe}         {Octagon Deposited Energy}
\acro{PROOF}         {Parallel ROOT Facility}
\acro{ROOT}          {An Object-Oriented Data Analysis Tool}
\acro{SpecN}         {the inner-ring Spectrometer arm}
\acro{SpecP}         {the outer-ring Spectrometer arm}
\acro{TAM}           {Tree-Analysis Modules}
\acro{TPC}           {Time Projection Chamber}
\acro{ZPC}           {Zhang's Parton Cascade}
\end{acronym}

\noindent%
Mathematical Terminology:

\begin{acronym}[dAuMinBias]
\acro{RMS}           {Root-Mean-Square}
\end{acronym}

\noindent%
Physics Terminology:

\begin{acronym}[dAuMinBias]
\acro{MIP}           {Minimum Ionizing Particle}
\acro{ONO}           {Oxide-Nitride-Oxide}
\acro{QCD}           {Quantum Chromodynamics}
\acro{QGP}           {Quark-Gluon Plasma}
\end{acronym}

%% file: srcs/acknowledge.tex
% $Id: acknowledge.tex,v 1.3 2006/09/05 08:10:24 cjreed Exp $

\chapter*{Acknowledgments}

When work on this thesis began, crude oil cost \$15 per barrel, a group
of about ten people was beta-testing its web site ``google.com'' and no
one knew whether neutrinos had mass. It should come as no surprise that
during the years it took for these things to change, I have become
indebted to so many people for so many reasons.

George Stephans has been the quintessential adviser. His patience and
liberalism -- not only in the political sense, but also in his ability
to grant professional freedom -- allowed me to take on many different
projects and learn about scientific techniques, detectors, computing
and, of course, nuclear physics. Yet this freedom was tempered with
enough guardianship that I did not stray too far from the path for too
long. This thesis is a direct result of his hard work and guidance,
and I cannot overstate my gratitude.

As a member of the {\phob} collaboration, I had the great privilege of
working alongside some of the best nuclear physicists in the
world. Wit Busza has shown an amazing ability to understand
complicated problems by describing them in simple terms, a technique I
now find invaluable.  Gunther Roland has been an essential part of
{\phob}, and I have been lucky to learn heavy ion physics from
him. His nearly limitless drive for knowledge and ability to break
down a problem to its most important aspects has been an
inspiration. The physics discussions I had with Peter Steinberg and
Mark Baker were absolutely critical to the understanding of heavy ion
collisions that I have so far attained, and as there is much I still
do not comprehend, I look forward to many more such
conversations. Nigel George provided daily consultation, and his work
on the peripheral trigger and proton calorimeters made this thesis
possible. I am confident that his students are now benefiting as much
from his guidance as I did.

Being present while heavy ion collisions were being observed at
40~{\tev} for the first time in history was an unforgettable
experience.  This would not have been possible without the dedicated
work of all of my colleagues. Bolek Wyslouch provided much needed
guidance, both for the {\phob} experiment and for myself. Christof
Roland has been both a role model and a friend. His ability to work at
the level of excellence while still living a well-rounded life is a
rare quality in our field.  Gerrit van Nieuwenhuizen, Heinz Pernegger
and Rachid Nouicer have shown me what it takes to build an outstanding
experiment. Heinz provided me with my first lesson, which will not be
forgotten, ``This is a particle detector, my friend. It needs
particles.'' Gerrit and Rachid have always shown confidence in me,
which was a welcome boost at times when I needed it most. My friends
Maarten Ballintijn, Burt Holzman and Constantin Loizides have
encouraged, nurtured and relied upon my abilities to compose software
and solve problems; without this I would not have such
abilities. Gabor Veres helped to peak my interest in isospin studies,
and he was as good a postdoc as Mozart was a composer. Dave Hofman and
Steve Manly helped to provide a fun atmosphere, and demonstrated what
it takes to be a great young physics professor. Robert Pak and Andrei
Sukhanov kept the experiment running, and without them {\phob} would
have had no data to analyze. Andrzej Olszewski, Adam Trzupek and
Krzysztof Wozniak provided vital advice and detector
simulations. Finally, to all those that I have neglected to mention
here, let me first apologize and then thank you for all of your hard
work; {\phob} could not have succeeded without you.

It is perhaps an uncommon occurrence at large collaborations, but it
was truly a joy working on {\phob}, and there was not a single person
at our assemblies whom I did not look forward to meeting. This great
environment was due in no small part to the fantastic collection of
budding young physicists that {\phob} was able to obtain. I will miss
the good times I had with my friends and esteemed colleagues Jay Kane
and Conor Henderson.  I could not have asked for better
roommates. Kris Gulbrandsen and Patrick Decowski provided much needed
laughs, conversations and teachings. Carla Vale, Abby Bickley,
Chia-Ming Kuo and Marguerite Tonjes created the fun-loving,
hard-working fraternity that was the 214 office. Josh Hamblen and Erik
Johnson provided, among other things, several solid years of great
fantasy football play and fun. Rick Bindel, Zhengwei Chai, Vasu
Chetluru, Richard Hollis, Aneta Iordanova, Johannes
M\"{u}lmenst\"{a}dt, Joe Sagerer and Pete Walters made meetings fun
and freely donated their time to helping others (me not least of
all). My fellow officemates Burak Alver, Wei Li, Siarhei Vaurynovich
and Ed Wenger were kind enough to allow me to share their office.
They made 24-416 a fun and productive place to be. It was my pleasure
knowing all these outstanding young scientists, and I hope they will
remember me after they're famous.

While I could not have completed this thesis without the help of all
of my colleagues, I could not have begun it without the love and
support of my family. My Uncle Wood and Grandpa Elias inspired,
challenged and encouraged me to make a career out of discovery. My
parents, Amy and Dennis, provided me all of the love, patience,
freedom, guidance and support any child could have asked for. I was
given every opportunity to choose my own path through life, and was
applauded at each step. My brother Chris and sister Catie have
provided countless hours of comedy, companionship, controversy and
conflict; thereby making life compelling (with a capital 'c') and
worth living. It is a great source of pride for me that we can be such
different people, and yet be so alike. My Grandpa and Grandma Reed
have always been excessively caring, supportive people. I treasure the
years I was able to live nearby and visit often, and not just because
of the delicious food. My Uncle Bob and Aunt Jeanette have given me
more than my share of love and merriment. My Uncle Ted has always been
a source of indispensable humor, which has been particularly valuable
recently. I have truly been blessed with a wonderful family. My only
regret is that those who were with me when I started school could not
be here when I finished it. My Grandpa and Grandma Elias and my Aunt
Donna were some of the best people I had the pleasure of knowing, and
each had a profound impact on my life. It is to them that I have
dedicated this work.

Finally, I wish to thank all of the great friends I have made whose
camaraderie has helped me make it to, and make it through, all these
years of school. Carl has ever been an ally in our crusade through
life, and I know I can always count on him. Everyone back in
California -- Julie, Brett, Sean, Beth, you know who you are -- made
going home fun. I couldn't have stayed sane without your company on
those trips. Deborah gave me the love and support I needed to make it
through school, family tragedy and life in general. It was a great
stroke of luck that we met against all odds, and I will always
remember our years together in Boston. My time at UCI would have been
intolerable without the great people I met there. Steve, Rina and
Grant made college life entertaining. Last, but by no means least, I
could not have survived MIT without the good times and support that
all of my friends in Boston have given me. Brian, Cathal, Conor,
Deborah, Gary, Jay, Jie-Eun, Katie, Kevin, Ksusha, Luisa, Miranda,
Oliver, Peter, Sejal, Shane, Susan and Yaz all made living here
great fun, and have reminded me that I need to take it easy every
now and then. Well, The Dude abides.